\documentclass[structabstract]{aa}  

\usepackage{graphicx}
\usepackage{txfonts}
\usepackage{longtable}
\usepackage{longtable,lscape}
\usepackage{lscape}
\begin{document}
   \title{Evidence for ultra-fast outflows in radio-quiet AGNs: I - detection and statistical incidence of Fe K-shell absorption lines}


   \author{F. Tombesi\inst{1,2,3,4}, M. Cappi\inst{1}, J. N. Reeves\inst{5}, G.G.C. Palumbo\inst{2}, T. Yaqoob\inst{3,4}, V. Braito\inst{6} \and M. Dadina\inst{1}
          }

   \offprints{tombesi@iasfbo.inaf.it}

   \institute{INAF-IASF Bologna, Via Gobetti 101, I-40129 Bologna, Italy 
\and
Dipartimento di Astronomia, Universit\`a degli Studi di Bologna, Via Ranzani 1, I-40127 Bologna, Italy 
\and
Department of Physics and Astronomy, Johns Hopkins University, 3400, Baltimore, MD 21218, USA
\and
Laboratory for High Energy Astrophysics, NASA/Goddard Space Flight Center, Greenbelt, MD 20771, USA
\and
Astrophysics Group, School of Physical and Geographical Sciences, Keele University, Keele, Staffordshire ST5 5BG, UK
\and
Department of Physics and Astronomy, University of Leicester, University Road, Leicester LE1 7RH, UK
}

   \date{Received ; accepted }

 
  \abstract
   {Blue-shifted Fe K absorption lines have been detected in recent years between 7 and 10~keV in the X-ray spectra of several radio-quiet AGNs. The derived blue-shifted velocities of the lines can often reach mildly relativistic values, up to 0.2--0.4c.  These findings are important because they suggest the presence of a previously unknown massive and highly ionized absorbing material outflowing from their nuclei, possibly connected with accretion disk winds/outflows.}
   {The scope of the present work is to statistically quantify the parameters and incidence of the blue-shifted Fe K absorption lines through a uniform analysis on a large sample of radio-quiet AGNs.
This allows us to assess their global detection significance and to overcome any possible publication bias.}
   {We performed a blind search for narrow absorption features at energies greater than 6.4~keV in a sample of 42 radio-quiet AGNs observed with XMM-Newton. A simple uniform model composed by an absorbed power-law plus Gaussian emission and absorption lines provided a good fit for all the data sets. 
We derived the absorption lines parameters and calculated their detailed detection significance making use of the classical F-test and extensive Monte Carlo simulations.}
{{We detect 36 narrow absorption lines on a total of 101 XMM-Newton EPIC pn observations. The number of absorption lines at rest-frame energies higher than 7~keV is 22. Their global probability to be generated by random fluctuations is very low, less than $3\times 10^{-8}$, and their detection have been independently confirmed by a spectral analysis of the MOS data, with associated random probability $<10^{-7}$. We identify the lines as Fe XXV and Fe XXVI K-shell resonant absorption. They are systematically blue-shifted, with a velocity distribution ranging from zero up to $\sim$0.3c, with a peak and mean value at $\sim$0.1c. We detect variability of the lines on both EWs and blue-shifted velocities among different XMM-Newton observations even on time-scales as short as a few days, possibly suggesting somewhat compact absorbers. Moreover, we find no significant correlation between the cosmological red-shifts of the sources and the lines blue-shifted velocities, ruling out any systematic contamination by local absorption. 
If we define Ultra-fast Outflows (UFOs) those highly ionized absorbers with outflow velocities higher than $10^4$~km/s, then the majority of the lines are consistent with being associated to UFOs and the fraction of objects with detected UFOs in the whole sample is at least $\sim$35\%. This fraction is similar for Type 1 and Type 2 sources. The global covering fraction of the absorbers is consequently estimated to be in the range $C$$\sim$0.4--0.6, thereby implying large opening angles.}}
   {{From our systematic X-ray spectral analysis on a large sample of radio-quiet AGNs we have been able to clearly assess the global veracity of the blue-shifted Fe K absorption lines at E$>$7~keV and to overcome their publication bias.
These lines indicate that UFOs are a rather common phenomenon observable in the central regions of these sources and they are probably the direct signature of AGN accretion disk winds/ejecta. The detailed photo-ionization modeling of these absorbers is presented in a companion paper.}}

\authorrunning{F. Tombesi et al.}

\titlerunning{ultra-fast outflows in radio-quiet AGNs}

   \maketitle
%

\section{Introduction}

Thanks to the high throughput of recent X-ray observatories, there is increasing evidence for the presence of narrow blue-shifted absorption lines at rest-frame energies higher than 7~keV in the spectra of a number of radio-quiet AGNs (see review by Cappi 2006).
These features are commonly identified with Fe XXV and/or Fe XXVI K-shell resonant absorption from a highly ionized, log$\xi$$\simeq$3--6~erg~s$^{-1}$~cm, zone of circumnuclear gas, with column densities as large as $N_H$$\simeq$$10^{22}$--$10^{24}$~cm$^{-2}$.
The blue-shifted velocities of the lines are also often quite large, reaching mildly relativistic values, up to $\sim$0.2--0.4c (e.g., Pounds et al.~2003a; Dadina et al.~2005; Markowitz et al.~2006; Braito et al.~2007; Cappi et al.~2009; Reeves et al.~2009). In some cases also short term variability has been reported (e.g, Dadina et al.~2005; Braito et al.~2007; Cappi et al.~2009 and references therein).
These findings suggest the presence of previously unknown highly ionized, massive and fast outflows from the innermost regions of radio-quiet AGNs, possibly connected with accretion disk winds/ejecta.

Detailed theoretical studies of accretion disk winds in AGNs showed that the inner regions of the outflowing material can be highly ionized by the intense nuclear radiation and can have large outflow velocities (e.g., King \& Pounds 2003; Proga \& Kallman 2004; Ohsuga et al.~2009; King 2010).
Therefore, when the observer line of sight intercepts the flow, considerable absorption features from highly ionized species can be imprinted in the X-ray spectrum.
In particular, the extreme values of the ionization parameter\footnote{Throughout this paper we use the definition of the ionization parameter $\xi=L_{ion}/nr^2$ (erg~s$^{-1}$~cm), where $L_{ion}$ is the source X-ray ionizing luminosity integrated between 1 and 1000 Ryd (1Ryd$=$13.6~eV), $n$ is the density of the material and $r$ the distance of the absorber from the central source (e.g., Tarter, Tucker \& Salpeter 1969).} strongly suggest 
that the absorption material could be preferentially observed in the spectrum 
through blue-shifted Fe K-shell absorption lines from Fe XXV and/or Fe
XXVI at energies greater than $\sim$7~keV (e.g., Sim et al.~2008; Schurch et al.~2009; King 2010; Sim et al.~2010).

However, these features are in a region of the spectrum where instrumental resolution and signal-to-noise (S/N) are usually much lower than in the soft X-rays (e.g., Sim et al.~2008; Schurch et al.~2009). Moreover, this highly ionized gas is hard to observe in soft X-ray spectra because all the elements lighter than Fe are almost completely ionized and this may be the reason why it has not been reported previously. 

In fact, extensive studies of local Seyfert 1 galaxies, which constitute the majority of the local radio-quiet AGN population, in the soft X-rays (E$\simeq$0.1--2~keV) demonstrated the presence of layers of ionized absorbing gas in at least 50\% of the sources, the so-called warm absorbers (WAs) (e.g., Blustin et al.~2005; McKernan et al.~2007). This gas has values of the ionization parameter (log$\xi$$\sim$0--2~erg~s$^{-1}$~cm), column density ($N_H$$\sim$$10^{20}$--$10^{22}$~cm$^{-2}$) and outflow velocity (v$\sim$100--1000~km/s) that are far less extreme than those of the highly ionized absorbers previously discussed. 
Hence, this can possibly suggest a different physical origin of the WAs, potentially connected with the optical-UV BLR or torus winds (Blustin et al.~2005; McKernan et al.~2007), rather than being directly linked to the accretion disk.

For clarity, we thereby define here Ultra-fast Outflows (UFOs) those highly ionized absorbers detected essentially through Fe XXV and Fe XXVI K-shell absorption lines with blue-shifted velocities v$\geq$10$^4$~km/s ($\simeq$0.033c), i.e., much greater than the maximum value for typical warm absorbers (e.g., Blustin et al.~2005; McKernan et al.~2007).

The detection of these UFOs is consistent with the observation of fast outflows in different classes of AGNs also in other wavebands, from the relativistic jets in radio-loud AGNs to the broad-absorption lines (BAL) in the UV and X-ray spectra of distant quasars (e.g., Chartas et al.~2002; Chartas et al.~2003).

There have been a number of paper reporting the evidence that UFOs from the central regions of radio-quiet AGNs might have the possibility to bring outward a significant amount of mass and energy, which can have an important influence on the surrounding environment (e.g., see review by Cappi 2006). In fact, feedback from AGNs is expected to have a significant role in the evolution of the host galaxy, such as the enrichment of the ISM or the reduction of star formation, and could also explain some fundamental relations (e.g., see review by Elvis 2006 and Fabian 2009).
The ejection of a huge amount of mass from the central regions of AGNs can also inhibit the growth of super-massive black holes (SMBHs), potentially affecting  their evolution.
The study of UFOs gives also important insights on the working of AGNs and on the black hole accretion/ejection physics in general (e.g., see review by Cappi 2006). 

However, even if in recent years there have been several papers in the literature reporting the detection of blue-shifted Fe K absorption lines in radio-quiet AGNs (e.g., Chartas et al.~2002; Chartas et al.~2003; Pounds et al.~2003a; Dadina et al.~2005; Markowitz et al.~2006; Braito et al.~2007; Cappi et al.~2009; Reeves et al.~2009), there is still some debate on their real statistical significance and therefore there are doubts on the effective existence of UFOs. 
In fact, the detections of these blue-shifted absorption features in the spectra of different sources have been published by several authors using distinct instruments, analysis and statistical techniques. 
Moreover, the fact that the detection significance of each single line can be weak and the lack of a global analysis on a complete sample of sources led Vaughan \& Uttley (2008) to claim the presence of a publication bias. 
Thus, there is the need to put a solid observational basis on the global veracity of such blue-shifted Fe K absorption lines.
The aim of the present work is to address this point. Therefore, we statistically quantify the incidence and characteristics of UFOs with a comprehensive and uniform analysis on a large sample of radio-quiet AGNs. We perform a blind search for blue-shifted Fe K absorption lines at E$>$6.4~keV, derive their observable parameters (centroid energy, equivalent width, blue-shifted velocity) and assess their statistical significance making use also of extensive Monte Carlo (MC) simulations. 
A detailed physical modeling of the absorbers using the Xstar (Kallman \& Bautista 2001) photo-ionization code and a curve of growth analysis is presented in a companion paper (Tombesi et al. in prep.). 

\section{Sample selection and data reduction}  

The sample of radio-quiet AGNs has been drawn from the RXTE All-Sky Slew Survey Catalog (XSS; Revnivtsev et al.~2004), which provides a list of 294 sources serendipitously detected in the hard X-rays. The survey is 90\% complete to a 4$\sigma$ limiting flux of $\simeq$$10^{-11}$~erg~s$^{-1}$~cm$^{-2}$ in the 4--10~keV band. From this catalog we extracted all the sources thereby classified as Type 1 (NLSy1, Sy1) and Type 2 (Sy2).
The detailed sources classification and redshifts were then cross-checked with the accurate database of NED\footnote{http://nedwww.ipac.caltech.edu} (NASA/IPAC Extragalactic Database).

In order to assure the direct observation of the nuclear continuum in the Fe K band (E$\simeq$4--10~keV), we selected only the Type 2 sources whose X-ray spectra were affected by intrinsic neutral absorption with a column density lower than $N_H$$\sim$$10^{24}$~cm$^{-2}$.   
Thus, the preliminary sample contains a total of 55 sources. These were divided into 42 Type 1 (including NLSy1, Sy1, Sy1.2 and Sy1.5) and 13 Type 2 (including Sy1.8, Sy1.9 and Sy2).
Given the limited RXTE sensitivity, the selected sources are mostly X-ray bright (4--10~keV flux $\ga 10^{-12}$~erg~s$^{-1}$~cm$^{-2}$) and local ($z\la0.1$).

The high effective area between 4--10~keV, coupled with the moderate spectral resolution (FWHM$\sim$150~eV at 6.4~keV) of the EPIC pn detector aboard XMM-Newton make this one of the best available instrument to perform a detailed spectral study in the Fe K band. Moreover, after about a decade of operation, the XMM-Newton catalogue contains a large number of pointed AGN observations suitable for a statistical study. 
Therefore, we cross-correlated the RXTE selected sample with the XMM-Newton Accepted Targets Catalog and downloaded all the pointed observations publicly available at the date of October 2008.

We limited our analysis to the mean spectra of the XMM-Newton EPIC pn observations of the sample sources. The data reduction of the XMM-Newton observations was performed following the standard procedure with the XMM-SAS v~8.0.1 package. We used the calibration files updated to October 2008.

   \begin{figure}[t]
   \centering
    \includegraphics[width=6.5cm,height=8cm,angle=270]{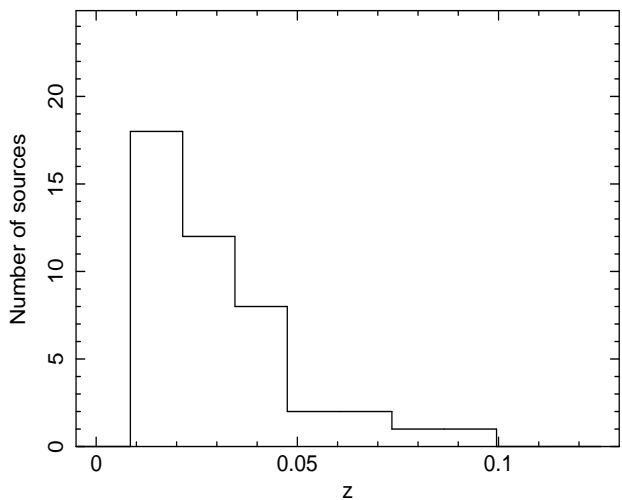}
   \caption{Distribution of the cosmological red-shifts of the sample sources.}
    \end{figure}

We checked the observations for high background contamination, looking for flares in the light curves at energies greater than 10~keV. We excluded these bad intervals from the successive analysis. 
We then selected only those observations with resultant net exposure time greater than 10~ks. Considering that the minimum 4--10~keV flux of the observations in the sample is $\sim10^{-12}$~erg~s$^{-1}$~cm$^{-2}$, this assures to have a minimal spectral sampling in this energy band (that is, to have at least 25 counts per $\sim$100~eV energy resolution element).
After these filtering processes, the total number of objects with at least one good XMM-Newton EPIC pn observation is 42. This constitutes about 80\% of the original 55 radio-quiet AGNs directly selected from the XSS catalog. The total number of observations is instead 101.
In particular, the number of Type 1 sources is 35 with 87 observations and that of Type 2 is 7 sources with 14 observations. These sources and the parameters of their XMM-Newton EPIC pn observations are reported in Table~A.1 in Appendix A.
It can be seen from the histograms in Fig.~1 and Fig.~2 that the sources in the final sample are local, with cosmological redshifts $z$$\le$$0.1$, and relatively X-ray bright, with 4--10~keV fluxes in the range $10^{-12}$--$10^{-10}$~erg~s$^{-1}$~cm$^{-2}$ and a mean value $\simeq$$2\times$$10^{-11}$~erg~s$^{-1}$~cm$^{-2}$.

\section{Data analysis}

We then proceeded with the uniform EPIC pn spectral analysis of the sample observations. We extracted the source photons from a circular region of 40$\arcsec$ radius, while the background ones were collected from an adjacent source free circular region of the same size. Only single and double events were selected. Using the SAS task \emph{epatplot} we checked that the Pile-up fraction was always negligible ($<$1\%).
Then, we extracted the source spectrum for all the observations, subtracted the correspondent background and grouped the data to a minimum of 25 counts per energy bin (to enable the use of the $\chi^2$ statistics when performing spectral fitting).

\subsection{Continuum and emission lines modeling}

The analysis of the EPIC pn observations was carried out using the \emph{heasoft} v.~6.5.1 package and XSPEC v.~11.3.2.
We limited our study to the Fe K band, between 3.5~keV and 10.5~keV. 

The large number of observations required a straightforward uniform analysis on the whole sample using a phenomenological baseline model composed of: a power-law continuum absorbed by neutral material intrinsic to the sources (\emph{zwabs} in XSPEC) and narrow Gaussians to model the expected neutral/ionized Fe K-shell emission lines in the interval E$\simeq$6.4--7~keV.

   \begin{figure}[t]
   \centering
    \includegraphics[width=6.5cm,height=8cm,angle=270]{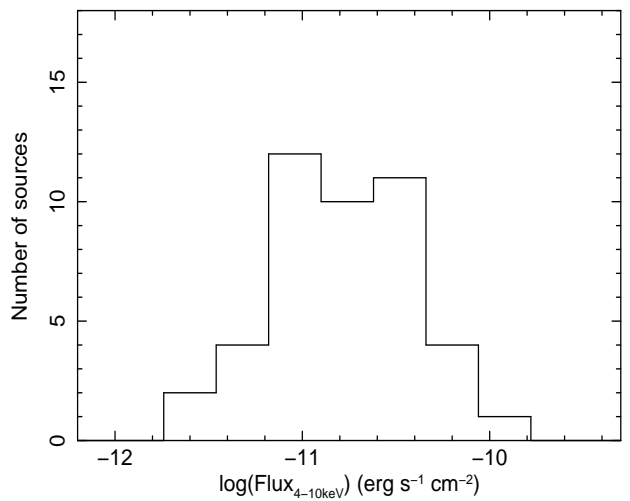}
   \caption{Distribution of the 4--10~keV flux among the sample sources. The values refer to the mean flux among the XMM-Newton observations for each source.}
    \end{figure}

Given the narrow energy band considered (E$=$3.5--10.5~keV), this modeling provides an excellent characterization of the data with the lowest number of free parameters. The validity of this choice is confirmed in \S3.4, where we report a detailed discussion of the effects of possible continuum modeling complexities.

The spectral fitting process was carried out in a uniform way for all the observations, starting from a simple power-law continuum and adding new components whenever they were required at more than the 99\% confidence level from the F-test. We first checked for the presence of power-law continuum curvature due to absorption intrinsic to the sources and then added Gaussians to model the expected Fe K fluorescence emission lines. We followed the method of adding consecutive spectral components to the model with decreasing $\chi^2$ improvement.
The ratios of the spectra against an absorbed, if required, power-law continuum model for all the observations of the sample sources are shown in the upper panels of the figures reported in Appendix C.

We confirm the ubiquitous presence of a neutral or mildly ionized Fe K$\alpha$ emission line at rest-frame energy E$\simeq$6.4--6.5~keV. 
In several cases we also detected emission lines at higher energies, possibly ascribable to highly ionized iron K$\alpha$ lines, such as Fe~XXV at E$\simeq$6.7~keV and Fe~XXVI at E$\simeq$7~keV, and/or to the neutral Fe K$\beta$ at E$\simeq$7.06~keV.
These emission lines turned out to be always marginally or not resolved at the energy resolution of the EPIC pn.
 We fixed their widths to 100~eV or 10~eV, depending on the higher $\chi^2$ improvement, in almost all the spectra. The values of the parameters of the baseline models for all the observations of the sample are reported in Table~A.2 in Appendix A for both Type 1 and Type 2 sources.

\subsection{Absorption lines search}

We carried out a blind search for absorption features in the 4--10~keV band looking at the deviations in the $\Delta\chi^2$ with respect to the baseline model. We applied a technique for the search and visualization of features in the data similar to the widely used contour plot method for the determination of the error contours of spectral features in the energy-intensity plane (e.g., Miniutti \& Fabian 2006; Miniutti et al.~2007; Cappi et al.~2009).
This calculation has been carried out as follows: 
1) we fitted the 3.5--10.5~keV data with the baseline model (without any absorption line) and stored the resulting $\chi^2$ value; 
2) we then added a further narrow (unresolved, $\sigma=10$~eV) Gaussian line to the model and searched for the presence of both emission and absorption features by making a series of fits stepping the line energy in the 4--10~keV band at intervals of 100~eV and its normalization in positive or negative values, each time storing the new $\chi^2$;
3) in this way we derived a grid of $\chi^2$ values and then made a plot of the contours with the same $\Delta\chi^2$ level relative to the baseline model fit. These levels are $\Delta\chi^2=-2.3$, $-4.61$ and $-9.21$, which can be translated using the F-test in statistical confidence levels for the addition of two more parameters of 68\%, 90\% and 99\%, respectively. 
Even if we were mainly interested in the 7--10~keV band, we performed this check on the whole 4--10~keV interval in order to have a general view of the Fe K band. 

In Fig.~3 we report the case of the first observation of PG~1211+143 (obs. 0112610101). This example illustrates the procedure used for the search of blue-shifted absorption lines, going from the actual spectral data modeling to the contour plots. The ratios against a simple (absorbed, if required) power-law continuum and the contour plots with respect to the baseline models for each XMM-Newton observation are displayed in Appendix C.

   \begin{figure}
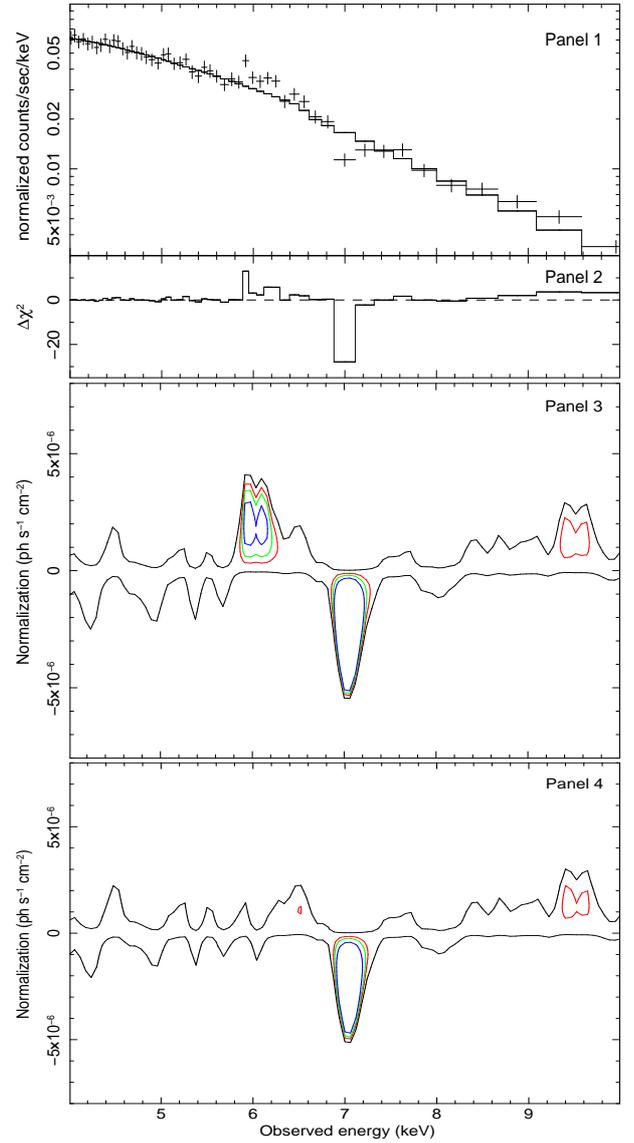

   \centering
    \includegraphics[width=5cm,height=8cm,angle=270]{pg1211_panel12.ps}
    \includegraphics[width=5cm,height=8cm,angle=270]{pg1211_panel3_obs.ps}
    \includegraphics[width=5cm,height=8cm,angle=270]{pg1211_panel4_obs.ps}
    \caption{\emph{Panel 1:} EPIC pn spectrum of PG~1211+143 (obs. 0112610101) fitted with a simple absorbed power-law model in the 4--10~keV band; \emph{Panel 2:} $\Delta\chi^2$ residuals; \emph{Panel 3:} confidence contour plot with respect to the simple absorbed power-law model (68\% (red), 90\% (green), 99\% (blue) levels); \emph{Panel 4:} contour plot after the inclusion of the Gaussian emission line in the model. The blue-shifted absorption line at the observed energy of $\sim$7~keV is clearly visible with high significance. The contours in black (calculated with $\Delta\chi^2=+0.5$) indicate the baseline model reference level.}
    \end{figure}

It should be noted that the negative $\Delta\chi^2$ values with respect to the baseline model indicate that a better fit would be reached with the inclusion of a line, reducing the $\chi^2$ value. 
In fact, the relative confidence levels do not depend on the $\Delta\chi^2$ sign but only on its absolute value. This method is similar to the one obtained with the \emph{steppar} command in XSPEC, but in this way the contours are inverted, \emph{which means that inner contours indicate higher significance than the outer ones}. This is indeed a powerful technique to visualize the presence of spectral structures in the data and simultaneously have an idea of their energy, intensity and confidence levels. However, it gives only a semi-quantitative indication and the absorption line parameters have been then determined by a direct spectral fitting. 

Therefore, the blind line search was performed as follows: 
1) we considered the 3.5--10.5~keV spectrum and fitted it with a simple power-law continuum model; 2) we checked for the presence of continuum curvature at energies below $\sim$6~keV due to intrinsic neutral/mildly-ionized absorption and approximated this component with \emph{zwabs} in XSPEC; 3) we looked at the spectral ratios, included the almost ubiquitous narrow Fe K$\alpha$ emission line at E$\simeq$6.4~keV and calculated the contour plots using this intermediate model; 
4) then, we adopted an iterative process at each step adding further narrow ionized emission lines (E$\simeq$6.4--7~keV) if required by the data at more than 99\% with the F-test (this corresponds to $\Delta\chi^2 \ge 9.21$ for two additional model parameters) and checked for the presence of possible blue-shifted absorption lines in the relative contour plots;
5) if there was evidence for a possible absorption feature at E$>$6.4~keV with confidence level $\ge$99\% in the contour plots, we checked for the presence of other intense ionized Fe K emission lines in the interval E$\simeq$6.4--7~keV that could influence the absorption line parameters. If further ionized emission lines were present, we included them in the baseline model, calculated again the energy-intensity contour plot using this refined baseline model and checked if the absorption feature with $\ge$99\% confidence contours was still present. Otherwise, if there was no evidence for further ionized emission lines, we stopped the process here, reported the baseline model in Table~A.2 and parameterized the absorption line with an inverted Gaussian, with width fixed to either 10~eV or 100~eV depending on the higher $\chi^2$ improvement;
6) if there was no evidence for blue-shifted absorption features we did not include additional emission lines besides the neutral Fe K$\alpha$ and directly reported the baseline model in Table~A.2.

It is important to note that in Table~A.2 we included only narrow ($\sigma$$\le$100~eV) absorption lines detected in the interval E$=$6.4--7.1~keV with F-test confidence levels $\ge$99\% and those detected at E$>$7.1~keV with, in addition, a Monte Carlo derived confidence level $\ge$95\% (see \S3.3).
In \S3.4 we assure that the inclusion of possible more complex spectral components to the baseline model yields completely consistent results.

It should also be noted that in the contour plots of some sources (see Appendix C) there is evidence for the presence of sporadic narrow absorption lines at energies lower than 6.4~keV (e.g., NGC~4151, NGC~3783, NGC~3516, MCG-6-30-15, Mrk~335, ESO~198-G024, NGC~7582), with F-test detection confidence levels $\sim$99\%.
Moreover, in some cases there is also evidence for broad/blended absorption features or possible edges at energies greater than 7~keV with F-test confidence level $\sim$99\% (e.g., IC4329A, Mrk~766, 1H419-577).
An interpretation of these possible additional spectral features is briefly discussed in \S4.5 and \S4.1, respectively, and in Appendix B for each source.

\subsection{Absorption lines detection significance}

Protassov et al.~(2002) pointed out that the standard likelihood ratio tests such as the $\chi^2$, Cash and related F-test can provide misleading results when applied to highly structured models, such as those used in the fitting of X-ray spectra.
Sometimes, in fact, the basic regularity conditions required by these standard tests are not met and the statistics do not follow the assumed null distributions. In particular, the F-test method can overestimate the actual detection significance for a blind search of emission/absorption lines as it does not take into account the possible range of energies where a line might be expected to occur, nor does it take into account the number of bins (resolution elements) present over that energy range. The F-test can yield the probability of finding a feature at a given energy if the line energy is known in advance from theory or laboratory results. This problem requires an additional test on the blue-shifted lines significance and can be solved by determining the unknown underlying statistical distribution by performing extensive Monte Carlo simulations (Porquet et al.~2004; Yaqoob \& Serlemitsos 2005; Miniutti \& Fabian 2006; Markowitz et al.~2006; Cappi et al.~2009).

The most intense atomic transitions expected in the 4--10~keV interval are the inner K-shell resonances of iron and, in particular, the 1s--2p. The energy of this transition depends on the ionization state of iron: E$\simeq$6.4--6.5~keV for Fe I--XVIII, E$\simeq$6.5--6.6~keV for Fe XIX--XXIV, E$\simeq$6.7~keV for Fe XXV and E$\simeq$6.97~keV for Fe XXVI (e.g., Kallman et al.~2004). Therefore, it is expected to find the presence of lines at least in this energy range.
However, it is important to note that K-shell 1s--2p absorption lines can actually occur only for ionization states of iron higher than Fe XVII/XVIII, with rest-frame energies E$\ga$6.6~keV, for which the L-shell is not fully occupied and there is vacancy for the 1s electron.

It is well established that at least half of the X-ray spectra of Seyfert galaxies are affected by intrinsic absorption from nuclear photo-ionized gas, the so-called warm absorber. This material is systematically found in outflow, with velocities lower than $\sim$1000~km/s (e.g., Blustin et al.~2005; McKernan et al.~2007).
Therefore, if we allow also for this possible low blue-shift, we would preferentially expect lines to be present in the energy interval E$\simeq$6.4--7.1~keV. Therefore, we can take a conservative approach assuming that the absorption lines in this energy band are rest-frame or lowly shifted Fe K-shell transitions and the standard F-test alone is adequate to estimate their confidence level.
The 14 absorption lines which satisfy this condition and which have F-test confidence levels greater than 99\% (see contour plots in Appendix C) are reported in Table~A.2 (in Appendix A). 
As we can see, their energies are indeed consistent with the Fe XXV and Fe XXVI 1s--2p transitions within $\sim$1--2\%, which correspond to velocity shifts smaller than $\sim$5000~km/s, that are known to dominate the Fe ions distribution for highly ionized absorbers (e.g., Kallman et al.~2004).

Instead, there is no a priori expectation to observe single narrow resonant Fe K absorption lines in the energy range 7.1--10~keV, if not consistent with higher order K-shell series lines.
Given that there is no preference in finding the line at a particular energy in this interval, we have to estimate the probability distribution of random generated lines in the whole energy range and compare this with that of the observed line to properly assess its detection significance.
Therefore, even if their F-test derived confidence levels are greater than 99\%, we performed extensive Monte Carlo simulations to clearly assess the significance of the narrow absorption lines detected in the E$=$7.1--10~keV band.

We tested the null hypothesis that the spectra were adequately fitted by a model that did not include the absorption line. The analysis has been carried out as follows:
1) we simulated a spectrum (with the \emph{fakeit} command in XSPEC) using the baseline model (listed in Table~A.2) without any absorption line and considered the same exposure as the real data. 
We subtracted the appropriate background and grouped the spectral data to a minimum of 25 counts per energy bin; 
2) we fitted again this spectrum with the baseline model in the 3.5--10.5~keV band, stored the new parameters values, and generated another simulated spectrum as before but using this refined model. This additional step is executed in order to account for the uncertainty in the null hypothesis model itself (Markowitz et al.~2006);
3) this new simulated spectrum was fitted again with the baseline model in the 3.5--10.5~keV and the resultant $\chi^2$ was stored; 
4) then, a new Gaussian line (unresolved, $\sigma=10$~eV) was added to the model, with its normalization initially set to zero and let free to vary between positive and negative values. In order to account for the possible range of energies in which the line could be detected in a blind search, we then stepped its centroid energy between 7.1 and 10~keV at intervals of 100~eV, to adequately sample the EPIC pn spectral resolution at these energies, each time making a fit and stored only the maximum of the resultant $\Delta\chi^2$ values; 5) this procedure was repeated $S=1000$ times and consequently a distribution of simulated $\Delta\chi^2$ values was generated. This would indicate the fraction of random generated narrow emission/absorption features in the 7.1--10~keV band that are expected to have a $\Delta\chi^2$ greater than a threshold value. In particular, if $N$ of these simulated $\Delta\chi^2$ values are greater or equal to the real value, then the estimated detection confidence level from Monte Carlo simulations is simply $1-{N/S}$. 

We note here that neglecting to include the step 2) of the Monte Carlo procedure does not affect the final probabilities for narrow emission/absorption lines.
This additional step is executed only in order to formally take into account the uncertainty in the baseline model itself, which is derived from a direct fit to the data, and therefore each time a new spectral simulation is initialized with a slightly different realization of the null hypothesis model (as in Markowitz et al.~2006).

We decided to select only those features with Monte Carlo derived confidence levels $\ge$95\%. 
Therefore, the number of narrow absorption lines at E$\ge$7.1~keV that are beyond this detection threshold is 22. 
We reported these lines and their parameters in Table~A.2 (in Appendix A). As expected, the Monte Carlo detection confidence levels are slightly lower than those derived from the F-test because they effectively take into account the random generated lines in the whole 7.1--10~keV energy interval. 
In \S3.4 we assess that the inclusion of possible more complex spectral components to the baseline models used in spectral simulations yields completely consistent results.

It is important to note here that we can also estimate the global probability to find such narrow absorption features in the 7.1--10~keV energy interval in the whole sample using the Binomial distribution. 
We have 21 observations with at least one absorption line detected at energies greater than 7.1~keV with Monte Carlo confidence level $\ge$95\%.
We can apply a conservative approach saying that the random probability of finding such features in that energy band for each observation is then $p<0.05$. Therefore, we can estimate the probability that the line detections in 21 out of a total of 101 observations are merely due to random fluctuations. This turned out to be less than $3\times 10^{-8}$. Therefore, even if some of the individual line detections can be weak, the global probability for these features to be generated by random fluctuations is very low.

\subsection{On the continuum modeling complexities}

There is the possibility that our modeling simplification (see \S3.1) could generate model dependent systematics that, in principle, can either favour or disfavour the detection of blue-shifted absorption lines at E$>$7~keV. This can be illustrated here with a few potential pitfalls.

Some sources are known to have broad, relativistically iron lines. Where the broad red or blue wing of such a line drops off rapidly, the continuum (power-law plus broad line) bends and the fit might improve by slightly modifying normalization and slope of the power-law, while adding an additional feature to compensate for the local bending. The same might occur at E$>$7~keV, where a hardening reflection component would emerge. At lower energies, improper modeling of additional curvature due to the presence of a warm absorber that is more complex than the simple cold absorber model used here to parametrize the continuum may also potentially lead to artifacts that might be reduced by including additional absorption lines. Moreover, not accounting for possible further weak emission lines may also potentially generate weak local drops, that could be alleviated by including absorption features. Therefore, we performed several additional checks to assure that our search for blue-shifted Fe K absorption lines is not globally affected by any bias due to a wrong continuum modeling.  

Some sources of the sample can in fact present rather complex 4--10~keV spectra (see figures in Appendix C) and their detailed modeling could require the inclusion of possible additional spectral components, such as a reflection component, a broad relativistic Fe line and/or ionized absorption.

A proper modeling of the possible reflection component requires a high energy coverage, from E$\simeq$10~keV up to E$\simeq$100--200~keV, and it can not be constrained with XMM-Newton. 
Therefore, we considered the average parameters from Dadina~2008, who performed a systematic broad-band spectral analysis (E$\simeq$1--100~keV) of local Seyfert galaxies ($z$$<$0.1) observed with BeppoSAX, i.e. a power-law photon index $\Gamma$$\sim$2, a reflection fraction R$\simeq$1 and a high energy cut-off E$_\textrm{c}$$\simeq$250~keV. We approximated this component with the model \emph{pexrav} in XSPEC.

The possible presence of a broad Fe K emission line, with a profile distorted by relativistic effects, has been reported in a number of AGN X-ray spectra. This could suggest an origin of the line from reflection on the inner regions of the putative accretion disk (e.g., Fabian et al.~1995).
In particular, Nandra et al.~(2007) recently performed a survey of broad iron lines in a large sample of Seyfert galaxies observed with XMM-Newton and found an incidence of $\sim$30\%. However, the strength of this component resulted much weaker than previously reported (e.g., Nandra et al.~1997). 
We modeled the line profile with \emph{diskline} in XSPEC and considered the typical parameters from Nandra et al.~(2007), i.e. a line energy E$\simeq$6.4~keV,  EW$\sim$80~eV, a disk inclination $i$$\simeq$$40\degr$ and emission radii between $r$$=$6--100~$r_g$.

Furthermore, it is well established that at least 50\% of local Seyfert galaxies show the presence of X-ray warm absorbers (e.g., Blustin et al.~2005; McKernan et al.~2007).
A proper modeling of WAs requires high energy resolution and low energy coverage (E$\simeq$0.1--2~keV), such as that offered by the gratings on board XMM-Newton and Chandra. However, these instruments have a lower sensitivity compared to the EPIC pn and require much longer exposures to reach comparable S/N. A systematic analysis of the soft X-ray spectra of Seyfert galaxies is beyond the scope of the present work.  
Therefore, given that the warm absorber parameters can not be constrained with the 4--10~keV EPIC pn spectra, we used the typical parameters already reported in the literature, i.e. a column density in the range $N_H$$\simeq$$10^{21}$--$10^{22}$~cm$^{-2}$ and an ionization parameter in the interval log$\xi$$\simeq$0--3~erg~s$^{-1}$~cm (Blustin et al.~2005; McKernan et al.~2007). 
We did not use the accurate photo-ionization code Xstar to model the WA because it would have introduced also possible narrow resonance absorption lines in the model. Instead, we used the \emph{absori} model in XSPEC because it accounts only for photo-electric absorption and it is sufficient for the modeling of the continuum curvature due to bound-free transitions from the WA.

Therefore, firstly, we performed several spectral simulations using the \emph{fakeit} command in XSPEC to check the general accuracy of our continuum modeling.
We considered the typical 4--10~keV spectrum of Seyfert galaxies in our sample, i.e. a power-law continuum with  $\Gamma$$\sim$2, a 4--10~keV flux of $10^{-11}$~erg~s$^{-1}$~cm$^{-2}$ and a 50~ks net exposure, which yields a total of $\sim$$5\times 10^4$ counts in the 4--10~keV band. We also added a typical narrow Fe K$\alpha$ emission line at E$=$6.4~keV with width $\sigma$$=$100~eV and EW$=$100~eV.
We then simulated several spectra iteratively including each of the possible additional spectral elements, i.e. the reflection component, the broad Fe K line, the warm absorber and all three together, and fitted them with our absorbed power-law plus Gaussian emission line baseline model and the more complex models.
It turned out that the difference in the estimated 4--10~keV continuum slope in the two cases is always less than $\sim$1--2\%, completely within the 1~$\sigma$ errors.

We did not find necessary to include neutral absorption from our own Galaxy because the low column densities involved ($N_H$$\la$$10^{21}$~cm$^{-2}$) have no effects in this energy band. 
However, we warn that even if the continuum curvature induced by the possible WAs in the limited 4--10~keV band is well approximated by a simple neutral absorption component, the derived column densities (see Table~A.2) should not be considered for any further physical interpretation.

Then, we performed the same spectral simulations, but this time including also narrow ($\sigma$$=$10~eV) absorption lines at the energies of 7, 8 and 9~keV with EW$=$$-50$~eV. We then fitted the spectra with both our phenomenological baseline model and the more complex models and derived that the absorption line parameters (energy and EW) were always consistent within the relative 1~$\sigma$ errors.

We also performed a more specific check considering two representative cases of complex 4--10~keV spectra among our sample sources. We chose NGC~3783 (obs. 0112210501) and MCG-5-23-16 (obs. 0302850201), which are among the brightest sources in our sample and have very high S/N spectra. 
The detailed analysis of the same XMM-Newton data sets in the Fe K band has already been reported by Reeves et al.~(2004) and Braito et al.~(2007), respectively. 
The authors used complex models which incorporate all the main spectral components previously discussed, i.e. a power-law continuum plus narrow Fe K emission lines, a broad relativistic iron line, a reflection component and intrinsic neutral/ionized absorption.
However, we compared the direct spectral fit with these complex models or with our phenomenological baseline models (see Table~A.2 in Appendix A) and found that the continuum parameters were always consistent within the 1~$\sigma$ errors.

Moreover, from the energy-intensity contour plots of these sources (see Fig.~C.1 and Fig.~C.7 in Appendix C) there is evidence for narrow absorption lines at rest-frame energies of E$\simeq$6.67~keV and E$\simeq$7.84~keV, respectively. The detection of these absorption lines in the same XMM-Newton data sets has already been reported by Reeves et al.~(2004) and Braito et al.~(2007), respectively.
Therefore, we directly fitted the data using our baseline models (see Table~A.2 in Appendix A) and also their more complex models and derived that the absorption line parameters (energy and EW) were consistent within the 1~$\sigma$ errors in both cases.  

   \begin{figure}[t]
   \centering
    \includegraphics[width=8cm,height=10.5cm,angle=0]{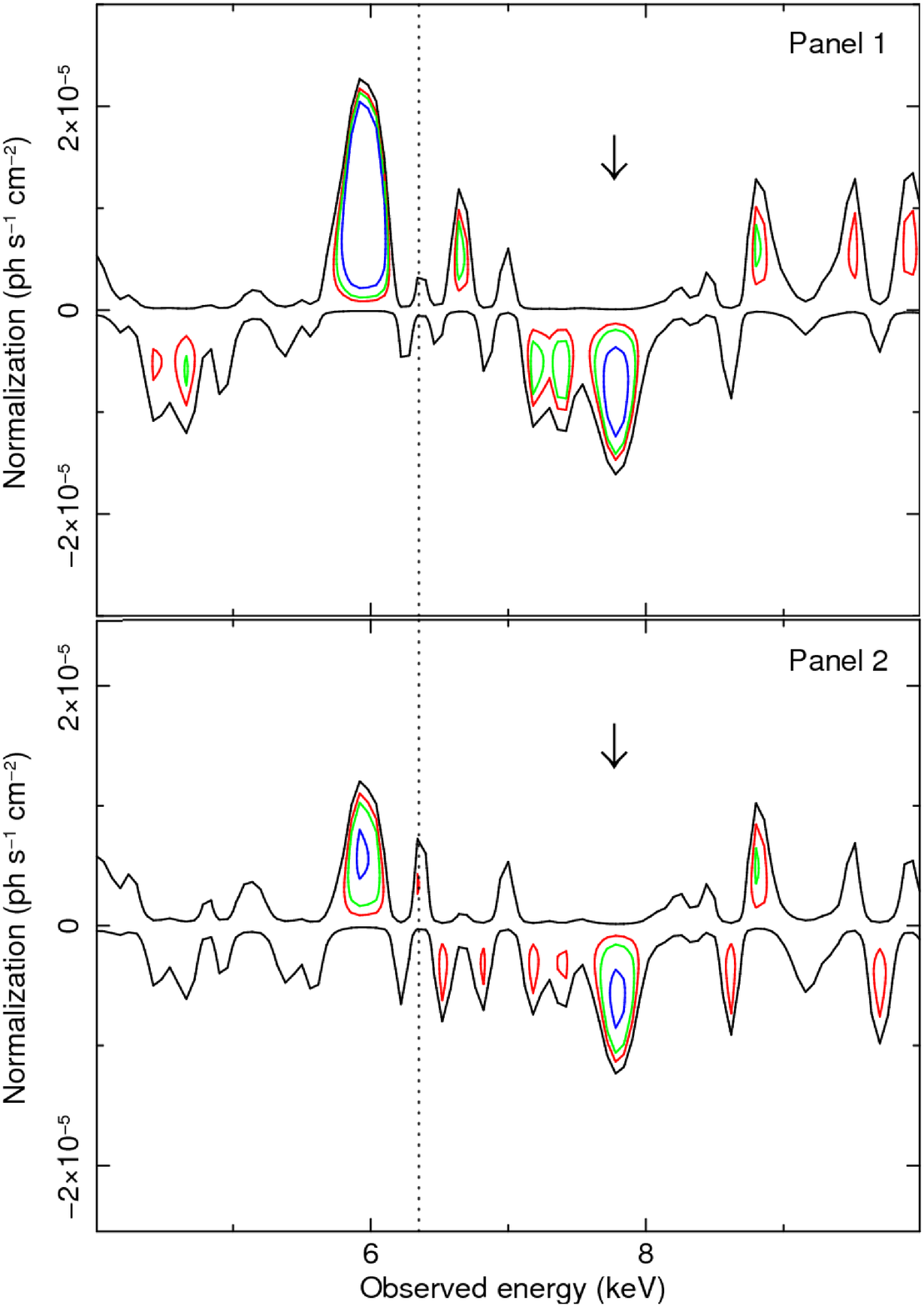}
   \caption{Comparison of the energy-intensity contour plots for MCG-5-23-16 (obs. 0302850201) calculated using our baseline model (\emph{Panel 1}) and the more complex model of Braito et al.~(2007) (\emph{Panel 2}). The blue-shifted absorption line at the observed energy of E$\simeq$7.77~keV (rest-frame E$\simeq$7.84~keV) is pointed by the arrow. The location of the neutral Fe K$\alpha$ emission line (rest-frame energy E$=$6.4~keV) is marked by the vertical line for reference.}
    \end{figure}

In Fig.~4 we report a comparison of the energy-intensity contour plots for MCG-5-23-16 (obs. 0302850101) calculated using our phenomenological baseline model (Panel 1) or the more complex model of Braito et al.~(2007) (Panel 2). It can be seen that, even if the model of Braito et al.~(2007) provides a somewhat better overall characterization of the 4--10~keV spectrum, the blue-shifted absorption line at the observed energy of E$\simeq$7.77~keV (rest-frame E$\simeq$7.84~keV) is indeed present in both cases, with equivalent parameters.

It is worth noting that in some spectra there is also evidence for sporadic weak line-like emission features in the $\sim$5--6~keV energy band (e.g., IC4329A, NGC~4151, NGC~3783, MCG-6-30-15), that could be associated with relativistically red-shifted lines (see figures in Appendix C).   
However, we did not find necessary to include also these features because, as previously demonstrated, our phenomenological model already provides a good continuum characterization. 
We refer the reader to the recent work by De Marco et al.~(2009) who performed a comprehensive study of these red-shifted features in the X-ray spectra of a sample of Seyfert 1 galaxies.  

However, for each spectrum with at least a detected narrow absorption line in the Fe K band, we assured that the parametrization of the possible remaining weak emission features with simple Gaussian lines (with free width), after fitting with our baseline models, gives absorption line parameters always consistent within the 1~$\sigma$ errors.

Therefore, we conclude that the possible additional spectral features that can complicate the analysis of the 4--10~keV spectra are typically not intense enough to require their detailed treatment and to induce possible systematics. Our phenomenological baseline model is indeed adequate to provide a suitable estimate of the continuum level in the limited 3.5--10.5~keV band.

Finally, we performed also a sanity check on the possible dependence of the Monte Carlo probability on the assumed baseline models.
We considered again NGC~3783 (obs. 0112210501) and MCG-5-23-16 (obs. 0302850201) as representative cases of the complex 4--10~keV spectra of some sample sources. We used the more complex models reported respectively by Reeves et al.~(2004) and Braito et al.~(2007) to simulate the spectra at the first step of the Monte Carlo procedure (see \S3.3), then fitted them with the same models and with our phenomenological baseline models and directly compared the resultant $\Delta\chi^2$ distributions for random generated lines. 
We derived that the difference in the confidence levels of possible absorption lines at E$>$7~keV in the two cases can be at maximum of the order of $\pm$0.5\%.

Therefore, the various tests described in this section clearly assure that our absorption line search is not globally biased and that the model dependence of our results is only very marginal.

\subsection{EPIC pn background and calibration checks}

We checked that the background level of the observations of the sample in the 
7--10~keV band is negligible, always less than $\sim$10\% of the source 
counts. 
However, it is important to note that the EPIC pn background has intense instrumental emission lines from Ni K$\alpha$ and Cu K$\alpha$ at the energies of 7.48~keV and 8.05~keV, respectively (Katayama et al.~2004). Their intensity has been found to be space dependent on the detector, increasing with off-axis angles. Therefore, an incorrect selection of the background extraction region could induce fake absorption features when subtracted from the source spectrum. This effect is more relevant for low flux or extended sources. We checked that none of the reported blue-shifted absorption lines were induced by an erroneous subtraction of the background spectrum.
Moreover, the detected features cannot be attributed to some sort of EPIC pn calibration artifact because they have been detected always at different energies and are narrow. 
Instead, if due to instrumental calibration problems, they would be expected to be observed always at some specific energies or, if due to effective area calibration problems, to induce broad continuum distortions.

\subsection{Consistency check with the EPIC MOS cameras}

We performed a consistency check of the EPIC pn spectral results using the data from the MOS1 and MOS2 cameras.
These instruments have a lower effective area compared to the EPIC pn ($\sim$30\% at 6~keV). However, they provide simultaneous observations of the same pointed source in the 0.5--10~keV band and therefore they help to independently confirm the veracity of the narrow absorption lines detected in the EPIC pn spectra. 
We reduced the MOS1 and MOS2 data of the 36 XMM-Newton observations in which absorption lines at E$>$6.4~keV were detected in their EPIC pn spectra (see Table~A.2 in Appendix A). We followed the standard reduction procedure and used the XMM-SAS v.8.0.1 package. Initially, we checked for the presence of high background flares looking at the light curves at energies greater than 10~keV and excluded the relative bad time intervals from the analysis. Then, we extracted the source photons from a circular region of 40$\arcsec$ radius centered on the source and the background ones from an adjacent source free circular region of the same size.
The pile-up fraction for observations performed with the MOS cameras can be high, especially for the brightest sources in our sample (even greater than 10\%). However, we expect such effect to have negligible influence on the narrow absorption lines we are interested in. In fact, the main result of the pile-up is to induce an overall distortion of the observed continuum, as it basically consists in counting two or more low-energy photons as a single high-energy photon.
Then, we extracted the MOS1 and MOS2 source and background spectra and grouped them to a minimum of 25 counts per energy bin (in order to allow the use of the $\chi^2$ statistics in spectral fitting). We limited the analysis to the 3.5--10.5~keV energy band.

   \begin{figure}[!t]
   \centering
    \includegraphics[width=6.5cm,height=8cm,angle=270]{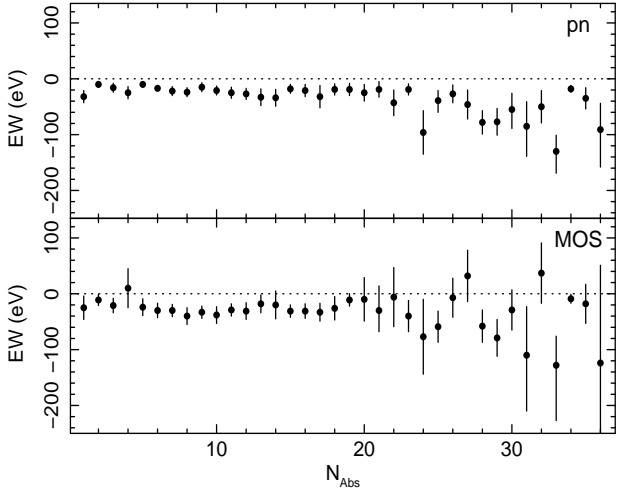}
   \caption{Consistency check of the EWs of the absorption lines detected in the pn spectra at E$>$6.4~keV (\emph{upper panel}) with respect to the MOS cameras (\emph{lower panel}). The EW errors are at the 90\% level. In the x axis we numerate each absorption line, for a total of 36. The horizontal dotted lines refer to the zero level. See \S3.6 for more details.}
    \end{figure}

The main purpose of this test is not to find new absorption lines with the MOS, but to confirm (or deny) features found with the pn at E$>$6.4~keV. Therefore, the line detections in the MOS cameras must be linked to the detections in the pn.
We fitted each MOS spectrum with the baseline model derived from the pn, with the inclusion of the absorption lines (see Table~A.2). We fixed the energy and width of the lines to the values determined with the pn and derived only the best-fit EW and associated errors, allowing the line normalization to be positive or negative. When possible, the MOS1 and MOS2 data have been combined in order to increase the statistics. The EWs of the absorption lines along with their 1$\sigma$ errors derived from the MOS cameras are reported in Table~A.2 (in Appendix A), after the pn results. The significance of the lines can be simply estimated comparing the values of the EWs with respect to the associated errors. 

In Fig.~5 we show a direct comparison of the EWs from the pn (upper panel) and the MOS (lower panel) for all the 36 absorption lines detected in the pn spectra at E$>$6.4~keV. In this plot the error bars are reported at the 90\% level and the horizontal dotted lines refer to the zero EW level. We can see that there is an overall consistency of the results, even if the MOS are systematically slightly less significant due to the lower statistics available.  
In fact, all the 36 line EWs from pn are not consistent with zero at $>$90\%. For the MOS, 26 are not consistent with zero at the 90\% level, which corresponds to a fraction of 26/36 ($\sim$70\%). For the remaining 11 we can only place lower limits.

Moreover, it is important to note that the EWs derived from the pn and MOS are consistent at the 90\% level for 34 out of 36 features ($\sim$94\% of the cases). 
In particular, if we restrict only to the 22 absorption lines detected at E$\ge$7.1~keV, 20 ($\sim$91\%) of them have EWs consistent at the 90\% level between pn and MOS. Furthermore, for 12/22 ($\sim$55\%) of them the EWs derived from the MOS are not consistent with zero at $>$90\%.

Looking at the right part of Fig.~5 (lower panel), we can note a discrepancy of the results between pn and MOS in only two cases, in which the EWs from MOS are slightly not consistent with the pn at the 90\%, i.e., Mrk~841 (obs. 0205340401; number 27 in Fig.~5) and Mrk~205 (obs. 0124110101; number 32 in Fig.~5).  
In fact, there is actually an indication for possible emission features in the MOS where absorption was expected. However, the lines are consistent with zero at the 1$\sigma$ level. Both these two lines have been detected in the pn at E$>$7.1~keV. This suggests that only these two features detected in the EPIC pn data might be fake. This is consistent with the Monte Carlo confidence level threshold of 95\%, for which on 22 absorption lines detected at E$\ge$7.1~keV we would expect that at maximum one or two of them might be generated by random fluctuations.

As already discussed in \S3.3, we can estimate the global probability to detect narrow absorption lines in the MOS spectra at energies fixed from the pn using the Binomial distribution. We have 26 cases out of 36 with a detected absorption line at $\ge$90\% in the MOS. We can apply a conservative approach saying that the random probability of finding each of these absorption lines in the MOS at the energies determined from the pn is $p\le0.1$. Therefore, we can estimate the probability that the line detections in 26 out of 36 cases are due to random fluctuations. This turns out to be extremely low, less than $\sim$$10^{-18}$. Moreover, if we resctrict only to the absorption features detected in the pn at E$>$7.1~keV, we have 12 detections out of 22. Also in this case the random probability is very low, $\la$$10^{-7}$. 

Therefore, we conclude that even if there are two dubious cases, the global consistency of the pn and MOS results is strong and the probability that the absorption lines in MOS are generated by random fluctuations is extremely low.
The fact that we have been able to independently confirm the detection of several absorption lines observed in the pn also with the MOS cameras places an additional very strong point in favour of the veracity of the lines, without relying on any statistical method.

\section{Results}

As already introduced in \S1, here we focus our analysis on the blue-shifted Fe K absorption lines because their less ambiguous interpretation and modeling make them a fundamental diagnostic tool for the study of accretion disk winds/outflows in AGNs (e.g., Sim et al.~2008; Schurch et al.~2009; King 2010; Sim et al.~2010).

\subsection{On the identification of absorption lines at E$>$6.4~keV}

The total number of narrow absorption lines detected at rest-frame energies greater than 6.4~keV in all the observations is 36. In particular, 14 are in the interval E$\simeq$6.4--7.1~keV and 22 at energies greater than 7.1~keV  (see \S3.3 and Table~A.2).
We identified these features with highly ionized iron absorption lines, specifically associated with Fe XXV and Fe XXVI K-shell transitions. The lines parameters have been taken from the NIST\footnote{http://physics.nist.gov/PhysRefData/ASD/index.html.} atomic database, if not otherwise stated.

The most prominent Fe XXV lines are due to the He$\alpha$ (1s$^2$--1s2p) and He$\beta$ (1s$^2$--1s3p) transitions, which further subdivide into inter-combination and resonance lines. At the moderate energy resolution of the EPIC pn instrument, these fine structure line components are not distinguishable and we can measure only a blend. Therefore, we averaged their energies, weighted for the respective oscillator strengths. The consequent line energies are 6.697~keV for the Fe XXV He$\alpha$ and 7.880~keV for the Fe XXV He$\beta$, respectively.
The most intense lines from Fe XXVI are instead due to Ly$\alpha$ (1s--2p) and Ly$\beta$ (1s--3p) transitions, each of which further subdivide into two resonance doublets. The average energies of these lines, weighted for their oscillator strengths, are 6.966~keV for the Fe XXVI Ly$\alpha$ and 8.250~keV for the Fe XXVI Ly$\beta$, respectively. 
From a simple comparison of the oscillator strengths of the lines, the equivalent widths of the  1s--3p transitions are expected to be $\sim$20\% of the 1s--2p. However, this is only a lower limit because their ratios can increase up to $\sim$1 if saturation effects are significant, i.e. for high column densities or low line velocity broadening (e.g., Bianchi et al.~2005; Risaliti et al.~2005; Tombesi et al. in prep.).
If the measured line energies were not consistent with these expected values, we calculated the relative velocity shifts. 
We adopted the convention to use positive and negative signs for blue-shifted and red-shifted line velocities, respectively.

The best-fit parameters and identification of the absorption lines are listed in Table~A.2 (in Appendix A). It is important to note that in this table we reported only the narrow ($\sigma$$\le$100~eV) absorption lines detected in the 6.4--7.1~keV interval with F-test confidence levels $\ge$99\% and those detected in the 7.1--10~keV energy band with additional Monte Carlo probability $\ge$95\% (see \S3.3).
The detailed physical modeling of these absorption lines with the photo-ionization code Xstar is presented in a companion paper (Tombesi et al. in prep.).
There we clearly show that the identification of the lines with K-shell transitions from highly ionized iron assumed here is indeed correct and the general conclusions are completely consistent with the somewhat qualitative discussion presented here.

In the majority of the cases we detected only a single narrow absorption line.  If the line energy was between E$\simeq$6.4--7.1~keV, we employed a conservative approach assuming that the line was rest-frame or low blue/red-shifted Fe XXV He$\alpha$ or Fe XXVI Ly$\alpha$, depending on the transition closer to the measured centroid energy of the line.
In particular, from Table~A.2 we can see that the energy of the lines in the interval E$\simeq$6.4--7~keV are indeed consistent with our identification as Fe XXV/XXVI 1s--2p transitions within $\sim$1--2\%, which correspond to velocity shifts smaller than $\sim$5000~km/s.
However, it should be noted that the almost ubiquitous presence of Fe K emission lines at E$\simeq$6.4--7~keV can potentially fill up possible absorption features, hampering their detection in this energy interval.

Instead, if a single line was detected at E$\ga$7~keV, we opted for the identification as Fe XXV He$\alpha$ or Fe XXVI Ly$\alpha$ following a different approach.
If the line detection and its detailed modeling was already reported in other papers in the literature, we used the same identification from these studies; this is the case for IC4329A, NGC~3783, Mrk~509, PG~1211$+$143 and MCG-5-23-16 (see Appendix B for a detailed description of each source).
On the other hand, if there were no other indications, we preferentially identified the line with Fe XXVI Ly$\alpha$ and calculated the relative blue-shifted velocity.
This choice is supported by fact that the EW of the Fe XXV He$\alpha$ transition with respect to the ionization parameter is sharp and peaks around log$\xi$$\sim$3~erg~s$^{-1}$~cm, instead that for Fe XXVI Ly$\alpha$ is much broader and spans from log$\xi$$\sim$4~erg~s$^{-1}$~cm up to log$\xi$$\sim$6~erg~s$^{-1}$~cm (e.g., Bianchi et al.~2005). 
Therefore, the identification of all single lines at E$>$7~keV with blue-shifted Fe XXV He$\alpha$ absorption is less probable because it is most intense only for a narrow range of ionization parameters, log$\xi$$\sim$3--4~erg~s$^{-1}$~cm; a slight lower value of $\xi$ would imply a significant low-energy curvature and would be accompanied by lines and edges from low-Z ions as well, instead a higher value of $\xi$ would give rise to lines from also Fe XXVI (e.g., Bianchi et al.~2005; Risaliti et al.~2005; Cappi et al.~2009 and references therein).
Instead, the EW of the Fe XXVI Ly$\alpha$ transition has a less limited interval of ionization parameters in which its intensity is strong, from log$\xi$$\sim$4~erg~s$^{-1}$~cm up to log$\xi$$\sim$6~erg~s$^{-1}$~cm (the exact value of the EW depending on the assumed total column density). At these high ionization states there are no additional spectral features from other ionic species imprinted in the spectrum at energies below $\sim$7~keV.
This solution might turn out not be adequate for one or few cases, but its global inadequacy is contrasted by the fact that otherwise it would require a systematic fine-tuning of the absorber parameters in several different sources.
Moreover, this is a conservative approach, because the association of the single lines at E$>$7~keV with Fe XXV He$\alpha$ would imply systematically larger velocity shifts.

Finally, there are a six observations in which we have two or more simultaneous absorption line detections at energies E$>$6.4~keV. These occurrences depend on the different configuration of the absorber parameters, i.e. ionization state, column density and turbulent velocity (e.g., Bianchi et al.~2005; Risaliti et al.~2005; Sim et al.~2008; Sim et al.~2010; Tombesi et al. in prep.).
In these cases we started with the identification of the lower energy lines with 1s--2p Fe XXV and/or Fe XXVI transitions, which are the most intense, and then considered the 1s--3p, taking also into account their expected energy spacing.  
For instance, in the spectra of two sources, namely NGC~3783 and ESO~323-G77, we detected two narrow absorption lines with the same common blue-shift and in NGC~3516 we detected two lines in three different spectra and even four lines with the same common blue-shift in another observation (see Table~A.2).

The identification of spectral features in the Fe K band is relatively secure since only K-shell features of heavy ions are expected at energies $\ga$6~keV. At lower energies, where several K-shell transitions of light or intermediate-mass elements might arise, line identification is more ambiguous, particularly if large velocity shifts are considered. However, several objections might be raised.

For instance, we should consider the possibility of the lines to be associated with blue-shifted 1s--2p transitions from lowly ionized iron.
We note however that K-shell ($n$$=$1) 1s--2p absorption lines can occur only for ionization states of iron higher than Fe XVII/XVIII, with rest-frame energies of the lines E$\ga$6.6~keV, which have the L-shell ($n$$=$2) level not fully occupied.
However, the column densities required to explain the EWs of the lines detected in the Fe K band are typically large, in the range $N_H$$\sim$$10^{22}$--$10^{24}$~cm$^{-2}$ (e.g., Pounds et al.~2003a; Dadina et al.~2005; Markowitz et al.~2006; Braito et al.~2007; Cappi et al.~2009; Reeves et al.~2009), and if the ionization state of the absorbing material is low, this would cause a strong systematic depression of the spectra up to $\sim$6~keV in both Type 1 and Type 2 sources, which is not seen.

The only case in which a potential contamination from lower ionization species of iron can be observable is for NCG~3783. 
In fact, as Reeves et al.~(2004) already discussed, there is evidence for a blend of the Fe XXV He$\alpha$ line with lower ionization components from Fe XXIII (6.630~keV) and Fe XXIV (6.659~keV). This would explain why its centroid energy is measured between E$\simeq$6.61--6.67~keV, resulting in an apparent low red-shifted velocity, not consistent with the expected rest-frame energy of Fe XXV He$\alpha$ at the 90\% level.
The possible contribution from lower ionization species of iron is consistently taken into account in the physical modeling of the Fe K absorbers with the photo-ionization code Xstar presented in the companion paper (Tombesi et al. in prep.).

We find also extremely unlikely the identification of absorption lines at E$>$6.4~keV with blue-shifted K-shell resonances from ionized low-Z elements. For instance, if associated with Ar XVIII Ly$\alpha$ at E$\simeq$3.3~keV or Ca XX Ly$\alpha$ at E$\simeq$4.4~keV, the blue-shifted velocities of the lines would be much greater than Fe, of the order of $\sim$0.5c.
Moreover, in this case also extremely blue-shifted lines and edges from several low-Z elements should be observed in the soft X-rays (E$\la$4~keV), resulting in a completely distorted spectrum.
Instead, only spectral features due to warm absorbers are detected in this energy band, with blue-shifted velocities always lower than $\sim$1000~km/s (e.g., Blustin et al.~2005; McKernan et al.~2007).
However, this might provide a possible explanation for some of the narrow absorption lines observable in a few energy-intensity contour plots at energies lower than 6.4~keV (see \S3.2). This possibility is discussed in more detail in \S4.5.

It should be noted that, in principle, also less intense (equivalent width $\sim$5\% of the Fe XXVI Ly$\alpha$, but can be comparable if saturation effects are significant) higher order Fe XXVI Lyman series lines could be observed in some spectra, such as: the Ly$\gamma$ (1s--4p) at E$=$8.700~keV and the Ly$\delta$ (1s--5p) at E$=$8.909~keV. However, their detection is hampered by the fact that they are intrinsically weak and their rest-frame energies are in a region of the spectrum in which the instrumental effective area and energy resolution are worse and the S/N is lower.
Nevertheless, we found that in the XMM-Newton observation of IC4329A (obs. 0147440101) the broad absorption trough observable at an energy greater than $\simeq$9~keV (see Fig.~C.1 in Appendix C) can be well described by a blend of unresolved higher order Fe XXVI Lyman series lines with a blue-shift consistent with that of the narrow Fe XXVI Ly$\alpha$ line reported in Table~A.2 (see Appendix B for more details).
Moreover, indication for a possible additional narrow absorption feature at rest-frame energy E$\simeq$8.4~keV in the first XMM-Newton observation of 1H419-577 (obs. 0148000201) can be seen in the energy-intensity contour plot in Fig.~C.5 (top right panel). However, this feature has not been included in Table~A.2 and is not further considered here because the relative Monte Carlo confidence level was less than the threshold value of 95\% (see note on this source in Appendix B for more details).

Moreover, even if the cosmic abundance of nickel is negligible with respect to that of iron ($\sim$5\%, from Grevesse et al.~1996), the K-shell transitions of this element are distributed at energies greater than 7~keV and could complicate our line identification. 
The contamination by mildly ionized Ni K$\alpha$ lines is very unlikely, as it would require extremely high column densities ($N_H$$>$$10^{24}$--$10^{25}$~cm$^{-2}$) for these lines to be intense enough to be observable, which would consequently generate strong absorption lines and edges from all the other elements as well.
The only possible contamination could be due to He/H-like nickel, whose 1s--2p transitions are at rest-frame energies of E$\simeq$7.8~keV and E$\simeq$8.1~keV, respectively. 
Also in this case the column densities required to have lines with measurable intensities would be extremely high ($N_H$$>$$10^{24}$--$10^{25}$~cm$^{-2}$). 
However, the very high ionization level required to have significant columns of these ions are so extreme (log$\xi$$\ga$6~erg~s$^{-1}$~cm) that all the lighter elements would be completely ionized (with iron being the only possible exception) and therefore they will not contribute with other absorption features. 
We found that only 4 over 22 absorption lines detected at E$\ge$7.1~keV have energies consistent at the 90\% level with rest-frame He/H-like nickel transitions. 
Therefore, we conclude that even if the unlikely interpretation as rest-frame highly ionized Ni K lines turned out not to be a mere coincidence, this would have a negligible effect on the global results anyway.

The search for narrow absorption lines at energies greater than 7~keV could be also complicated by the presence of ionized Fe K edges at energies in the range from E$\simeq$7.1~keV to E$\simeq$9.3~keV, depending on the ionization state of iron (from neutral to H-like). 
Hence, one could object that some of the spectral structures we identified as blue-shifted absorption lines could actually be interpreted equally well as ionized Fe K edges. 
As a sanity check, we tested that the alternative modeling of the Gaussian absorption lines with simple sharp absorption edges (\emph{zedge} in XSPEC) did not significantly improve the spectral fits, as expected from the narrowness of the observed spectral features. 
Moreover, it is important to note that the commonly held view of sharp Fe K edges is an oversimplification of the real process and could lead to misleading results. 
In fact, it has been demonstrated that if the adequate treatment of the decay pathways of resonances converging to the K threshold is properly taken into account, the resulting edges are not sharp but smeared and broadened (Palmeri et al.~2002; Kallman et al.~2004). This effect can be negligible for neutral or extremely ionized iron (He/H-like) but is quite relevant for intermediate states, with energies in the range E$\simeq$7.2--9~keV. 
Furthermore, intense Fe K resonance absorption lines from different ionization states would be expected to accompany the edges. 
A proper modeling of the Fe K edges would need to be performed only using more sophisticated photo-ionization codes (such as Xstar), which is presented in a companion paper (Tombesi et al. in prep.).
The only case in which a significant broad absorption trough is observed at the energies of $\sim$8--9~keV is in one observation of Mrk~766 (0109141301, see Fig.~C.4 in Appendix C). As already reported by Pounds et al.~(2003c), this feature can only be well modeled by a rest-frame ionized Fe K edge (see Appendix B for more detailed information) and therefore it has not been included in Table~A.2.

\subsection{Global incidence of blue-shifted Fe K absorption lines}

We can divide the detected absorption lines into two groups: at rest, if their blue-shifted velocity is consistent with or slightly less than zero at the 90\% level and blue-shifted, if their velocity is greater than zero at the 90\% level. The first group is composed of only 8 features, the second instead of 28. Therefore, the majority of the detected features are consistent with the presence of outflows.
The total number of objects with at least one absorption line detected in the Fe K band is 19 over 42, that translates into a frequency of $\sim$45\%.  
If we further consider the fraction of objects having at least one feature with blue-shifted velocity, the fraction is instead 17/42, which corresponds to about 40\% of the cases. 

In \S1 we defined Ultra-fast Outflows (UFOs) those highly ionized absorbers detected essentially through Fe XXV and Fe XXVI K-shell absorption lines with blue-shifted velocities v$\geq$10$^4$~km/s ($\simeq$0.033c), i.e., much greater than the maximum value for typical warm absorbers (e.g., Blustin et al.~2005; McKernan et al.~2007).

The ratio of sources with at least one detection of a narrow absorption feature with blue-shifted velocity v$\ge$$10^4$~km/s, thus identifiable with UFOs, is 15/42 ($\simeq$35\%) (see Table~A.2 in Appendix A).
This fraction is high. If we consider this with respect to the number of sources with blue-shifted lines it corresponds to 15/17 ($\simeq$90\%).
Therefore, the majority of the sources with detected lines show the presence of UFOs. 
If we further consider the lines with mildly relativistic velocities v$\geq$0.1c, these fractions are instead 11/42 ($\simeq$25\%) and 11/17 ($\simeq$65\%).

   \begin{figure}[!t]
   \centering
    \includegraphics[width=6.5cm,height=8cm,angle=270]{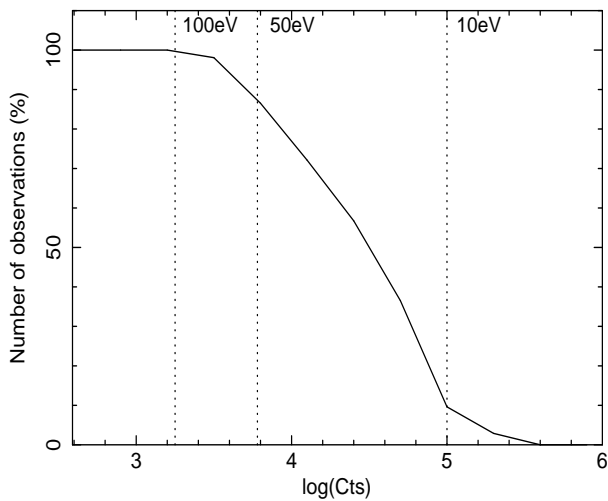}
   \caption{Percentage of XMM-Newton EPIC pn observations of the whole sample with total 4--10~keV counts greater than a fixed value. The vertical lines indicate the counts level over which it would be possible to detect a narrow emission/absorption line at 7~keV at the 3$\sigma$ level for EWs of 100~eV, 50~eV and 10~eV, respectively. It should be noted that the statistics would allow to detect the presence of such line in the 100\%, 80-90\% and only in the 10\% of the observations, respectively.}
    \end{figure}

However, we warn that the measured blue-shifted velocities are only lower limits because they depend on the unknown inclination angle of the outflow with respect to the line of sight (e.g. Elvis 2000). 
Therefore, the obtained fractions are only conservative estimates and the number of objects showing UFOs can be even higher, potentially extending to all the sources with detected Fe K absorption lines (i.e. 19/42).

Moreover, it is important to note that these fractions are also only conservative estimates, because they do not take into account the number of observations that have low counts levels and therefore do not have enough statistics for the detection of lines even if they were present. 

The total 4--10~keV counts level of each observation can be regarded as an indication of their statistics.
The percentage of observations of the whole sample with total 4--10~keV counts greater than a fixed value is shown in Fig.~6. 
To estimate the effect of the different statistics in the XMM-Newton observations on the detectability of the absorption lines we calculated the counts levels needed for a 3$\sigma$ detection of a narrow emission/absorption feature at the indicative energy of 7~keV (we assumed a power-law continuum with $\Gamma=2$) for different EWs. 
As it can be seen from Fig.~6, 100\% of the observations have a 4--10~keV counts level greater than $\simeq$$10^{3}$, that would allow the detection of a line of EW$=$100~eV if present. For lower EWs, however, the counts levels needed for the detection consequently increase. For instance, for a line of EW$=$50~eV the limit is at $\simeq$$6\times 10^3$ counts, which implies that it could be detected only in the $\sim$80--90\% of the observations. 
Finally, only $\sim$10\% of the sources have total 4--10~keV counts greater than $\sim$$10^5$, that would allow the detection of a line with EW$=$10~eV at 7~keV. 
If instead we consider lines at energies greater than 7~keV, the counts needed increase further because the intrinsic power-law spectrum has less photons at those energies and also the effective area of the instrument drops. 
For instance, the number of 4--10~keV counts needed for a 3$\sigma$ detection of a narrow absorption line of EW$=$100~eV (10~eV) are $\sim$$2\times 10^3$ ($\sim$$10^5$) at 7~keV and $\sim$$5\times 10^3$ ($\sim$$4\times 10^5$) at 9~keV. 
This means that from a statistical point of view we would be able to detect the presence of narrow absorption features with EW$=$100~eV (10~eV) in $\sim$100\% ($\sim$10\%) of the observations at 7~keV and in $\sim$90\% ($\leq1$\%) of the sample observations at 9~keV.
  
This clearly indicates that there is a bias against the detection of narrow absorption lines at higher energies/velocities with respect to those at lower energies/velocities.
However, we can derive a rough estimate of the global effect of the statistics available in the spectra of the whole sample considering that for a mean EW of $\sim$50~eV and a mean line energy of $\sim$8~keV, that corresponds to a blue-shifted velocity of about 0.1c for Fe XXVI Ly$\alpha$, the 4--10~keV counts level needed for a 3$\sigma$ detection is $\sim$$10^4$ counts. From Fig.~6 we can note that only about 80\% of the available XMM-Newton observations have enough counts and instead for about 20\% of them there is not enough statistics for a proper detection of a narrow absorption line if present. Hence, we can estimate that the total fraction of sources with blue-shifted absorption features (40\%) could actually be larger, up to a maximum of 60\%.

Finally, if we consider only the sources classified as Type 1, the ratio of objects with detected absorption lines is 16/35 ($\simeq$46\%). The ratio of sources having blue-shifted lines is 14/35 ($\simeq$40\%). 
If we consider those with velocities v$\geq$$10^4$~km/s, the number is 12. Therefore, the fraction of Type 1 sources with detected UFOs is 12/35 ($\simeq$35\%). If we limit to those with mildly relativistic velocities (v$\ge$0.1c) the fraction reduces to 8/35 ($\simeq$23\%). The remaining fraction of objects with line velocities consistent with zero (when negative they are probably indicating a blending with lower ionization Fe species, see \S4.1) is 2/35 ($\simeq$6\%).  
Instead, only absorption lines with blue-shifted velocities larger than $10^4$~km/s (actually, larger than 0.1c) have been detected in Type 2 sources.
Therefore, the fraction of UFOs in this class of objects is of 3/7 ($\simeq$43\%), similar to the Type 1s.

\subsection{Variability of the Fe K absorption lines}

Variability of the Fe K absorption lines detected in the X-ray spectra of some radio-quiet AGNs have been reported by several authors, even on time-scales shorter than $\sim$100~ks (e.g., Dadina et al.~2005; Risaliti et al.~2005; Braito et al.~2007; Reeves et al.~2008; Cappi et al.~2009 and references therein).

In Table~A.2 (in Appendix A) we can note that there are indeed some sources with multiple XMM-Newton observations in which Fe K absorption lines have been detected. Therefore, here we derive a rough estimate of the variability time-scales of the EWs and velocity shifts of the lines among the different observations.

A proper analysis of the variability patterns of these lines would require a detailed case by case study and a comparison with other source characteristics, such as flux history, cross-correlations among light curves, time-resolved and broader spectral analysis and so on, which are beyond the scope of the present work (e.g., Dadina et al.~2005; Braito et al.~2007 and references therein). 

To derive the variability time-scales for EW or velocity shift we must follow two different methods.
One can first consider one observation with a detected Fe K absorption line and look at the closest one with another line detection, checking if the two EWs are different at $\ge$90\% level. Otherwise, if there are no line detections in the other observations, we added a Gaussian line with energy and width fixed to the best-fit values (see Table~A.2) and derived the 90\% lower limits on the EW, letting the line normalization free to be positive or negative.

In the latter case, the variability in velocity, there is the need to have at least two different observations with detected absorption lines. Therefore, when possible, we checked whether the measured energy shifts were not consistent at $\ge$90\% level (see Table~A.2). We compared only lines with the same identification.

We estimated upper limits on the variability time-scales as the spacing between the two closest XMM-Newton observations in which the EWs or energy shifts of the lines were inconsistent at $\ge$90\% level. However, these are only conservative estimates because they depend on the arbitrary time spacing among the XMM-Newton observations.

herefore, we detected significant variability of the Fe K absorption lines in these cases: NGC~3516 in EW on $\Delta t \la 5$ yrs (e.g., Turner et al.~2008); Mrk~509 in velocity shifts $\Delta t \la 6$ months (e.g., Cappi et al.~2009); Mrk~79 in EW on $\Delta t \la 1$ month; NGC~4051 in both EW and velocity on $\Delta t \la 1.5$ yrs; Mrk~766 in EW on $\Delta t \la 2$ days ($\la$ 150--200~ks) and in velocity on $\la$ 4 days ($\la$ 300--400~ks); Mrk~841 in EW on $\la$ 6 months; PG~1211$+$143 in EW $\Delta t \la 3$ yrs (e.g., Reeves et al.~2008); MCG-5-23-16 in EW on $\Delta t \la 4$ yrs.

All the lines that have been found to be significantly variable are at rest-frame energies E$\ge$7.1~keV. The associated blue-shifted velocities are typically larger than $10^4$ km/s, indicating a classification as UFOs (see \S1). 

The only cases in which there is clear evidence for constancy of the Fe K absorption lines among different observations are for the two bright sources NGC~3783 and NCG~3516 (see Table~A.2 in Appendix A). In particular, for NGC~3783 the lines are constant over a period of $\sim 1$ year. Instead, for NGC~3516, the line parameters are consistent among four observations performed over a short interval of $\sim 7$ days and are variable with respect to an older observation performed $\sim$ 5 years before (see Table~A.2). In both cases the most intense lines have been detected in the interval E$\simeq$6.4--7.1~keV, with low or even slightly negative blue-shifts, and therefore can not be classified as UFOs.

As already noted by several authors (e.g, Dadina et al.~2005; Braito et al.~2007; Cappi et al.~2009 and references therein), the variability patterns of the highly blue-shifted absorption lines classified as UFOs suggest a transient nature for these features, which can appear and disappear in time in the same source. This might then be linked to the intrinsic physical phenomena that generate them, such as the sporadic ejection of blobs of material with different densities and velocities from the central regions of AGNs, which are observable in absorption only when crossing the line of sight (e.g., Dadina et al.~2005; Cappi et al.~2009 and references therein).
However, in the cases in which we did not detect significant line variability, we can not clearly distinguish from the fact that the lines were actually not present in that observation or their EWs were intrinsically less intense and could not be measured properly because the S/N was lower.

\subsection{Average parameters of the Fe K absorption lines}

The topic of this work is to assess the statistical significance and incidence the Fe K absorption lines and derive their general mean parameters on the whole sample of radio-quiet AGNs. 
Therefore, we then estimated also the average EWs and blue-shifted velocities of the lines.
We did not consider the line detections for each XMM-Newton observation separately because we would have introduced a bias. In fact, there are different numbers of observations for each source and consequently there are sources with more line detections (e.g. NGC~3783 and NGC~3516). In this way, the sources with more observations would have had a higher weight.
Thus, for each source we calculated the mean EW and blue-shifted velocity of the lines with the same identification among all the relative XMM-Newton observations.

However, as already discussed before, there are several cases with evidence for variability, especially for the lines at E$>$7.1~keV, and therefore care must be taken when averaging the data from multiple observations.
For instance, only the reported detections are averaged here. The non-detections might be due to a real non-existence of the lines or to an intrinsically lower EW. Hence, in the latter case, if they are included, they would systematically decrease the mean values. 
However, the variability patterns discussed earlier seem to prefer the former case, in which transient lines can appear and disappear in time. In this scenario, we can treat each line detection as a different realization of the phenomenon and the averaging process would not be biased.

   \begin{figure}
   \centering
    \includegraphics[width=6.5cm,height=8cm,angle=270]{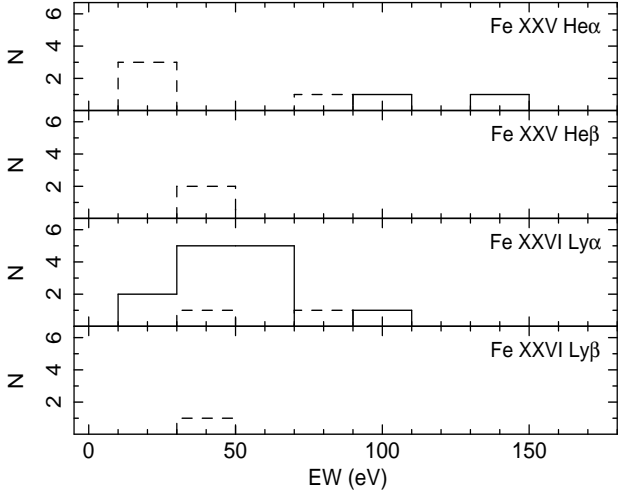}
   \caption{Histograms representing the mean EW distribution of each Fe K absorption line type with respect to the number of sources in which the lines have been detected. The plot has been divided in four panels according to the line classification in Fe XXV, He$\alpha$ and He$\beta$, and Fe XXVI, Ly$\alpha$ and Ly$\beta$, respectively. The solid lines refer to the absorption lines with blue-shifted velocities larger than $10^4$~km/s, here classified as UFOs (see \S1), and the dotted lines to the others. The mean values have been taken from Table~A.3.}
    \end{figure}

The list of sources with their mean EWs and blue-shifted velocities of the Fe K absorption lines among different XMM-Newton observations are reported in Table~A.3 (in Appendix A). In the cases with more line detections in the same source we reported also the lowest and highest values of the parameters. This has been divided according to the line identification as Fe XXV, He$\alpha$ and He$\beta$, and Fe XXVI, Ly$\alpha$ and Ly$\beta$, respectively. 

The distribution of mean EWs of the lines with respect to the number of sources in which they have been detected is shown in the histograms of Fig.~7.
We used the positive sign for the absorption lines EWs. 
It can be seen that the majority of the lines have associated blue-shifted velocities larger than $10^4$~km/s (solid lines in Fig.~7), suggesting the identification as UFOs (see \S1). 
In particular, the distribution in EW is dominated by Fe XXVI Ly$\alpha$ lines, with values ranging from $\sim10$~eV up to $\sim100$~eV and a peak and mean value around $\sim$50~eV.

Instead, the distribution of mean blue-shifted velocities of the Fe K absorption lines for each source is shown in the histogram in Fig.~8. As it can be seen, the distribution spans from about zero up to $\sim$0.3c, with a strong tendency toward high velocities.
In fact, there is a clear peak at v$\simeq$0.1c, which also coincides with the mean value (in this calculation we did not consider negative values). 
The majority of the outflow velocities are higher than v$=$$10^4$~km/s (marked with a red vertical dashed line in Fig.~8), which indicates when the absorbers can be associated with UFOs (see \S1).
However, we warn again that these velocities are actually only lower limits, because they depend on the unknown inclination angle of the outflows with respect to the line of sight (e.g. Elvis 2000). Therefore, their intrinsic values could be even higher and potentially all the Fe K absorbers detected in the sample could be identified with UFOs.

Finally, the fraction of sources with blue-shifted velocities of the Fe K absorption lines greater than a fixed value with respect to the total number of sample sources is reported in Fig.~9. We can note that these fractions are consistent with those already given in \S4.2.

   \begin{figure}[t]
   \centering
    \includegraphics[width=6.5cm,height=8cm,angle=270]{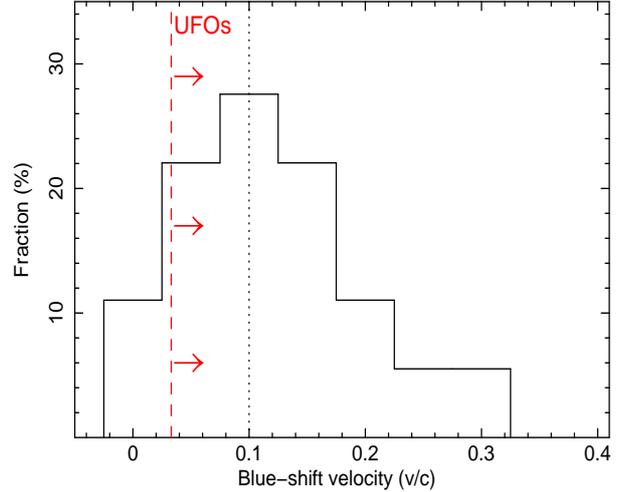}
   \caption{Histogram representing the distribution of mean blue-shifted velocities for each source with detected Fe K absorption lines with respect to the total number of sources. The mean velocities have been taken from Table~A.3 (column 10). The vertical dashed line (in red) indicates the threshold velocity value of v$=$$10^4$~km/s ($\simeq$0.033c) over which the absorbers can be associated with UFOs. The vertical dotted line (in black) indicates the peak and mean value of the velocity distribution at v$\simeq$0.1c.}
    \end{figure}

\subsection{On the evidence for absorption lines at E$<$6.4~keV}

As already introduced in \S3.2, there are 10 spectra (over a total of 101) which show possible narrow absorption lines at energies lower than 6.4~keV with F-test confidence levels $\sim$99\% in their energy-intensity contour plots (see figures in Appendix C), i.e., NGC~4151 (obs. 0112830201 and 0402660201), NGC~3783 (obs. 0112210201), NGC~3516 (obs. 0107460701), MCG-6-30-15 (obs. 0029740101, 0029740701 and 0029740801), Mrk~335 (obs. 0510010701), ESO~198-G024 (obs. 0305370101) and NGC~7582 (0112310201).  We did not include these lines in Table~A.2 and we did not further investigate their detection significance with extensive MC simulations (see \S3.3) because in this work we focused only on the study of absorption lines at energies greater than 6.4~keV. 
However, it is important to discuss here at least some possible explanations for their presence.

The rest-frame energies of the lines observed at E$<$6.4~keV are typically in the range E$\sim$4--5~keV, where no intense transitions from cosmically abundant elements are expected.
It is possible that at least part of them can be associated with Doppler or gravitationally red-shifted Fe XXV He$\alpha$ or Fe XXVI Ly$\alpha$ lines.
In fact, there have been several papers in the literature reporting the detection of such red-shifted Fe K absorption lines in the X-ray spectra of Seyfert galaxies or quasars (e.g., Nandra et al.~1999; Dadina et al.~2005; Reeves et al.~2005; Yaqoob \& Serlemitsos 2005; Longinotti et al.~2007a), with velocities in the range $\sim$0.1--0.4c. In our case this identification would imply substantial red-shifted velocities of $\sim$0.4--0.7c.

If the ionization parameter of the absorbing material is not extreme or if there are different components with intermediate ionization states (log$\xi$$\sim$1--3~erg~s$^{-1}$~cm), some of the lines could be alternatively identified with blue-shifted K-shell absorption lines from ionized elements lighter than iron, such as Si, S, Ar, Ca. 
The most intense 1s--2p transitions of their He/H-like ions are expected in the energy range E$\simeq$2--4~keV, in particular: Si XIII at E$\simeq$1.87~keV, Si XIV at E$\simeq$2~keV, S XV at E$\simeq$2.46~keV, S XVI at E$\simeq$2.62~keV, Ar XVII at E$\simeq$3.14~keV, Ar XVIII at E$\simeq$3.32~keV, Ca XIX at E$\simeq$3.9~keV and Ca XX at E$\simeq$4.11~keV (from Verner et al.~1996).
Therefore, the possible blue-shifted velocities would be in the range $\sim$0.1--0.5c.

   \begin{figure}[t]
   \centering
    \includegraphics[width=6.5cm,height=8cm,angle=270]{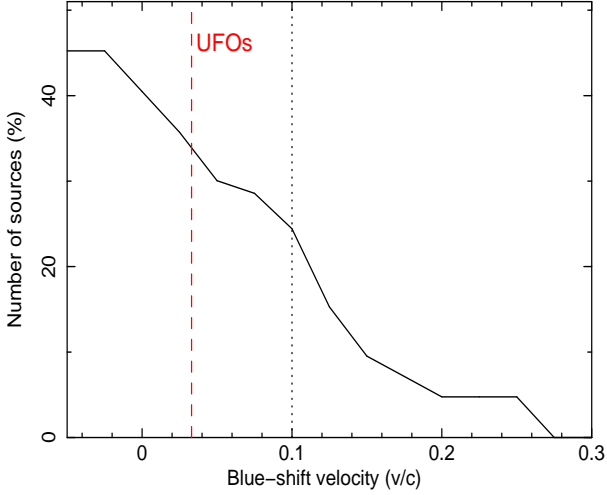}
   \caption{Fraction of sources with mean blue-shifted velocity of the Fe K absorption lines greater than a fixed value with respect to the total number of sources in the sample. The mean velocities have been taken from Table~A.3 (column 10). The vertical dashed line (in red) indicates the threshold velocity value of v$=$$10^4$~km/s ($\simeq$0.033c) over which the absorbers can be associated with UFOs. The dotted line (in black) refers the mean value of the velocity distribution at v$\simeq$0.1c.
}
   \end{figure}

A related intriguing possibility is to check whether these spectral features are consistent with absorption lines from ionized low-Z elements with the same blue-shift inferred from the Fe K lines.
In fact, if the ionization state of the material is not extremely high, log$\xi$$\sim$3~erg~s$^{-1}$~cm, there is the possibility to find K-shell absorption lines from He/H-like ions of Si, S, Ar or Ca, in conjunction with Fe.
This finding has already been reported by Pounds \& Page (2006) in one XMM-Newton observation of PG~1211+143 (obs. 0112610101).
Combining the EPIC pn based results with the analysis of the lower energy (E$\la$3~keV, below our low energy boundary) MOS and RGS data, the authors have been able to detect several K-shell absorption lines from highly ionized Ne, Mg, Si, S, Ar and Fe, all consistently blue-shifted with the same mildly-relativistic velocity of $\simeq$0.13c.
  
There are three observations which simultaneously show a narrow absorption line at E$>$7~keV and another in the interval E$\simeq$4--5~keV with F-test confidence level $\sim$99\% in their contour plots in Appendix C, i.e., two observations of NGC~4151 (obs. 0112830201 and 0402660201, see Fig.~C.1) and one of NGC~7582 (obs. 0112310201, see lower panel of Fig.~C.7).
For the former, we found that the absorption lines at energies lower than 5~keV are consistent with being Ca XIX He$\alpha$ (1s$^2$--1s2p; at E$\simeq$3.9~keV) and Ca XX Ly$\alpha$ (1s--2p; at E$\simeq$4.11~keV), blue-shifted with the same velocity ($\simeq$0.1c) of the higher energy Fe XXVI Ly$\alpha$ lines. However, we did not include the detection of an absorption line at E$\simeq$7.8~keV in the first observation of NGC~4151 in Table~A.2 because the resultant Monte Carlo confidence level was lower than the threshold value of 95\%.

The latter case, NGC~7582, is even more interesting. There we find the presence of three narrow absorption lines between 4 and 5.3~keV, one with an F-test confidence level $\ga$99\% and the other two $\simeq$95\%.
If the absorption line at E$\simeq$9~keV is identified with Fe XXV He$\alpha$ (as reported in Table~A.2), these lines are consistent with being absorption from Ar XVII He$\alpha$ (1s$^2$--1s2p; at E$\simeq$3.14~keV), Ar XVIII Ly$\alpha$ (1s--2p; at E$\simeq$3.3~keV) and Ca XIX He$\alpha$ (1s$^2$--1s2p; at E$\simeq$3.9~keV) blue-shifted with the same common velocity of $\simeq$0.255c (see Table~A.2). 
The probability for random fluctuations to give rise to this series of lines with the exact energy spacing and common blue-shift is very low, $\simeq$$2\times 10^{-6}$. 
This is an important result, which contributes to strengthen our conclusions on the veracity of the blue-shifted Fe K absorption lines.
We refer the reader to Appendix B for more detailed information regarding the spectral fitting and possible identification of these low energy absorption lines for each source. A detailed photo-ionization modeling of these absorbers is presented in a companion paper (Tombesi et al. in prep.).

\section{Discussion}

Despite an increasing number of works showing evidence for the presence of blue-shifted Fe K absorption lines in the X-ray spectra of radio-quiet AGNs (e.g., Chartas et al.~2002; Chartas et al.~2003; Pounds et al.~2003a; Dadina et al.~2005; Markowitz et al.~2006; Braito et al.~2007; Cappi et al.~2009; Reeves et al.~2009), there is still much debate on their physical interpretation and even on their real statistical significance.

For instance, it has been argued by Vaughan \& Uttley (2008) that several of the published detections of narrow red/blue-shifted emission and absorption lines in the Fe K band of AGNs could actually be falsified by the presence of a publication bias. 
In fact, only the observations with detected features have been reported in the literature and we actually do not know the fraction of detections/non-detections on the full population. Therefore, it is rather difficult to estimate the global significance of any individual case. 
Moreover, the fact that the significance of some of the published lines can be weak could suggest that some of them are simply the most significant from a distribution of random fluctuations. 

The authors stated that the presence of this bias can be shown by simply plotting the EWs of the lines with respect their relative 90\% errors (see their Fig.~1). They restricted to the published red/blue-shifted lines with velocities $\ge$0.05c.
There seems to be a tendency for the data points to lay close and parallel to the border line between detection and non-detection. In other words, the EWs and their associated errors seem to follow a linear relation. This means that lines with higher EWs have consequently higher error bars. Moreover, the fact that none of the data display in the top left corner of the diagram, which would indicate that more intense lines have smaller error bars, could suggest that the lines with higher EWs are preferentially detected in low quality spectra. 

Vaughan \& Uttley (2008) claim that these properties are actually indicative of a publication bias and several, if not all, of the reported red/blue-shifted lines could actually be false detections.
As also recognized by Vaughan \& Uttley (2008), the only way out to overcome this bias would be to perform a uniform and systematic analysis on a complete sample of sources and directly report the ratio of line detections over non-detections.
This is actually what has been done independently in this work, as far as blue-shifted Fe K absorption lines are concerned.
To clearly define the statistical veracity of the blue-shifted lines detections one must carry out extensive Monte Carlo simulations. This should be done in a uniform and comprehensive way on a complete sample of sources, as shown here. Only in this way the line detections can be directly tested against random noise fluctuations and consequently they can be corrected for possible biases (see \S3.3).

We have performed such Monte Carlo tests for the blue-shifted lines at energies greater than 7.1~keV (corresponding to blue-shifted velocities $\ga$0.01c if identified as Fe XXVI Ly$\alpha$) and we placed a lower detection confidence level threshold of 95\%. 
Consequently, we have been able also to estimate the global probability for these features to be generated by random fluctuations. It is very low, less than $3\times 10^{-8}$ (see \S3.3). 
Therefore, even if some single line detection can be statistically poor, the global probability (note that this is a conservative estimate) for these blue-shifted features to be generated by random fluctuations is very low.
Last but not least, we have been able to independently confirm the detection of several features with the MOS cameras. Also in this case their global random probability is very low, less than $10^{-7}$ (see \S3.6). The consistency between the pn and MOS results places an additional very strong point in favour of the veracity of the lines, without relying on any statistical method.

   \begin{figure}[!t]
   \centering
    \includegraphics[width=6.5cm,height=8cm,angle=270]{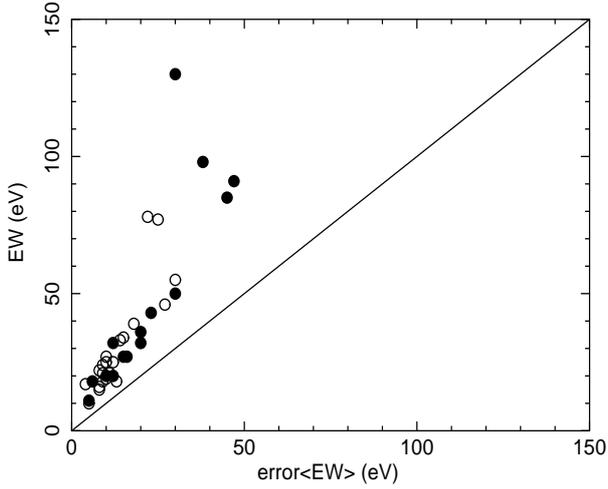}
   \caption{Plot of the EW of the blue-shifted absorption lines detected in the sample with respect to their 90\% errors. The open circles refer to those with blue-shifted velocities v$<$0.05c and the filled circles instead to those with v$\ge$0.05c. The diagonal line indicates the border region in which EW$=$error$<$EW$>$. The plot axes are in linear scale.}
    \end{figure}

For a comprehensive comparison with the Vaughan \& Uttley (2008) work, we reproduced their EW--error$<$EW$>$ plot with the lines detected in our sample. 
This is shown in Fig.~10. We used the positive sign for the absorption lines EWs. 
As it can be seen, some of the lines with more intense EWs indeed populate the upper left part of the graph and the global trend seems to diverge from the 90\% detection border line. 
Moreover, lines with blue-shifted velocities both lower or greater than $0.05$c do follow the same distribution. 
It should be noted that the diagram in Fig.~10 has been plotted with the axes in linear scale.
Using logarithmic axes and a wider scale (as in Vaughan \& Uttley 2008) would tend to visually compress more the data points to the detection border line.
 
We also performed a further sanity check by plotting together the EWs of the blue-shifted absorption lines together with the EWs of the neutral Fe K$\alpha$ emission lines at $\sim$6.4~keV detected in the sample, for which there is no doubt on the veracity. 
This is shown in Fig.~11. For a direct comparison of the EWs we used the positive sign for both emission and absorption lines. Interestingly enough, both the narrow absorption lines and the Fe K$\alpha$ emission lines do follow the same trend. 
They actually seem to be drawn from the same distribution, with the narrow absorption lines having only systematically lower EWs. 
This difference is of course due to a distinct physical origin.
Therefore, the data trend does seem to be rather general. 
The effect of closeness of the data points to the ``detection border line'' is again visually magnified by plotting the data in a logarithmic scale.
We therefore argue that the fact that none of the lines populate the upper left corner of the diagram does not directly tell us that the lines are fake, but that the capabilities of the X-ray instruments to detect spectral lines are intrinsically limited. 
In fact, to populate the upper left corner of the graph it would require, as an example, for a line of EW$\simeq$1~keV to have an error of less than 10~eV (at 90\%), therefore indicating a detection confidence level at about $200\sigma$! A measure with such accuracy has never been possible regardless of the X-ray instrument flown.  

The EWs of the blue-shifted lines (with 90\% errors) with respect to the 4--10~keV counts levels of the associated XMM-Newton EPIC pn observations of our sample is plotted in Fig.~12. We used the positive sign for the absorption line EWs. 
As it can be seen, there is a slight trend for the EWs of the lines and the associated error bars to increase for lower counts levels. 
However, this trend does not seem to be dependent on the intrinsic line detection significance, as demonstrated by the fact that it is followed by the lines detected at both 95--99\% and $\ge99$\% confidence levels (see Fig.~12).

   \begin{figure}[!t]
   \centering
    \includegraphics[width=6.5cm,height=8cm,angle=270]{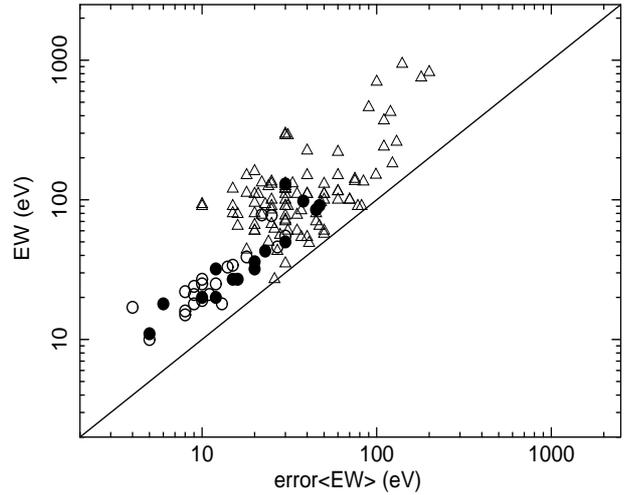}
   \caption{EWs with respect to the 90\% errors for the narrow blue-shifted absorption features (open circles v$<0.05$c and filled circles v$\ge0.05$c, respectively) and the neutral Fe K$\alpha$ emission lines (open triangles) detected in the sample. The plot axes are in logarithmic scale.}
    \end{figure}

We performed a simple additional test making use of some spectral simulations. We assumed to have a narrow ($\sigma=10$~eV) absorption feature at 8~keV with fixed intensity of $-5\times 10^{-6}$~ph~s$^{-1}$~cm$^{-2}$ and a simple power-law continuum with $\Gamma=2$ and 4--10~keV flux of $10^{-11}$~erg~s$^{-1}$~cm$^{-2}$. The resultant EW of the line is $\simeq$30~eV. 
Then, we simulated different EPIC pn spectra, each time changing only the exposure time in order to have different total 4--10~keV counts. The resulting spectra have been fitted with the same model and the relative EWs and errors have been determined.
The simulated EWs (with 90\% errors) with respect to the different 4--10~keV count levels are plotted in Fig.~11. 
We note that the trend of the EW values to increase with decreasing total counts is reproduced by the simulations. However, it is followed by an increase in the error bars as well. Therefore, the EW estimates are still consistent with the assumed constant value of $\simeq$30~eV (the small scattering is due to the randomization process in spectra simulation). 

The increase in the estimated EW values followed by corresponding bigger error bars seems to be due to the fact that for decreasing counts levels (decreasing S/N) there is an expansion of the error contours and therefore the location of the true minimum of the $\chi^2$ distribution is less constrained. 
For instance, if the EW best fit value is chosen as the average between the error bar limits, it is then expected to increase following the expansion of the available parameter space when the statistics is lower.
This has nothing to do with random fluctuations.
This can explain the trend followed by the data points in Fig.~10 and Fig.~11, that is, higher EWs have systematically higher associated error bars. 
Therefore, the trend does not state that the lines are fake (as we demonstrated including the line in the spectral simulations) but that we are approaching the detection limit of the instrument. 
It only indicates that the accuracy with which we are able to measure a certain parameter is worse when the available statistics is lower.

   \begin{figure}[!t]
   \centering
    \includegraphics[width=6.5cm,height=8cm,angle=270]{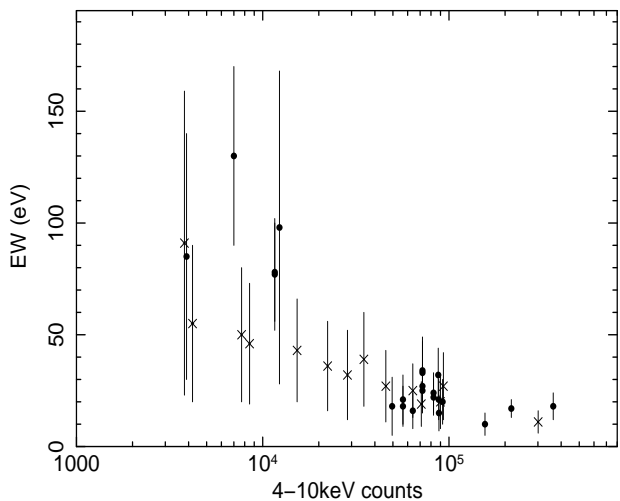}
   \caption{EWs of the Fe K absorption lines detected in the sample with respect to the 4--10~keV counts of the relative EPIC pn observation. The crosses refer to those with detection probabilities in the 95--99\% range and the filled circles instead to those with $\ge$99\%. The errors are at the 90\% level.}
    \end{figure}

We have derived the blue-shifted velocities of the narrow absorption lines (see Table~A.2) assuming that they were due to H/He-like Fe K-shell absorption intrinsic to the sources.
However, it has been debated that some, or all, of these features could instead be indicative of absorption from local or intermediate red-shift material ($z$$\simeq$0), due to the fact that some of them have blue-shifted velocities comparable to the sources cosmological red-shifts (McKernan et al.~2004; McKernan et al.~2005). 
To test such eventuality, we also plot the cosmological red-shifts of the sources ($z$) with respect to the blue-shifted velocities of the absorption lines (in units of $c$) from our sample. 

As it can be seen from Fig.~14, apart from a few data points (relative to NGC~3516, Mrk~841 and ESO~323-G77), the values do not match the simple linear relation expected if the lines were due to local absorption. Furthermore, the blue-shifted velocities are systematically higher compared to the cosmological red-shifts of the sources. 
This clearly demonstrates that the detected blue-shifted absorption lines in the Fe K band are indeed due to genuine absorption intrinsic to the sources.
 
The same conclusion using velocity information have been reached by Reeves et al.~(2008), comparing some observations recently reported in the literature.
Moreover, Reeves et al.~(2008) also suggested further physical arguments supporting this thesis. 
These narrow blue-shifted absorption features have been found to be variable on time-scales ranging from $\sim$years to less than $\sim$100~ks (e.g., Risaliti et al.~2005; Braito et al.~2007; Reeves et al.~2008; Cappi et al.~2009 and references therein) therefore implying sub-pc scale absorbers. The variability of the lines even on time-scales as short has a few days among different XMM-Newton observations is also confirmed by our analysis (see \S4.3).
These somewhat compact absorbers are at odds with the expected kpc scale extension of diffuse Galactic halos or Warm-Hot Intergalactic Medium (WHIM).    
Moreover, the local hot gas is expected to be collisionally ionized, instead of being photo-ionized by the AGN continuum. Therefore, the temperature required to have a substantial He/H-like iron population would be very high, greater than T$\sim$$10^{7}$--$10^{8}$~K.
Finally, the huge column densities of gas ($N_H$$\ga$$10^{23}$--$10^{24}$~cm$^{-2}$) required to reproduce the observed features are too high to be associated with any reasonable hot diffuse local gas ($N_H$$\la$$10^{21}$~cm$^{-2}$). 
Therefore, we are confident enough that the Fe K absorption lines detected at E$>$6.4~keV in our sample are indeed produced by absorption intrinsic to the sources.

From the fraction of sources showing Fe K absorption lines with blue-shifted velocities higher than $10^4$~km/s, $f$$\sim$0.4--0.6 (see \S4.2), we can derive a rough estimate of the global covering fraction of the UFOs, averaged over all lines of sight: $C=\Omega /4\pi \simeq f$, where $\Omega$ is the solid angle (Crenshaw et al.~1999).
Thus, as an ensemble, the UFOs might cover even about one-half of the sky as seen by the central continuum source.
This provides an important geometric constraint indicating that the distribution of the absorbing material cannot be very collimated, thereby implying large opening angles.
Overall this is consistent with what derived for the UV absorbers and classical X-ray warm absorbers detected in Seyfert 1 galaxies (George et al.~1998; Crenshaw et al.~1999). 

   \begin{figure}[!t]
   \centering
    \includegraphics[width=6.5cm,height=8cm,angle=270]{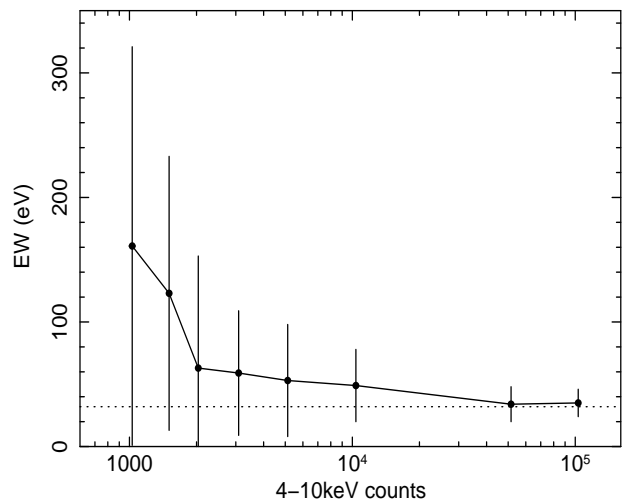}
   \caption{Simulated trend of the EWs of the narrow absorption lines with respect to different 4--10~keV counts levels. The horizontal line refer to the simulated line with EW$\simeq$30~eV (see \S5 for details). The error bars are at the 90\% level.}
    \end{figure}

\section{Conclusions}

We performed a search for narrow blue-shifted absorption lines in the Fe K band of a complete sample of 42 radio-quiet AGNs observed with XMM-Newton.
We detect 36 absorption lines in a total of 101 EPIC pn observations. 
Instead, the number of narrow absorption lines detected at rest-frame energies greater than 7.1~keV is 22. 
The detailed confidence levels of these features have been determined using the F-test and extensive Monte Carlo simulations.
Their global probability to be generated by random fluctuations is very low ($<$$3 \times 10^{-8}$).
Furthermore, the detection of several of these lines has been independently confirmed using the MOS cameras on board XMM-Newton and their random probability is also very low ($<$$10^{-7}$).
These results allow to overcome the publication bias that has been claimed for the blue-shifted Fe K absorption lines and to clearly assess their global veracity.

We assumed the most likely identification of the lines as due to Fe XXV/XXVI K-shell resonant absorption.
We did not find any significant correlation between the blue-shifted velocity of the lines and the cosmological red-shifts of the sources. This evidence, strengthened also by other physical arguments and the detection of short time-scale variability, rule out any possible interpretation of these features in term of absorption by local ($z\simeq0$) diffuse material or WHIM.
Therefore, all the detected lines can be genuinely attributed to Fe K-shell absorption intrinsic to the sources.

The majority of the lines are blue-shifted with respect to the expected energies of the relative atomic transitions.
The distribution of blue-shifted velocities ranges from about zero up to $\sim$0.3c, with a peak and mean value at $\sim$0.1c.
In particular, the fraction of objects of the sample with at least one Fe K absorption line with blue-shifted velocity higher than $10^4$~km/s, identified here with UFOs, is at least 35\%. This value is similar for Type 1 and Type 2 sources. 
Therefore, we conclude that UFOs are a rather common phenomenon observable in radio-quiet AGN and they can be the direct signature of AGN accretion disk winds/ejecta.
 The global covering fraction of these absorbers is consequently estimated to be in the range $C$$\sim$0.4--0.6, implying large opening angles.
The detailed modeling and physical interpretation of the detected Fe K absorption lines through the Xstar photo-ionization code and a curve of growth analysis is reported in a companion paper (Tombesi et al. in prep.).

   \begin{figure}[!t]
   \centering
    \includegraphics[width=6.5cm,height=8cm,angle=270]{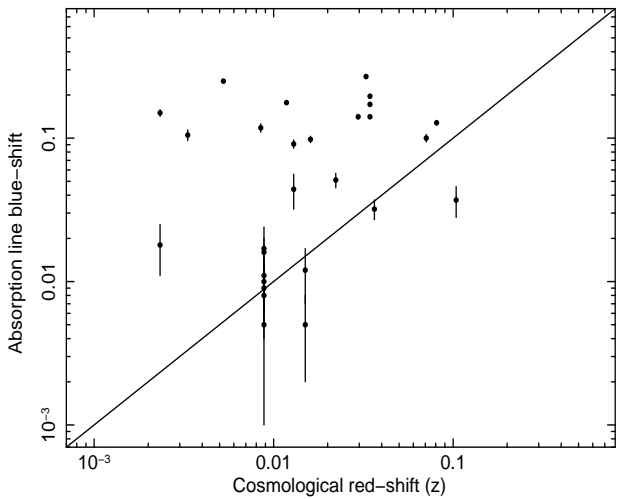}
   \caption{Cosmological red-shifts of the sample sources with respect to the blue-shifted velocities (in units of c) of the narrow Fe K absorption lines detected in their XMM-Newton observations. Errors are at the 90\% level. Only positive blue-shifted velocities have been reported.}
    \end{figure}

These results indicate that UFOs might likely have a strong impact on the overall AGN geometry/energetic and consequently also on the overall X-ray spectrum. 
In fact, several authors have demonstrated that accretion disk winds/outflows might well imprint also other spectral signatures in the X-ray spectra of AGNs besides blue-shifted Fe K absorption lines (e.g., Pounds \& Reeves 2009 and references therein).
For instance, Sim et al.~(2008) and Sim et al.~(2010) have been able to successfully reproduce the 2--10~keV spectra of two bright sources of our sample in which the detection of strong blue-shifted Fe K absorption lines in their XMM-Newton spectra were already reported, namely Mrk~766 (from Miller et al.~2007) and PG~1211+143 (from Pounds et al.~2003a), using their AGN accretion disk wind model. Notably, the authors have been able to account for both emission and absorption features in a physically self-consistent way. 
Therefore, in the near future we expect to obtain promising results in this field from the improvement of theoretical modeling of accretion disk winds/outflows in AGN.

Finally, we note that present X-ray telescopes (such as XMM-Newton, Chandra and Suzaku) are able to detect blue-shifted Fe K absorption lines in the spectra of AGNs with just a sufficient level of accuracy.
A substantial improvement is expected from the higher effective area and supreme energy resolution (down to $\sim$2--5~eV) in the Fe K band offered by the calorimeters on board the future Astro-H and IXO missions (e.g., Tombesi et al.~2009).

\begin{acknowledgements}

This paper is part of the PhD thesis in Astronomy of FT from the University of Bologna. We refer the reader to that document for more detailed information. This paper is based on observations obtained with the \emph{XMM-Newton} satellite, an ESA funded mission with contributions by ESA member states and USA. FT thank Prof. K. A. Pounds for useful discussion. MC acknowledge financial support from ASI under contract I/088/06/0. The authors thank the anonymous referee for suggestions that led to important improvements in the paper.
      
\end{acknowledgements}

\onecolumn{

\begin{appendix}

\section{List of tables}

\begin{small}

{

\begin{longtable}{l c c c c c c}

\caption{List of Type 1 and Type 2 sources with relative XMM-Newton observations.}\\

\hline\hline       
Source & $z$ & OBSID & Date & Counts & Expo & Flux\\ 
       &     &       &      & \scriptsize{($10^3$~cts)}& \scriptsize{(ks)} & \scriptsize{($10^{-12}$~erg~cm$^{-2}$~s$^{-1}$)}\\
\scriptsize{(1)} & \scriptsize{(2)} & \scriptsize{(3)} & \scriptsize{(4)} & \scriptsize{(5)} & \scriptsize{(6)} & \scriptsize{(7)}\\
\hline   
&\\ 
\endfirsthead   

\caption{Continued.}\\           
\hline\hline       
Source & $z$ & OBSID & Date & Counts & Expo & Flux\\ 
       &     &       &      & \scriptsize{($10^3$~cts)}& \scriptsize{(ks)} & \scriptsize{($10^{-12}$~erg~cm$^{-2}$~s$^{-1}$)}\\
\scriptsize{(1)} & \scriptsize{(2)} & \scriptsize{(3)} & \scriptsize{(4)} & \scriptsize{(5)} & \scriptsize{(6)} & \scriptsize{(7)}\\
\hline   
&\\ 
\endhead
\endfoot

& & & & & &\\[-16pt] 
\multicolumn{7}{c}{Type 1 sources}\\[1.2pt]
\hline
& & & & & &\\ 
NGC~4151 & 0.00332 & 0112310101 & 2000-12-21 & 51.1 & 21.0 & 37.9\\
 & & 0112830501 & 2000-12-22 & 43.4 & 17.6 & 38.5\\
 & & 0112830201 & 2000-12-22 & 124.0 & 50.9 & 38.7\\
 & & 0143500201 & 2003-05-26 & 139.0 & 12.7 & 170.0\\
 & & 0143500301 & 2003-05-27 & 176.0 & 13.0 & 209.0\\
 & & 0402660201 & 2006-11-29 & 87.5 & 19.6 & 70.7\\
IC4329A & 0.01605 & 0147440101 & 2003-08-06 & 301.0 & 71.3 & 61.9\\
NGC~3783 & 0.00973 & 0112210101 & 2000-12-28 & 63.9 & 26.0 & 35.8\\
 & & 0112210201 & 2001-12-17 & 156.0 & 80.0 & 28.9\\
 & & 0112210501 & 2001-12-19 & 216.0 & 84.0 & 38.1\\
MCG+8-11-11 & 0.02048 & 0201930201 & 2004-04-10 & 54.2 & 26.5 & 29.7\\
NGC~5548 & 0.01717 & 0109960101 & 2000-12-24 & 23.1 & 16.0 & 20.8\\
 & & 0089960301 & 2001-07-09 & 97.9 & 56.0 & 25.6\\
 & & 0089960401 & 2001-07-12 & 37.8 & 17.5 & 31.4\\
NGC~3516 & 0.00883 & 0107460701 & 2001-11-09 & 66.0 & 84.0 & 11.8\\
 & & 0401210401 & 2006-10-06 & 82.6 & 35.0 & 33.2\\
 & & 0401210501 & 2006-10-08 & 88.0 & 42.0 & 29.6\\
 & & 0401210601 & 2006-10-10 & 72.0 & 42.0 & 24.6\\
 & & 0401211001 & 2006-10-12 & 56.6 & 28.0 & 28.3\\
NGC~4593 & 0.00900 & 0059830101 & 2002-06-23 & 93.2 & 52.1 & 26.2\\
Mrk~509 & 0.03440 & 0130720101 & 2000-10-25 & 28.5 & 20.7 & 20.8\\
 & & 0130720201 & 2001-04-20 & 41.7 & 23.1 & 25.1\\
 & & 0306090201 & 2005-10-18 & 92.4 & 59.8 & 22.6\\
 & & 0306090301 & 2005-10-20 & 50.9 & 32.4 & 22.7\\
 & & 0306090401 & 2006-04-25 & 89.9 & 48.6 & 26.8\\
MCG-6-30-15 & 0.00775 & 0111570101 & 2000-07-11 & 34.6 & 28.8 & 16.7\\
 & & 0111570201 & 2000-07-11 & 57.6 & 35.1 & 22.8\\
 & & 0029740101 & 2001-07-31 & 85.4 & 47.6 & 25.2\\
 & & 0029740701 & 2001-08-02 & 166.0 & 83.6 & 27.8\\
 & & 0029740801 & 2001-08-04 & 149.0 & 83.9 & 24.9\\
Ark~120 & 0.03271 & 0147190101 & 2003-08-24 & 93.3 & 58.4 & 22.7\\
Mrk~110 & 0.03529 & 0201130501 & 2004-11-15 & 41.8 & 32.8 & 18.0\\
NGC~7469 & 0.01632 & 0112170101 & 2000-12-26 & 13.8 & 12.3 & 15.9\\
 & & 0112170301 & 2000-12-26 & 19.0 & 16.2 & 16.7\\
 & & 0207090101 & 2004-11-30 & 72.4 & 59.2 & 17.5\\
 & & 0207090201 & 2004-12-03 & 70.5 & 55.0 & 18.4\\
IRAS~5078+1626 & 0.01788 & 0502090501 & 2007-08-21 & 42.7 & 38.5 & 16.2\\
Mrk~279 & 0.03045 & 0302480401 & 2005-11-15 & 51.3 & 41.5 & 16.8\\
 & & 0302480501 & 2005-11-17 & 49.5 & 41.4 & 15.8\\
 & & 0302480601 & 2005-11-19 & 23.0 & 21.0 & 15.4\\
NGC~526A & 0.01910 & 0150940101 & 2003-06-21 & 36.9 & 35.5 & 15.5\\
NGC~3227 & 0.00386 & 0101040301 & 2000-11-28 & 14.3 & 30.4 & 7.1\\
 & & 0400270101 & 2006-12-03 & 142.0 & 92.4 & 23.2\\
NGC~7213 & 0.00584 & 0111810101 & 2001-05-28 & 28.2 & 29.6 & 13.6\\
ESO~511-G30 & 0.02239 & 0502090201 & 2007-08-05 & 65.0 & 76.2 & 12.2\\
Mrk~79 & 0.02219 & 0400070201 & 2006-09-30 & 15.3 & 14.4 & 14.5\\
 & & 0400070301 & 2006-11-01 & 12.6 & 14.0 & 13.0\\
 & & 0400070401 & 2007-03-19 & 11.5 & 14.0 & 11.9\\
NGC~4051 & 0.00233 & 0109141401 & 2001-05-16 & 71.0 & 67.9 & 14.2\\
 & & 0157560101 & 2002-11-22 & 12.3 & 41.3 & 4.3\\
Mrk~766 & 0.01293 & 0096020101 & 2000-05-20 & 16.2 & 24.8 & 8.9\\
 & & 0109141301 & 2001-05-20 & 75.6 & 76.8 & 13.5\\
 & & 0304030101 & 2005-05-23 & 19.3 & 54.5 & 5.1\\
 & & 0304030301 & 2005-05-25 & 34.9 & 68.9 & 6.9\\
 & & 0304030401 & 2005-05-27 & 39.3 & 65.6 & 8.2\\
 & & 0304030501 & 2005-05-29 & 45.8 & 63.5 & 10.1\\
 & & 0304030601 & 2005-05-31 & 37.4 & 59.1 & 8.6\\
 & & 0304030701 & 2005-06-03 & 8.3 & 15.5 & 7.3\\
Mrk~841 & 0.03642 & 0205340201 & 2005-01-16 & 14.7 & 29.9 & 7.1\\
 & & 0205340401 & 2005-07-17 & 8.5 & 15.8 & 7.6\\
Mrk~704 & 0.02923 & 0300240101 & 2005-10-21 & 8.1 & 14.9 & 7.4\\
Fairall~9 & 0.04702 & 0101040201 & 2000-07-05 & 13.3 & 25.8 & 7.4\\
ESO~323-G77 & 0.01501 & 0300240501 & 2006-02-07 & 11.6 & 22.5 & 7.3\\
1H419-577 & 0.10400 & 0148000201 & 2002-09-25 & 4.2 & 11.0 & 5.7\\
 & & 0148000401 & 2003-03-30 & 5.3 & 10.8 & 7.2\\
 & & 0148000601 & 2003-09-16 & 5.0 & 11.4 & 6.5\\
Mrk~335 & 0.02578 & 0101040101 & 2000-12-25 & 14.8 & 27.2 & 7.5\\
 & & 0306870101 & 2006-01-03 & 56.9 & 76.9 & 10.3\\
 & & 0510010701 & 2007-07-10 & 2.8 & 15.5 & 2.5\\
ESO~198-G024 & 0.04550 & 0305370101 & 2006-02-04 & 37.3 & 80.5 & 6.6\\
Mrk~290 & 0.02958 & 0400360201 & 2006-04-30 & 6.3 & 13.9 & 6.3\\
 & & 0400360301 & 2006-05-02 & 4.8 & 10.5 & 5.8\\
 & & 0400360601 & 2006-05-04 & 3.9 & 10.5 & 5.3\\
 & & 0400360801 & 2006-05-06 & 4.7 & 13.4 & 5.1\\
Mrk~205 & 0.07084 & 0124110101 & 2000-05-07 & 7.7 & 31.3 & 3.4\\
 & & 0401240201 & 2006-10-18 & 10.3 & 28.2 & 5.1\\
 & & 0401240501 & 2006-10-22 & 17.1 & 34.6 & 6.9\\
Mrk~590 & 0.02638 & 0201020201 & 2004-07-04 & 23.7 & 70.7 & 4.5\\
H~557-385 & 0.03387 & 0404260101 & 2006-08-11 & 5.1 & 24.5 & 3.3\\
 & & 0404260301 & 2006-11-03 & 9.3 & 52.5 & 2.8\\
TON~S180 & 0.06198 & 0110890401 & 2000-12-14 & 3.9 & 20.6 & 2.6\\
 & & 0110890701 & 2002-06-30 & 2.2 & 12.6 & 2.3\\
PG~1211+143 & 0.08090 & 0112610101 & 2001-06-15 & 7.0 & 49.5 & 2.0\\
 & & 0208020101 & 2004-06-21 & 4.7 & 32.3 & 1.9\\
 & & 0502050101 & 2007-12-21 & 7.2 & 42.6 & 2.2\\
 & & 0502050201 & 2007-12-23 & 4.8 & 25.4 & 2.5\\
Ark~564 & 0.02468 & 0206400101 & 2005-01-05 & 42.1 & 69.0 & 8.0\\
\hline\\[-6.5pt] 
\multicolumn{7}{c}{Type 2 sources}\\[1.2pt]
\hline
& & & & & &\\ 
MCG-5-23-16 & 0.00848 & 0112830401 & 2001-12-01 & 65.5 & 19.5 & 50.5\\
 & & 0302850201 & 2005-12-08 & 362.7 & 96.2 & 57.7\\
NGC~5506 & 0.00618 & 0013140101 & 2001-02-02 & 39.3 & 13.8 & 42.3\\
 & & 0201830201 & 2004-07-11 & 48.3 & 14.8 & 48.1\\
 & & 0201830301 & 2004-07-14 & 44.3 & 14.0 & 46.2\\
 & & 0201830401 & 2004-07-22 & 40.2 & 13.9 & 41.9\\
 & & 0201830501 & 2004-08-07 & 64.9 & 14.0 & 67.4\\
NGC~7172 & 0.00868 & 0147920601 & 2002-11-18 & 13.1 & 10.8 & 18.5\\
 & & 0414580101 & 2007-04-24 & 63.4 & 27.7 & 35.7\\
NGC~7314 & 0.00476 & 0111790101 & 2001-05-02 & 52.4 & 30.1 & 24.6\\
NGC~2110 & 0.00779 & 0145670101 & 2003-03-05 & 37.2 & 29.7 & 19.0\\
NGC~4507 & 0.01180 & 0006220201 & 2001-01-04 & 22.3 & 32.3 & 12.1\\
NGC~7582 & 0.00525 & 0112310201 & 2001-05-25 & 3.8 & 17.5 & 3.7\\
 & & 0204610101 & 2005-04-29 & 7.8 & 62.0 & 2.1\\
\hline   
\end{longtable}
  \tablefoot{
(1) Source name. (2) Cosmological red-shift. (3) XMM-Newton observation ID. (4) Observation starting date. (5) EPIC pn 4--10~keV total counts. (6) EPIC pn net exposure, after the subtraction of high background intervals and the CCD dead time fraction. (7) Observed flux in the 4--10~keV band.}
}

\end{small}

\begin{scriptsize}

\longtabL{2}{
\begin{landscape}
\begin{longtable}{l c||c c c c c c| c c c c c c c c c}

\caption{Best-fit parameters of the baseline models and absorption lines for the Type 1 and Type 2 sources.}\\

\hline\hline 
\small{Source} & \small{OBSID} & \multicolumn{6}{c}{\small{Baseline model}} & \multicolumn{9}{|c}{\small{Absorption line}} \\
\hline
 & & $\Gamma$ & $N_H$ & E & $\sigma$ & EW & $\chi^2/\nu$ & E & $\sigma$ & EW & $v_{out}$ & $\Delta\chi^2$ & F-test & M.C. & ID & EW$_{\mathrm{MOS}}$\\ 
 & & & \scriptsize{($10^{22}$~cm$^{-2}$)} & \scriptsize{(keV)} & \scriptsize{(eV)} & \scriptsize{(eV)} & & \scriptsize{(keV)} & \scriptsize{(eV)} & \scriptsize{(eV)} & \scriptsize{($c$)} & & & & & \scriptsize{(eV)}\\
\scriptsize{(1)} & \scriptsize{(2)} & \scriptsize{(3)} & \scriptsize{(4)} & \scriptsize{(5)} & \scriptsize{(6)} & \scriptsize{(7)}& \scriptsize{(8)} & \scriptsize{(9)} & \scriptsize{(10)} & \scriptsize{(11)} & \scriptsize{(12)} & \scriptsize{(13)} & \scriptsize{(14)} & \scriptsize{(15)} & \scriptsize{(16)} & \scriptsize{(17)}\\
\hline   
\endfirsthead   

\caption{Continued.}\\    

\hline\hline 
\small{Source} & \small{OBSID} & \multicolumn{6}{c}{\small{Baseline model}} & \multicolumn{9}{|c}{\small{Absorption line}} \\
\hline
 & & $\Gamma$ & $N_H$ & E & $\sigma$ & EW & $\chi^2/\nu$ & E & $\sigma$ & EW  & $v_{out}$ & $\Delta\chi^2$ & F-test & M.C. & ID & EW$_{\mathrm{MOS}}$\\ 
 & & & \scriptsize{($10^{22}$~cm$^{-2}$)} & \scriptsize{(keV)} & \scriptsize{(eV)} & \scriptsize{(eV)} & & \scriptsize{(keV)} & \scriptsize{(eV)} & \scriptsize{(eV)} & \scriptsize{($c$)} & & & & \scriptsize{(eV)}\\
\scriptsize{(1)} & \scriptsize{(2)} & \scriptsize{(3)} & \scriptsize{(4)} & \scriptsize{(5)} & \scriptsize{(6)} & \scriptsize{(7)}& \scriptsize{(8)} & \scriptsize{(9)} & \scriptsize{(10)} & \scriptsize{(11)} & \scriptsize{(12)} & \scriptsize{(13)} & \scriptsize{(14)} & \scriptsize{(15)} & \scriptsize{(16)} & \scriptsize{(17)}\\
\hline   
& & & & & & & & & & & & & & & &\\

\endhead
\endfoot

\multicolumn{17}{c}{Type 1 Sources}\\[1.2pt]
\hline
& & & & & & & & & & & & & & & &\\ 
NGC~4151 & 0112310101 & $1.46\pm0.04$ & $8.2\pm0.5$ & $6.398\pm0.004$ & 100 & $290\pm18$ & 1134/1064 & & & & & & & & &\\
 & 0112830501 & $1.43\pm0.04$ & $7.9\pm0.5$ & $6.397\pm0.005$ & 100 & $290\pm19$ & 1067/1021 & & & & & & & &\\
 & 0112830201 & $1.37\pm0.02$ & $7.4\pm0.1$ & $6.391\pm0.002$ & $62\pm4$ & $240\pm67$ & $1290/1309$ & & & & & & & & & \\
 & & & & $6.94\pm0.02$ & $100\pm24$ & $50\pm12$ & & & & & & & & & &\\
 & 0143500201 & $1.64\pm0.02$ & $7.2\pm0.2$ & $6.40\pm0.01$ & $79\pm10$ & $90^{+12}_{-9}$ & 1250/1317 & & & & & & & & &\\
 & 0143500301 & $1.66\pm0.02$ & $6.4\pm0.2$ & $6.40\pm0.01$ & $80\pm9$ & $80\pm10$ & 1362/1354 & & & & & & & & &\\
 & 0402660201 & $1.58\pm0.03$ & $11.0\pm0.4$ & $6.399^{+0.005}_{-0.002}$ & $79\pm8$ & $160\pm12$ & 1260/1226 & $7.74\pm0.05$ & 100 & $-32\pm7$ & $+0.105\pm0.005$ & 18 & 99.99\% & 99.7\% & Ly$\alpha$ & $-25\pm13$\\
IC4329A & 0147440101 & $1.65\pm0.01$ & & $6.40\pm0.01$ & 100 & $92\pm7$ & $1421/1388$ & $7.69\pm0.03$ & 10 & $-10\pm3$ & $+0.098\pm0.003$ & 11 & 99.5\% & 96\% & Ly$\alpha$ & $-11\pm6$\\
 & & & & $6.95\pm0.02$ & 100 & $30\pm7$ & & & & & & & & & &\\
NGC~3783 & 0112210101 & $1.77\pm0.04$ & $2.0\pm0.4$ & $6.36\pm0.01$ & 100 & $135^{+18}_{-15}$ & 1131/1076 & $6.61\pm0.02$ & 10 & $-16\pm5$ & $-0.013\pm0.004$ & 15.4 & 99.94\% & & He$\alpha$ & $-21\pm8$\\
 & & & & $7.01\pm0.02$ & 100 & $39\pm13$ & & $7.74\pm0.03$ & 10 & $-25\pm7$ & $-0.018\pm0.004$ & 12 & 99.7\% & 98.2\% & He$\beta$ & $+10\pm21$\\ 
 & 0112210201$^b$ & $1.66\pm0.02$ & $1.9\pm0.2$ & $6.39\pm0.01$ & 100 & $150^{+10}_{-11}$ & 1290/1309 & $6.63\pm0.03$ & 10 & $-10\pm3$ & $-0.010\pm0.005$ & 13 & 99.87\% &  & He$\alpha$ & $-24\pm9$\\
 & & & & $7.01\pm0.01$ & 100 & $57\pm11$ & & & & & & & & & &\\
 & 0112210501$^b$ & $1.74\pm0.02$ & $2.2\pm0.2$ & $6.39\pm0.01$ & 100 & $120\pm9$ & 1578/1363 & $6.67\pm0.02$ & 10 & $-17\pm2$ & $-0.004\pm0.002$ & 51 & $>99.99$\% & & He$\alpha$ & $-30\pm8$\\
 & & & & $7.05\pm0.02$ & 100 & $37\pm9$ & & & & & & & & & &\\
MCG+8-11-11 & 0201930201 & $1.57\pm0.01$ & & $6.42\pm0.01$ & 100 & $130\pm18$ & 981/1051 & & & & & & & & &\\
NGC~5548 & 0109960101 & $1.59\pm0.02$ & & $6.41\pm0.01$ & 10  & $72\pm18$ & 968/756 & & & & & & & & &\\
 & 0089960301 & $1.60\pm0.01$ & & $6.41\pm0.01$ & 10 & $60\pm12$ & 1221/1225 & & & & & & & & &\\
 & 0089960401 & $1.69\pm0.01$ & & $6.37\pm0.02$ & 100 & $77\pm18$ & 922/919 & & & & & & & & &\\
NGC~3516 & 0107460701 & $1.76\pm0.03$ & $8.1\pm0.4$ & $6.41\pm0.01$ & 10 & $297\pm18$ & 1194/1106 & & & & & & & & &\\
 & 0401210401 & $2.07\pm0.03$ & $3.6\pm0.4$ & $6.39\pm0.01$ & 100 & $93\pm14$ & 1285/1138 & $6.73\pm0.02$ & 10 & $-22\pm5$ & $+0.005\pm0.002$ & 27.4 & $>99.99$\% & & He$\alpha$ & $-30\pm7$\\
 & & & & & & & & $7.02\pm0.02$ & 10 & $-24\pm5$ & $+0.008\pm0.002$ & 21.6 & 99.99\% & & Ly$\alpha$ & $-40\pm9$\\
 & 0401210501 & $2.06\pm0.03$ & $3.8\pm0.4$ & $6.39\pm0.01$ & 100 & $125\pm15$ & 1226/1146 & $6.69\pm0.02$ & 10 & $-15\pm5$ & $-0.001\pm0.004$ & 14 & 99.6\% & & He$\alpha$ & $-33\pm7$\\
 & & & & & & & & $7.04\pm0.02$ & 10 & $-21\pm5$ & $+0.010\pm0.002$ & 20.2 & 99.99\% & & Ly$\alpha$ & $-38\pm9$\\
 & 0401210601 & $2.01\pm0.03$ & $4.9\pm0.4$ & $6.38\pm0.01$ & 100 & $133^{+16}_{-13}$ & 1170/1095 & $6.77\pm0.02$ & 10 & $-25\pm6$ & $+0.011\pm0.002$ & 26.2 & $>99.99$\% & & He$\alpha$ & $-29\pm7$\\
 & & & & & & & & $7.08\pm0.02$ & 10 & $-27\pm6$ & $+0.016\pm0.002$ & 17 & 99.96\% & & Ly$\alpha$ & $-31\pm9$\\
 & & & & & & & & $7.95\pm0.02$ & 10 & $-33\pm9$ & $+0.009\pm0.002$ & 24.2 & $>99.99$\% & $>99.9$\% & He$\beta$ & $-18\pm10$\\
 & & & & & & & & $8.28\pm0.03$ & 10 & $-34\pm9$ & $+0.004\pm0.004$ & 14.6 & 99.9\% & 99.5\% & Ly$\beta$ & $-20\pm15$\\
 & 0401211001 & $2.05\pm0.04$ & $3.5\pm0.4$ & $6.42\pm0.01$ & 100 & $131\pm20$ & 1067/1024 & $6.71\pm0.04$ & 10 & $-18\pm5$ & $+0.002\pm0.006$ & 10.4 & 99.3\% & & He$\alpha$ & $-31\pm7$\\
 & & & & & & & & $7.09\pm0.03$ & 10 & $-21\pm7$ & $+0.017\pm0.004$ & 12.7 & 99.8\% & & Ly$\alpha$ & $-31\pm8$\\
NGC~4593 & 0059830101 & $1.68\pm0.01$ & & $6.40\pm0.01$ & 100 & $130\pm15$ & 1212/1186 & & & & & & & & &\\
Mrk~509 & 0130720101 & $1.54\pm0.02$ & & $6.38\pm0.02$ & 100 & $84\pm23$ & 837/855 & $8.29\pm0.03$ & 10 & $-32\pm12$ & $+0.172\pm0.003$ & 9.6 & $99.3\%$ & $95\%$ & Ly$\alpha$ & $-33\pm10$\\
 & 0130720201 & $1.60\pm0.02$ & & $6.43\pm0.02$ & 100 & $60\pm19$ & 979/981 & & & & & & & & &\\
 & 0306090201 & $1.70\pm0.01$ & & $6.43\pm0.01$ & 100 & $67\pm11$ & 1059/1192 & $8.03\pm0.02$ & 10 & $-19\pm6$ & $+0.141\pm0.002$ & 15 & 99.9\% & 99.4\% & Ly$\alpha$ & $-26\pm13$\\
 & & & & $7.02\pm0.04$ & 100 & $24\pm13$ & & & & & & & & & &\\ 
 & 0306090301 & $1.72\pm0.02$ & & $6.43\pm0.02$ & 100 & $62\pm16$ & 967/1019 & & & & & & & & &\\
 & 0306090401 & $1.68\pm0.01$ & & $6.44\pm0.02$ & 100 & $58\pm12$ & 1147/1187 & $8.51\pm0.04$ & 10 & $-19\pm7$ & $+0.196\pm0.003$ & 10 & 99.5\% & 95.3\% & Ly$\alpha$ & $-11\pm7$\\
 & & & & $6.85\pm0.08$ & 100 & $22\pm12$ & & & & & & & & & &\\ 
MCG-6-30-15 & 0111570101 & $2.09\pm0.04$ & $3.4\pm0.5$ & $6.35\pm0.03$ & 100 & $78^{+24}_{-21}$ & 895/876 & & & & & & & & &\\
 & 0111570201 & $2.25\pm0.04$ & $4.1\pm0.4$ & $6.39\pm0.02$ & 100 & $67\pm15$ & 1028/1013 & & & & & & & & &\\
 & 0029740101 & $2.26\pm0.03$ & $4.4\pm0.3$ & $6.43\pm0.01$ & 100 & $78\pm13$ & 1154/1127 & & & & & & & & &\\
 & 0029740701 & $2.16\pm0.02$ & $2.8\pm0.2$ & $6.42\pm0.01$ & 100 & $65\pm10$ & 1381/1289 & & & & & & & & &\\
 & 0029740801 & $2.28\pm0.02$ & $3.8\pm0.2$ & $6.43\pm0.01$ & 100 & $79^{+16}_{-10}$ & 1338/1253 & & & & & & & & &\\
Ark~120 & 0147190101$^b$ & $1.86\pm0.01$ & & $6.39\pm0.01$ & 100 & $90\pm15$ & 1148/1163 & $9.18\pm0.03$ & 10 & $-25\pm9$ & $+0.269\pm0.002$ & 11 & 99.6\% & 97.1\% & Ly$\alpha$ & $-10\pm24$\\
 & & & & $6.71\pm0.03$ & 100 & $43\pm13$ & & & & & & & & & &\\
 & & & & $7.02\pm0.02$ & 10 & $24\pm12$ & & & & & & & & & &\\
Mrk~110 & 0201130501 & $1.68\pm0.02$ & & $6.45\pm0.02$ & 100 & $56\pm17$ & 931/967 & & & & & & & & & \\
NGC~7469 & 0112170101 & $1.69\pm0.03$ & & $6.40\pm0.01$ & 10 & $110\pm30$ & 467/509 & & & & & & & & & \\
 & 0112170301 & $1.75\pm0.02$ & & $6.43\pm0.02$ & 100 & $96\pm29$ & 611/653 & & & & & & & & &\\
 & 0207090101 & $1.75\pm0.01$ & & $6.41\pm0.01$ & 100 & $110\pm12$ & 1145/1119 & & & & & & & & &\\
 & 0207090201 & $1.76\pm0.01$ & & $6.42\pm0.01$ & 100 & $110\pm13$ & 1131/1104 & & & & & & & & &\\
IRAS~5078+1626 & 0502090501 & $1.55\pm0.02$ & & $6.39\pm0.01$ & 100 & $120\pm18$ & 1026/978 & & & & & & & & & \\
Mrk~279 & 0302480401 & $1.74\pm0.02$ & & $6.43\pm0.01$ & 100 & $100\pm18$ & 1037/1028 & & & & & & & & & \\
 & 0302480501$^a$ & $1.71\pm0.02$ & & $6.43\pm0.01$ & 100 & $115\pm23$ & 976/1022 & $6.69\pm0.02$ & 10 & $-19\pm9$ & $-0.001\pm0.004$ & 10.5 & 99.6\% & & He$\alpha$ & $-30^{+27}_{-23}$\\
 & & & & $7.00\pm0.06$ & 10 & $29\pm16$ & & & & & & & & & &\\
 & 0302480601 & $1.70\pm0.02$ & & $6.42\pm0.01$ & 100 & $130\pm30$ & 731/753 & & & & & & & & &\\
NGC~526A & 0150940101 & $1.43\pm0.02$ & & $6.40\pm0.02$ & 100 & $90\pm19$ & 969/945 & & & & & & & & & \\
NGC~3227 & 0101040301 & $1.50\pm0.06$ & $7.3^{+1.0}_{-0.7}$ & $6.38\pm0.01$ & 100 & $220^{+61}_{-36}$ & 514/484 & & & & & & & & &\\
 & 0400270101 & $1.55\pm0.01$ & & $6.39\pm0.01$ & 100 & $112\pm11$ & 1292/1297 & & & & & & & & &\\
NGC~7213 & 0111810101 & $1.71\pm0.02$ & & $6.42\pm0.02$ & 100 & $110\pm24$ & 786/824 & & & & & & & & &\\
ESO~511-G30 & 0502090201 & $1.71\pm0.01$ & & $6.39\pm0.01$ & 100 & $100\pm15$ & 1215/1093 & & & & & & & & &\\
Mrk~79 & 0400070201 & $1.74\pm0.03$ & & $6.43\pm0.02$ & 10 & $54\pm24$ & 586/568 & $7.63\pm0.03$ & 10 & $-43\pm14$ & $+0.091\pm0.004$ & 12 & 99.5\% & 97.7\% & Ly$\alpha$ & $+6\pm32$\\
 & 0400070301 & $1.64\pm0.03$ & & $6.42\pm0.01$ & 10 & $80\pm27$ & 411/467 & & & & & & & & &\\
 & 0400070401 & $1.60\pm0.03$ & & $6.37\pm0.02$ & 100 & $115\pm36$ & 441/439 & & & & & & & & &\\
NGC~4051 & 0109141401$^b$ & $2.04\pm0.03$ & $1.7\pm0.4$ & $6.39\pm0.01$ & 100 & $120\pm18$ & 1127/1105 & $7.10\pm0.03$ & 10 & $-19\pm6$ & $+0.018\pm0.004$ & 11.2 & 99.6\% & 97.5\% & Ly$\alpha$ & $-40\pm17$\\
 & 0157560101 & $1.77\pm0.04$ & $6.5\pm0.9$ & $6.42\pm0.01$ & 100 & $275\pm40$ & 507/436 & $8.10\pm0.05$ & 100 & $-96\pm24$ & $+0.150\pm0.005$ & 17 & 99.94\% & 99.7\% & Ly$\alpha$ & $-77\pm41$\\
 & & & & $7.08\pm0.02$ & 10 & $48\pm25$ & & & & & & & & & &\\
Mrk~766 & 0096020101 & $1.96\pm0.03$ & & $6.41\pm0.02$ & 10 & $65\pm27$ & 608/568 & & & & & & & & &\\
 & & & & $6.73\pm0.04$ & 100 & $74\pm34$ & & & & & & & & & &\\
 & 0109141301 & $2.02\pm0.02$ & & $6.39\pm0.02$ & 10 & $44\pm11$ & 1254/1089 & & & & & & & & &\\
 & & & & $6.73\pm0.04$ & 100 & $60\pm15$ & & & & & & & & & &\\
 & 0304030101 & $1.90\pm0.06$ & $6.6\pm0.8$ & $6.43\pm0.02$ & 10 & $57\pm30$ & 704/667 & & & & & & & & &\\
 & & & & $6.66\pm0.05$ & 10 & $57\pm29$ & & & & & & & & &\\
 & 0304030301 & $1.88\pm0.02$ & & $6.39\pm0.01$ & 10 & $60\pm21$ & 961/888 & $7.28\pm0.05$ & 100 & $-39^{+11}_{-13}$ & $+0.044\pm0.007$  & 12 & 99.6\% & 98.6\% & Ly$\alpha$ & $-59\pm17$\\
 & & & & $6.62\pm0.04$ & 100 & $63\pm24$ & & & & & & & & & &\\
 & 0304030401 & $1.88\pm0.02$ & & $6.43\pm0.01$ & 10 & $43\pm16$ & 993/920 & & & & & & & & &\\
 & & & & $6.79\pm0.04$ & 100 & $65\pm21$ & & & & & & & & &\\
 & 0304030501 & $1.99\pm0.02$ & & $6.41\pm0.01$ & 10 & $50\pm15$ & 995/960 & $7.63\pm0.03$ & 10 & $-27\pm10$ & $+0.091\pm0.004$ & 12.2 & 99.7\% & 98.7\% & Ly$\alpha$ & $-7\pm21$\\
 & & & & $6.78\pm0.03$ & 100 & $77\pm21$ & & & & & & & & & &\\
 & 0304030601 & $1.89\pm0.02$ & & $6.35\pm0.04$ & 10 & $27\pm16$ & 1012/908 & & & & & & & & &\\
 & & & & $6.67\pm0.04$ & 100 & $52\pm23$ & & & & & & & & & &\\
 & 0304030701 & $1.70\pm0.06$ & & $6.49\pm0.04$ & 10 & $54\pm22$ & 360/329 & & & & & & & & &\\
Mrk~841 & 0205340201 & $1.80\pm0.06$ & $5.1^{+0.8}_{-1.0}$ & $6.46\pm0.02$ & 100 & $100^{+43}_{-30}$ & 491/522 & & & & & & & & &\\
 & 0205340401$^a$ & $1.61\pm0.04$ & & $6.49\pm0.03$ & 100 & $100^{+43}_{-36}$ & 321/337 & $7.19\pm0.02$ & 10 & $-46\pm16$ & $+0.034\pm0.003$ & 12 & 99.8\% & 98.4\% & Ly$\alpha$ & $+32\pm28$\\
Mrk~704 & 0300240101 & $1.70\pm0.10$ & $5.0\pm1.2$ & $6.36\pm0.03$ & 100 & $136\pm46$ & 340/320 & & & & & & & & &\\
Fairall~9 & 0101040201 & $1.64\pm0.03$ & & $6.39\pm0.02$ & 100 & $150\pm36$ & 451/493 & & & & & & & & &\\
ESO~323-G77 & 0300240501 & $2.36\pm0.09$ & $12\pm1$ & $6.40\pm0.03$ & 100 & $115^{+49}_{-36}$ & 503/417 & $6.73\pm0.01$ & 10 & $-78\pm13$ & $+0.005\pm0.002$ & 46.2 & $>99.99$\% & & He$\alpha$ & $-58\pm18$\\
 & & & & & & & & $7.02\pm0.02$ & 10 & $-77\pm15$ & $+0.008\pm0.003$ & 30 & $>99.99$\% & & Ly$\alpha$ & $-79\pm20$\\
1H419-577 & 0148000201 & $1.21\pm0.05$ & & $6.33\pm0.05$ & 100 & $90^{+56}_{-47}$ & 197/176 & $7.23\pm0.04$ & 10 & $-55^{+18}_{-21}$ & $+0.037\pm0.005$ & 10.3 & 99\% & 95.8\% & Ly$\alpha$ & $-29\pm22$ \\
 & 0148000401 & $1.55\pm0.05$ & 198/225 & & & & & & & & & & & & &\\
 & 0148000601 & $1.49\pm0.05$ & 214/213 & & & & & & & & & & & & &\\
Mrk~335 & 0101040101 & $2.00\pm0.03$ & & $6.46\pm0.08$ & 100 & $60\pm30$ & 542/543 & & & & & & & & &\\
 & 0306870101 & $1.97\pm0.01$ & & $6.41\pm0.02$ & 100 & $90\pm18$ & 1019/1018 & & & & & & & & &\\
 & & & & $6.96\pm0.02$ & 100 & $64\pm18$ & & & & & & & & & &\\
 & 0510010701 & $2.5\pm0.2$ & $13\pm2$ & $6.39\pm0.03$ & 100 & $182^{+80}_{-75}$ & 135/121 & & & & & & & & &\\
ESO~198-G024 & 0305370101 & $1.62\pm0.02$ & & $6.43\pm0.02$ & 100 & $90\pm18$ & 944/937 & & & & & & & & &\\
Mrk~290 & 0400360201 & $1.54\pm0.05$ & & & & & 281/265 & & & & & & & & &\\
 & 0400360301 & $1.50\pm0.6$ & & & & & 195/203 & & & & & & & & &\\
 & 0400360601 & $1.61\pm0.05$ & & & & & 157/173 & $8.03\pm0.03$ & 10 & $-85^{+27}_{-33}$ & $+0.141\pm0.003$ & 13.4 & 99.96\% & 99\% & Ly$\alpha$ & $-110^{+53}_{-61}$\\
 & 0400360801 & $1.54\pm0.05$ & & $6.42\pm0.03$ & 100 & $151^{+63}_{-60}$ & 181/203 & & & & & & & & &\\
Mrk~205 & 0124110101 & $1.71\pm0.04$ & & $6.37\pm0.04$ & 100 & $100\pm43$ & 309/314 & $7.70\pm0.03$ & 10 & $-50\pm18$ & $+0.100\pm0.004$ & 11 & 99.7\% & 96.9\% & Ly$\alpha$ & $+37\pm33$\\
 & & & & $6.86\pm0.05$ & 100 & $90\pm49$ & & & & & & & & & &\\
 & 0401240201 & $1.79\pm0.04$ & & $6.34\pm0.02$ & 10 & $70\pm27$ & 465/392 & & & & & & & & &\\
 & 0401240501 & $1.85\pm0.02$ & & $6.55\pm0.04$ & 100 & $65\pm29$ & 553/603 & & & & & & & & &\\
Mrk~590 & 0201020201 & $1.52\pm0.03$ & & $6.42\pm0.01$ & 10 & $110\pm21$ & 739/776 & & & & & & & & &\\
H~557-385 & 0404260101 & $1.20\pm0.06$ & $21\pm2$ & $6.42\pm0.01$ & 100 & $370\pm67$ & 231/202 & & & & & & & & & \\
 & 0404260301 & $1.14\pm0.05$ & $21\pm2$ & $6.44\pm0.01$ & 100 & $460\pm55$ & 409/337& & & & & & & & &\\
 & & & & $6.90\pm0.02$ & 100 & $189^{+47}_{-41}$ & & & & & & & & & &\\
TON~S180 & 0110890401 & $2.05\pm0.05$ & & & & & 174/178 & & & & & & & & &\\
 & 0110890701 & $1.96\pm0.08$ & & & & & 123/104 & & & & & & & & &\\
PG~1211+143 & 0112610101 & $2.80\pm0.12$ & $11.7\pm1.5$ & $6.52\pm0.05$ & 100 & $100\pm43$ & 307/273 & $7.62\pm0.02$ & 100 & $-130^{+18}_{-24}$ & $+0.128\pm0.003$ & 56 & $>99.99$\%  & $>99.9$\%  & He$\alpha$ & $-128^{+32}_{-60}$\\
 & 0208020101 & $1.71\pm0.05$ & & $6.52\pm0.03$ & 100 & $135^{+60}_{-51}$ & 219/200 & & & & & & & & &\\
 & 0502050101 & $2.07\pm0.04$ & & $6.36\pm0.03$ & 100 & $141\pm46$ & 258/291 & & & & & & & & &\\
 & 0502050201 & $2.06\pm0.05$ & & $6.36\pm0.06$ & 100 & $90^{+53}_{-50}$ & 237/211 & & & & & & & & & \\
Ark~564 & 0206400101 & $2.45\pm0.02$ & & $6.34\pm0.04$ & 100 & $35\pm18$ & 829/900 & & & & & & & & &\\
 & & & & $6.71\pm0.03$ & 100 & $50\pm18$ & & & & & & & & & &\\
& & & & & & & & & & & & & & & &\\ 
\hline\\[-6.5pt] 
\multicolumn{17}{c}{Type 2 Sources}\\[1.2pt]
\hline
& & & & & & & & & & & & & & & &\\ 
MCG-5-23-16 & 0112830401 & $1.61\pm0.03$ & $2.5\pm0.4$ & $6.37\pm0.01$ & 100 & $90\pm15$ & 1081/1123 & & & & & & & & &\\
 & 0302850201 & $1.53\pm0.01$ & $1.2\pm0.2$ & $6.41\pm0.01$ & 100 & $90\pm6$ & 1554/1391 & $7.84\pm0.04$ & 100 & $-18\pm4$ & $+0.118\pm0.005$ & 22.6 & 99.996\% & $>99.9$\% & Ly$\alpha$ & $-9\pm5$\\
 & & & & $6.93\pm0.02$ & 100 & $26\pm7$ & & & & & & & & & &\\
NGC~5506 & 0013140101 & $1.68\pm0.04$ & $2.8\pm0.5$ & $6.44\pm0.01$ & 100 & $125^{+21}_{-18}$ & 1049/960 & & & & & & & & &\\
 & 0201830201 & $1.79\pm0.04$ & $3.4\pm0.4$ & $6.39\pm0.01$ & 100 & $110\pm18$ & 1060/1002 & & & & & & & & &\\
 & 0201830301 & $1.88\pm0.03$ & $4.7\pm0.5$ & $6.39\pm0.01$ & 100 & $120\pm18$ & 1116/984 & & & & & & & & &\\
 & 0201830401 & $1.93\pm0.04$ & $4.8\pm0.5$ & $6.41\pm0.01$ & 100 & $130\pm18$ & 931/948 & & & & & & & & &\\
 & 0201830501 & $1.93\pm0.03$ & $4.2\pm0.4$ & $6.41\pm0.02$ & 100 & $85\pm15$ & 1115/1082 & & & & & & & & &\\
NGC~7172 & 0147920601 & $1.70\pm0.06$ & $9.9\pm0.9$ & $6.40\pm0.02$ & 100 & $110^{+36}_{-30}$ & 411/455 & & & & & & & & &\\
 & 0414580101 & $1.72\pm0.03$ & $9.0\pm0.4$ & $6.39\pm0.01$ & 100 & $80\pm12$ & 1135/1128 & & & & & & & & &\\
NGC~7314 & 0111790101 & $1.97\pm0.04$ & $1.8\pm0.4$ & $6.42\pm0.02$ & 100 & $70\pm18$ & 916/1006 & & & & & & & & &\\
NGC~2110 & 0145670101 & $1.59\pm0.04$ & $4.7\pm0.5$ & $6.43\pm0.01$ & 100 & $150^{+18}_{-24}$ & 903/954 & & & & & & & & &\\
NGC~4507 & 0006220201 & $1.50\pm0.06$ & $38.0\pm0.5$ & $6.37\pm0.01$ & $76^{+10}_{-12}$ & $227\pm26$ & 649/682 & $8.32\pm0.02$ & 10 & $-35\pm12$ & $+0.177\pm0.002$ & 12.5 & 99.9\% & 98.7\% & Ly$\alpha$ & $-18\pm21$\\
 & & & & $6.90\pm0.04$ & 100 & $36\pm20$ & & & & & & & & & &\\
NGC~7582 & 0112310201 & $0.80\pm0.06$ & $19.0\pm2.4$ & $6.39\pm0.01$ & 100 & $425^{+113}_{-73}$ & 168/149 & $8.99\pm0.04$ & 10 & $-91^{+29}_{-41}$ & $+0.255\pm0.003$ & 12 & 99.5\% & 98.2\% & He$\alpha$ & $-124^{+107}_{-119}$\\
 & & & & $7.02\pm0.05$ & 100 & $143^{+74}_{-68}$ & & & & & & & & & &\\
 & 0204610101 & $0.50\pm0.06$ & $9.6^{+1.2}_{-2.1}$ & $6.41\pm0.05$ & 100 & $940^{+122}_{-85}$ & 317/287 & & & & & & & & &\\
 & & & & $6.91\pm0.02$ & 100 & $290\pm61$ & & & & & & & & & &\\
 & & & & & & & & & & & & & & & &\\
\hline\\
\end{longtable}
  \tablefoot{
The baseline models are composed by a simple absorbed power-law continuum with narrow Gaussian emission lines. 
The narrow absorption lines refer to those that have been detected in the E$=$6.4--7.1~keV band with F-test confidence levels greater than 99\% and to those detected at energies greater than 7.1~keV with additional Monte Carlo probability $\ge$95\%. The errors are at the 1$\sigma$ level. (1) Source name. (2) XMM-Newton observation ID. (3) Power-law continuum photon index. (4) Neutral absorber equivalent Hydrogen column density. (5) Gaussian emission line rest-frame energy. (6) Emission line width. (7) Emission line equivalent width. (8) Best-fit $\chi^2$ and degrees of freedom $\nu$ of the baseline model. (9) Absorption line rest-frame energy. (10) Absorption line width. (11) Absorption line equivalent width. (12) Absorption line blue-shifted velocity. (13) $\Delta\chi^2$ relative to the addition of the absorption line. (14) Detection confidence level from the F-test. (15) Detection confidence level from extensive Monte Carlo simulations. (16) Absorption line identification as Fe XXV 1s$^2$--1s2p (He$\alpha$), Fe XXV 1s$^2$--1s3p (He$\beta$), Fe XXVI 1s--2p (Ly$\alpha$) or Fe XXVI 1s--3p (Ly$\beta$). (17) Consistency check of the EW of the absorption line with the MOS cameras.\\
\tablefoottext{a}{Only MOS1 data available.}
\tablefoottext{b}{Only MOS2 data available.}
}

\end{landscape}
}

\end{scriptsize}

\clearpage

\begin{table*}[!t]

\caption{Estimates of the mean Fe K absorption line parameters for the sample sources among the XMM-Newton observations in which they have been detected.}
\begin{scriptsize}
\begin{center}
\begin{tabular}{l |c c |c c |c c |c c |c}
\hline\hline\\[-6pt]     
 Source & Fe XXV He$\alpha$ & $v_{out}$ & Fe XXV He$\beta$ & $v_{out}$ & Fe XXVI Ly$\alpha$ & $v_{out}$ & Fe XXVI Ly$\beta$ & $v_{out}$ & $<v_{out}>$\\  
 & (eV) & (c) & (eV) & (c) & (eV) & (c) & (eV) & (c) & (c)\\
\scriptsize{(1)} & \scriptsize{(2)} & \scriptsize{(3)} & \scriptsize{(4)} & \scriptsize{(5)} & \scriptsize{(6)} & \scriptsize{(7)} & \scriptsize{(8)} & \scriptsize{(9)} & \scriptsize{(10)}\\
\hline   
 & & & & & & & & &\\ 
NGC~4151 & & & & & $-32\pm7$ & $+0.105\pm0.005$ & & & $+0.105$\\[+3pt]
IC4329A & & & & & $-10\pm3$ & $+0.098\pm0.003$ & & & $+0.098$\\[+3pt]
NGC~3783 & $-14(^{-10}_{-17})$ & $-0.009(^{-0.004}_{-0.013})$ & $-25\pm7$ & $-0.018\pm0.004$ & & & & & $-0.013$\\[+3pt]
NGC~3516 & $-20(^{-15}_{-25})$ & $+0.004(^{+0.011}_{-0.001})$ & $-33\pm9$ & $+0.009\pm0.002$ & $-23(^{-21}_{-27})$ & $+0.013(^{+0.017}_{+0.008})$ & $-34\pm9$ & $+0.004\pm0.004$ & $+0.008$\\[+3pt]
Mrk~509 & & & & & $-23(^{-19}_{-32})$ & $+0.170(^{+0.196}_{+0.141})$ & & & $+0.170$\\[+3pt]
Ark~120 & & & & & $-25\pm9$ & $+0.269\pm0.002$ & & & $+0.269$\\[+3pt]
Mrk~279 & $-19\pm9$ & $-0.001\pm0.004$ & & & & & & & $-0.001$\\[+3pt]
Mrk~79 & & & & & $-43\pm14$ & $+0.091\pm0.004$ & & & $+0.091$\\[+3pt]
NGC~4051 & & & & & $-58(^{-19}_{-96})$ & $+0.084(^{+0.150}_{+0.018})$ & & & $+0.084$\\[+3pt]
Mrk~766 & & & & & $-33(^{-27}_{-39})$ & $+0.067(^{+0.091}_{+0.044})$ & & & $+0.067$\\[+3pt]
Mrk~841 & & & & & $-46\pm16$ & $+0.034\pm0.003$ & & & $+0.034$\\[+3pt]
ESO~323+G77 & $-78\pm13$ & $+0.005\pm0.002$ & & & $-77\pm15$ & $+0.008\pm0.003$ & & & $+0.007$\\[+3pt]
1H419-577 & & & & & $-55^{+18}_{-21}$ & $+0.037\pm0.005$ & & & $+0.037$\\[+3pt]
Mrk~290 & & & & & $-85^{+27}_{-33}$ & $+0.141\pm0.003$ & & & $+0.141$\\[+3pt]
Mrk~205 & & & & & $-50\pm18$ & $+0.100\pm0.004$ & & & $+0.100$\\[+3pt]
PG~1211+143 & $-130^{+18}_{-24}$ & $+0.128\pm0.003$ & & & & & & & $+0.128$\\[+3pt]
MCG-5-23-16 & & & & & $-18\pm4$ & $+0.118\pm0.005$ & & & $+0.118$\\[+3pt]
NGC~4507 & & & & & $-35\pm12$ & $+0.177\pm0.002$ & & & $+0.177$\\[+3pt]
NGC~7582 & $-91^{+29}_{-41}$ & $+0.255\pm0.003$ & & & & & & & $+0.255$\\[+3pt]
\hline     
\end{tabular}
  \tablefoot{
 For the sources with multiple line detections in different XMM-Newton observations we report also the lowest and highest values of the EW and blue-shifted velocity within brackets. Instead, if there is only one line detection we report the associated 1$\sigma$ errors. (1) Source name. (2)-(3) Mean EW and blue-shifted velocity for Fe XXV He$\alpha$. (4)-(5) Mean EW and blue-shifted velocity for Fe XXV He$\beta$. (6)-(7) Mean EW and blue-shifted velocity for Fe XXVI Ly$\alpha$. (8)-(9) Mean EW and blue-shifted velocity for Fe XXVI Ly$\beta$. (10) Mean blue-shifted velocity from all the Fe K absorption lines.
}
\end{center}
\end{scriptsize}
\end{table*}

\end{appendix}

}

\clearpage

\newpage

\begin{appendix}

\section{Notes on single sources}

In this section we discuss and compare our work with the results already published in the literature. We payed particular attention to the works reporting a spectral analysis of the Fe K band of each sample source, especially if performed with the XMM-Newton EPIC pn instrument.

\emph{NGC~4151:} Piro et al.~(2005) reported the detection of an absorption feature around E$\simeq$8.5--9~keV with a statistical significance of 99.96\% in one out of five Beppo-SAX observations of this source. Due to the low energy resolution of that instrument, the feature could be fitted either with an absorption edge due to highly ionized iron at rest or with an absorption line due to Fe XXV/XXVI with blue-shifted velocity of $\sim$0.1--0.2c.  
The presence of a complex and possibly multi-phase ionized absorber in NGC~4151 has also been reported by other authors, such as Schurch et al.~(2003) using XMM-Newton data and Kraemer et al.~(2005) using simultaneous HST and Chandra observations. 
In the contour plots of the third XMM-Newton observation of the source (obs. 0112830201) we find evidence for a possible narrow absorption line at E$\simeq$7.8~keV (see Fig.~C.1). This line might be identified with blue-shifted Fe XXVI Ly$\alpha$. However, we did not include this line in Table~A.2 because the relative Monte Carlo confidence level was lower than 95\%. 
Instead, we clearly detected a narrow absorption line at rest-frame energy E$\simeq$7.74~keV in the last observation (obs. 0402660201). We identified the line with Fe XXVI Ly$\alpha$ with blue-shifted velocity $\simeq$0.1c (see Table~A.2). 
It is important to note that in the contour plots of both these observations we found also evidence for two possible narrow absorption lines at energies lower than 6.4~keV.
The parameters of these lines are E$=$$4.26^{+0.03}_{-0.06}$~keV, $\sigma$$=$100~eV and EW$=$$-15\pm4$~eV, in the former, and E$=$$4.84\pm0.03$~keV, $\sigma$$=$10~eV and EW$=$$-13\pm4$~eV, in the latter. In both cases the F-test detection confidence levels are $\ga$99.9\%.
If identified with blue-shifted absorption lines from Ca XIX He$\alpha$ (E$\simeq$3.9~keV) and Ca XX Ly$\alpha$ (E$\simeq$4.11~keV), their velocity is consistent with that inferred from the Fe K lines of $\simeq$0.1c.    

\emph{IC4329A:} the detection of a blue-shifted narrow absorption feature at E$\simeq$7.7~keV ascribable to Fe XXVI Ly$\alpha$ in the XMM-Newton spectrum of IC4329A has already been reported by Markowitz et al.~(2006). We confirm their results.
Moreover, from the contour plot in Appendix C (see Fig.~C.1), a broad absorption trough can be observed at an energy greater than $\simeq$9~keV. We tried to model it with an ionized Fe K edge (\emph{zedge} in XSPEC). However, we obtained a better fit with a model composed of a blend of further three unresolved Fe XXVI Lyman series lines (Ly$\beta$ at E$=$$8.250$~keV, Ly$\gamma$ at E$=$$8.700$~keV and Ly$\delta$ at E$=$$8.909$~keV) 
with a blue-shift consistent with that of the Fe XXVI Ly$\alpha$. 
In particular, we performed a fit adding to the baseline model (see Table~A.2) 
four additional narrow absorption lines with energies fixed to
the expected values for the Fe XXVI Lyman series and left their
common energy shift as a free parameter. This series of lines provide a very
good simultaneous modeling of all the absorption features, with a global
$\Delta\chi^2=30$ for five additional parameters. The
probability of having these four absorption lines at these exact
energies simply from random fluctuations is low, about $10^{-4}$.
Interestingly enough, their common blue-shifted velocity is
$+0.097\pm0.005$c, completely consistent with that estimated for the
 Fe XXVI Ly$\alpha$ line alone (see Table~A.2).
The resultant EWs of the four Fe XXVI Lymann series
lines are: EW$=$$-11\pm6$~eV for the Ly$\alpha$, EW$=$$-11\pm8$~eV
for the Ly$\beta$, EW$=$$-10\pm6$~eV for the Ly$\gamma$ and EW$=$$-15\pm7$~eV
for the Ly$\delta$. Their ratios are close to unity, which suggests 
possible saturation effects. A physically self-consistent modeling of these lines with the photo-ionization code Xstar is presented in a companion paper.

\emph{NGC~3783:} a strong absorption line at the rest-frame energy of E$\simeq$6.7~keV has been detected by Reeves et al.~(2004) in the XMM-Newton spectrum of this source. The authors pointed out that the line energy is consistent with rest-frame resonant absorption from a blend of different highly ionized iron ions, such as Fe XXIII (E$\simeq$6.62~keV), Fe XXIV (E$\simeq$6.66~keV) and Fe XXV (E$\simeq$6.70~keV) (see \S4.1). We confirm their detection and the line parameters here derived are in agreement with theirs.
Furthermore, we note that in the contour plot of one observation (0112210201) there is evidence for a narrow absorption line at E$<$6.4~keV. From a fit with an inverted Gaussian we derived E$=$$4.30\pm0.04$~keV, $\sigma$$=$10~eV and EW$=$$-6\pm3$~eV. The F-test confidence level of the line is $\sim$99\%, but it slightly decreases if all the remaining emission features redwards the neutral Fe K$\alpha$ line are modeled with Gaussians, indicating a possible slight model dependency.
 The identification of this line is ambiguous. However, as already discussed in \S4.5, the line could be possibly associated with blue-shifted Ca XX Ly$\alpha$ with v$\sim$$10^4$~km/s or red-shifted Fe XXV/XXVI 1s--2p with v$\sim$0.5--0.6c.

\emph{MCG+8-11-11:} a detailed analysis of the XMM-Newton spectrum of this source has already been published by Matt et al.~(2006). We confirm their overall results. We did not detect any narrow Fe K absorption line.

\emph{NGC~5548:} a detailed analysis of the XMM-Newton spectrum of this source has already been published by Pounds et al.~(2003b). Our analysis confirms their overall results. We did not detect any narrow Fe K absorption line.

\emph{NGC~3516:} the detection of narrow highly ionized absorption features in the Chandra HETG and XMM-Newton EPIC pn spectra of this source have already been  published by Turner et al.~(2005) and Turner et al.~(2008). In particular, the authors reported the presence of H/He-line resonant absorption lines from Mg, Si, S and Fe. Our detection of several Fe XXV/XXVI K-shell absorption lines at E$>$6.4 keV is in agreement with their results. 
It should be noted that in the energy-intensity contour plot of the first observation of the source (0107460701) there is evidence for a narrow absorption feature at E$<$6.4~keV (see Fig.~C.2) with F-test confidence level $\sim$99\%. The resultant line parameters when modeled with a simple inverted Gaussian are E$=$$4.75\pm0.04$~keV, $\sigma$$=$10~eV and EW$=$$-19\pm6$~eV. As discussed in \S4.5, some possible identifications for this feature are with blue-shifted Ca XX Ly$\alpha$ with v$\sim$0.1c or red-shifted Fe XXV/XXVI 1s--2p with v$\sim$0.4--0.5c.

\emph{NGC~4593:} a detailed analysis of the XMM-Newton spectrum of this source has already been published by Reynolds et al.~(2004). Our analysis confirms their overall results. We did not detect any narrow Fe K absorption line.

\emph{Mrk~509:} the detection of narrow blue-shifted absorption lines ascribable to the Fe XXVI Ly$\alpha$ resonant transition in the XMM-Newton spectra of this source have already been reported by Dadina et al.~(2005) and Cappi et al.~(2009). Our results are in complete agreement with theirs.

\emph{MCG-6-30-15:} the Fe K band of this source is known to be complex, with the overlapping of several spectral components, such as a broad relativistic emission line, narrow emission/absorption features and warm absorption. Therefore, even if the EPIC pn has a high effective area in this energy band, its moderate energy resolution can be not sufficient to unambiguously disentangle all the spectral features. For instance, Fabian et al.~(2002) and Vaughan \& Fabian (2004) found hints for a narrow absorption feature at $\sim$6.7~keV probably due to Fe XXV He$\alpha$ in the EPIC pn spectrum of this source. However, they stated that it is difficult to clearly discriminate between narrow absorption and emission features due to the spectral complexity.
Instead, from a deep spectral analysis of this source performed with the higher energy resolution Chandra HETG spectrometer, Young et al.~(2005) have been able to clearly detect two narrow, unresolved, absorption lines ascribable to Fe XXV He$\alpha$ and Fe XXVI Ly$\alpha$. The lines have blue-shifted velocities $\simeq$0.007c and EW$\simeq$$-20$~eV.
In our analysis we have not been able to unambiguously detect any blue-shifted narrow Fe K absorption features. This demonstrates the need to perform such systematic studies with other observatories as well, in order to exploit the different capabilities offered by each instrument.
However, it should be noted that from the energy-intensity contour plots in Appendix C there is evidence for narrow absorption lines at E$<$6.4~keV with F-test significance $\sim$99\% in three observations of the source (0029740101, 0029740701 and 0029740801). If modeled with inverted Gaussians, their parameters are: E$=$$4.20\pm0.05$~keV, $\sigma$$=$100~eV and EW$=$$-12\pm7$~eV, for the first, E$=$$4.49\pm0.05$~keV, $\sigma$$=$100~eV and EW$=$$-14\pm5$~eV, for the second, and E$=$$4.27\pm0.05$~keV, $\sigma$$=$10~eV and EW$=$$-10\pm5$~eV, for the third. As discussed in \S4.5, these lines could be possibly identified with blue-shifted Ca XX Ly$\alpha$ with velocities $\sim$0.02--0.08c or Fe XXV/XXVI 1s--2p with very large red-shifted velocities of $\sim$0.5--0.6c.

\emph{Ark~120:} a detailed analysis of the XMM-Newton spectrum of this source has already been published by Vaughan et al.~(2004). The authors argued that from a broad-band X-ray analysis of the combined RGS and EPIC pn data there is no evidence for an intrinsic warm absorber and placed upper limits on the ionic column densities that are substantially lower than those of more typical, absorbed Seyfert 1s. This led them to claim that Ark~120 could actually represent a  ``bare'' Seyfert 1 nucleus.\\
However, the reason for the lack of an X-ray warm absorber is quite unclear. It is plausible that a substantial column of ionized gas exists but it is either too highly ionized to show significant spectral features or lies out of the line of sight. These arguments are in agreement with our finding of a substantially blue-shifted Fe XXVI Ly$\alpha$ absorption line at the rest-frame energy of $\sim$9.18~keV.  

\emph{Mrk~110:} a deep analysis of the XMM-Newton spectrum of this source have been reported by Boller et al.~(2007). 
We did not detect any narrow blue-shifted Fe K absorption feature.

\emph{NGC~7469:} a detailed analysis of the XMM-Newton spectrum of this source has already been published by Blustin et al.~(2003). Our analysis confirms their overall results. We did not detect any significant Fe K absorption feature.

\emph{IRAS~05078+1626:} the analysis of the XMM-Newton EPIC pn spectrum of the source has never been published. We found that a simple power-law continuum plus a neutral Fe K$\alpha$ emission line at $\sim$6.4~keV provides a good fit to the spectral data in the 4--10~keV band. We did not find evidence for narrow Fe K absorption features. 

\emph{Mrk~279:} we did not find any published work on the XMM-Newton EPIC pn spectrum of the source. However, Fields et al.~(2007) found evidence for the presence of different layers of highly ionized absorbing material from the spectral analysis of the soft X-ray Chandra LETG data. The associated column densities are low, of the order of $N_H$$\sim 10^{20}$~cm$^{-2}$, and the outflow velocities are of few $\sim$1000~km/s. The authors also stated that the existence of a more ionized outflow component with iron ions from Fe XXIV to Fe XXVI cannot be ruled out. This is in agreement with our detection of a narrow absorption line ascribable to Fe XXV He$\alpha$ in the EPIC pn spectrum of the source, consistent with a rest-frame energy of $\simeq$6.69~keV.

\emph{NGC~526A:} the analysis of the XMM-Newton EPIC pn spectrum of the source has never been published. We found that a simple power-law continuum plus a neutral Fe K$\alpha$ emission line at $\sim$6.4~keV provides a good fit to the spectral data in the 4--10~keV band. We did not find evidence for narrow Fe K absorption features. 

\emph{NGC~3227:} a deep analysis of the EPIC pn spectrum of the source have been reported by Gondoin et al.~(2003) and Markowitz et al.~(2009). The 4--10~keV spectrum of the source is well modeled by a simple power-law continuum plus a neutral Fe K$\alpha$ emission line at $\sim$6.4~keV. However, in the first XMM-Newton observation there is the need for a substantial neutral absorption component with column density $N_H$$\sim10^{23}$~cm$^{-2}$. This has been reported to be consistent with an eclipsing event by a broad line region cloud (Lamer et al.~2003). Our overall results are consistent with these conclusions and we did not find evidence for narrow Fe K absorption features. 

\emph{NGC~7213:} a deep analysis of the XMM-Newton EPIC pn spectrum of the source in the Fe K band have been performed by Bianchi et al.~(2003a). We refer the reader to that paper for detailed information. However, our overall results are consistent with theirs and we did not find evidence for highly ionized Fe absorption features.

\emph{ESO~511-G030:} the analysis of the XMM-Newton EPIC pn spectrum of the source has never been published. We found that a simple power-law continuum plus a neutral Fe K$\alpha$ emission line at $\sim$6.4~keV provides a good fit to the spectral data in the 4--10~keV band. We did not find evidence for narrow Fe K absorption features. 

\emph{Mrk~79:} (or UGC3973) the spectral analysis of two snapshot XMM-Newton observations of the source have been reported by Gallo et al.~(2005). We did not consider those observations because of their too short exposures ($<10$~ks). Instead, here we report for the first time the spectral analysis of three new longer XMM-Newton EPIC pn observation of the source. In all the cases the baseline model is constituted by a simple power-law continuum plus a neutral Fe K$\alpha$ emission line at $\sim$6.4~keV, with parameters consistent with those of Gallo et al.~(2005). However, in one observation we have detected a narrow absorption feature ascribable to Fe XXVI Ly$\alpha$ at E$\simeq$7.63~keV, consistent with a blue-shifted velocity of $\sim$0.1c.    

\emph{NGC~4051:} a detailed analysis of the two XMM-Newton observations of this source has been published by Pounds et al.~(2004a). The overall spectral fit is consistent with ours. We confirm their detection of a narrow blue-shifted absorption line ascribable to Fe XXVI Ly$\alpha$ at the energy of $\sim$7.1~keV in the first observation. We detected a further absorption line at $\simeq$8.1~keV in the spectrum of the second observation. We interpreted this feature as a blue-shifted Fe XXVI Ly$\alpha$ absorption line. 

\emph{Mrk~766:} a detailed analysis of the XMM-Newton EPIC pn spectra of the source has already been reported by Pounds et al.~(2003c), Miller et al.~(2007) and Turner et al.~(2007). The spectral analysis of the Fe K band show the presence of both broad absorption troughs and narrow absorption line-like features at energies $\ga$7~keV. This complexity could indicate absorption from various layers of gas in different physical states.  
In fact, in two spectra we detected absorption features at energies greater than $7$~keV that could only be modeled by narrow, unresolved, absorption lines. We interpreted such features as blue-shifted Fe XXVI Ly$\alpha$ absorption lines. The presence of the first one at a rest-frame energy of $\sim7.3$~keV has already been suggested by Miller et al.~(2007) and Turner et al.~(2007). The detection of the second one at $\sim7.6$~keV has instead never been reported. 
However, in another observation of the source (0109141301) a significant broad absorption trough between E$\simeq$8--9~keV can be observed (see contour plot in Appendix B). As already reported by Pounds et al.~(2003c), this feature is well modeled by a photoelectric edge, whose energy is consistent with the rest-frame Fe XXV K edge (E$\simeq$8.8~keV). Therefore, we did not include this feature in Table~A.2.   

\emph{Mrk~841:} a detailed analysis of the XMM-Newton EPIC pn spectra of this source has already been published by Longinotti et al.~(2004) and Petrucci et al.~(2007). The authors tested different complex models in order to explain the broad-band X-ray spectral shape. However, our more simple spectral parametrization is in agreement with their results in the 4--10~keV band. Moreover, we detected a narrow absorption feature ascribable to a blue-shifted Fe XXVI Ly$\alpha$ resonance absorption line at the rest-frame energy of $\sim$7.2~keV.  

\emph{Mrk~704:} the XMM-Newton EPIC pn spectral analysis of this source has never been published in the literature. We have found the 4--10~keV spectrum to be well modeled by an absorbed power-law continuum plus a narrow neutral Fe K$\alpha$ emission line. We did not find evidence for additional narrow Fe K absorption features. 

\emph{Fairall~9:} the XMM-Newton EPIC pn spectral analysis of this source has already been published by Gondoin et al.~(2001). We confirm their overall results. We did not detect any narrow Fe K absorption line. 

\emph{ESO~323-G77:} the detailed XMM-Newton EPIC pn spectral analysis of this source has been published by Jim{\'e}nez-Bail{\'o}n et al.~(2008). We confirm their detection of a couple of narrow absorption features at the rest-frame energy of $\sim$6.7~keV and $\sim$7~keV ascribable to Fe XXV He$\alpha$ and Fe XXVI Ly$\alpha$ resonant absorption. 

\emph{1H419-577:} the detailed broad-band spectral analysis of the XMM-Newton EPIC pn observations of this source has been published by Pounds et al.~(2004b) and Fabian et al.~(2005). From their study, the authors argued that a good fit to the data can be provided either by an absorption dominated or by a reflection dominated model. However, these two models can not be distinguished in the narrow energy band we considered (E$\simeq$4--10~keV) and the continuum can be well approximated by a simple power-law. We detected a blue-shifted narrow absorption line ascribable to Fe XXVI Ly$\alpha$ at the rest-frame energy E$\simeq$7.23~keV in the first XMM-Newton observation of this source (obs. 0148000201), with F-test and Monte Carlo confidence levels of 99\% and 95.8\%, respectively. It should be noted that in the energy-intensity contour plot of this observation reported in Fig.~C.5 (upper right panel) there is evidence for a possible further narrow absorption line at rest-frame energy E$\simeq$8.4~keV with F-test confidence level $\sim$99\%. 
However, we did not include it in Table~A.2 and we did not consider it further here because the associated $\Delta\chi^2\simeq9.3$ yielded a Monte Carlo confidence level $\simeq$93\%, which is lower than the threshold value of 95\%. 

\emph{Mrk~335:} the detection of an absorption feature at the rest-frame energy of $\sim$5.9~keV in the first XMM-Newton observation of this source (obs. 0101040101) has been reported by Longinotti et al.~(2007a). If identified with Fe XXVI Ly$\alpha$ resonant absorption, this would indicate a red-shifted velocity for the line of $\sim$0.18~c. We find evidence for such a feature in the energy-intensity contour plot in Fig.~C.5. However, the measured detection confidence level is in the range $\sim$95--99\% using the F-test. 
Moreover, also in the third observation (obs. 0510010701) we find evidence for a narrow absorption line at E$<$6.4~keV with F-test confidence level $\sim$99\% (see Fig.~C.5). If modeled as an inverted Gaussian, the rest-frame energy is E$=$$4.58\pm0.06$~keV, $\sigma$$=$100~eV and EW$=$$-80\pm30$~eV. As already discussed in \S4.5, there are different possible identifications for the line, such as: blue-shifted Ca XX Ly$\alpha$ with velocity $\sim$0.1c or red-shifted Fe XXV/XXVI 1s--2p with velocity $\sim$0.4--0.5c. However, from a further detailed spectral analysis, we find evidence for a possible additional absorption line at rest-frame energy E$\simeq$3.9~keV with F-test confidence level $\simeq$97\%. Therefore, we could also speculate an association of both lines with rest-frame Ca XIX He$\alpha$ and Ca XIX He$\beta$, which are expected exactly at the energies E$\simeq$3.9~keV at E$\simeq$4.58~keV (Verner et al.~1996). Finally, we did not detect any significant narrow blue-shifted absorption lines at E$\ga$7~keV.

\emph{ESO~198-G024:} the analysis of the XMM-Newton EPIC pn spectrum of this source has already been reported by Guainazzi (2003) and Porquet et al.~(2004). We agree with their overall results in the 4--10~keV band and we did not find evidence for highly ionized Fe K absorption lines.
However, in the energy-intensity contour plot in Fig.~C.5 there is evidence for a possible narrow absorption line at E$<$6.4~keV with F-test confidence level $\sim$99\%. If modeled as an inverted Gaussian, the rest-frame energy is E$=$$4.59\pm0.04$~keV, $\sigma$$=$10~eV and EW$=$$-14\pm6$~eV.
As already discussed in \S4.5, the line could be possibly associated with blue-shifted Ca XX Ly$\alpha$ with velocity $\sim$0.1c or red-shifted Fe XXV/XXVI 1s--2p with velocity $\sim$0.4--0.5c.
However, as for the line in Mrk~335, the centroid energy of the feature is consistent with Ca XIX He$\beta$ at the rest-frame energy E$\simeq$4.58~keV. If associated with such a line, we would expect to observe also the Ca XIX He$\alpha$ at E$\simeq$3.9~keV. Instead, at this energy we detect only a narrow emission line with F-test confidence level $\simeq$98\%. Therefore, even if appealing, this last identification is not straightforward.

\emph{Mrk~290:} we did not find any published analysis of the XMM-Newton EPIC pn spectrum of the source. We have detected a narrow absorption line ascribable to Fe XXVI Ly$\alpha$ in one out of four observations. The line rest-frame energy is E$\simeq$8~keV, which would suggest a blue-shifted velocity of $\sim$0.14c.

\emph{Mrk~205:} the Fe K band analysis of the XMM-Newton EPIC pn spectrum of the source has already been published by Reeves et al.~(2001) and Page et al.~(2003). We found that the spectrum in this band can be well modeled by a simple power-law continuum plus narrow Gaussian emission lines. However, we detected an additional narrow absorption feature at the rest-frame energy of $\sim$7.7~keV in one observation. If identified with Fe XXVI Ly$\alpha$ resonant absorption this would suggest an outflow velocity of $\sim$0.1c.

\emph{Mrk~590:} the detailed analysis of the XMM-Newton EPIC pn spectrum of the source has already been published by Longinotti et al.~(2007b). We confirm their overall results. We did not detect any narrow Fe K absorption lines.    

\emph{H~577-385:} the detailed spectral analysis of the XMM-Newton EPIC pn observations of the source has already been reported by Longinotti et al.~(2009). We agree with their overall results. We did not find evidence for Fe K absorption features.

\emph{TON~S180:} we did not find any published analysis of the XMM-Newton EPIC pn spectrum of the source. The 4--10~keV band is well modeled by a simple power-law continuum without additional spectral features.

\emph{PG~1211+143:} a detailed analysis of the XMM-Newton EPIC pn spectra of the source has already been reported by Pounds et al.~(2003a), Pounds \& Page (2006) and Pounds \& Reeves (2009). In particular, Pounds et al~(2003a) clearly detected a narrow blue-shifted Fe K absorption line at the rest-frame energy of $\sim7.6$~keV in the pn spectrum of the first XMM observation. The identification of this line with Fe XXVI Ly$\alpha$ suggests a relativistic outflow velocity of $\sim0.08$c. Moreover, the presence of a highly ionized absorber in this source was confirmed also by the detection of several other blue-shifted narrow absorption lines due to lighter elements in the RGS as well. In a subsequent re-analysis of this observation using also the MOS data Pounds \& Page (2006) have been able to confirm the presence of additional absorption lines, yielding a revised outflow velocity of $\sim$0.13--0.15c. Furthermore, the authors stated that they removed the ambiguity in the identification of the $\sim7.6$~keV absorption feature, with a preference for Fe XXV He$\alpha$. This would imply a consequently higher outflow velocity of $\sim0.13$c. We adopted this latter line identification. 

\emph{Ark~564:} we did not find any published analysis of the XMM-Newton EPIC pn spectrum of the source. We found the 4--10~keV spectrum to be well modeled by a simple power-law continuum and Gaussian emission lines. We did not detect additional Fe K absorption features.

\emph{MCG-5-23-16:} a detailed spectral analysis of the XMM-Newton EPIC pn observations of this source has already been published by Dewangan et al.~(2003) and Braito et al.~(2007). We agree with their overall results in the 4--10~keV band. Moreover, we confirm the detection made by Braito et al.~(2007) of a narrow absorption feature at the rest-frame energy of $\sim$7.8~keV in the last observation. If identified with Fe XXVI Ly$\alpha$ resonant absorption, the feature indicates a substantial blue-shifted velocity of $\sim$0.12c.

\emph{NGC~5506:} the analysis of the XMM-Newton spectra of the source have been reported by Matt et al.~(2001) and Bianchi et al.~(2003b). We agree with their overall results and we did not find evidence for narrow ionized Fe K absorption lines.   

\emph{NGC~7172:} the 4--10~keV spectra of the XMM-Newton EPIC pn observations of the source have never been published. We found them to be well modeled by a simple absorbed power-law continuum plus a neutral Fe K$\alpha$ emission line. We did not detect additional Fe K absorption features.

\emph{NGC~7314:} we did not find any reported analysis of the XMM-Newton EPIC pn spectrum of the source. We found the 4--10~keV band to be well modeled by a simple absorbed power-law continuum plus a narrow neutral Fe K$\alpha$ emission line. We did not find evidence for narrow Fe K absorption lines.

\emph{NGC~2110:} a detailed analysis of the broad-band X-ray spectrum of the source observed by XMM-Newton has been reported by Evans et al.~(2007). We agree with their results in the 4--10~keV band. We did not detect any Fe K absorption line.

\emph{NGC~4507:} a detailed broad-band analysis of the XMM-Newton EPIC pn spectrum of the source has been reported by Matt et al.~(2004). We agree with their results in the 4--10~keV band. Furthermore, we have detected a narrow absorption feature at the rest-frame energy of $\sim$8.3~keV. If associated with Fe XXVI Ly$\alpha$ resonant absorption, the corresponding blue-shifted velocity is substantial, of the order of 0.18c.

\emph{NGC~7582:} a broad-band spectral analysis of the XMM-Newton observations of this source has been reported by Piconcelli et al.~(2007). 
However, we report for the first time the detection a narrow absorption feature at the rest-frame energy of E$\simeq$9~keV in the first observation (obs. 0112310201). The description of this line has been included in Table~A.2.
Moreover, in the contour plot in Fig.~C.7 there is evidence for a narrow absorption line at the rest-frame energy  E$\sim$4.5~keV with F-test confidence contours $\ga$99\%. 
If modeled with an inverted Gaussian, we derived E$=$$4.52\pm0.03$~keV, $\sigma$$=$10~eV, EW$=$$-76^{+18}_{-21}$~eV and F-test detection confidence level $\simeq$99.9\% ($\Delta\chi^2$$\simeq$13 for two more model parameters). 
Furthermore, from a detailed spectral analysis we also found the presence of two further narrow ($\sigma$$=$10~eV) absorption lines with F-test confidence levels $\simeq$95--98\% at E$=$$4.12\pm0.03$~keV, with EW$=$$-63^{+18}_{-25}$~eV, and E$=$$5.22\pm0.03$~keV, with EW$=$$-41^{+19}_{-22}$~eV.
If the absorption line at E$\simeq$9~keV is identified with the Fe XXV He$\alpha$ transition (E$\simeq$6.7~keV), these low energy lines turn out to be consistent with being absorption from Ar XVII He$\alpha$ (E$\simeq$3.14~keV), Ar XVIII Ly$\alpha$ (E$\simeq$3.3~keV) and Ca XIX He$\alpha$ (E$\simeq$3.9~keV), blue-shifted by the same common velocity of $\simeq$0.255c.
Moreover, we performed a test fitting the spectrum with the baseline model and adding four absorption lines with energies fixed to these expected values, letting the common blue-shift free to vary. 
We obtained a very good fit with a $\chi^2$ improvement of 35 for five additional parameters and the derived blue-shifted velocity is $+0.255\pm0.003$c, completely consistent with the value determined only from the Fe K line (see Table~A.2).
The probability for random fluctuations to give rise to this series of lines with the exact energy spacing and common blue-shift is very low, $\simeq$$2\times 10^{-6}$.

\end{appendix}

\newpage

\onecolumn{

\begin{appendix}
\section{Ratios and contour plots}

   \begin{figure*}[ht]
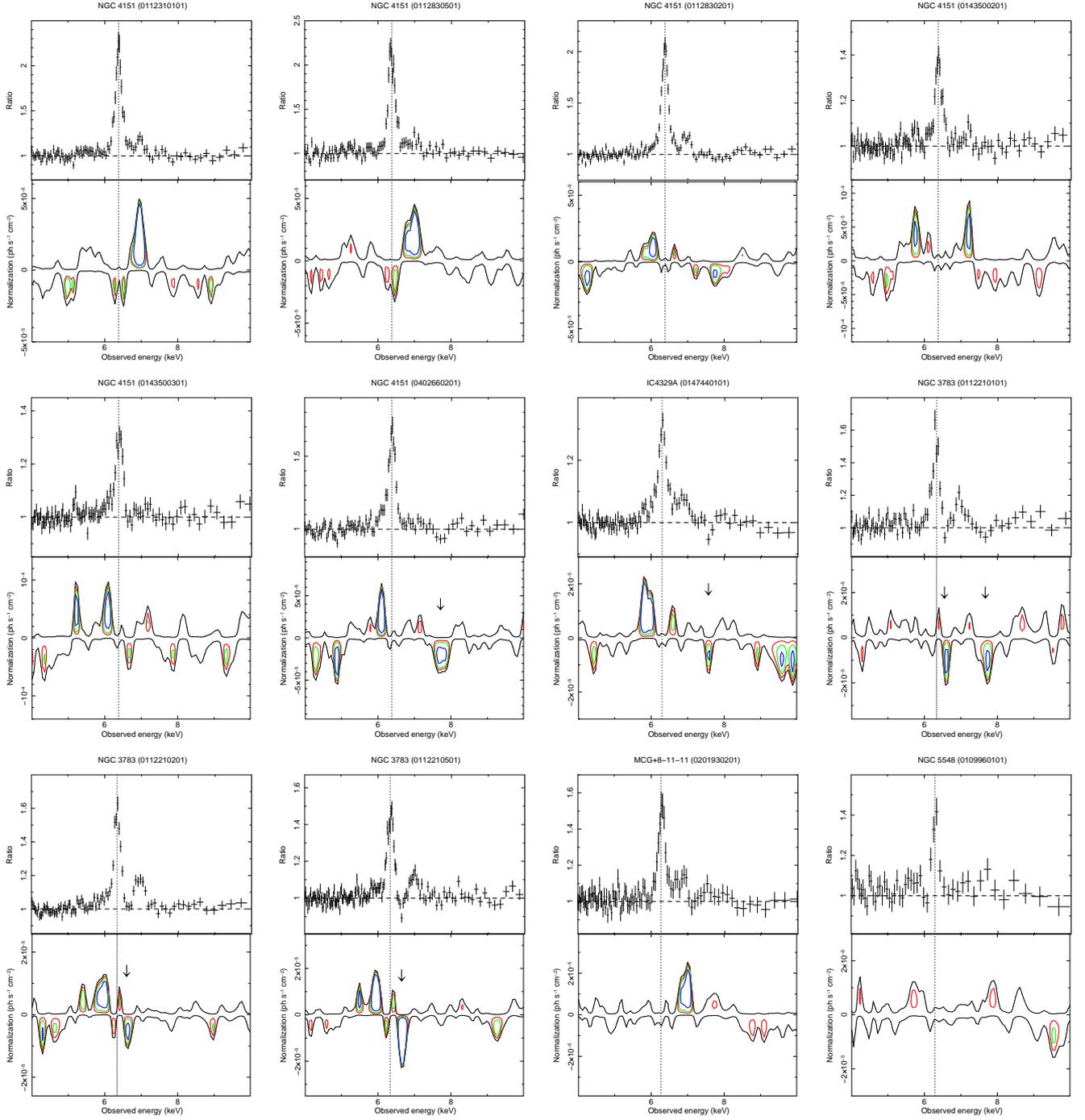

   \centering


    \includegraphics[width=3cm,height=4cm,angle=270]{ngc4151_0112310101_ratio.ps}
\hspace{0.3cm}
    \includegraphics[width=3cm,height=4cm,angle=270]{ngc4151_0112830501_ratio.ps}
\hspace{0.3cm}
    \includegraphics[width=3cm,height=4cm,angle=270]{ngc4151_0112830201_ratio.ps}
\hspace{0.3cm}
    \includegraphics[width=3cm,height=4cm,angle=270]{ngc4151_0143500201_ratio.ps}

\vspace{-0.07cm}

\hspace{-0.12cm}
    \includegraphics[width=3cm,height=4cm,angle=270]{ngc4151_0112310101_cont.ps}
\hspace{0.3cm}
    \includegraphics[width=3cm,height=4cm,angle=270]{ngc4151_0112830501_cont.ps}
\hspace{0.3cm}
    \includegraphics[width=3cm,height=4cm,angle=270]{ngc4151_0112830201_cont.ps}
\hspace{0.3cm}
    \includegraphics[width=3cm,height=4cm,angle=270]{ngc4151_0143500201_cont.ps}


\vspace{0.2cm}

    \includegraphics[width=3cm,height=4cm,angle=270]{ngc4151_0143500301_ratio.ps}
\hspace{0.3cm}
    \includegraphics[width=3cm,height=4cm,angle=270]{ngc4151_0402660201_ratio.ps}
\hspace{0.3cm}
    \includegraphics[width=3cm,height=4cm,angle=270]{ic4329a_0147440101_ratio.ps}
\hspace{0.3cm}
    \includegraphics[width=3cm,height=4cm,angle=270]{ngc3783_0112210101_ratio.ps}

\vspace{-0.07cm}

\hspace{-0.12cm}
    \includegraphics[width=3cm,height=4cm,angle=270]{ngc4151_0143500301_cont.ps}
\hspace{0.3cm}
    \includegraphics[width=3cm,height=4cm,angle=270]{ngc4151_0402660201_cont.ps}
\hspace{0.3cm}
    \includegraphics[width=3cm,height=4cm,angle=270]{ic4329a_0147440101_cont_BIS.ps}
\hspace{0.3cm}
    \includegraphics[width=3cm,height=4cm,angle=270]{ngc3783_0112210101_cont.ps}


\vspace{0.2cm}

    \includegraphics[width=3cm,height=4cm,angle=270]{ngc3783_0112210201_ratio.ps}
\hspace{0.3cm}
    \includegraphics[width=3cm,height=4cm,angle=270]{ngc3783_0112210501_ratio.ps}
\hspace{0.3cm}
    \includegraphics[width=3cm,height=4cm,angle=270]{mcg+8-11-11_0201930201_ratio.ps}
\hspace{0.3cm}
    \includegraphics[width=3cm,height=4cm,angle=270]{ngc5548_0109960101_ratio.ps}

\vspace{-0.07cm}

\hspace{-0.12cm}
    \includegraphics[width=3cm,height=4cm,angle=270]{ngc3783_0112210201_cont.ps}
\hspace{0.3cm}
    \includegraphics[width=3cm,height=4cm,angle=270]{ngc3783_0112210501_cont.ps}
\hspace{0.3cm}
    \includegraphics[width=3cm,height=4cm,angle=270]{mcg+8-11-11_0201930201_cont.ps}
\hspace{0.3cm}
    \includegraphics[width=3cm,height=4cm,angle=270]{ngc5548_0109960101_cont.ps}

   \caption{Ratio against the continuum (\emph{upper panel}) and contour plots with respect to the best-fit baseline model (\emph{lower panel}; 68\% (red), 90\% (green), 99\% (blue) levels) for the Type 1 sources. The Fe K absorption lines are pointed by arrows. The vertical line refer to Fe I K$\alpha$.}
    \end{figure*}

   \begin{figure*}[ht]
   \centering


    \includegraphics[width=3cm,height=4cm,angle=270]{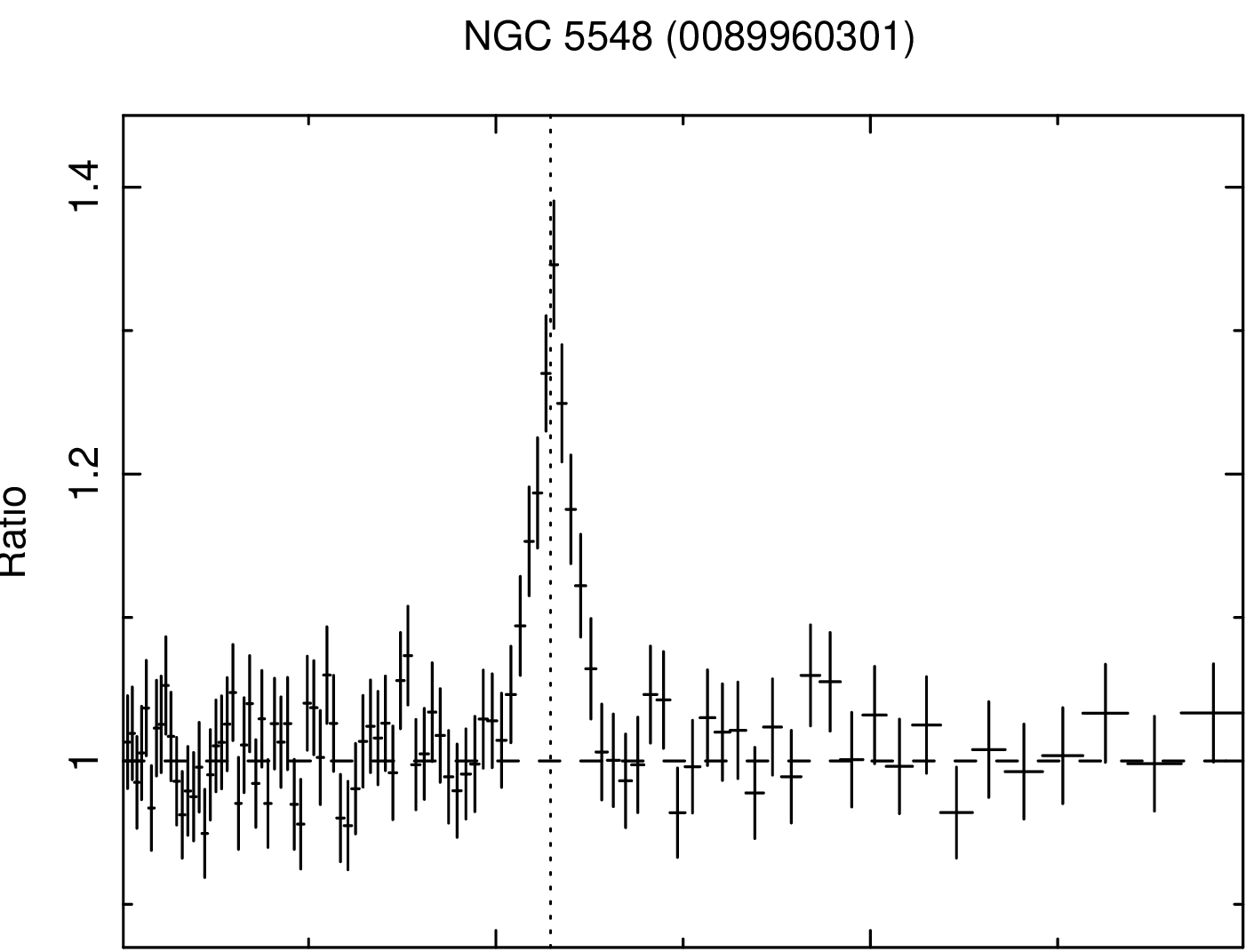}
\hspace{0.3cm}
    \includegraphics[width=3cm,height=4cm,angle=270]{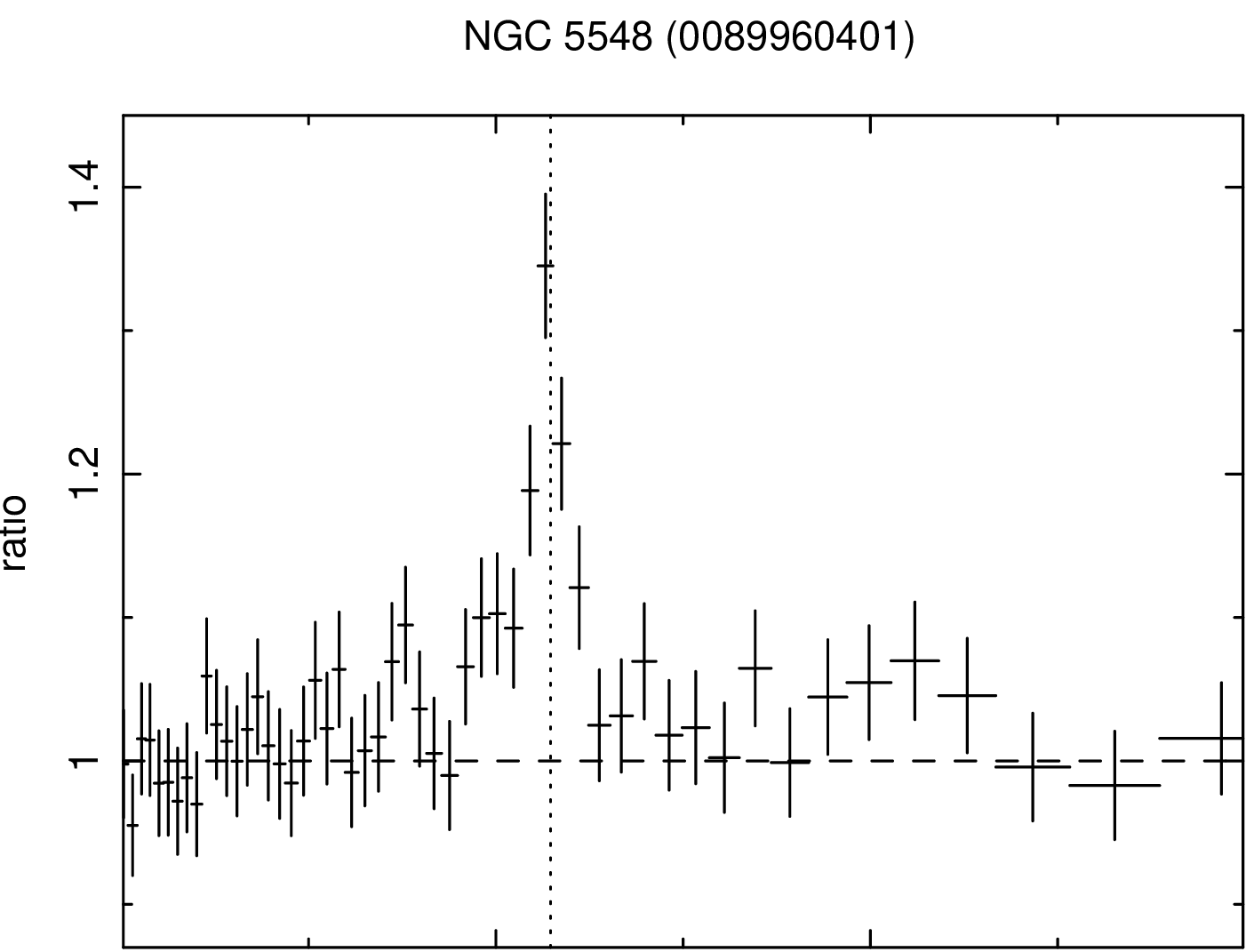}
\hspace{0.3cm}
    \includegraphics[width=3cm,height=4cm,angle=270]{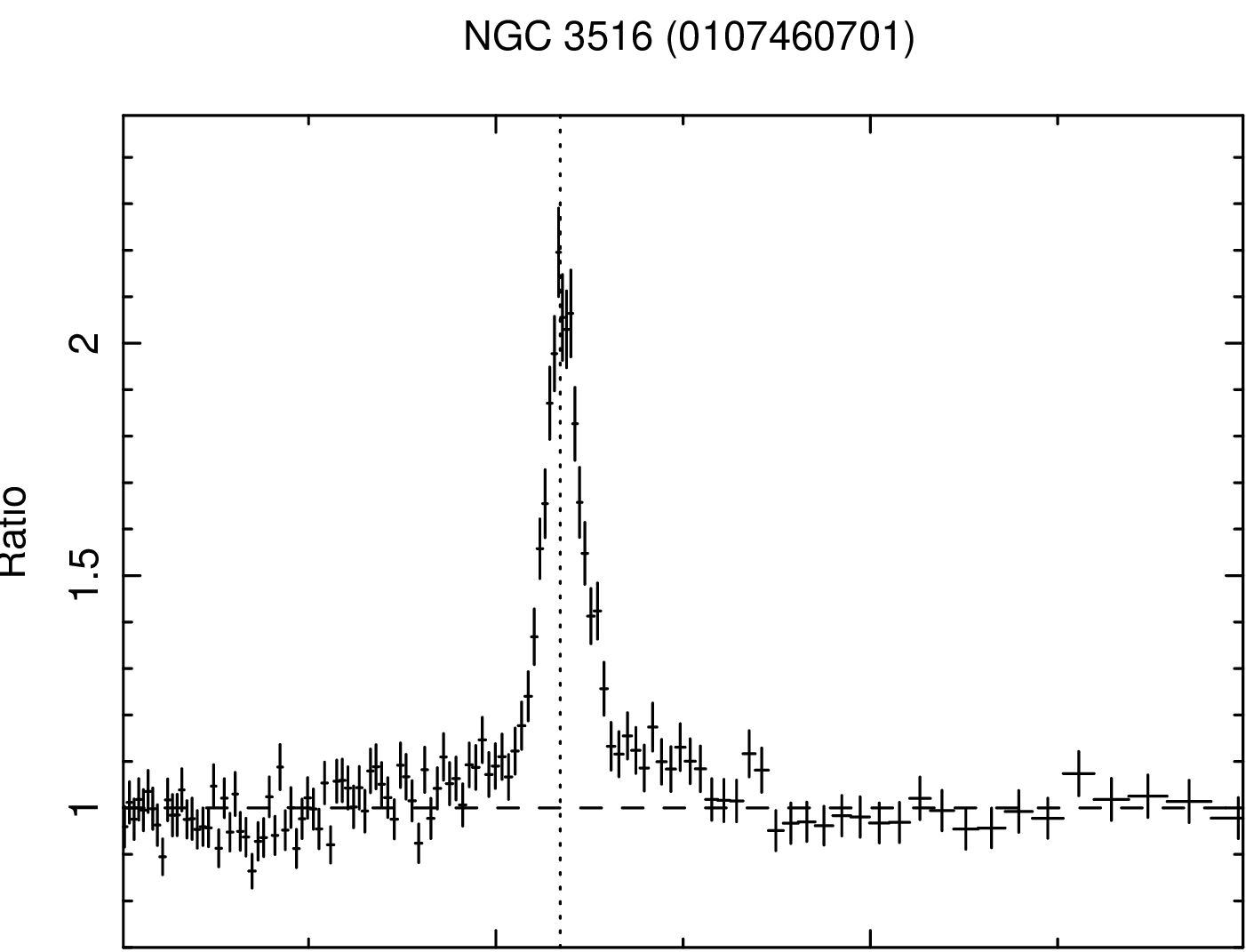}
\hspace{0.3cm}
    \includegraphics[width=3cm,height=4cm,angle=270]{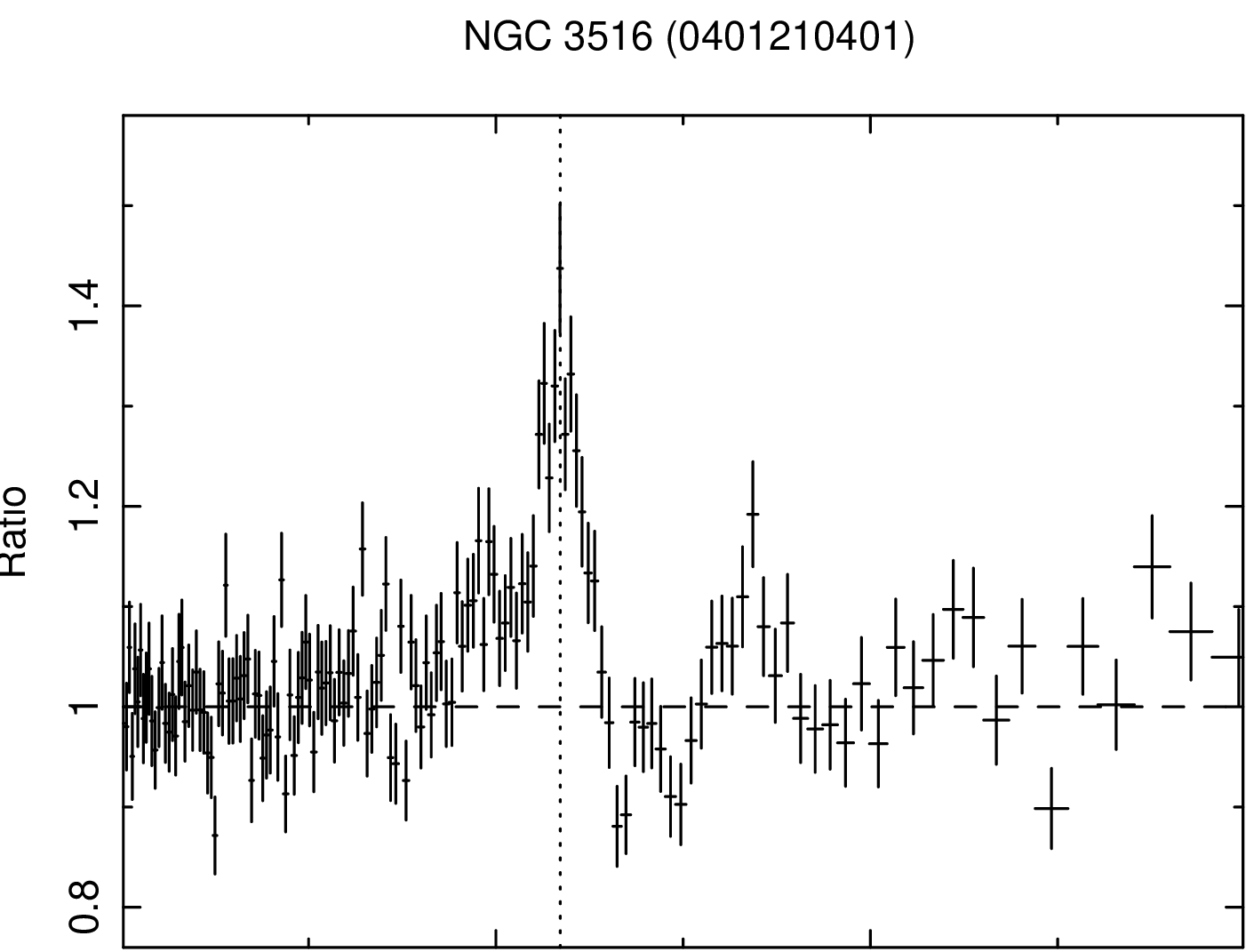}

\vspace{-0.07cm}

\hspace{-0.12cm}
    \includegraphics[width=3cm,height=4cm,angle=270]{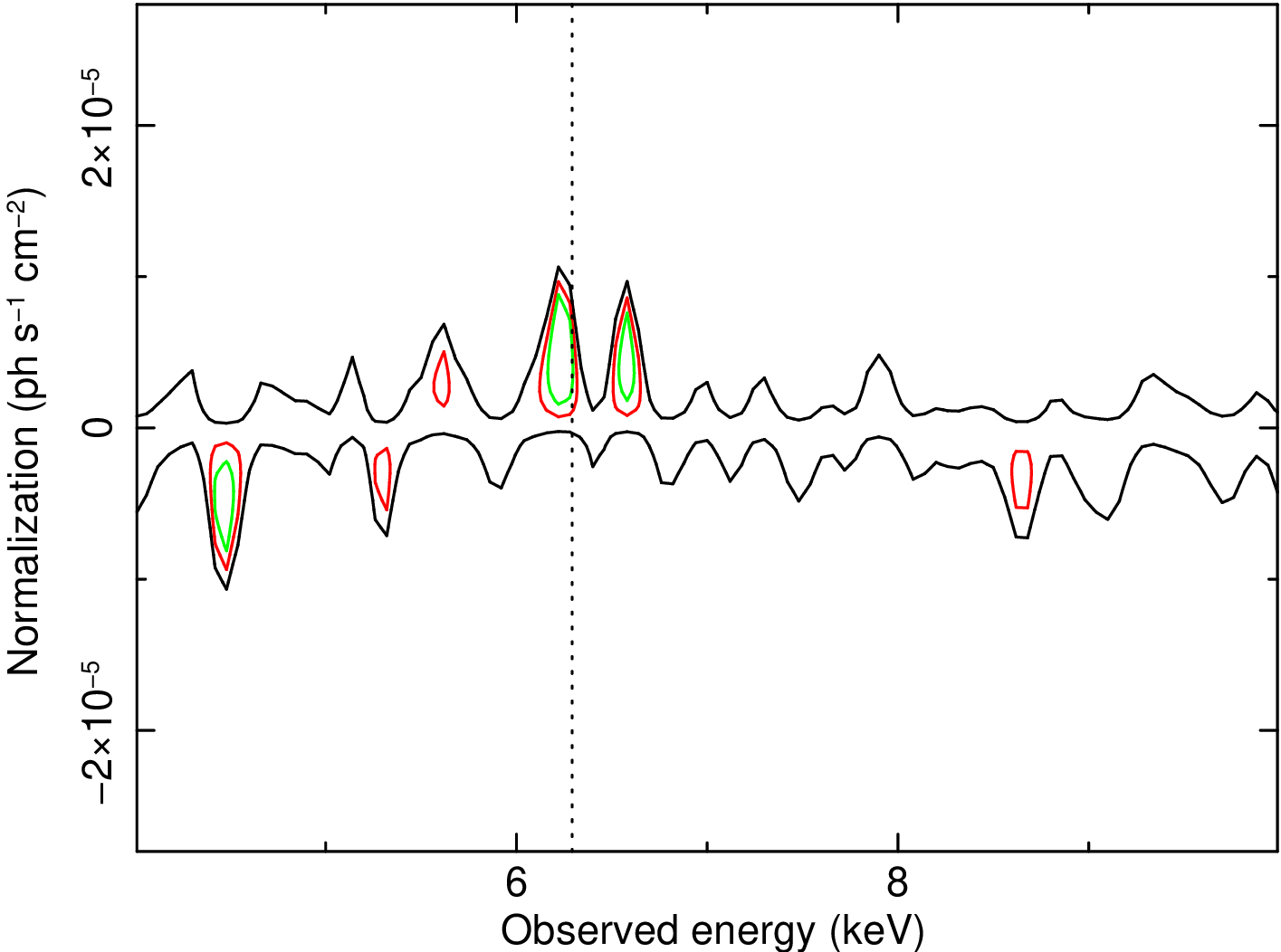}
\hspace{0.3cm}
    \includegraphics[width=3cm,height=4cm,angle=270]{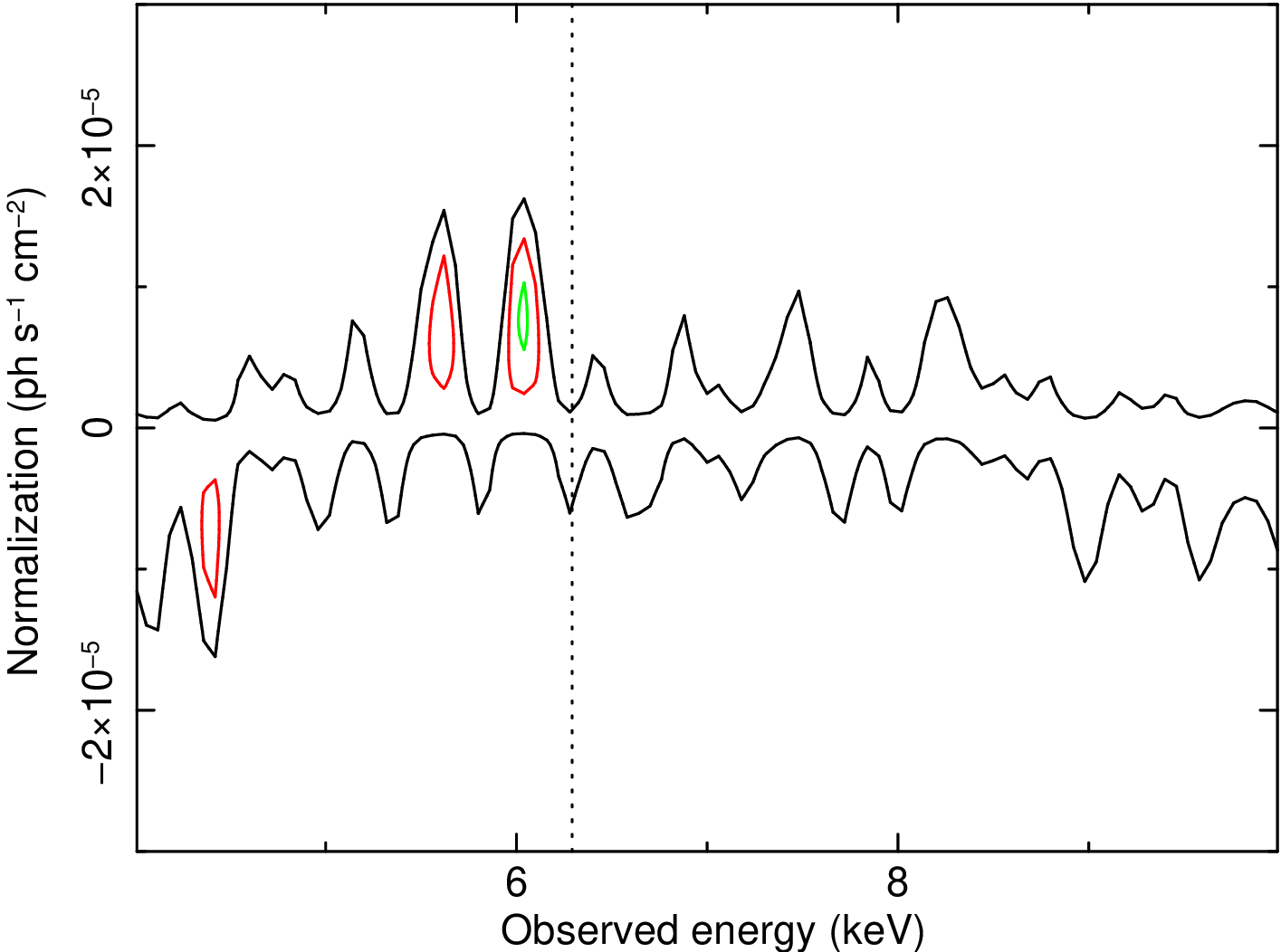}
\hspace{0.3cm}
    \includegraphics[width=3cm,height=4cm,angle=270]{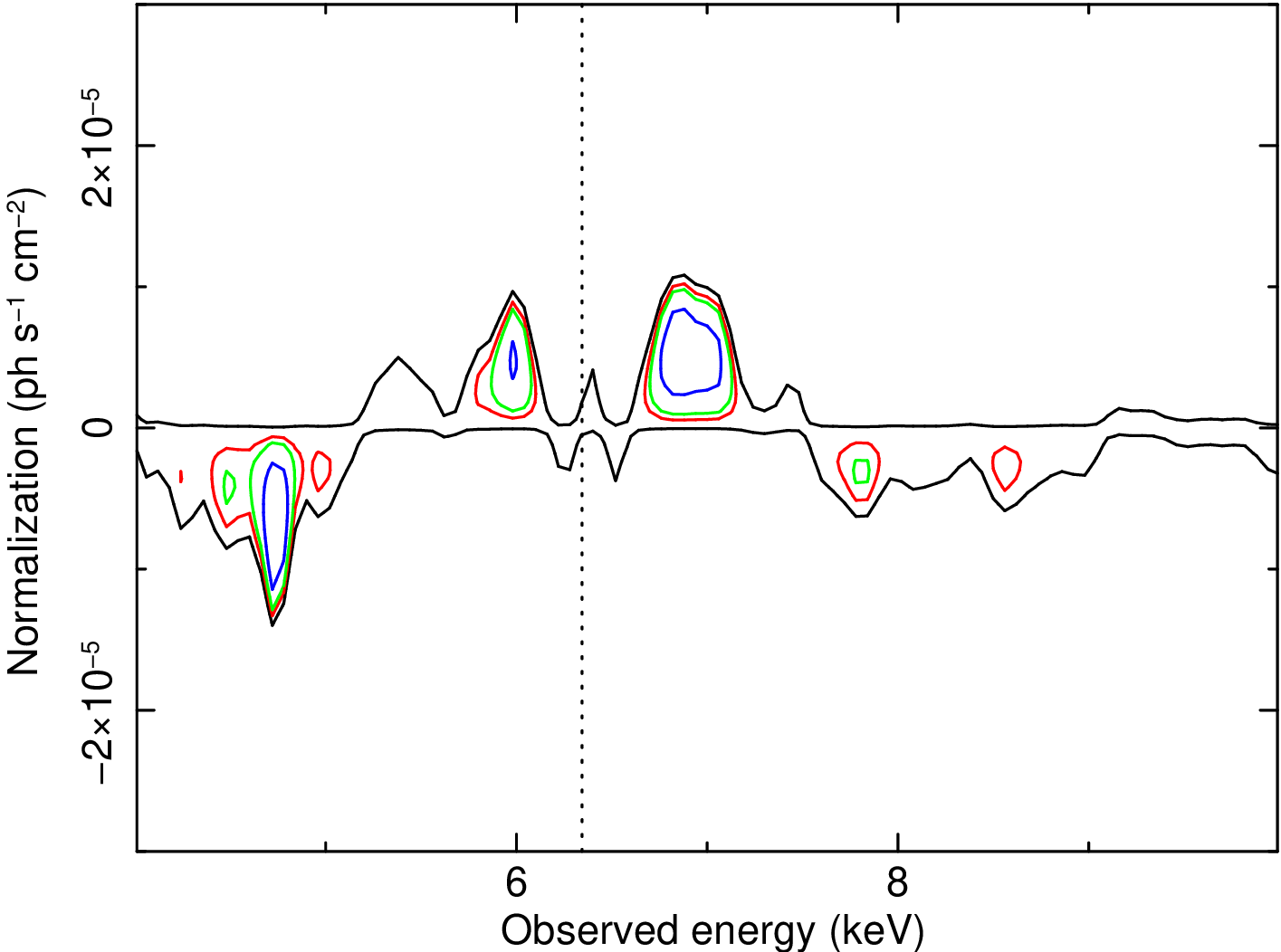}
\hspace{0.3cm}
    \includegraphics[width=3cm,height=4cm,angle=270]{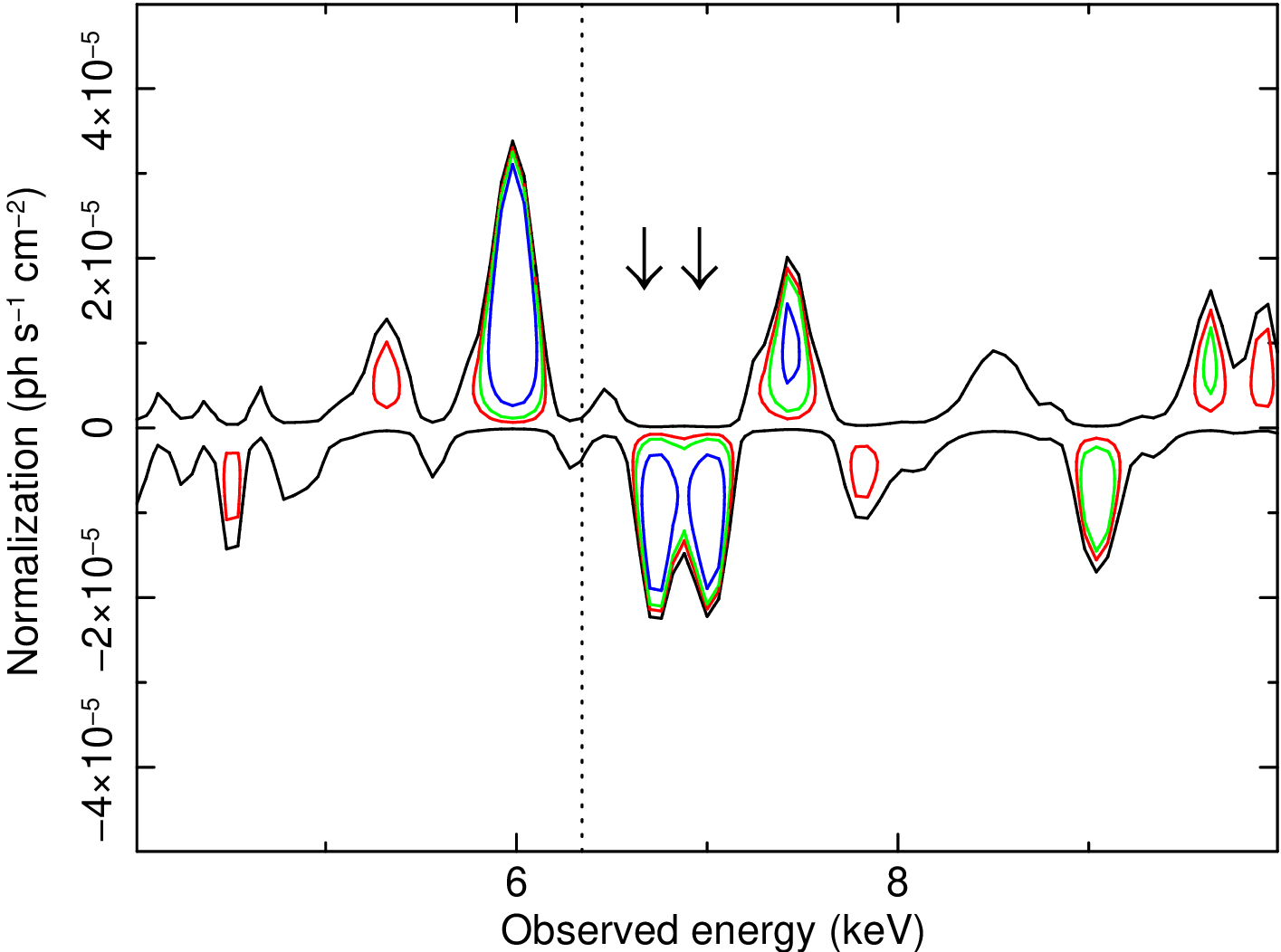}


\vspace{0.2cm}

    \includegraphics[width=3cm,height=4cm,angle=270]{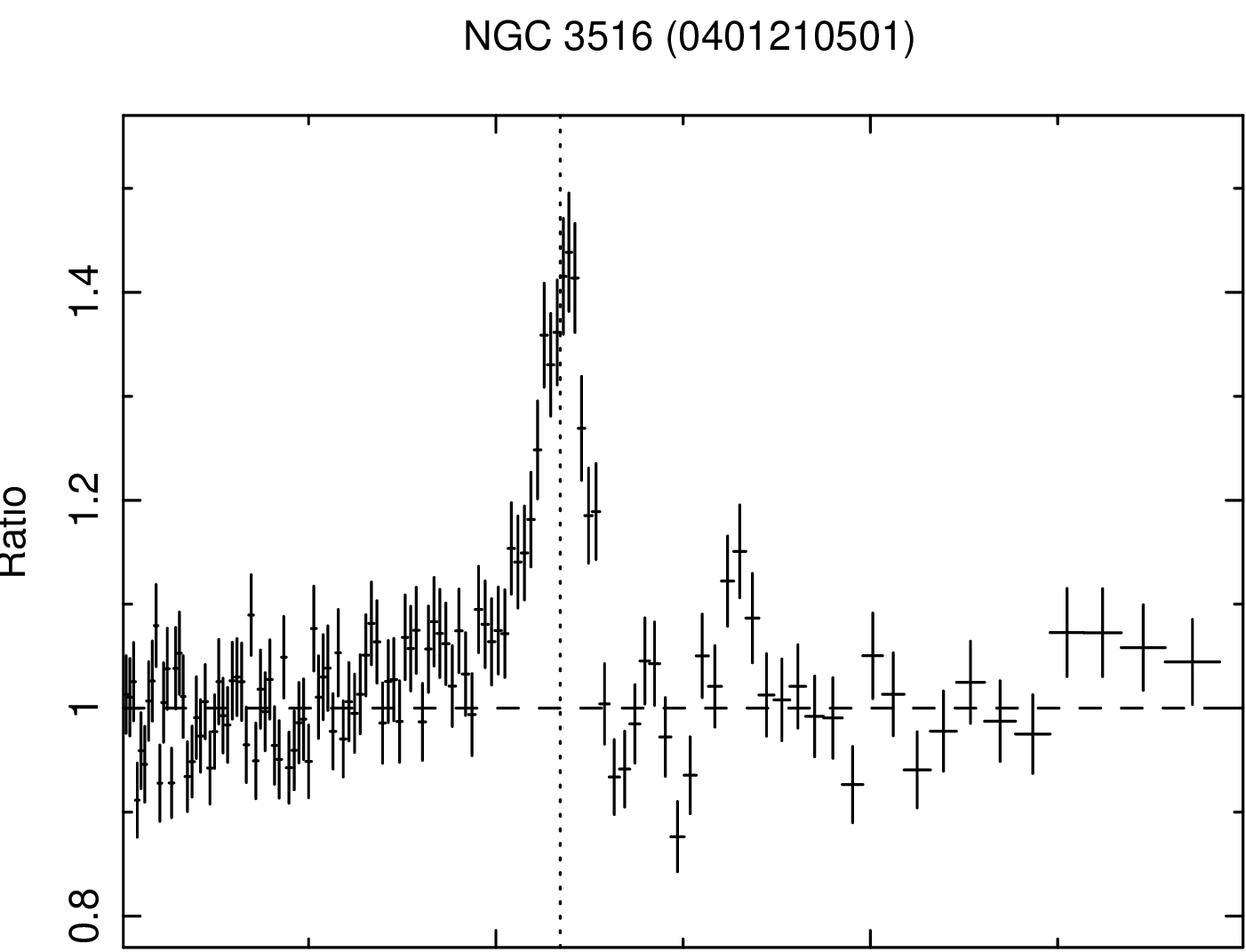}
\hspace{0.3cm}
    \includegraphics[width=3cm,height=4cm,angle=270]{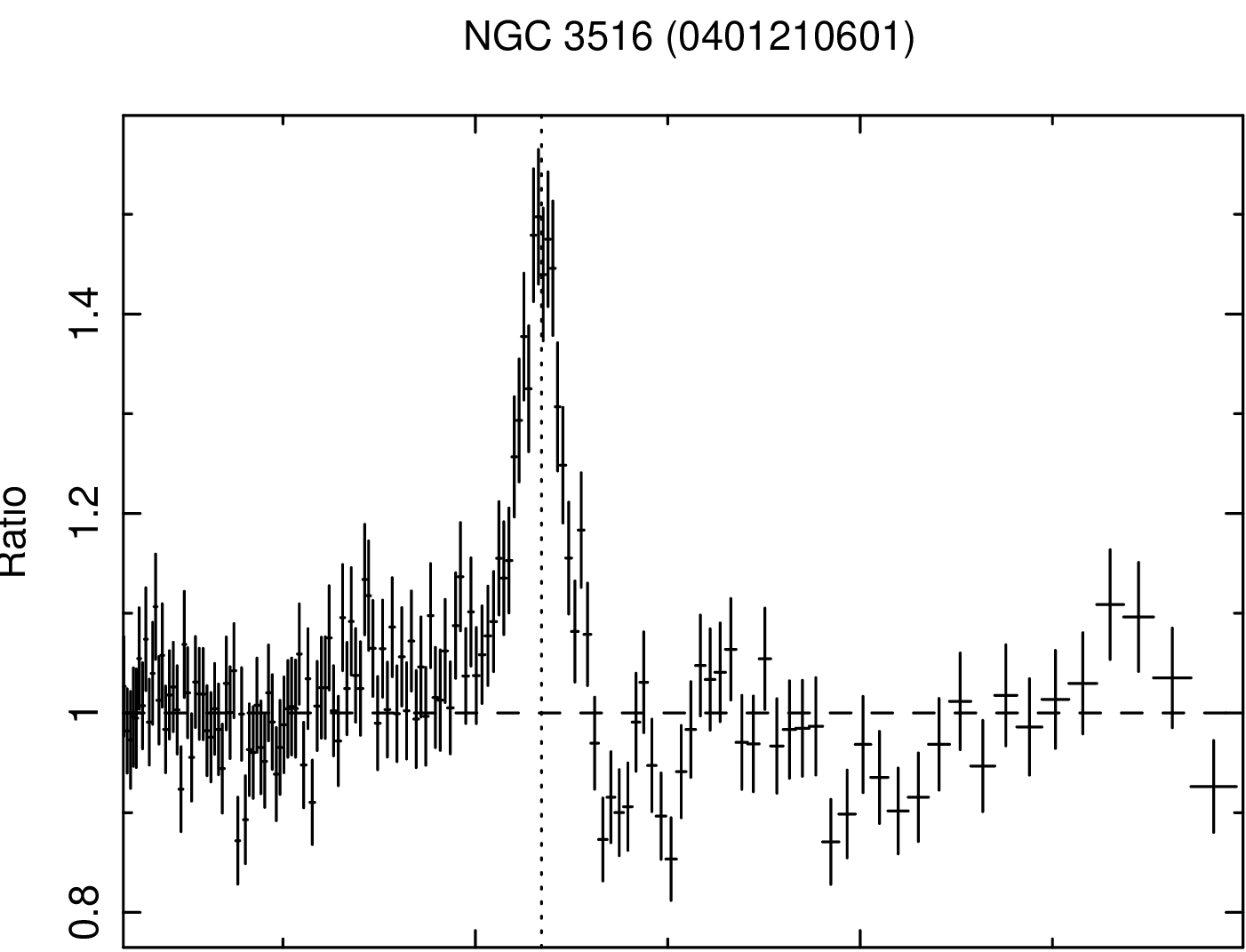}
\hspace{0.3cm}
    \includegraphics[width=3cm,height=4cm,angle=270]{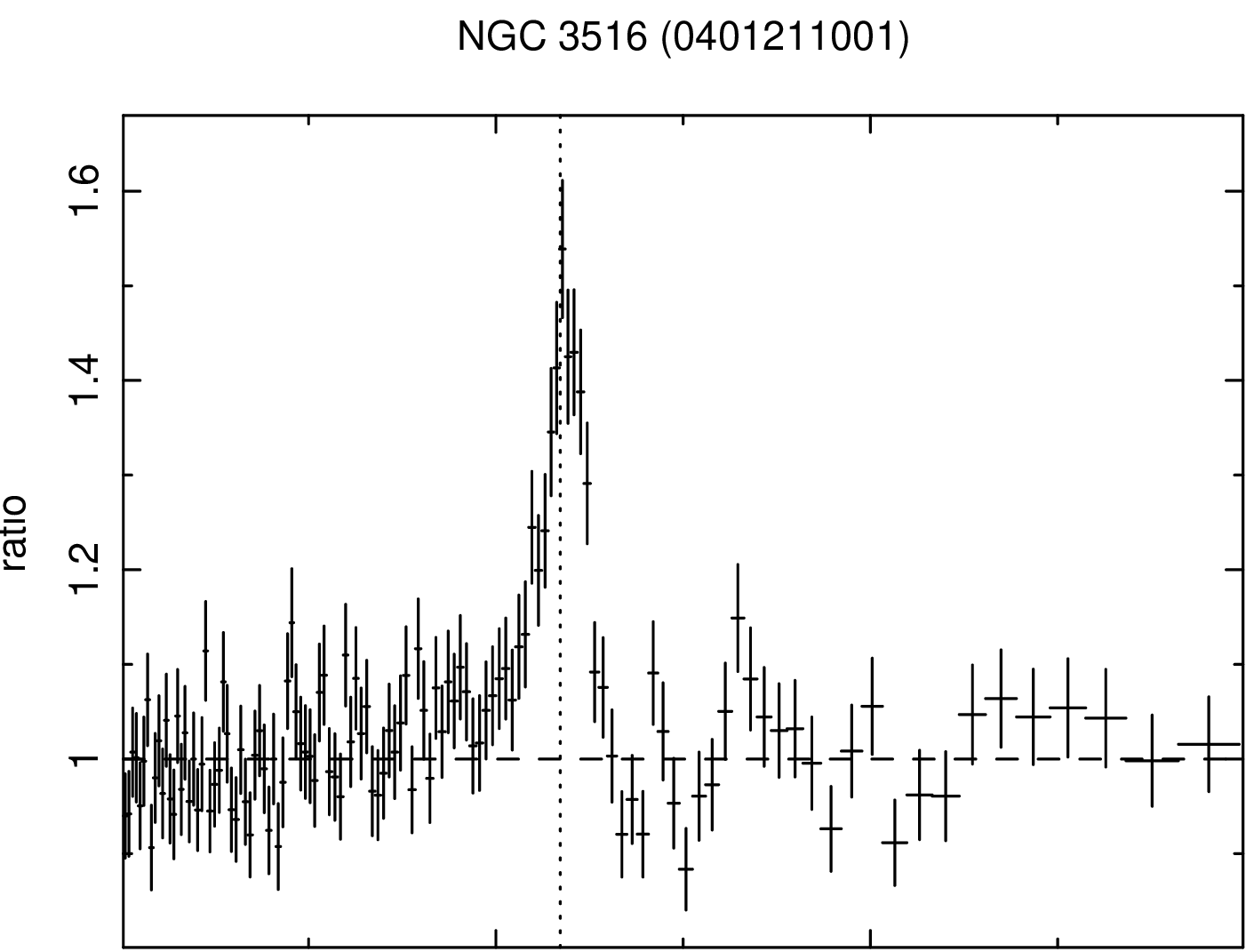}
\hspace{0.3cm}
    \includegraphics[width=3cm,height=4cm,angle=270]{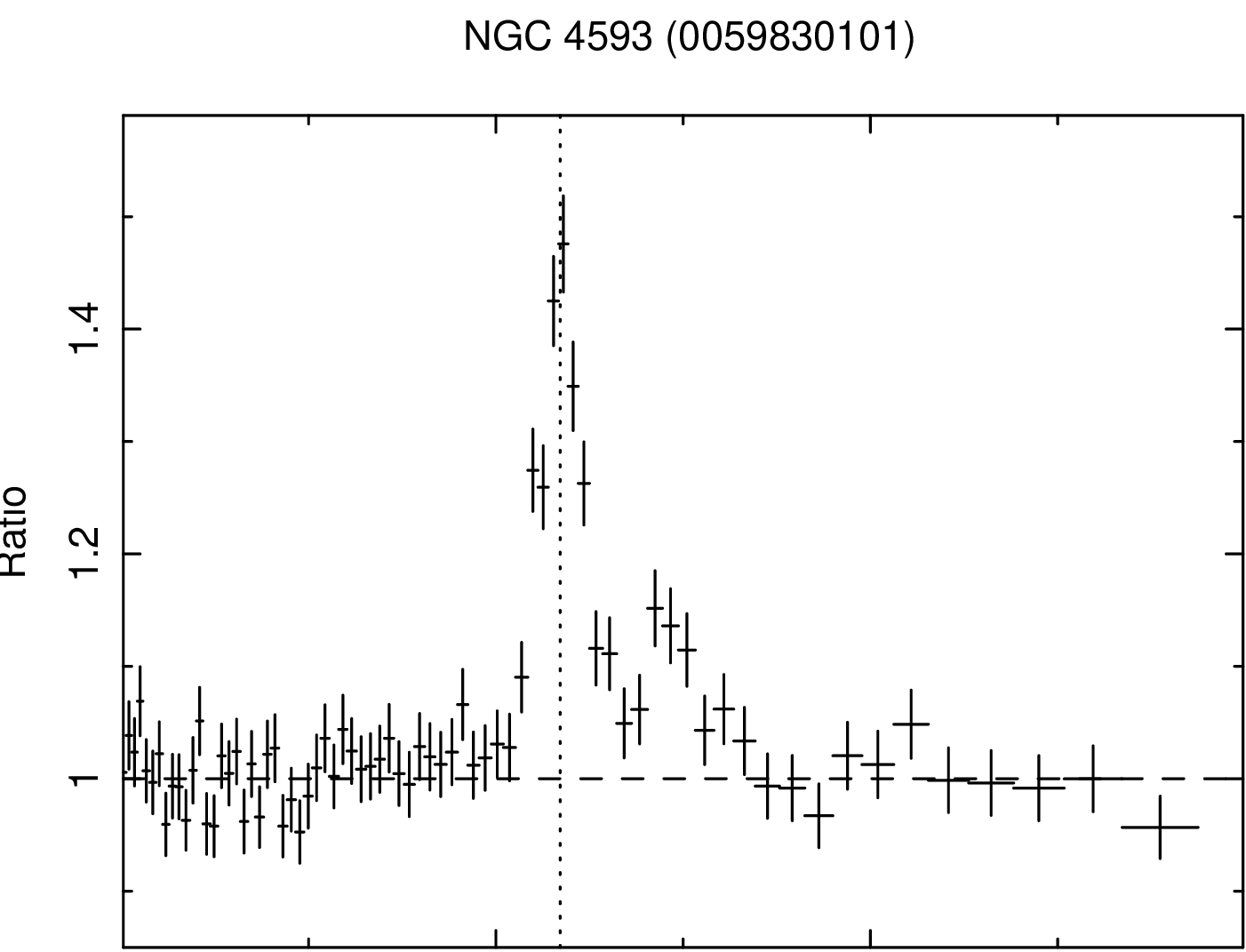}

\vspace{-0.07cm}

\hspace{-0.12cm}
    \includegraphics[width=3cm,height=4cm,angle=270]{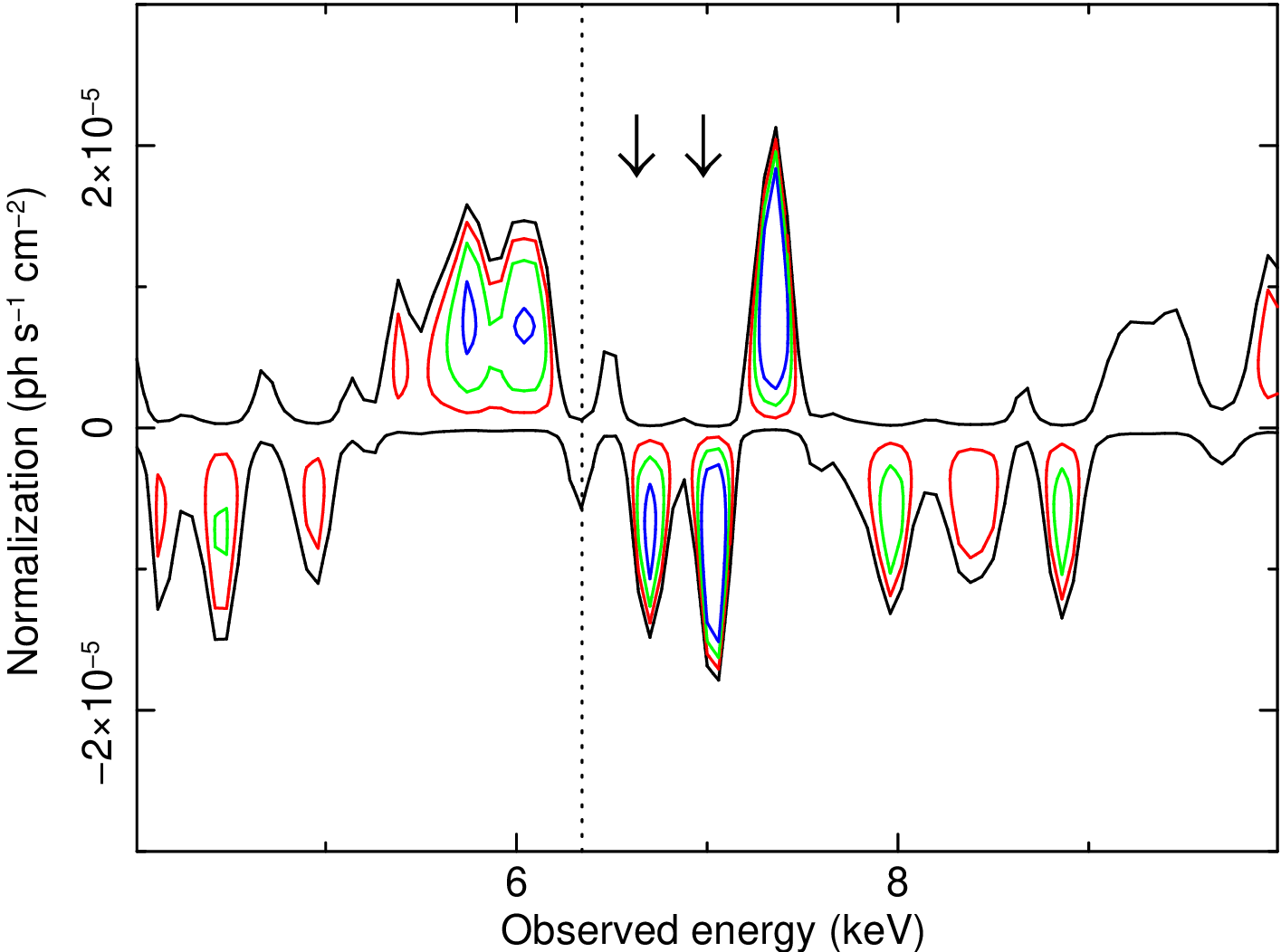}
\hspace{0.3cm}
    \includegraphics[width=3cm,height=4cm,angle=270]{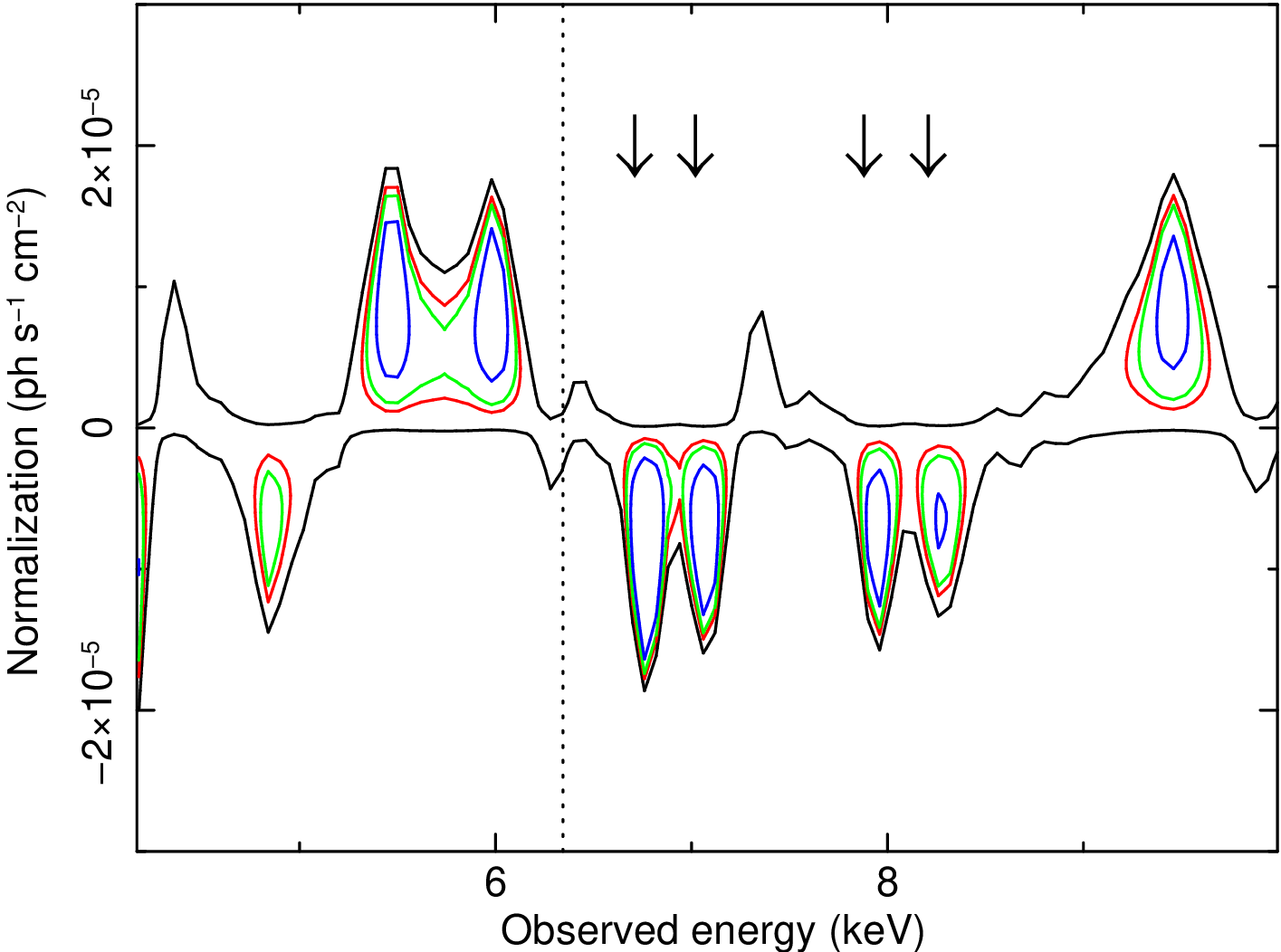}
\hspace{0.3cm}
    \includegraphics[width=3cm,height=4cm,angle=270]{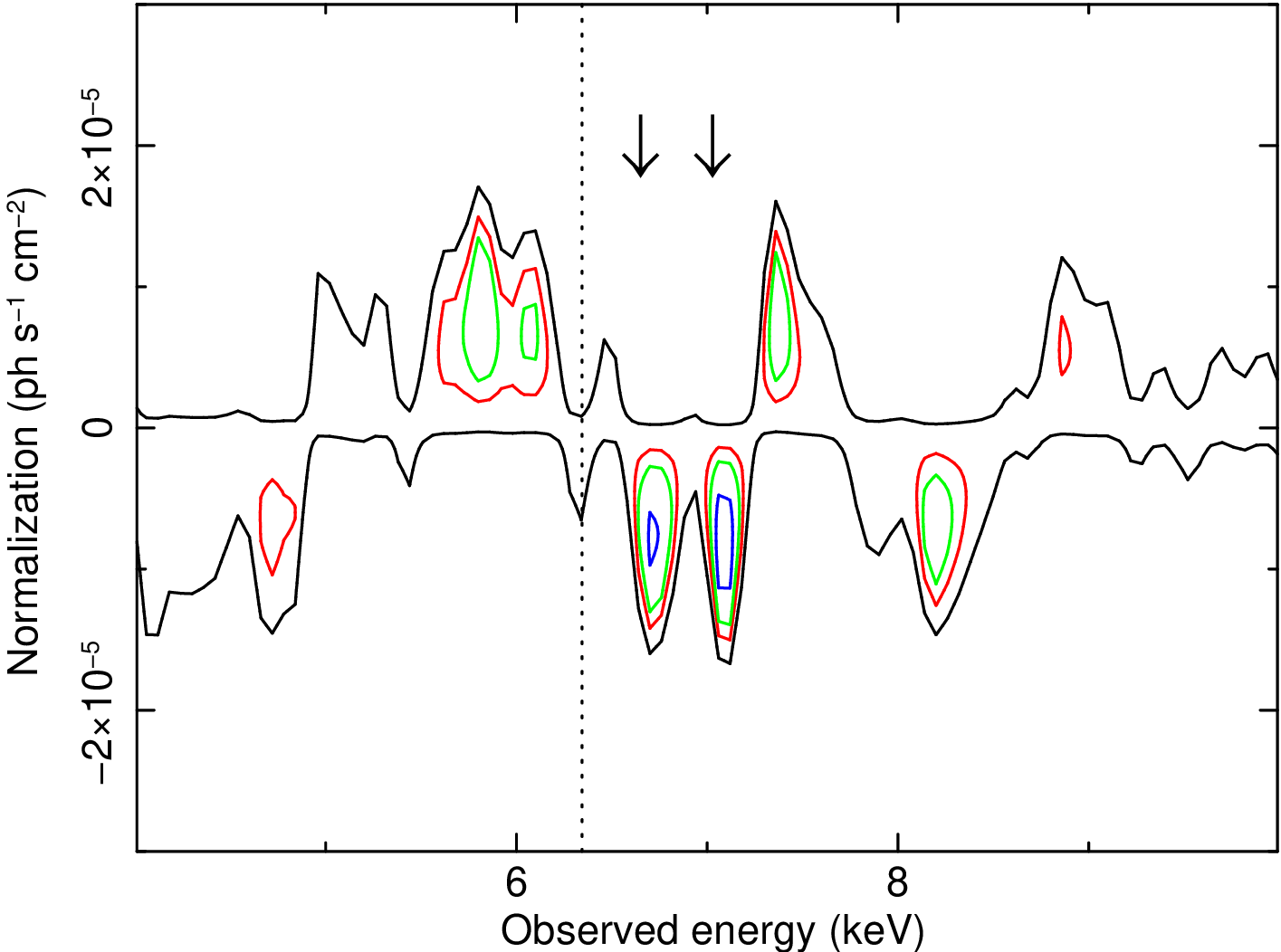}
\hspace{0.3cm}
    \includegraphics[width=3cm,height=4cm,angle=270]{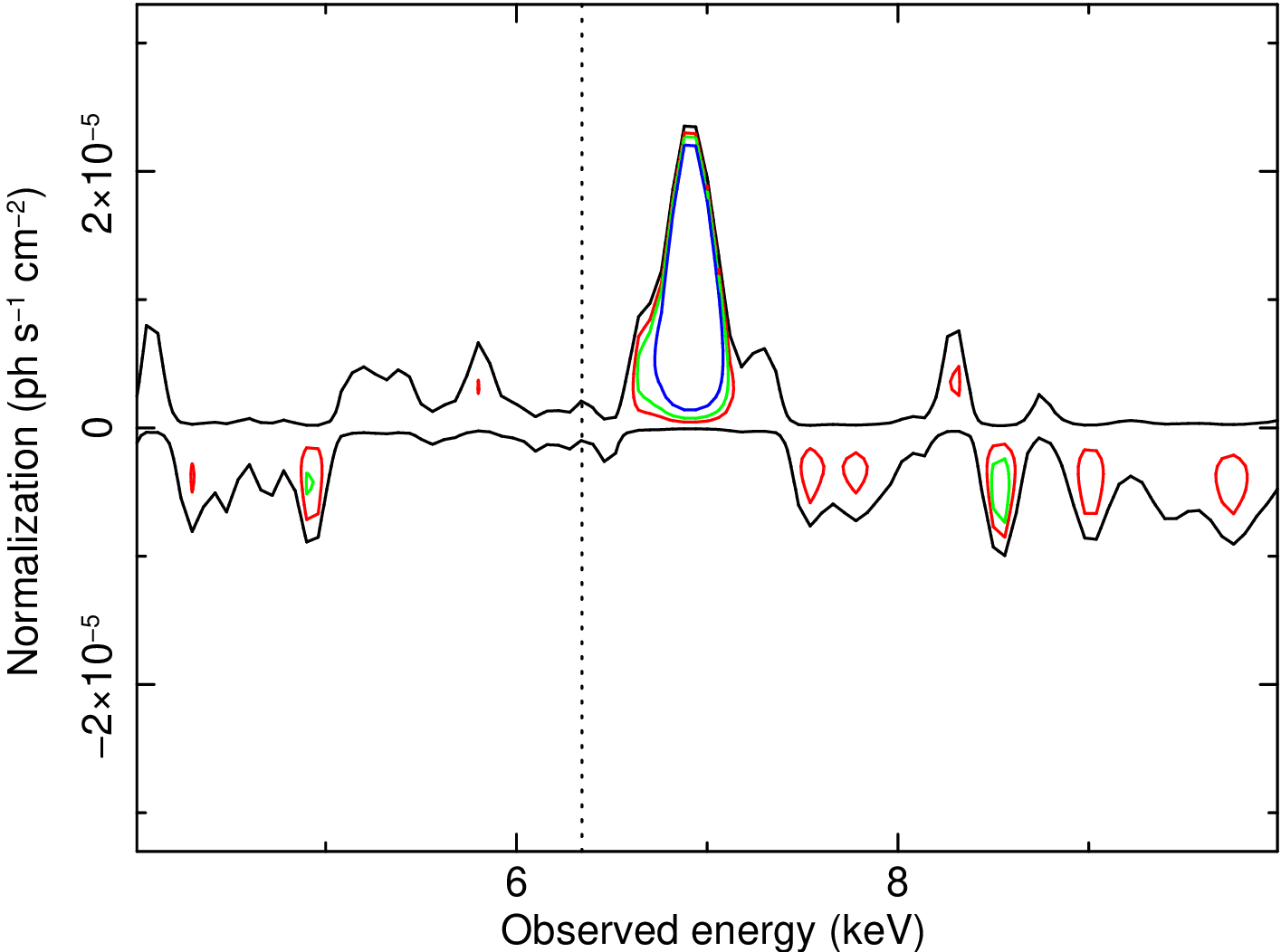}


\vspace{0.2cm}

    \includegraphics[width=3cm,height=4cm,angle=270]{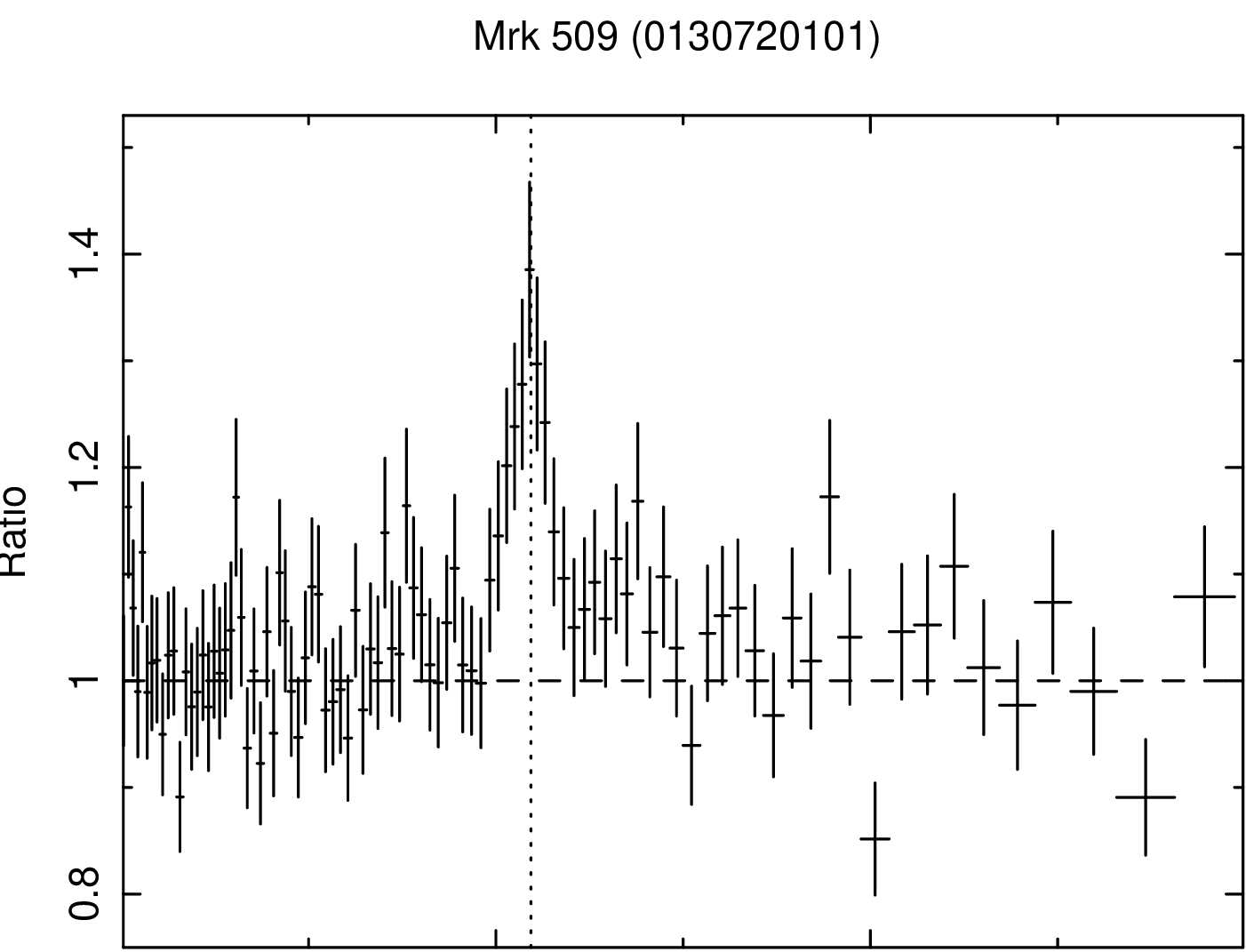}
\hspace{0.3cm}
    \includegraphics[width=3cm,height=4cm,angle=270]{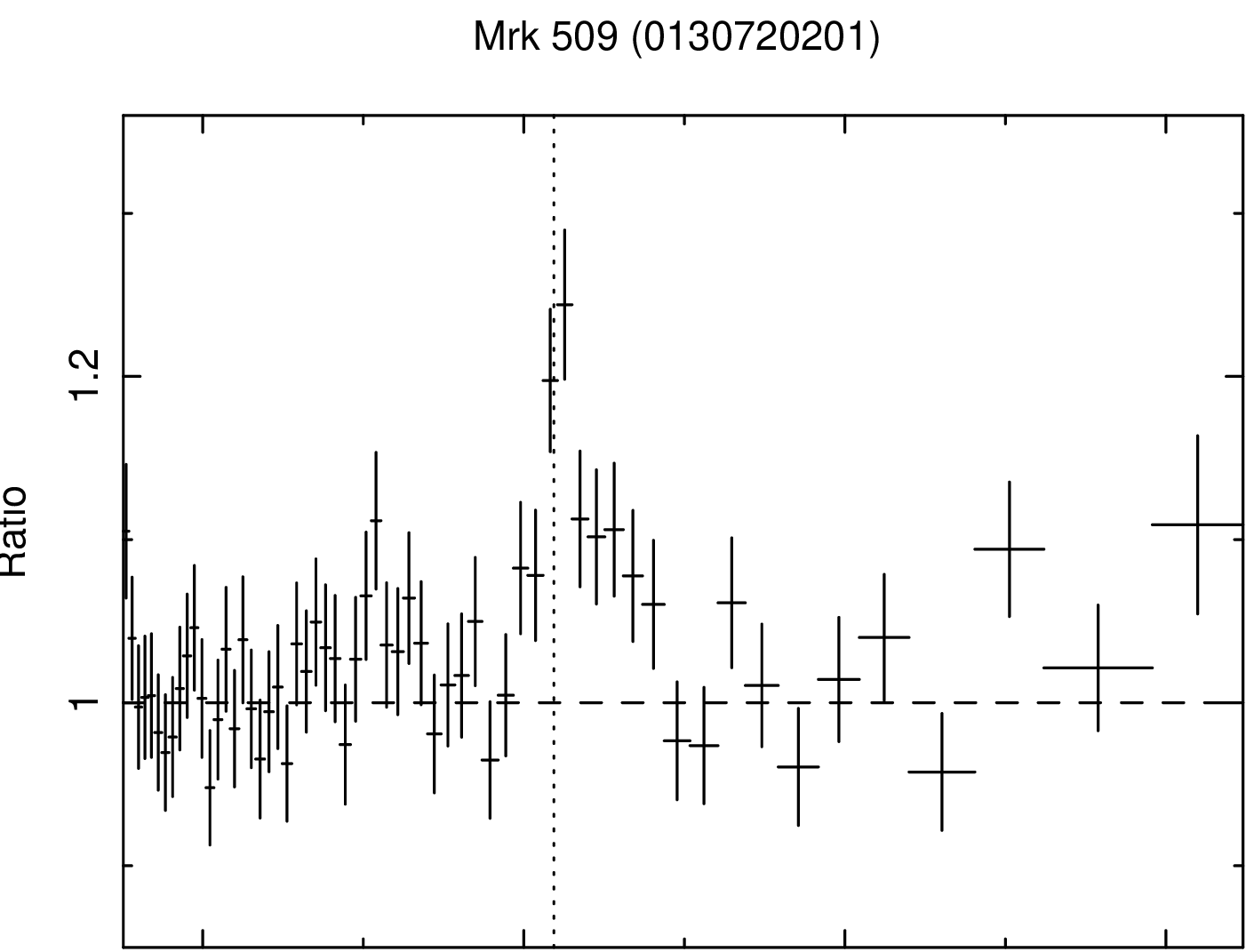}
\hspace{0.3cm}
    \includegraphics[width=3cm,height=4cm,angle=270]{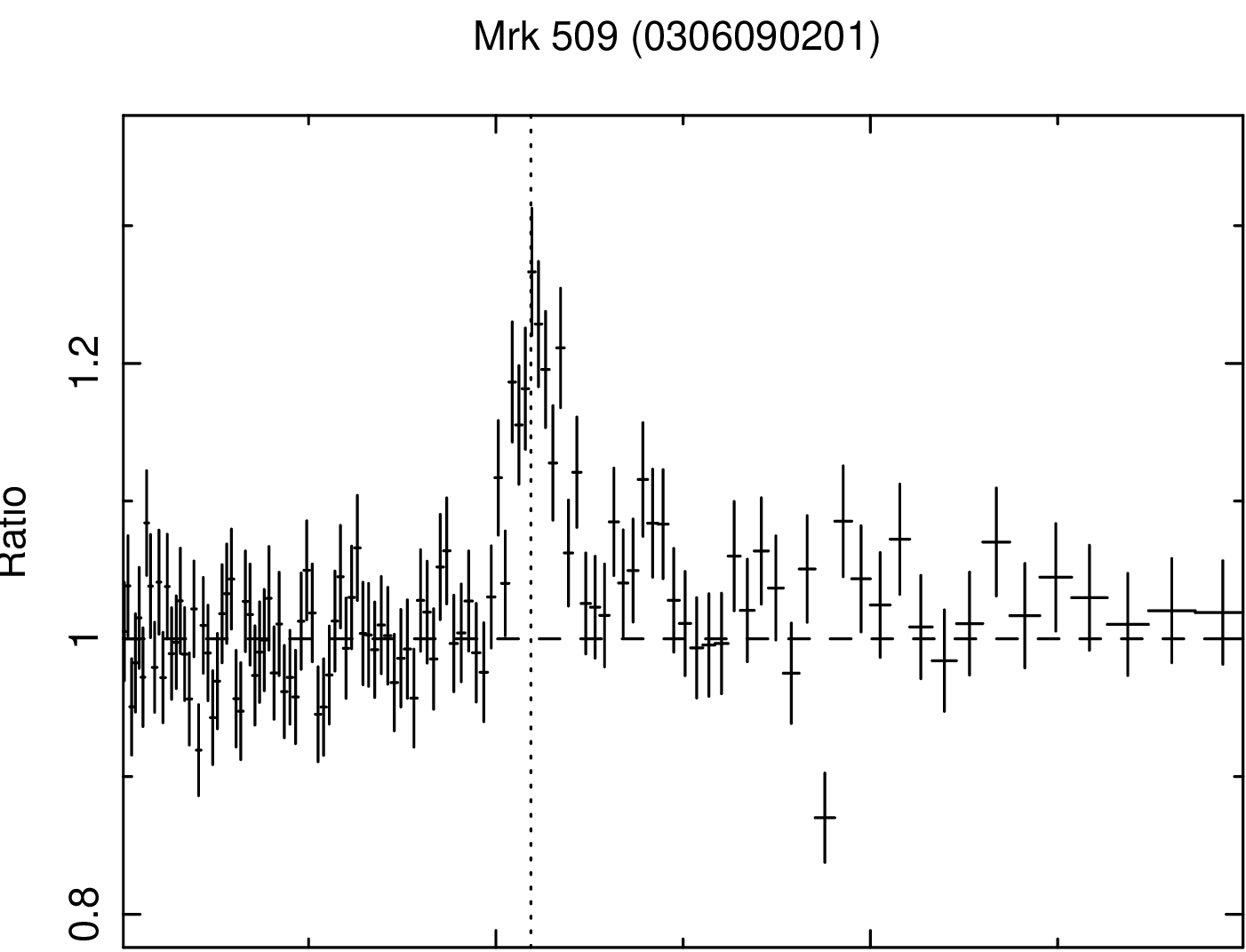}
\hspace{0.3cm}
    \includegraphics[width=3cm,height=4cm,angle=270]{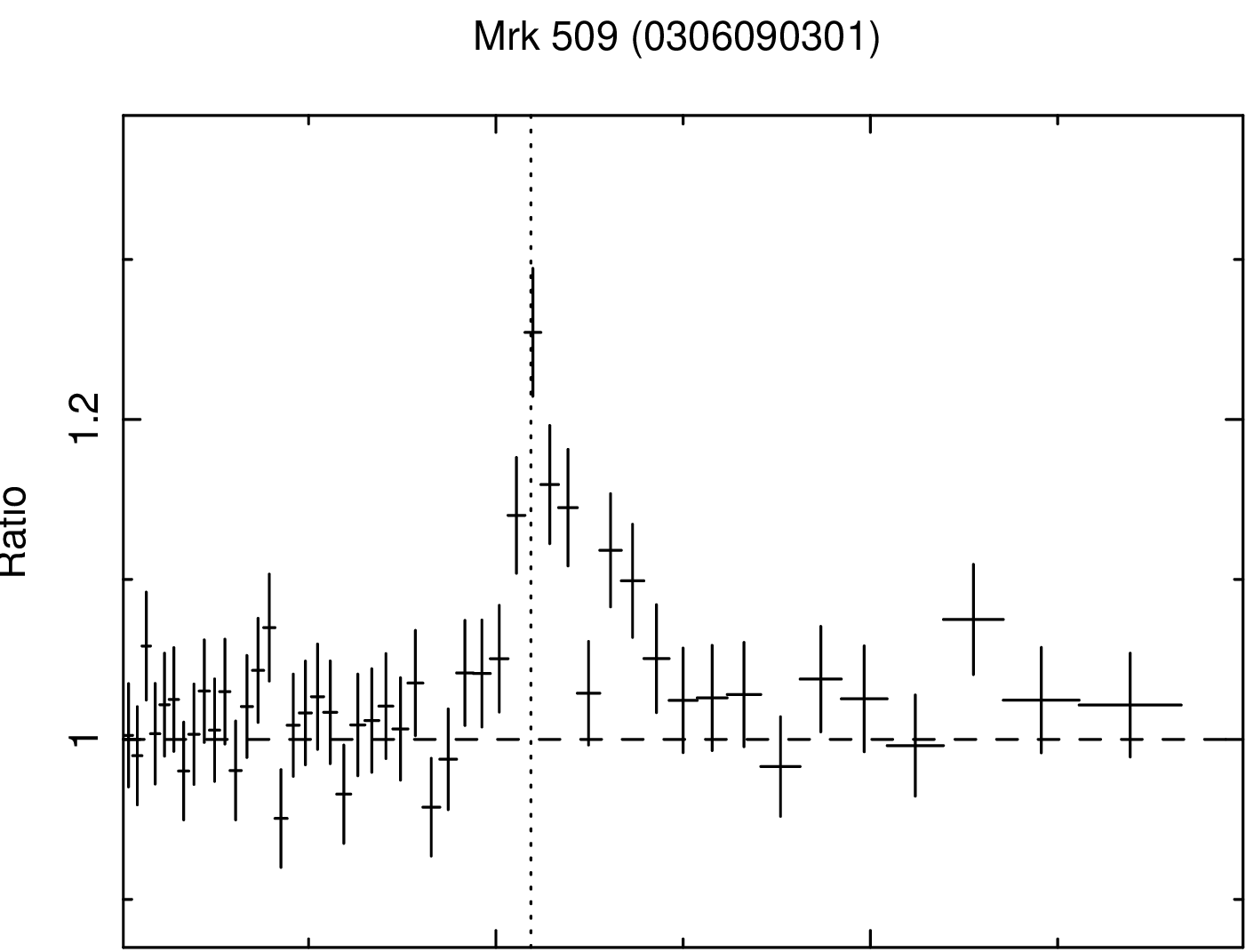}

\vspace{-0.07cm}

\hspace{-0.12cm}
    \includegraphics[width=3cm,height=4cm,angle=270]{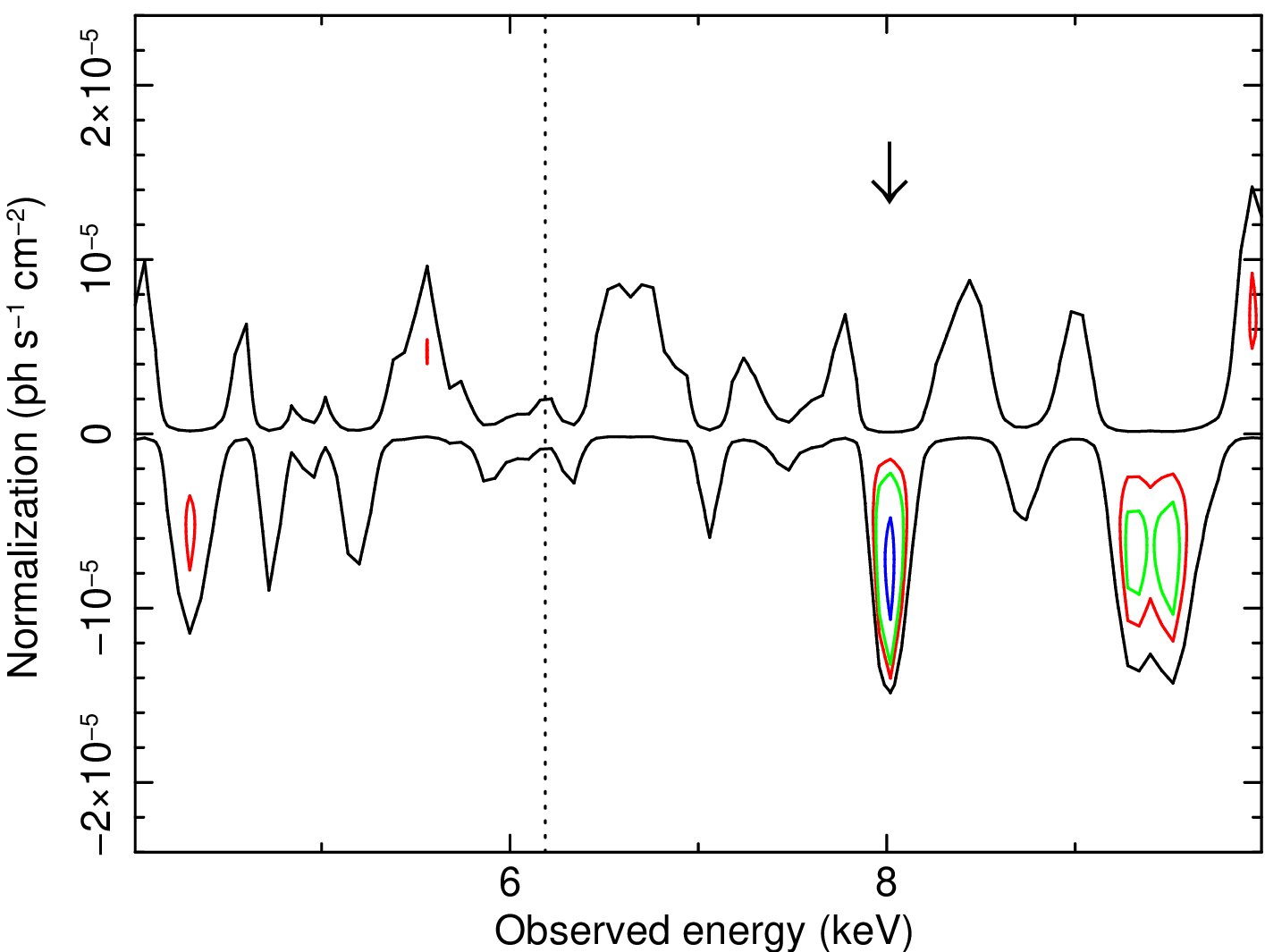}
\hspace{0.3cm}
    \includegraphics[width=3cm,height=4cm,angle=270]{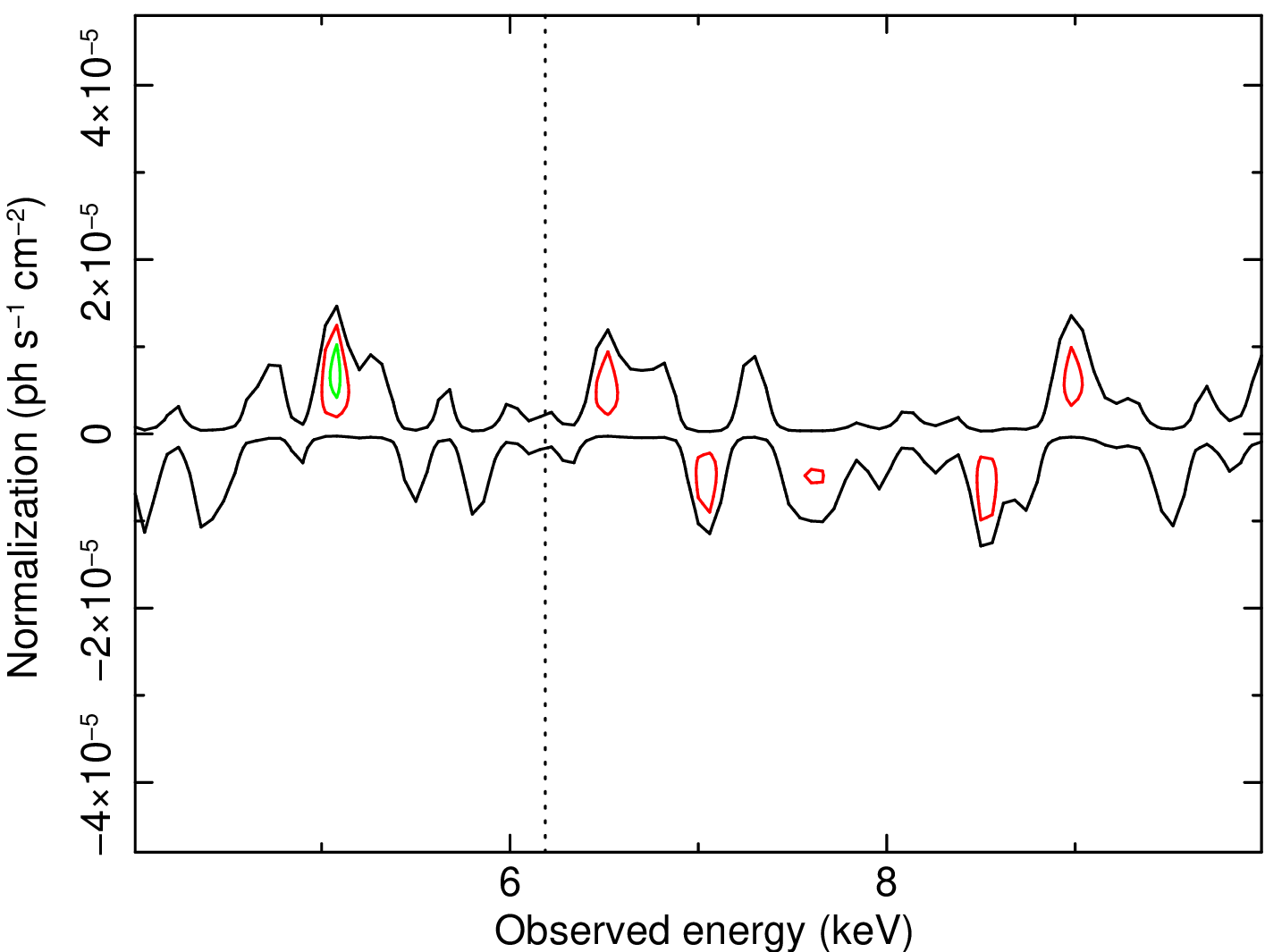}
\hspace{0.3cm}
    \includegraphics[width=3cm,height=4cm,angle=270]{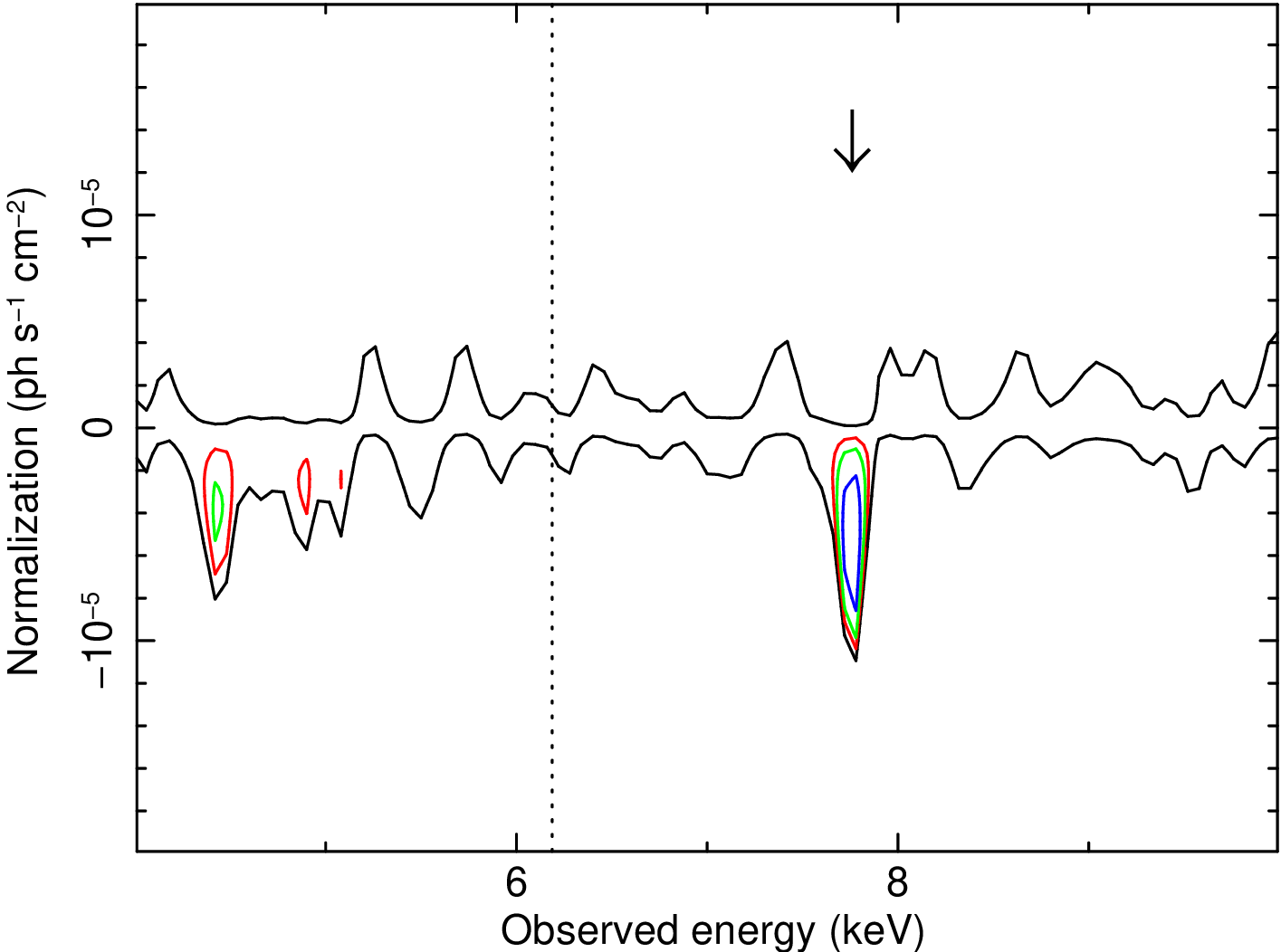}
\hspace{0.3cm}
    \includegraphics[width=3cm,height=4cm,angle=270]{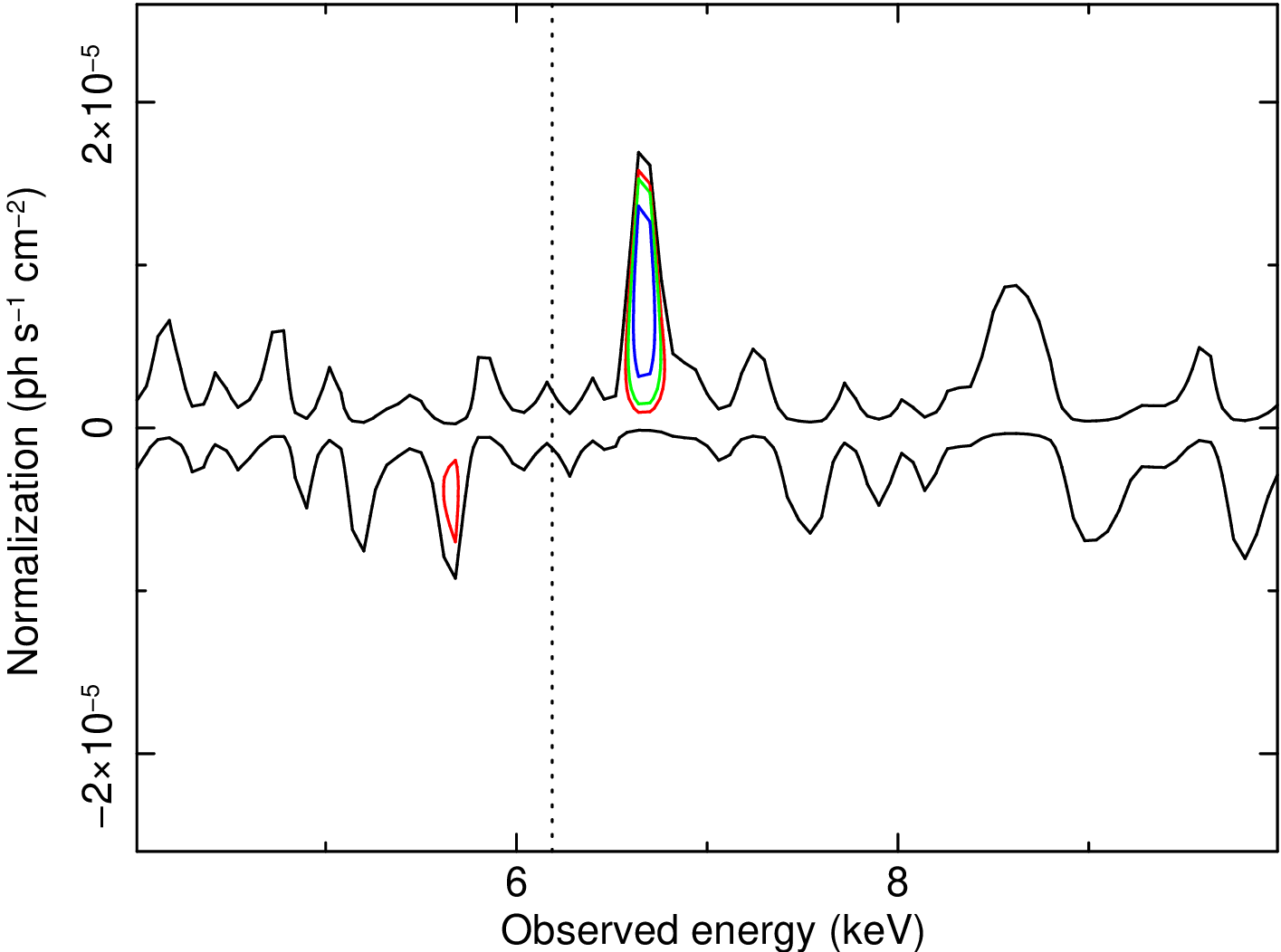}


\vspace{0.2cm}

    \includegraphics[width=3cm,height=4cm,angle=270]{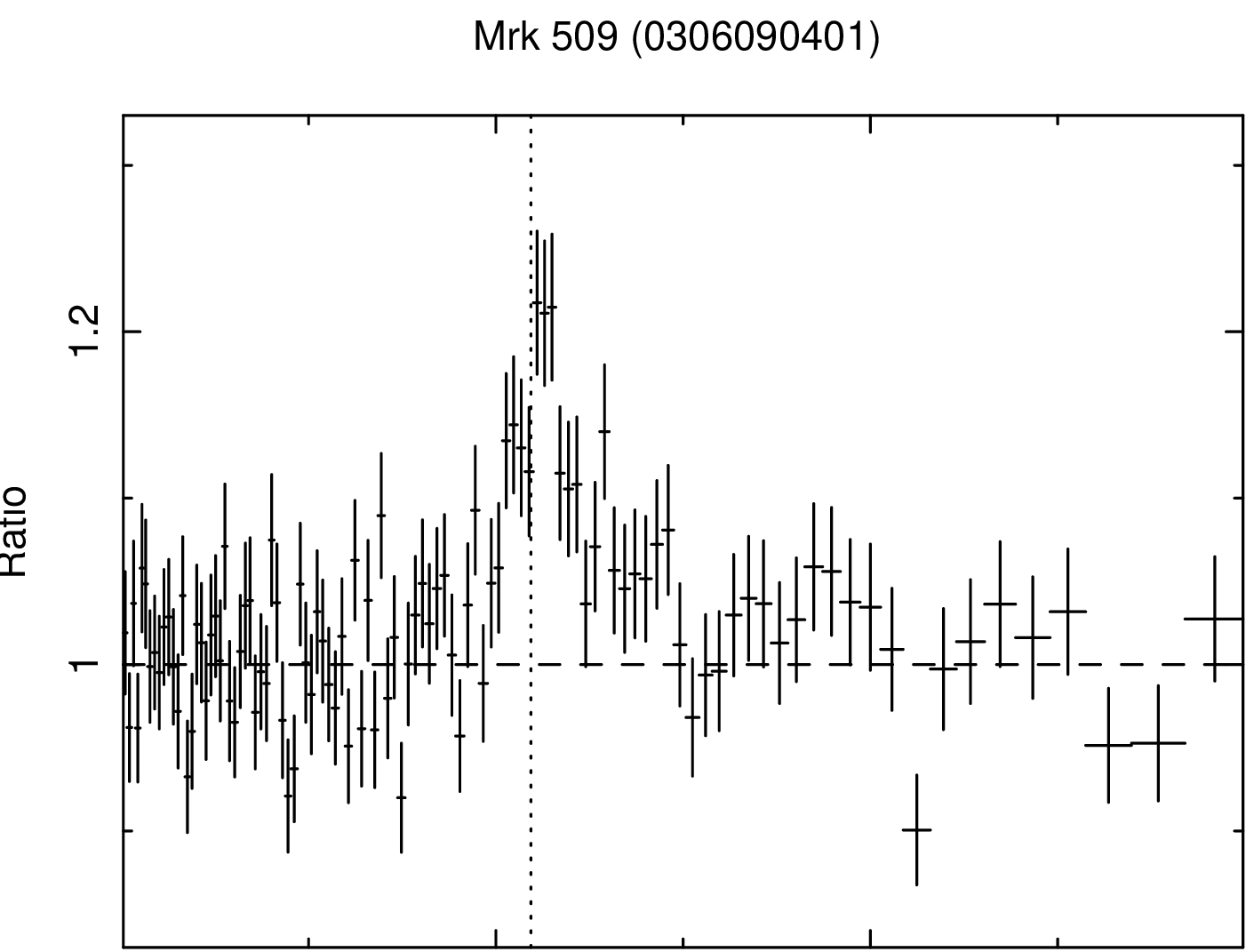}
\hspace{0.3cm}
    \includegraphics[width=3cm,height=4cm,angle=270]{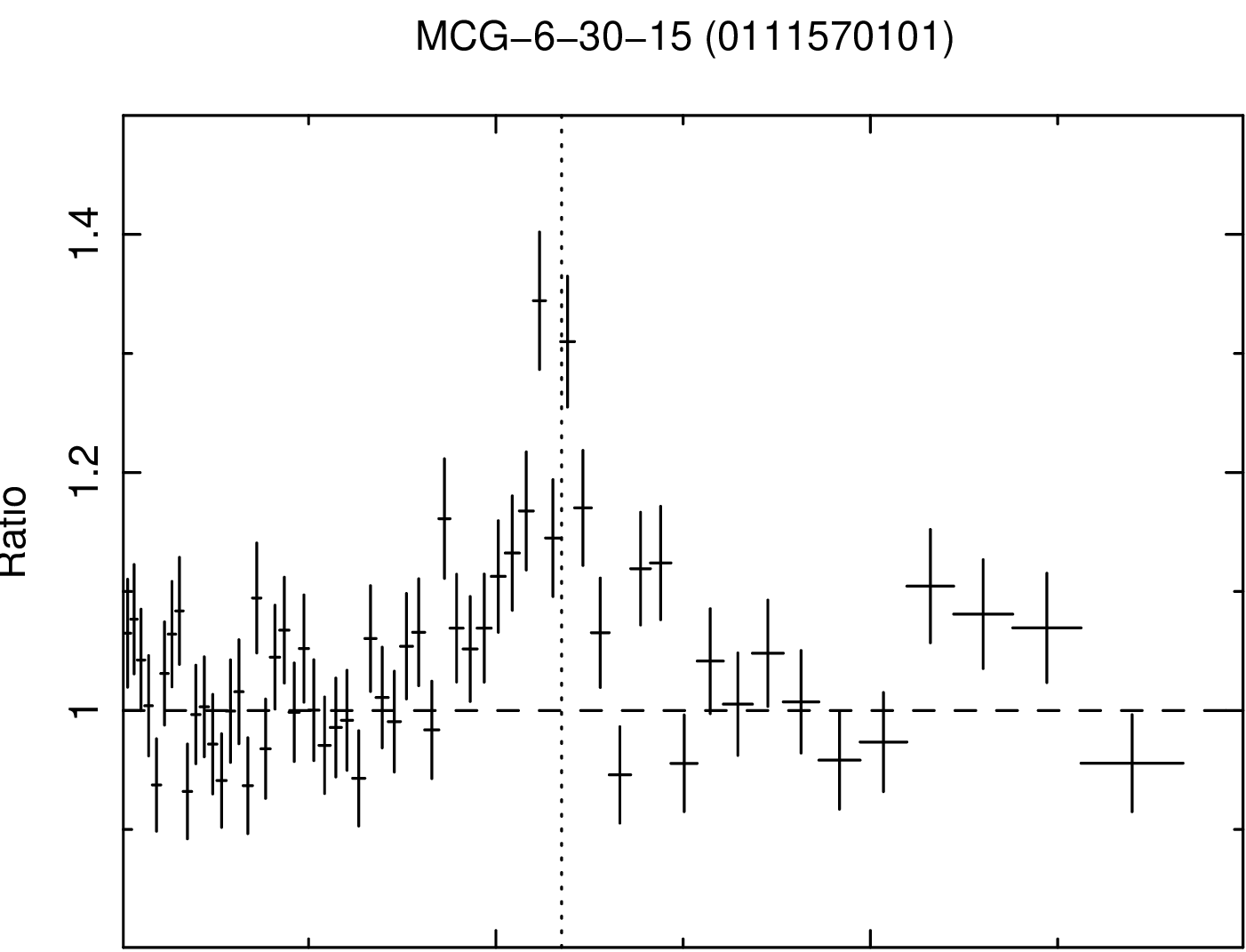}
\hspace{0.3cm}
    \includegraphics[width=3cm,height=4cm,angle=270]{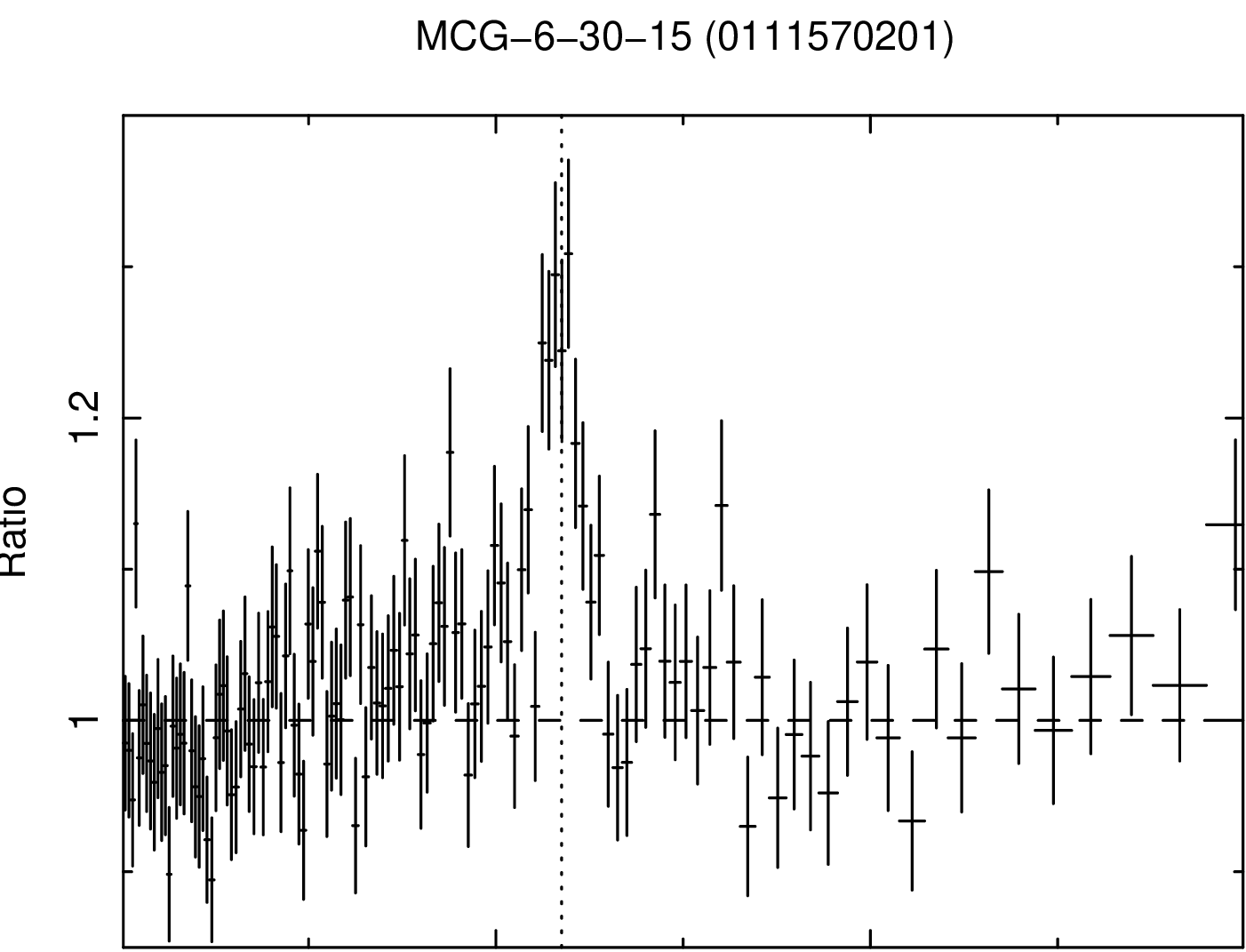}
\hspace{0.3cm}
    \includegraphics[width=3cm,height=4cm,angle=270]{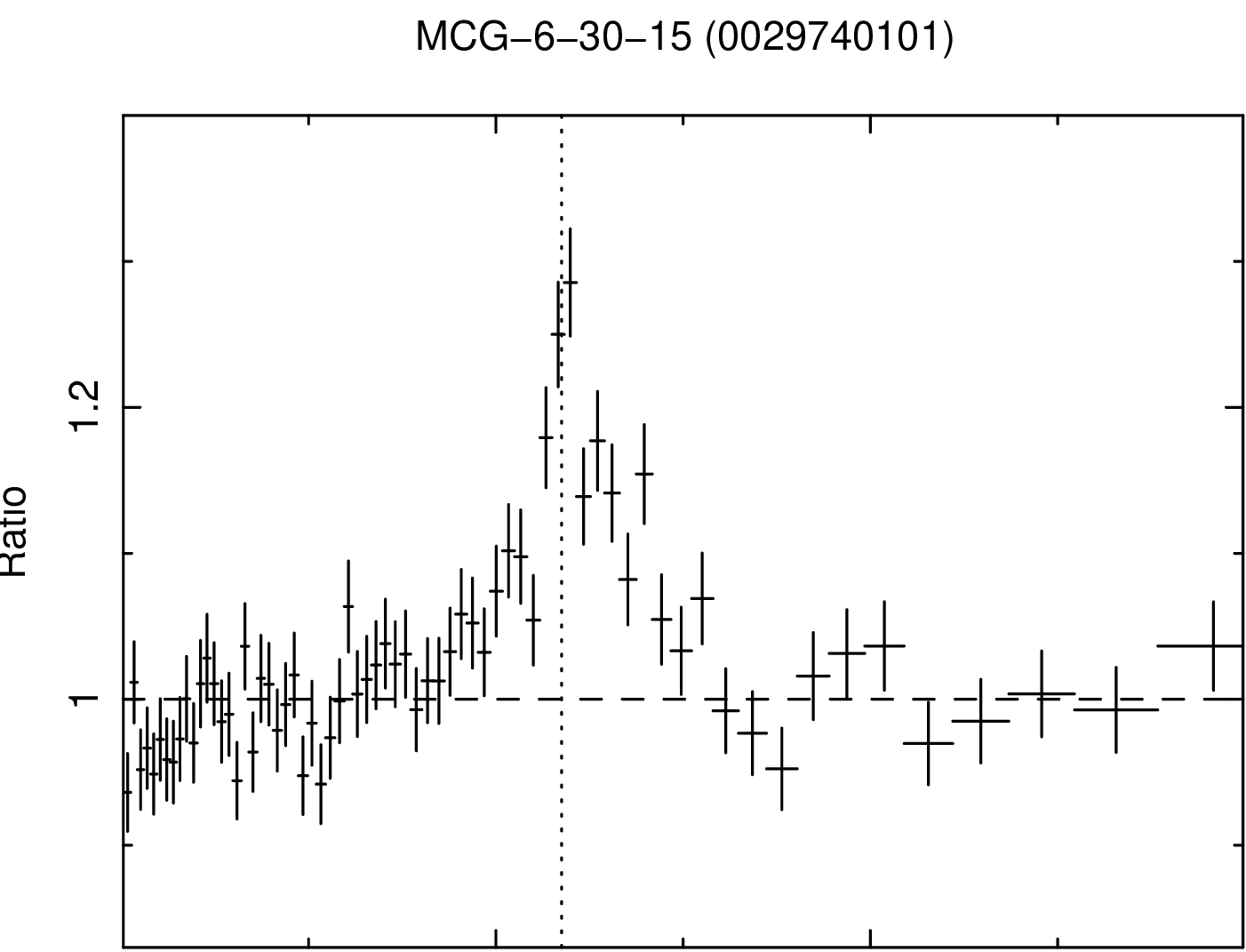}

\vspace{-0.07cm}

\hspace{-0.12cm}
    \includegraphics[width=3cm,height=4cm,angle=270]{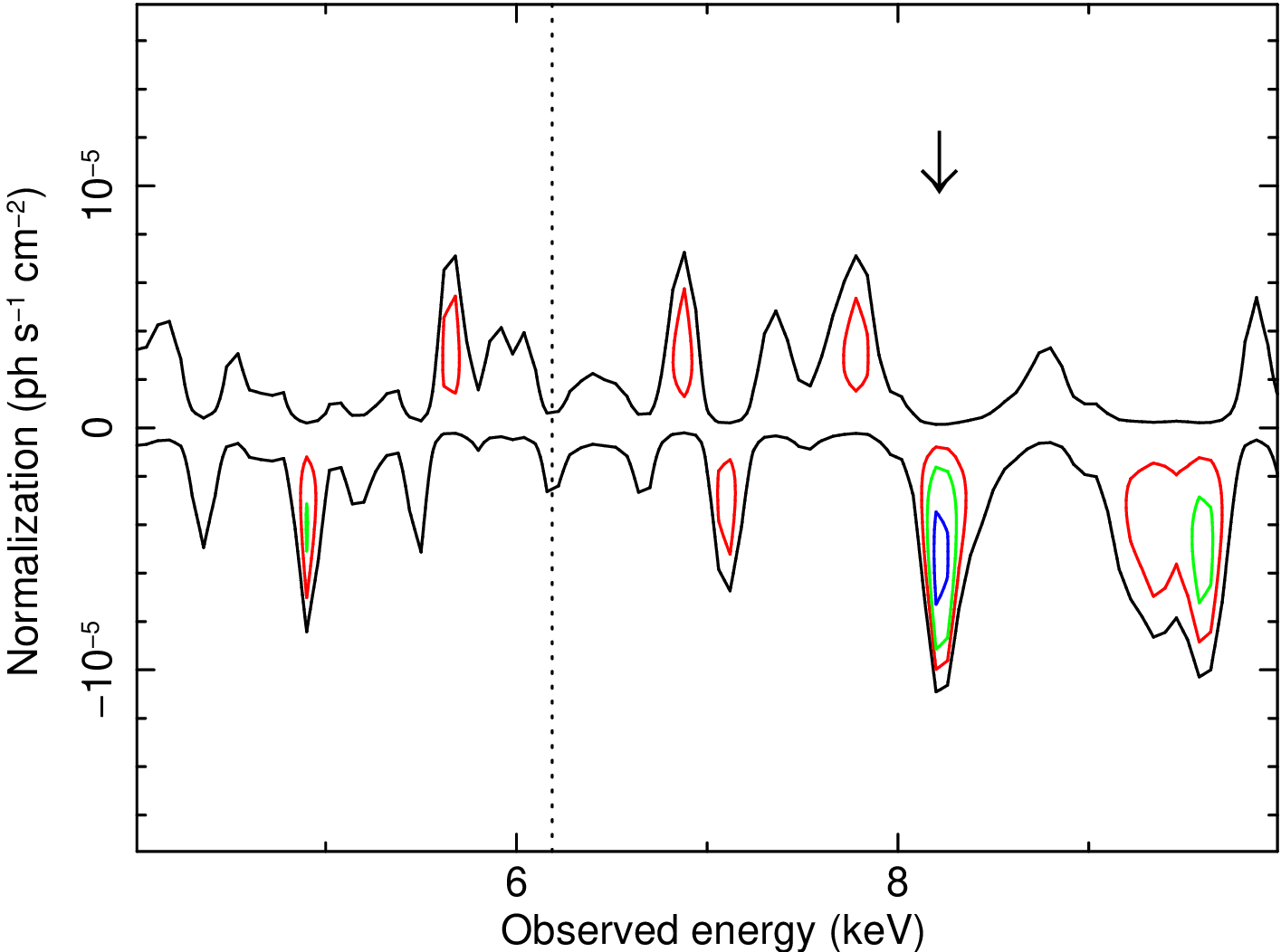}
\hspace{0.3cm}
    \includegraphics[width=3cm,height=4cm,angle=270]{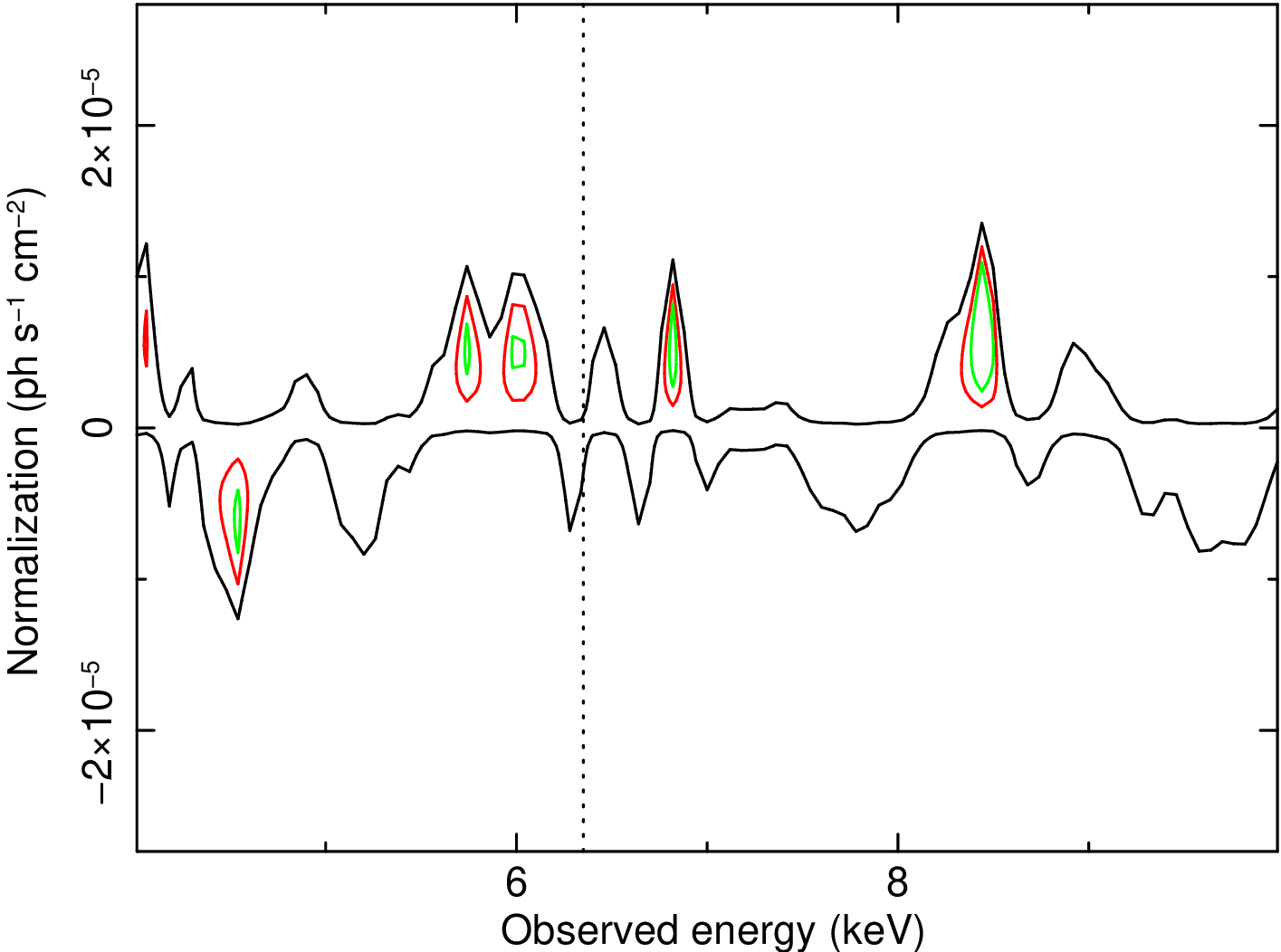}
\hspace{0.3cm}
    \includegraphics[width=3cm,height=4cm,angle=270]{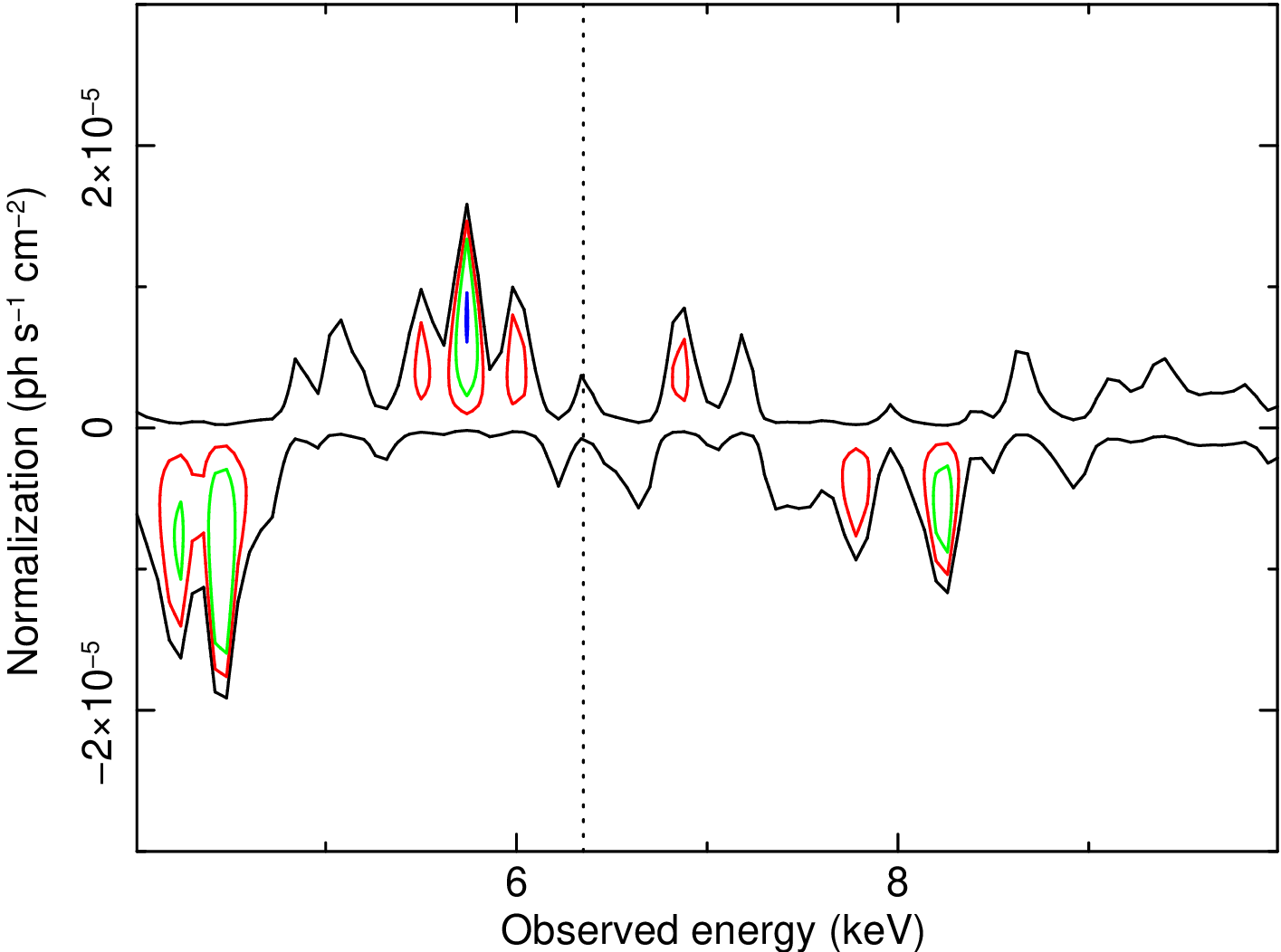}
\hspace{0.3cm}
    \includegraphics[width=3cm,height=4cm,angle=270]{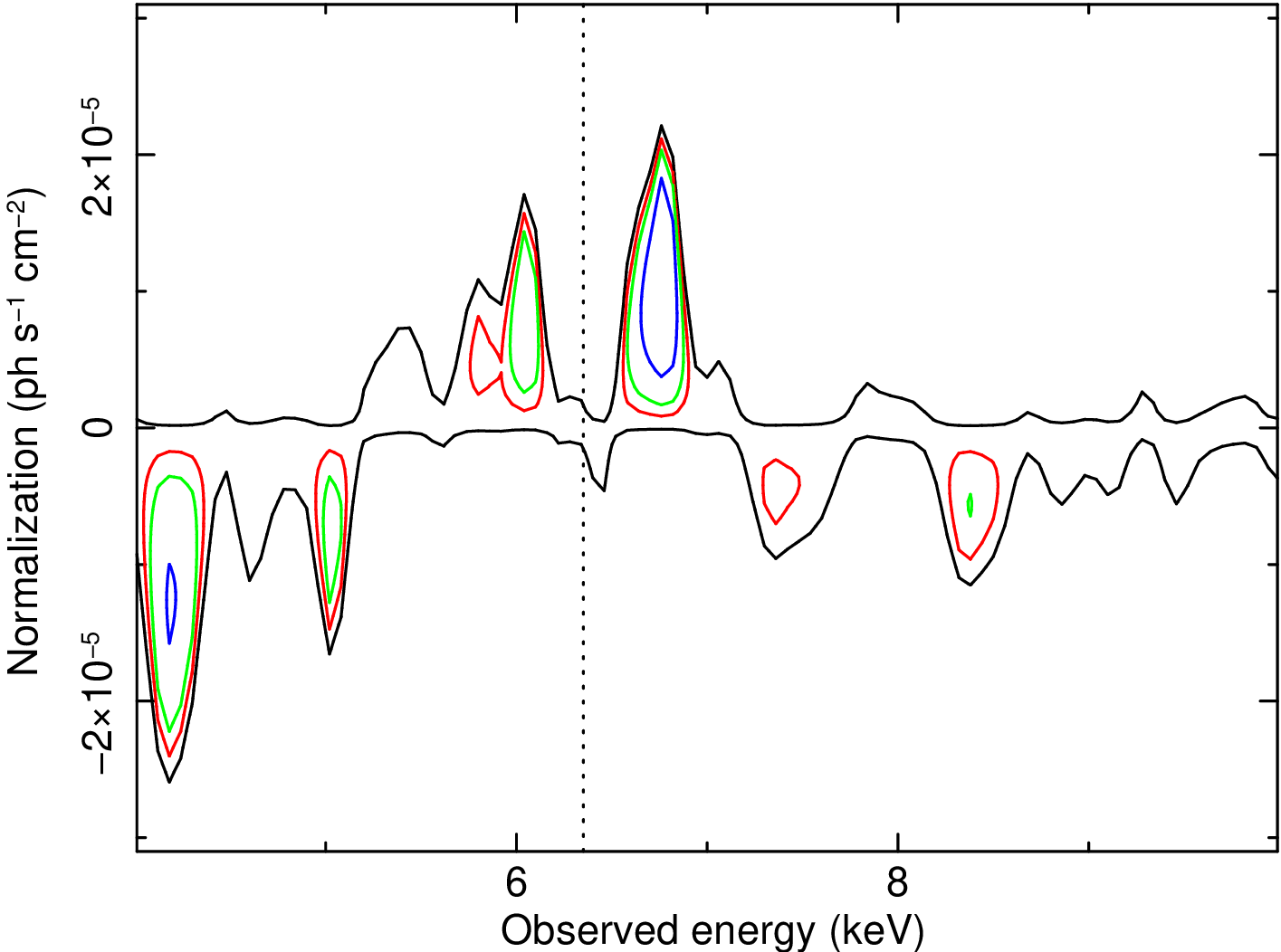}

   \caption{Ratio against the continuum (\emph{upper panel}) and contour plots with respect to the best-fit baseline model (\emph{lower panel}; 68\% (red), 90\% (green), 99\% (blue) levels) for the Type 1 sources. The Fe K absorption lines are pointed by arrows. The vertical line refer to Fe I K$\alpha$.}
    \end{figure*}

   \begin{figure*}[ht]
   \centering


    \includegraphics[width=3cm,height=4cm,angle=270]{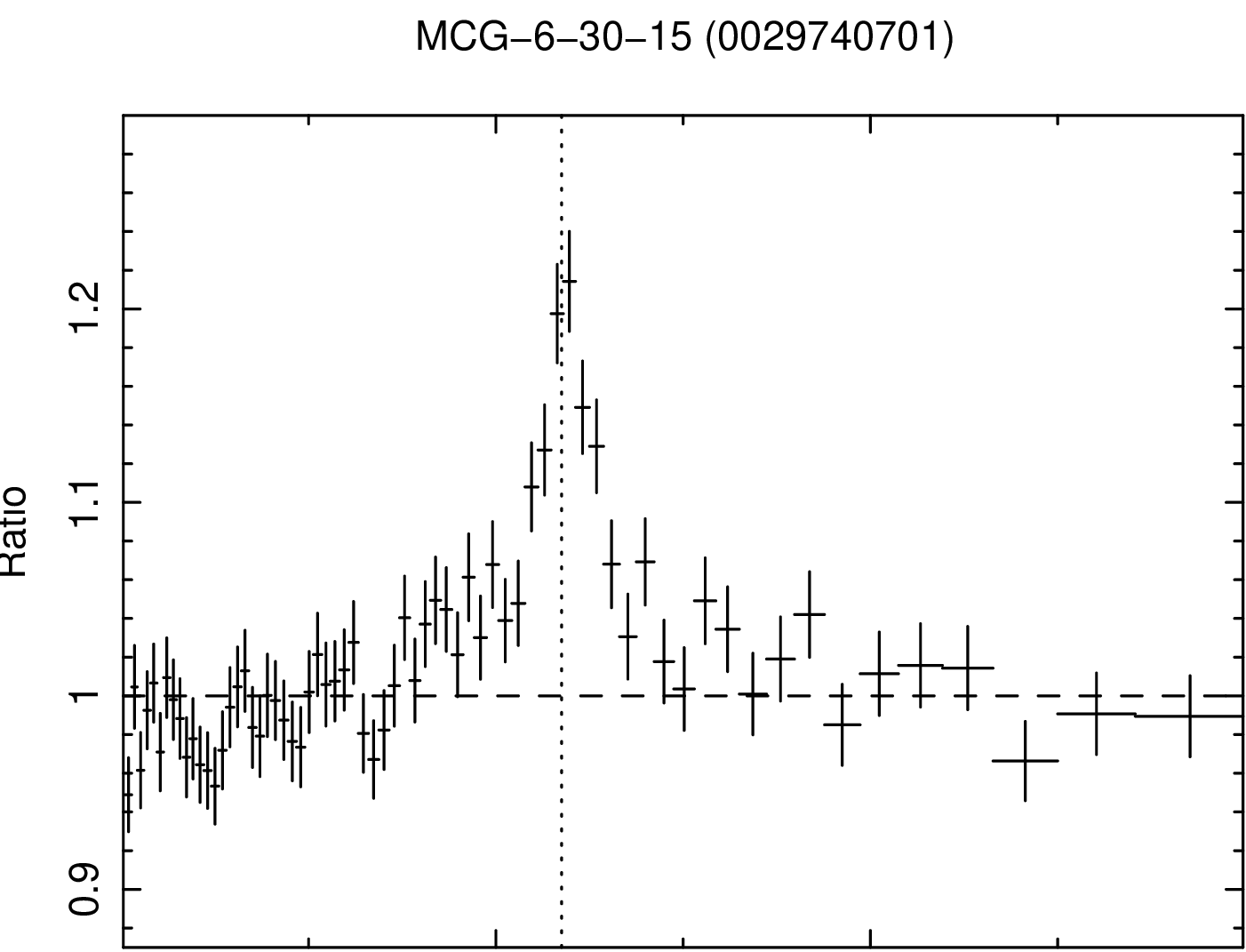}
\hspace{0.3cm}
    \includegraphics[width=3cm,height=4cm,angle=270]{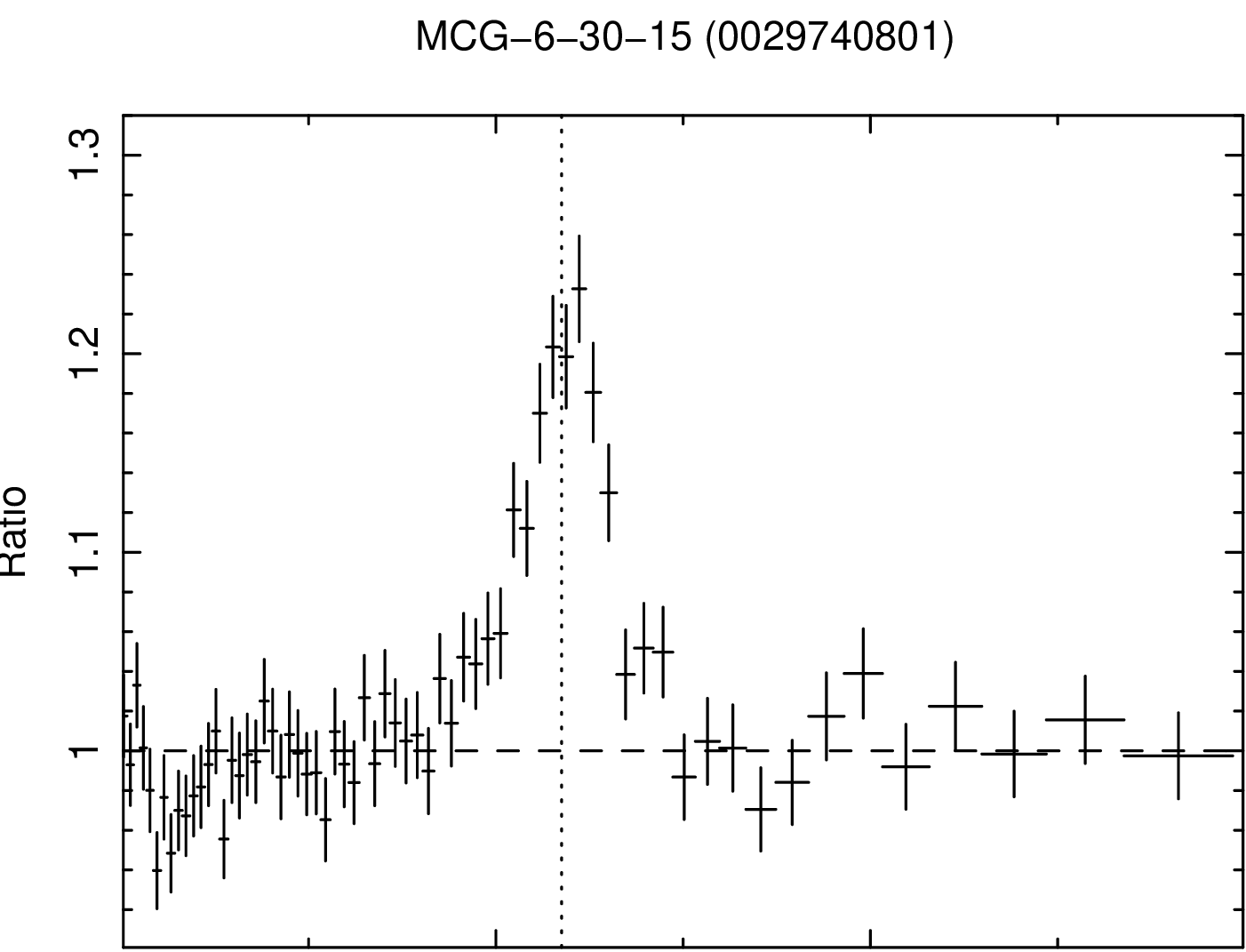}
\hspace{0.3cm}
    \includegraphics[width=3cm,height=4cm,angle=270]{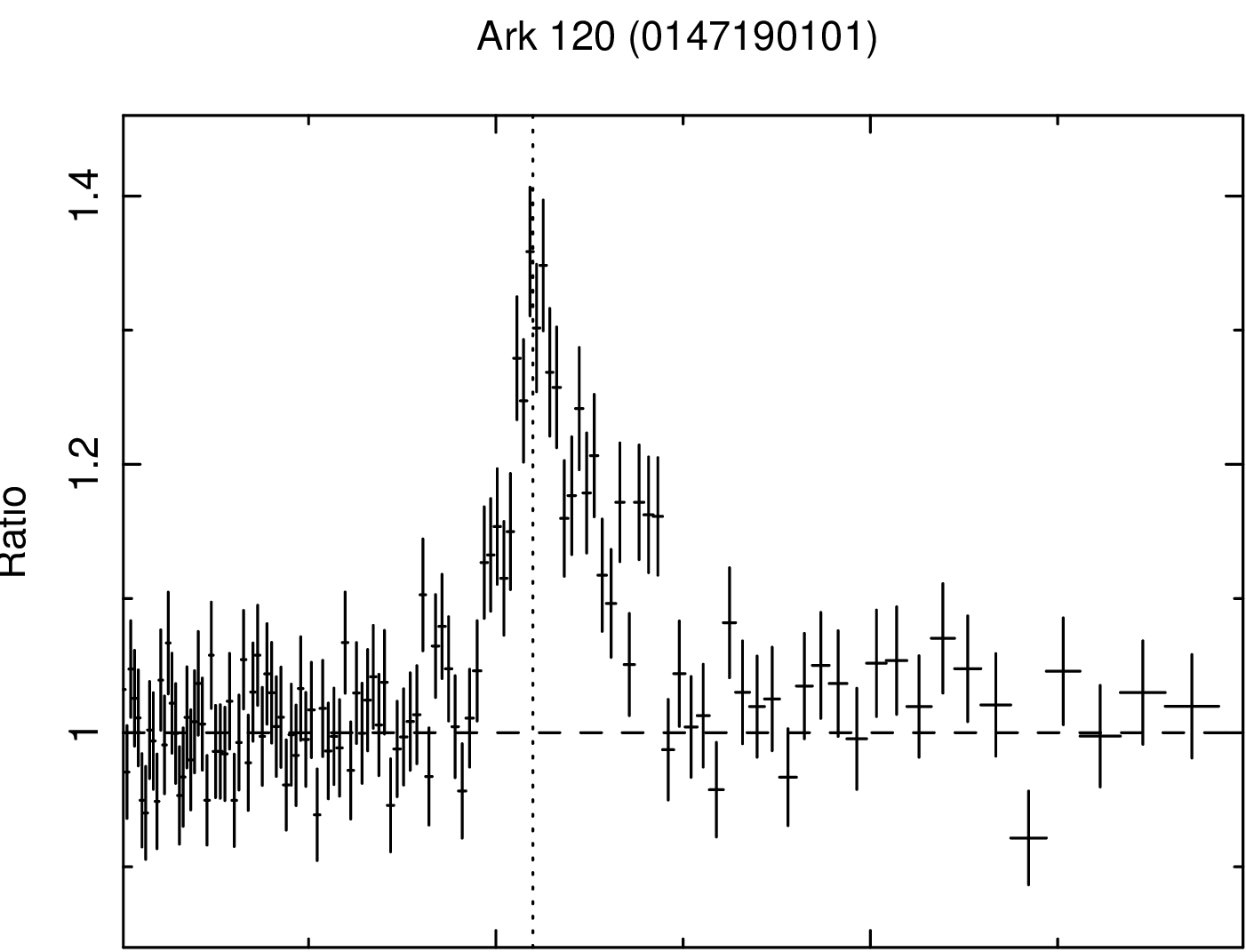}
\hspace{0.3cm}
    \includegraphics[width=3cm,height=4cm,angle=270]{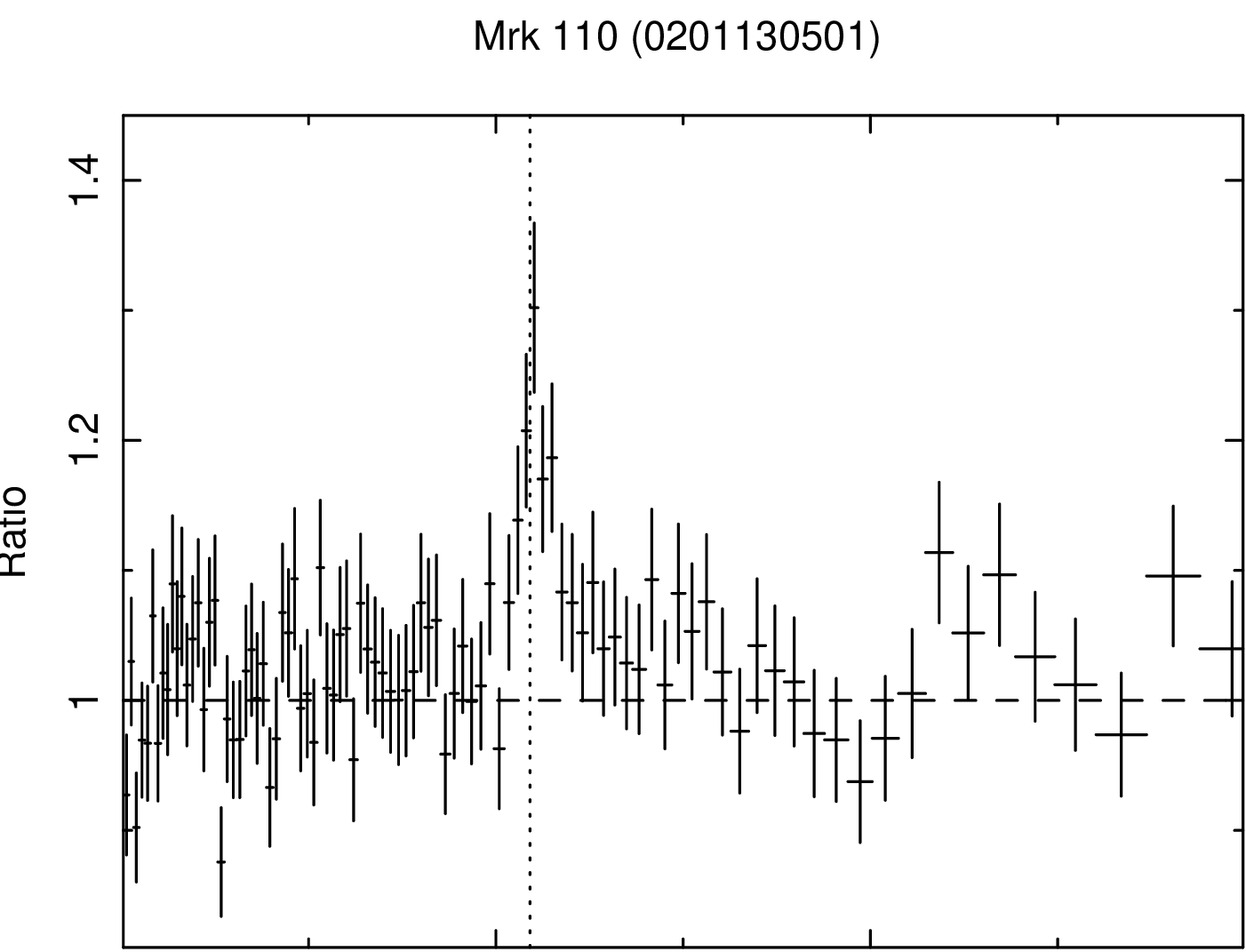}

\vspace{-0.07cm}

\hspace{-0.12cm}
    \includegraphics[width=3cm,height=4cm,angle=270]{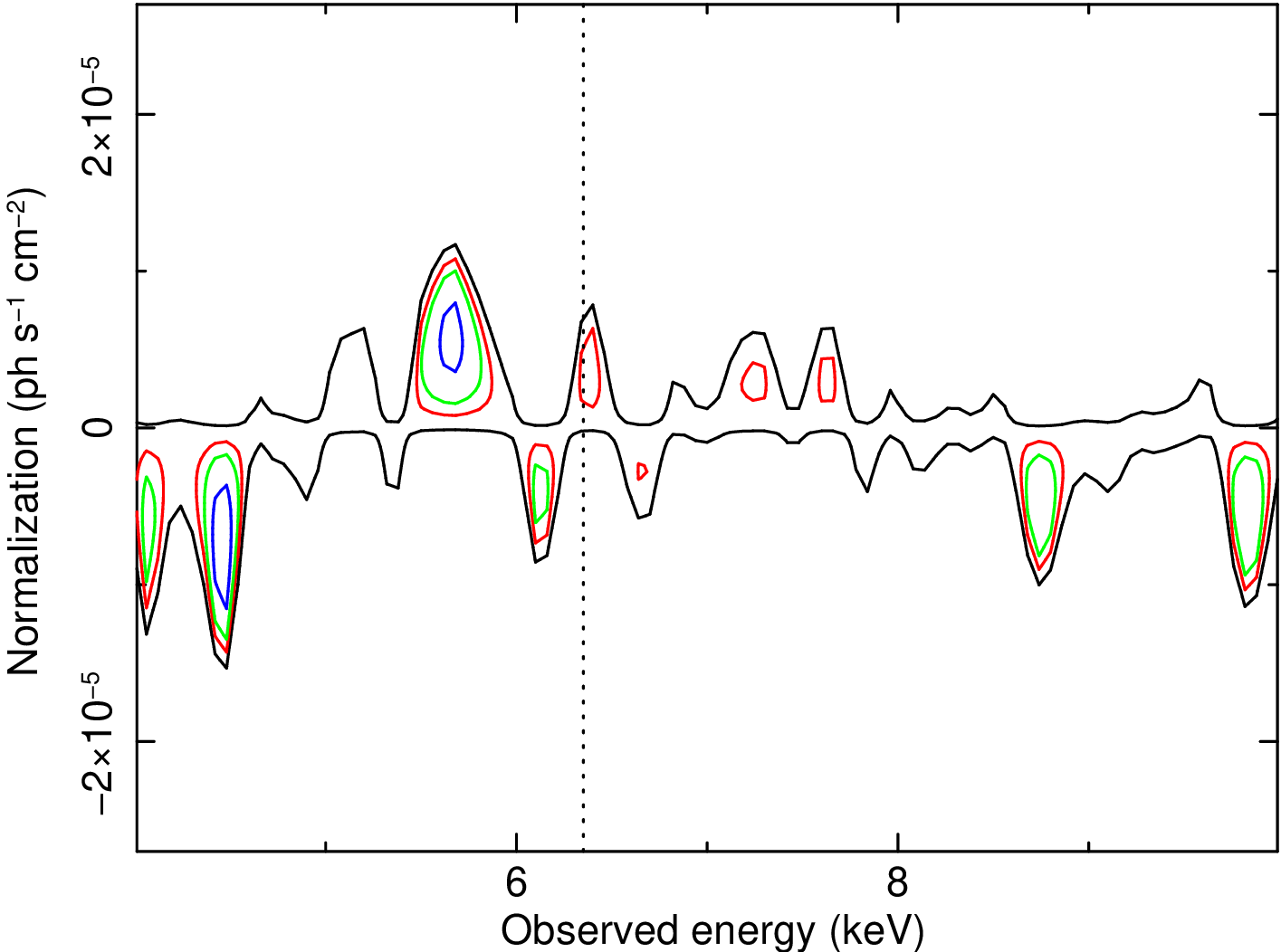}
\hspace{0.3cm}
    \includegraphics[width=3cm,height=4cm,angle=270]{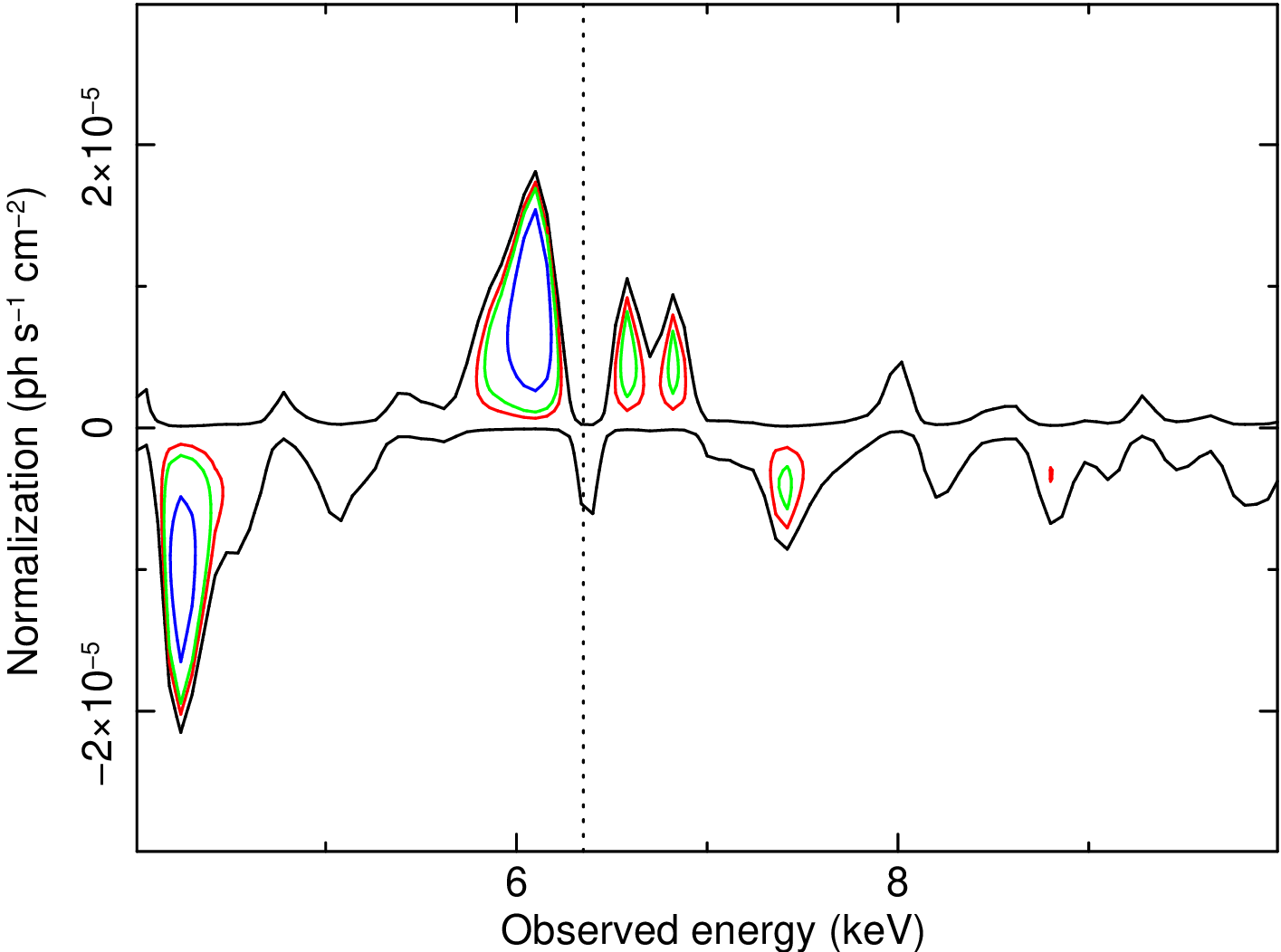}
\hspace{0.3cm}
    \includegraphics[width=3cm,height=4cm,angle=270]{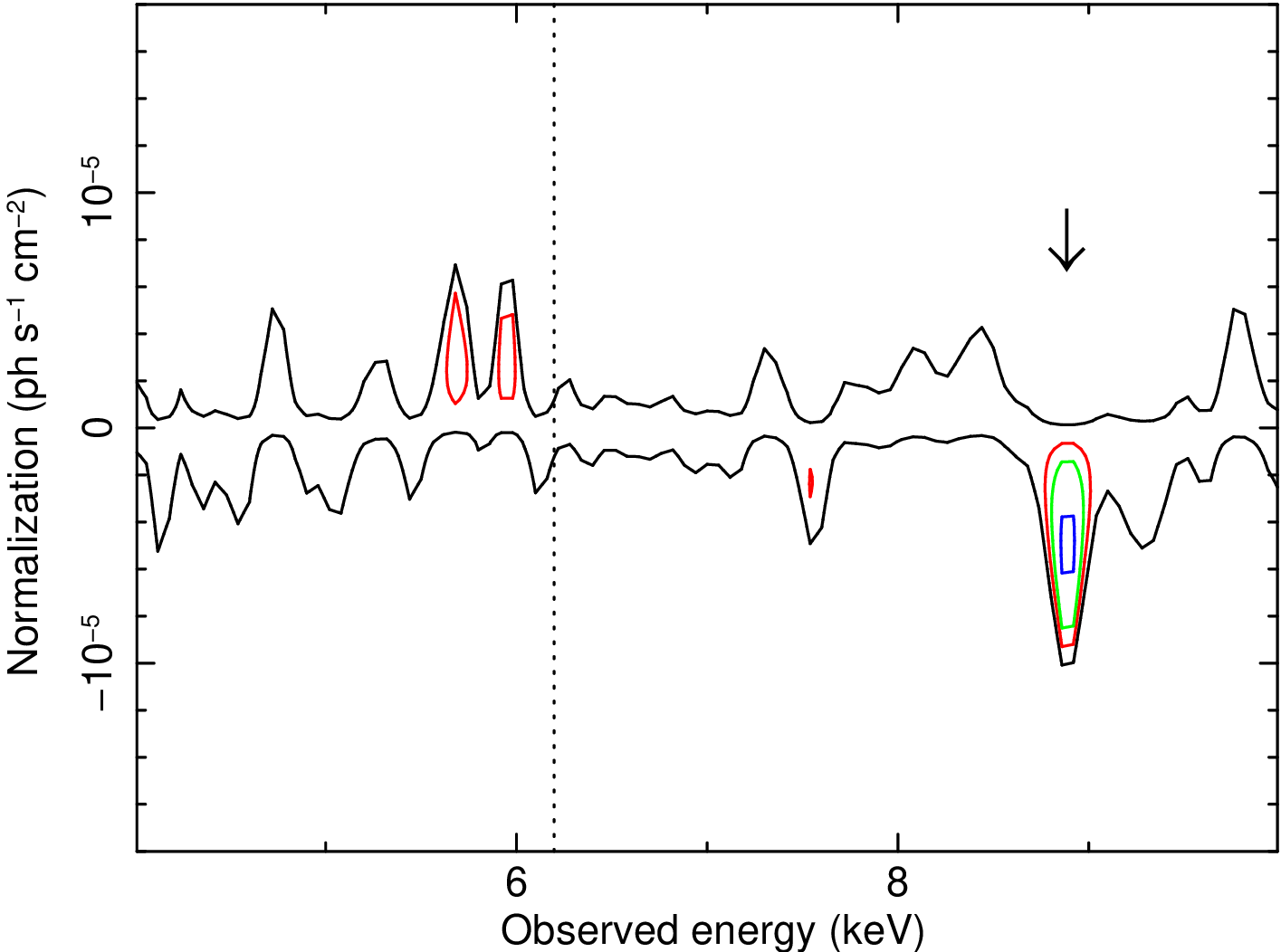}
\hspace{0.3cm}
    \includegraphics[width=3cm,height=4cm,angle=270]{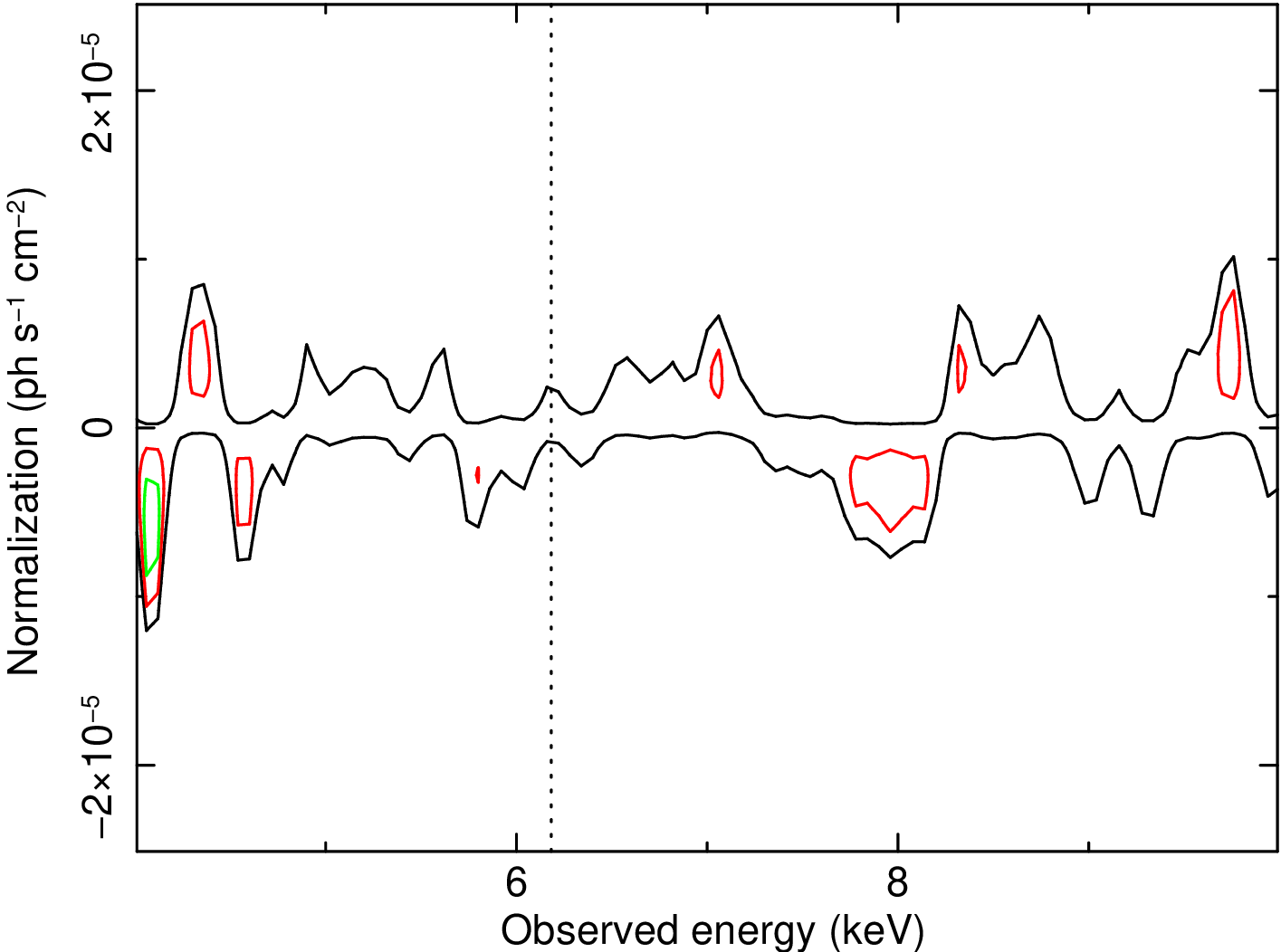}


\vspace{0.2cm}

    \includegraphics[width=3cm,height=4cm,angle=270]{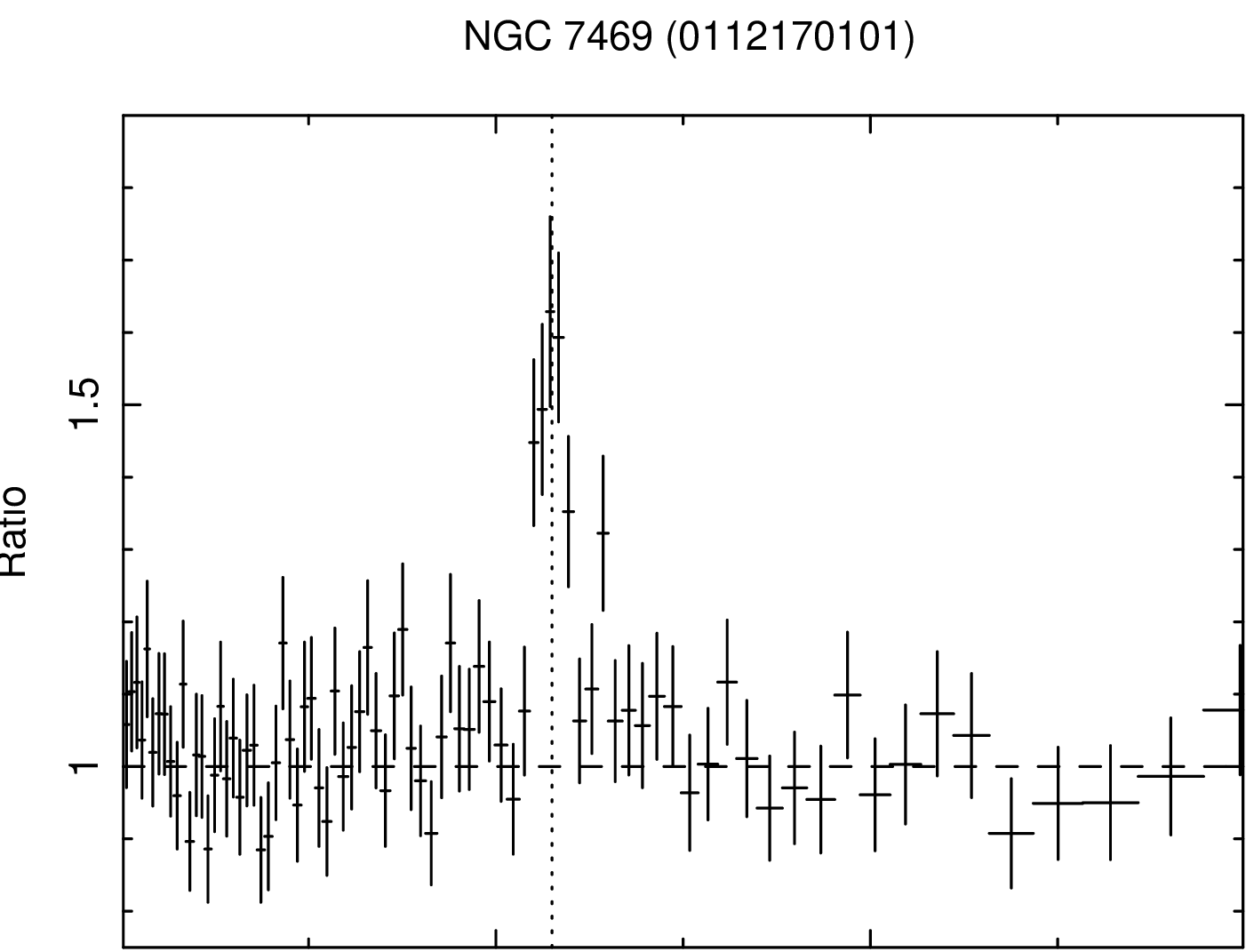}
\hspace{0.3cm}
    \includegraphics[width=3cm,height=4cm,angle=270]{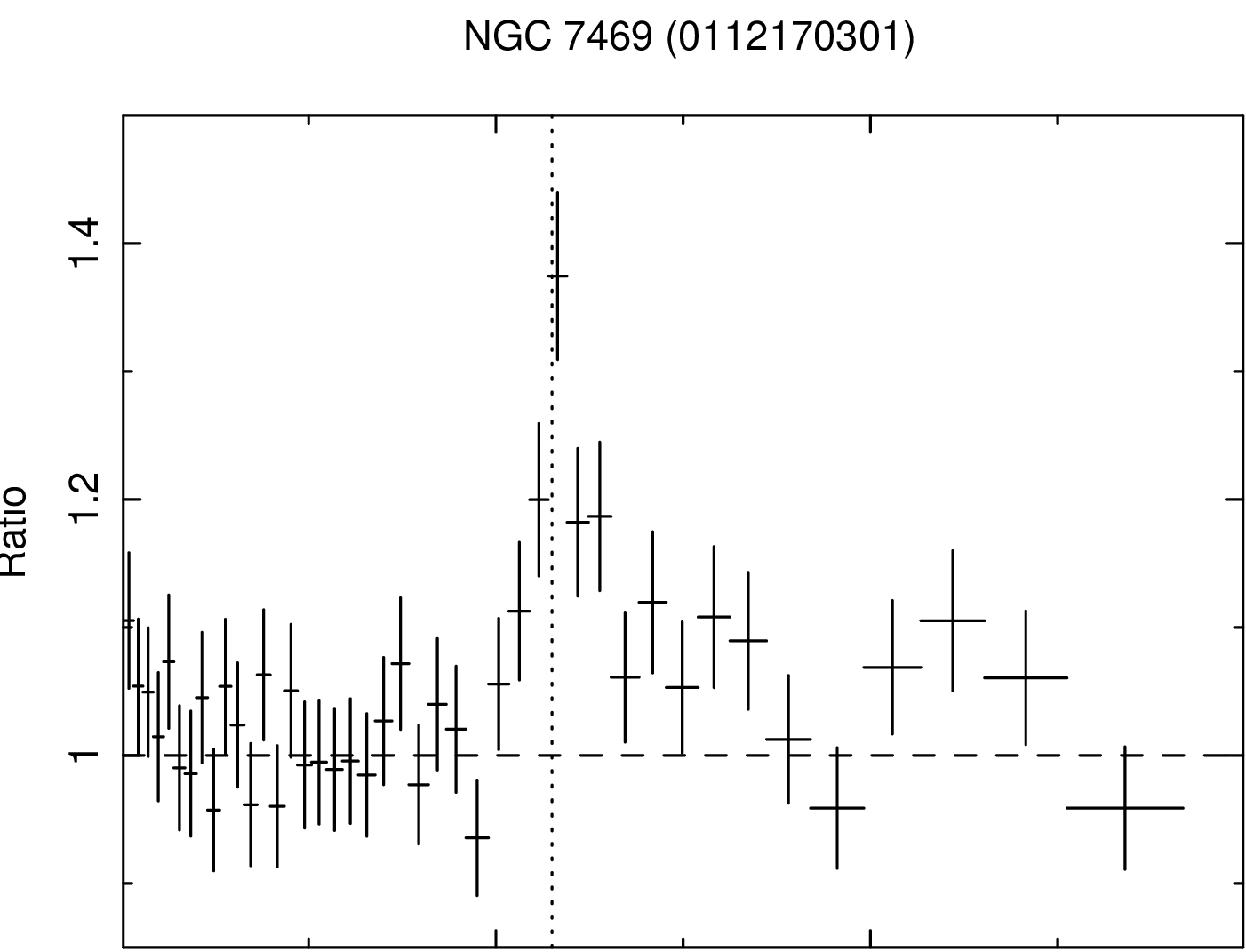}
\hspace{0.3cm}
    \includegraphics[width=3cm,height=4cm,angle=270]{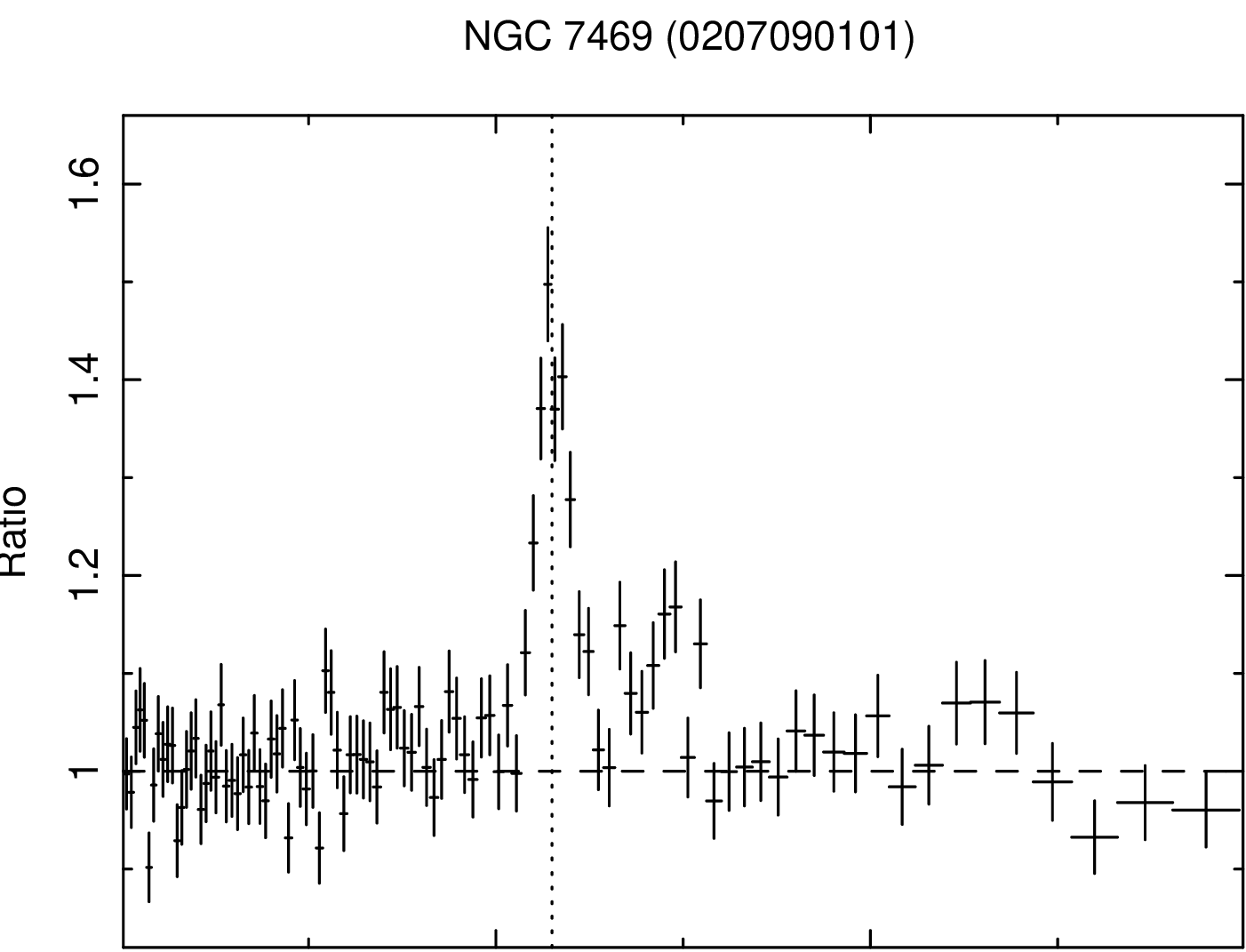}
\hspace{0.3cm}
    \includegraphics[width=3cm,height=4cm,angle=270]{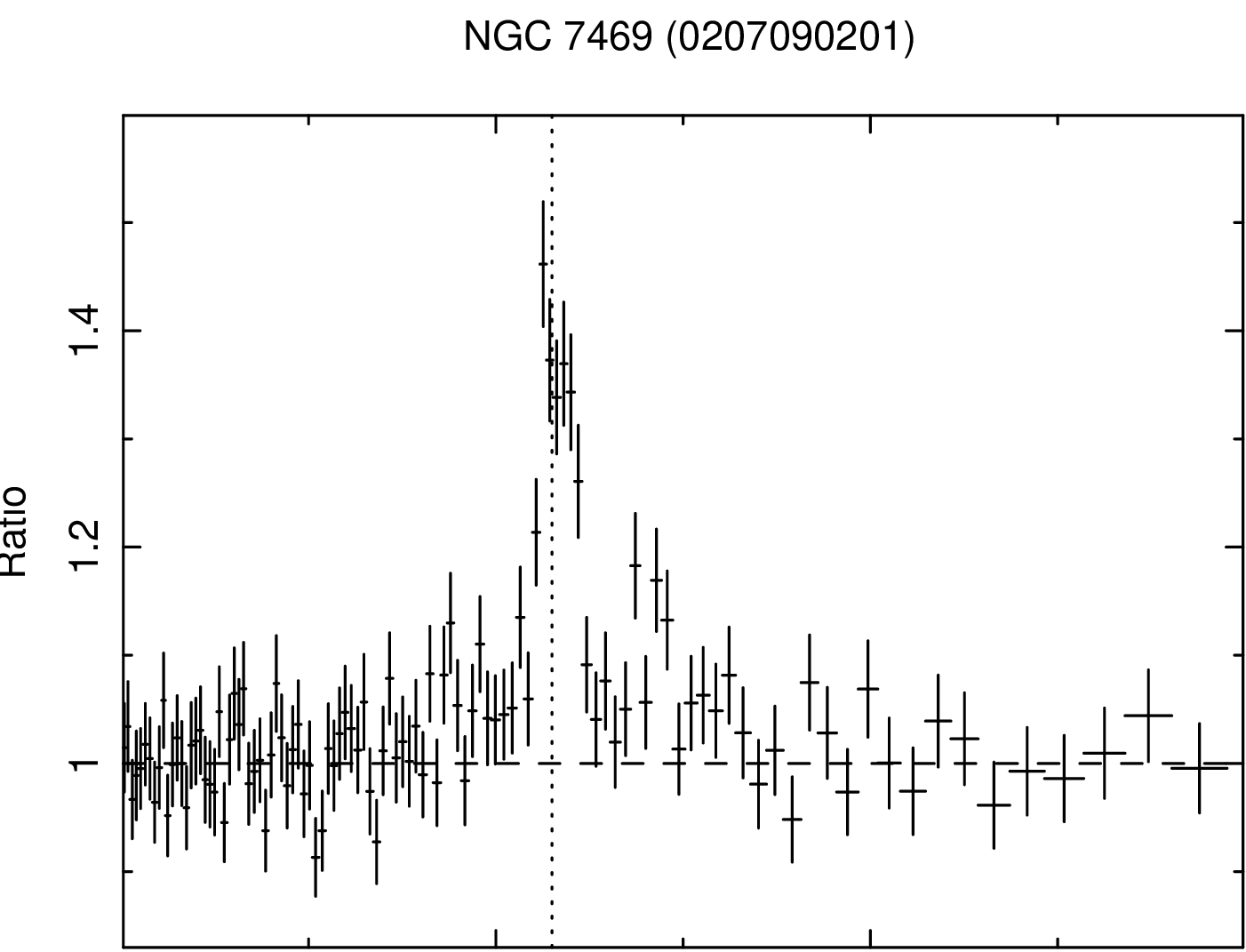}

\vspace{-0.07cm}

\hspace{-0.12cm}
    \includegraphics[width=3cm,height=4cm,angle=270]{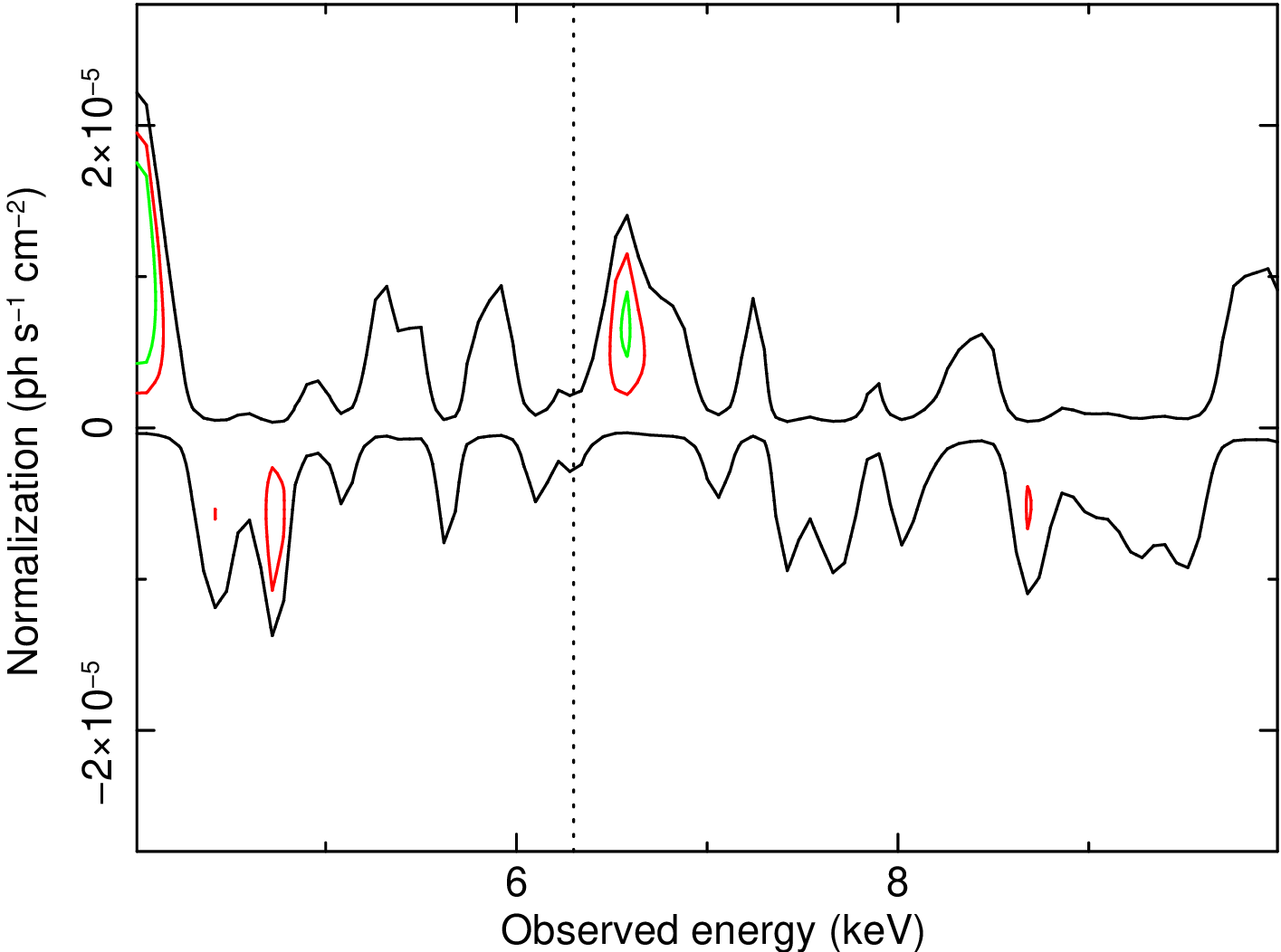}
\hspace{0.3cm}
    \includegraphics[width=3cm,height=4cm,angle=270]{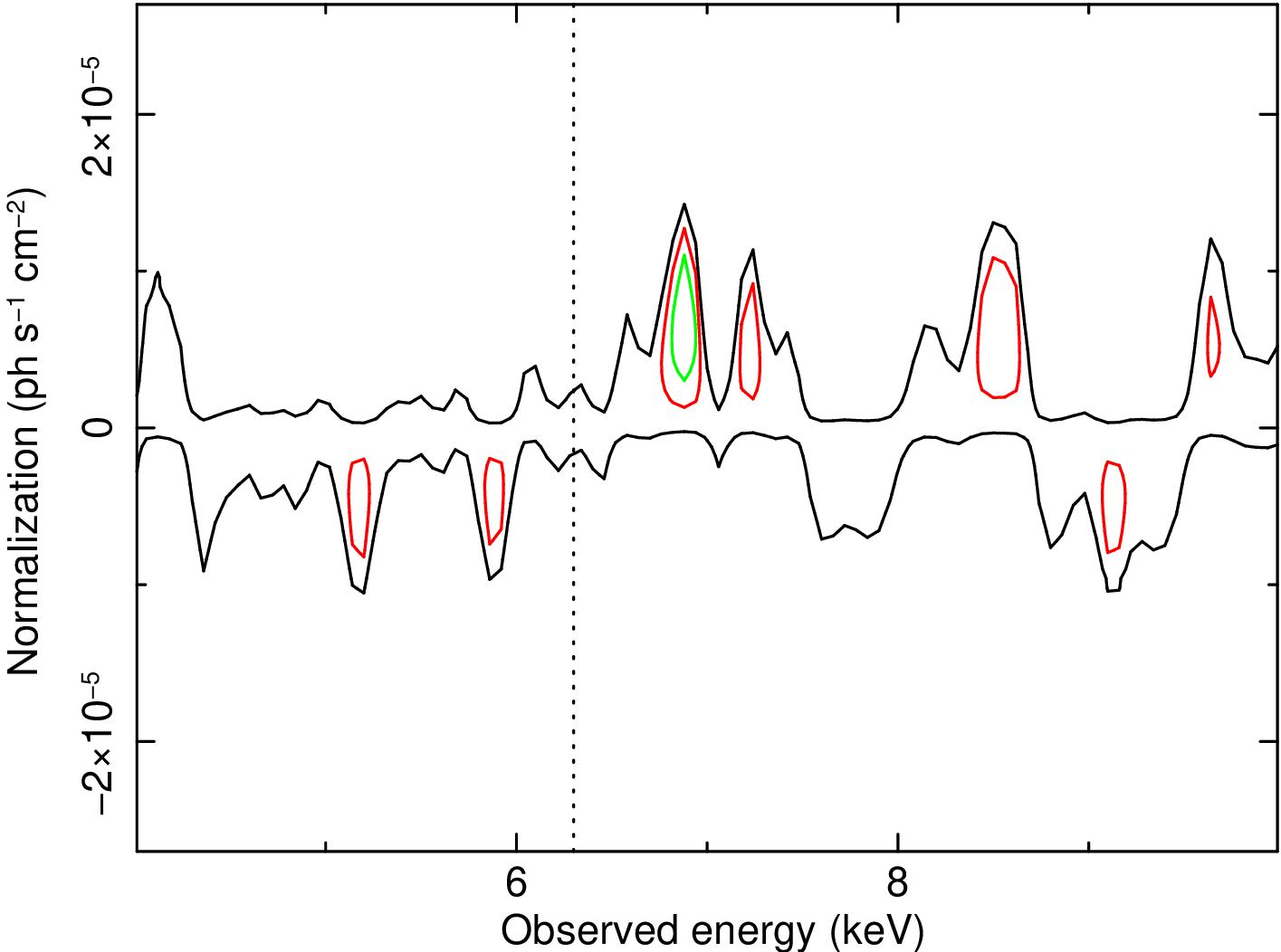}
\hspace{0.3cm}
    \includegraphics[width=3cm,height=4cm,angle=270]{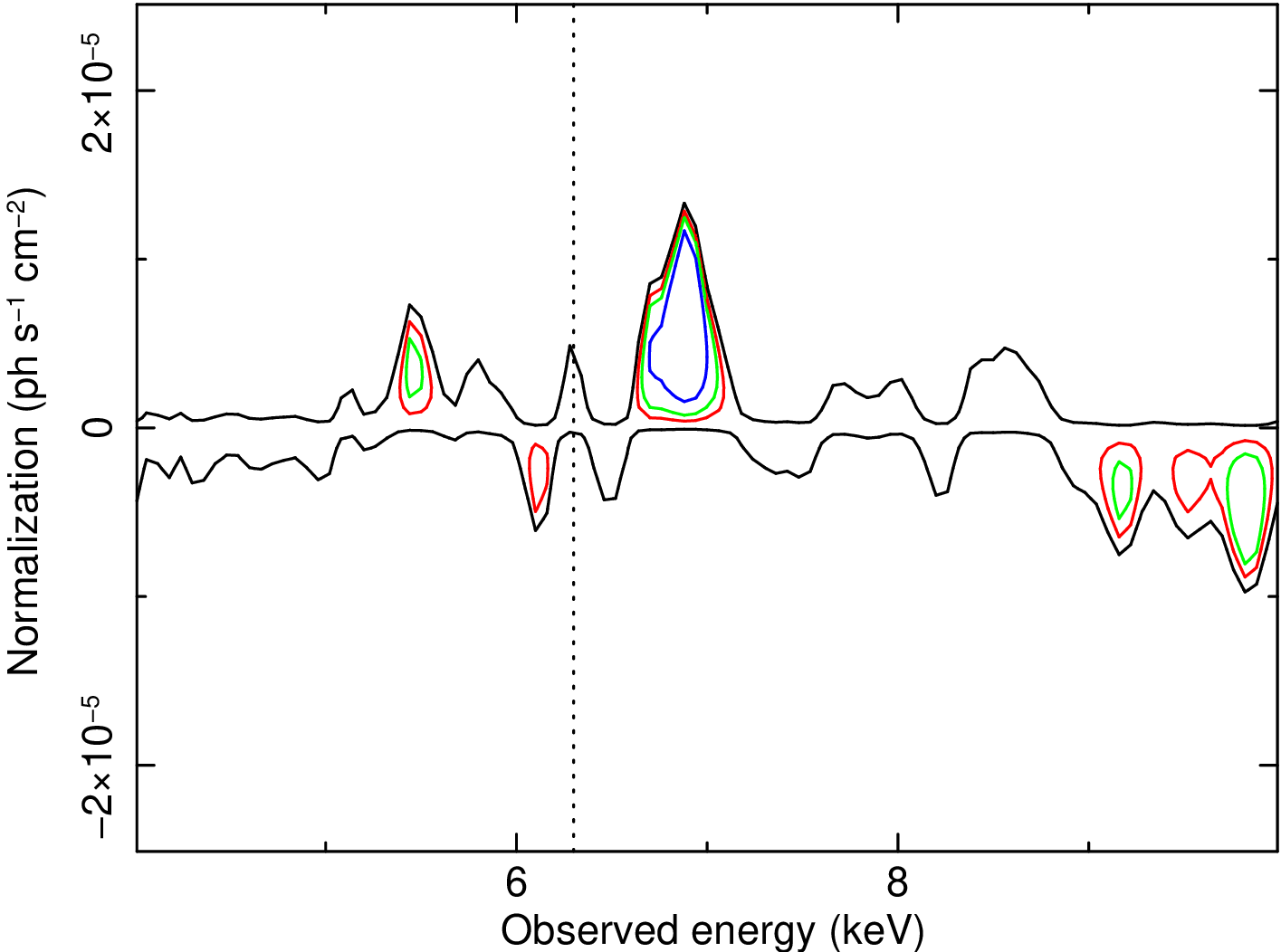}
\hspace{0.3cm}
    \includegraphics[width=3cm,height=4cm,angle=270]{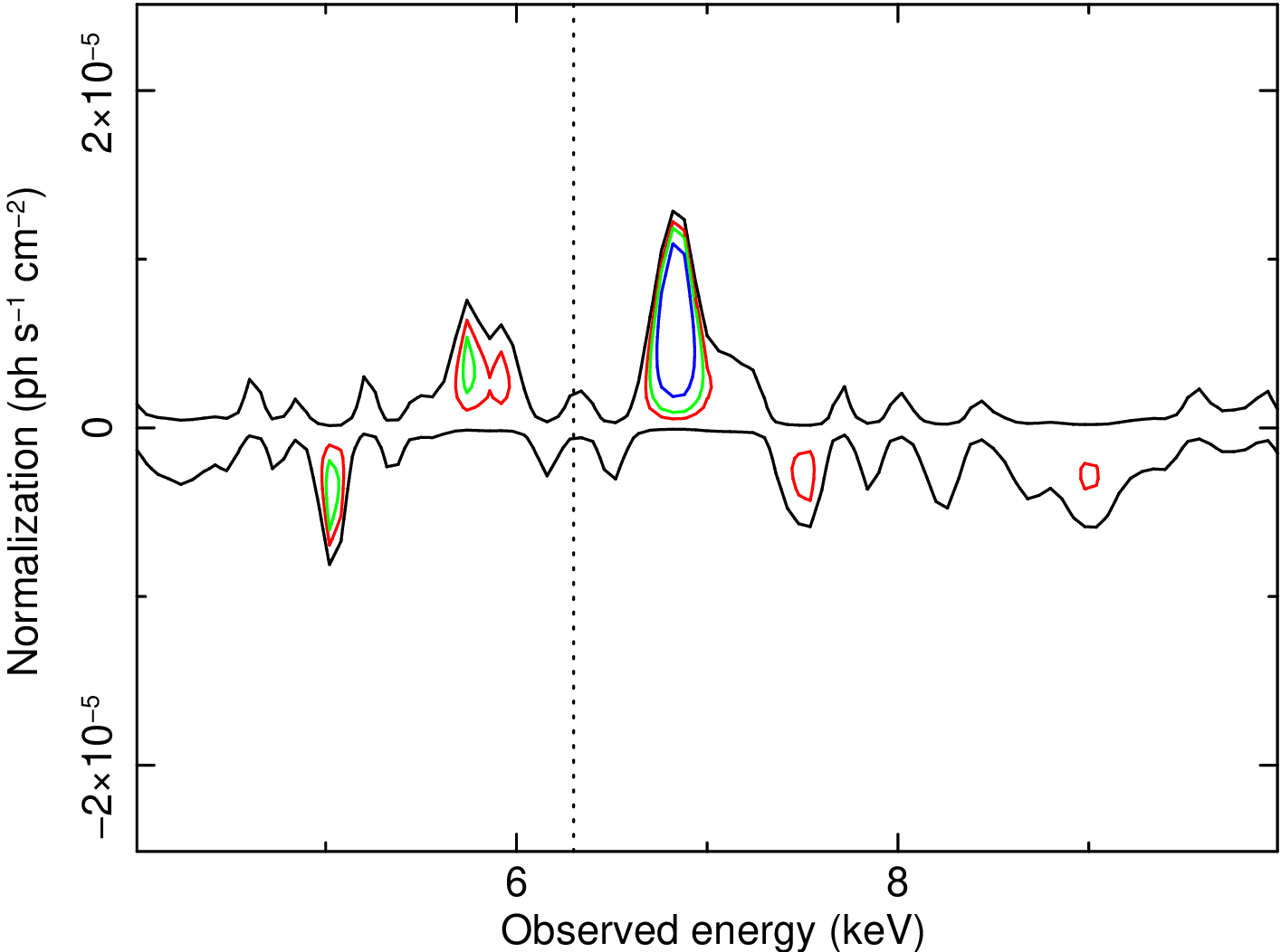}


\vspace{0.2cm}

    \includegraphics[width=3cm,height=4cm,angle=270]{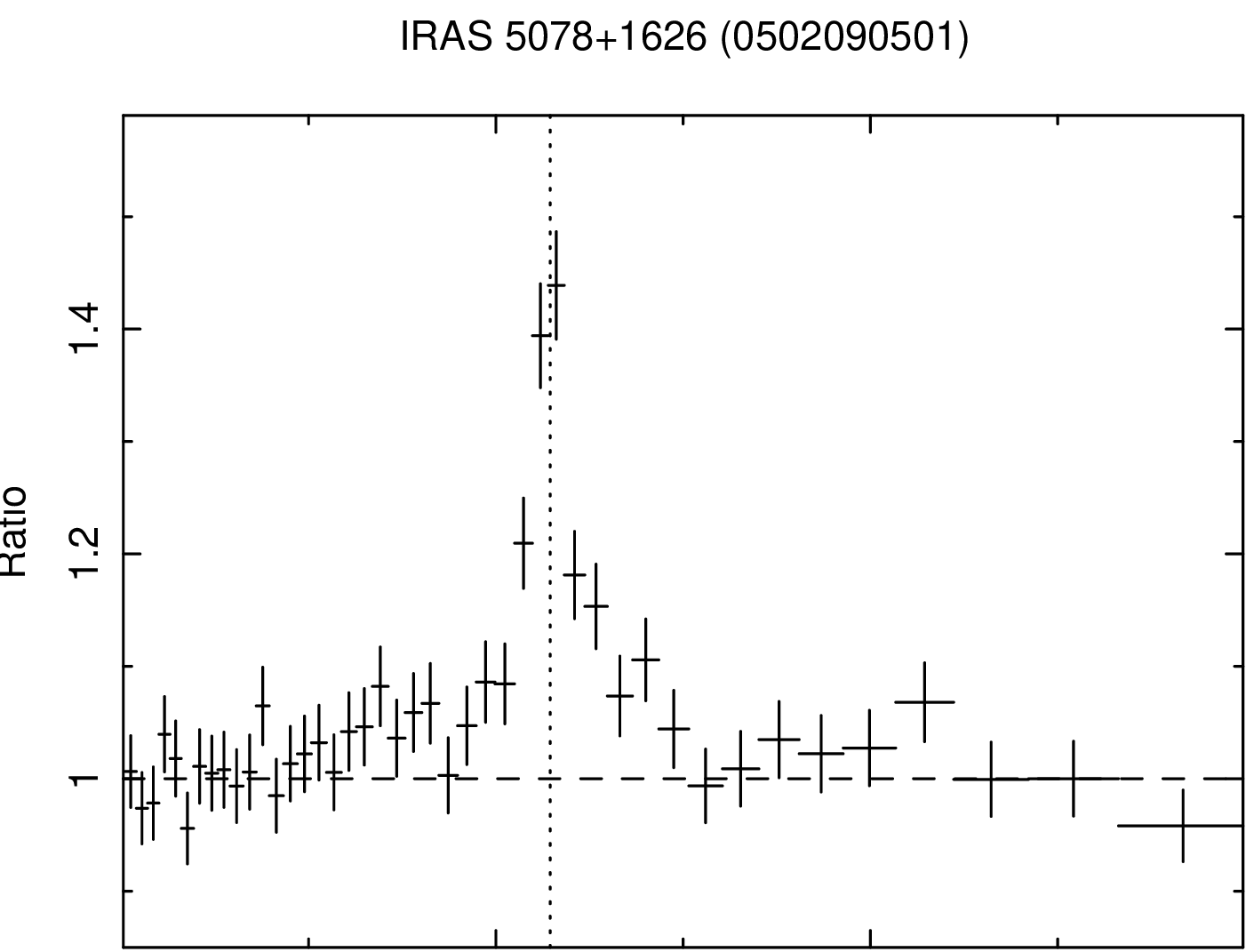}
\hspace{0.3cm}
    \includegraphics[width=3cm,height=4cm,angle=270]{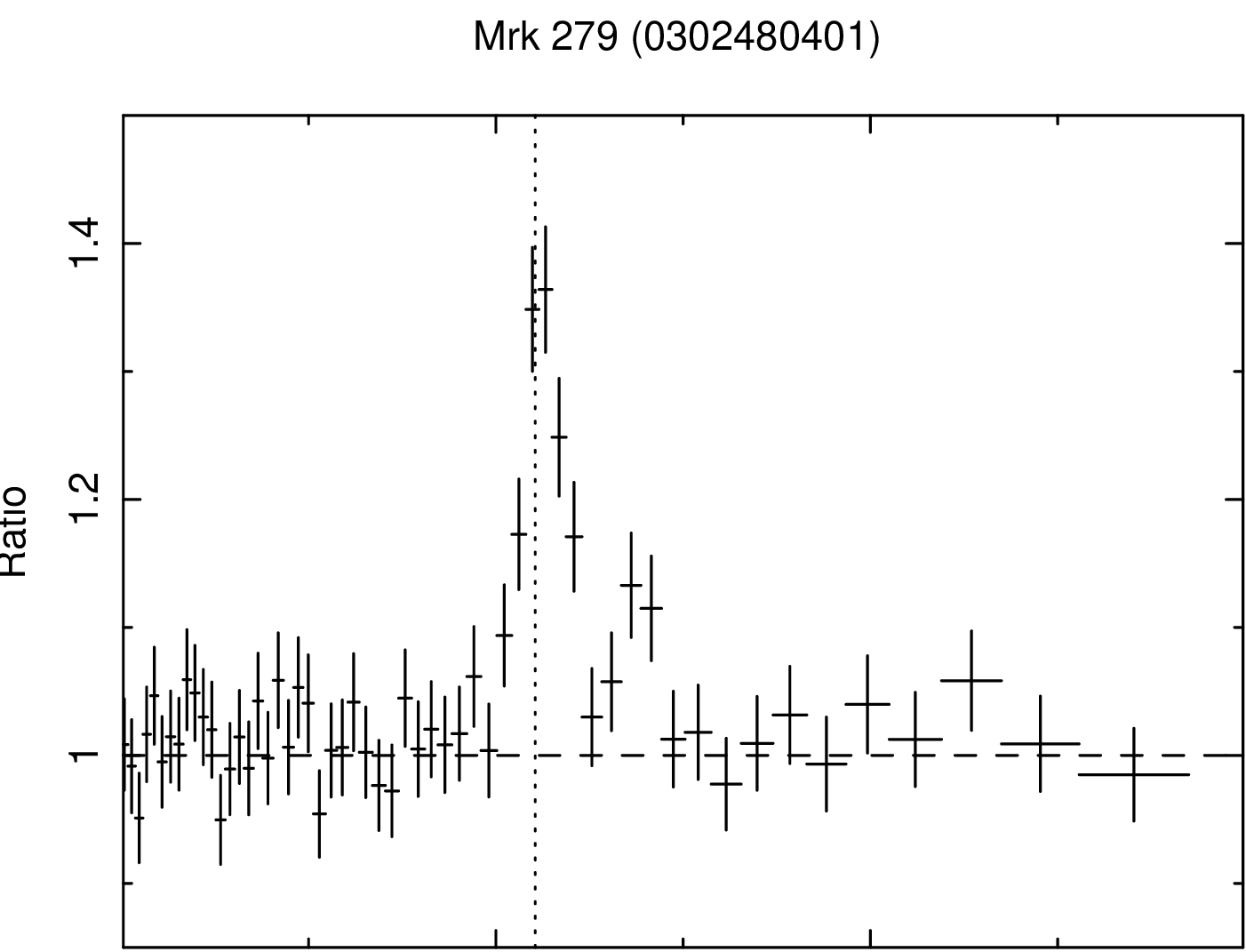}
\hspace{0.3cm}
    \includegraphics[width=3cm,height=4cm,angle=270]{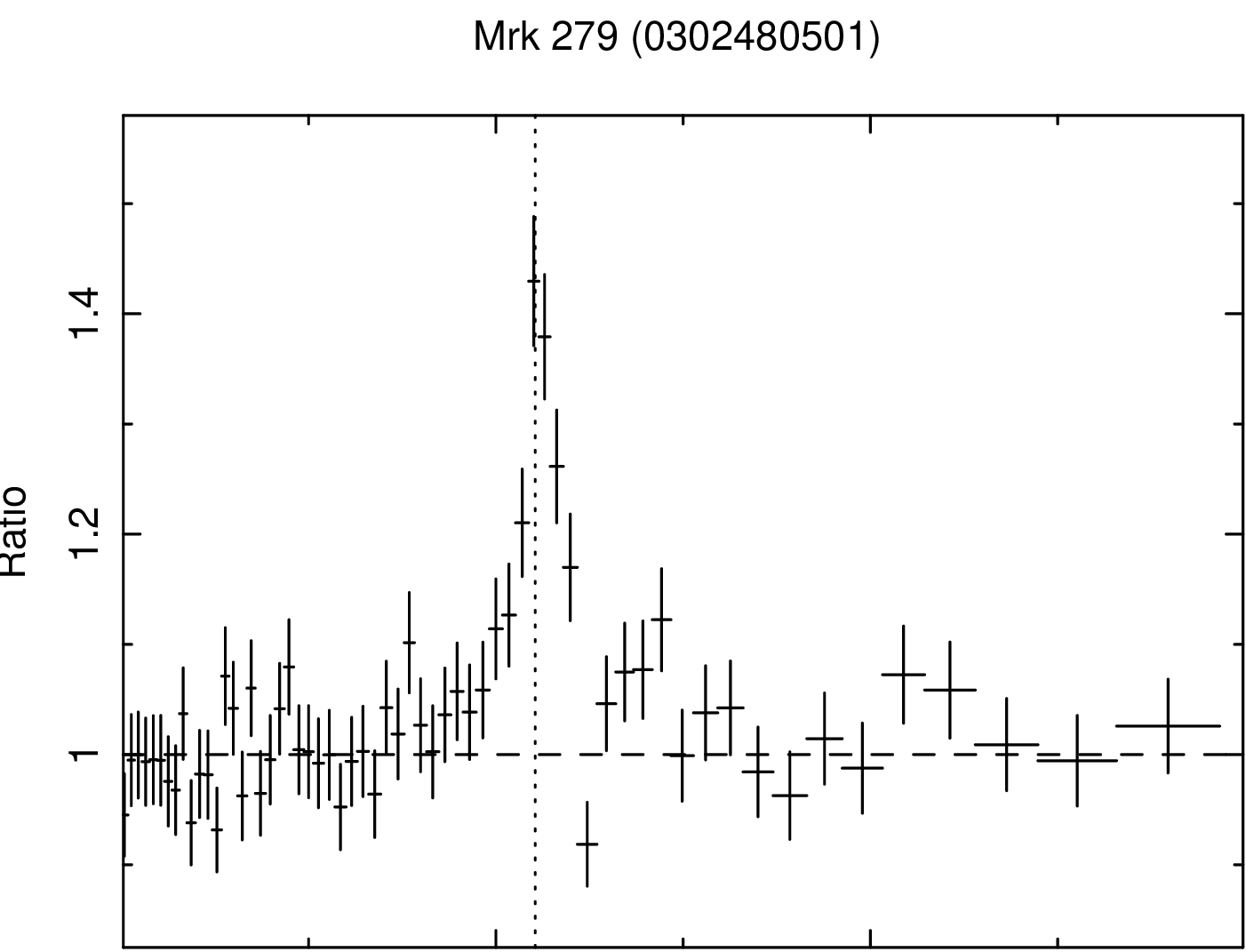}
\hspace{0.3cm}
    \includegraphics[width=3cm,height=4cm,angle=270]{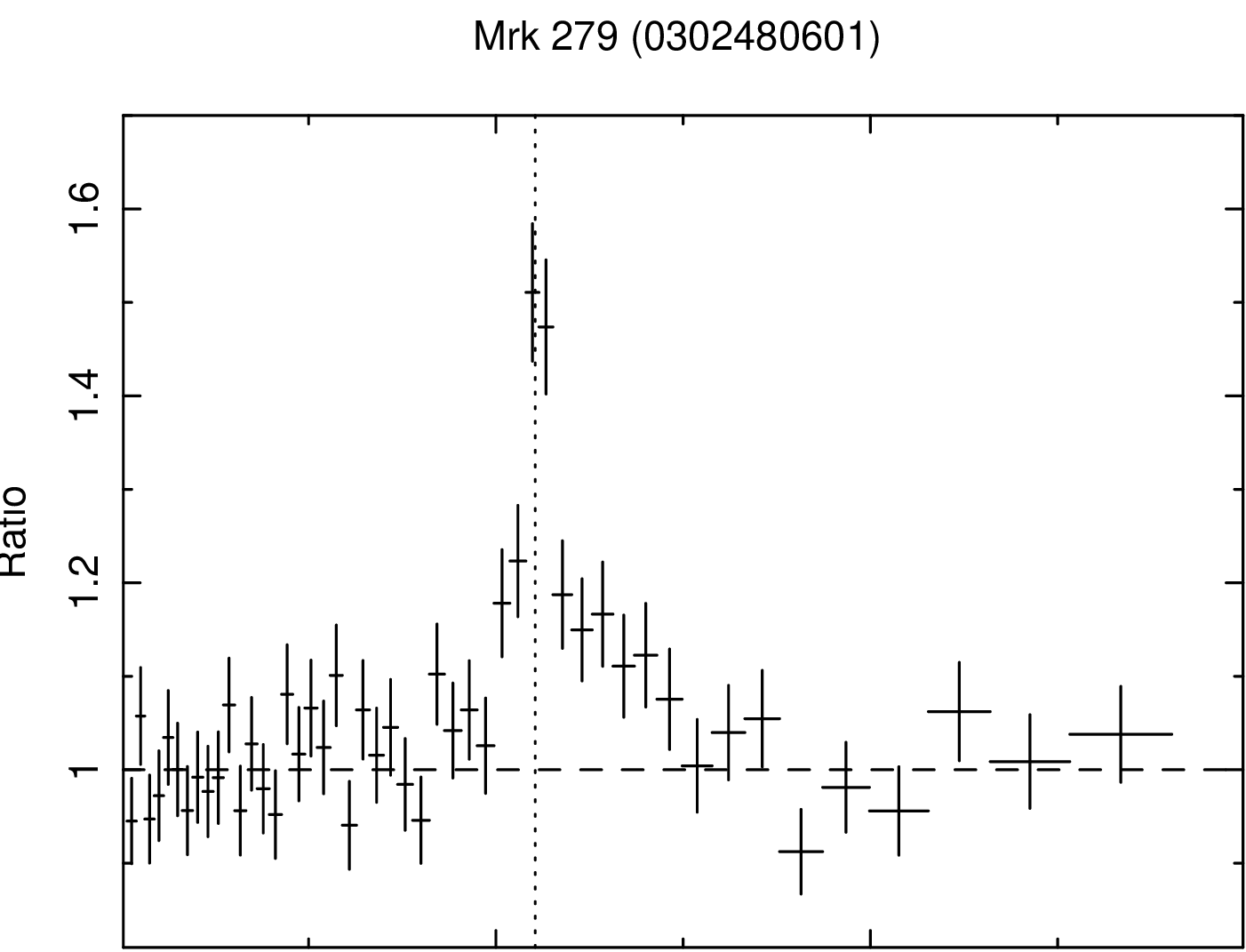}

\vspace{-0.07cm}

\hspace{-0.12cm}
    \includegraphics[width=3cm,height=4cm,angle=270]{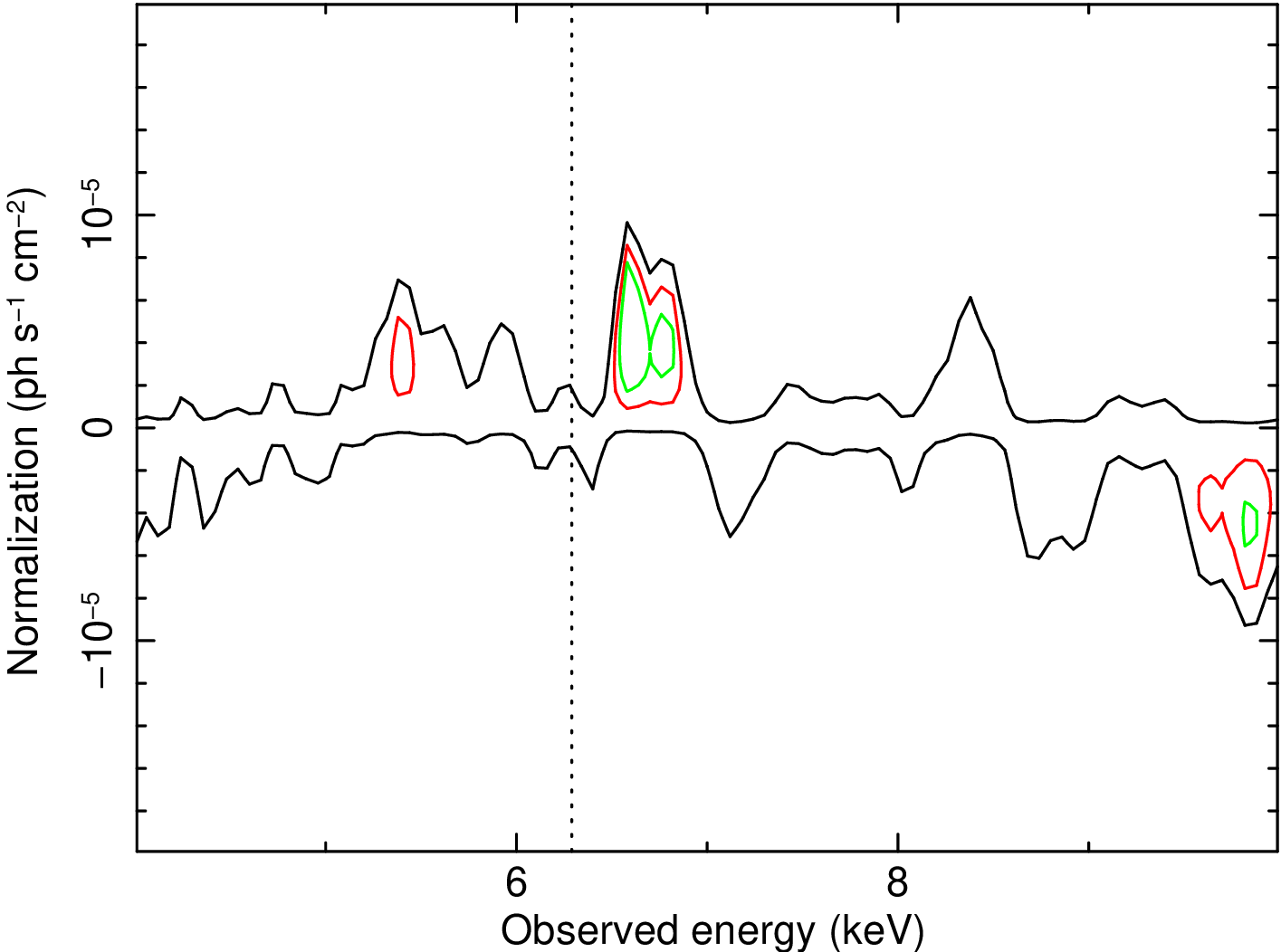}
\hspace{0.3cm}
    \includegraphics[width=3cm,height=4cm,angle=270]{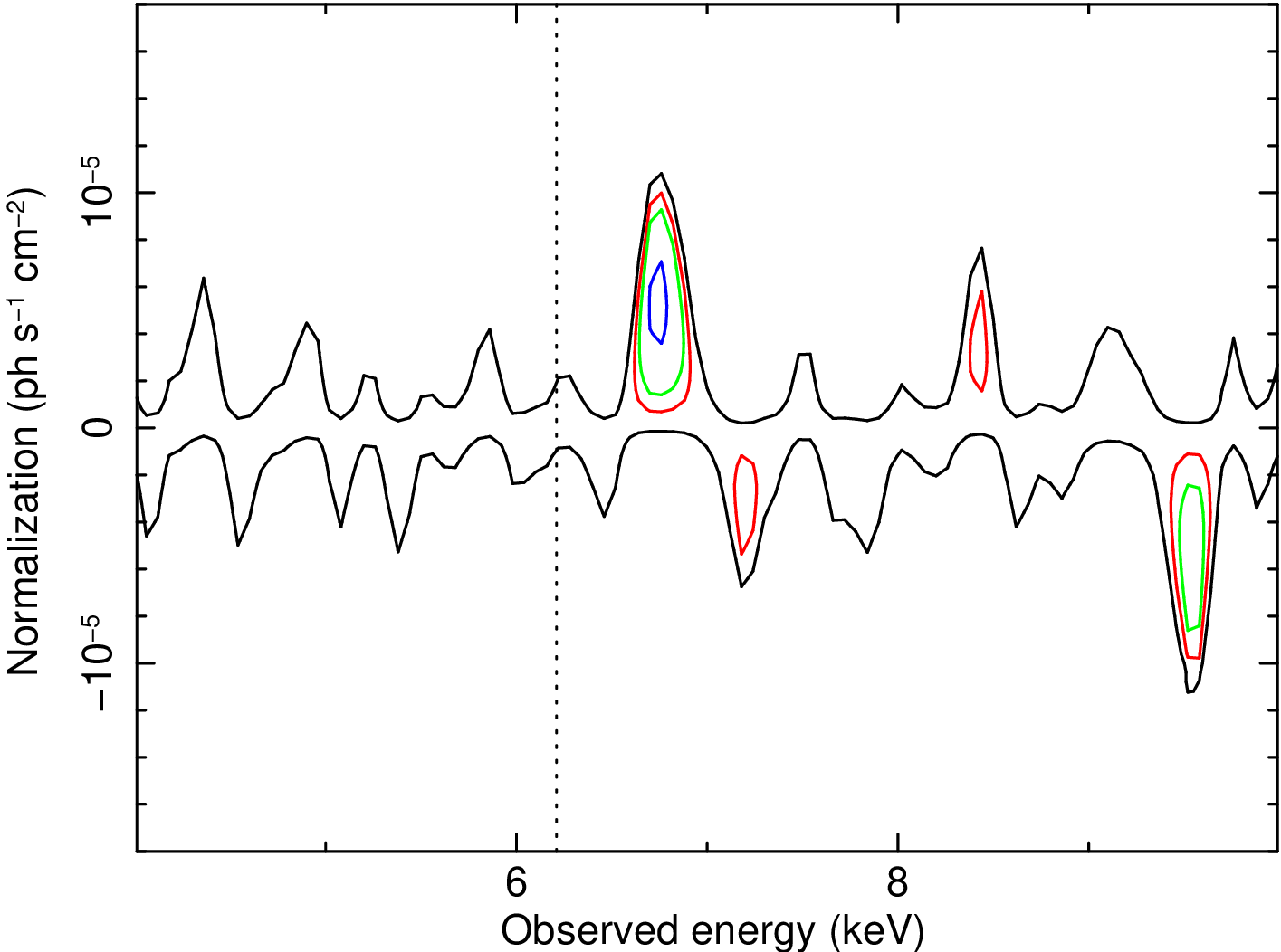}
\hspace{0.3cm}
    \includegraphics[width=3cm,height=4cm,angle=270]{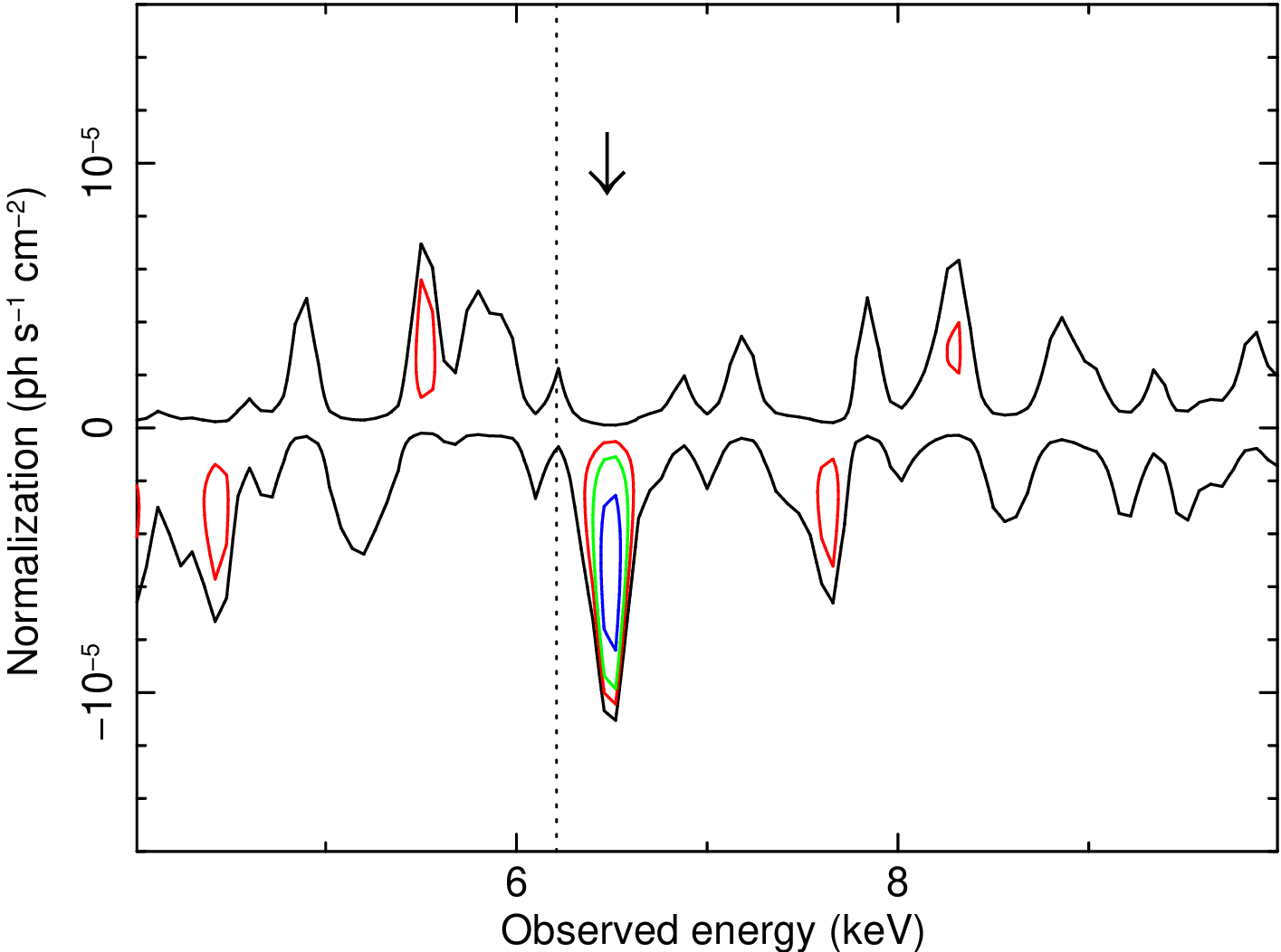}
\hspace{0.3cm}
    \includegraphics[width=3cm,height=4cm,angle=270]{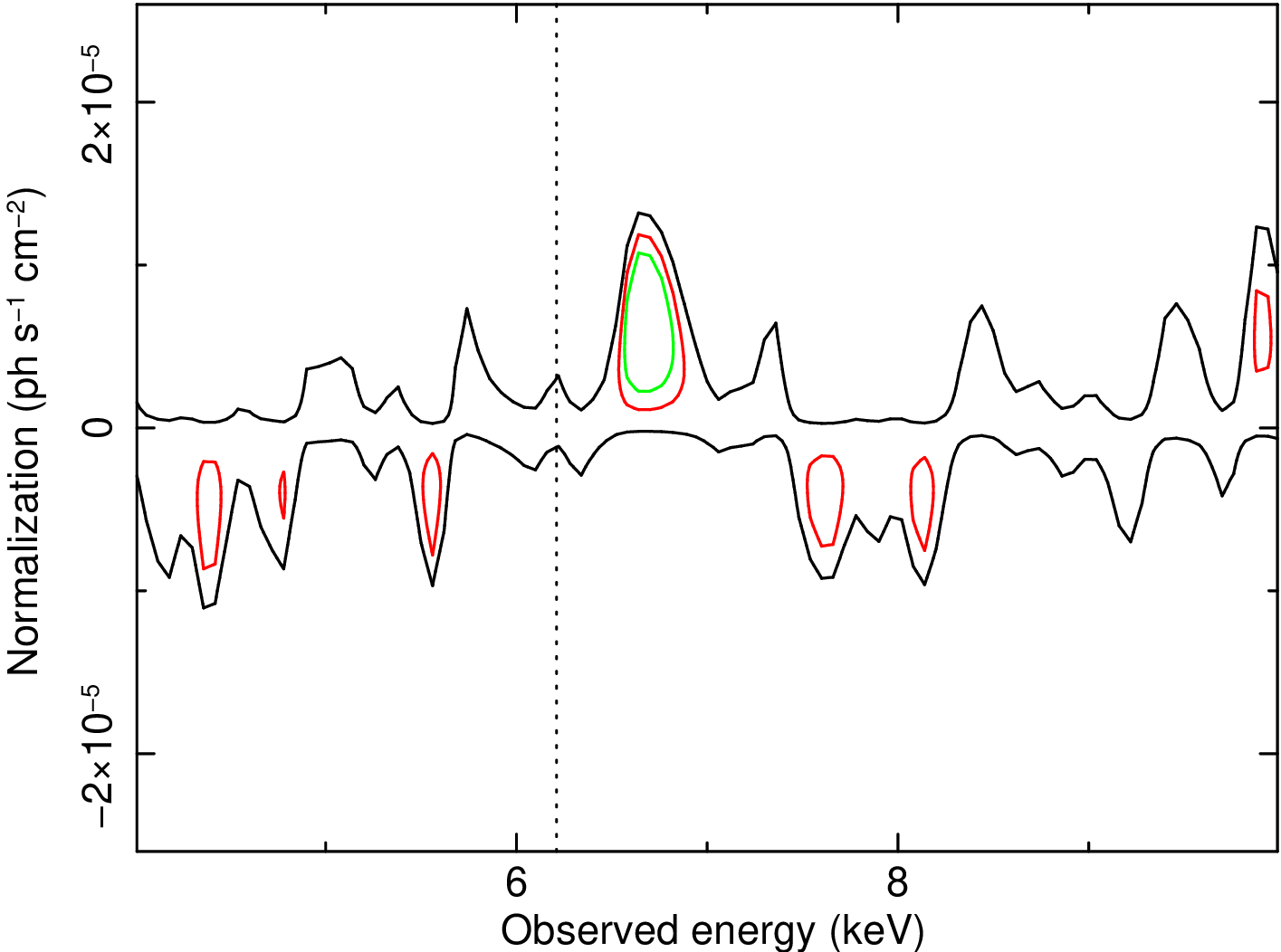}


\vspace{0.2cm}

    \includegraphics[width=3cm,height=4cm,angle=270]{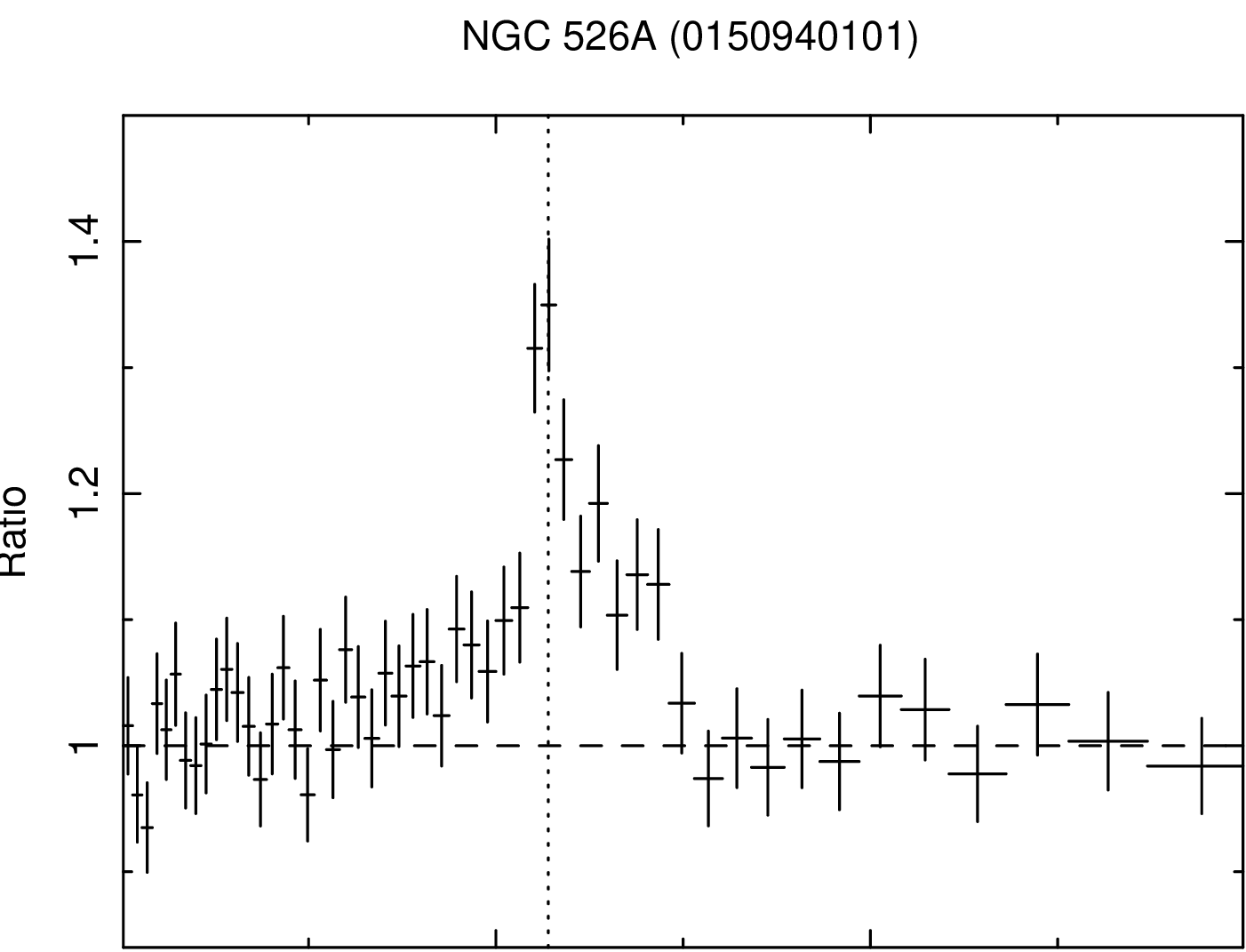}
\hspace{0.3cm}
    \includegraphics[width=3cm,height=4cm,angle=270]{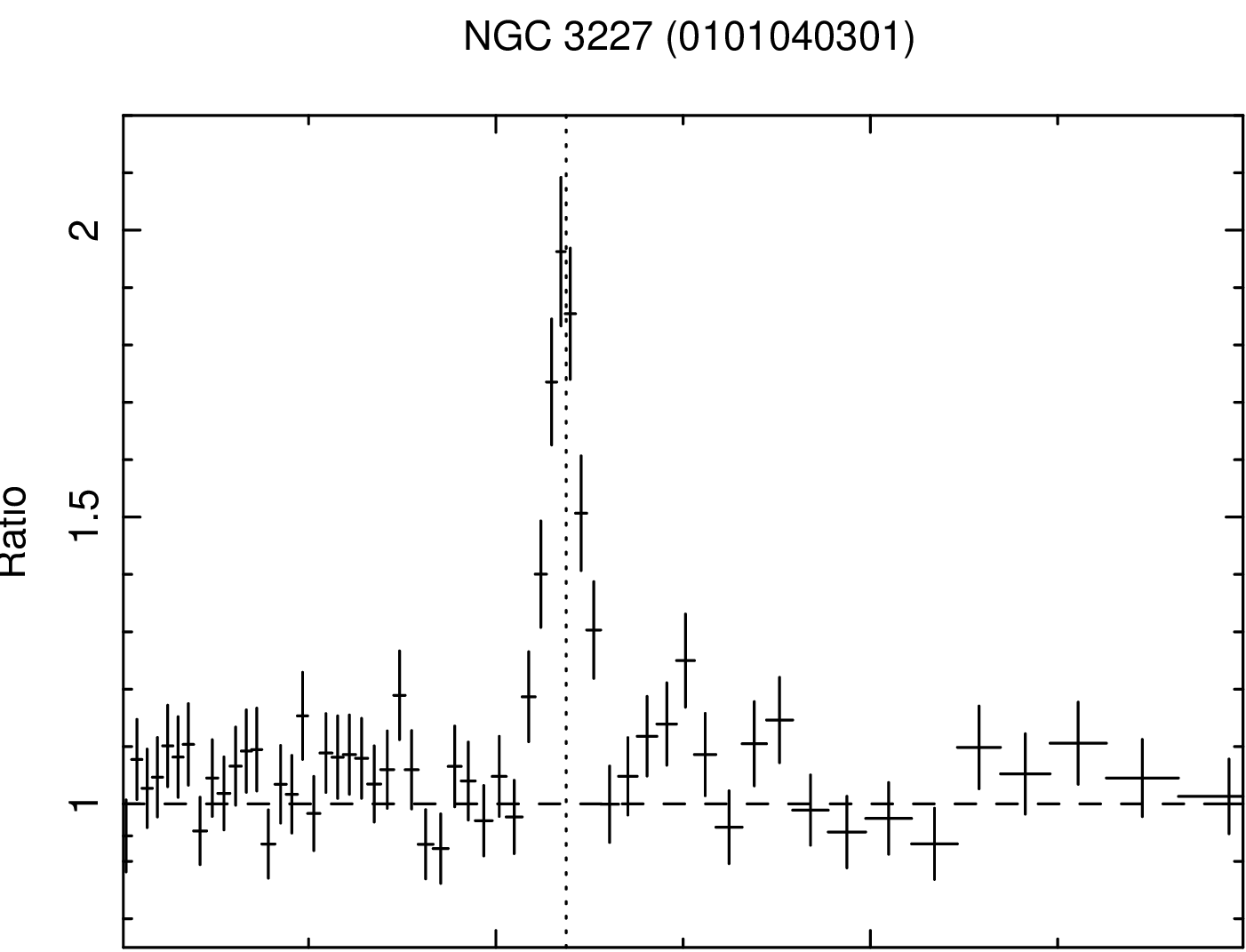}
\hspace{0.3cm}
    \includegraphics[width=3cm,height=4cm,angle=270]{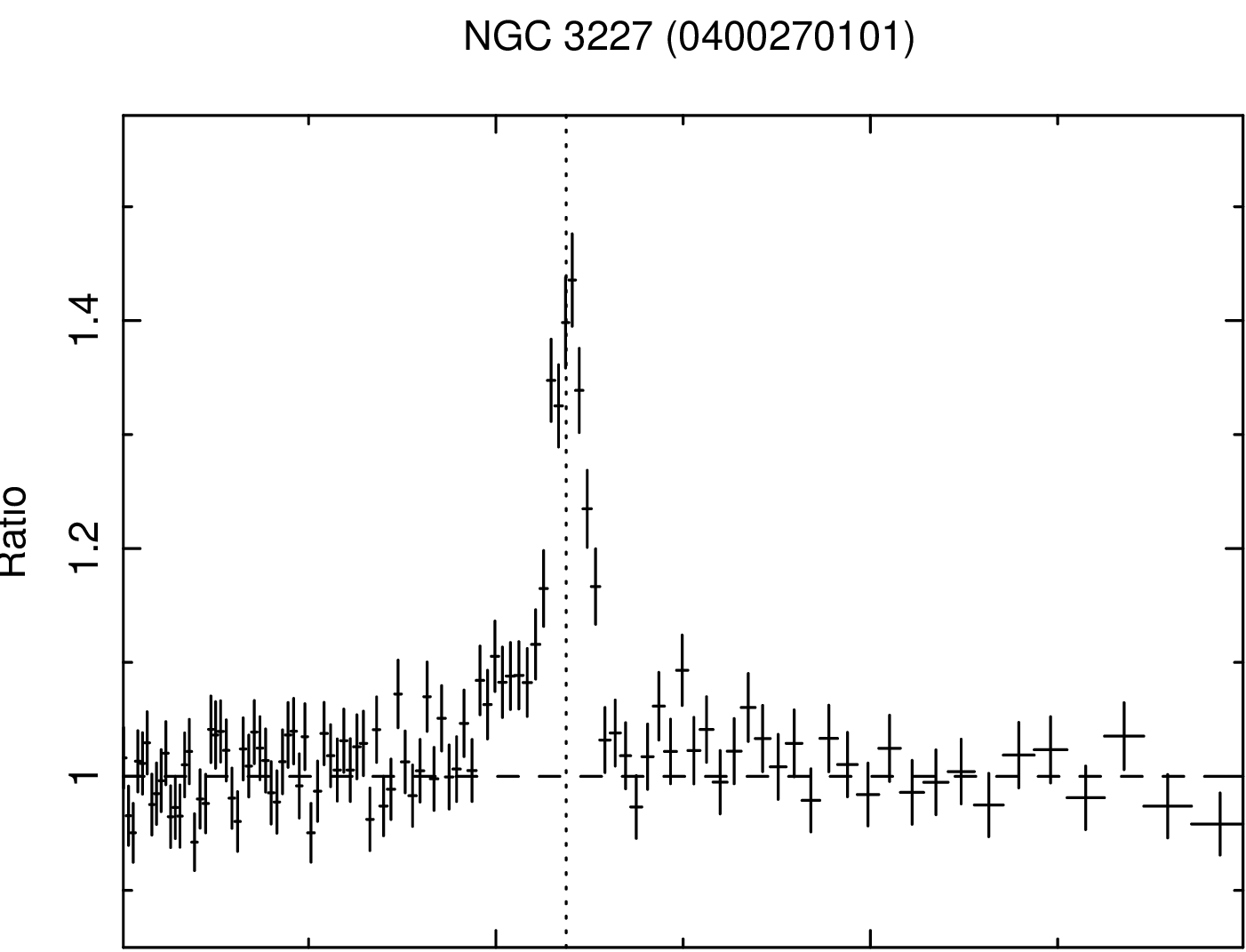}
\hspace{0.3cm}
    \includegraphics[width=3cm,height=4cm,angle=270]{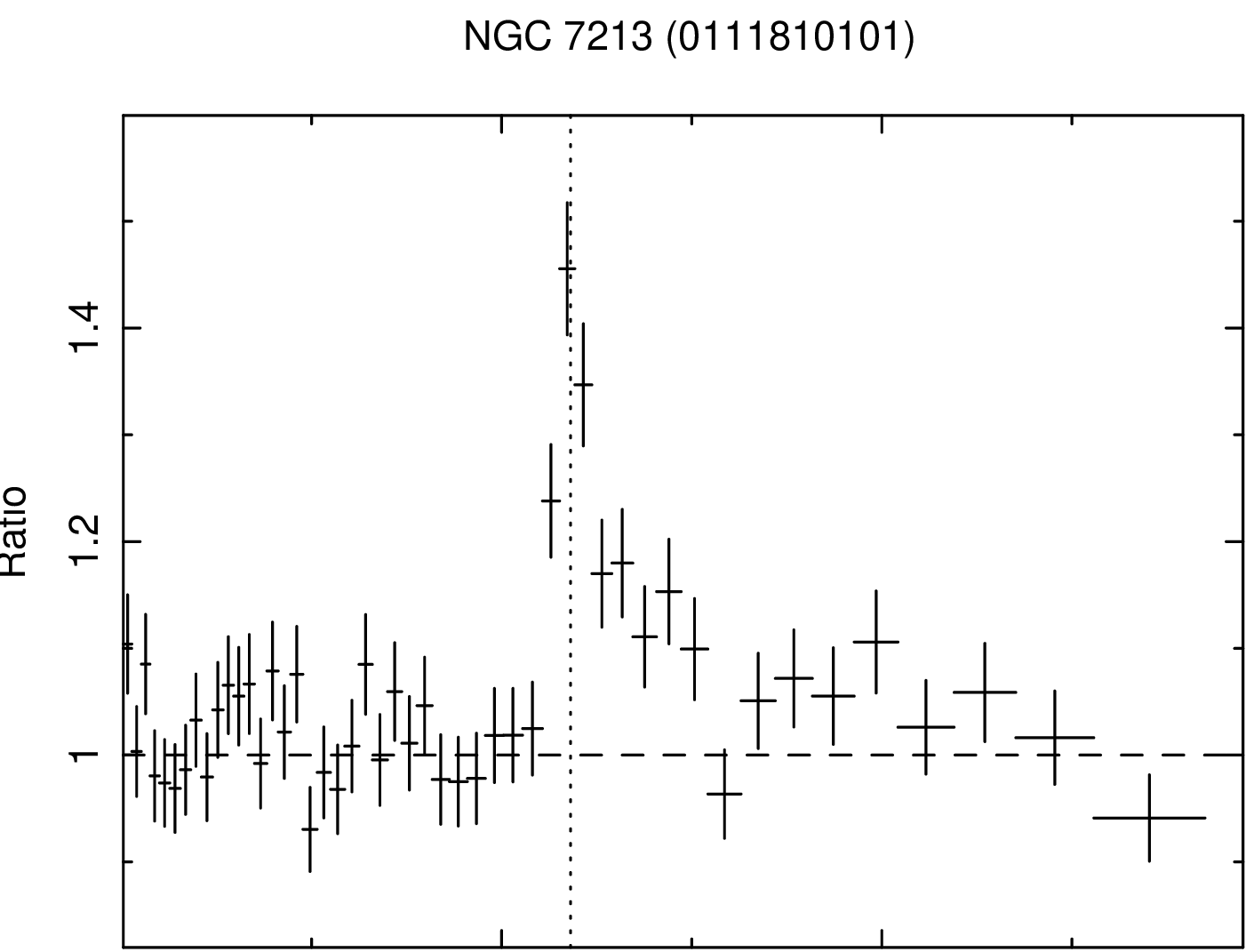}

\vspace{-0.07cm}

\hspace{-0.12cm}
    \includegraphics[width=3cm,height=4cm,angle=270]{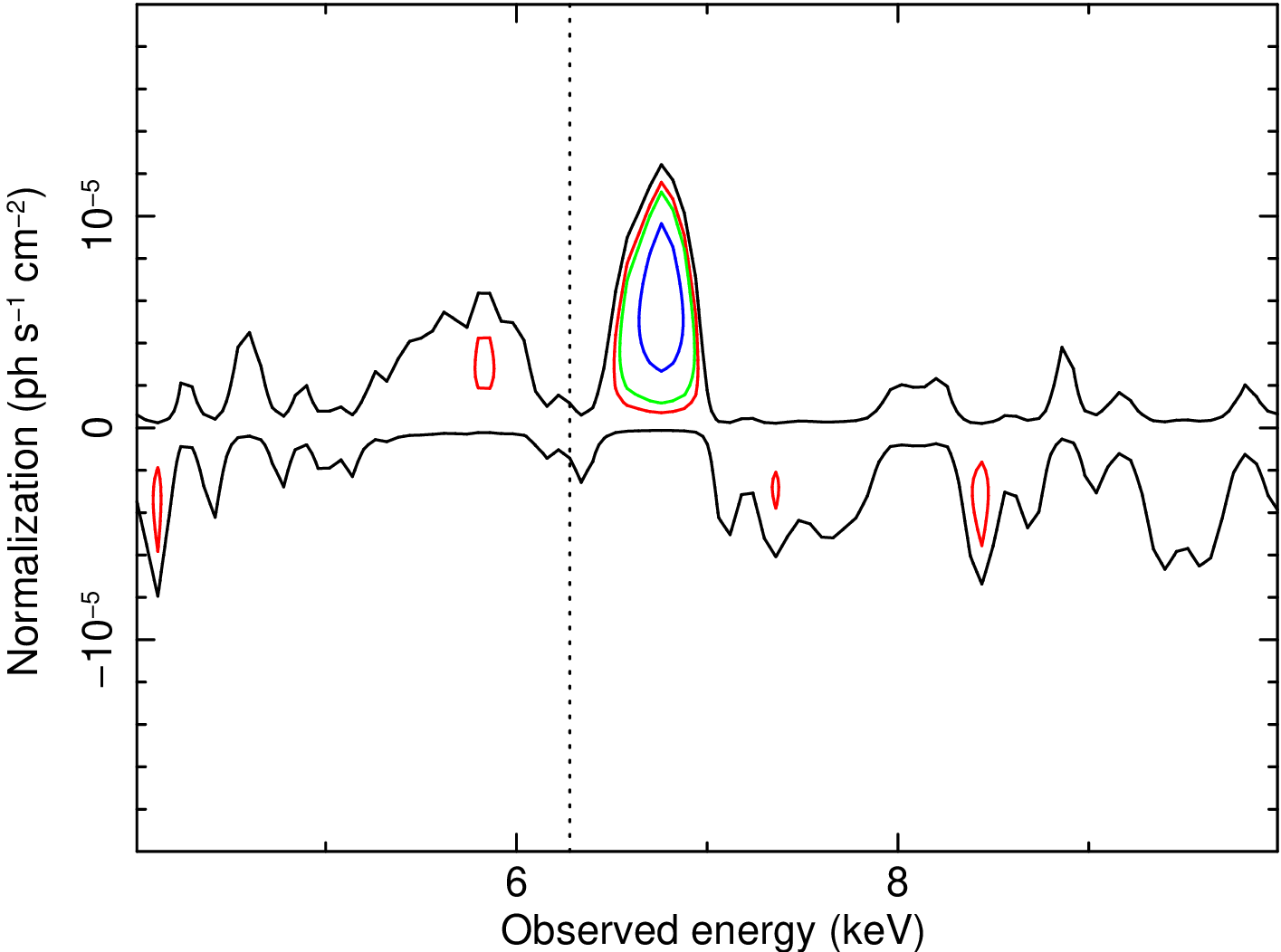}
\hspace{0.3cm}
    \includegraphics[width=3cm,height=4cm,angle=270]{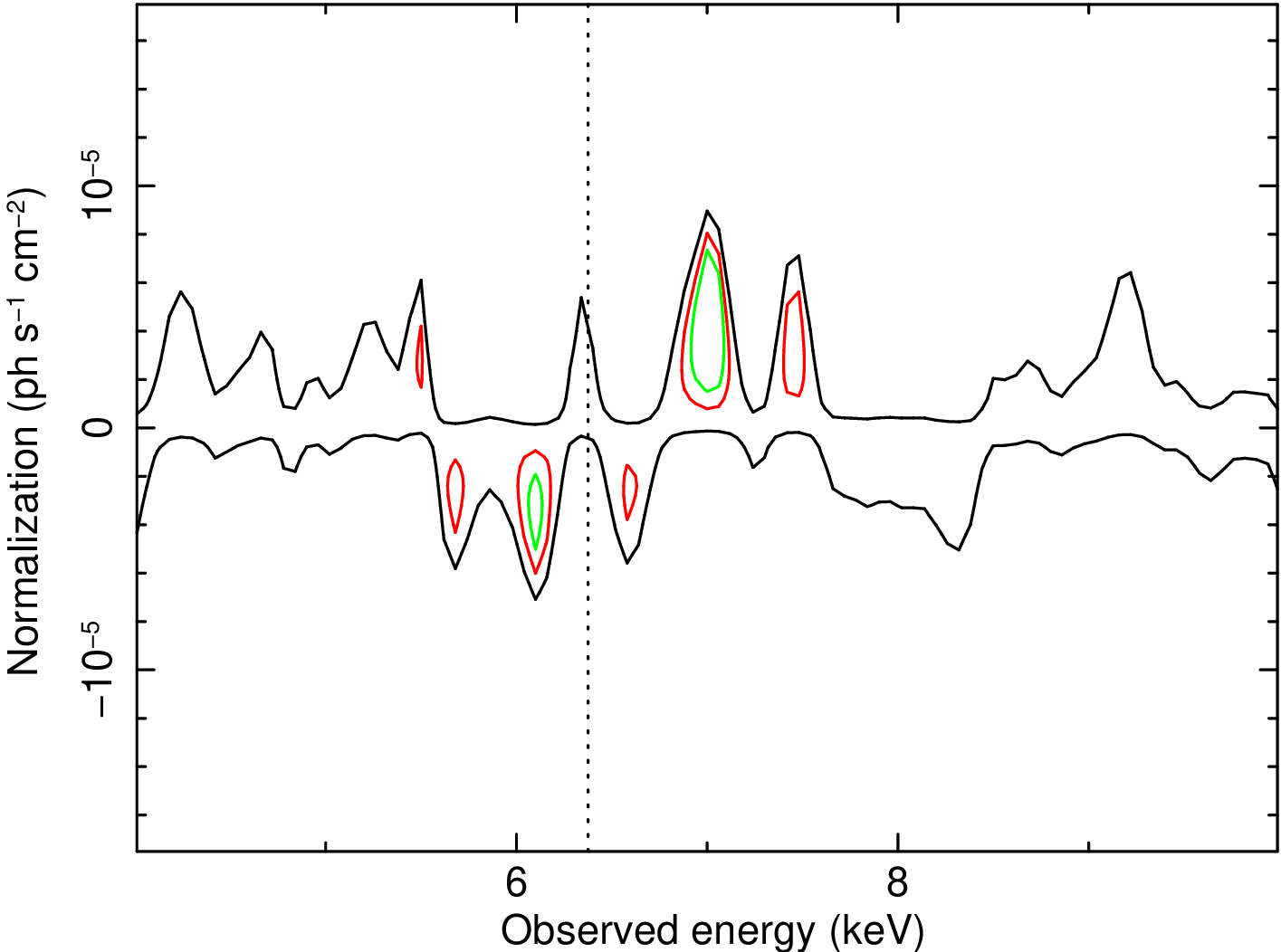}
\hspace{0.3cm}
    \includegraphics[width=3cm,height=4cm,angle=270]{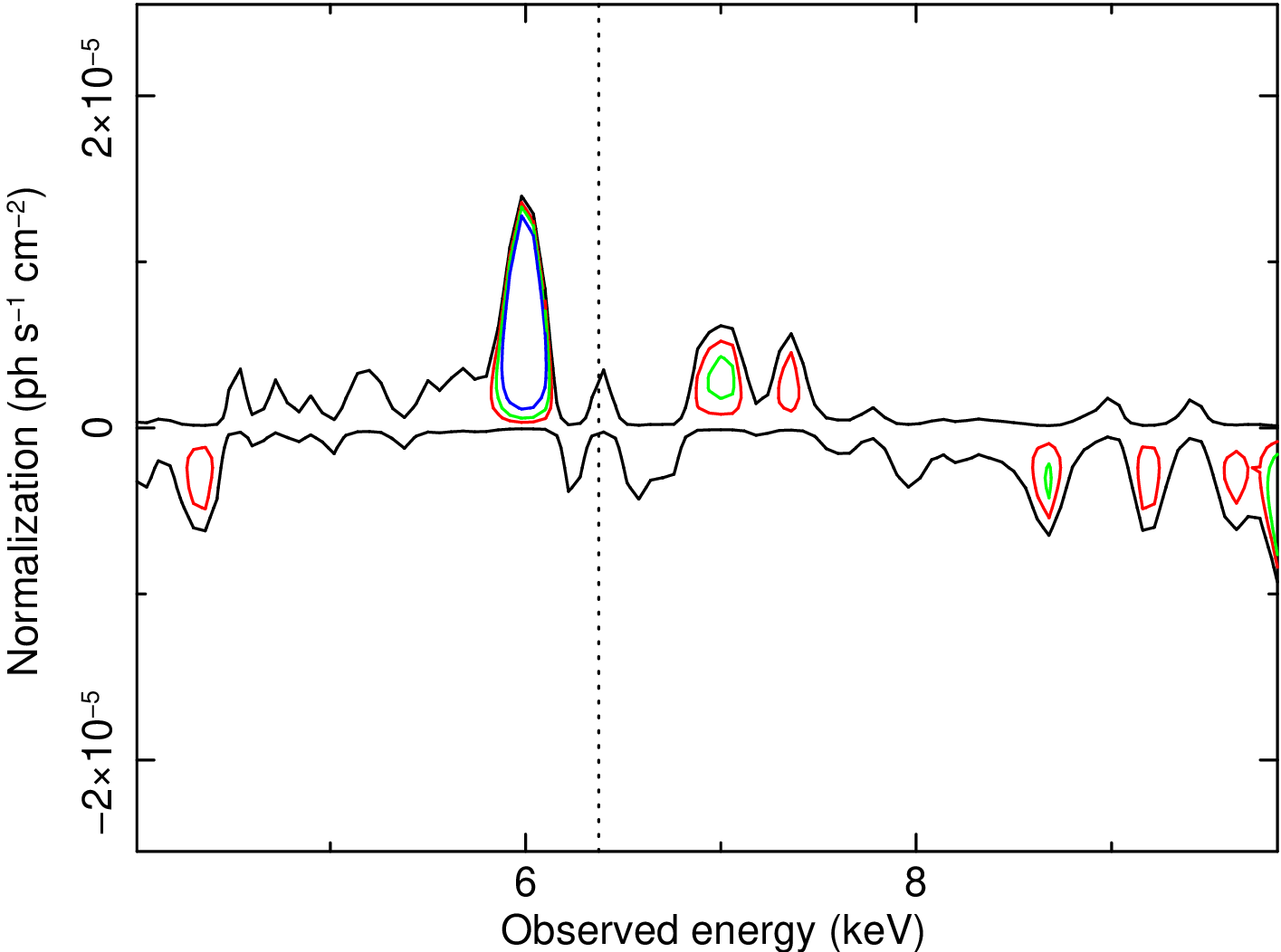}
\hspace{0.3cm}
    \includegraphics[width=3cm,height=4cm,angle=270]{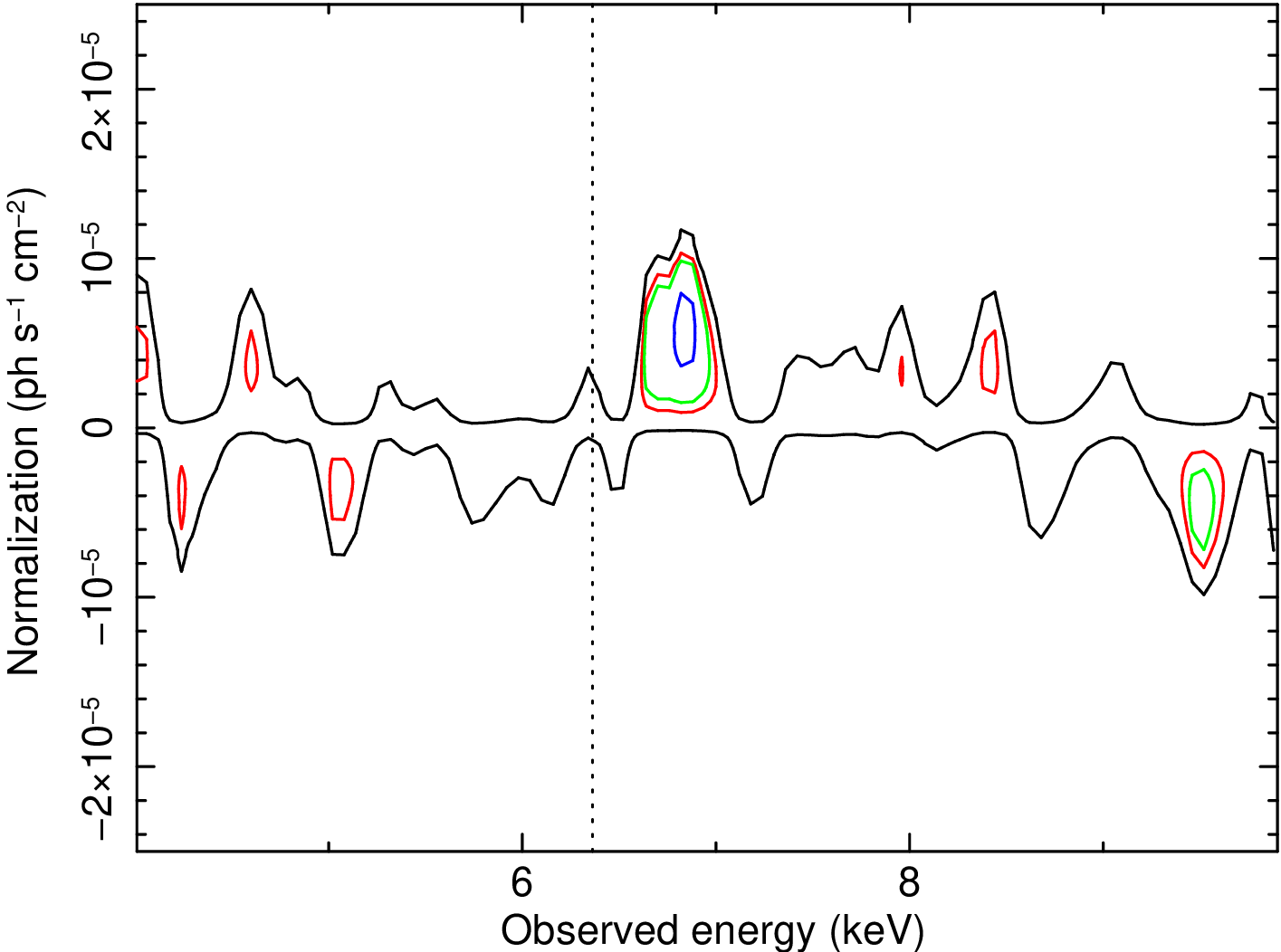}

   \caption{Ratio against the continuum (\emph{upper panel}) and contour plots with respect to the best-fit baseline model (\emph{lower panel}; 68\% (red), 90\% (green), 99\% (blue) levels) for the Type 1 sources. The Fe K absorption lines are pointed by arrows. The vertical line refer to Fe I K$\alpha$.}
    \end{figure*}

   \begin{figure*}[ht]
   \centering


    \includegraphics[width=3cm,height=4cm,angle=270]{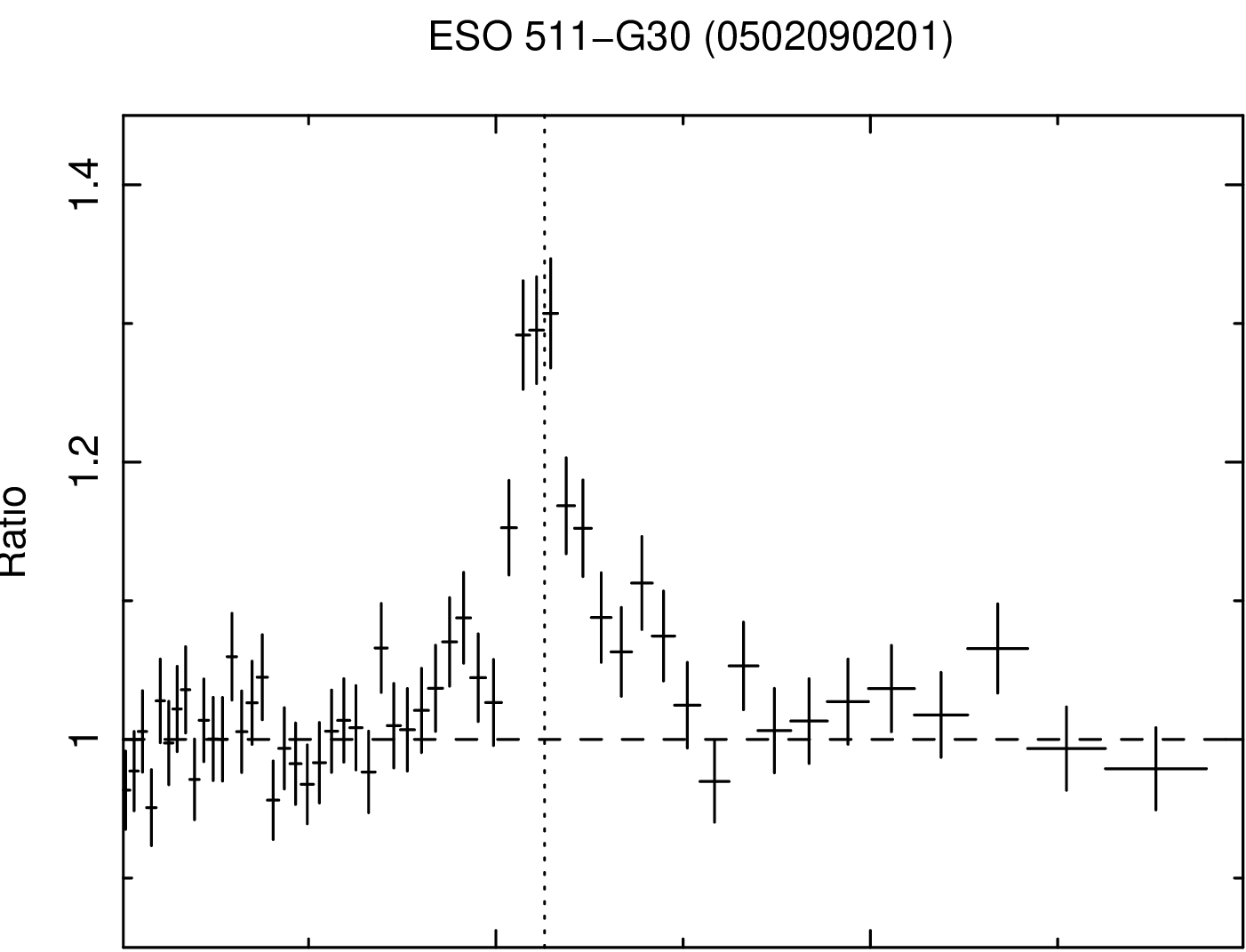}
\hspace{0.3cm}
    \includegraphics[width=3cm,height=4cm,angle=270]{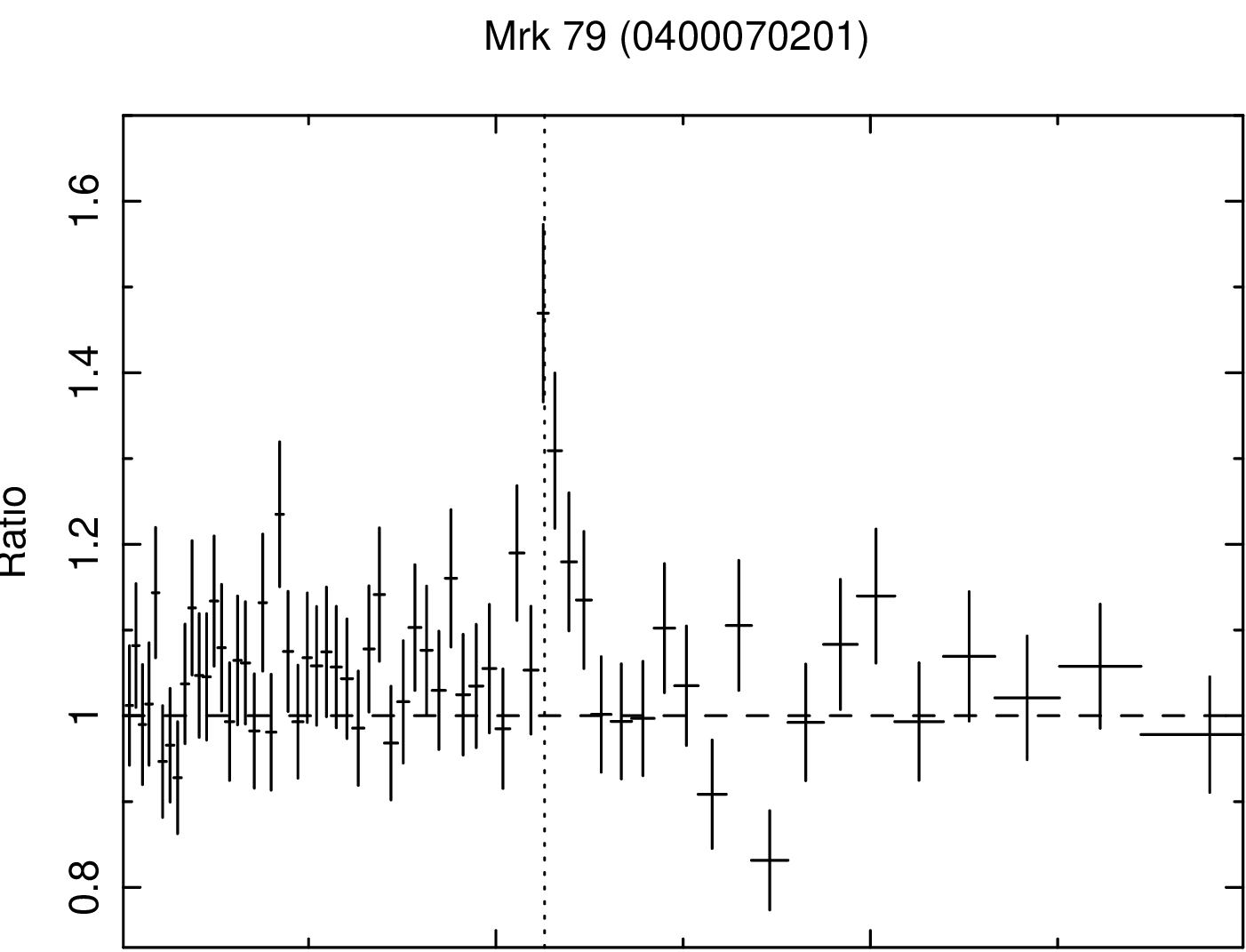}
\hspace{0.3cm}
    \includegraphics[width=3cm,height=4cm,angle=270]{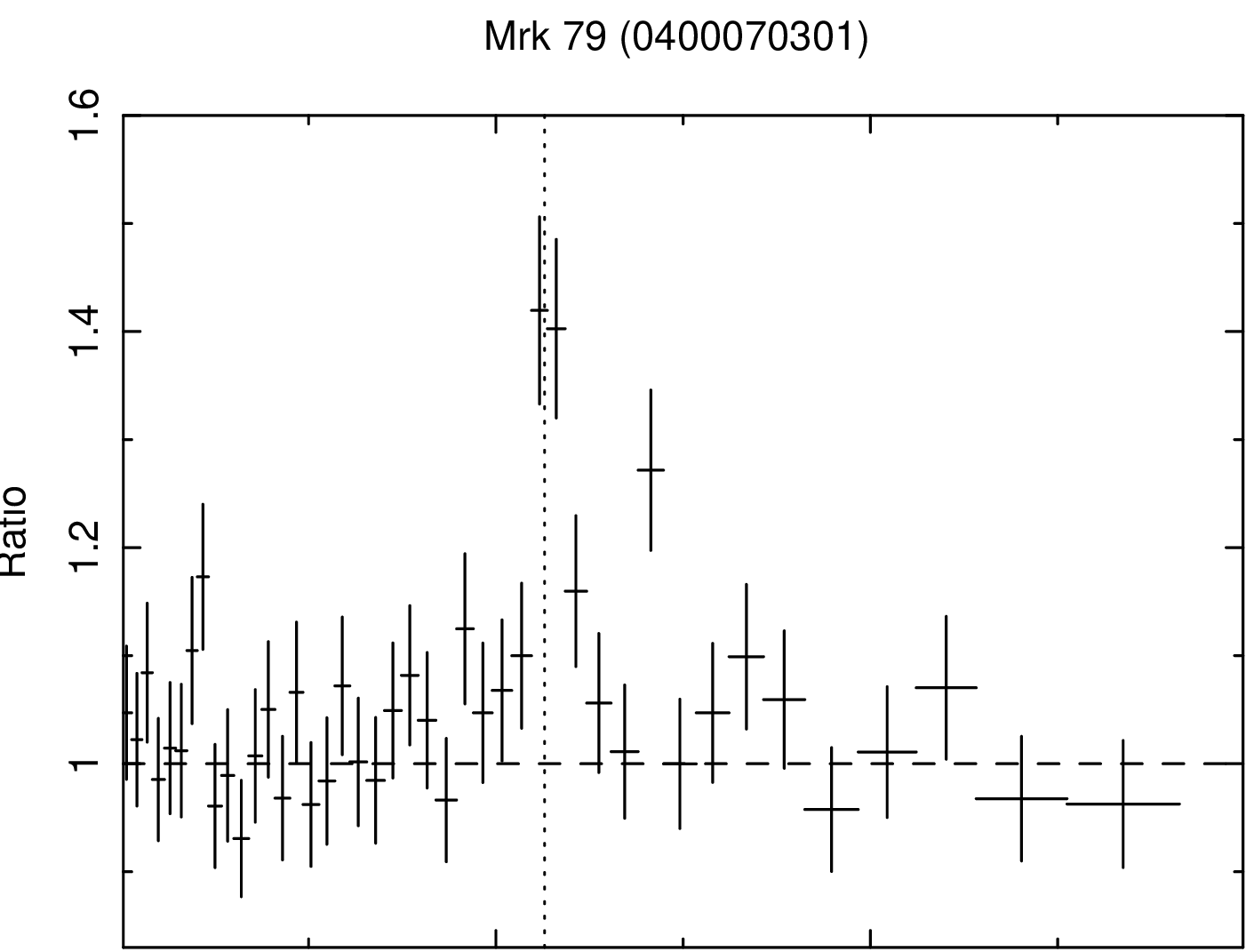}
\hspace{0.3cm}
    \includegraphics[width=3cm,height=4cm,angle=270]{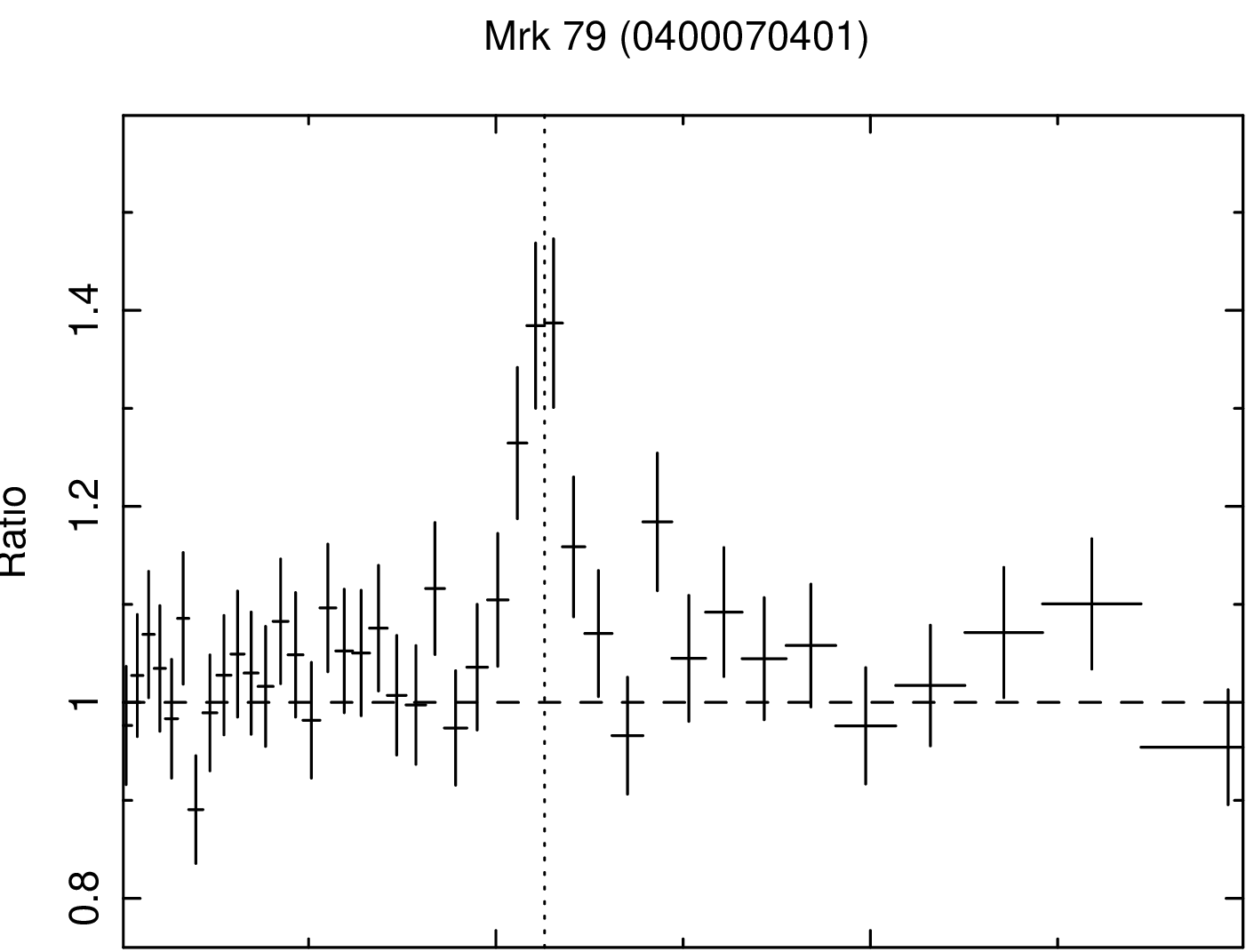}

\vspace{-0.07cm}

\hspace{-0.12cm}
    \includegraphics[width=3cm,height=4cm,angle=270]{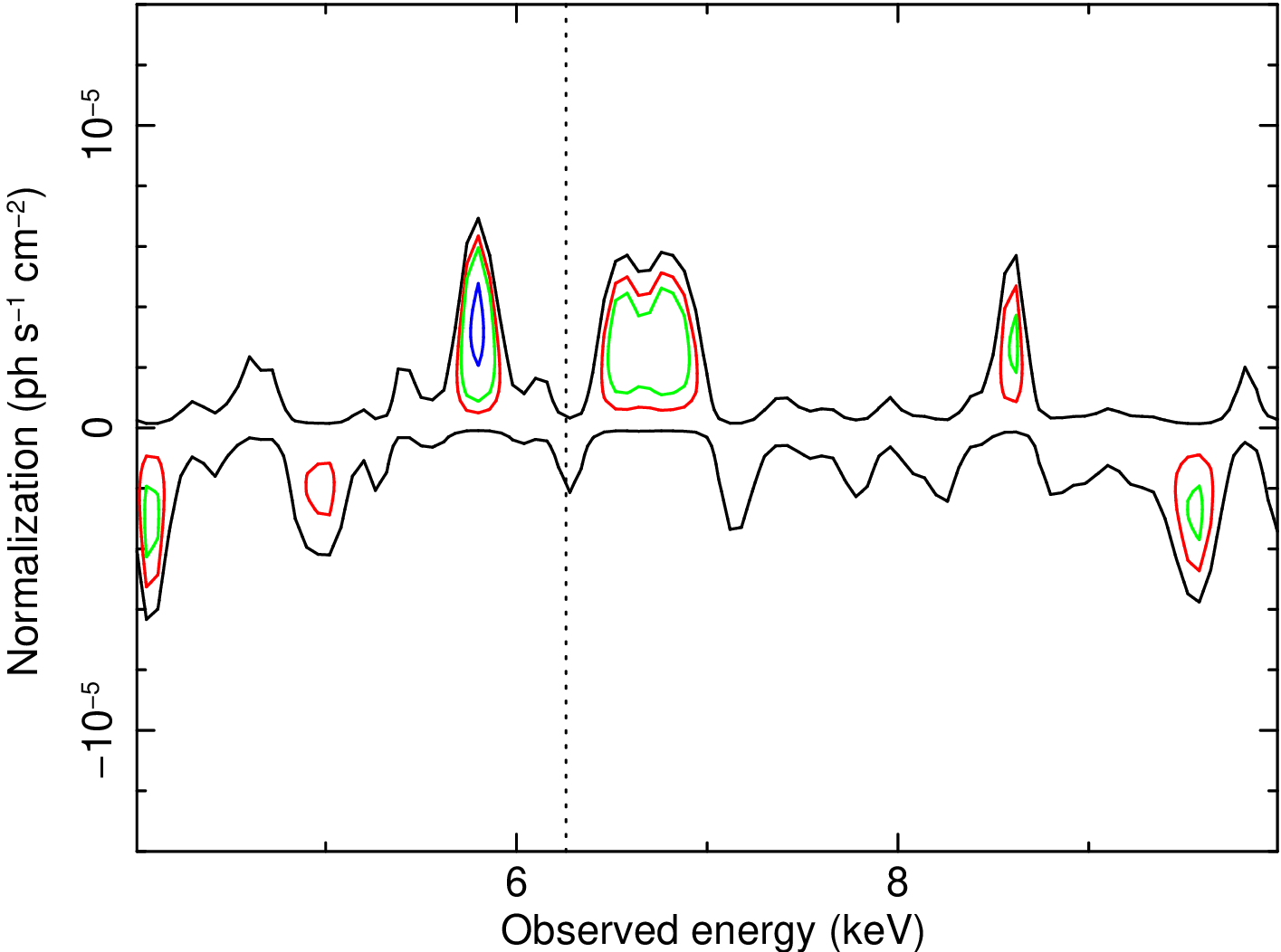}
\hspace{0.3cm}
    \includegraphics[width=3cm,height=4cm,angle=270]{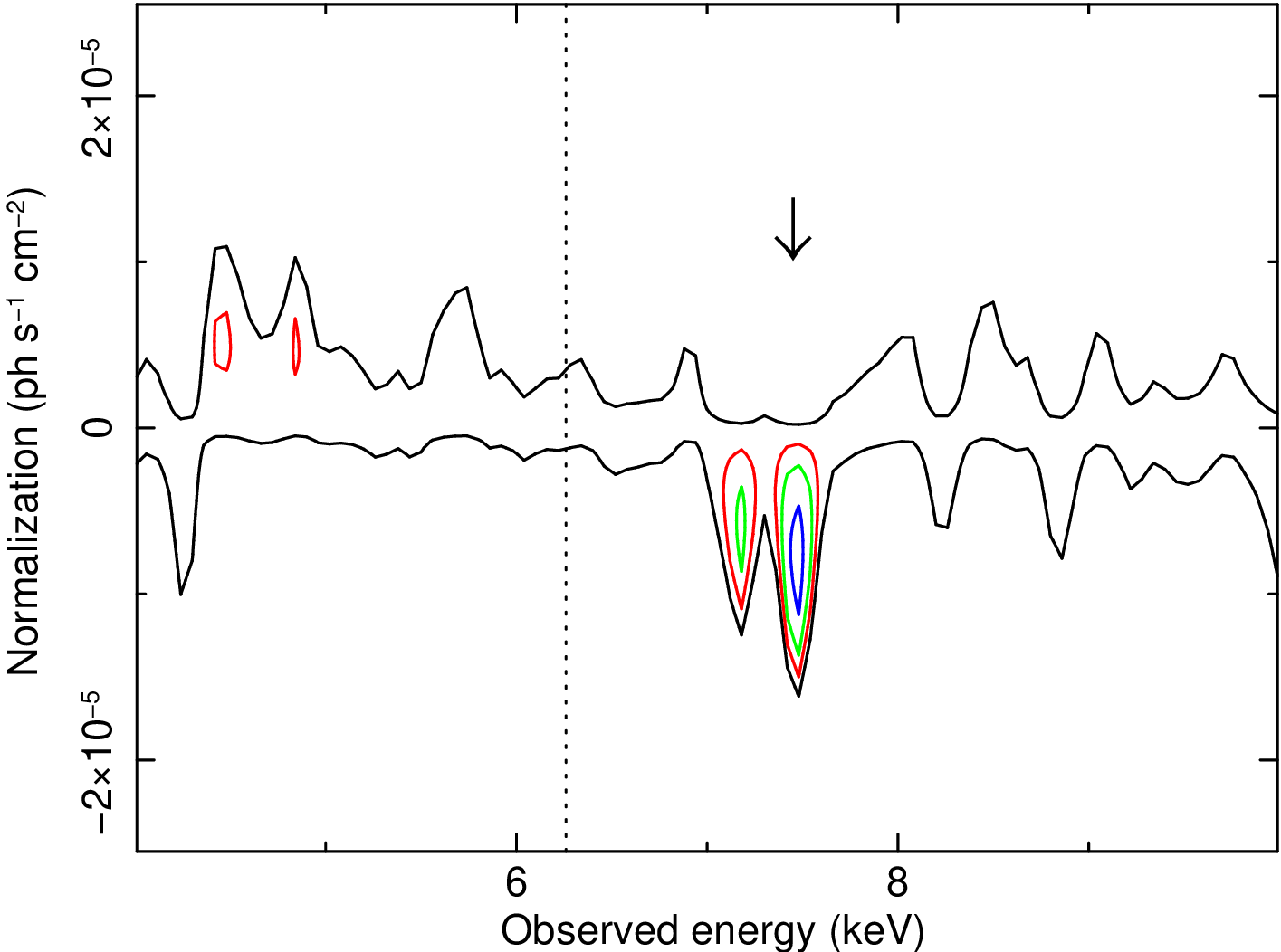}
\hspace{0.3cm}
    \includegraphics[width=3cm,height=4cm,angle=270]{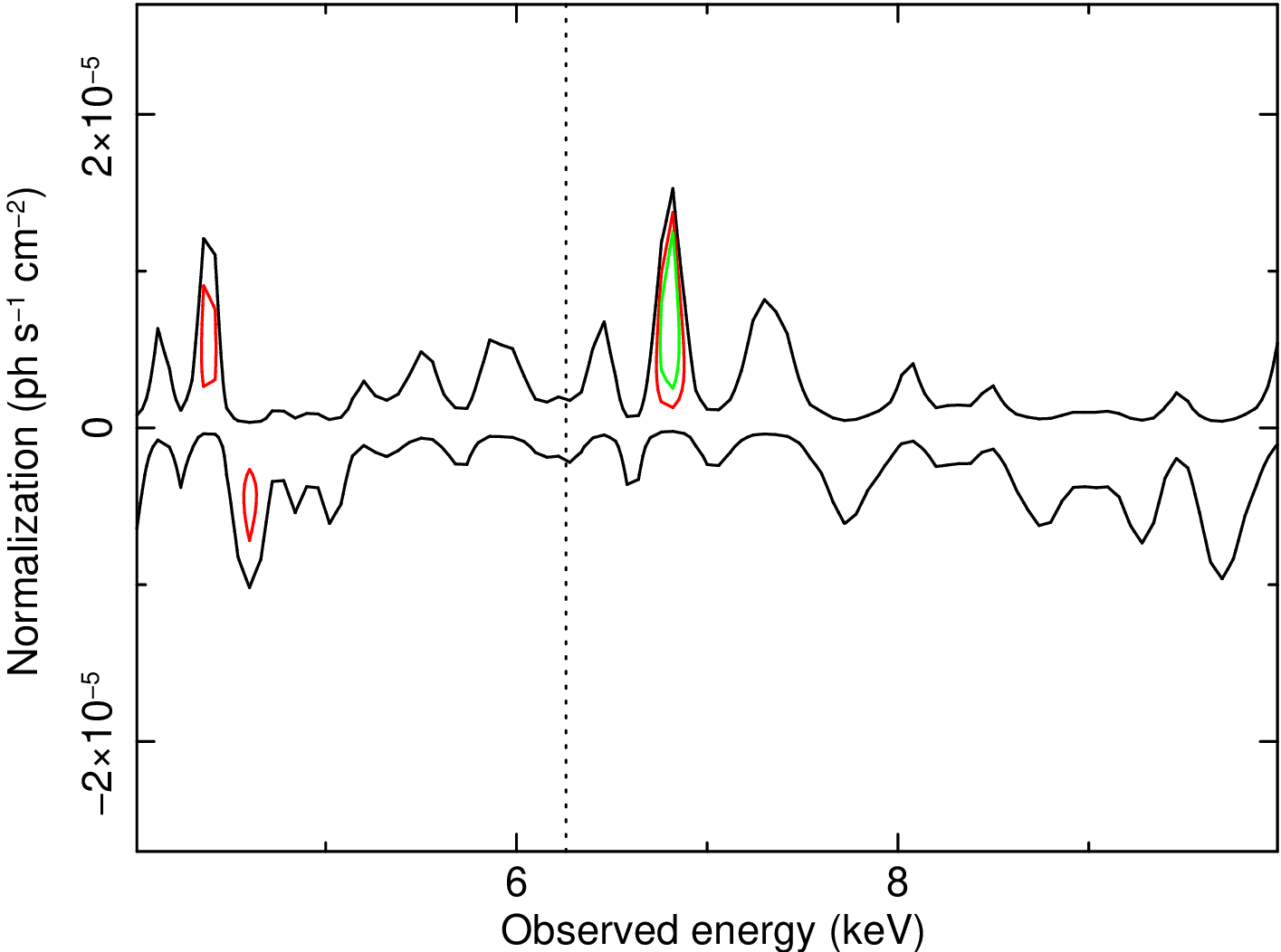}
\hspace{0.3cm}
    \includegraphics[width=3cm,height=4cm,angle=270]{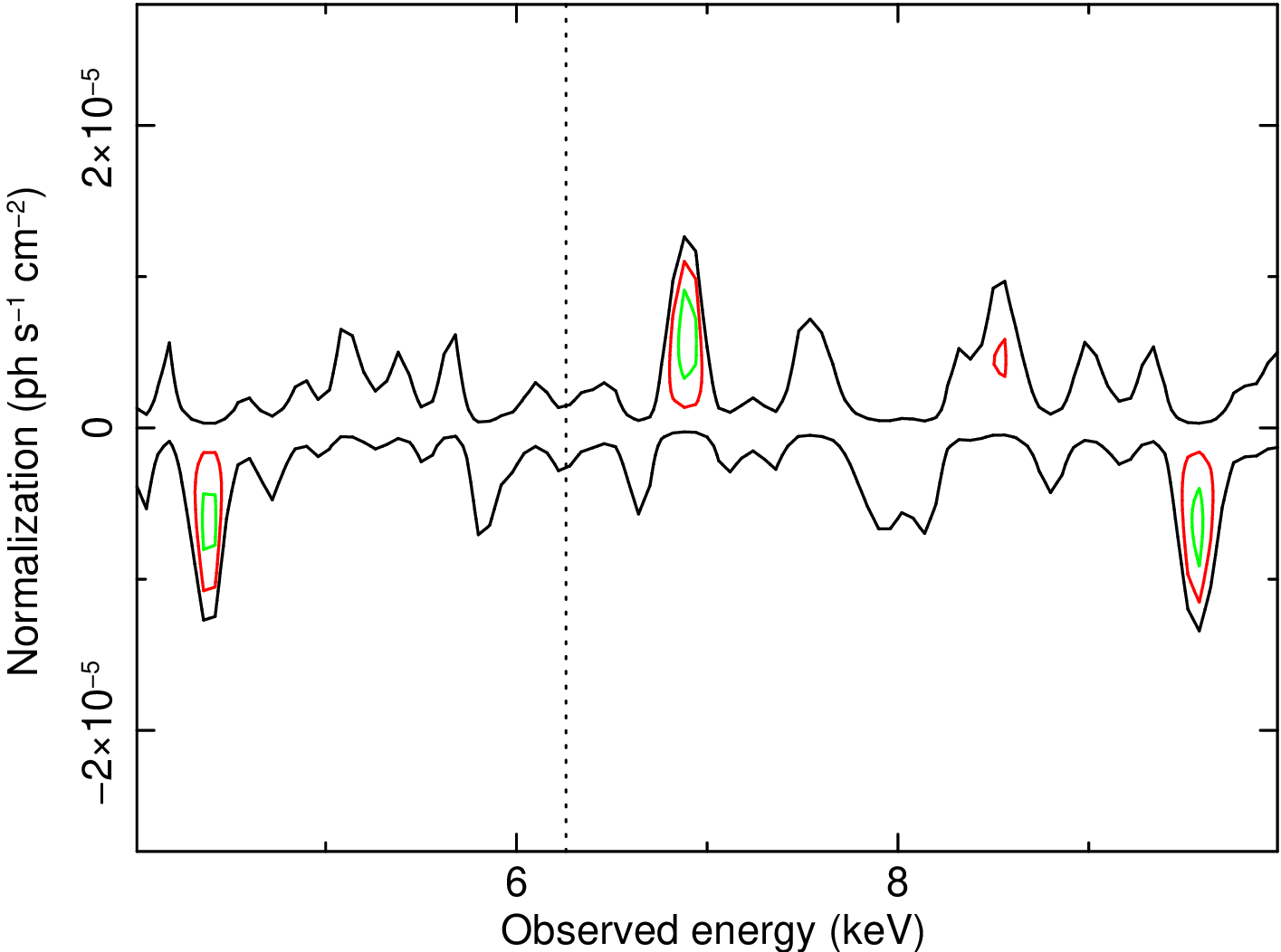}


\vspace{0.2cm}

    \includegraphics[width=3cm,height=4cm,angle=270]{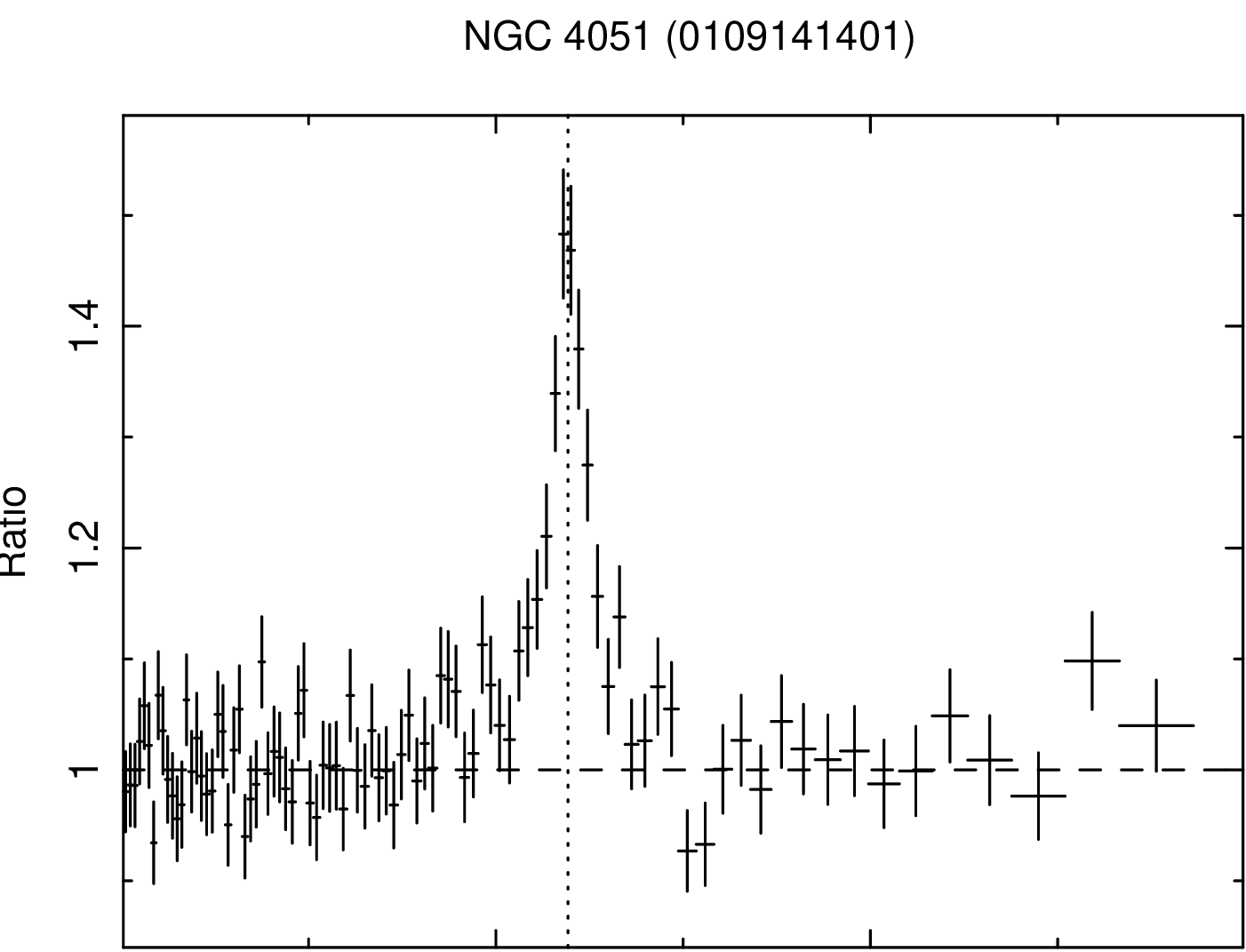}
\hspace{0.3cm}
    \includegraphics[width=3cm,height=4cm,angle=270]{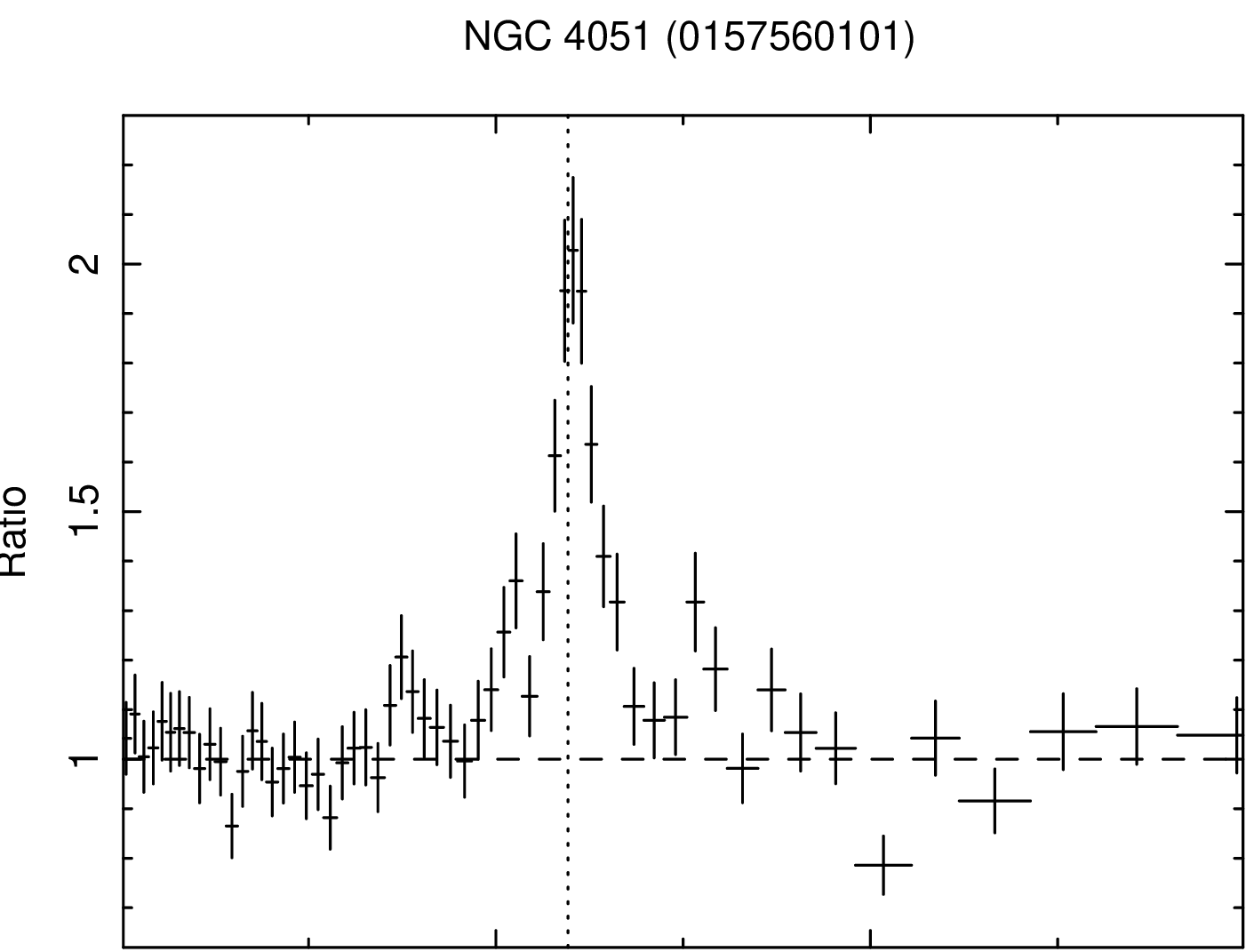}
\hspace{0.3cm}
    \includegraphics[width=3cm,height=4cm,angle=270]{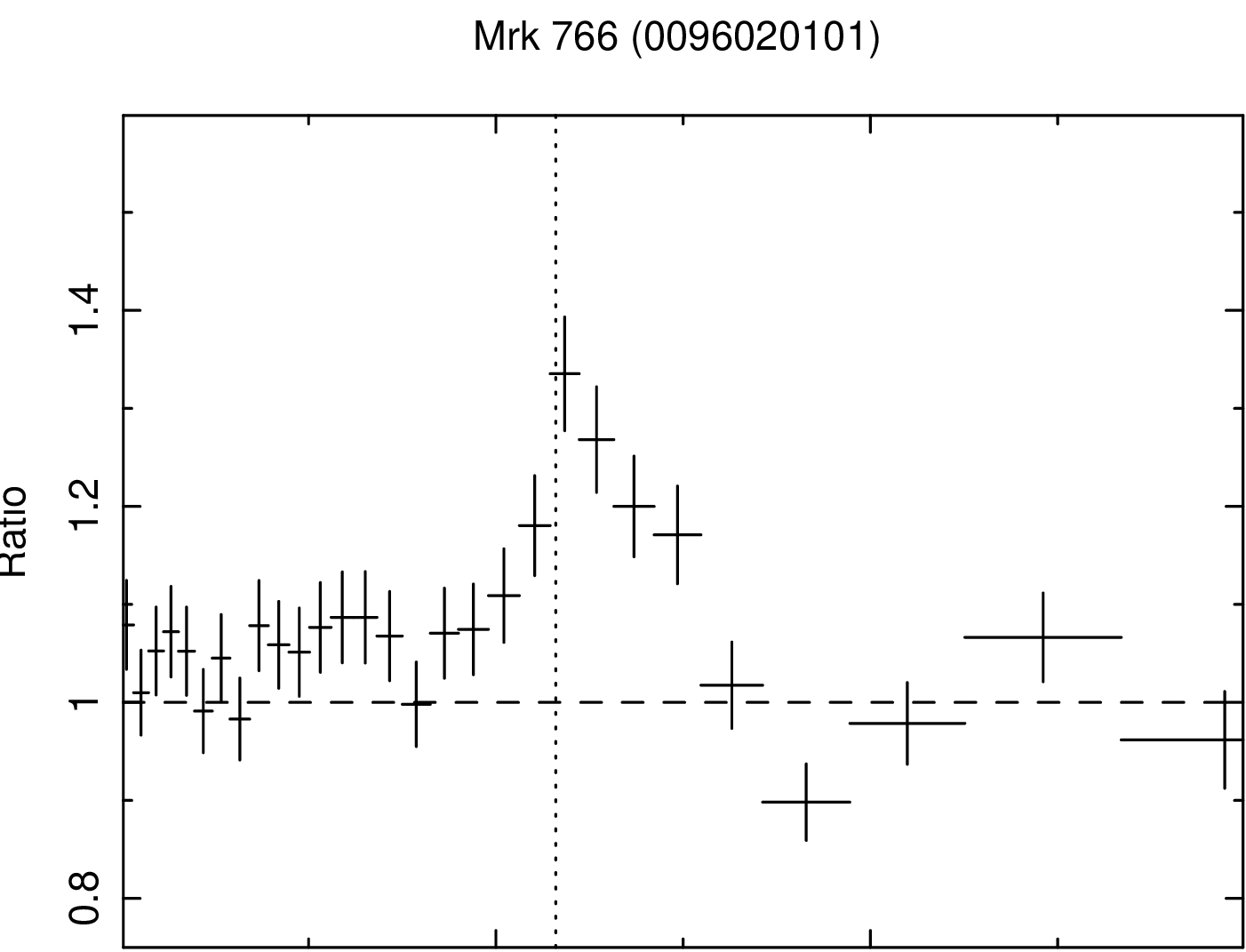}
\hspace{0.3cm}
    \includegraphics[width=3cm,height=4cm,angle=270]{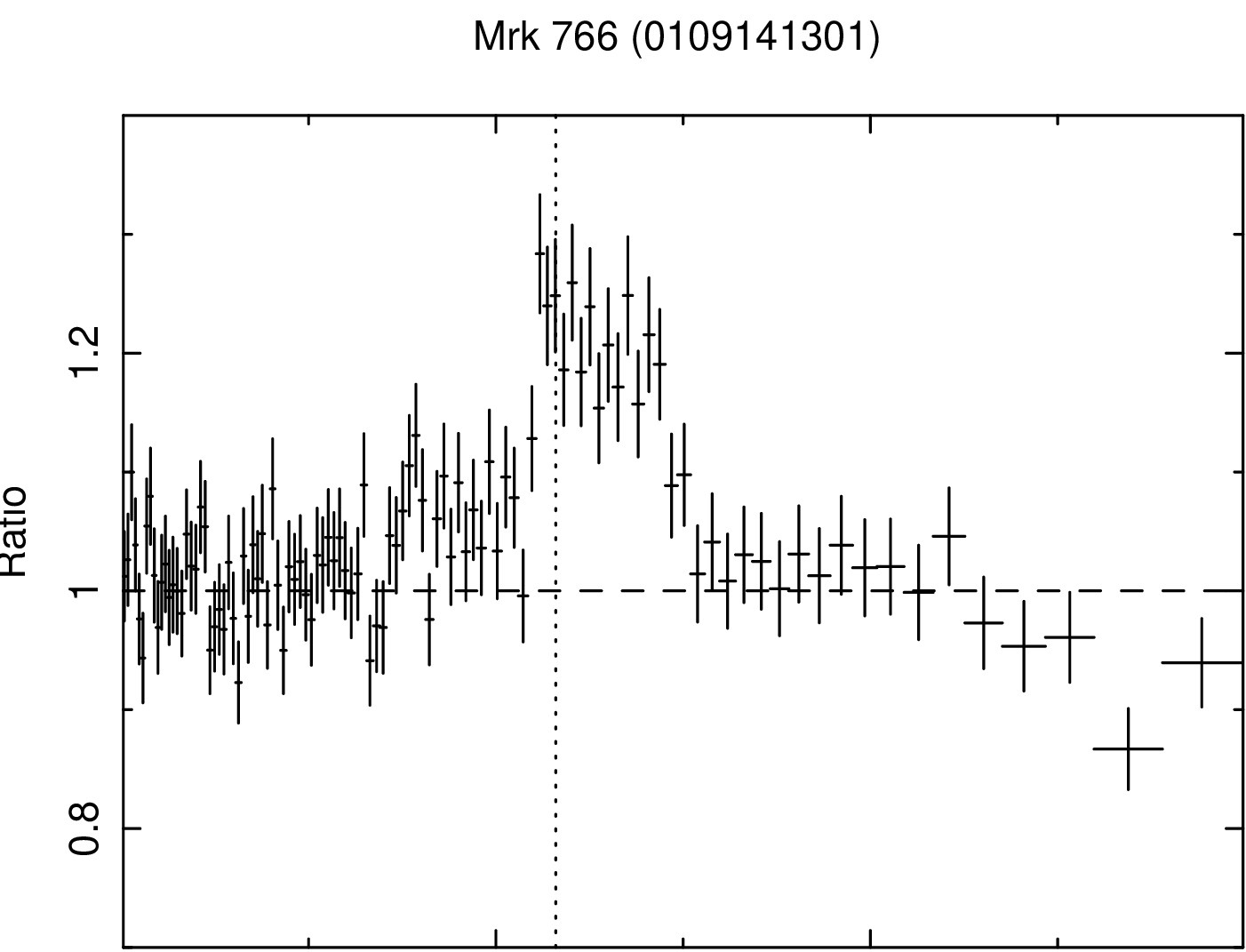}

\vspace{-0.07cm}

\hspace{-0.12cm}
    \includegraphics[width=3cm,height=4cm,angle=270]{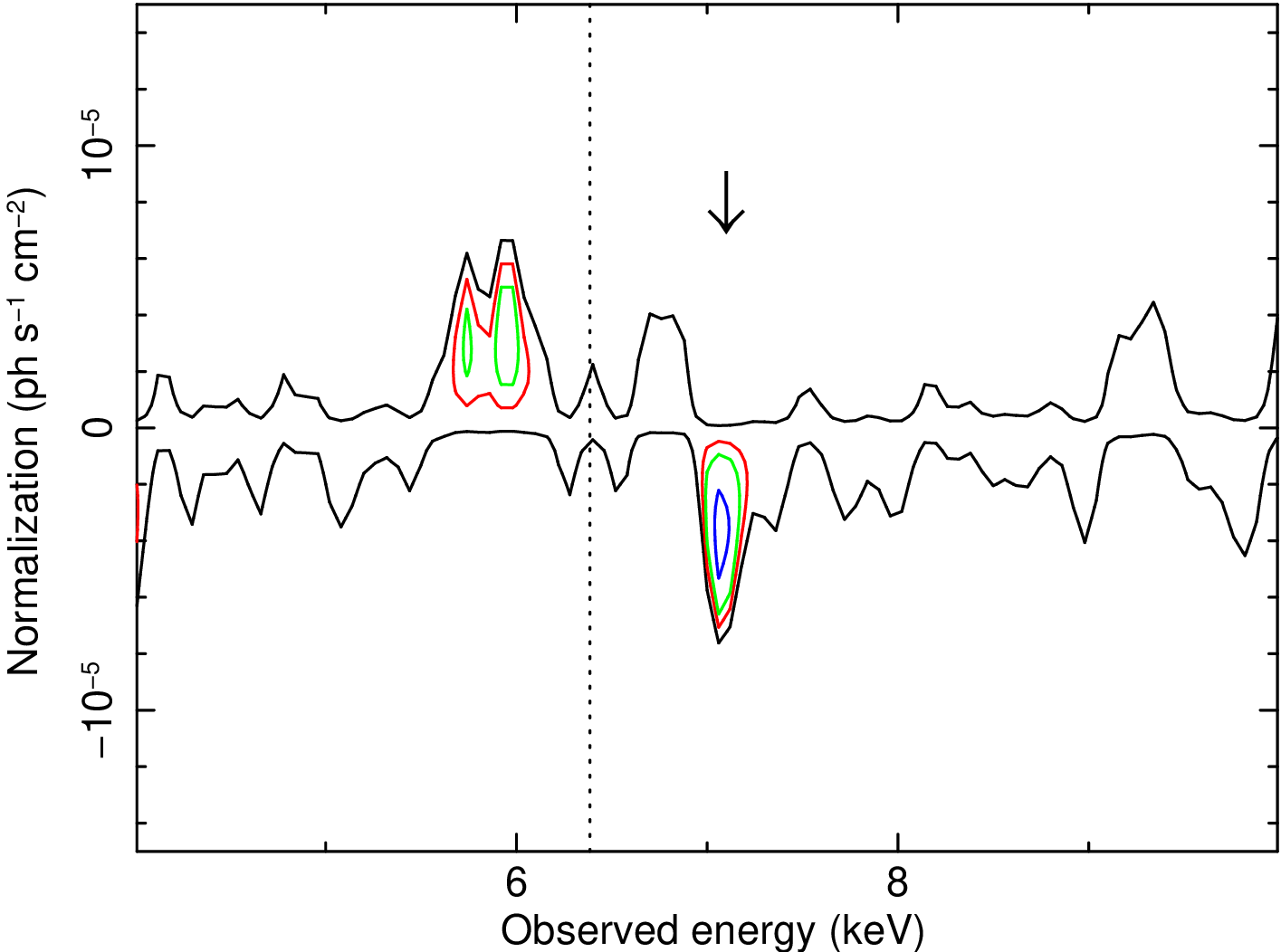}
\hspace{0.3cm}
    \includegraphics[width=3cm,height=4cm,angle=270]{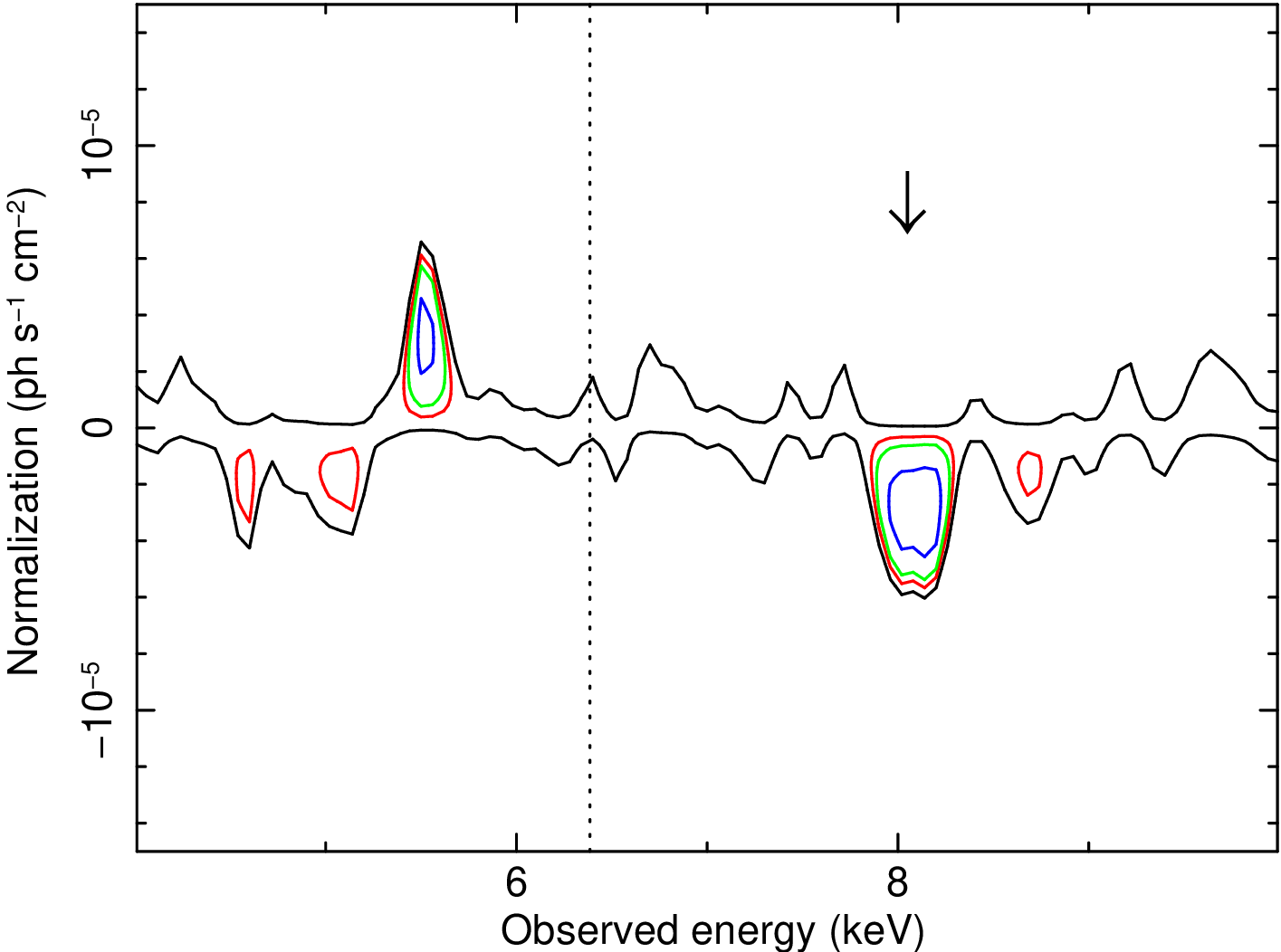}
\hspace{0.3cm}
    \includegraphics[width=3cm,height=4cm,angle=270]{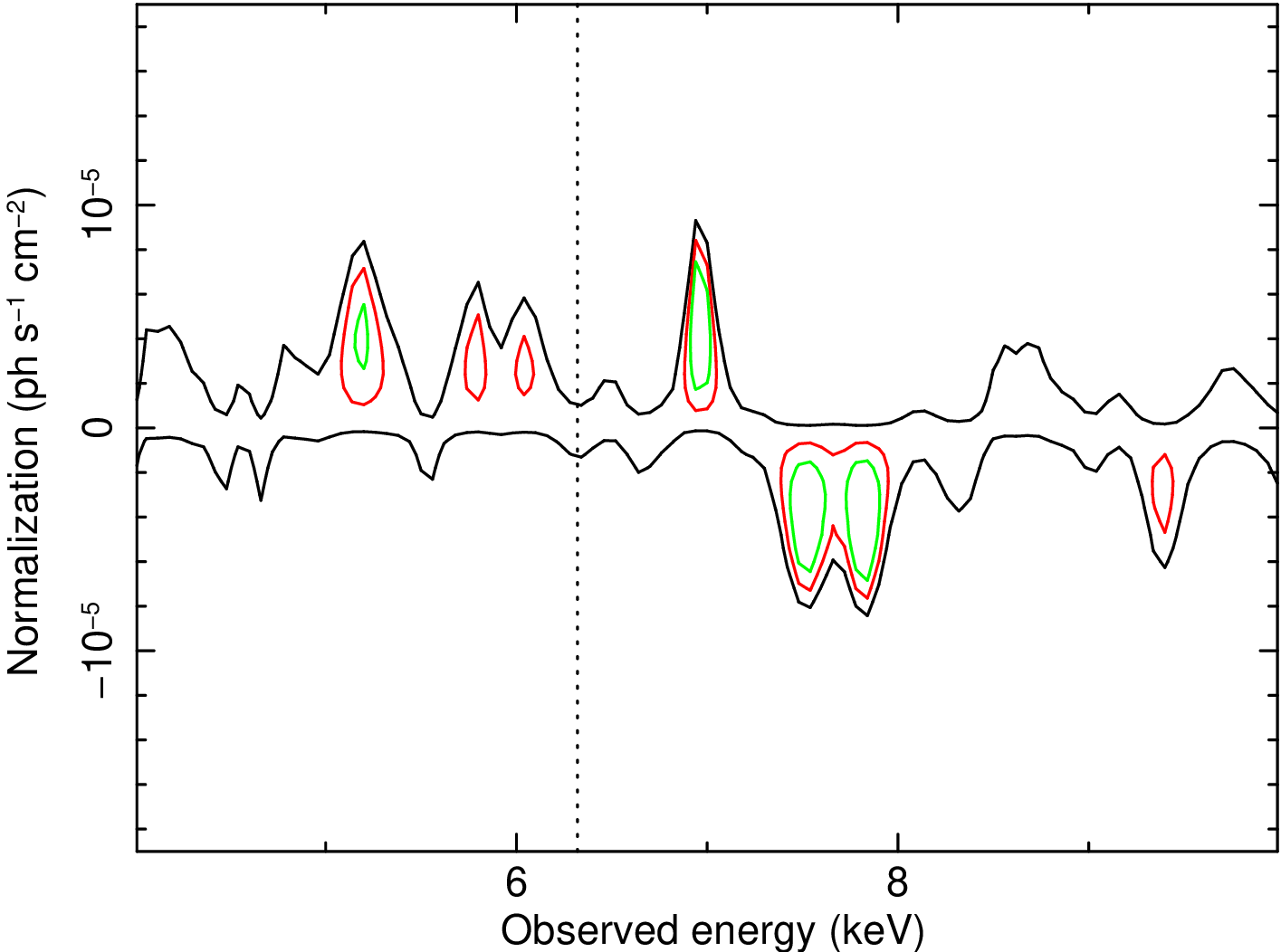}
\hspace{0.3cm}
    \includegraphics[width=3cm,height=4cm,angle=270]{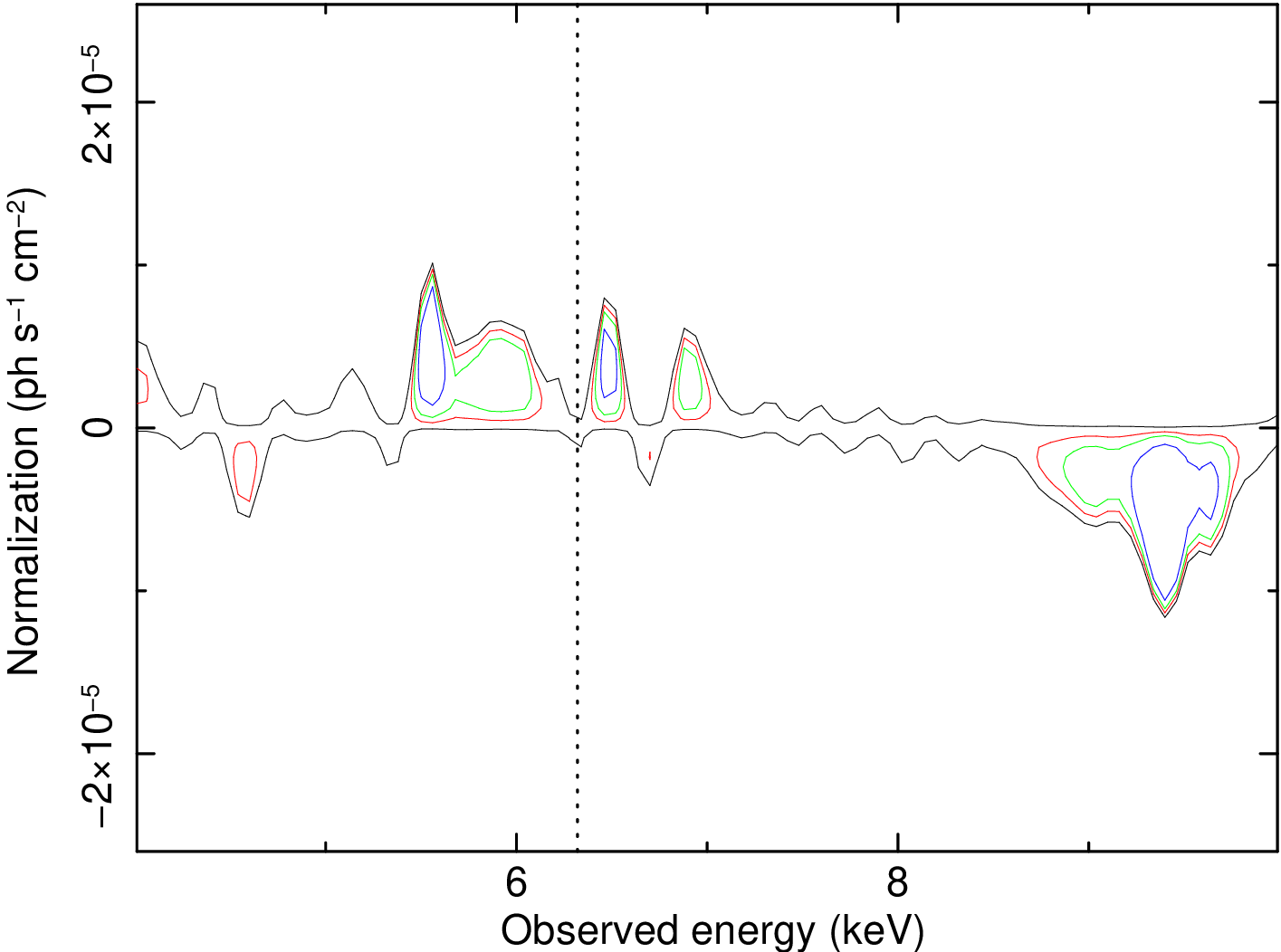}


\vspace{0.2cm}

    \includegraphics[width=3cm,height=4cm,angle=270]{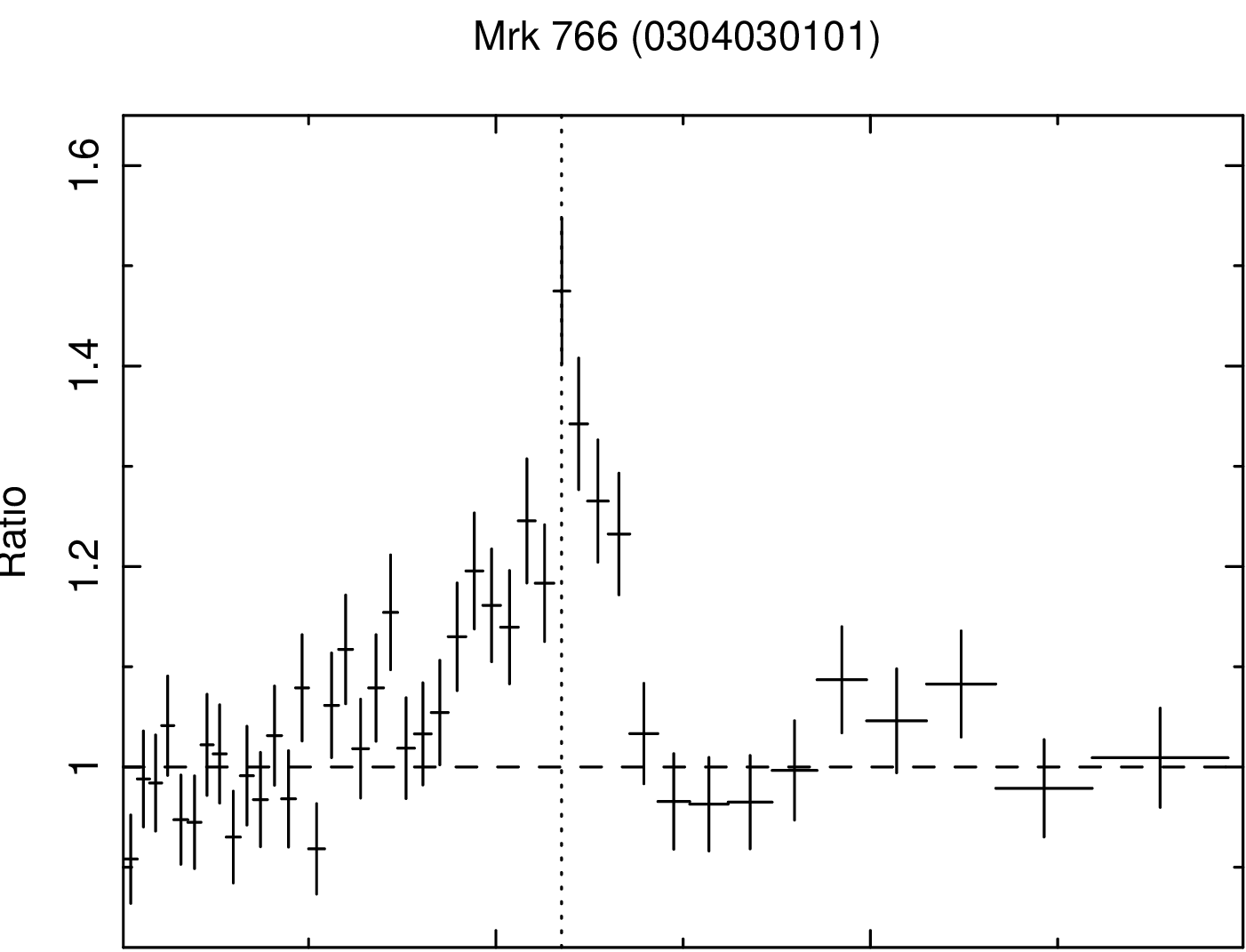}
\hspace{0.3cm}
    \includegraphics[width=3cm,height=4cm,angle=270]{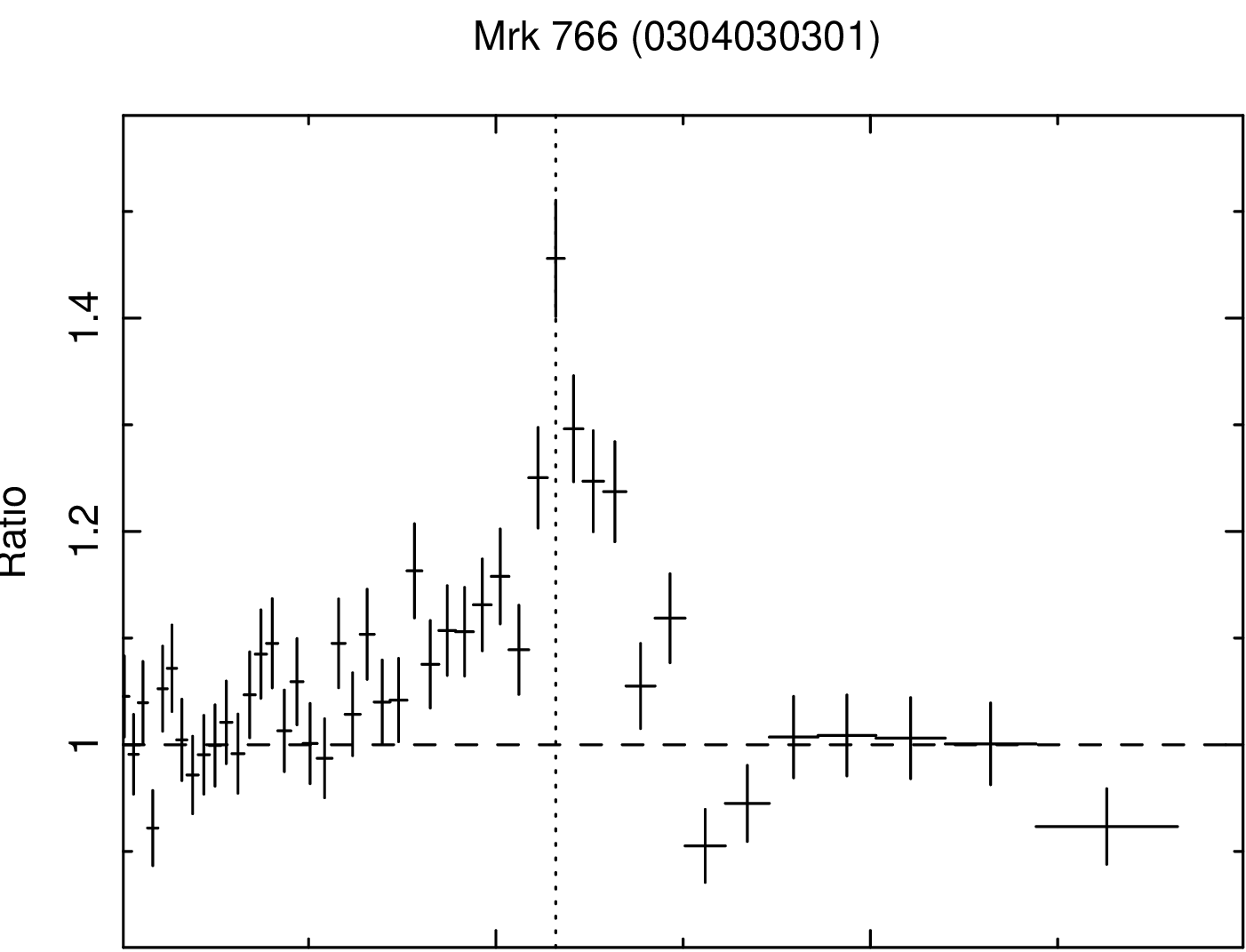}
\hspace{0.3cm}
    \includegraphics[width=3cm,height=4cm,angle=270]{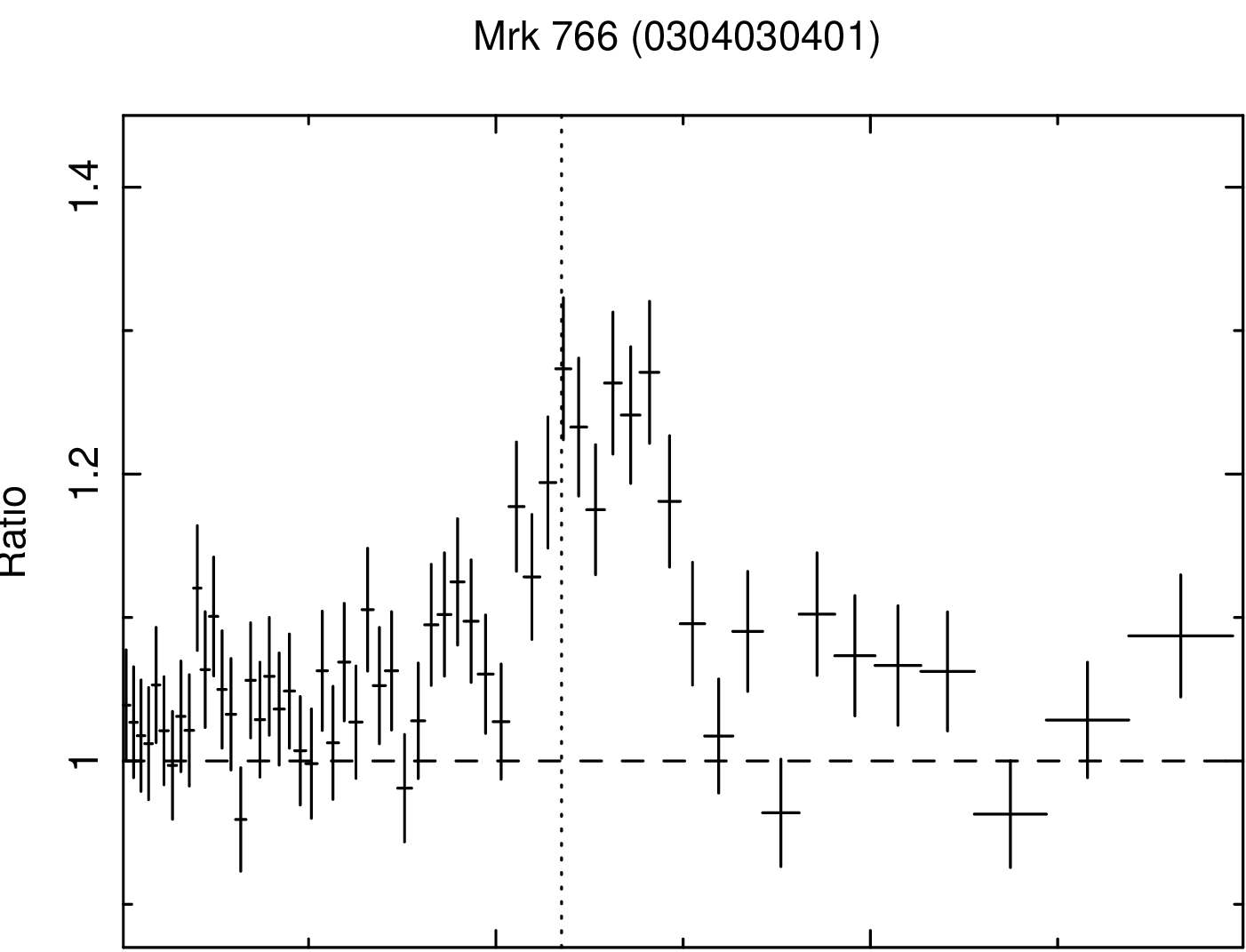}
\hspace{0.3cm}
    \includegraphics[width=3cm,height=4cm,angle=270]{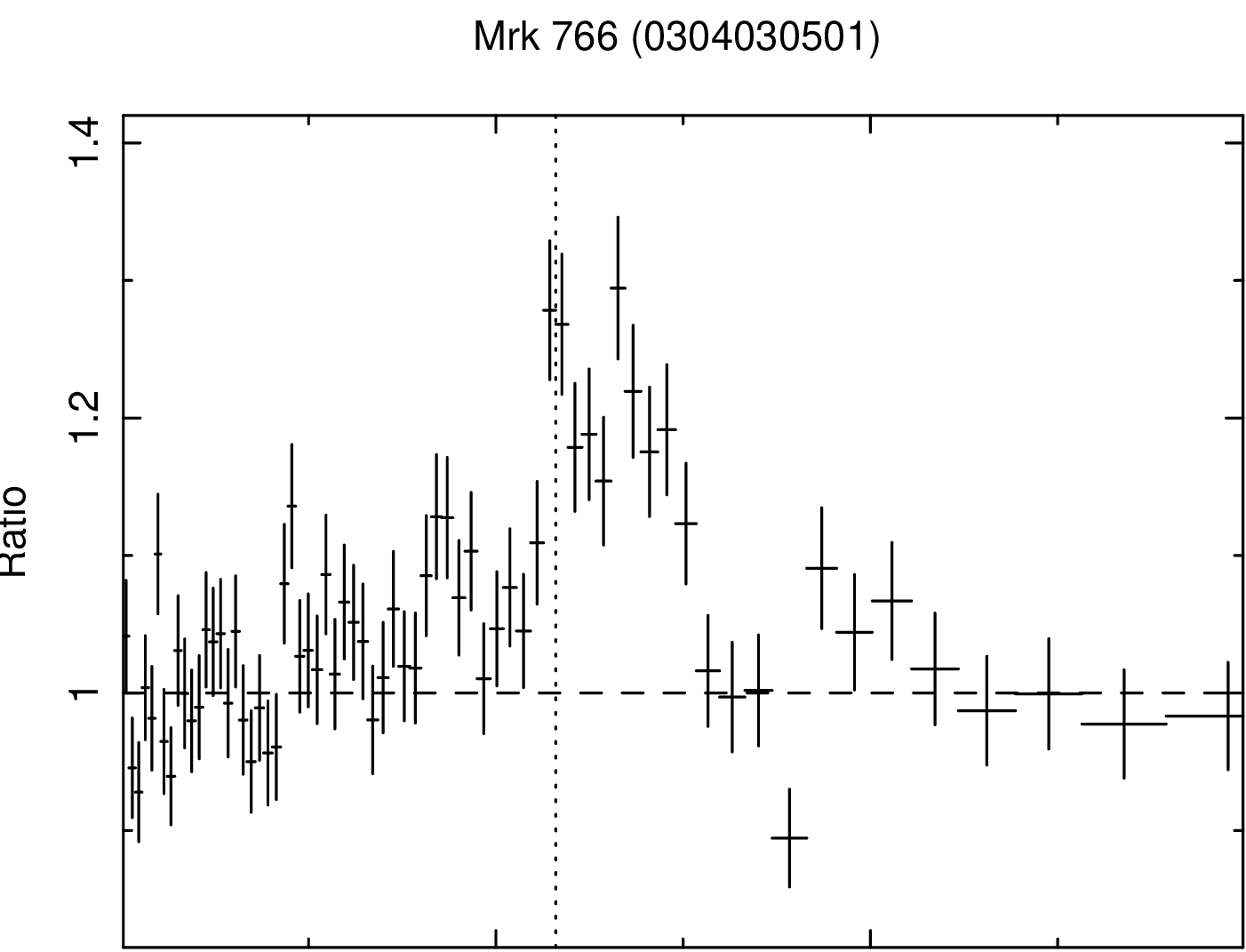}

\vspace{-0.07cm}

\hspace{-0.12cm}
    \includegraphics[width=3cm,height=4cm,angle=270]{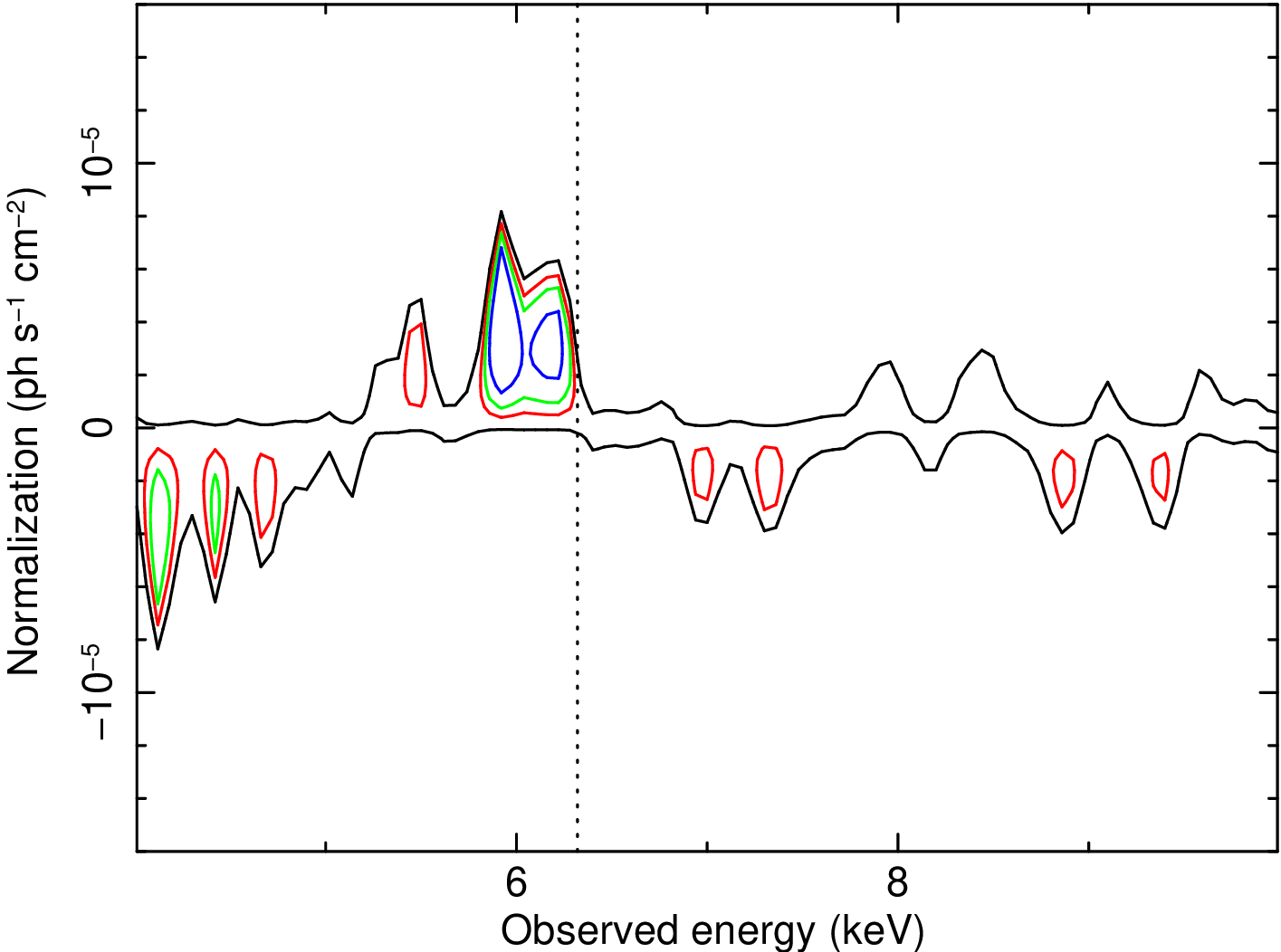}
\hspace{0.3cm}
    \includegraphics[width=3cm,height=4cm,angle=270]{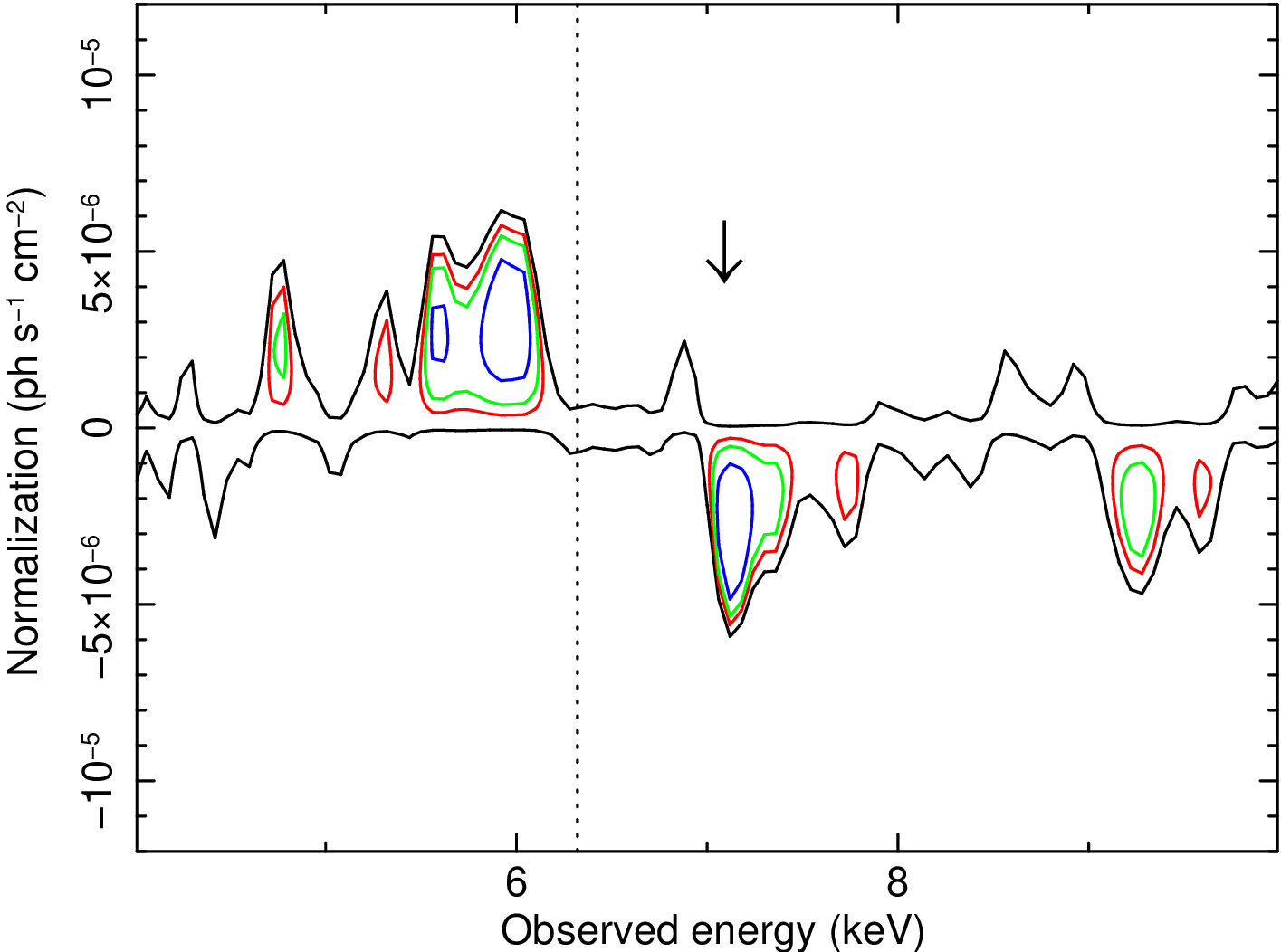}
\hspace{0.3cm}
    \includegraphics[width=3cm,height=4cm,angle=270]{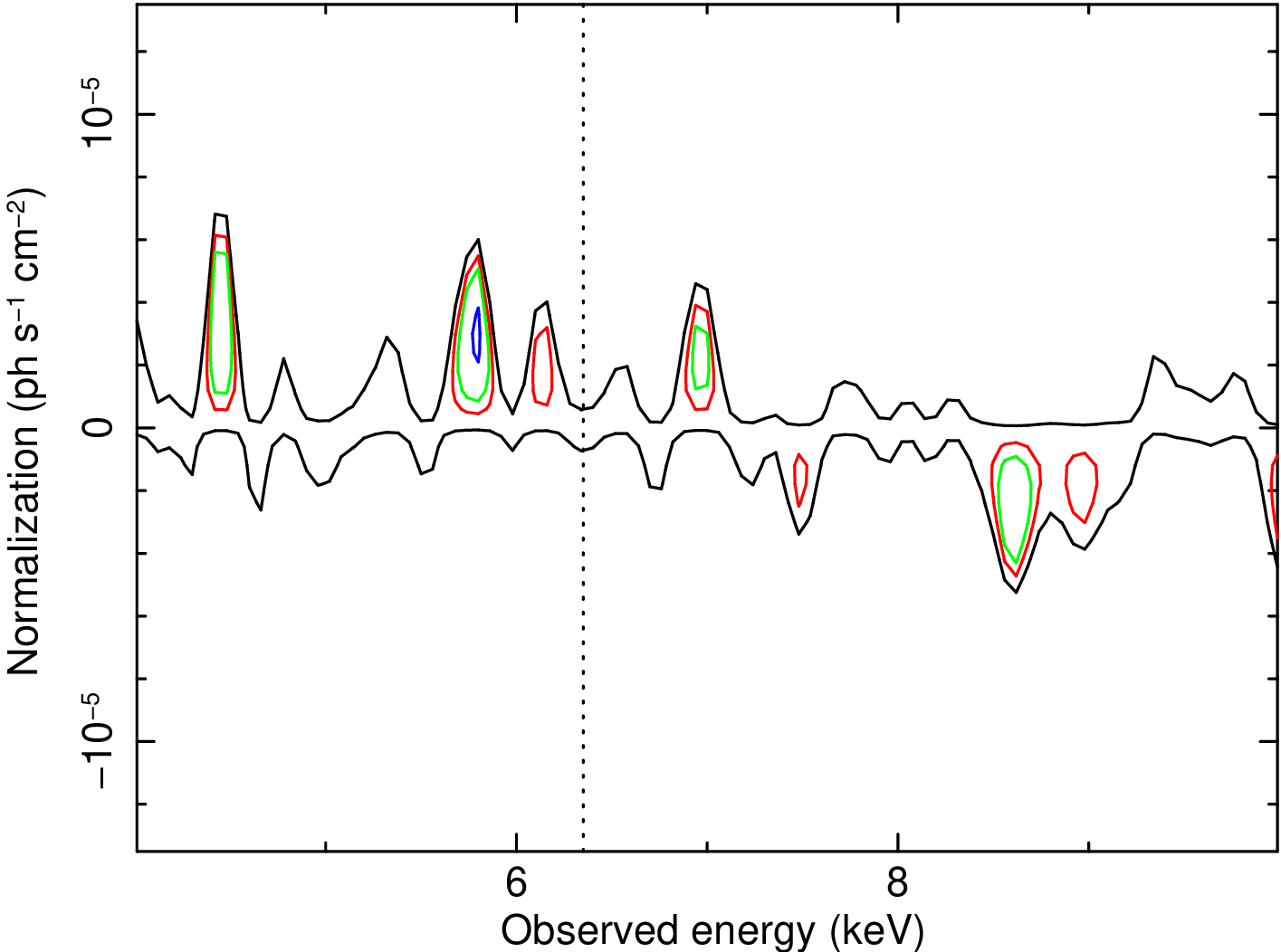}
\hspace{0.3cm}
    \includegraphics[width=3cm,height=4cm,angle=270]{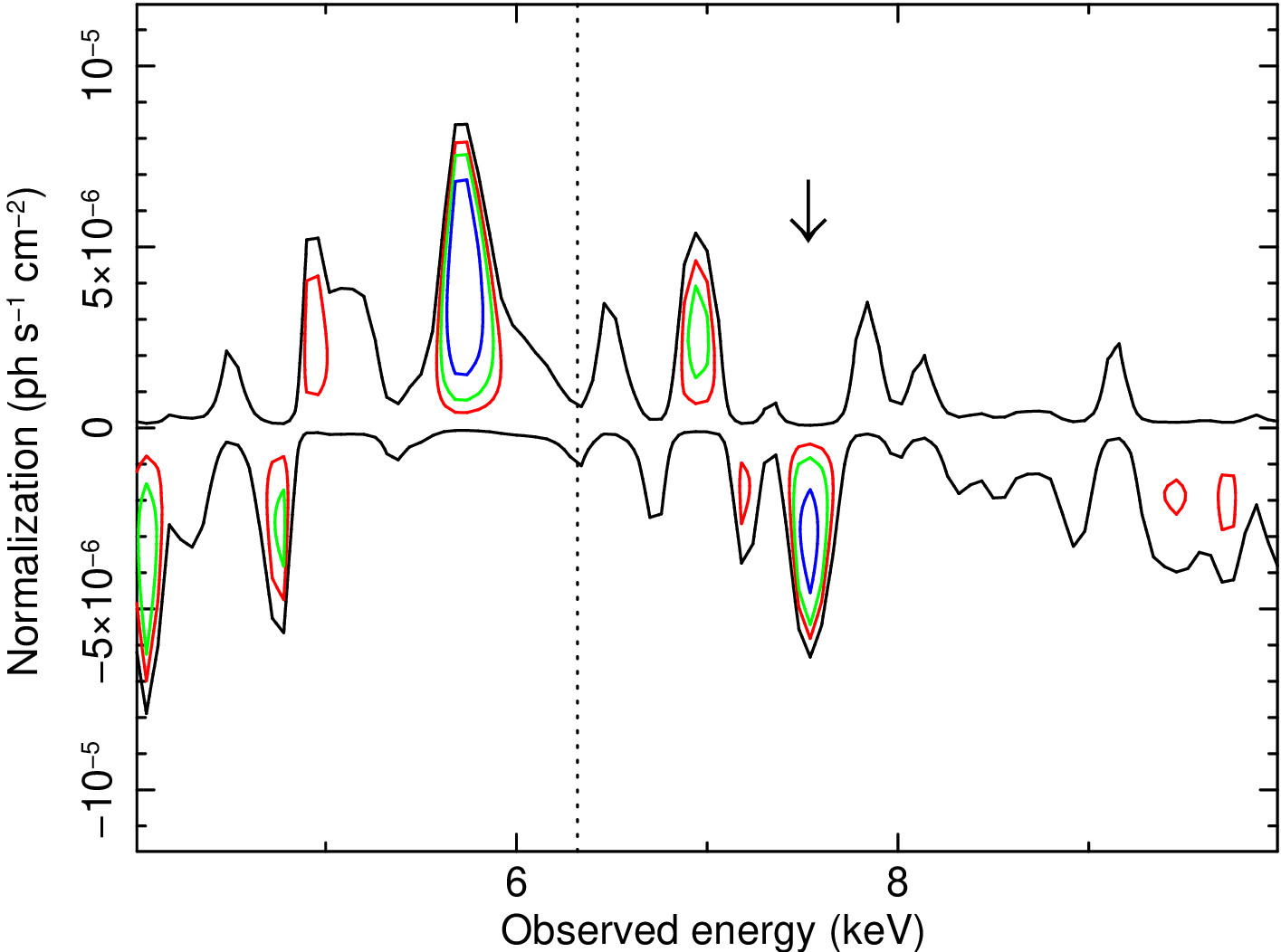}


\vspace{0.2cm}

    \includegraphics[width=3cm,height=4cm,angle=270]{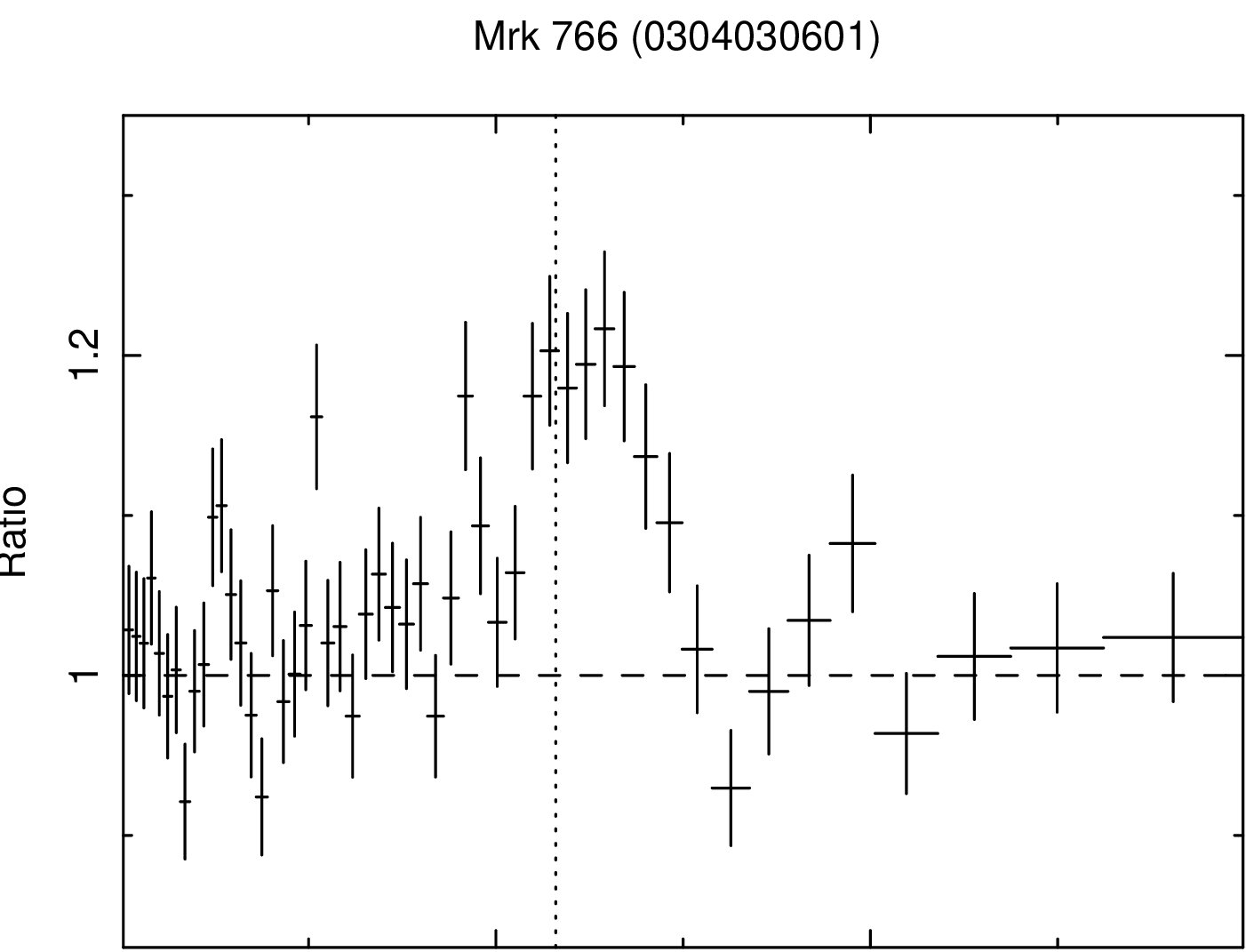}
\hspace{0.3cm}
    \includegraphics[width=3cm,height=4cm,angle=270]{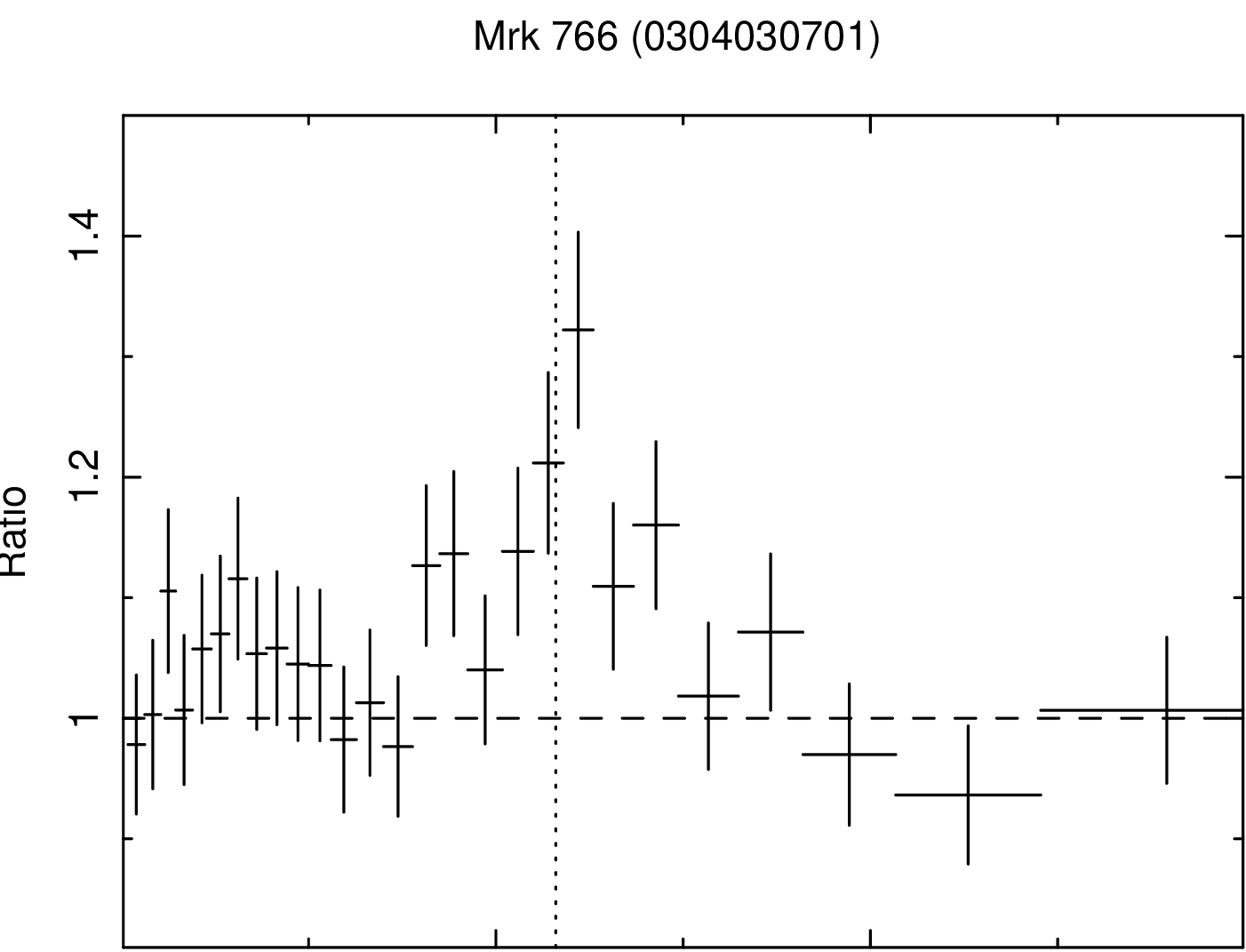}
\hspace{0.3cm}
    \includegraphics[width=3cm,height=4cm,angle=270]{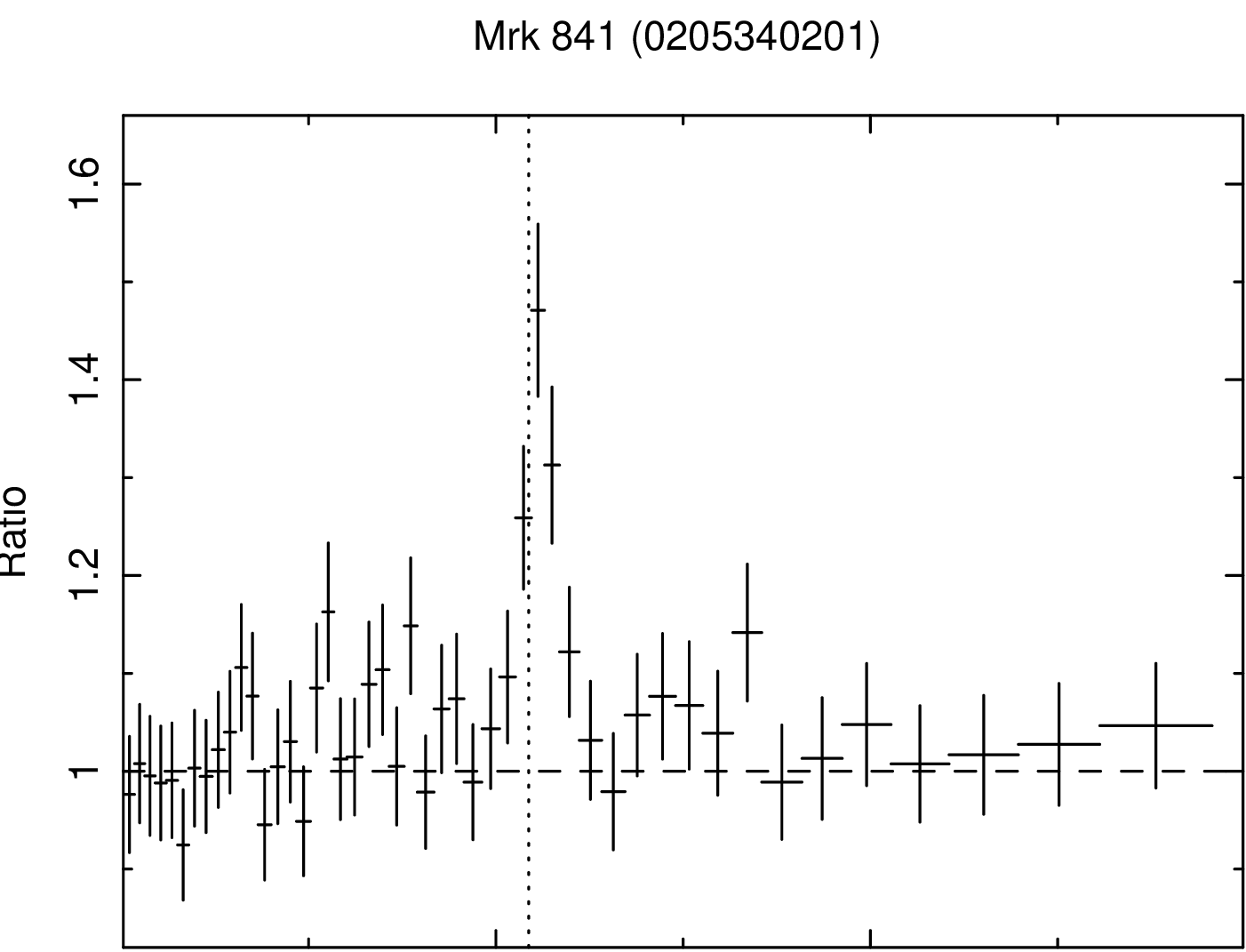}
\hspace{0.3cm}
    \includegraphics[width=3cm,height=4cm,angle=270]{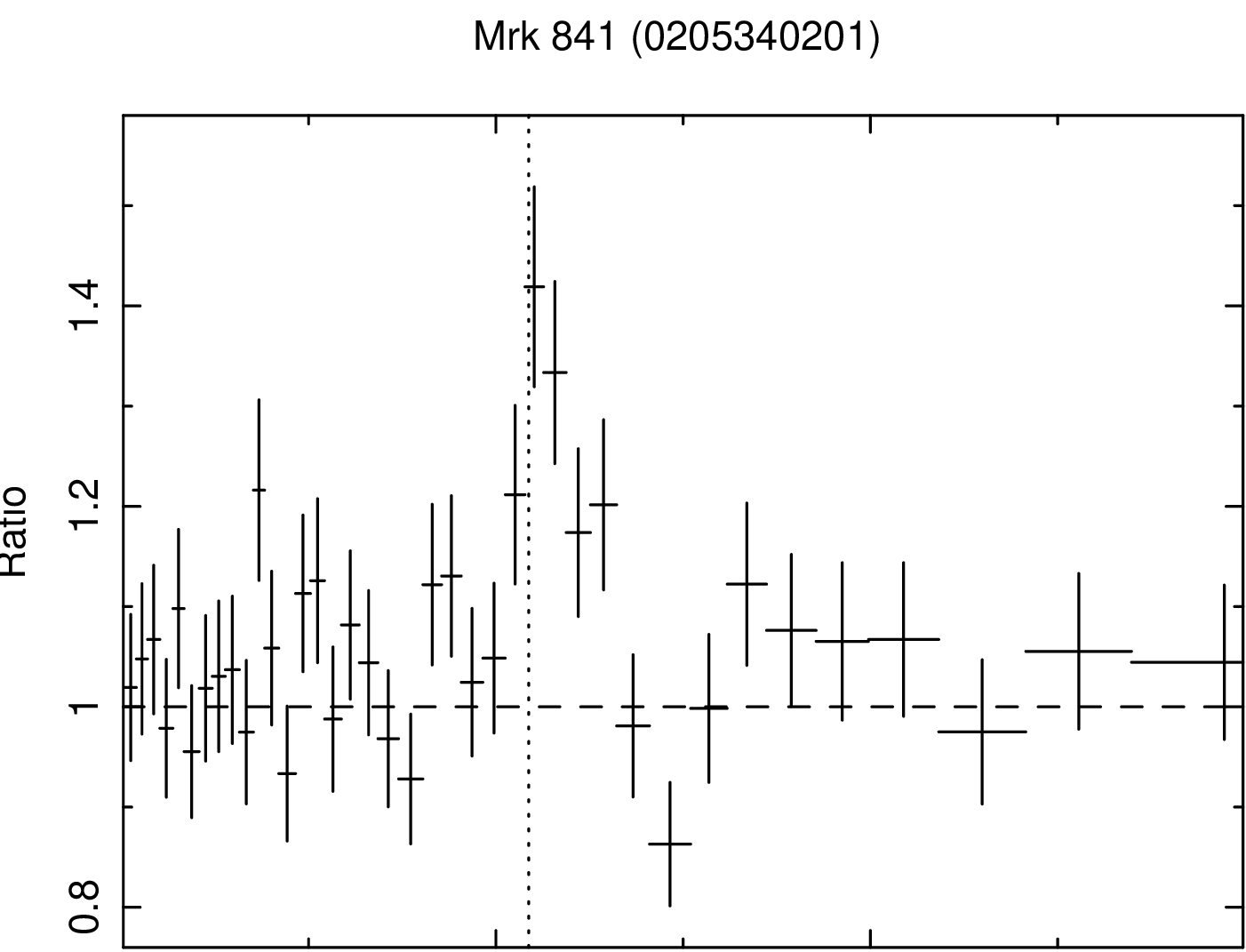}

\vspace{-0.07cm}

\hspace{-0.12cm}
    \includegraphics[width=3cm,height=4cm,angle=270]{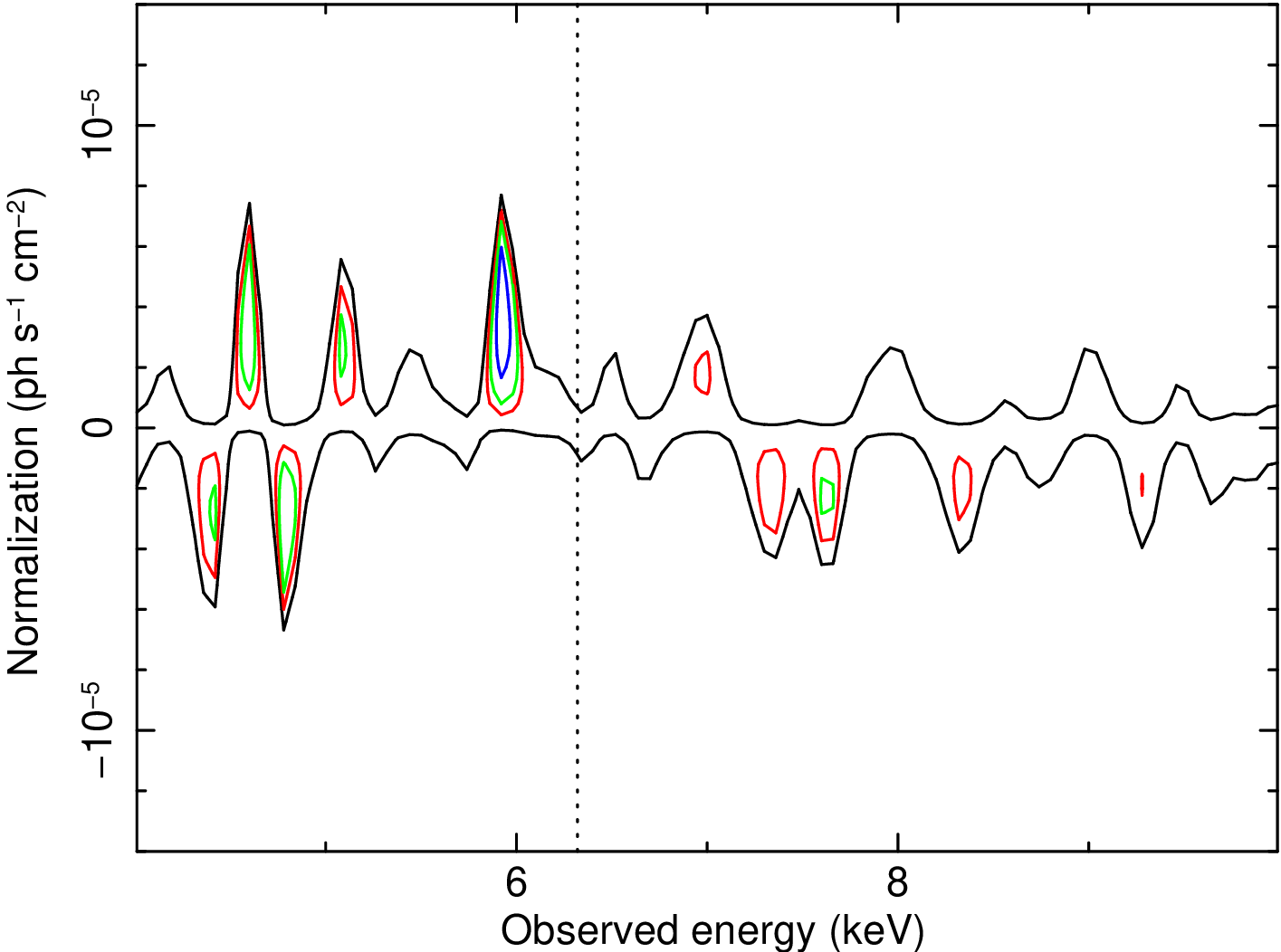}
\hspace{0.3cm}
    \includegraphics[width=3cm,height=4cm,angle=270]{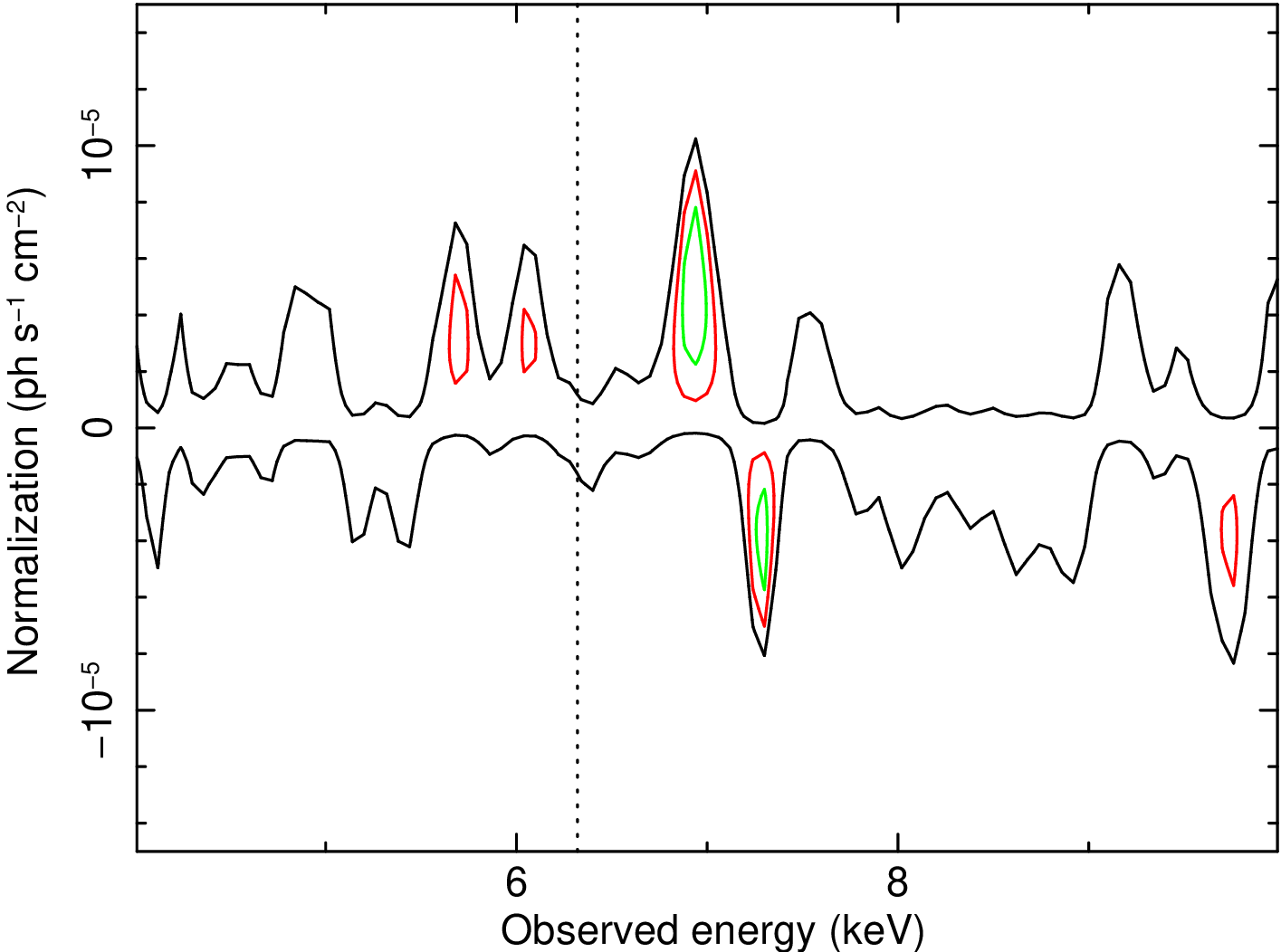}
\hspace{0.3cm}
    \includegraphics[width=3cm,height=4cm,angle=270]{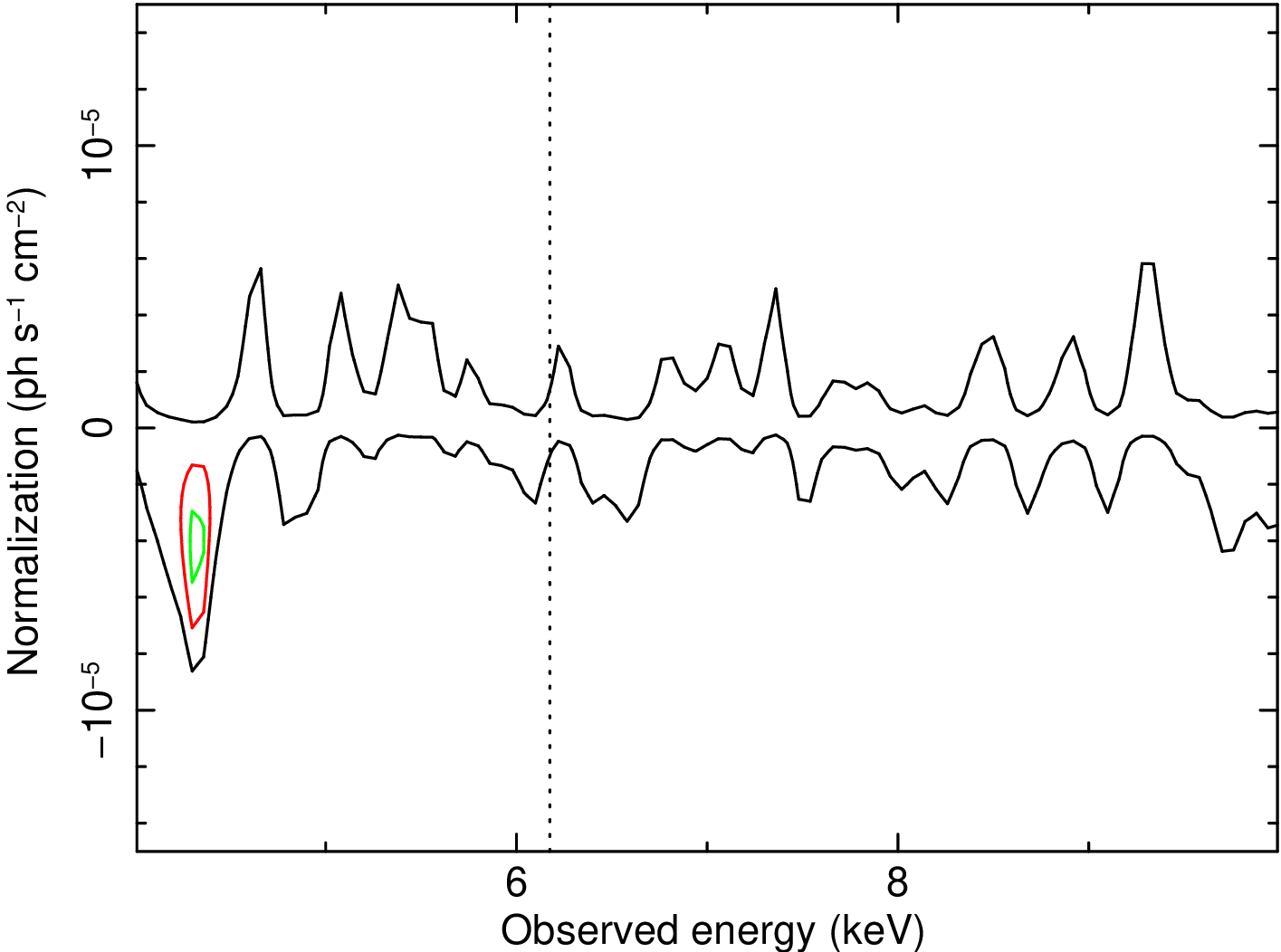}
\hspace{0.3cm}
    \includegraphics[width=3cm,height=4cm,angle=270]{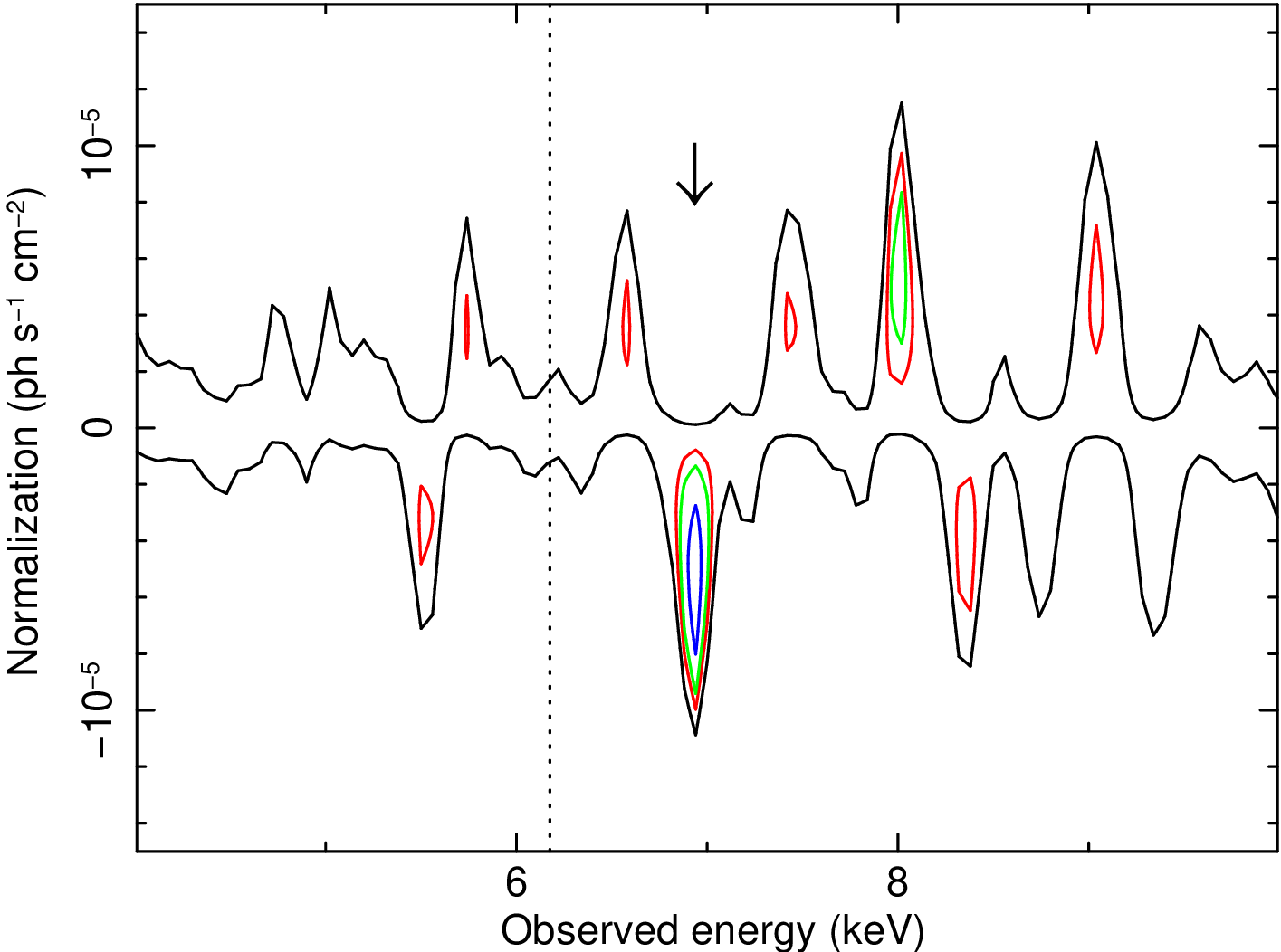}

   \caption{Ratio against the continuum (\emph{upper panel}) and contour plots with respect to the best-fit baseline model (\emph{lower panel}; 68\% (red), 90\% (green), 99\% (blue) levels) for the Type 1 sources. The Fe K absorption lines are pointed by arrows. The vertical line refer to Fe I K$\alpha$.}
    \end{figure*}

   \begin{figure*}[ht]
   \centering


    \includegraphics[width=3cm,height=4cm,angle=270]{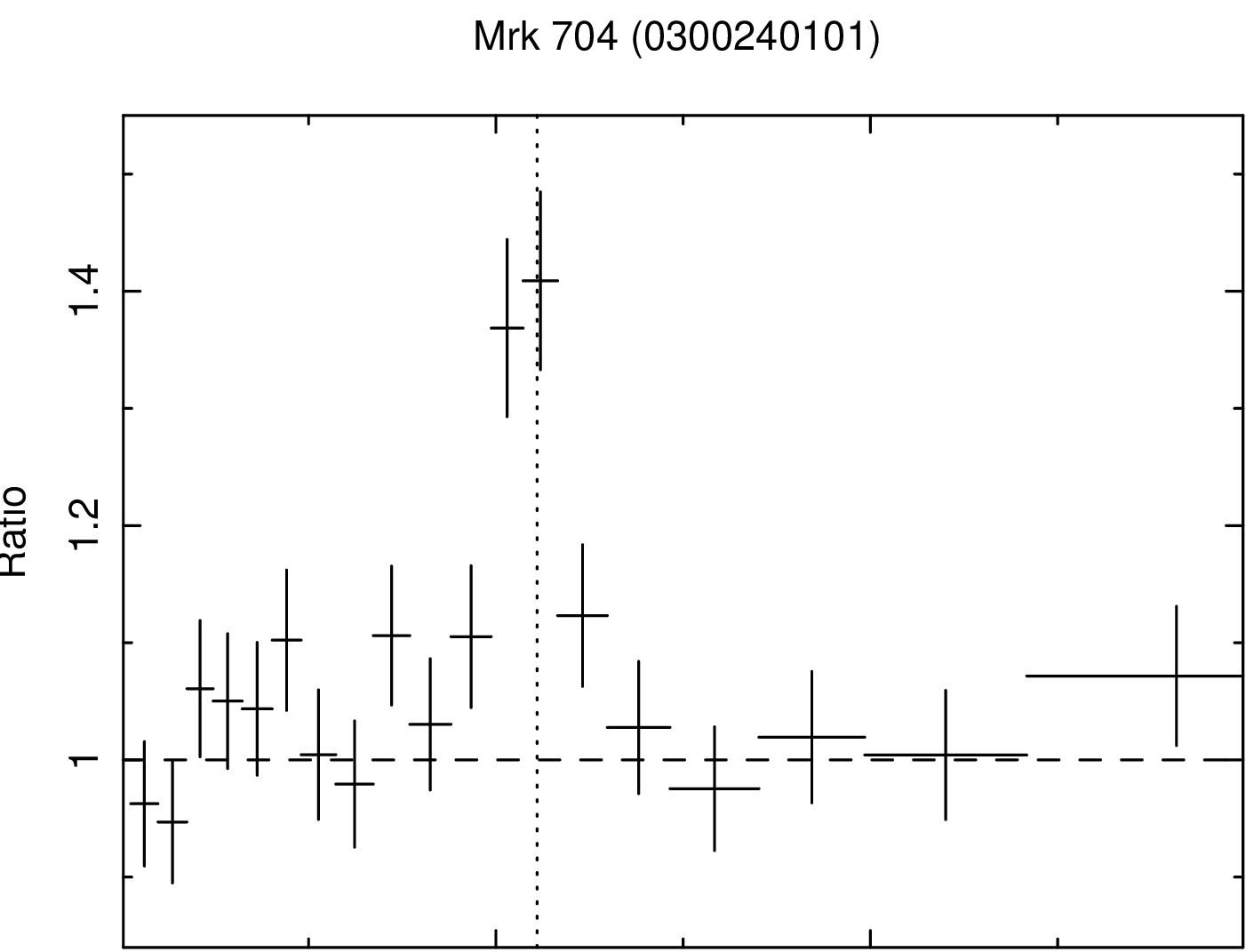}
\hspace{0.3cm}
    \includegraphics[width=3cm,height=4cm,angle=270]{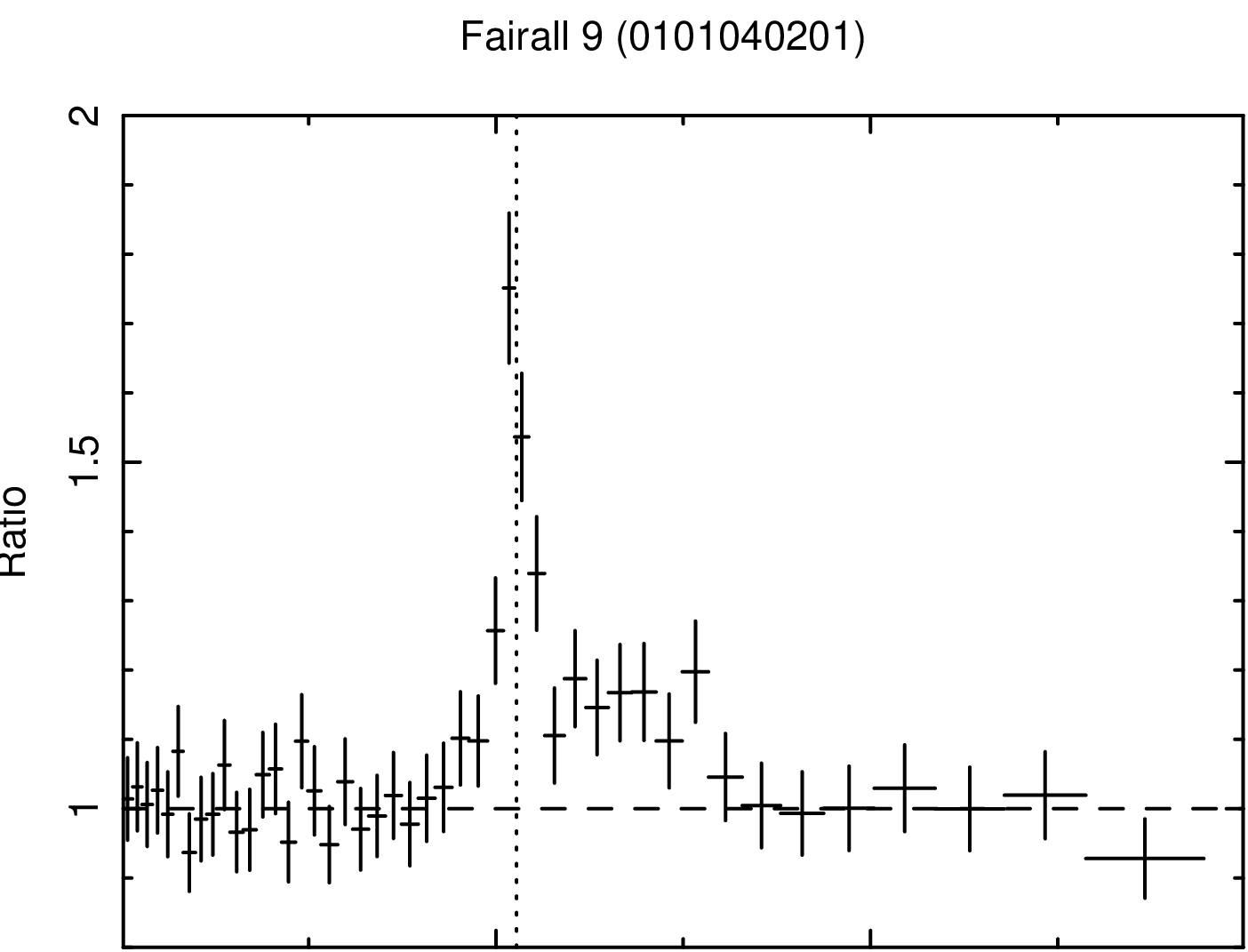}
\hspace{0.3cm}
    \includegraphics[width=3cm,height=4cm,angle=270]{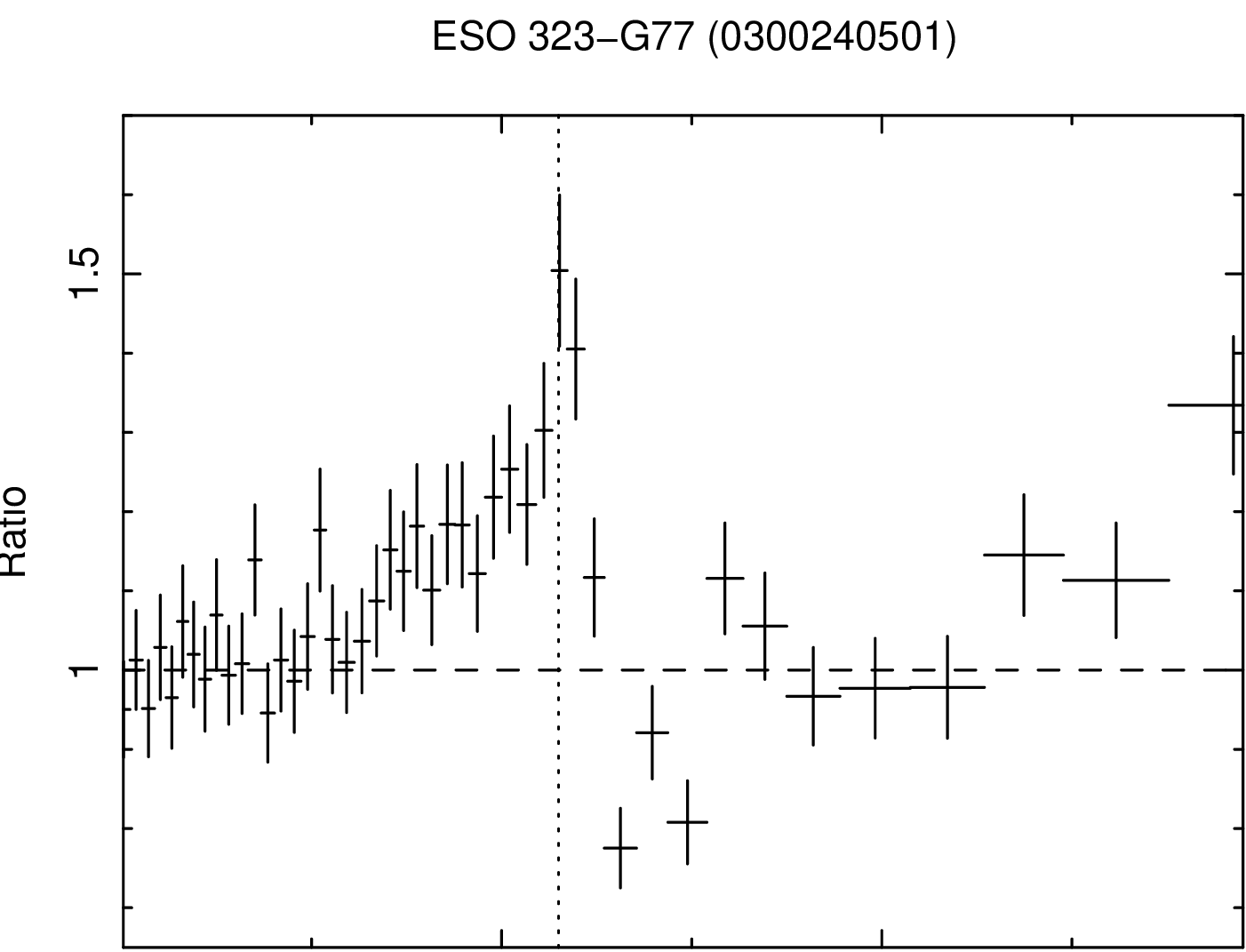}
\hspace{0.3cm}
    \includegraphics[width=3cm,height=4cm,angle=270]{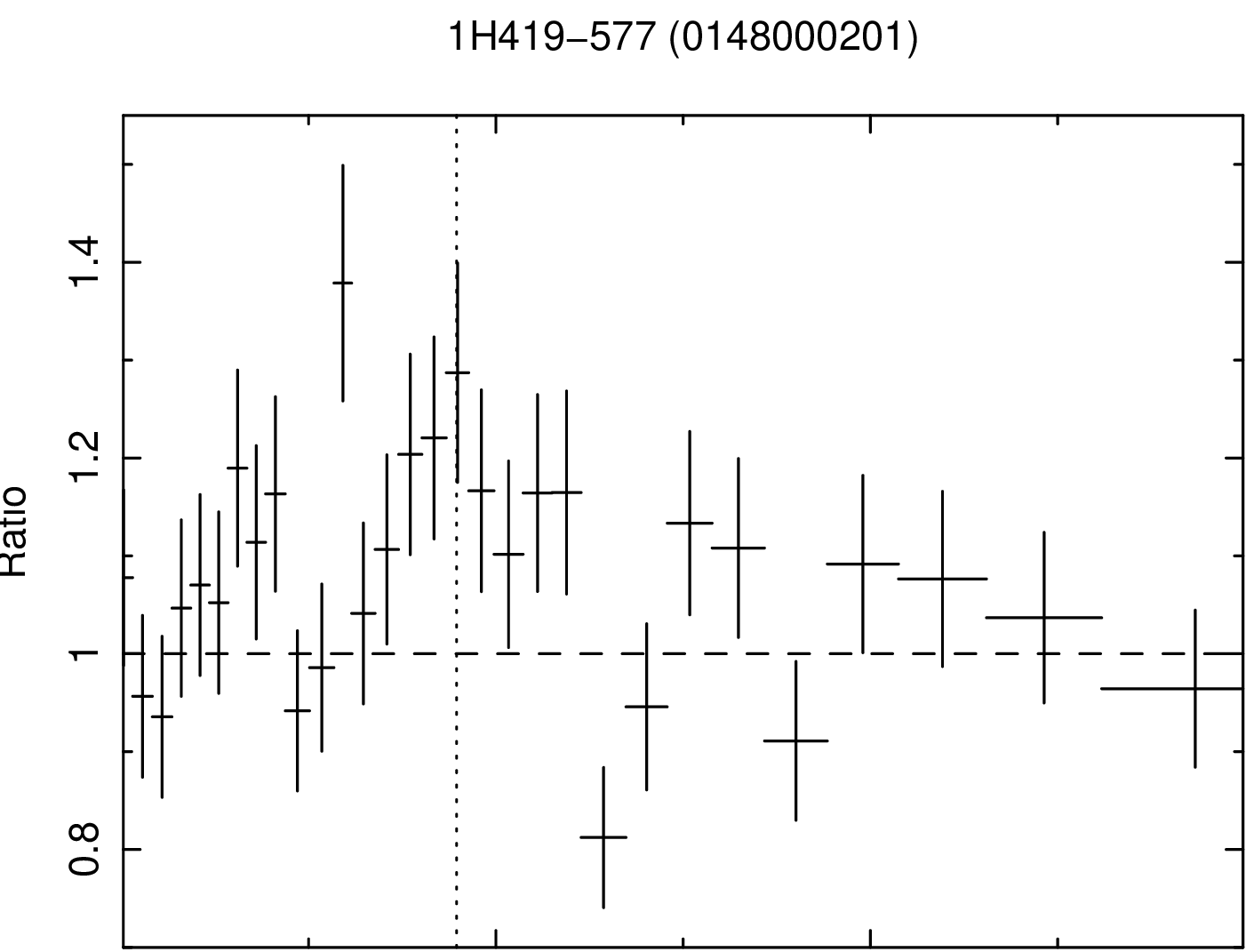}

\vspace{-0.07cm}

\hspace{-0.12cm}
    \includegraphics[width=3cm,height=4cm,angle=270]{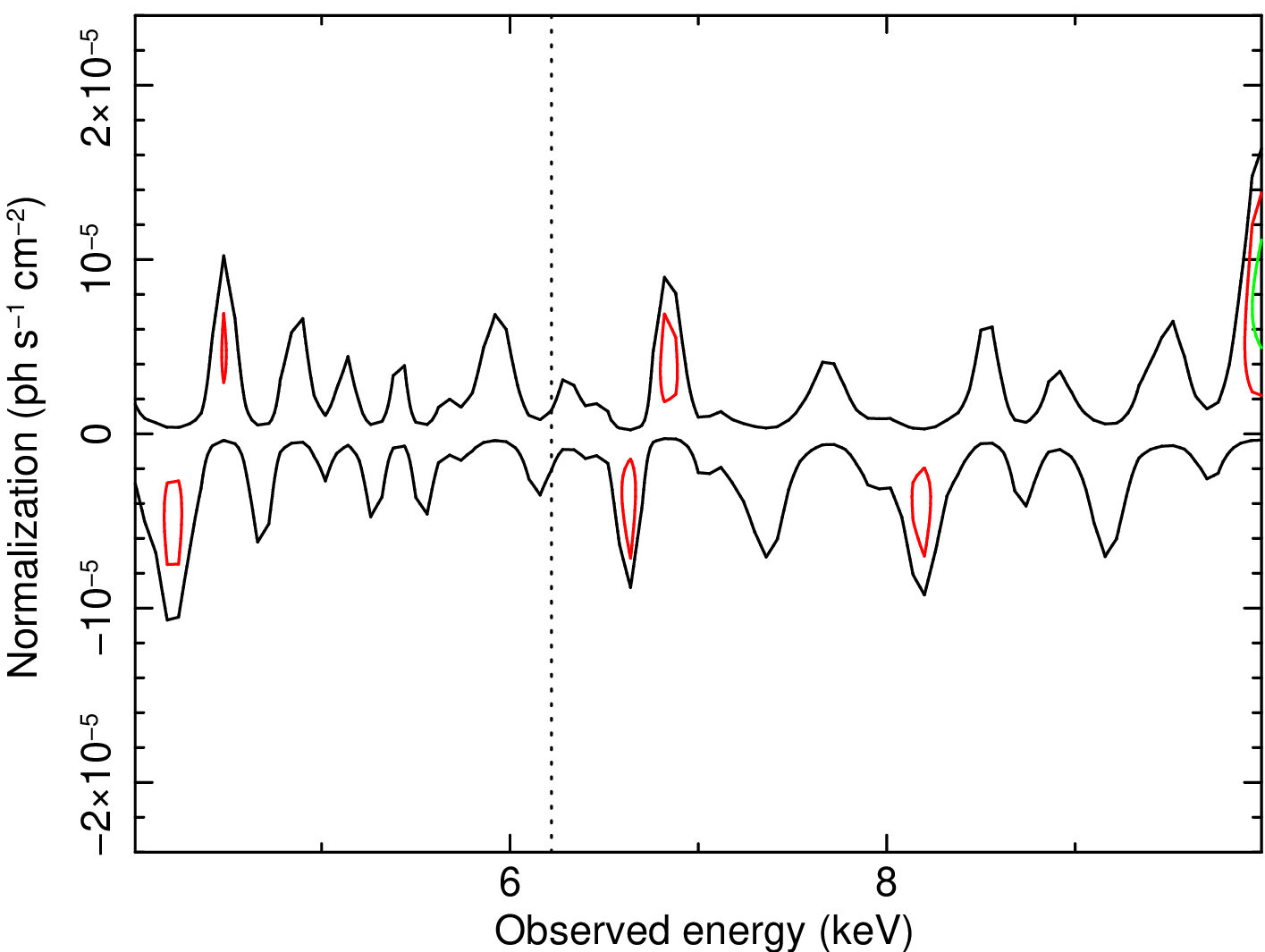}
\hspace{0.3cm}
    \includegraphics[width=3cm,height=4cm,angle=270]{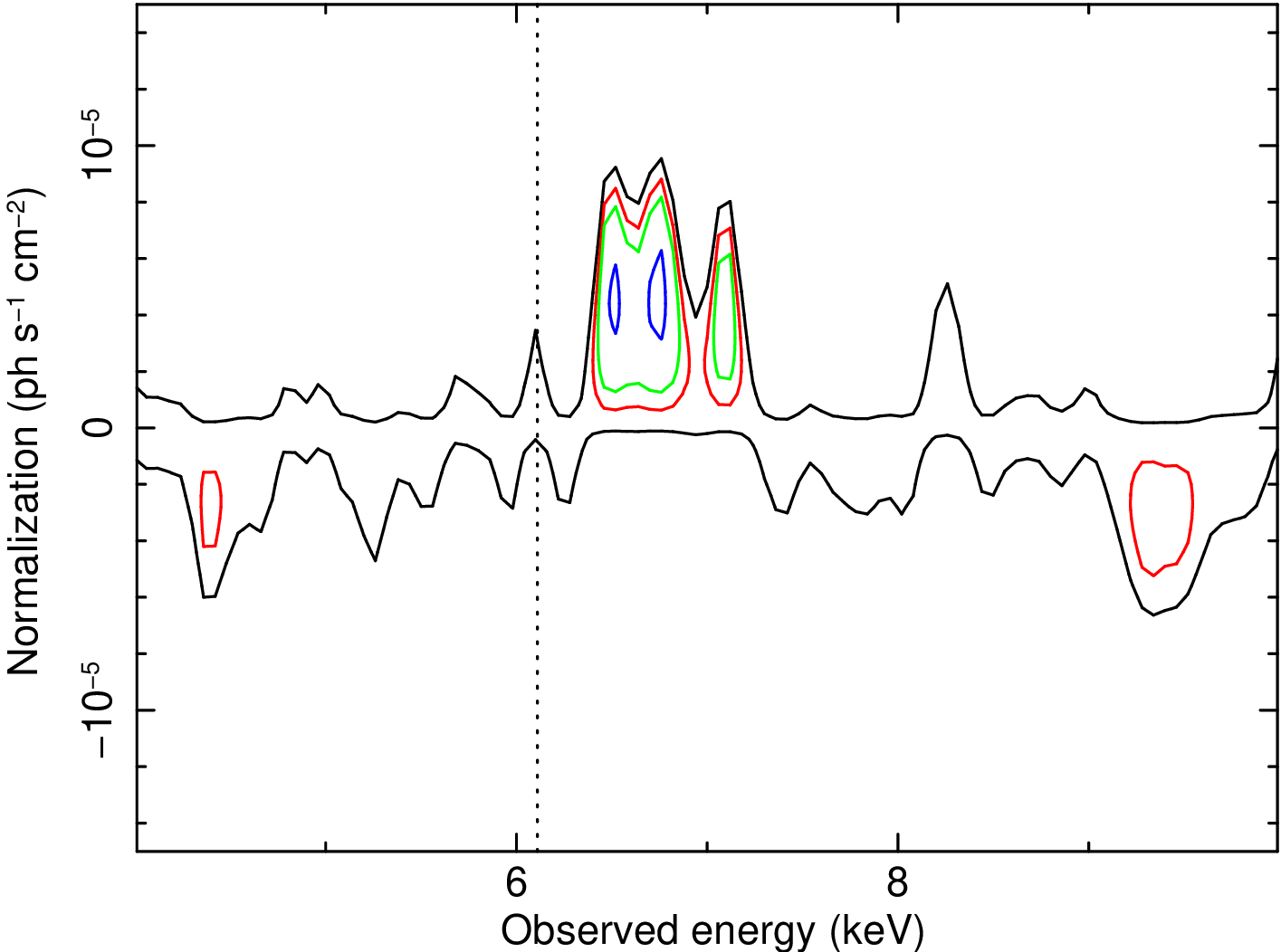}
\hspace{0.3cm}
    \includegraphics[width=3cm,height=4cm,angle=270]{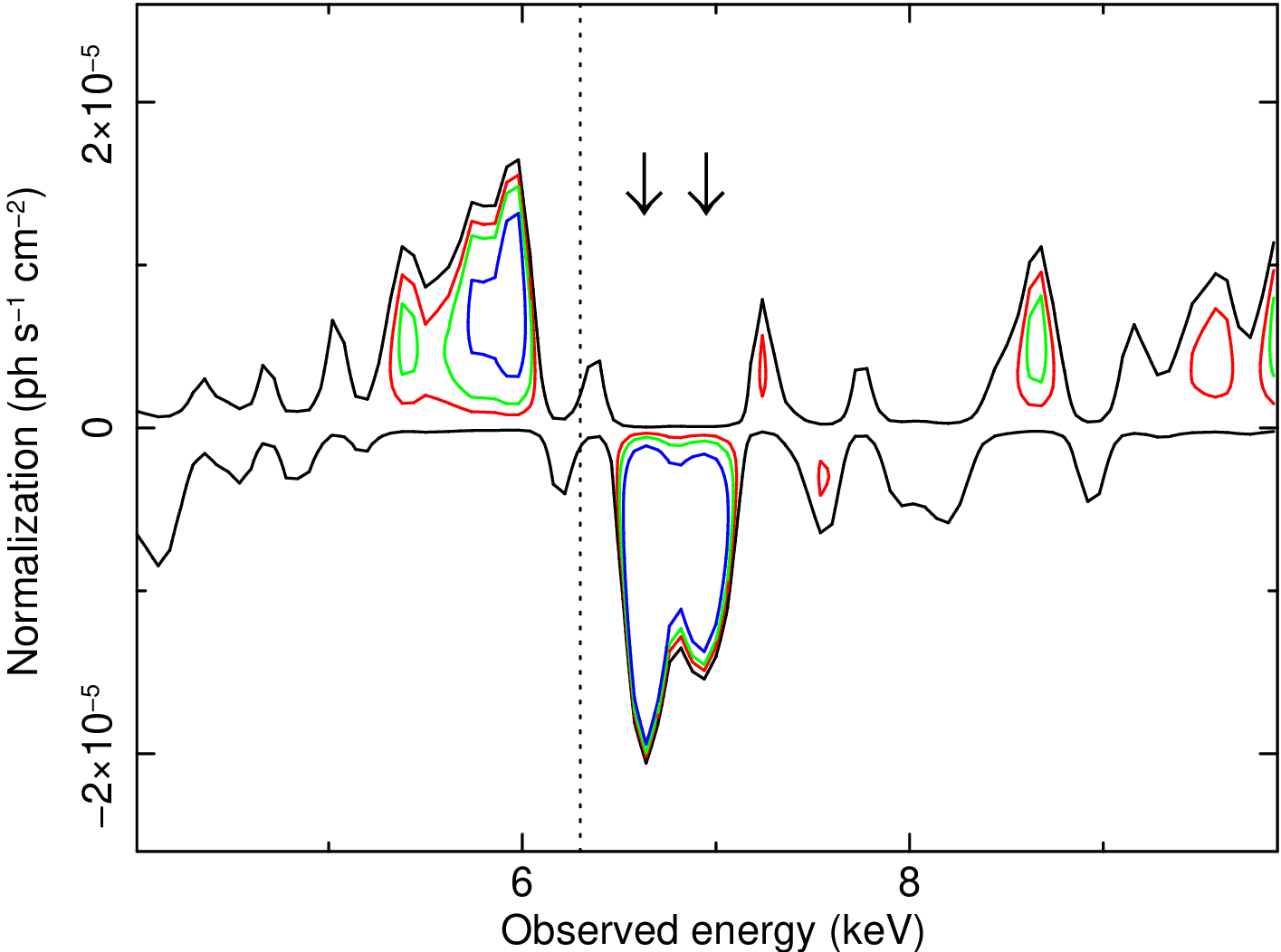}
\hspace{0.3cm}
    \includegraphics[width=3cm,height=4cm,angle=270]{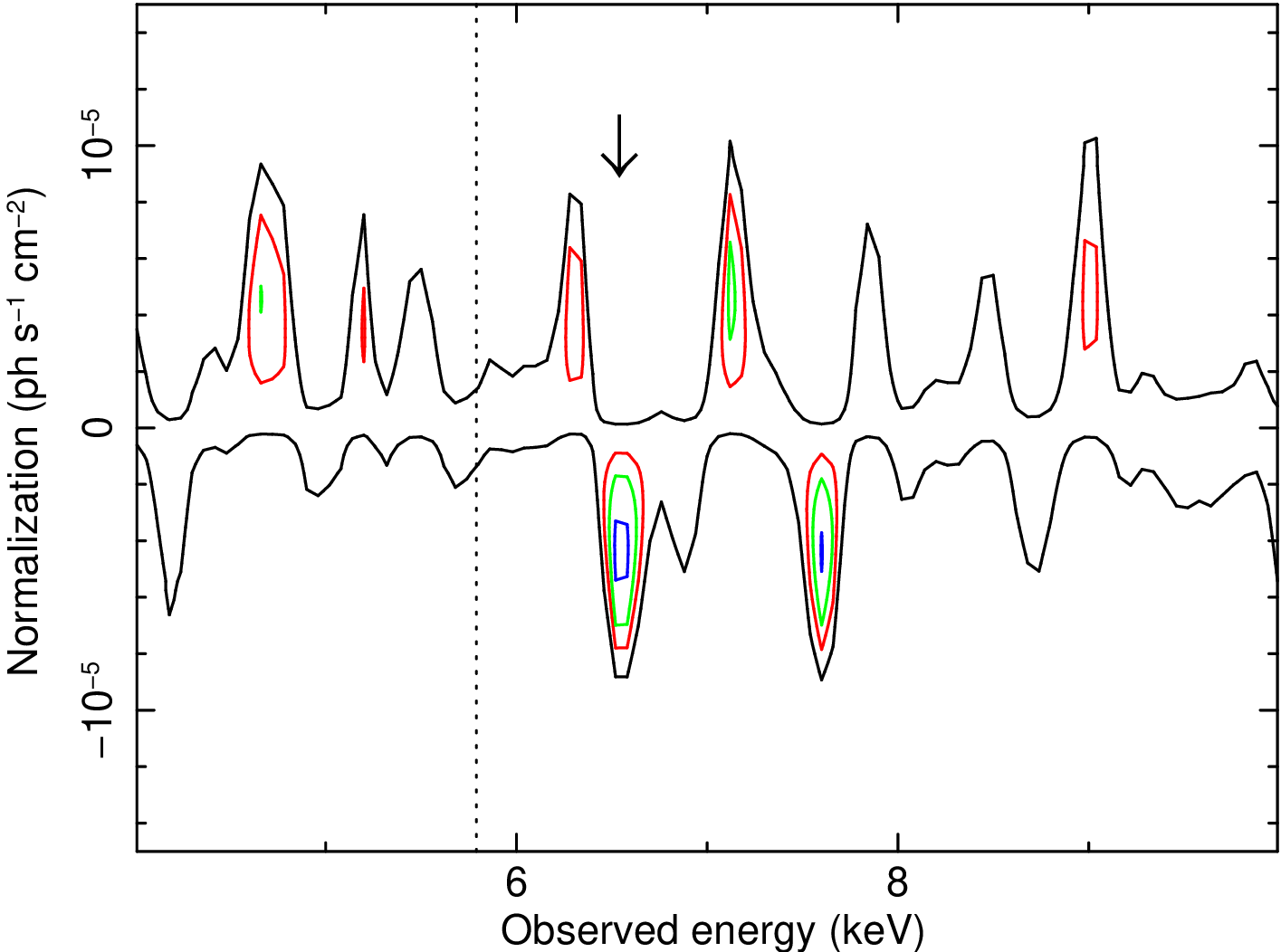}


\vspace{0.2cm}

    \includegraphics[width=3cm,height=4cm,angle=270]{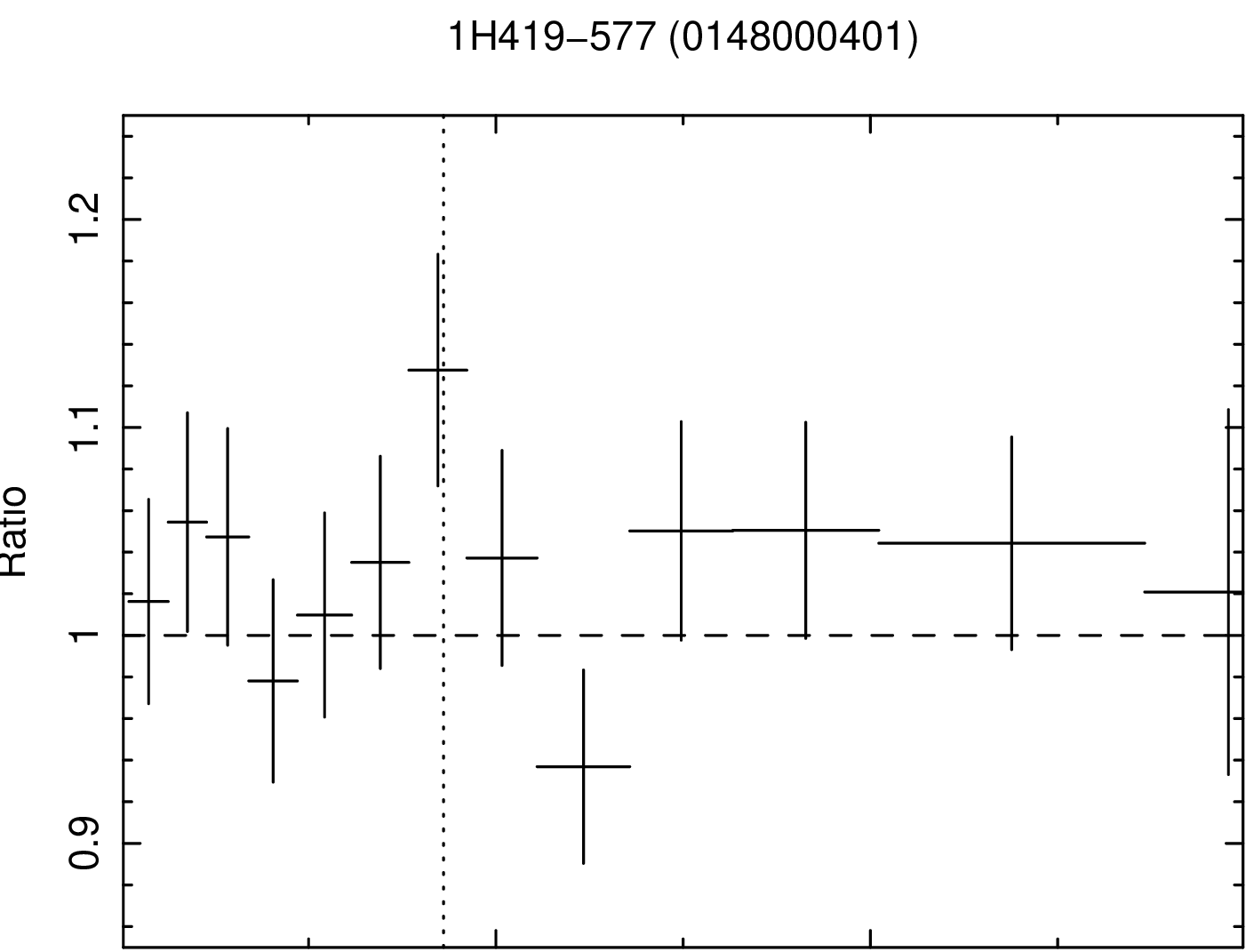}
\hspace{0.3cm}
    \includegraphics[width=3cm,height=4cm,angle=270]{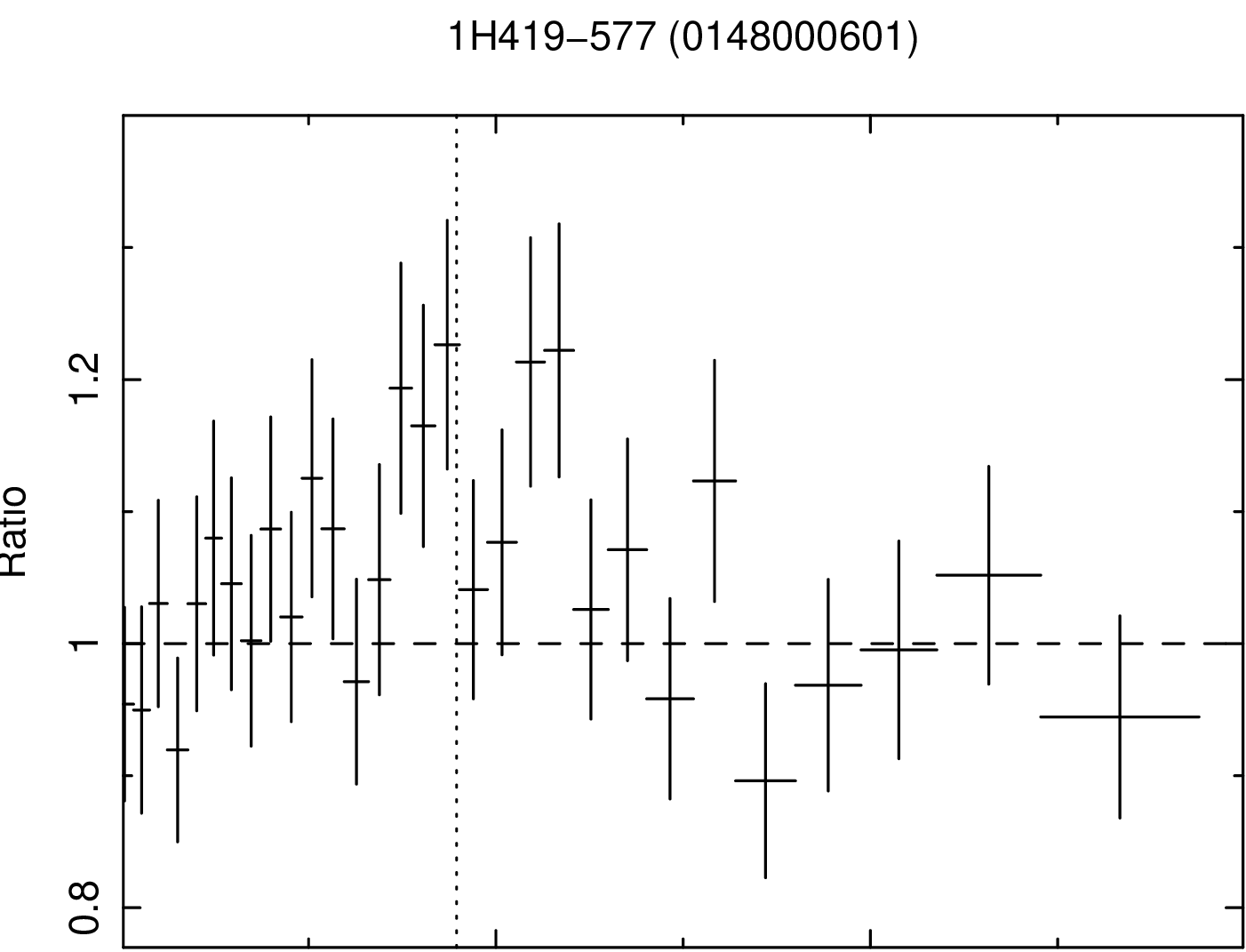}
\hspace{0.3cm}
    \includegraphics[width=3cm,height=4cm,angle=270]{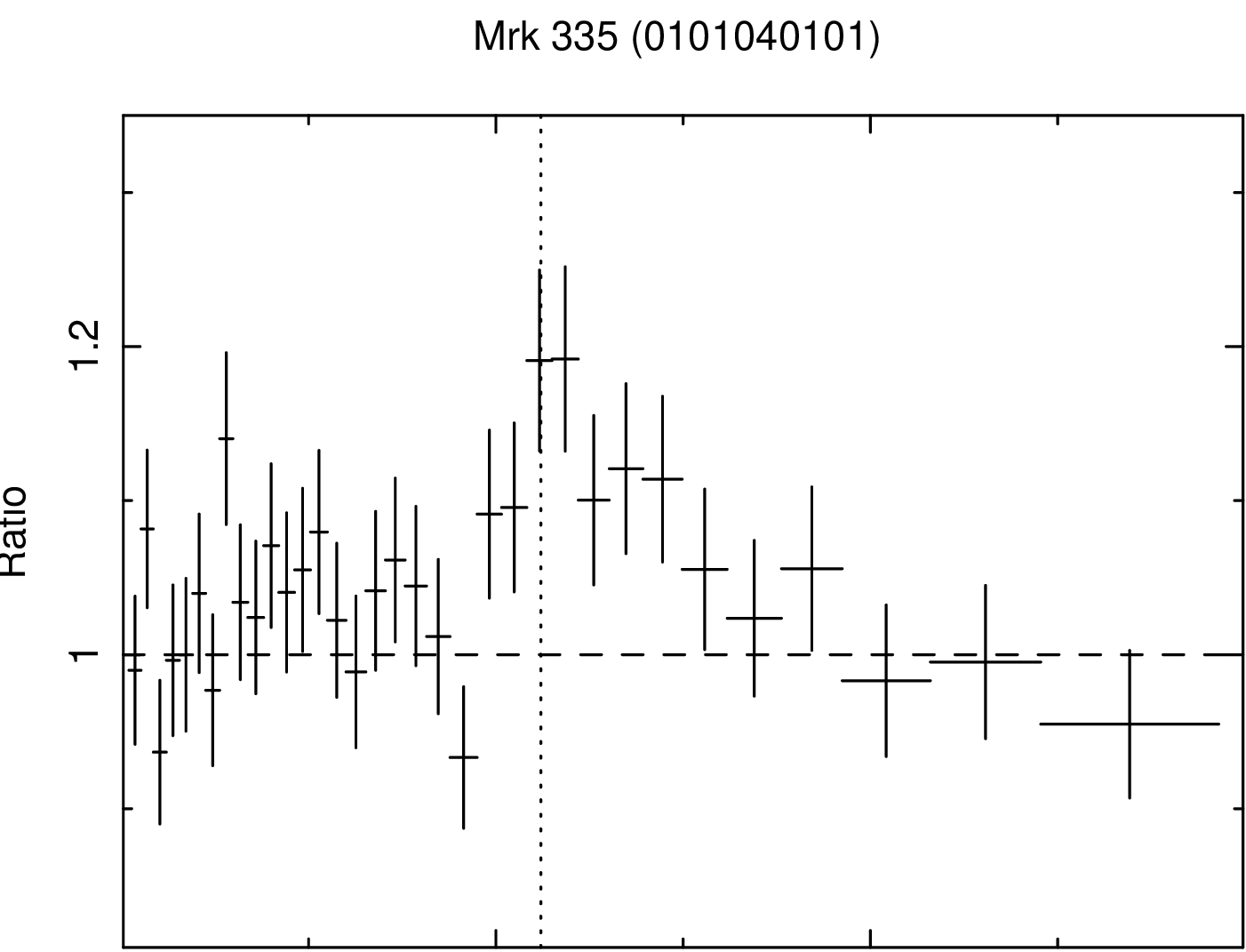}
\hspace{0.3cm}
    \includegraphics[width=3cm,height=4cm,angle=270]{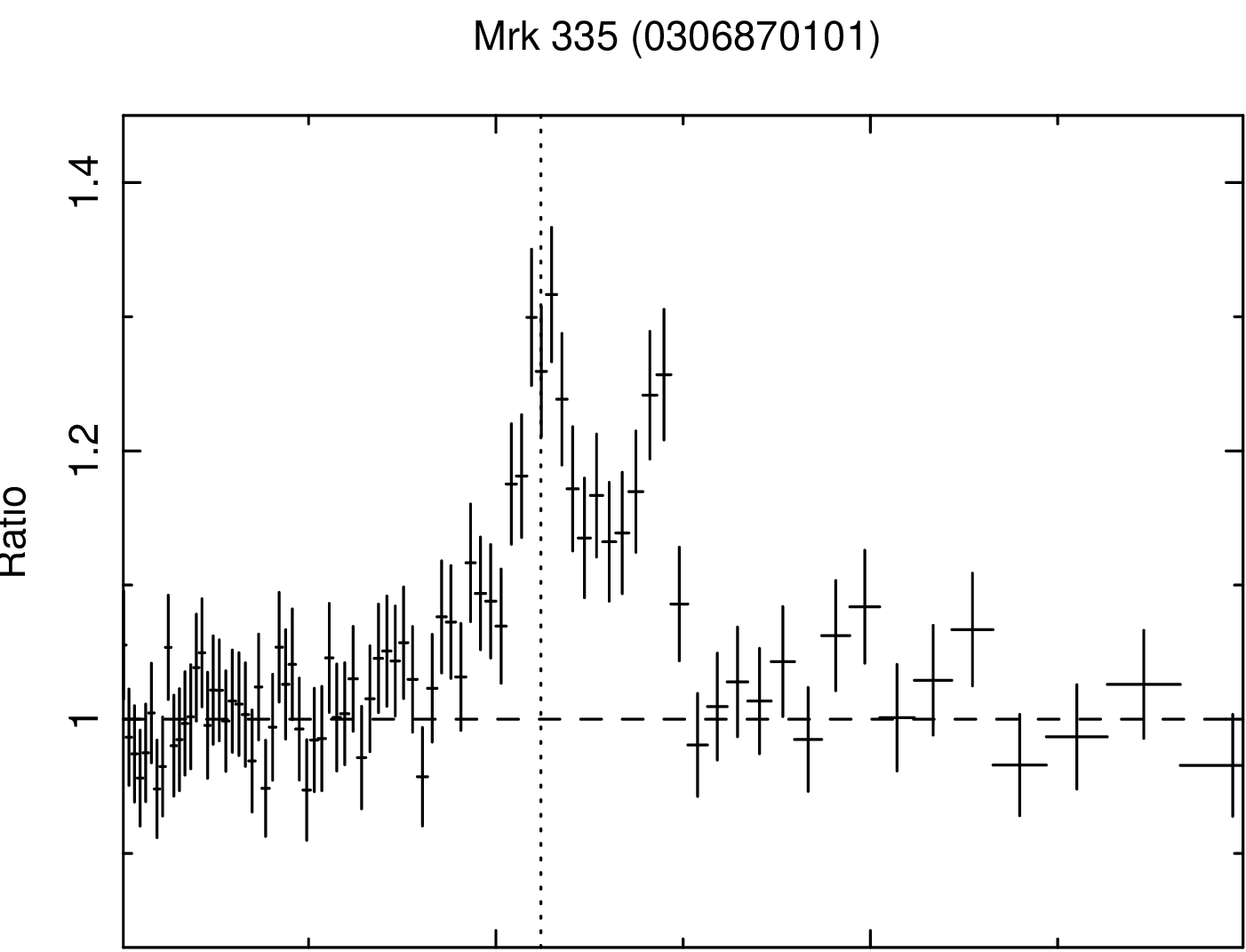}

\vspace{-0.07cm}

\hspace{-0.12cm}
    \includegraphics[width=3cm,height=4cm,angle=270]{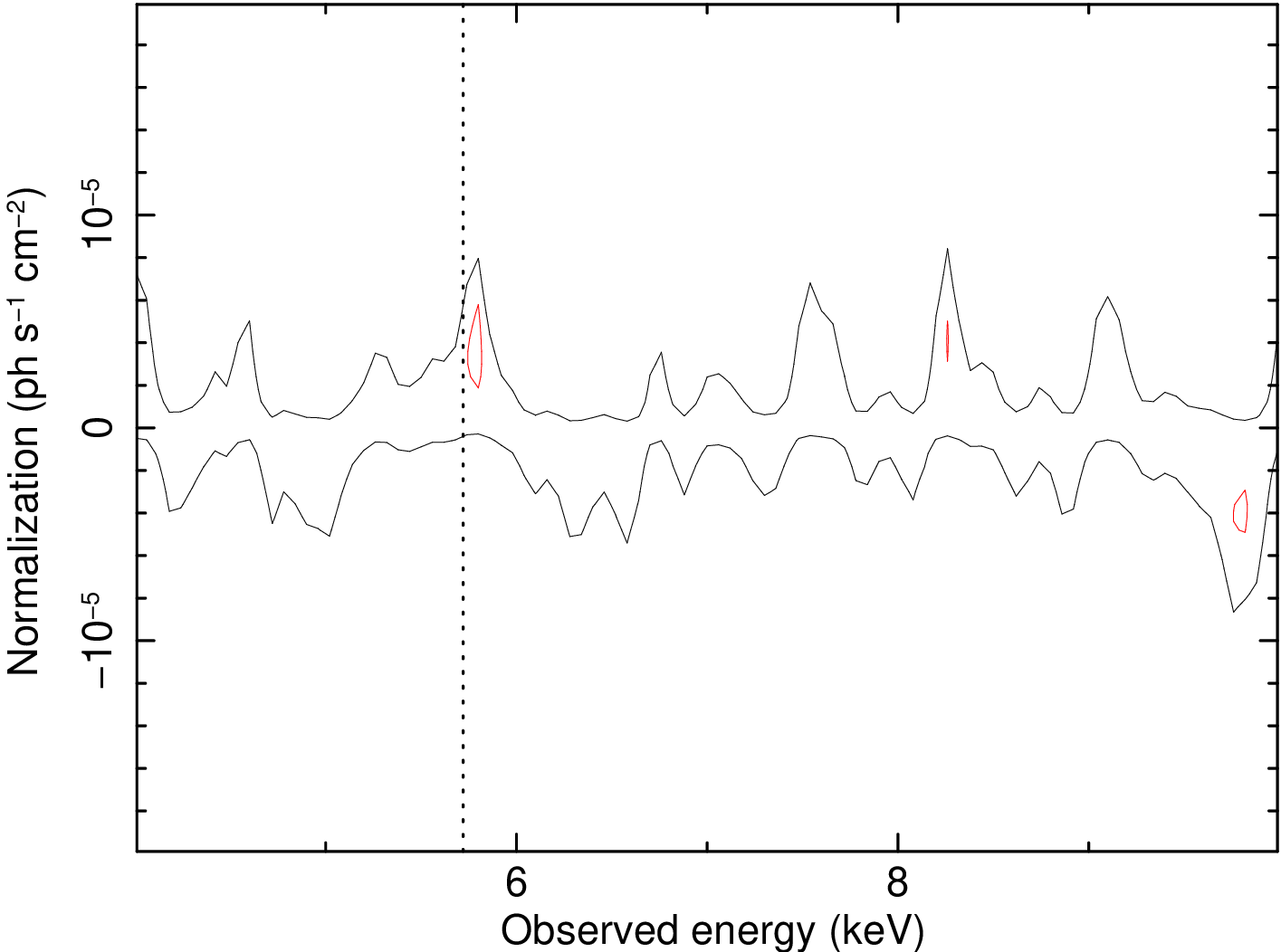}
\hspace{0.3cm}
    \includegraphics[width=3cm,height=4cm,angle=270]{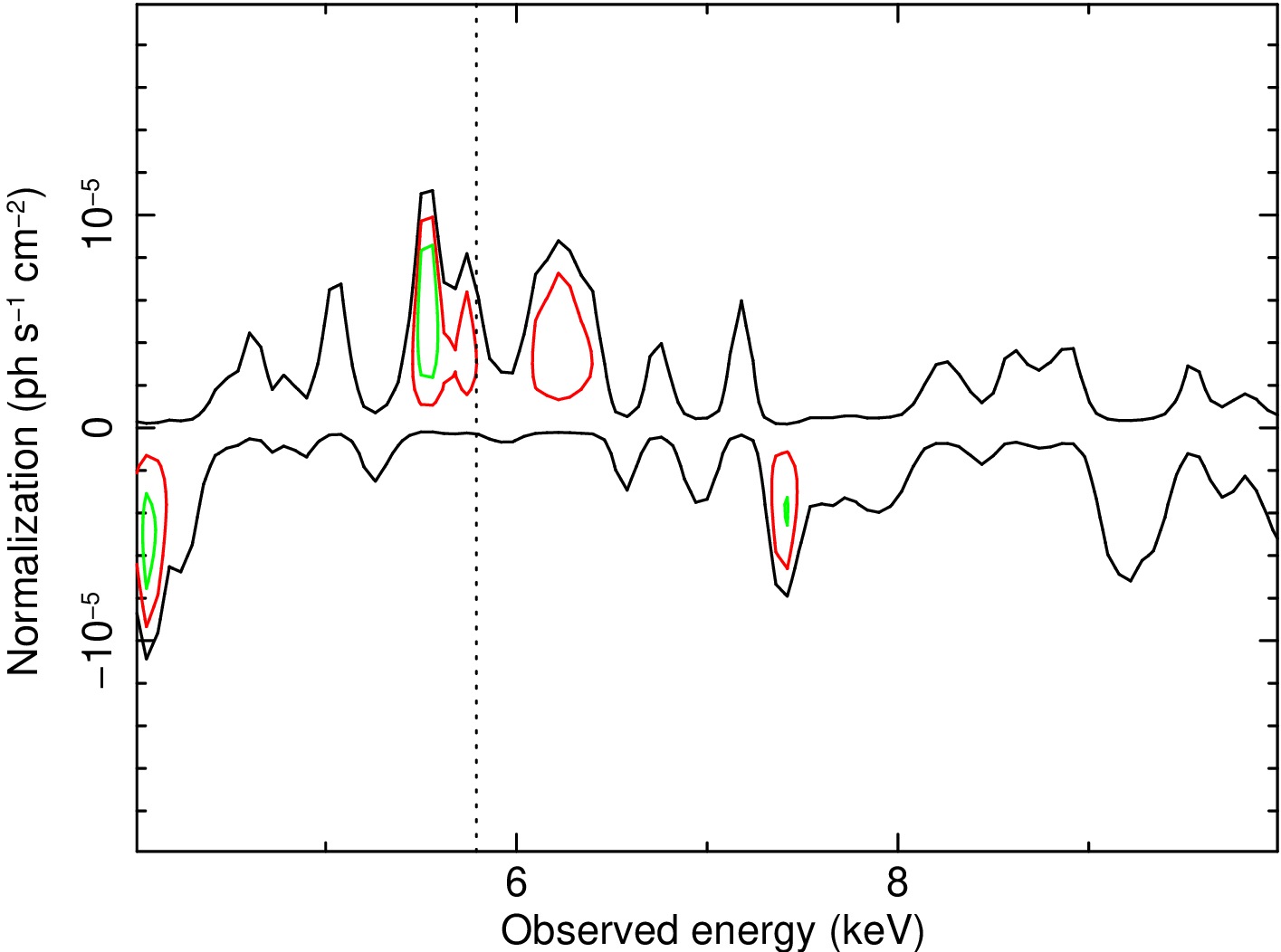}
\hspace{0.3cm}
    \includegraphics[width=3cm,height=4cm,angle=270]{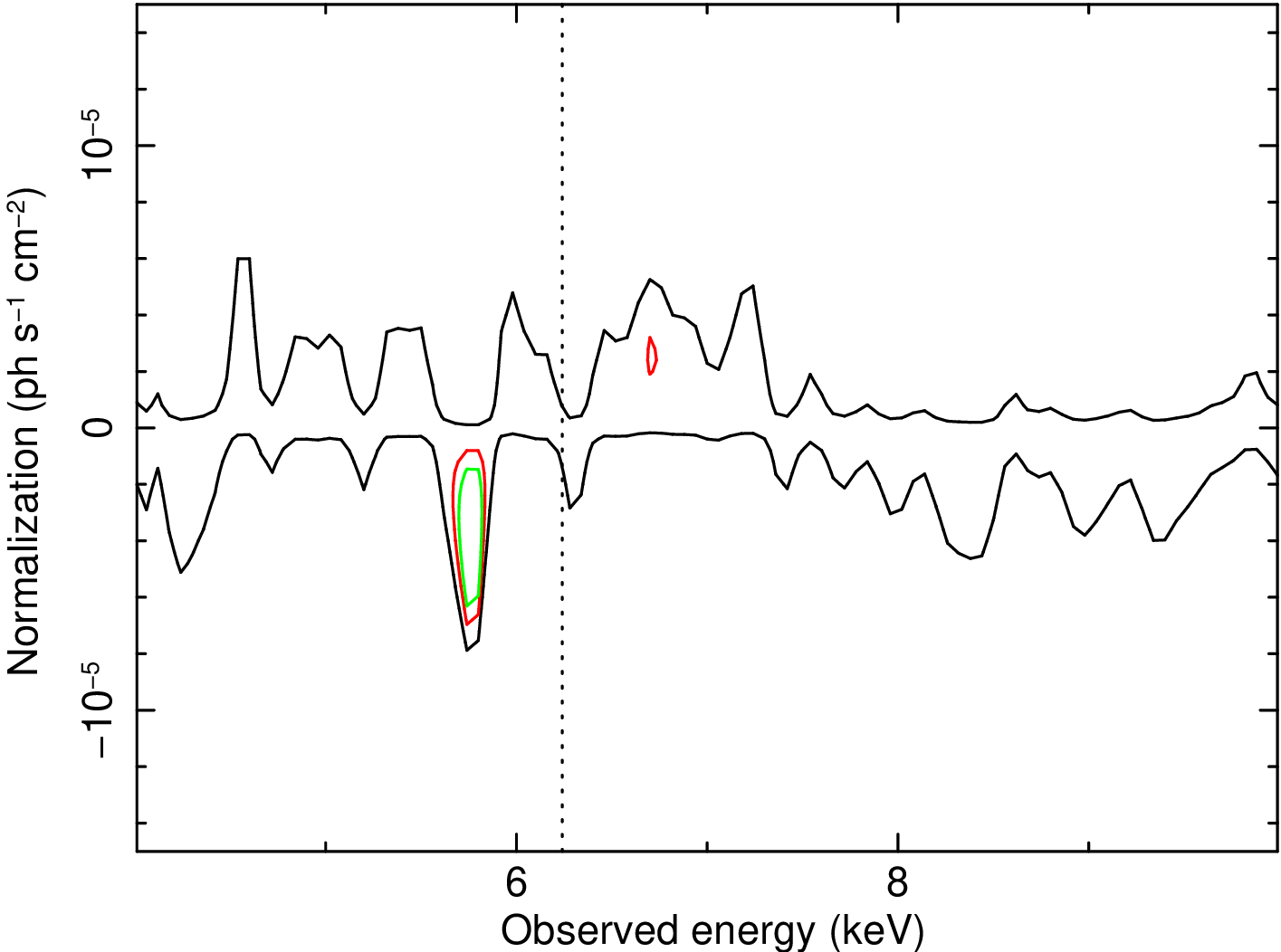}
\hspace{0.3cm}
    \includegraphics[width=3cm,height=4cm,angle=270]{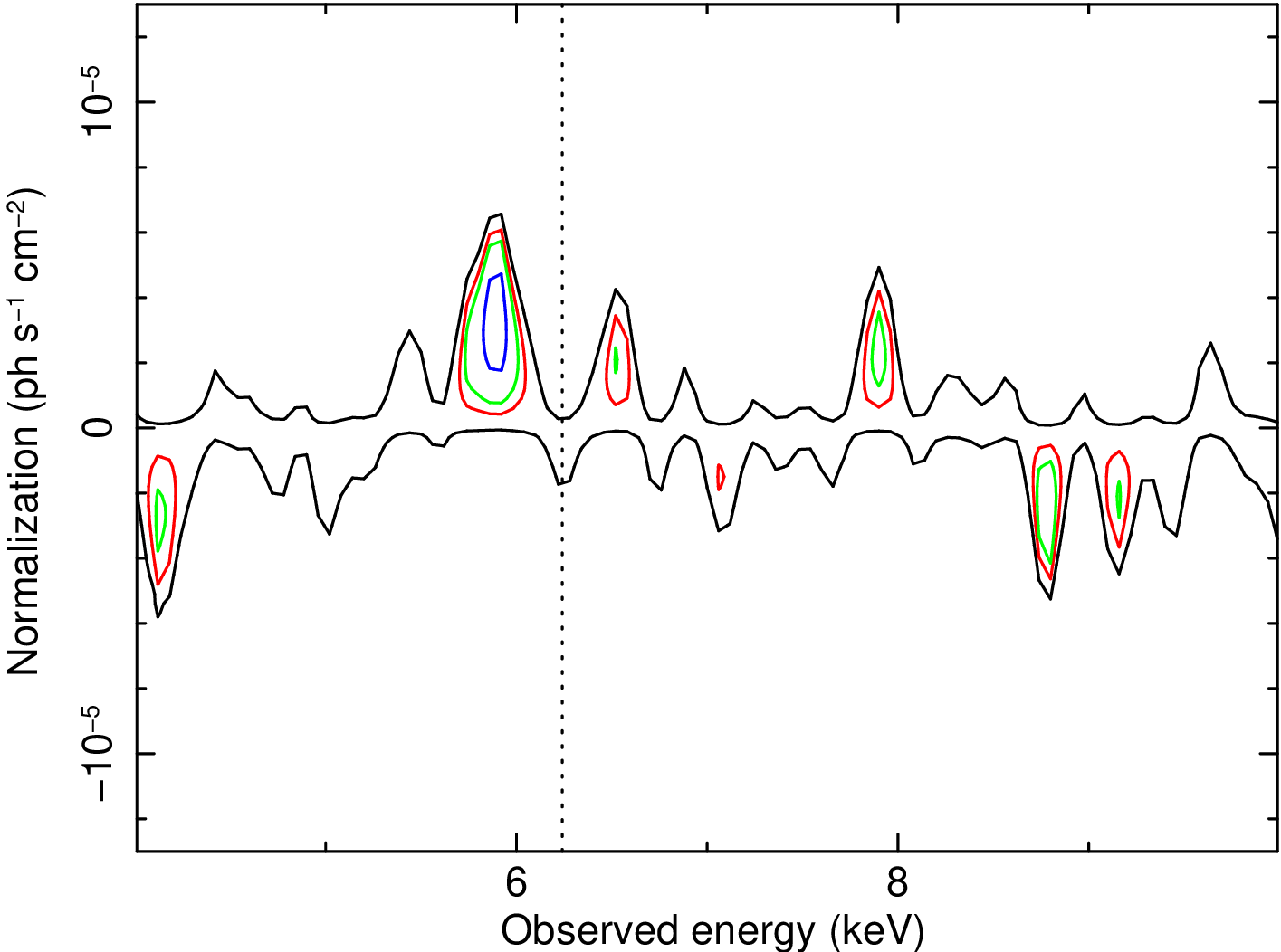}


\vspace{0.2cm}

    \includegraphics[width=3cm,height=4cm,angle=270]{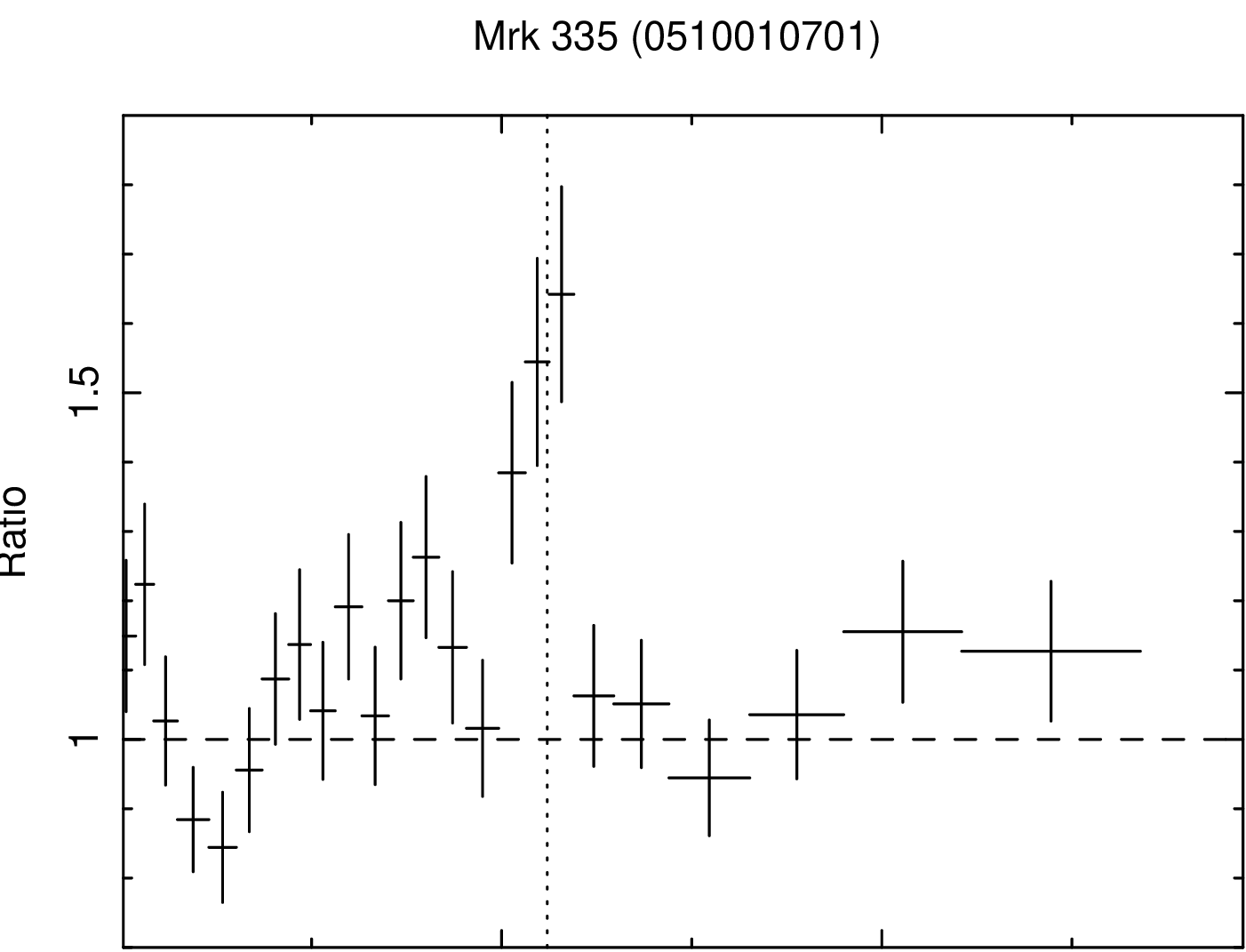}
\hspace{0.3cm}
    \includegraphics[width=3cm,height=4cm,angle=270]{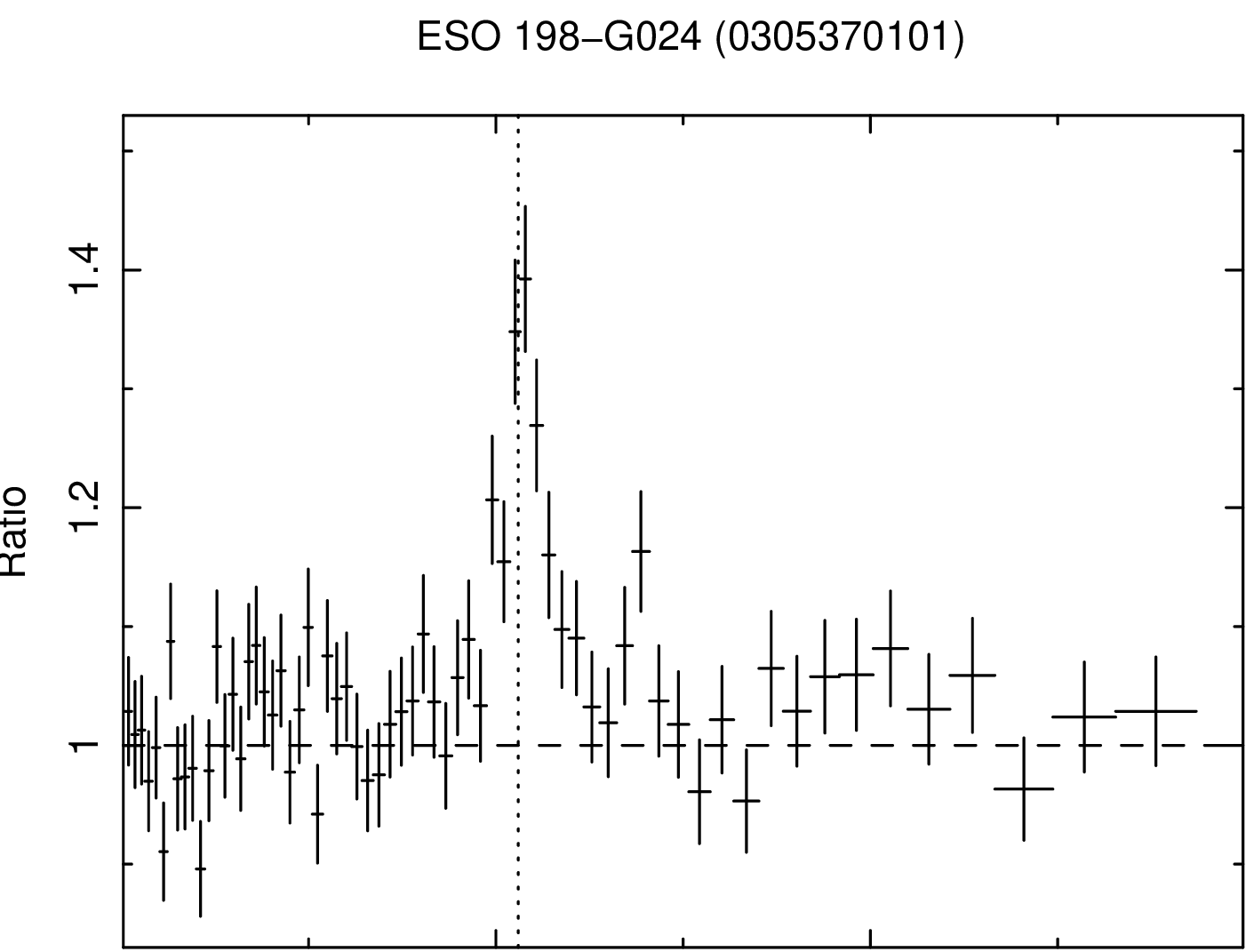}
\hspace{0.3cm}
    \includegraphics[width=3cm,height=4cm,angle=270]{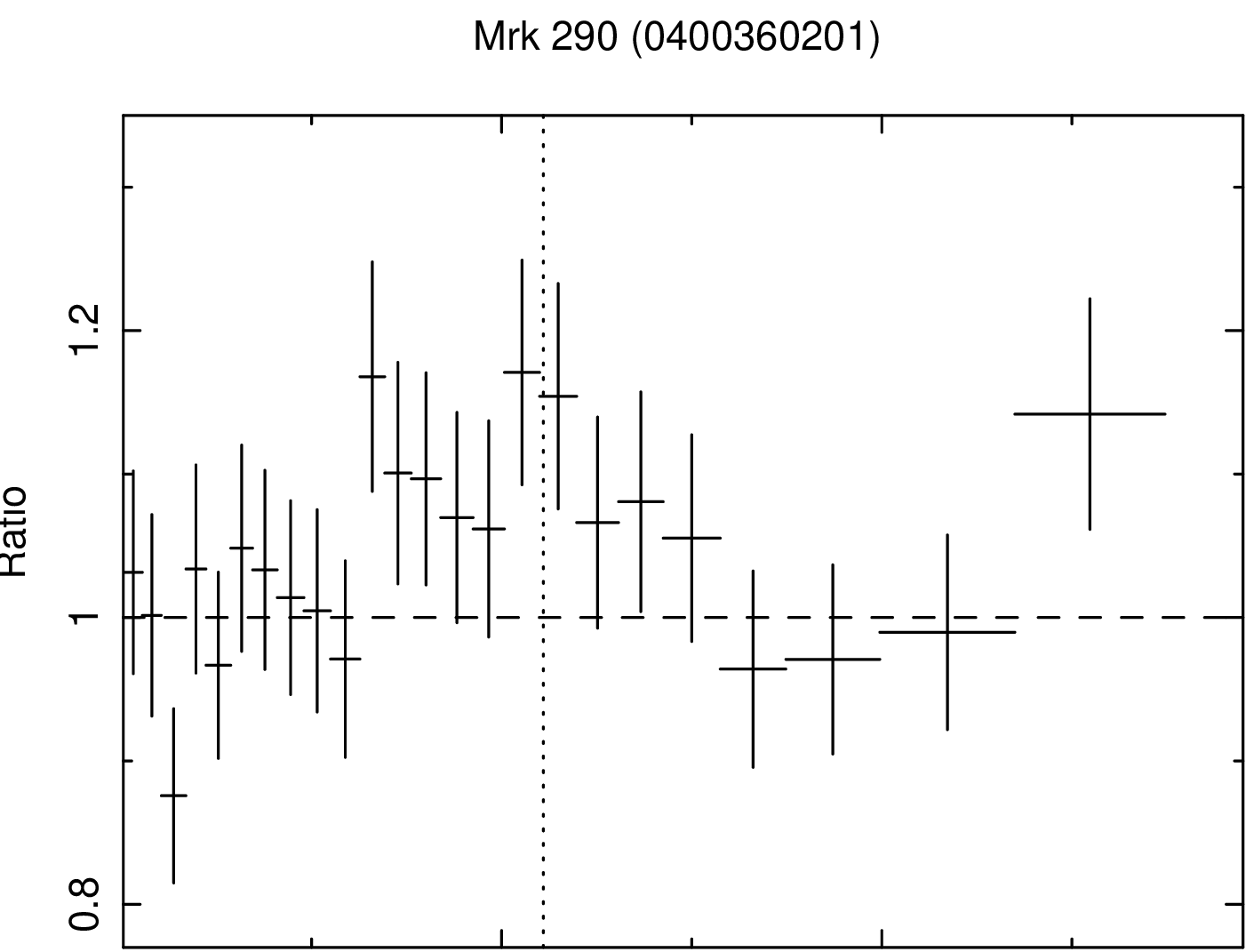}
\hspace{0.3cm}
    \includegraphics[width=3cm,height=4cm,angle=270]{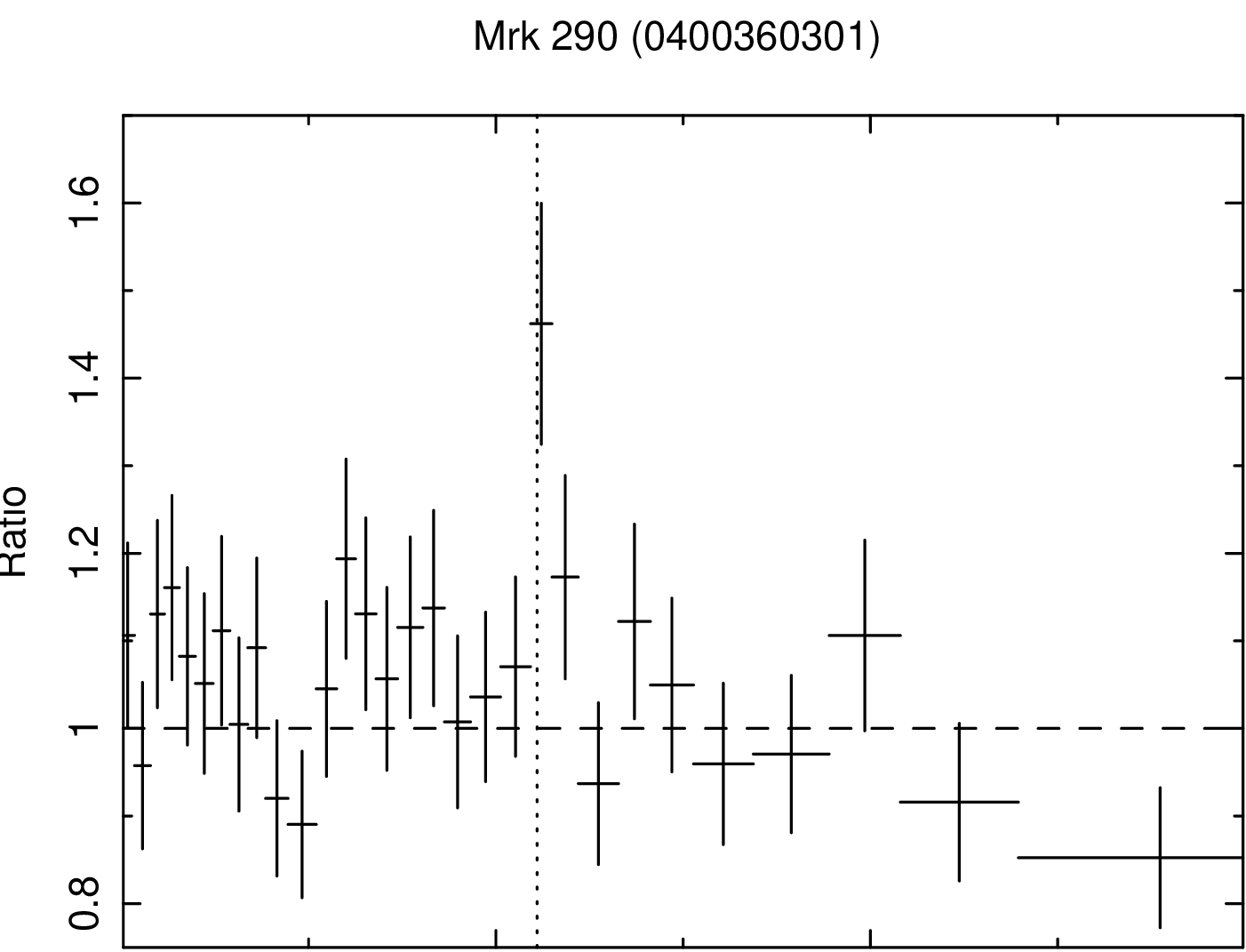}

\vspace{-0.07cm}

\hspace{-0.12cm}
    \includegraphics[width=3cm,height=4cm,angle=270]{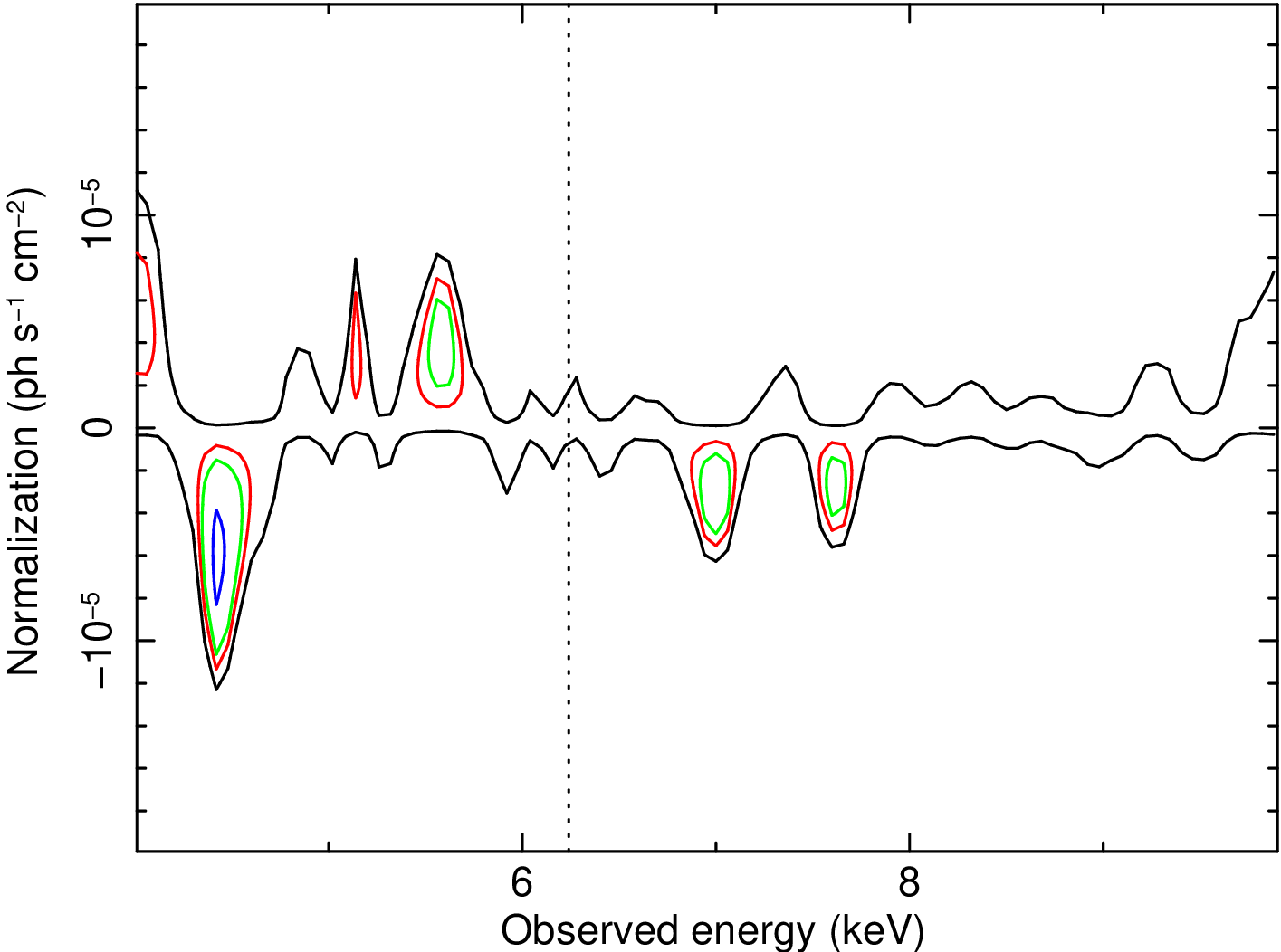}
\hspace{0.3cm}
    \includegraphics[width=3cm,height=4cm,angle=270]{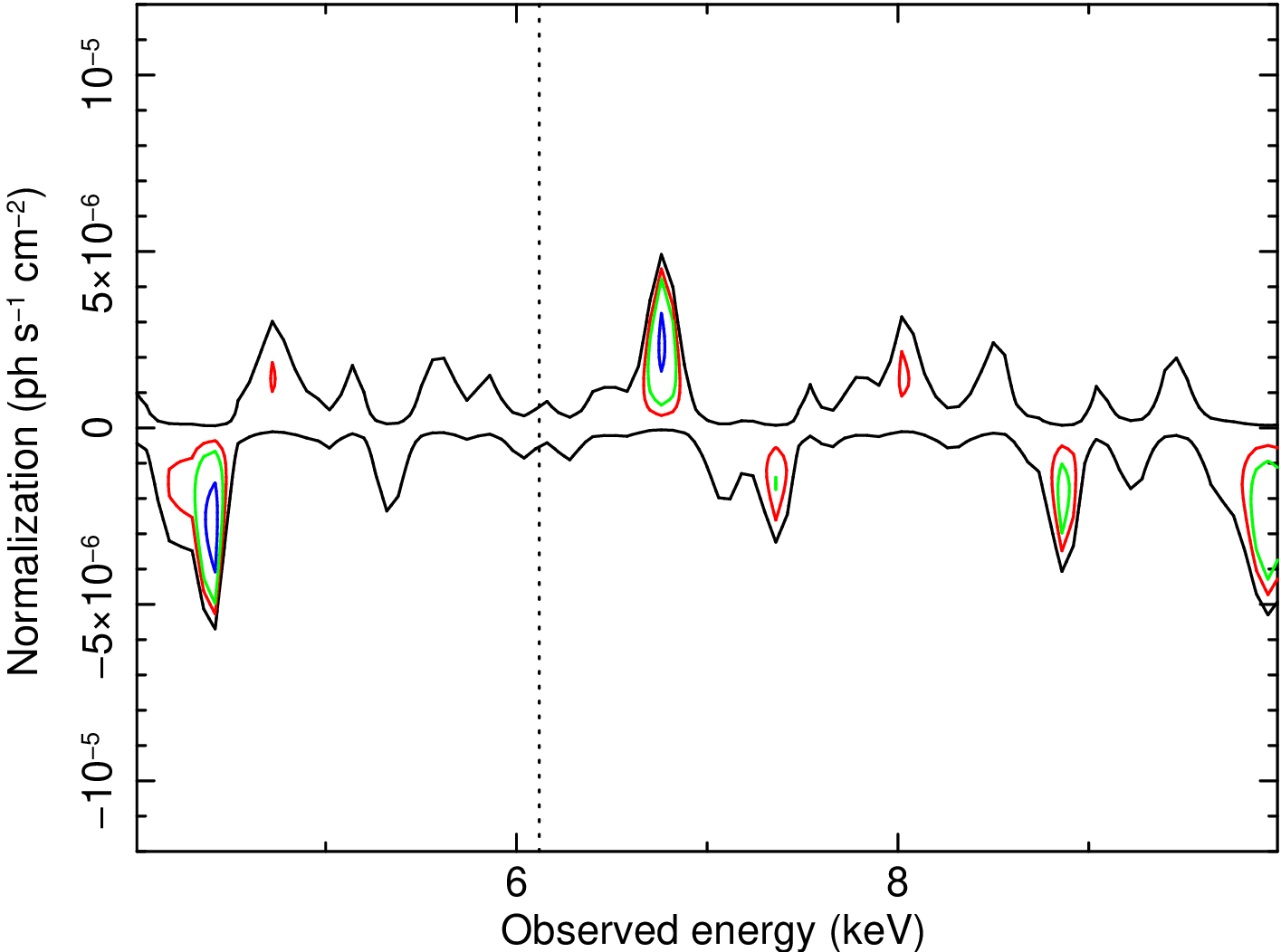}
\hspace{0.3cm}
    \includegraphics[width=3cm,height=4cm,angle=270]{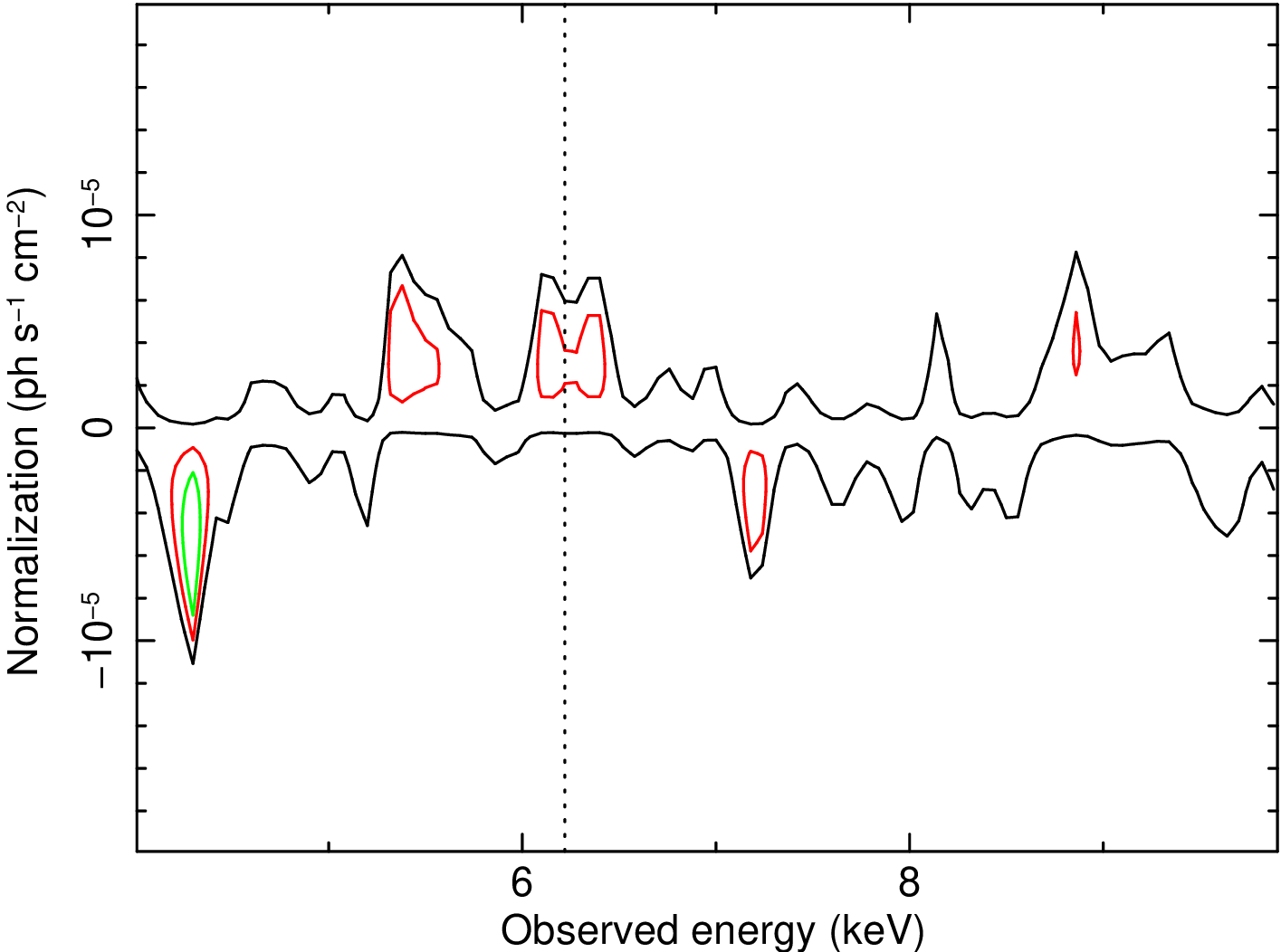}
\hspace{0.3cm}
    \includegraphics[width=3cm,height=4cm,angle=270]{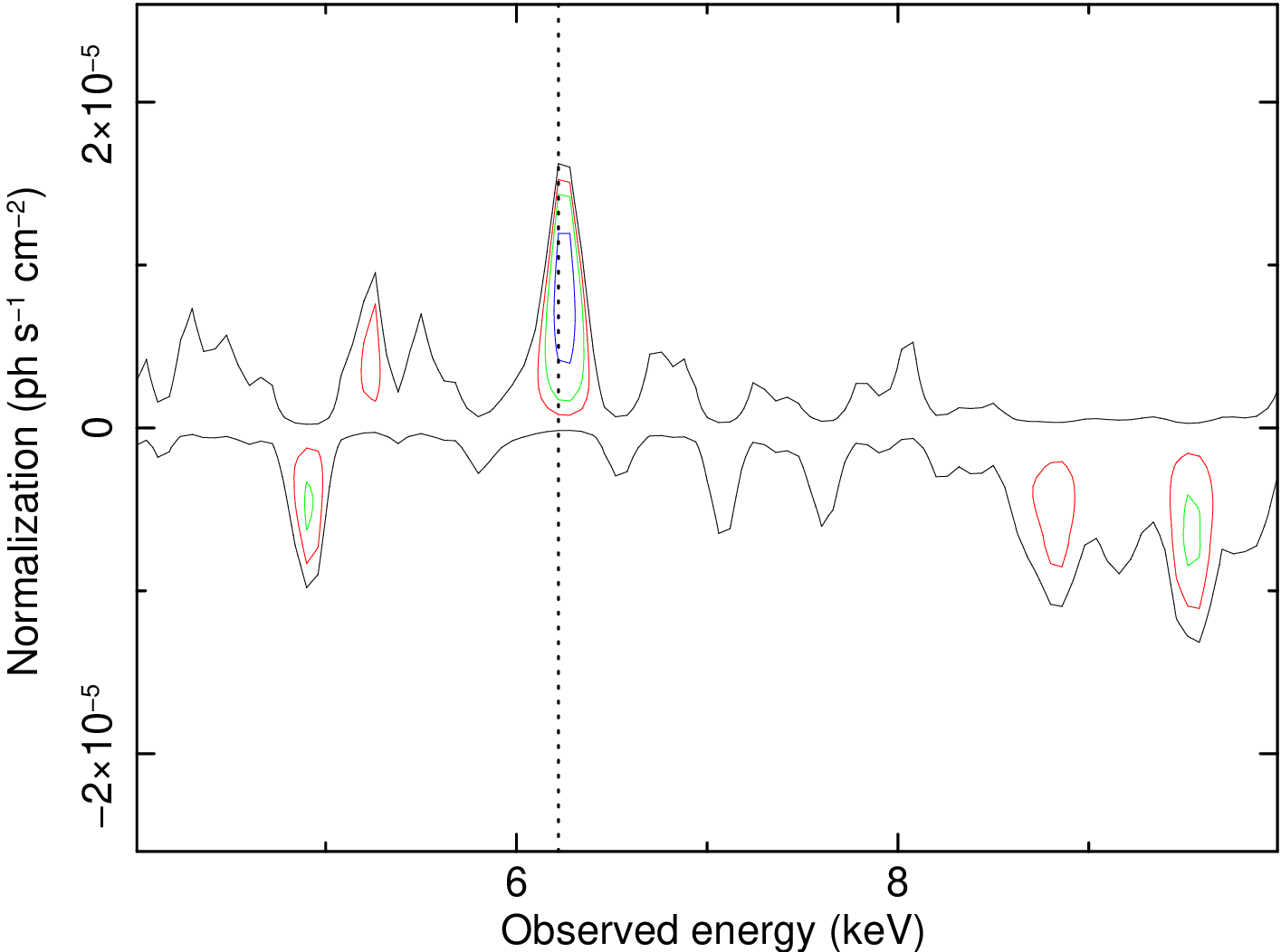}


\vspace{0.2cm}

    \includegraphics[width=3cm,height=4cm,angle=270]{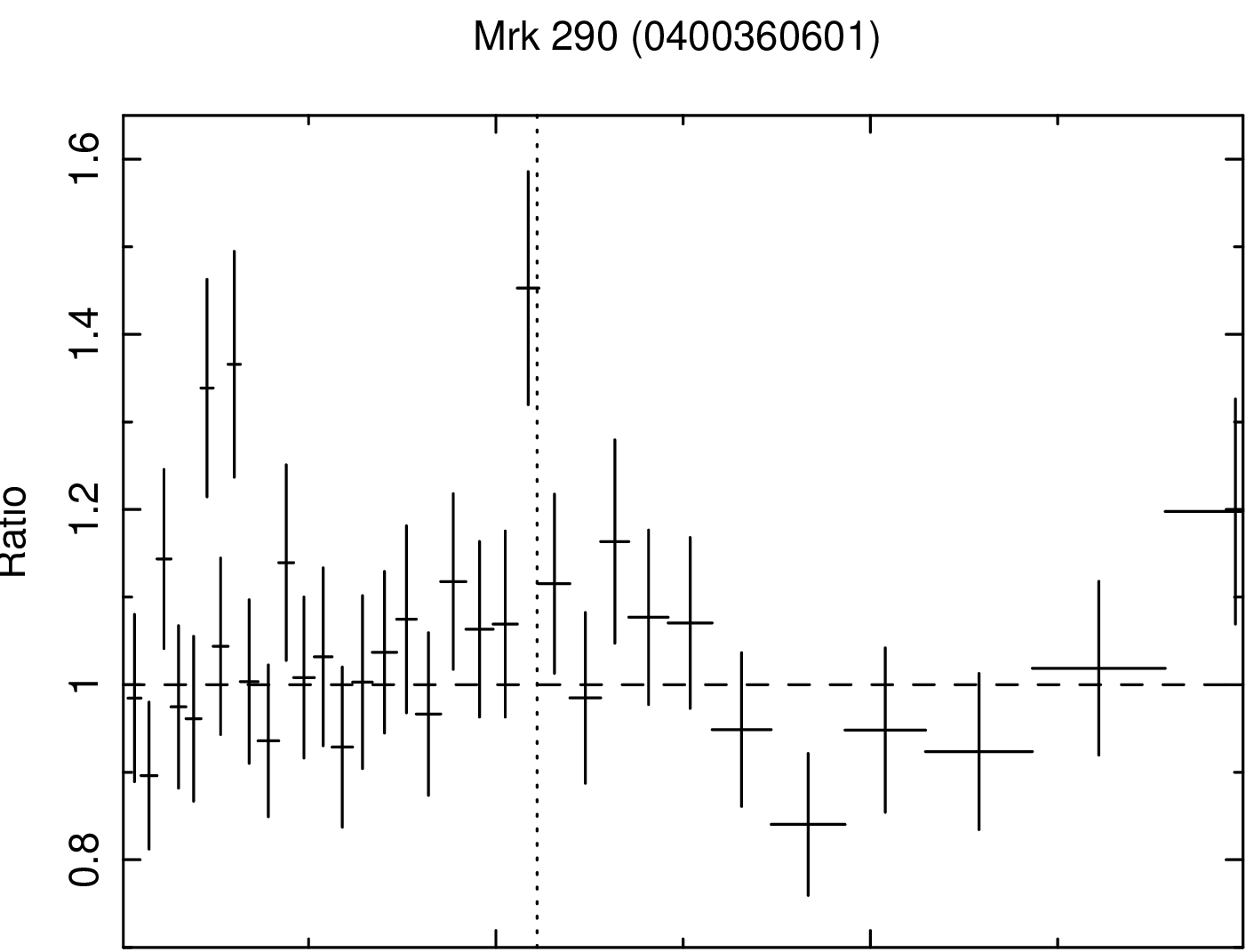}
\hspace{0.3cm}
    \includegraphics[width=3cm,height=4cm,angle=270]{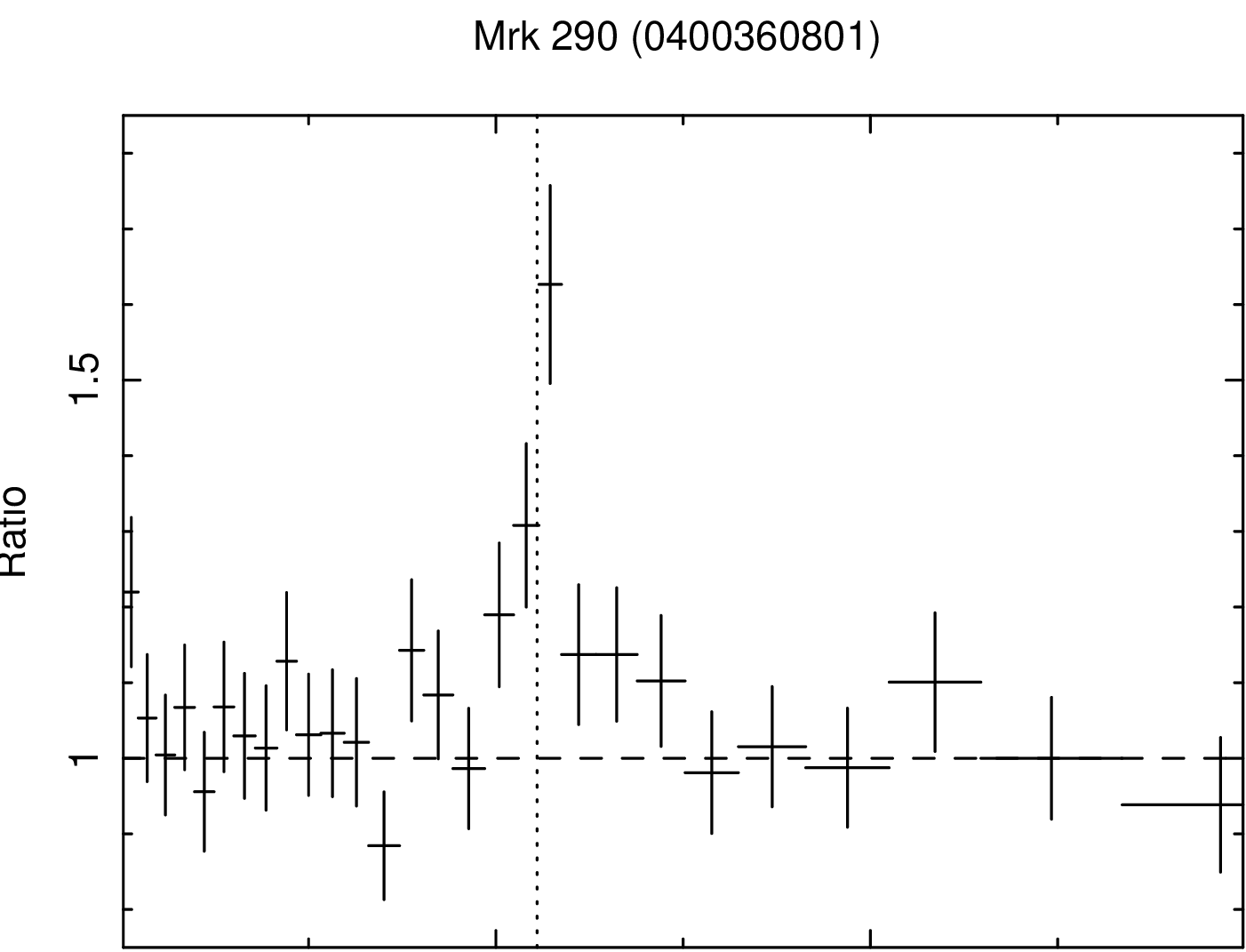}
\hspace{0.3cm}
    \includegraphics[width=3cm,height=4cm,angle=270]{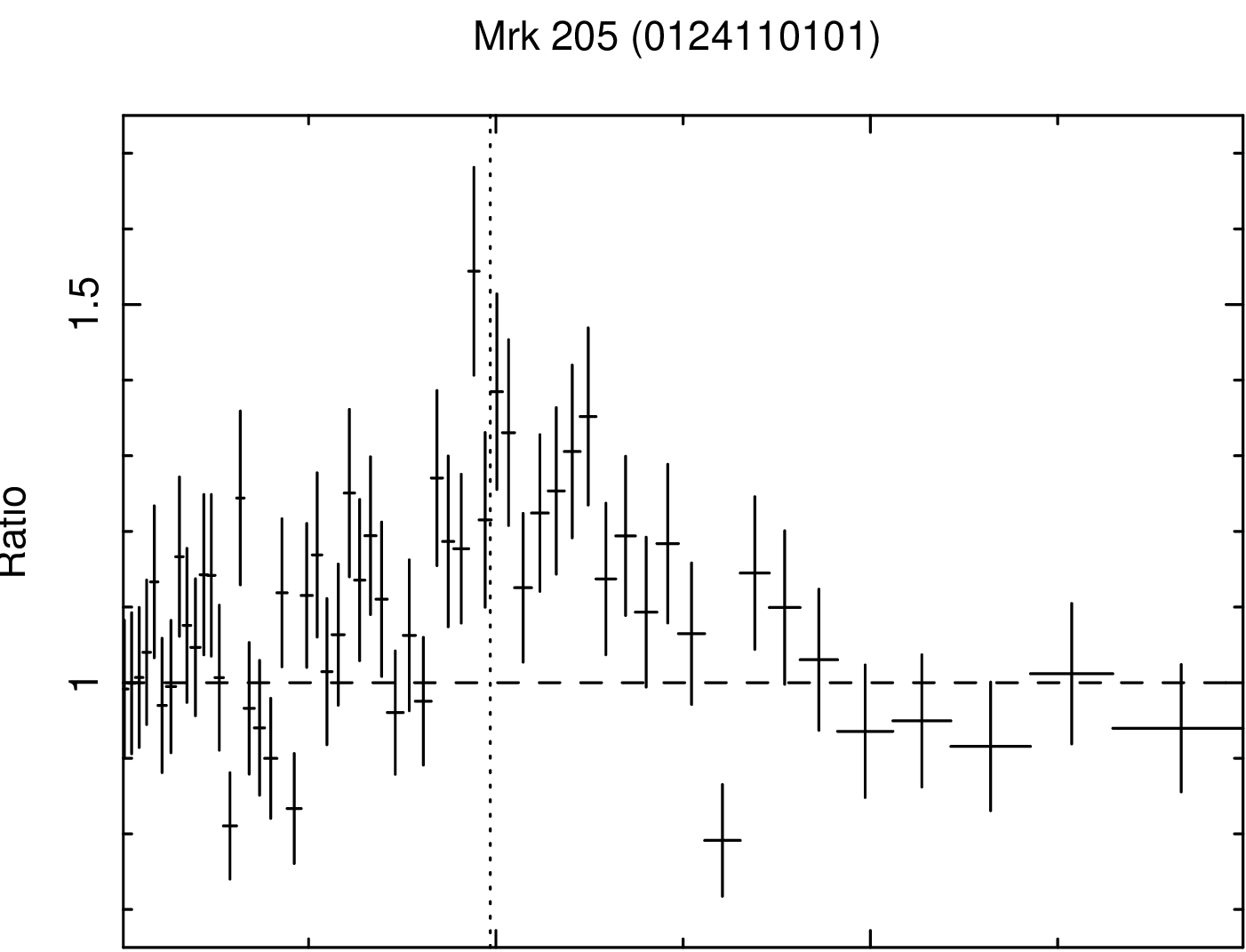}
\hspace{0.3cm}
    \includegraphics[width=3cm,height=4cm,angle=270]{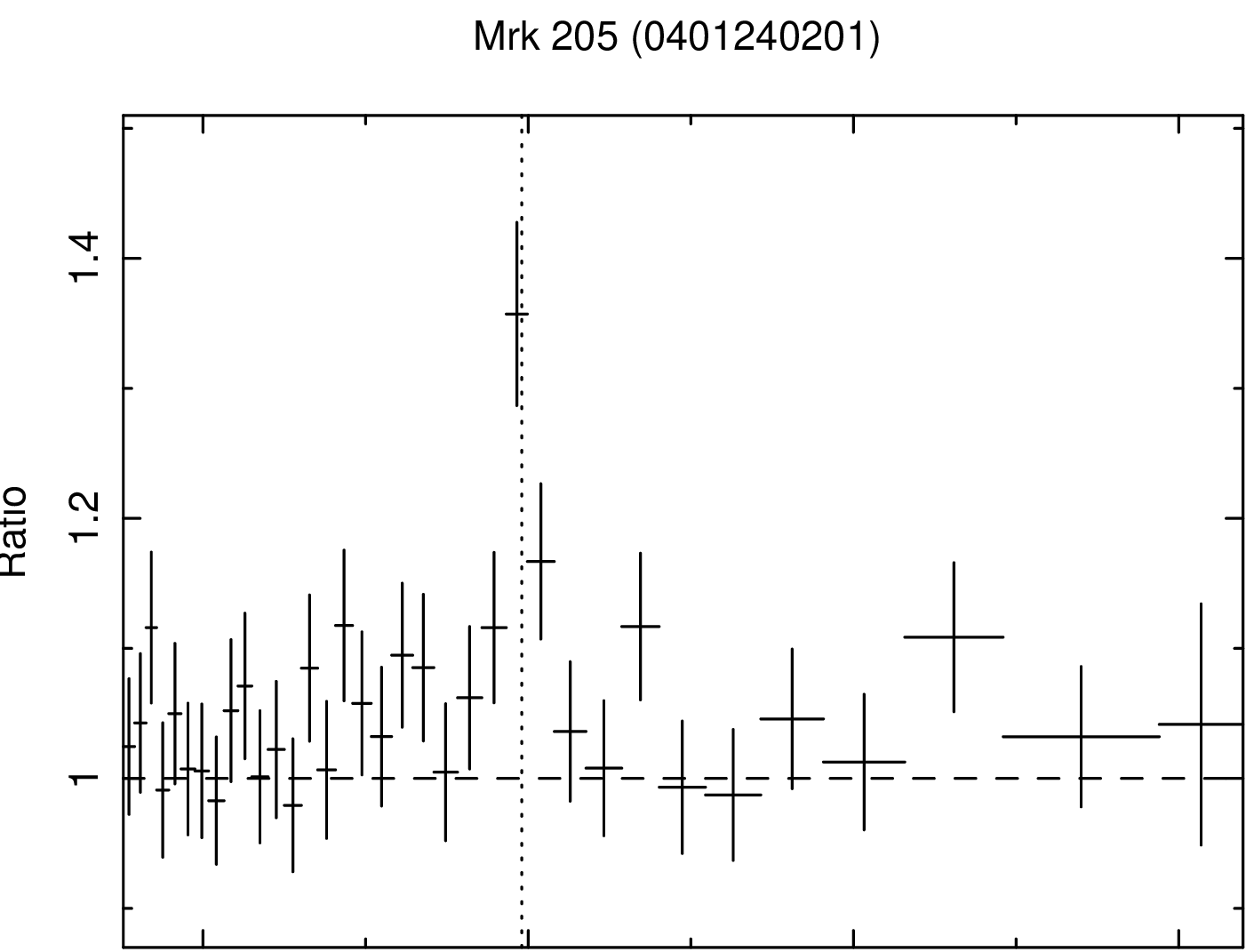}

\vspace{-0.07cm}

\hspace{-0.12cm}
    \includegraphics[width=3cm,height=4cm,angle=270]{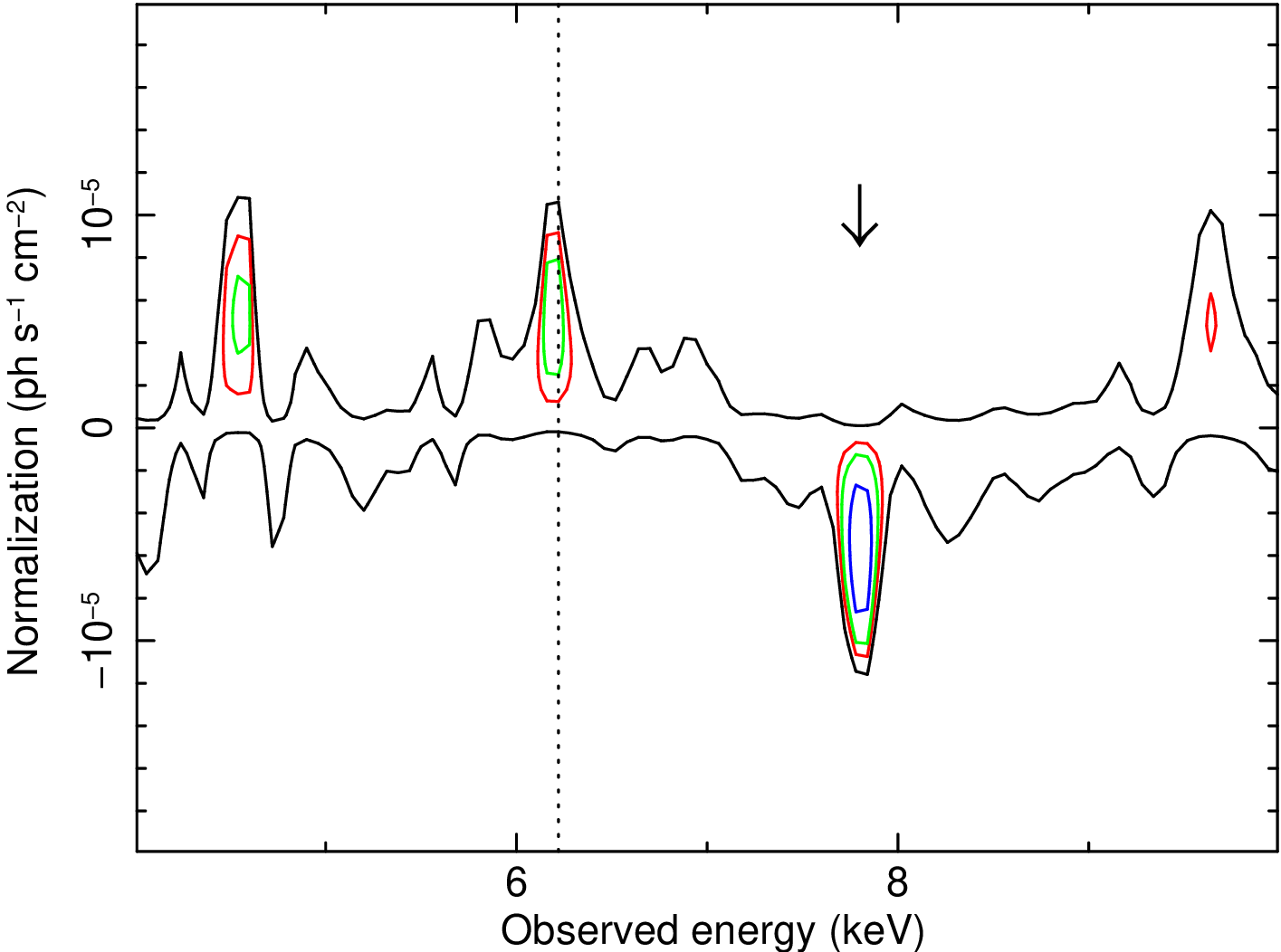}
\hspace{0.3cm}
    \includegraphics[width=3cm,height=4cm,angle=270]{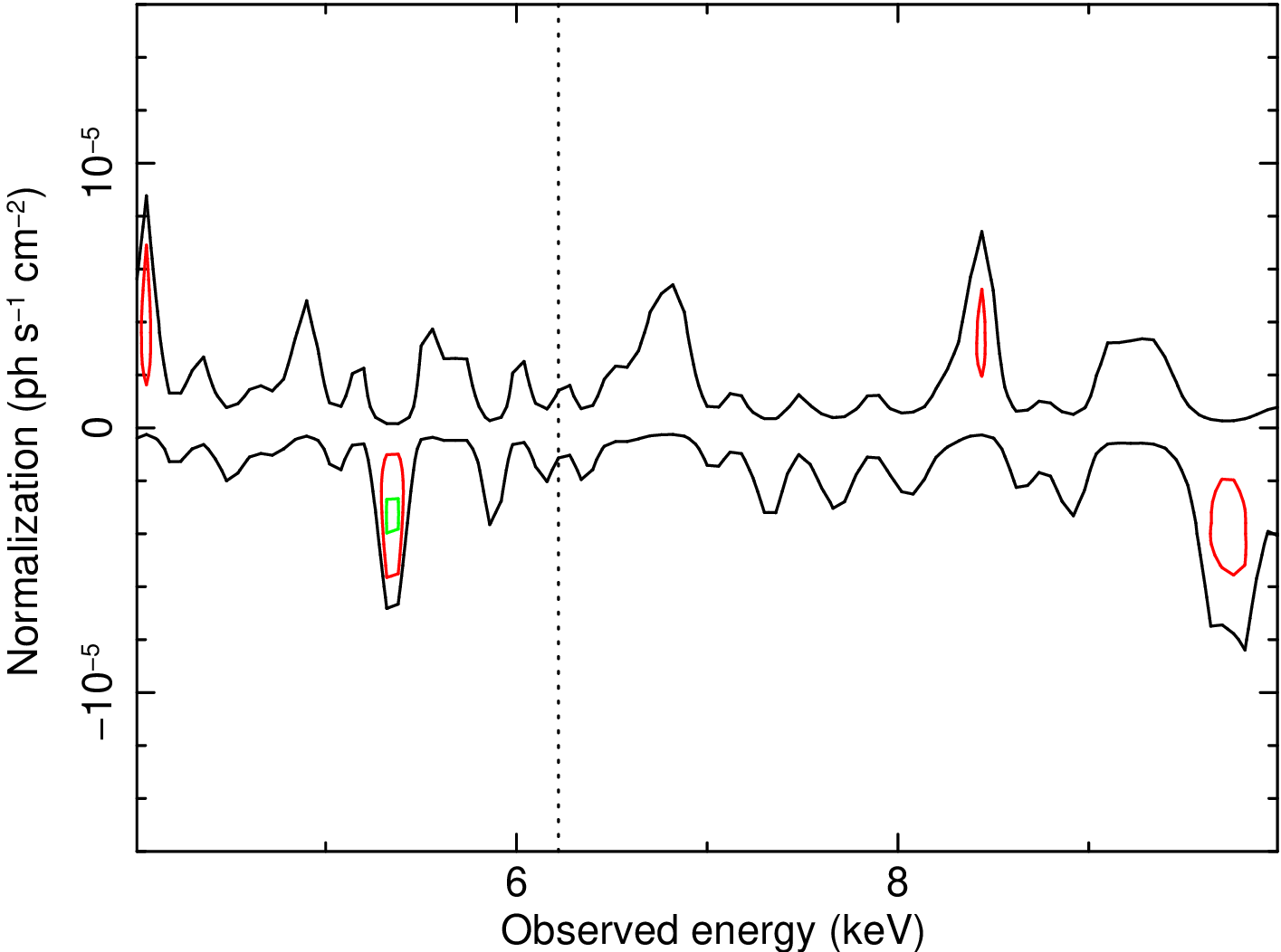}
\hspace{0.3cm}
    \includegraphics[width=3cm,height=4cm,angle=270]{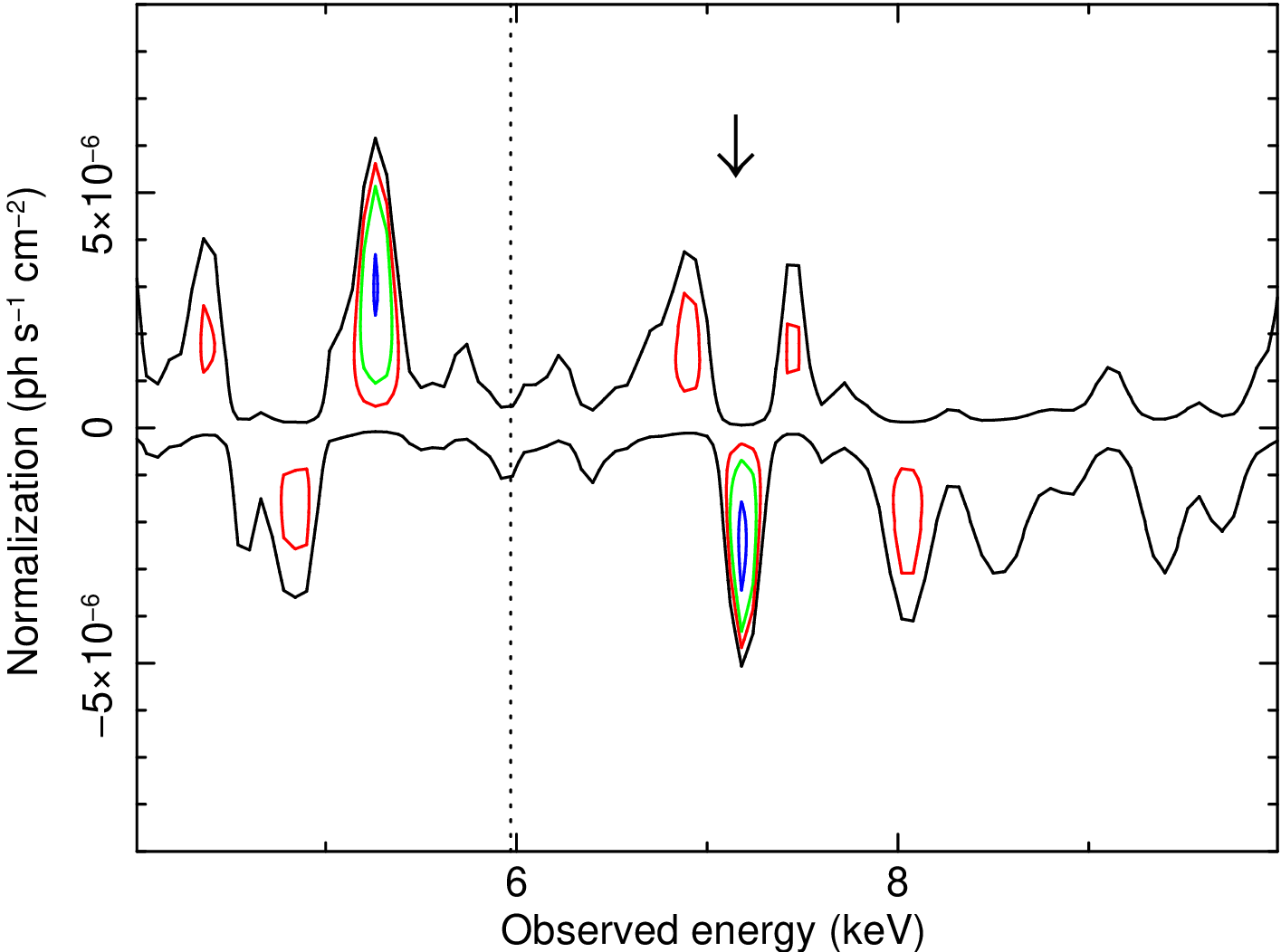}
\hspace{0.3cm}
    \includegraphics[width=3cm,height=4cm,angle=270]{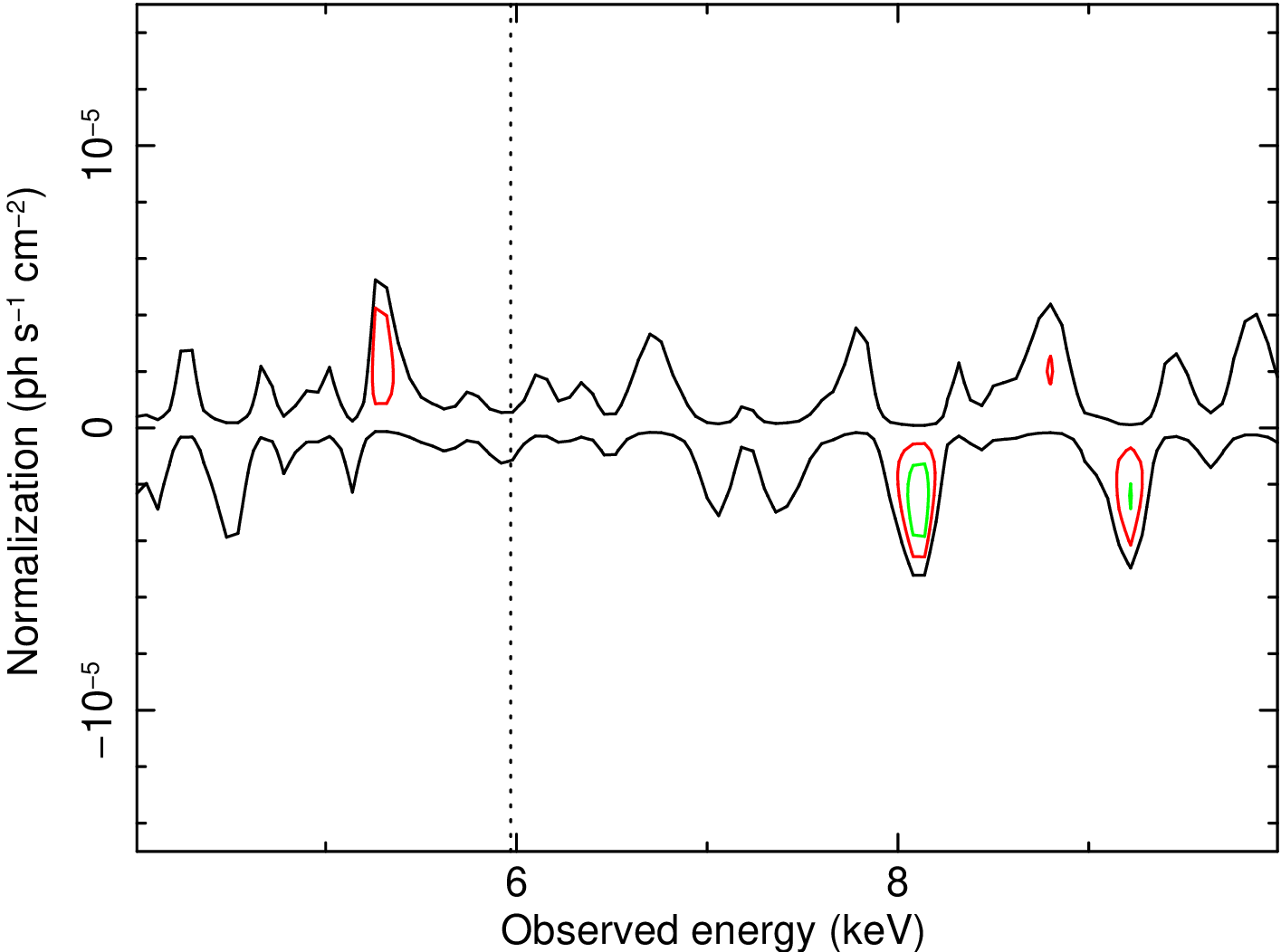}

   \caption{Ratio against the continuum (\emph{upper panel}) and contour plots with respect to the best-fit baseline model (\emph{lower panel}; 68\% (red), 90\% (green), 99\% (blue) levels) for the Type 1 sources. The Fe K absorption lines are pointed by arrows. The vertical line refer to Fe I K$\alpha$.}
    \end{figure*}

   \begin{figure*}[ht]
   \centering


    \includegraphics[width=3cm,height=4cm,angle=270]{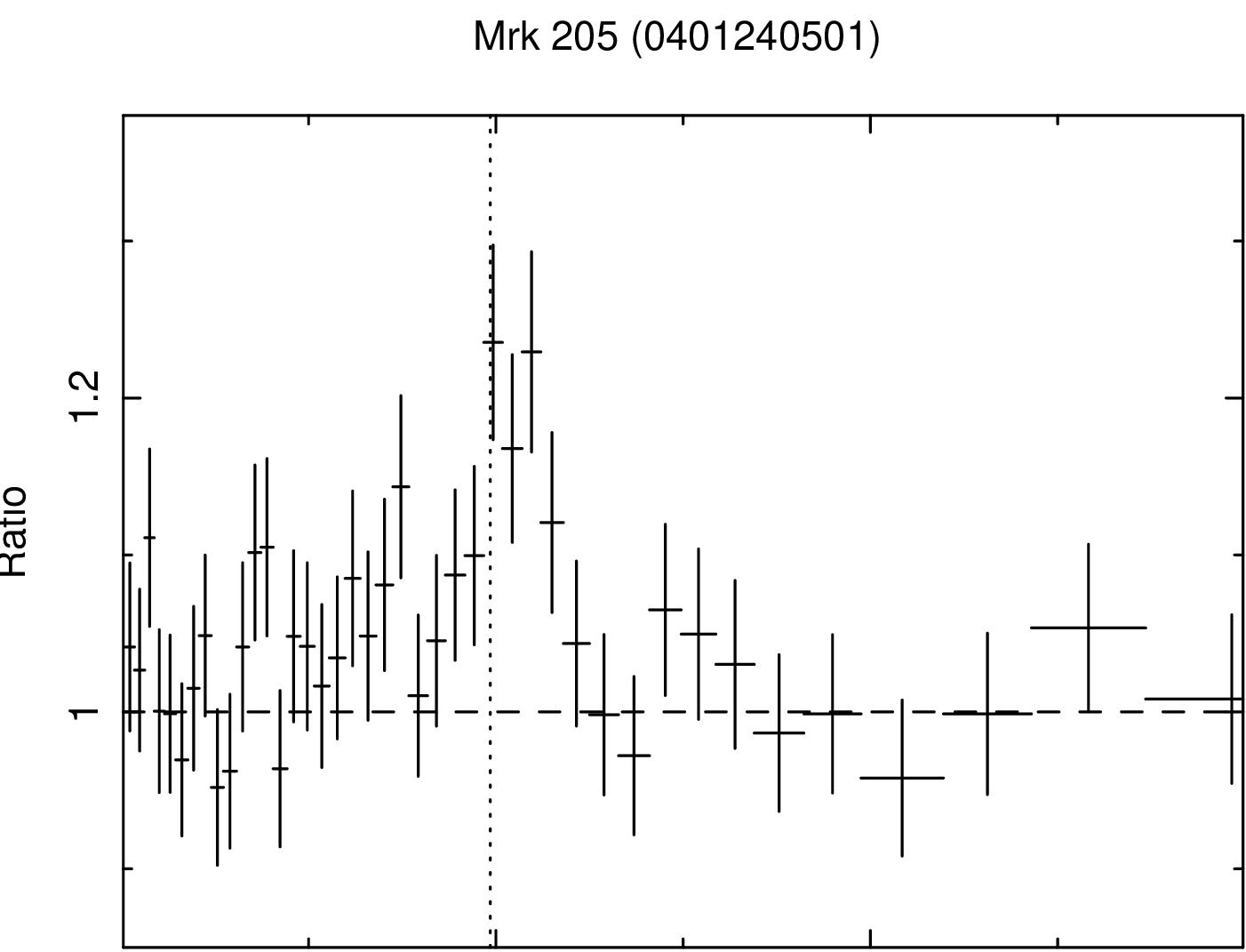}
\hspace{0.3cm}
    \includegraphics[width=3cm,height=4cm,angle=270]{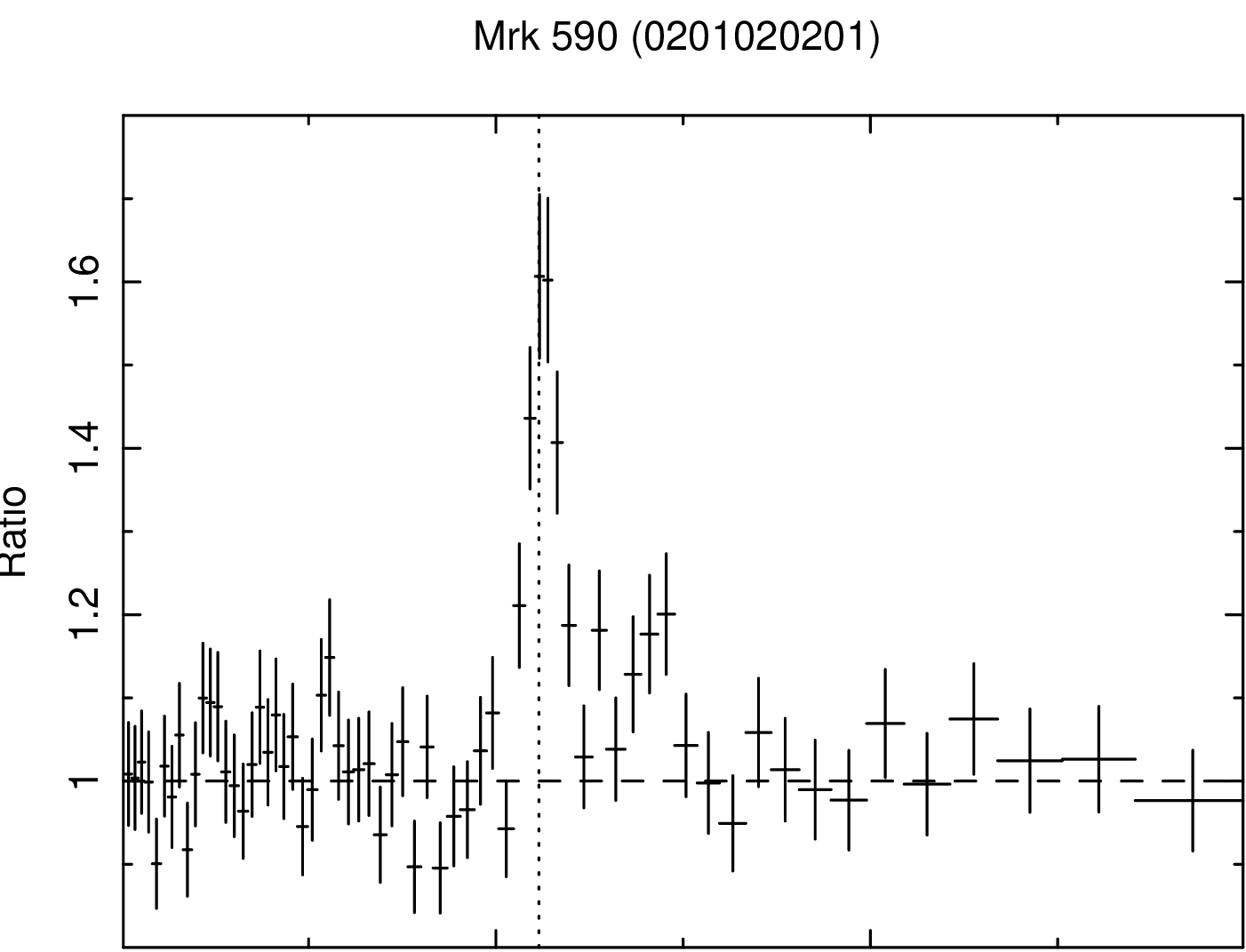}
\hspace{0.3cm}
    \includegraphics[width=3cm,height=4cm,angle=270]{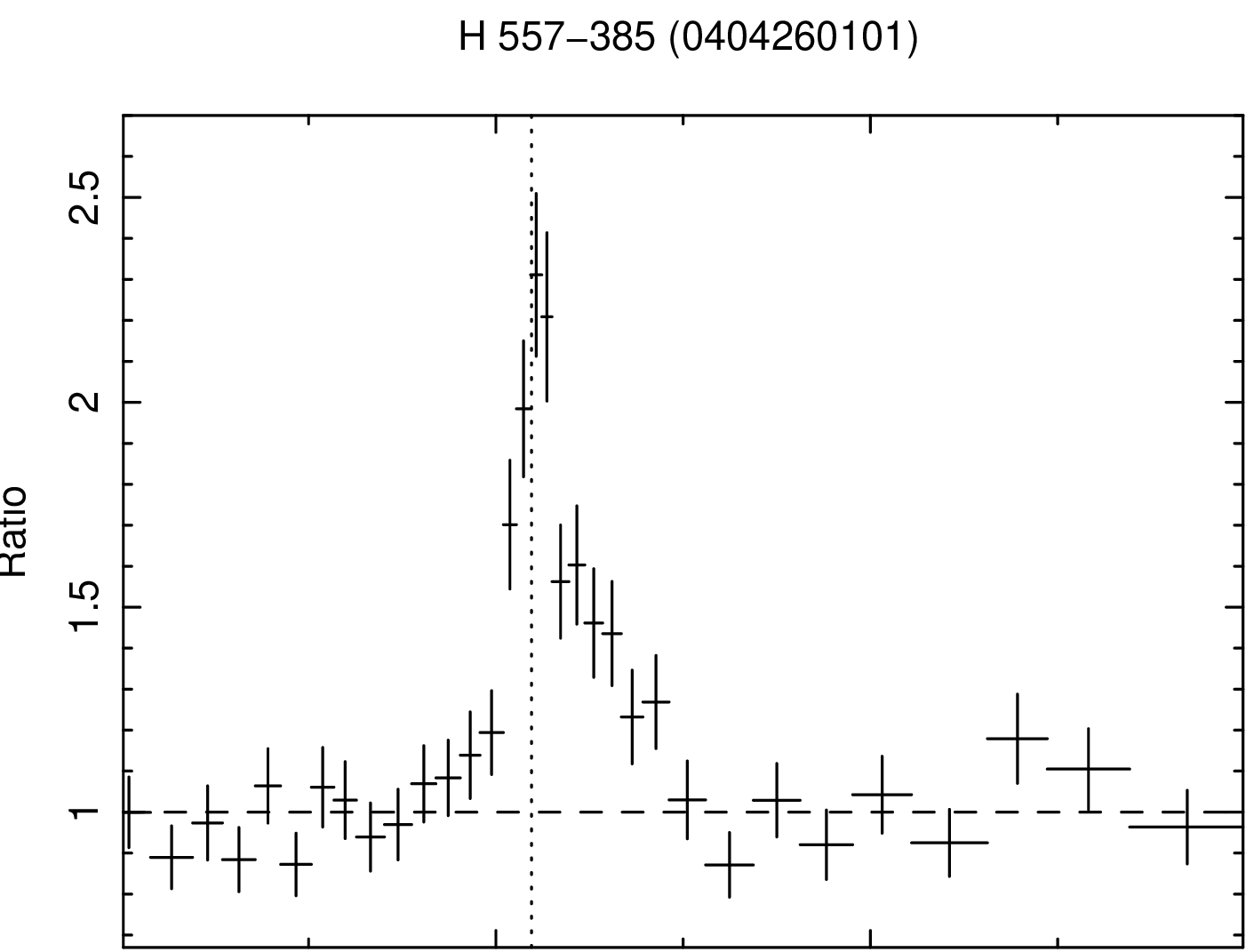}
\hspace{0.3cm}
    \includegraphics[width=3cm,height=4cm,angle=270]{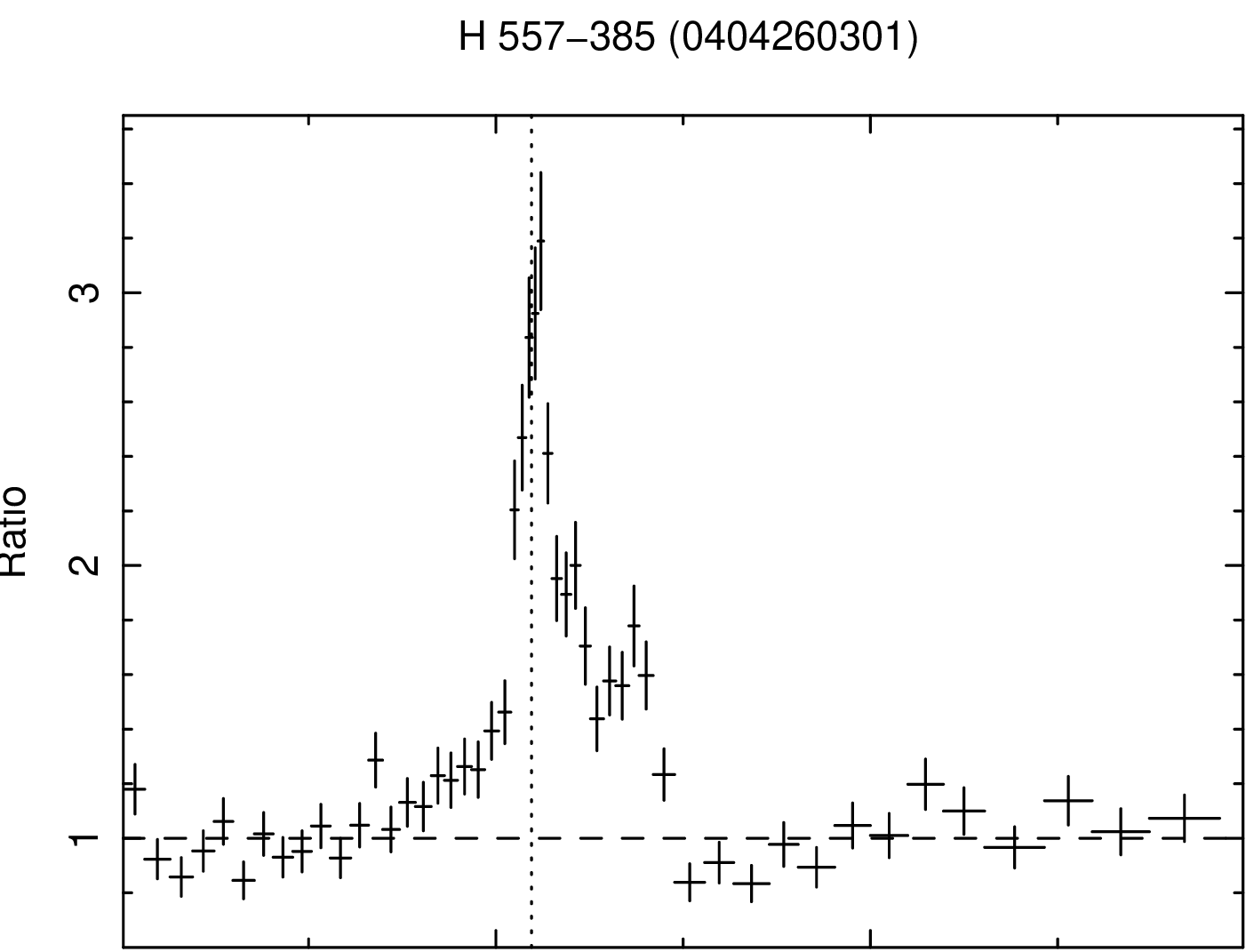}

\vspace{-0.07cm}

\hspace{-0.12cm}
    \includegraphics[width=3cm,height=4cm,angle=270]{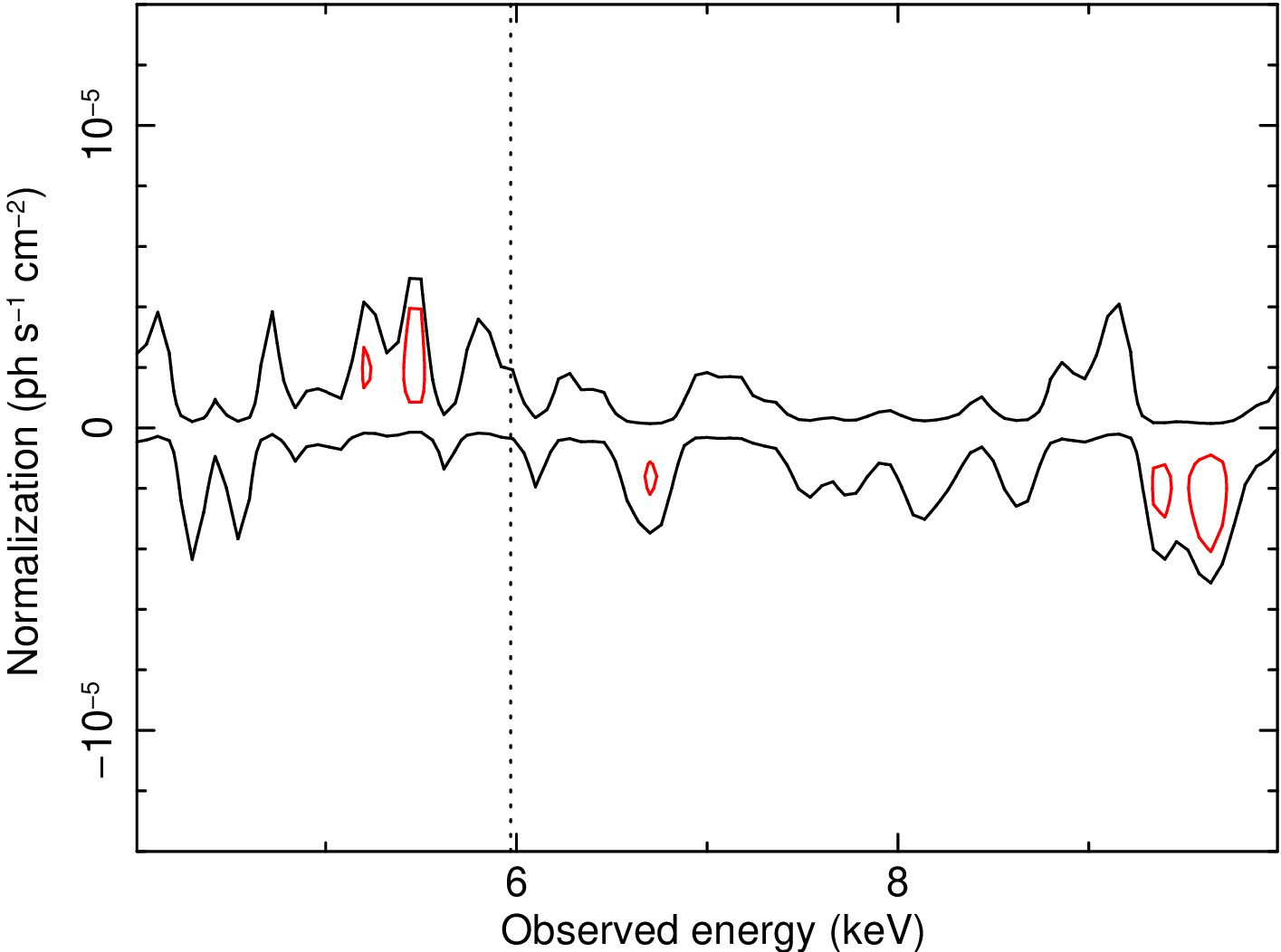}
\hspace{0.3cm}
    \includegraphics[width=3cm,height=4cm,angle=270]{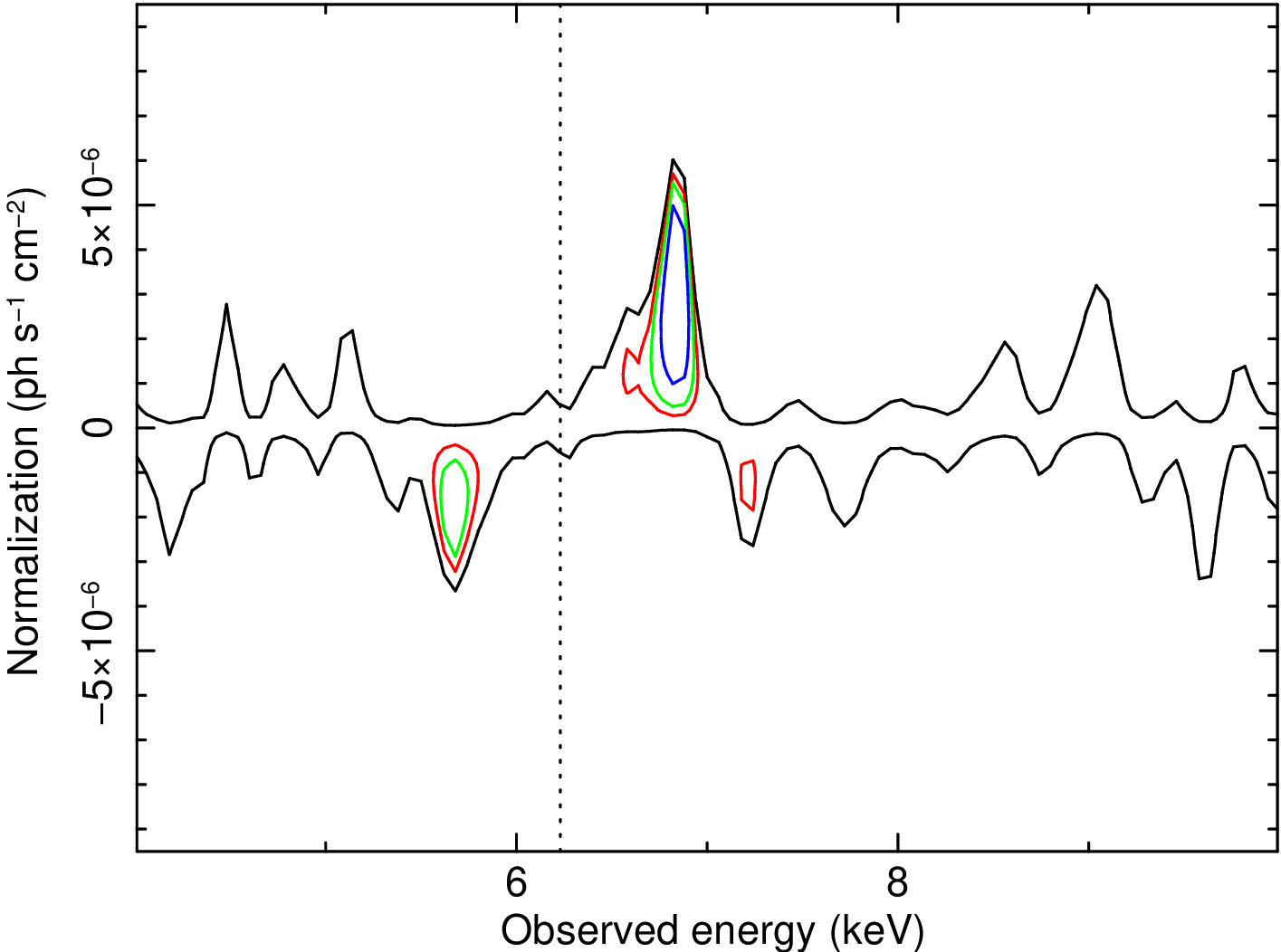}
\hspace{0.3cm}
    \includegraphics[width=3cm,height=4cm,angle=270]{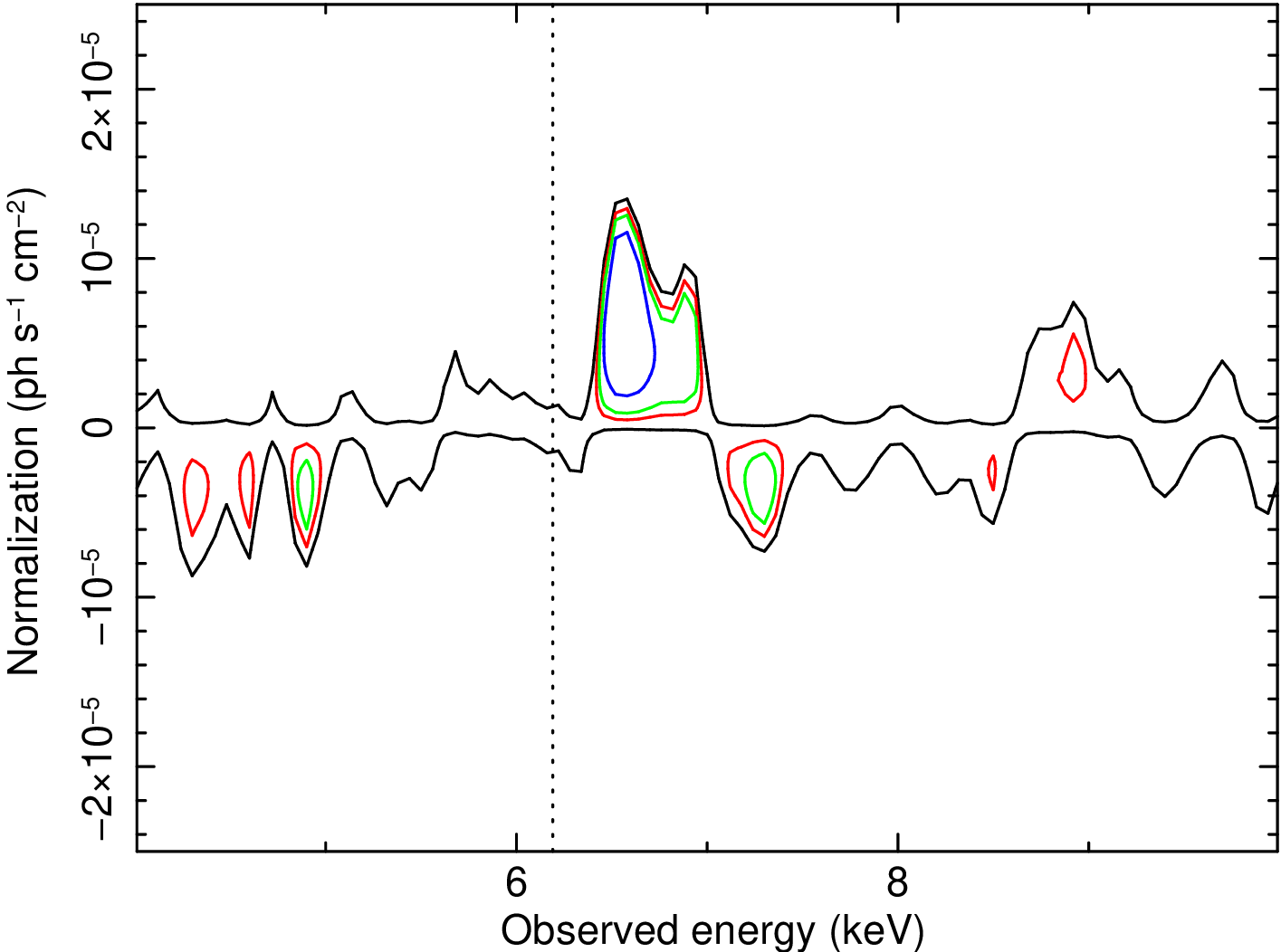}
\hspace{0.3cm}
    \includegraphics[width=3cm,height=4cm,angle=270]{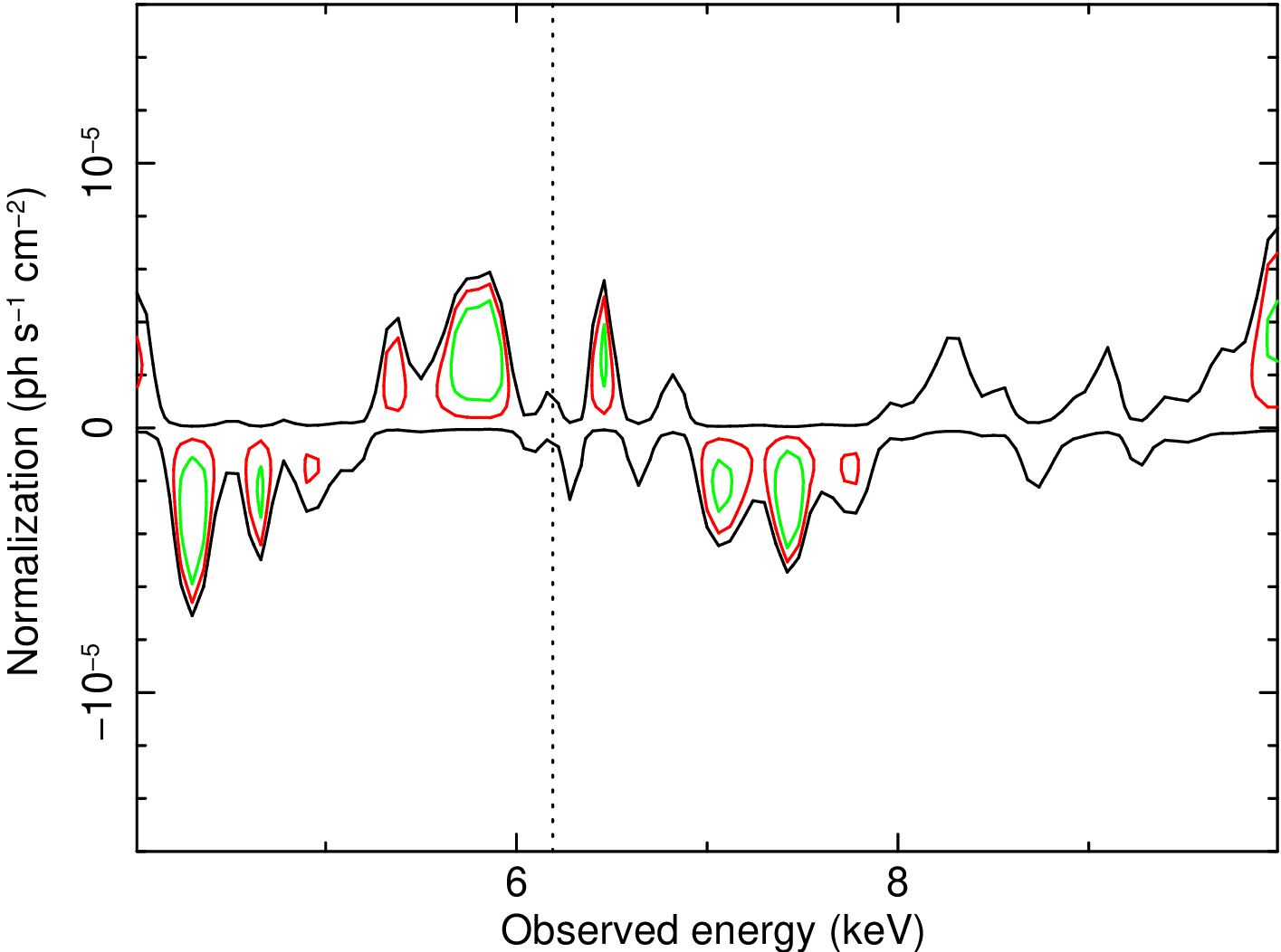}


\vspace{0.2cm}

    \includegraphics[width=3cm,height=4cm,angle=270]{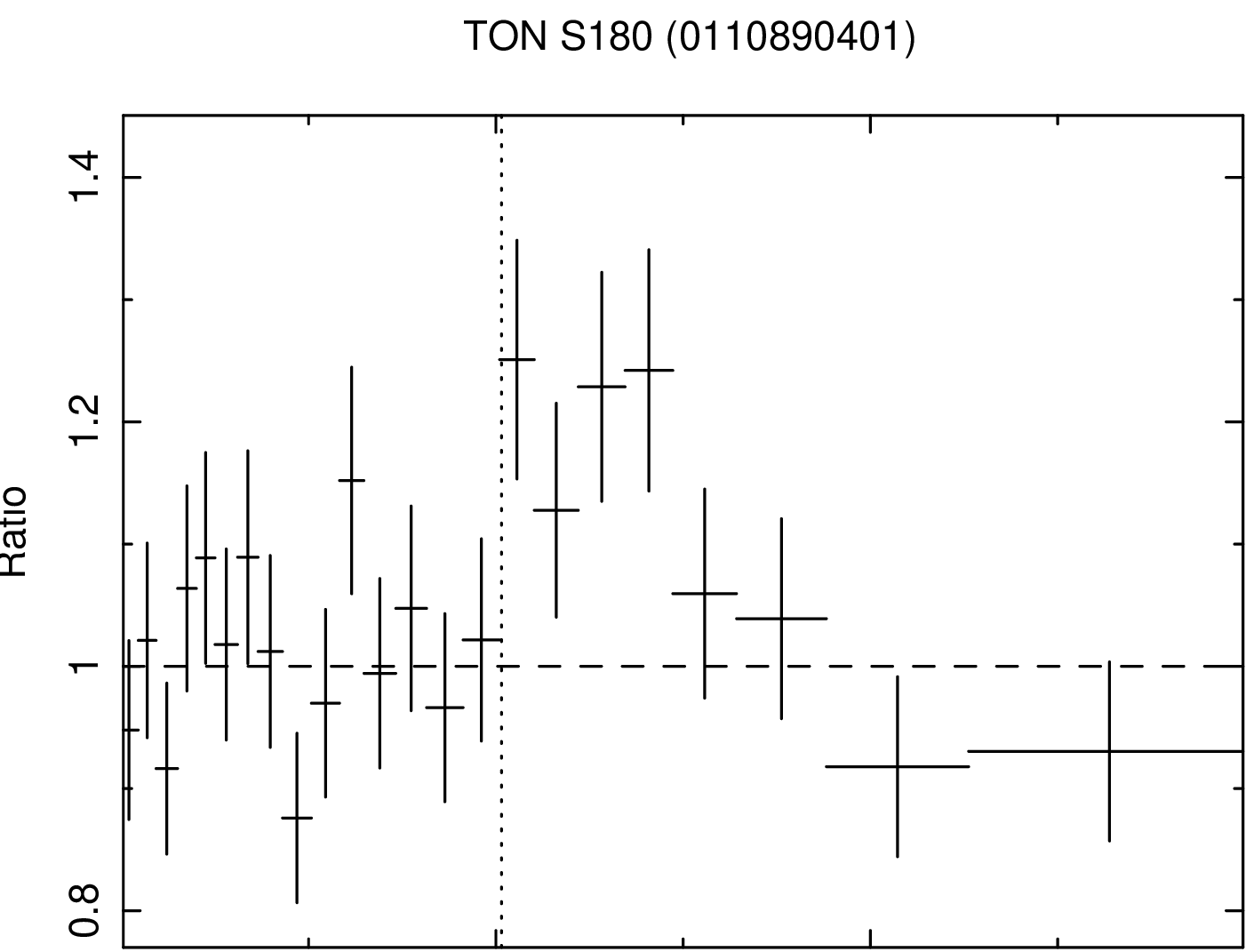}
\hspace{0.3cm}
    \includegraphics[width=3cm,height=4cm,angle=270]{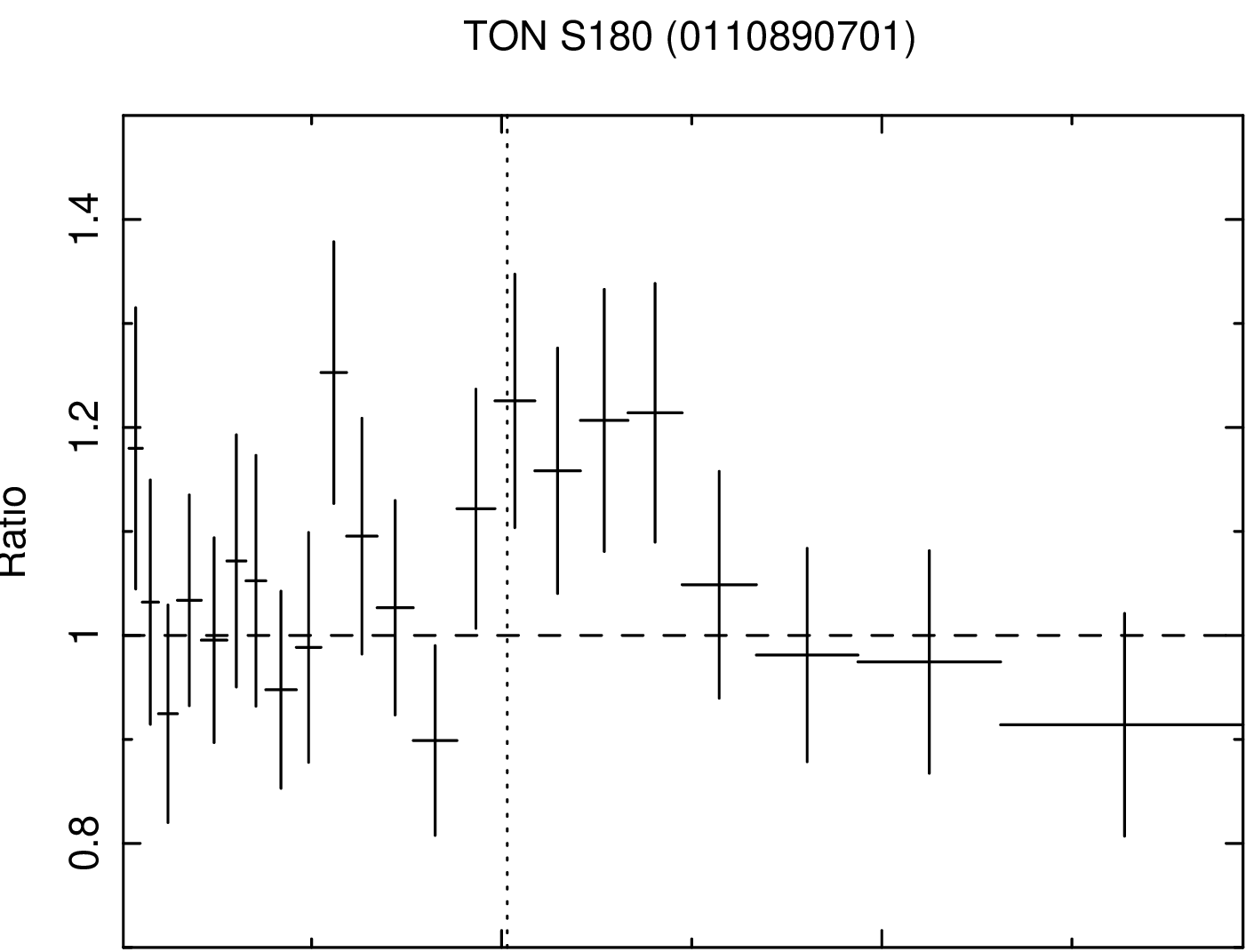}
\hspace{0.3cm}
    \includegraphics[width=3cm,height=4cm,angle=270]{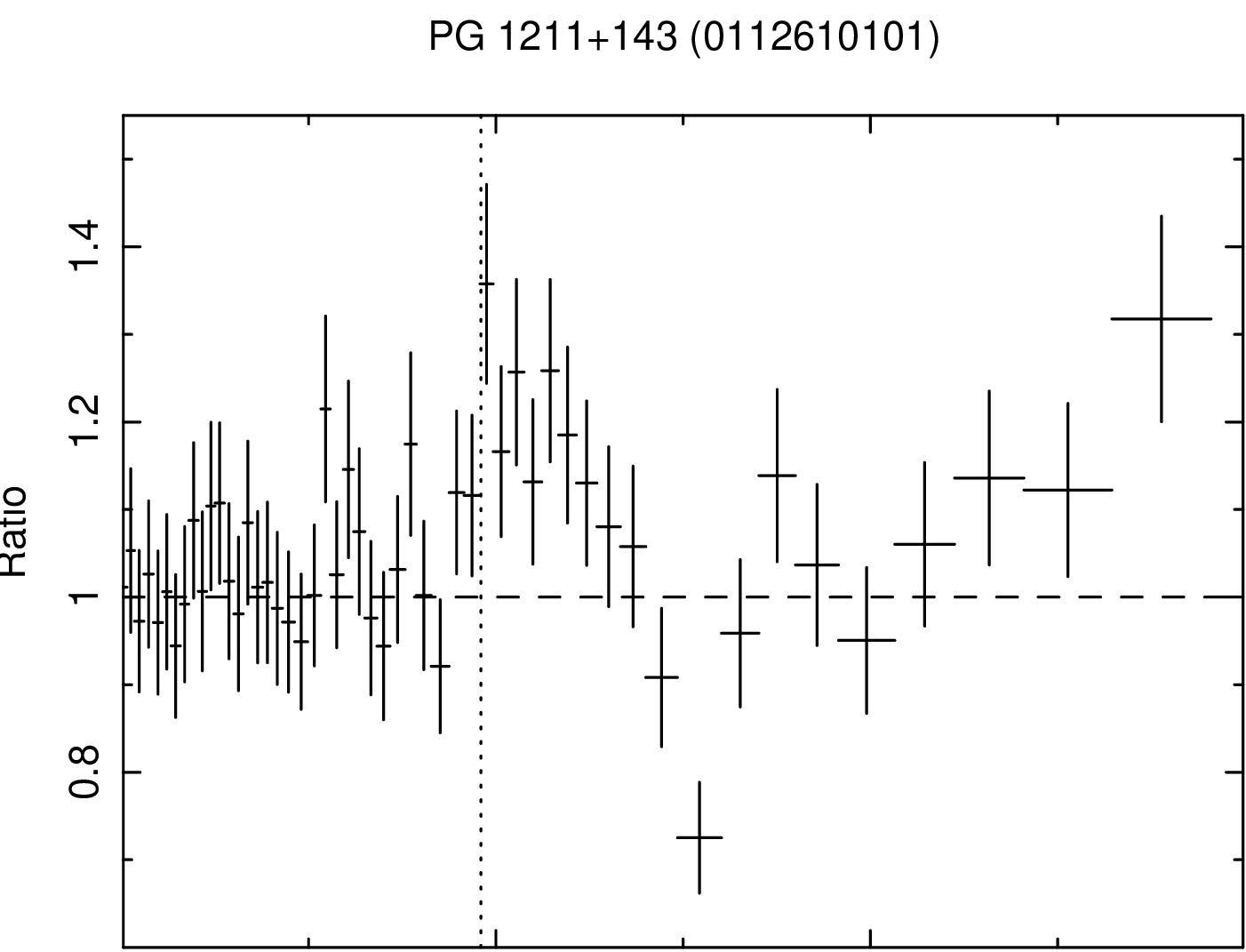}
\hspace{0.3cm}
    \includegraphics[width=3cm,height=4cm,angle=270]{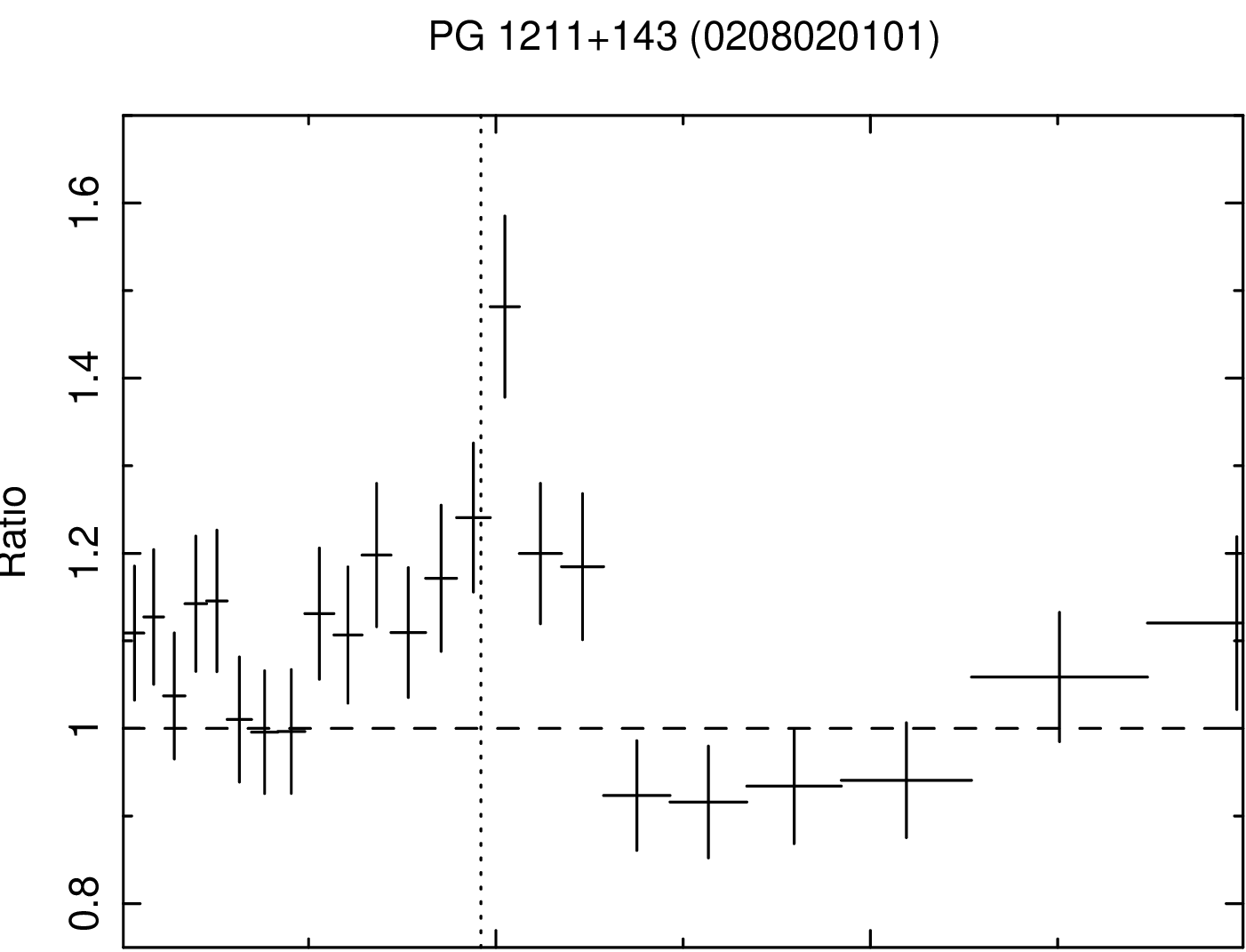}

\vspace{-0.07cm}

\hspace{-0.12cm}
    \includegraphics[width=3cm,height=4cm,angle=270]{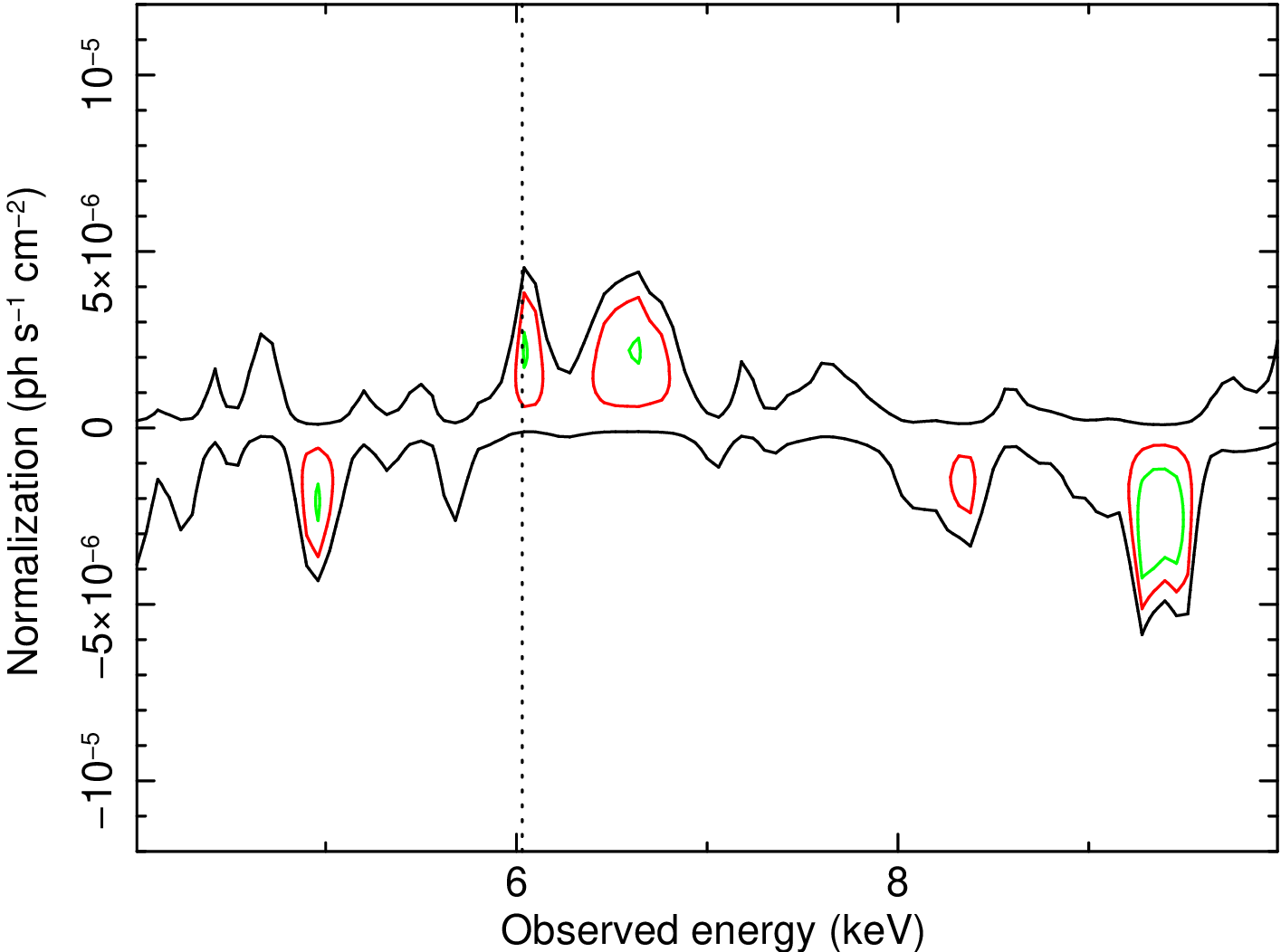}
\hspace{0.3cm}
    \includegraphics[width=3cm,height=4cm,angle=270]{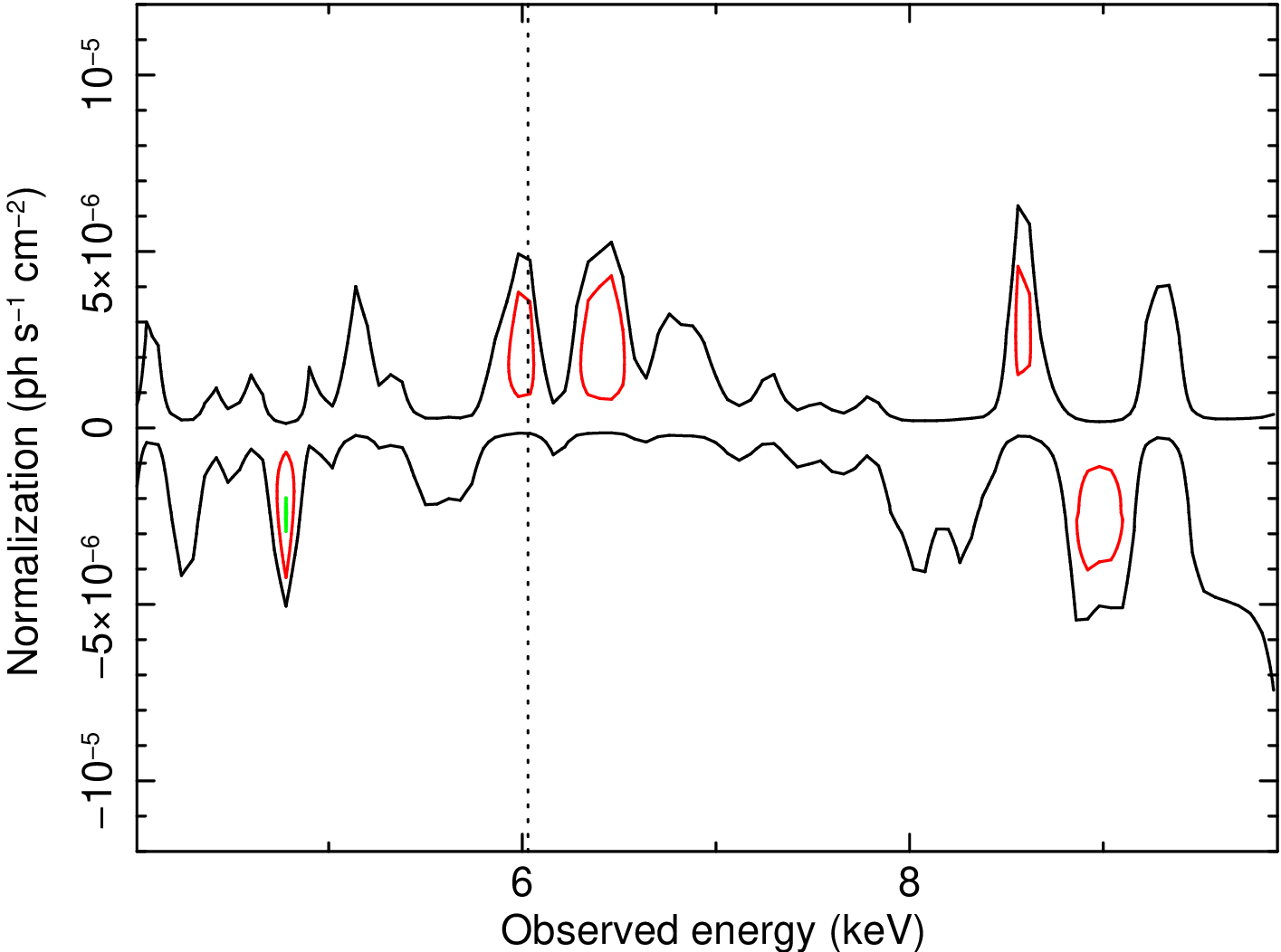}
\hspace{0.3cm}
    \includegraphics[width=3cm,height=4cm,angle=270]{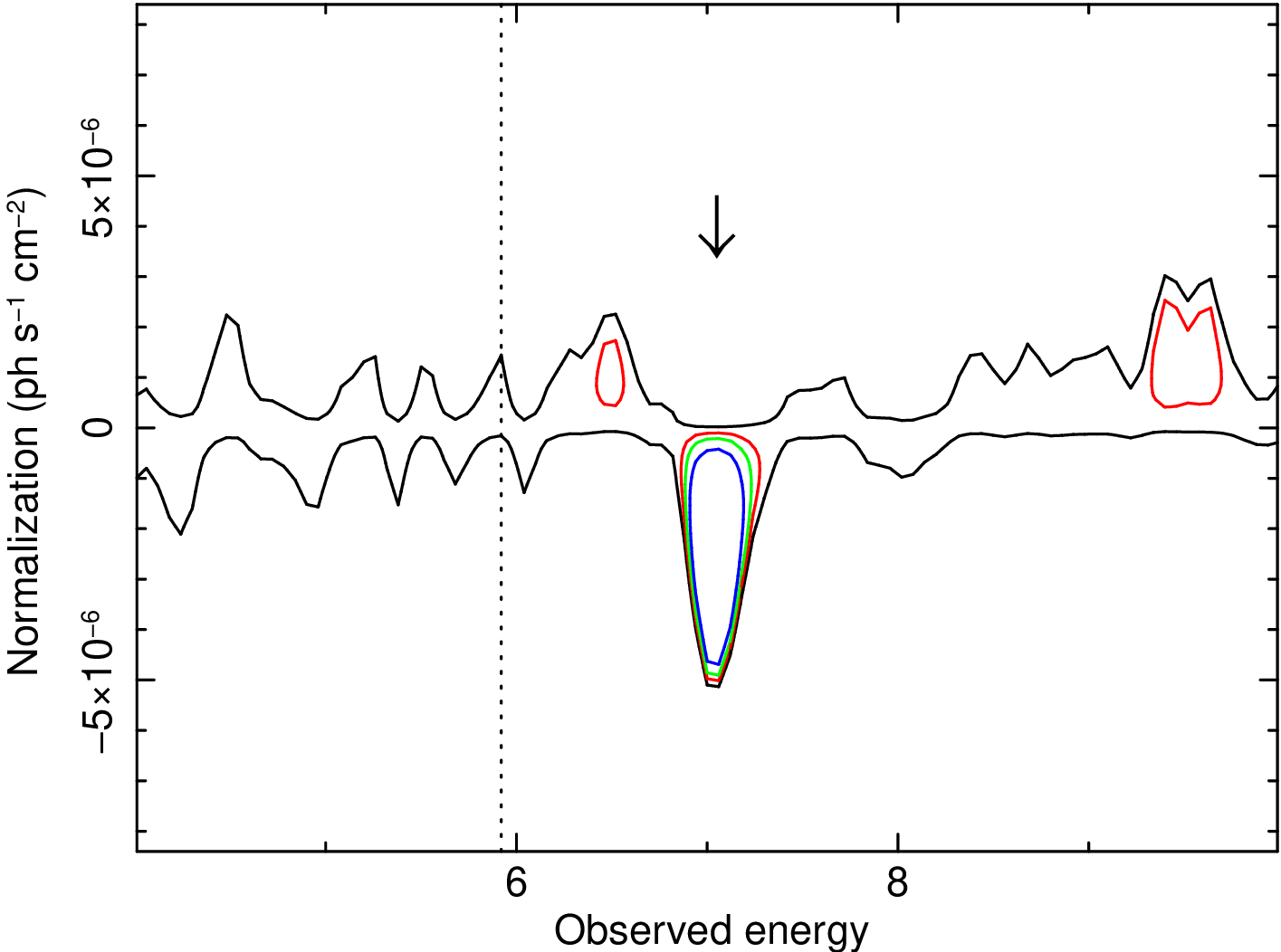}
\hspace{0.3cm}
    \includegraphics[width=3cm,height=4cm,angle=270]{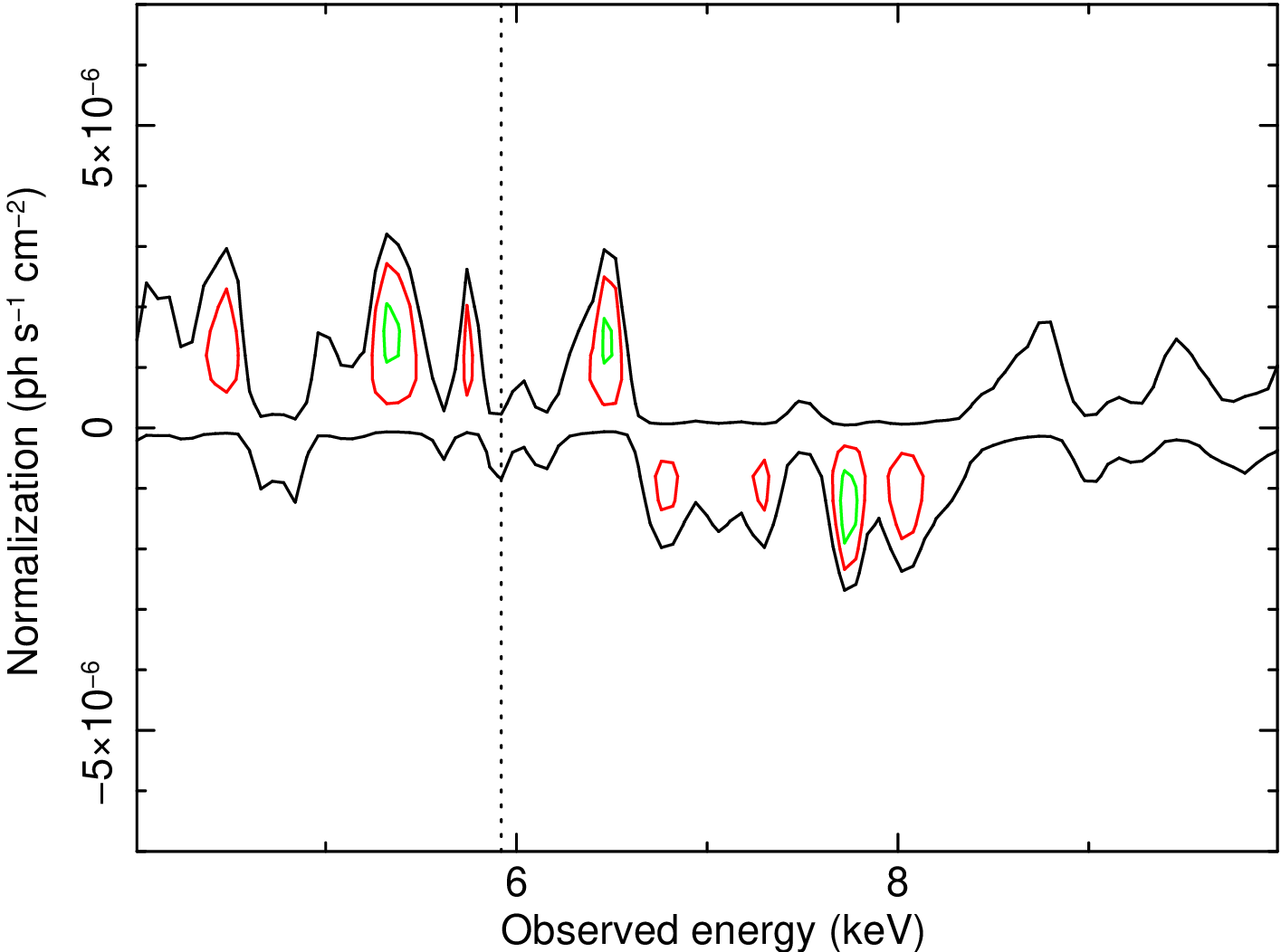}


\vspace{0.2cm}

    \includegraphics[width=3cm,height=4cm,angle=270]{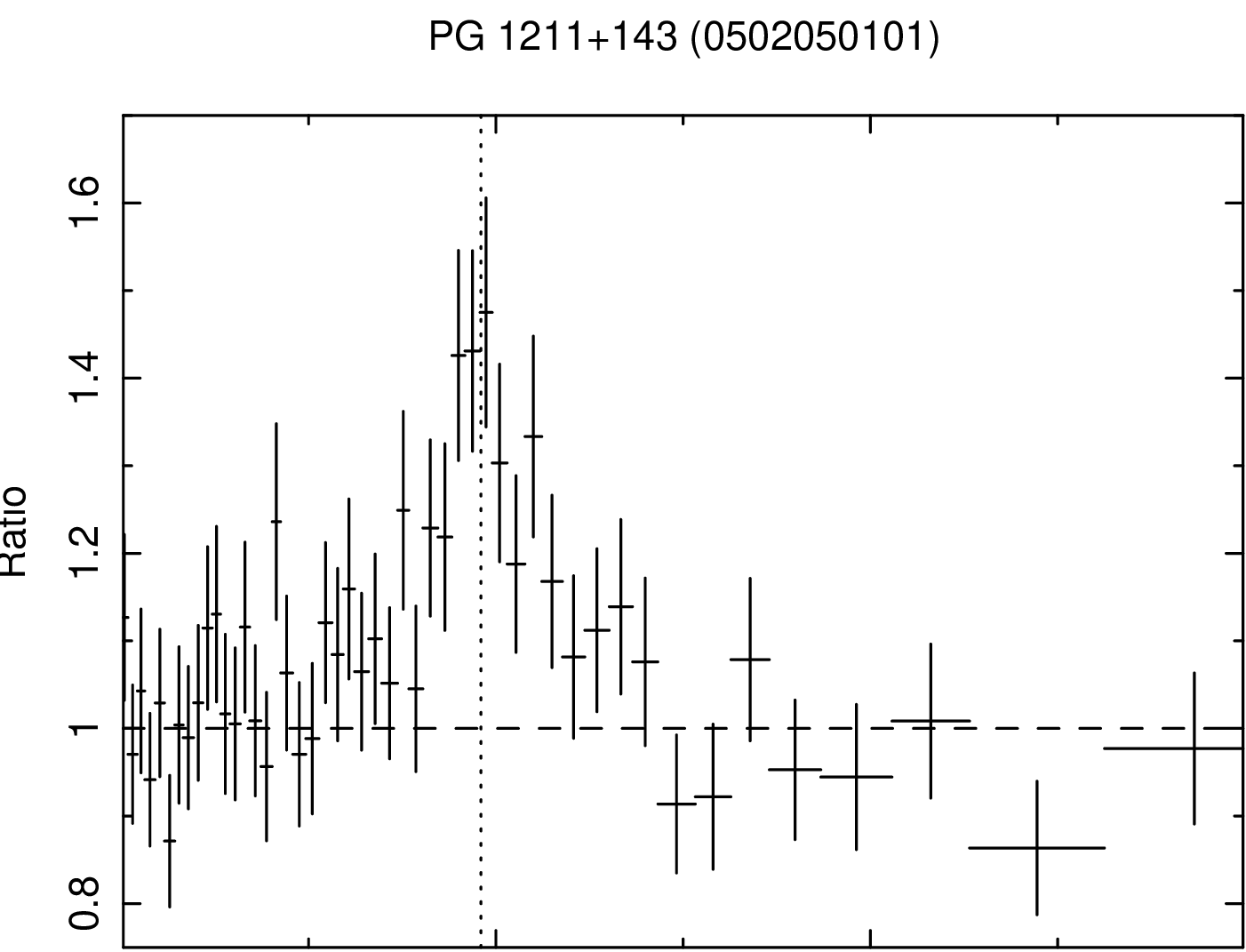}
\hspace{0.3cm}
    \includegraphics[width=3cm,height=4cm,angle=270]{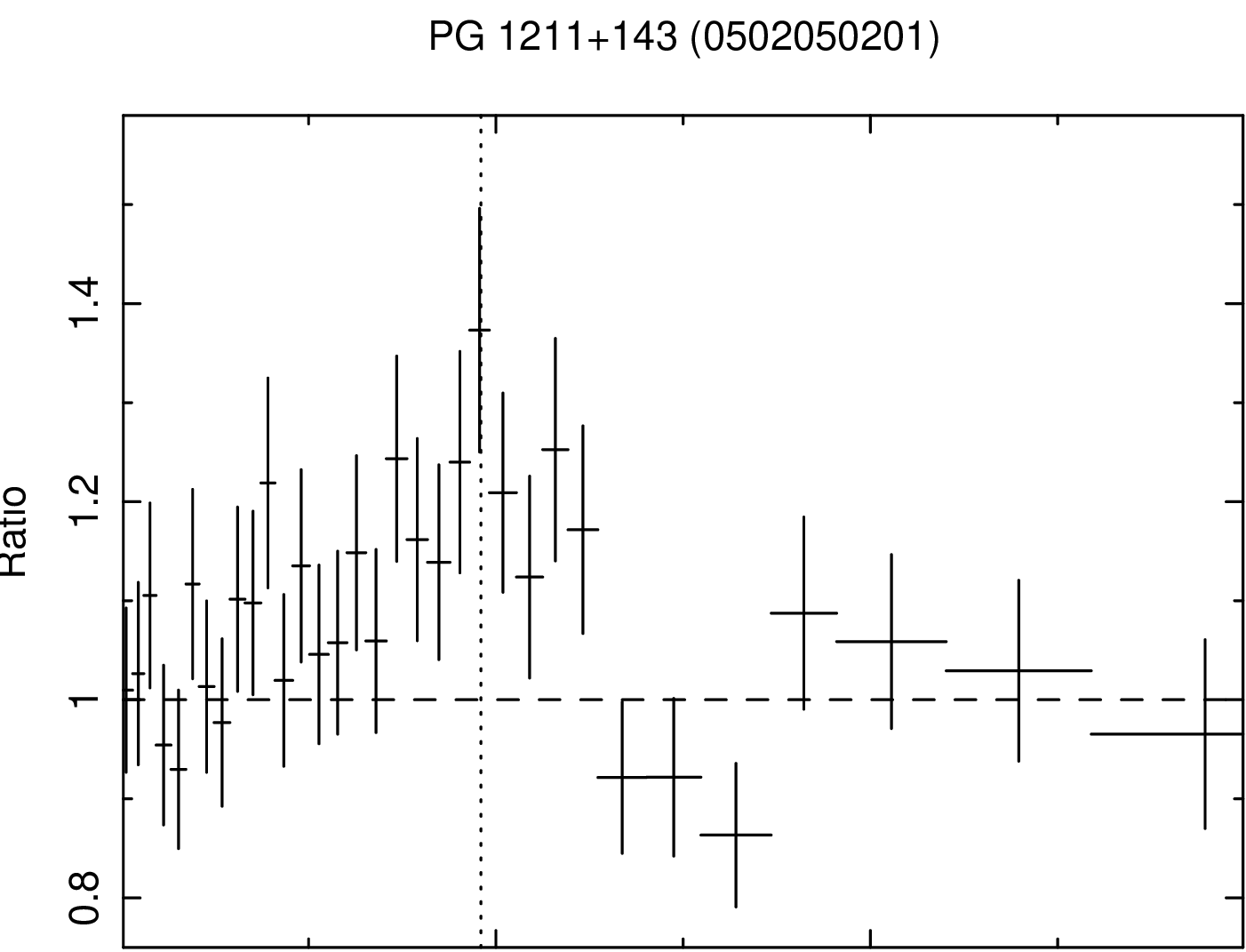}
\hspace{0.3cm}
    \includegraphics[width=3cm,height=4cm,angle=270]{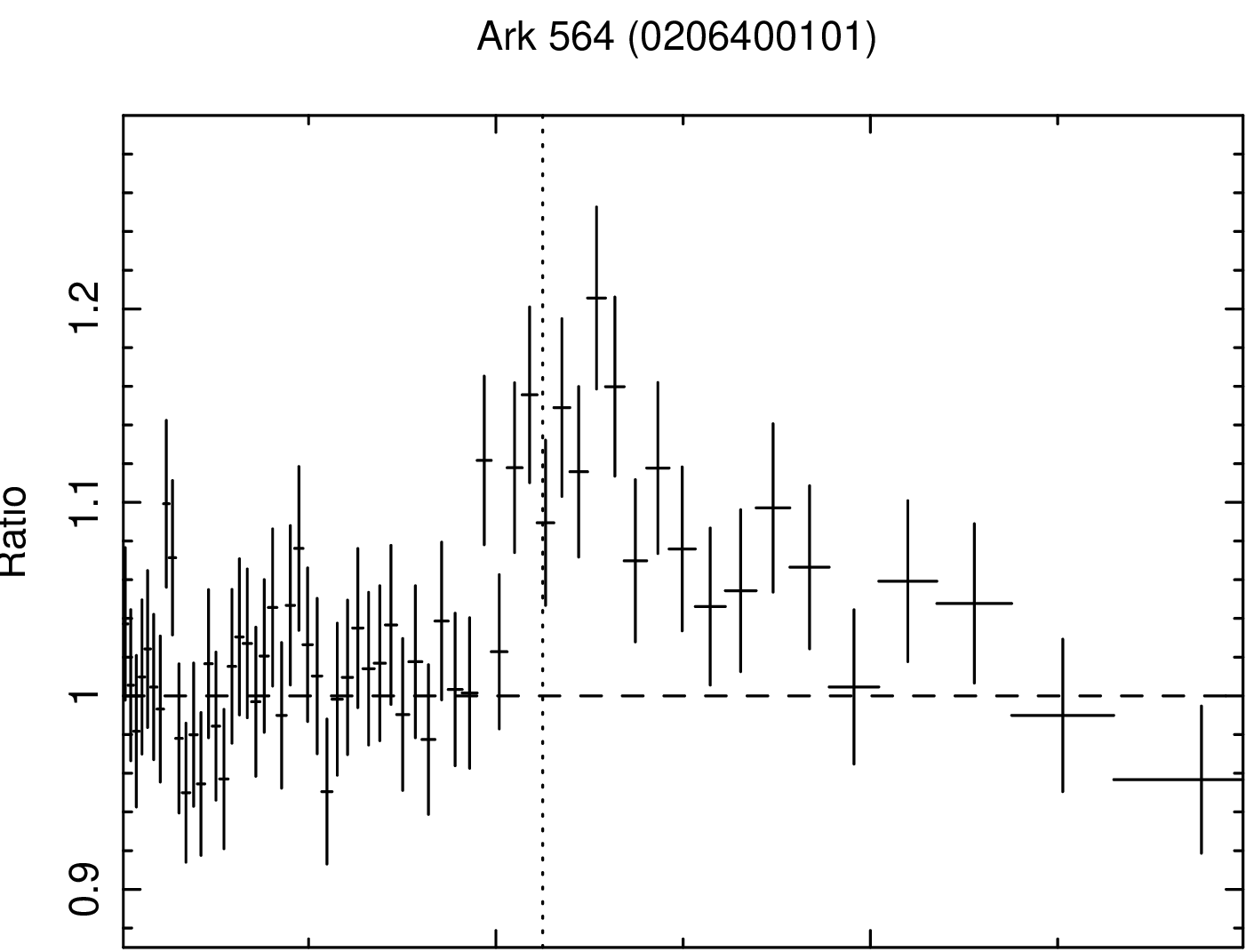}

\vspace{-0.07cm}

\hspace{-0.12cm}
    \includegraphics[width=3cm,height=4cm,angle=270]{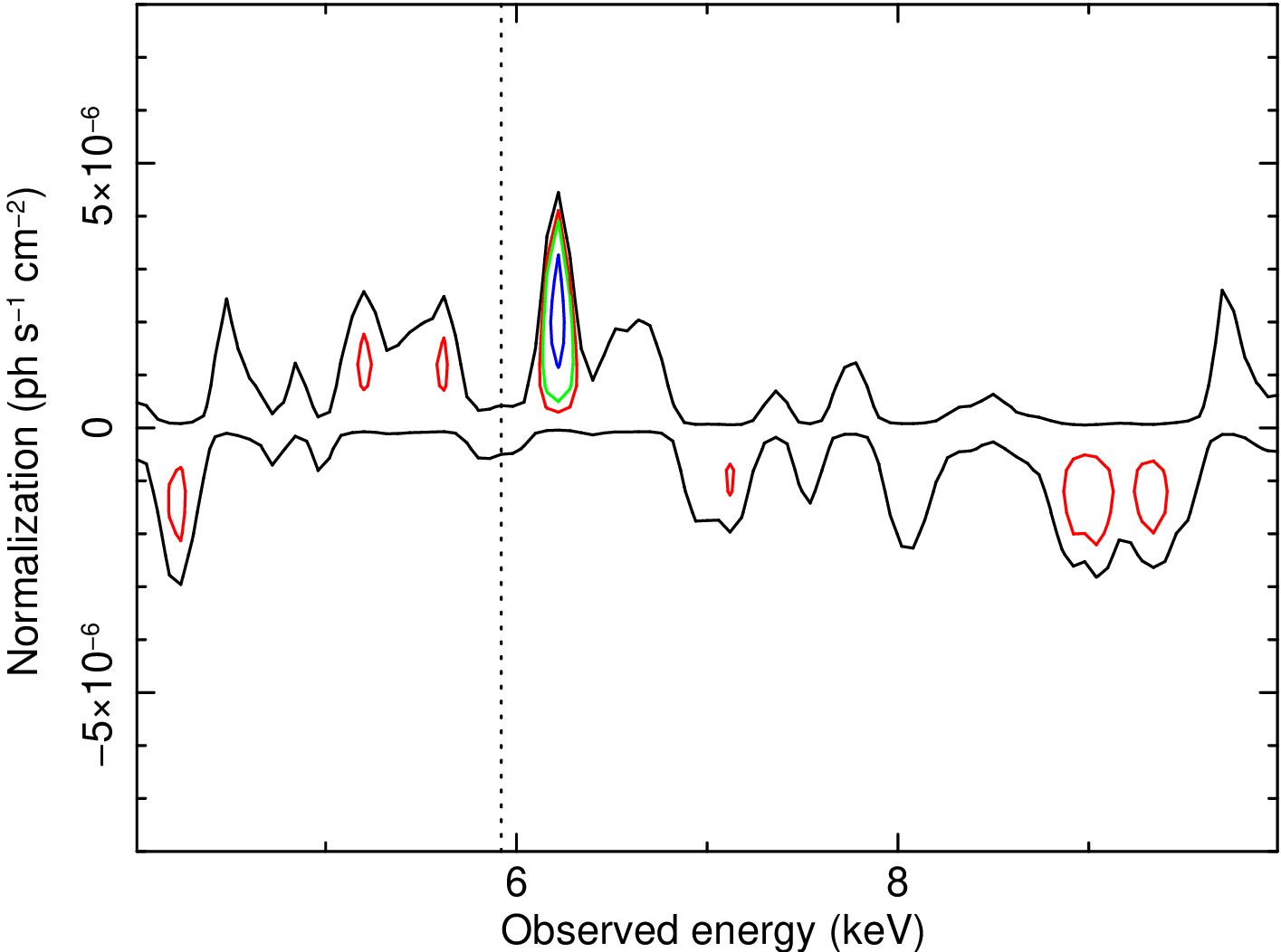}
\hspace{0.3cm}
    \includegraphics[width=3cm,height=4cm,angle=270]{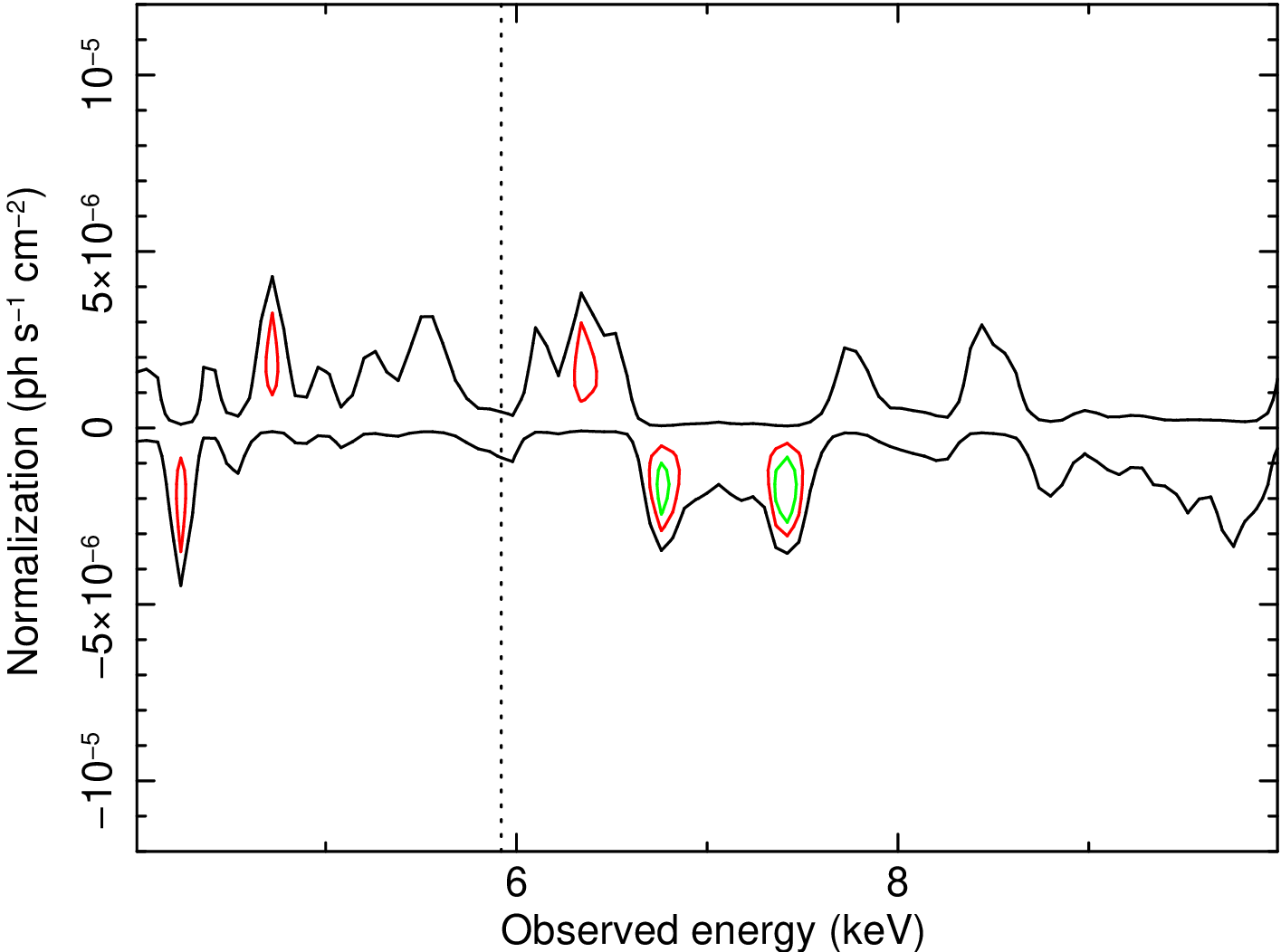}
\hspace{0.3cm}
\includegraphics[width=3cm,height=4cm,angle=270]{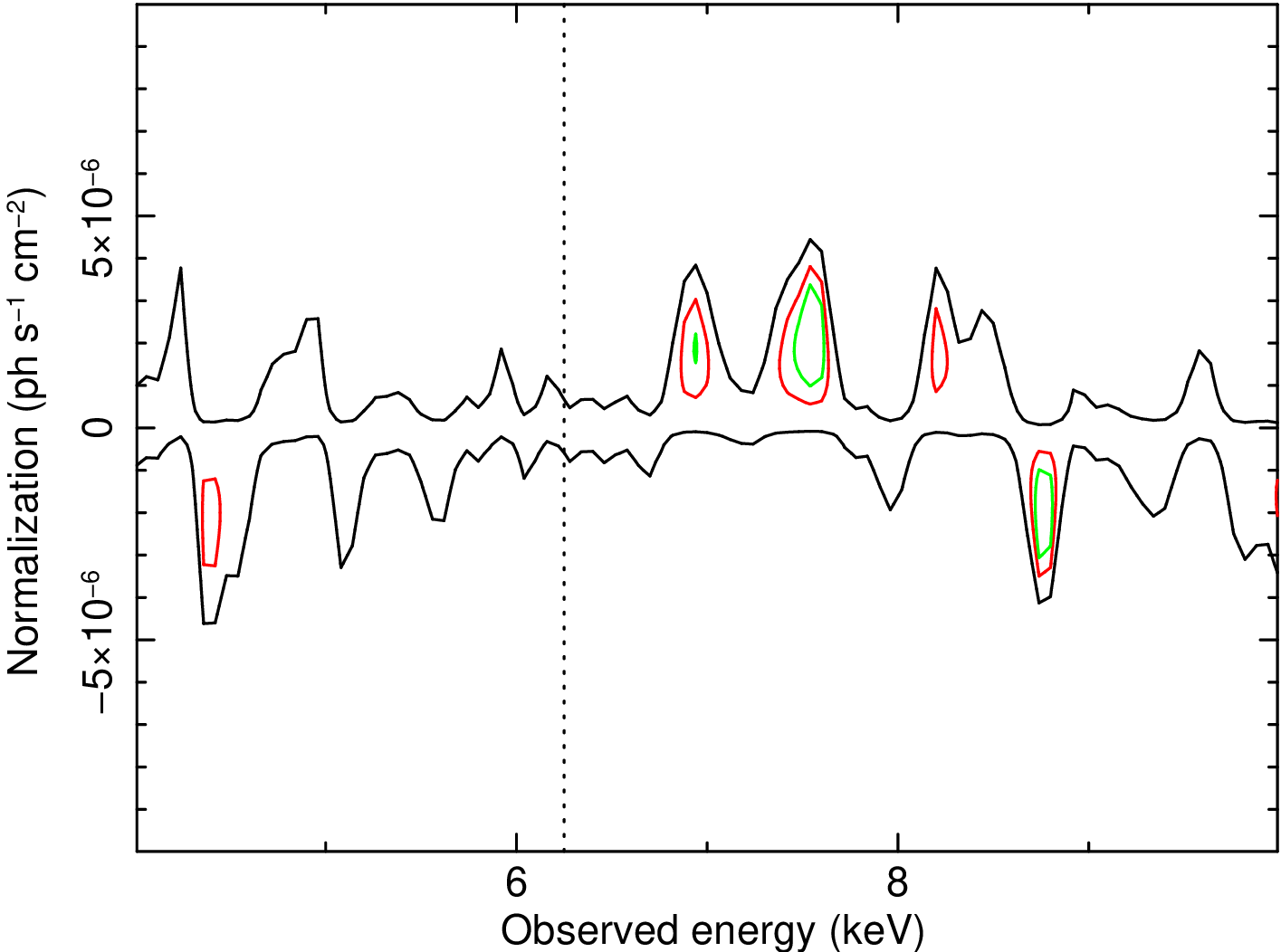}

   \caption{Ratio against the continuum (\emph{upper panel}) and contour plots with respect to the best-fit baseline model (\emph{lower panel}; 68\% (red), 90\% (green), 99\% (blue) levels) for the Type 1 sources. The Fe K absorption lines are pointed by arrows. The vertical line refer to Fe I K$\alpha$.}
    \end{figure*}

   \begin{figure*}[ht]
   \centering


    \includegraphics[width=3cm,height=4cm,angle=270]{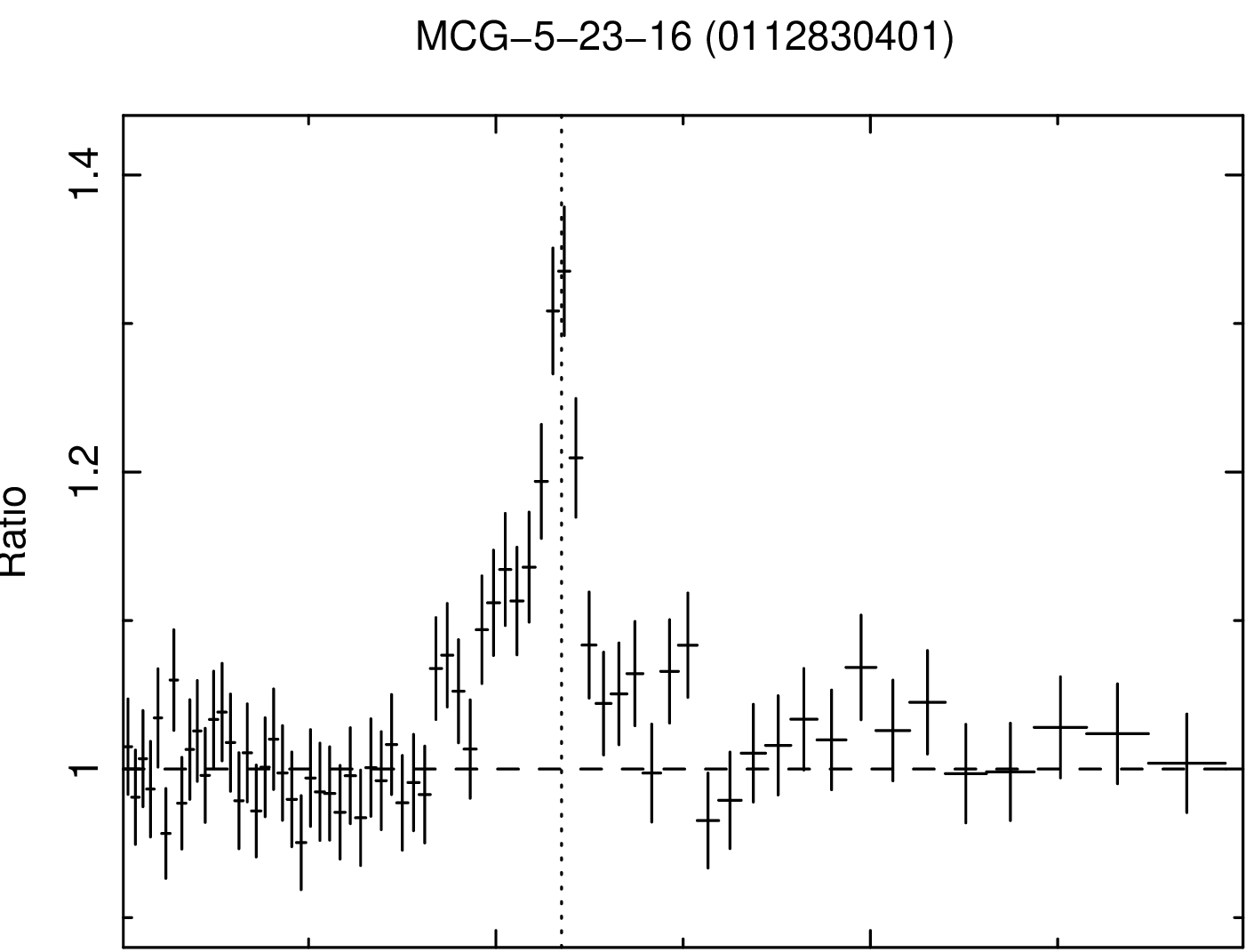}
\hspace{0.3cm}
    \includegraphics[width=3cm,height=4cm,angle=270]{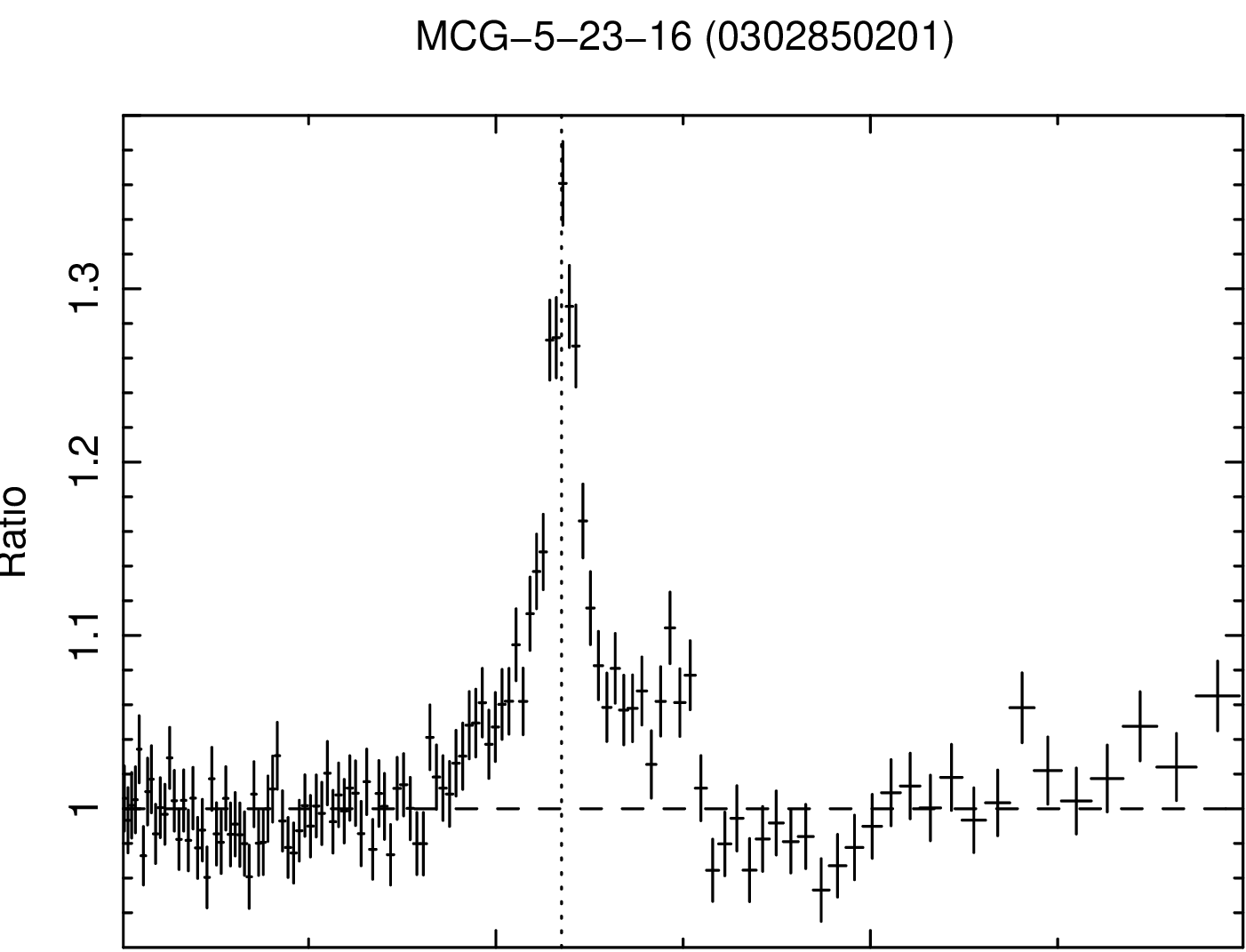}
\hspace{0.3cm}
    \includegraphics[width=3cm,height=4cm,angle=270]{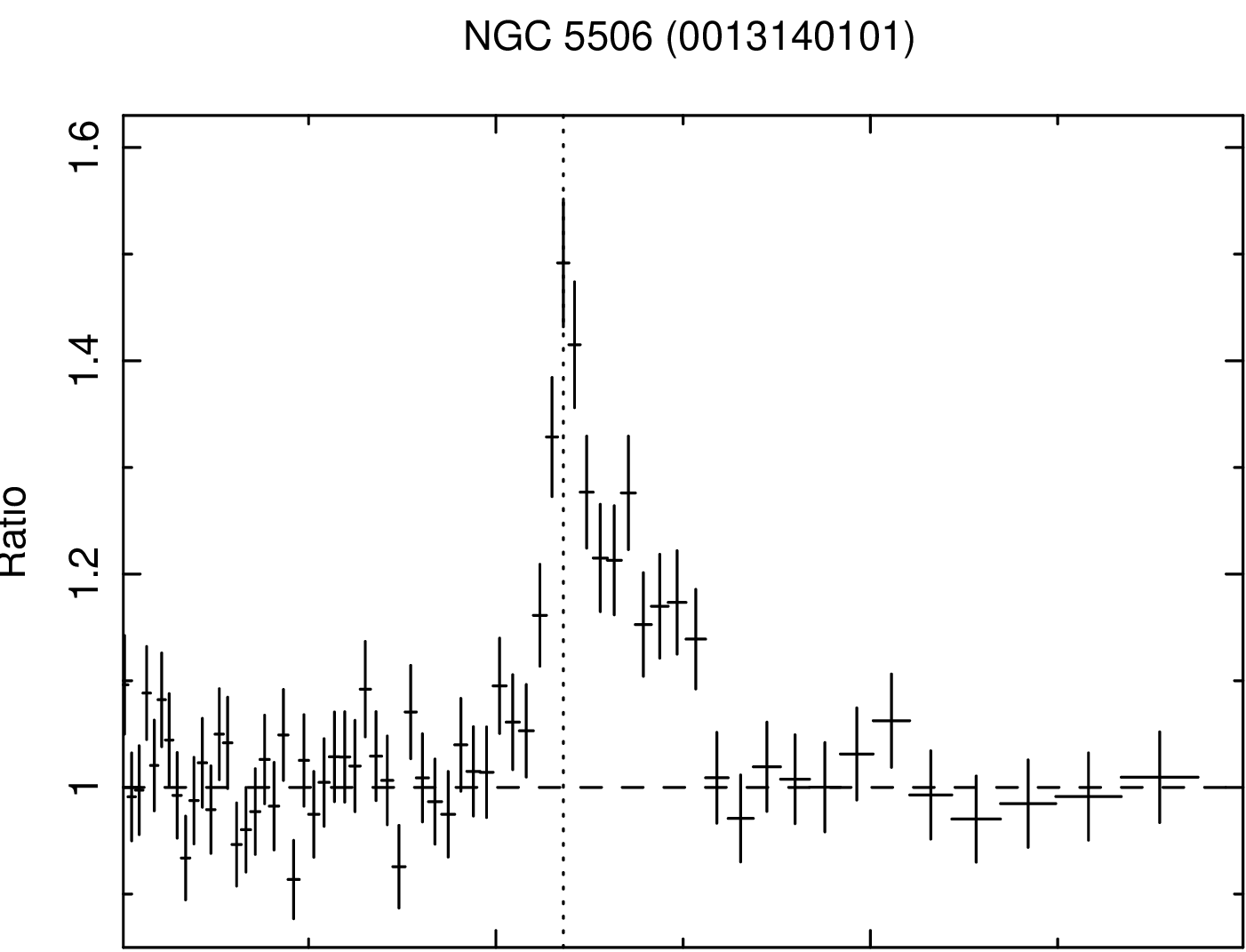}
\hspace{0.3cm}
    \includegraphics[width=3cm,height=4cm,angle=270]{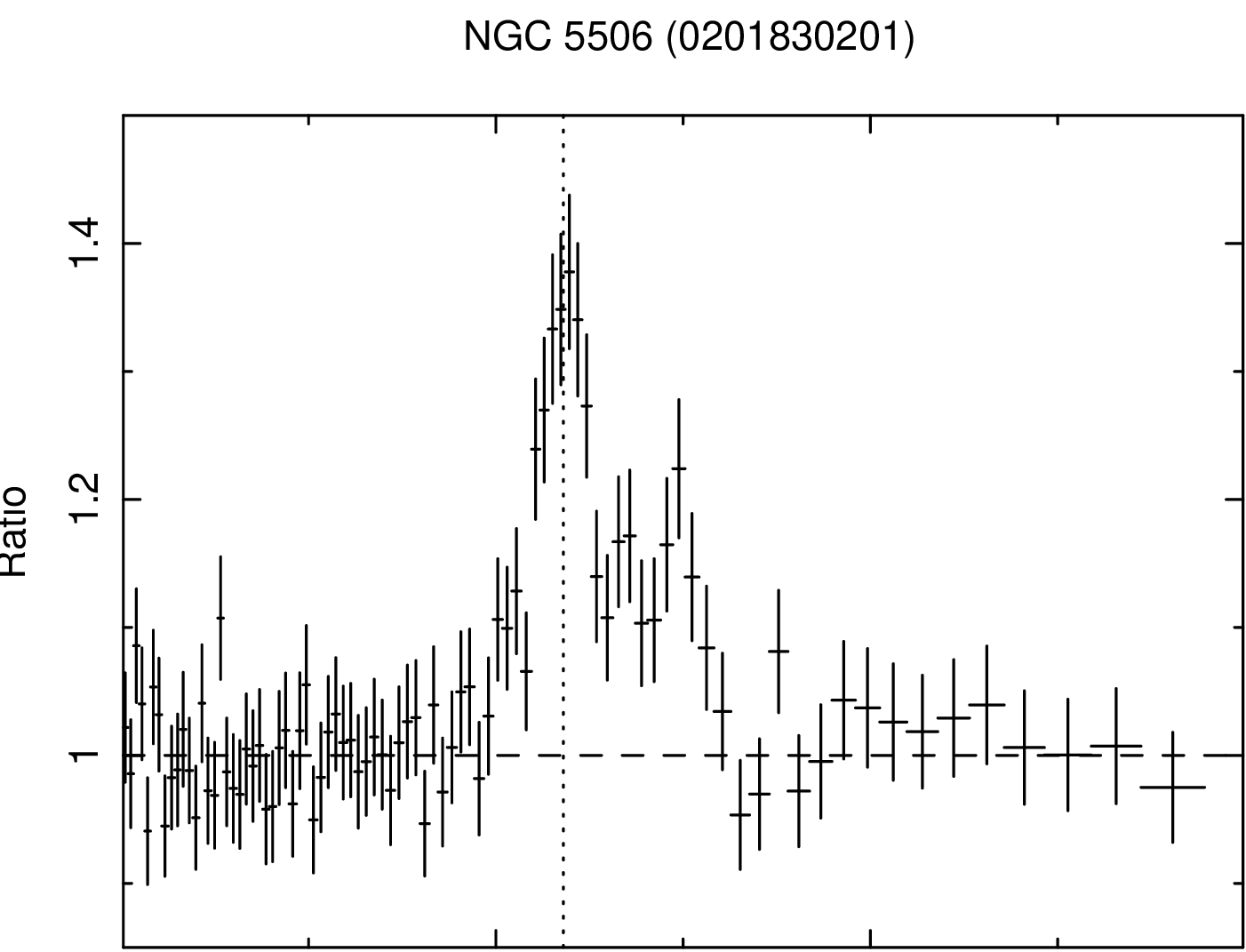}

\vspace{-0.07cm}

\hspace{-0.12cm}
    \includegraphics[width=3cm,height=4cm,angle=270]{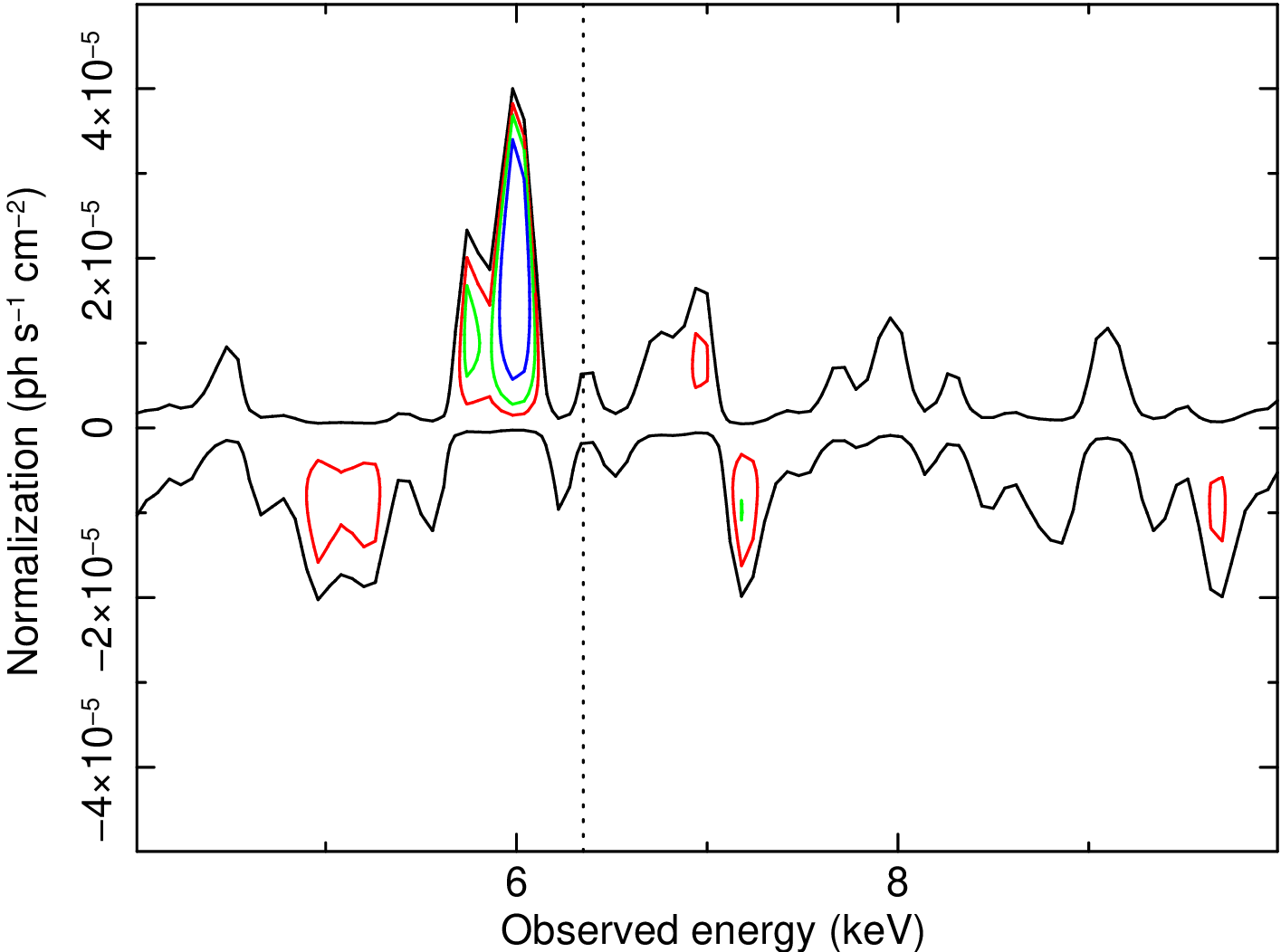}
\hspace{0.3cm}
    \includegraphics[width=3cm,height=4cm,angle=270]{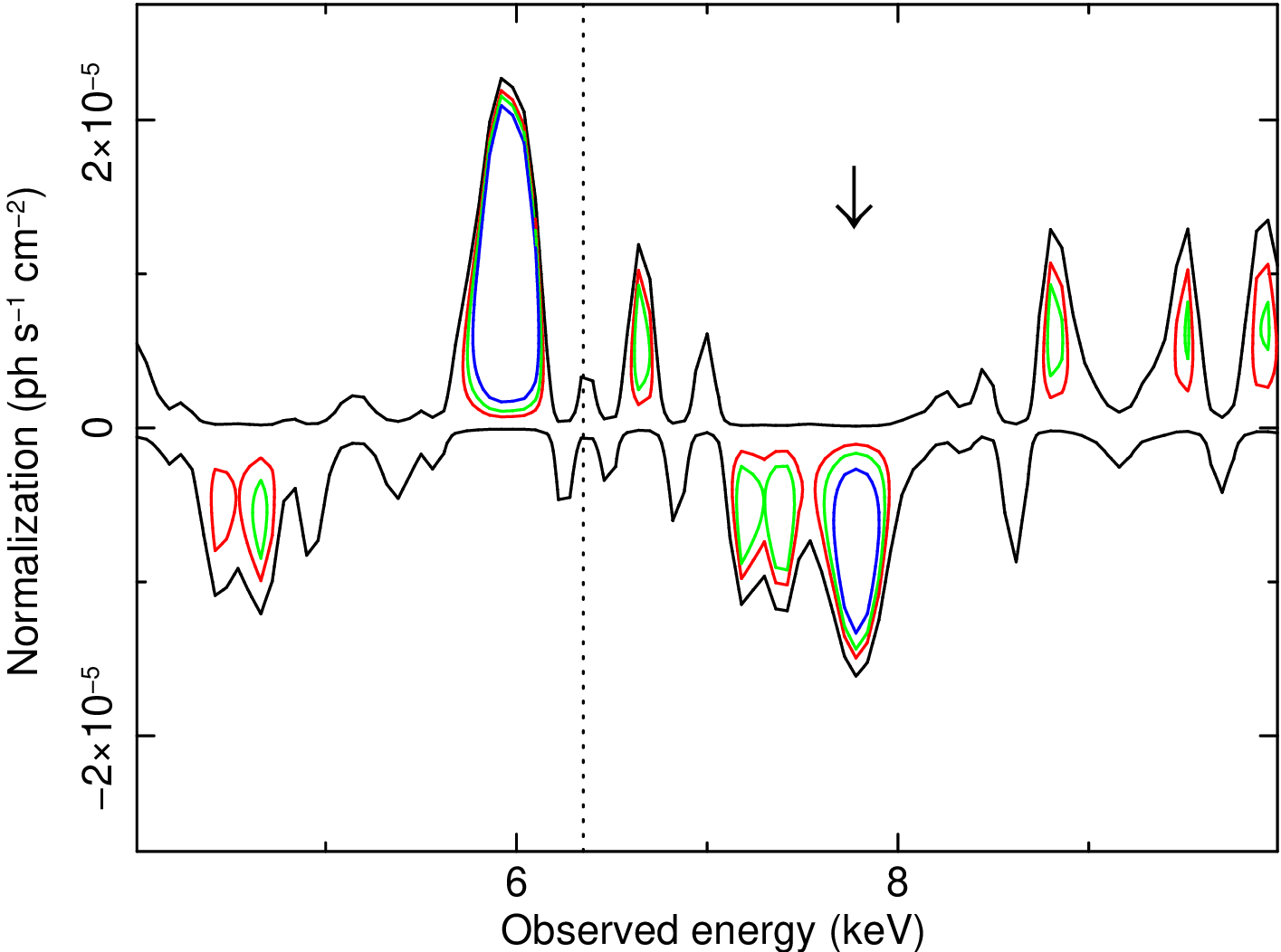}
\hspace{0.3cm}
    \includegraphics[width=3cm,height=4cm,angle=270]{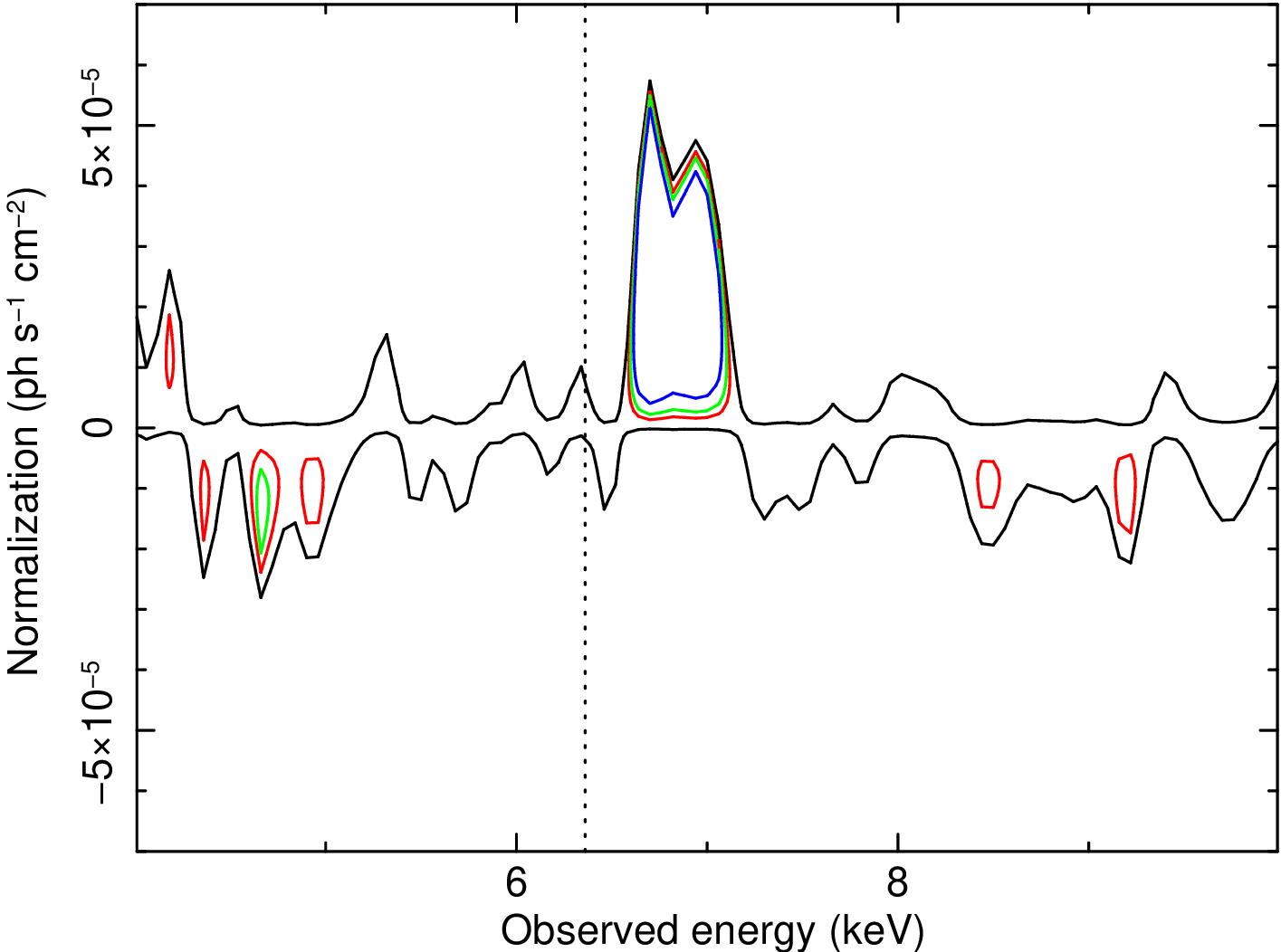}
\hspace{0.3cm}
    \includegraphics[width=3cm,height=4cm,angle=270]{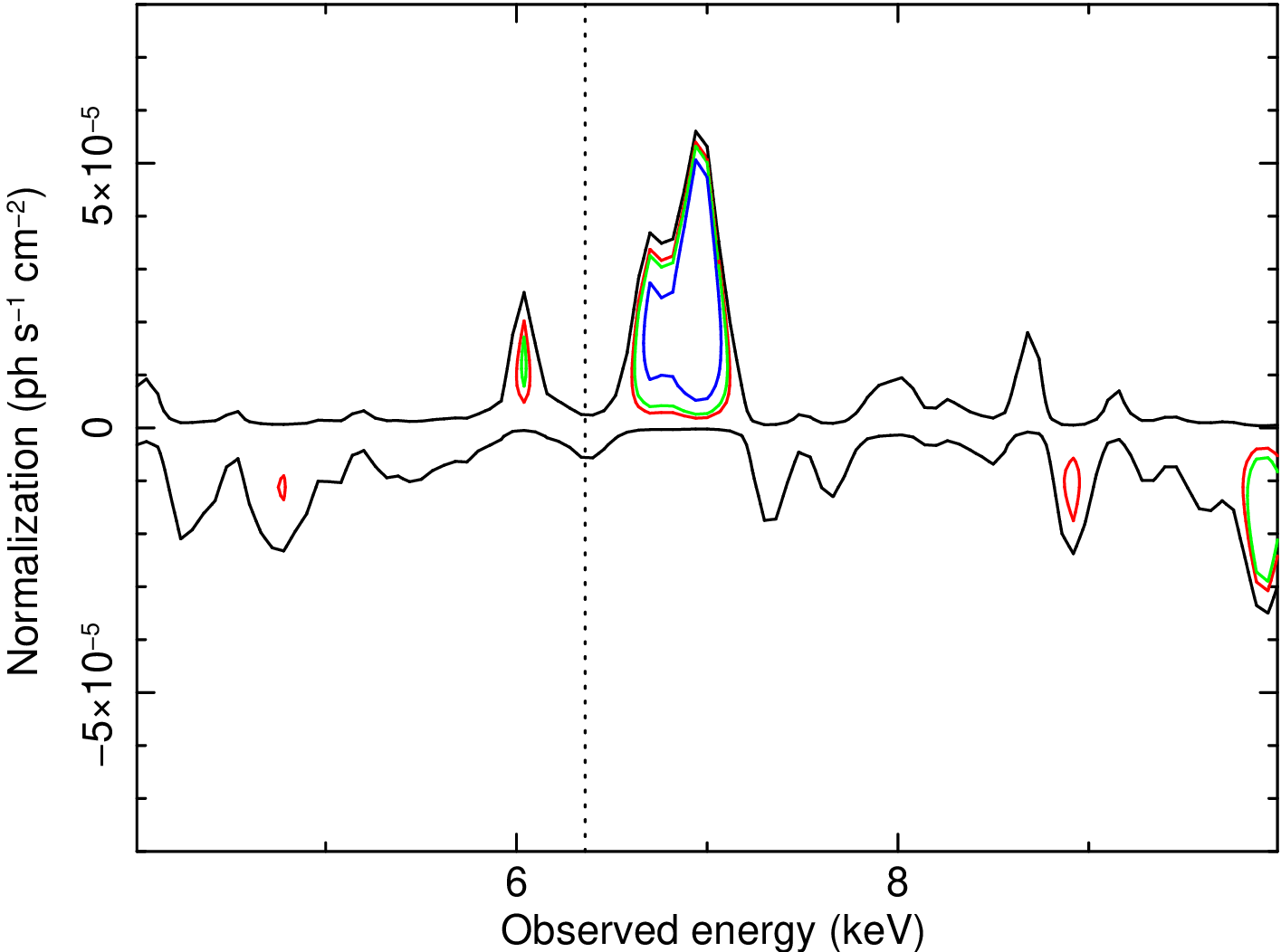}


\vspace{0.2cm}

    \includegraphics[width=3cm,height=4cm,angle=270]{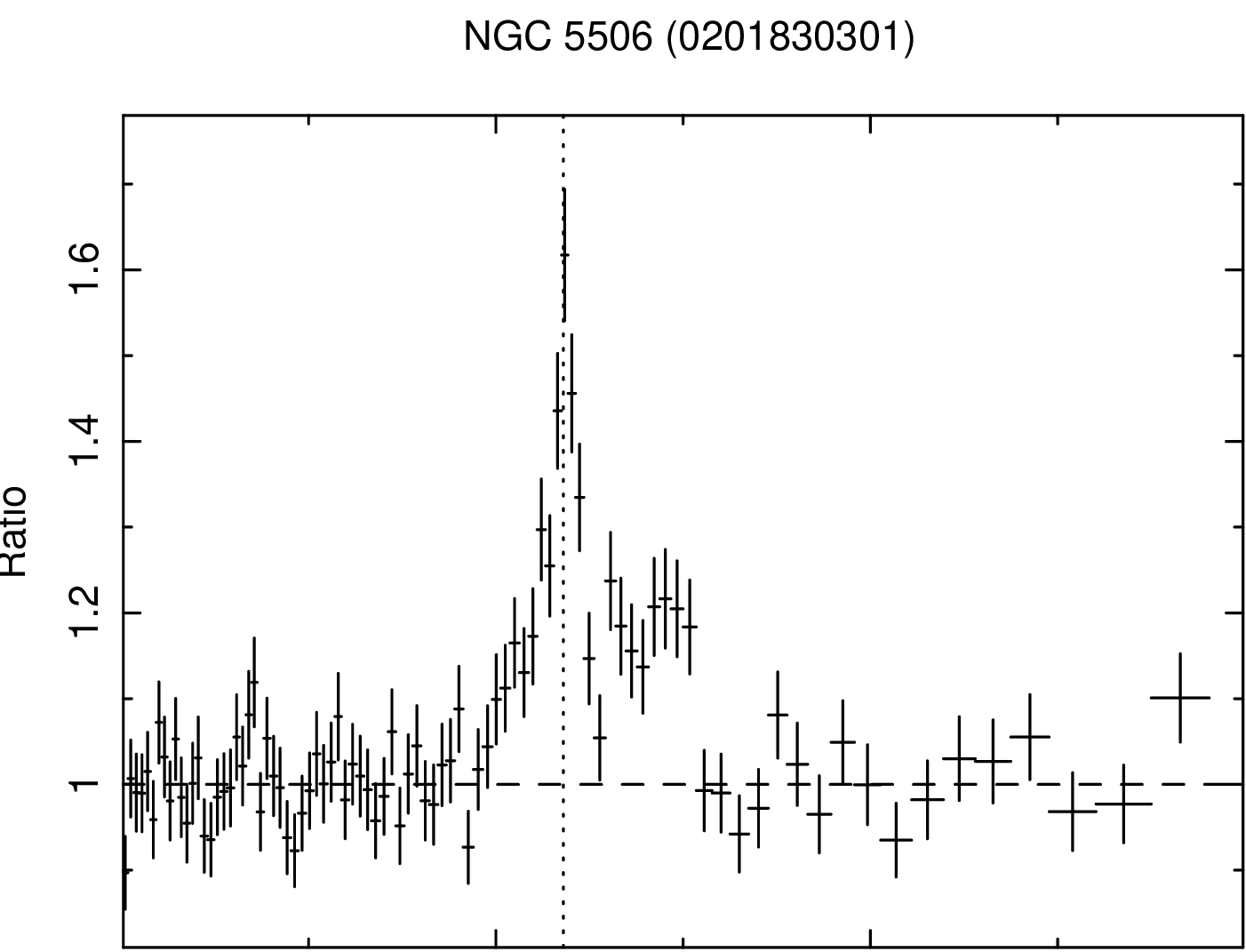}
\hspace{0.3cm}
    \includegraphics[width=3cm,height=4cm,angle=270]{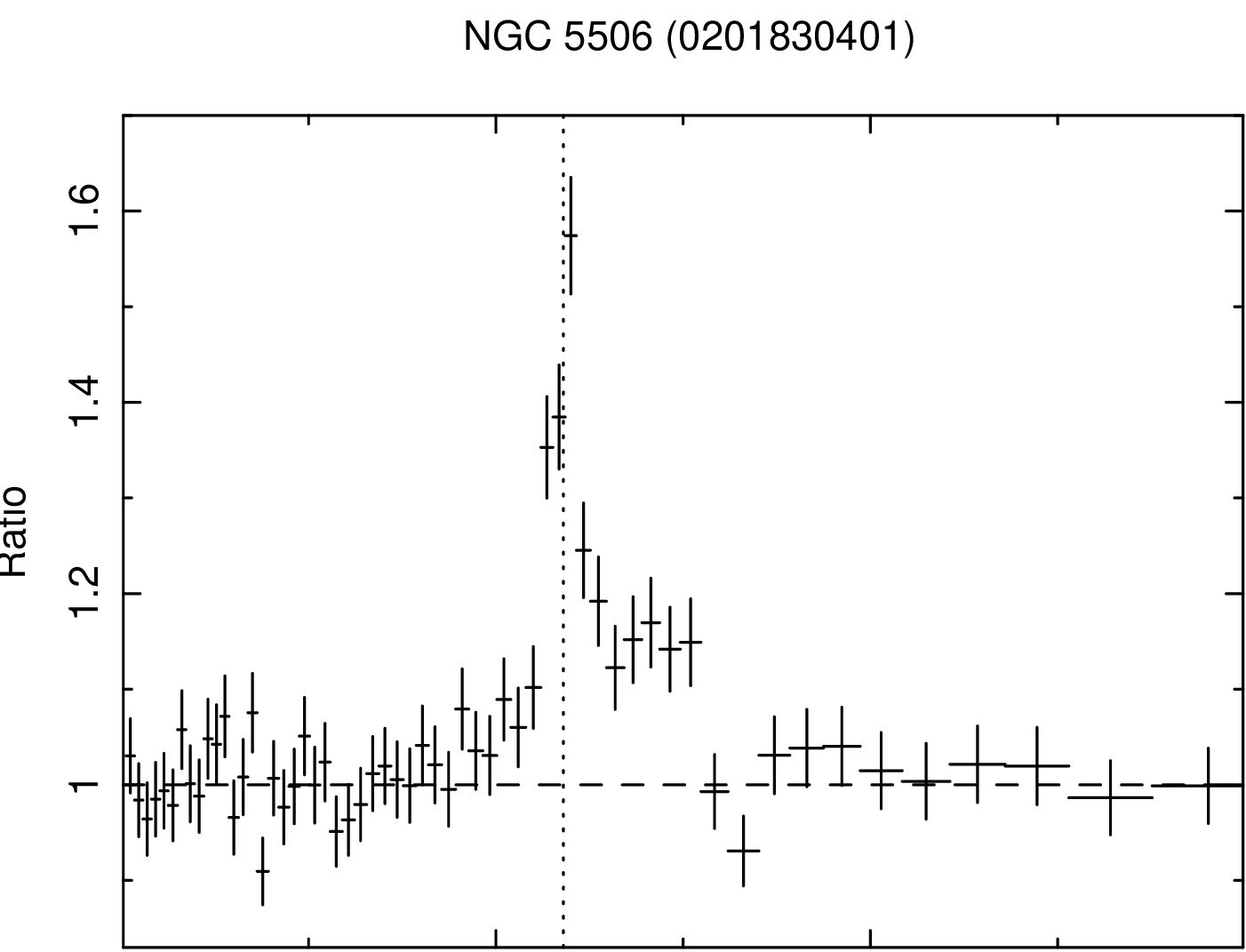}
\hspace{0.3cm}
    \includegraphics[width=3cm,height=4cm,angle=270]{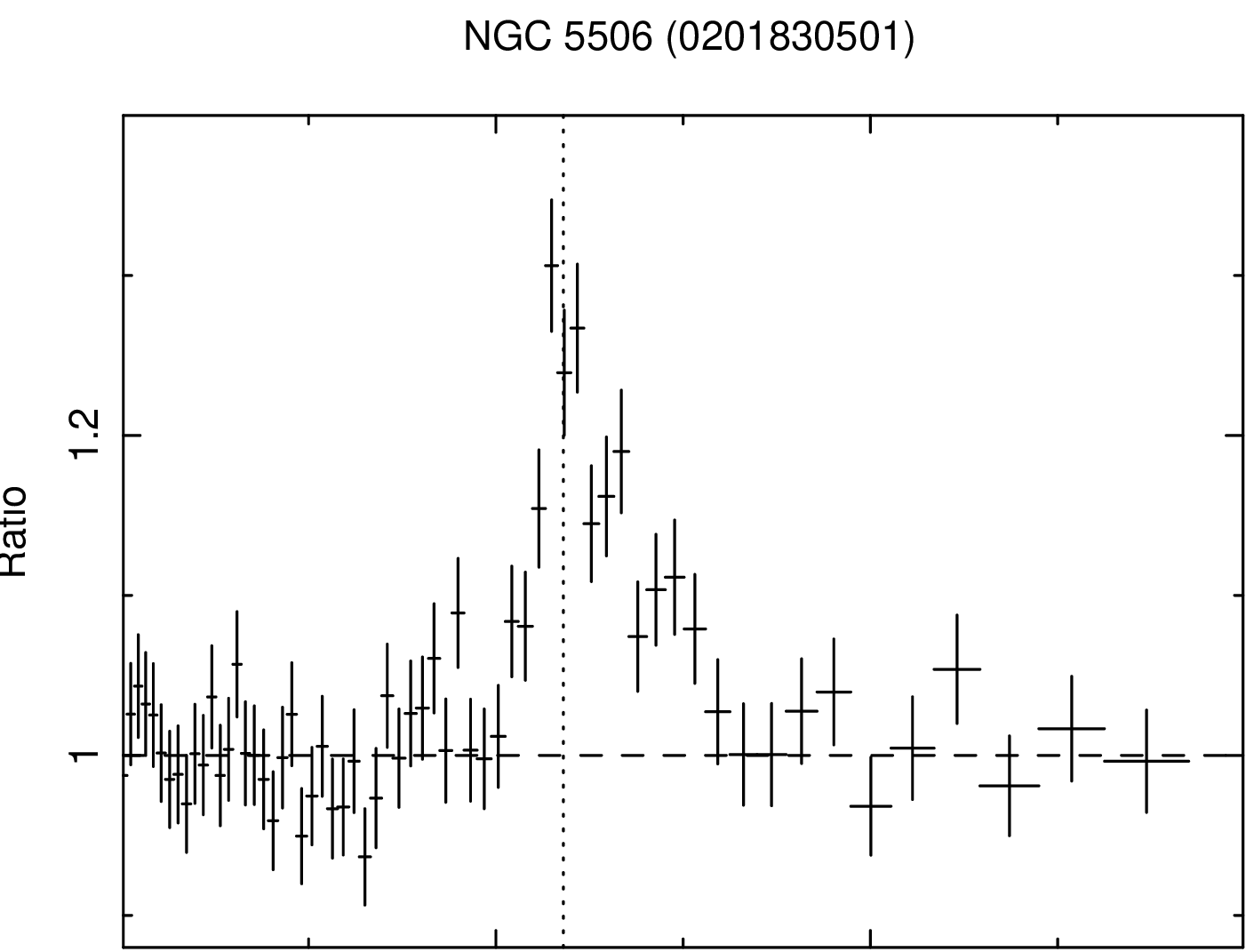}
\hspace{0.3cm}
    \includegraphics[width=3cm,height=4cm,angle=270]{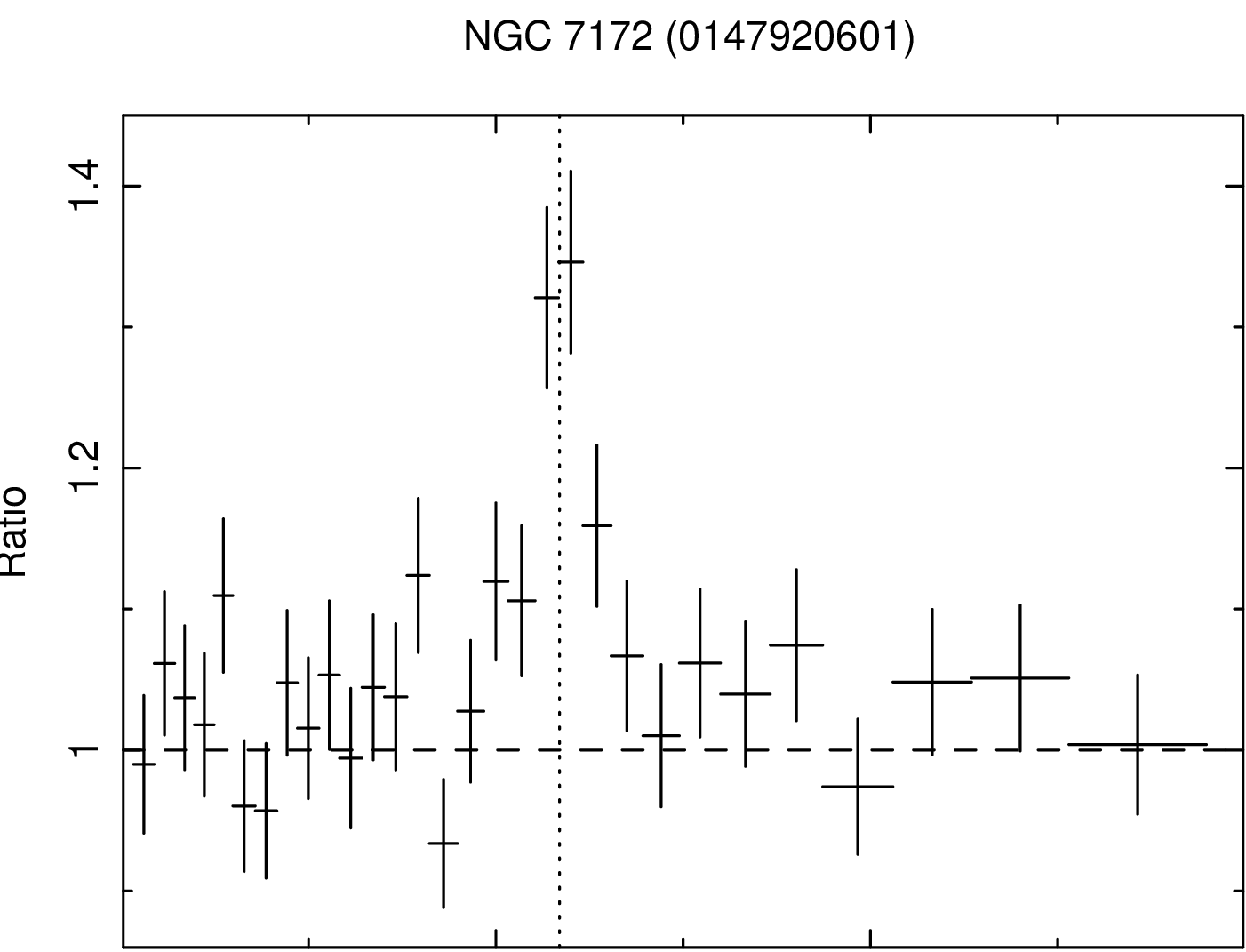}

\vspace{-0.07cm}

\hspace{-0.12cm}
    \includegraphics[width=3cm,height=4cm,angle=270]{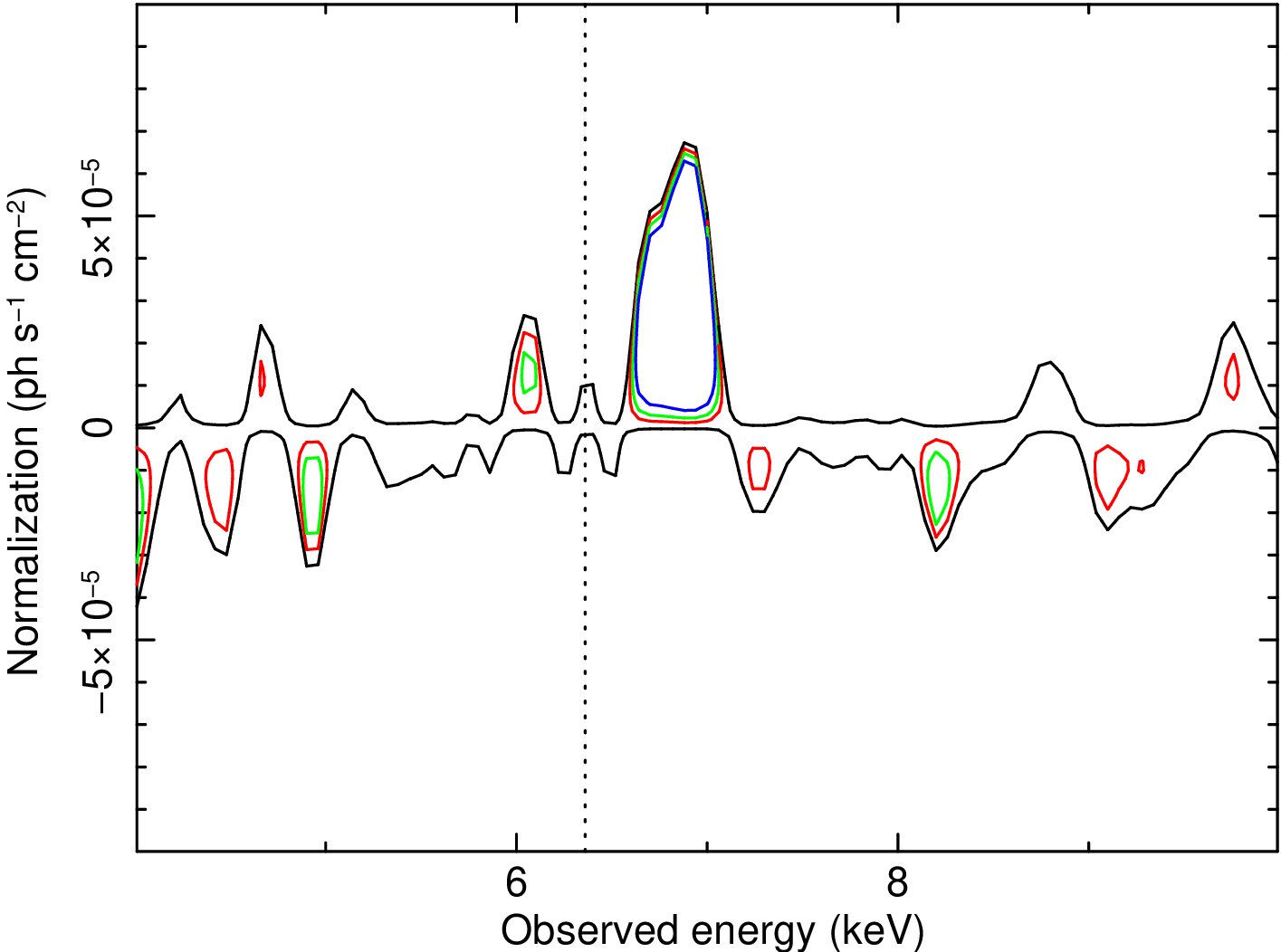}
\hspace{0.3cm}
    \includegraphics[width=3cm,height=4cm,angle=270]{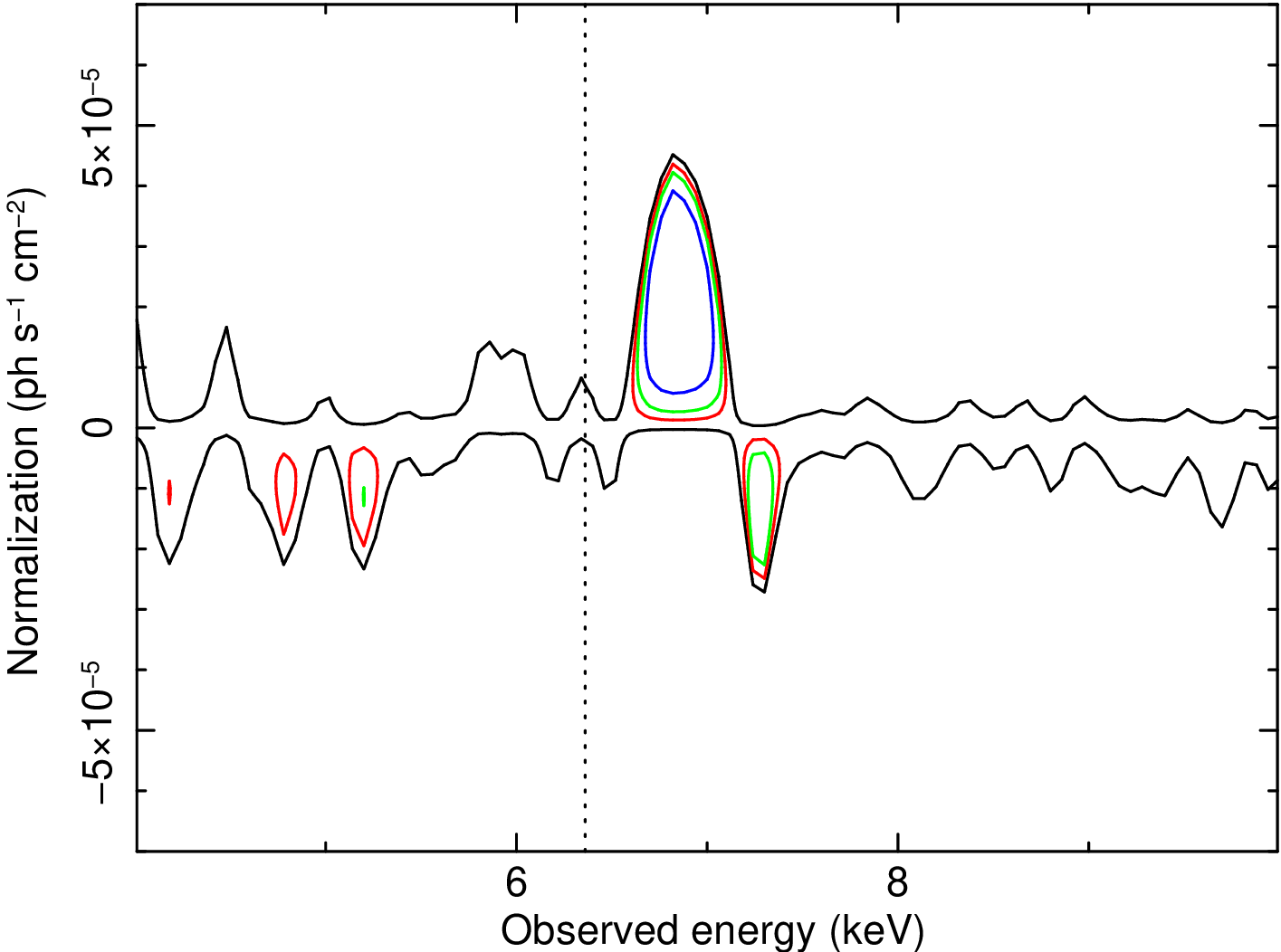}
\hspace{0.3cm}
    \includegraphics[width=3cm,height=4cm,angle=270]{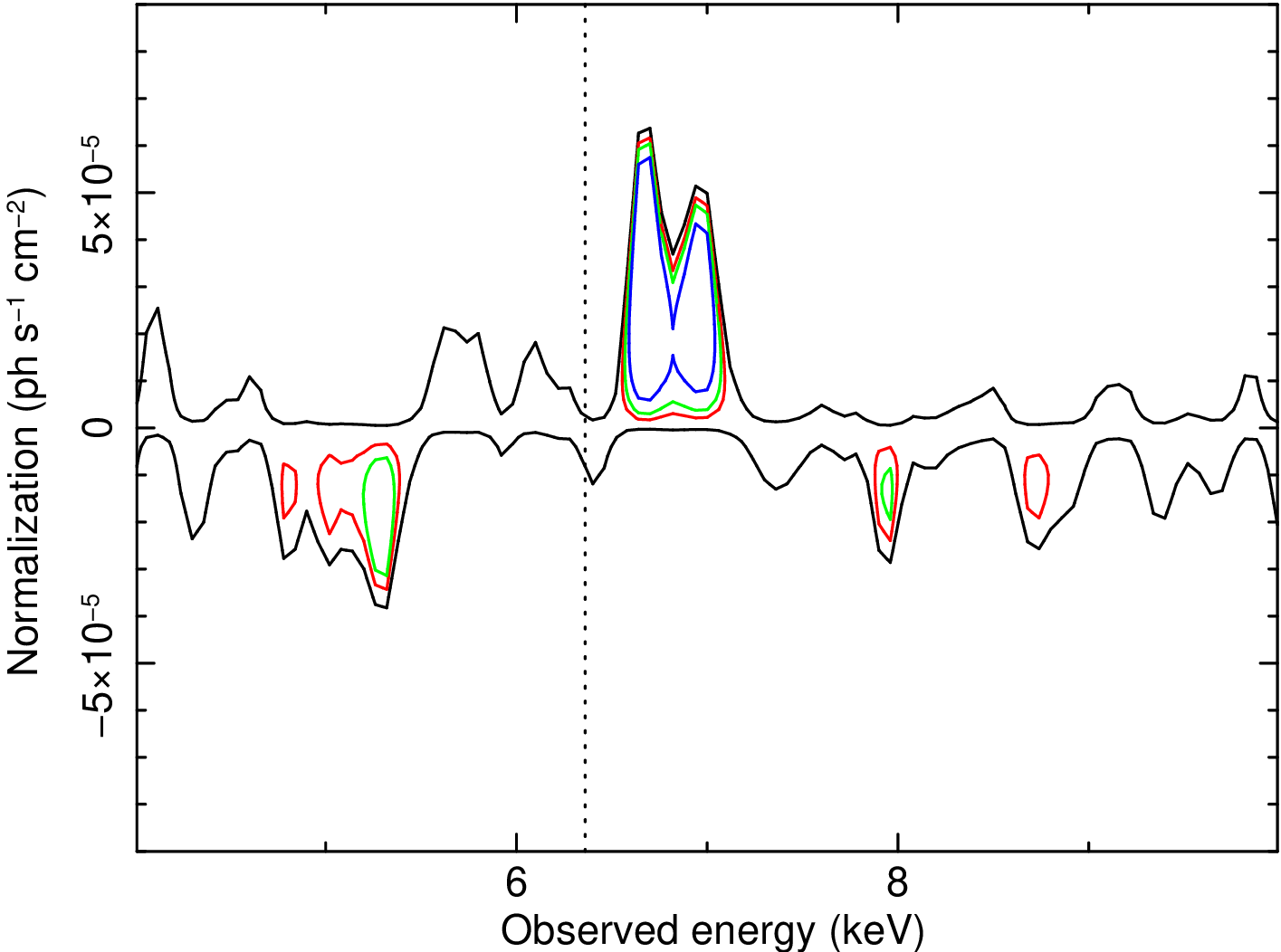}
\hspace{0.3cm}
    \includegraphics[width=3cm,height=4cm,angle=270]{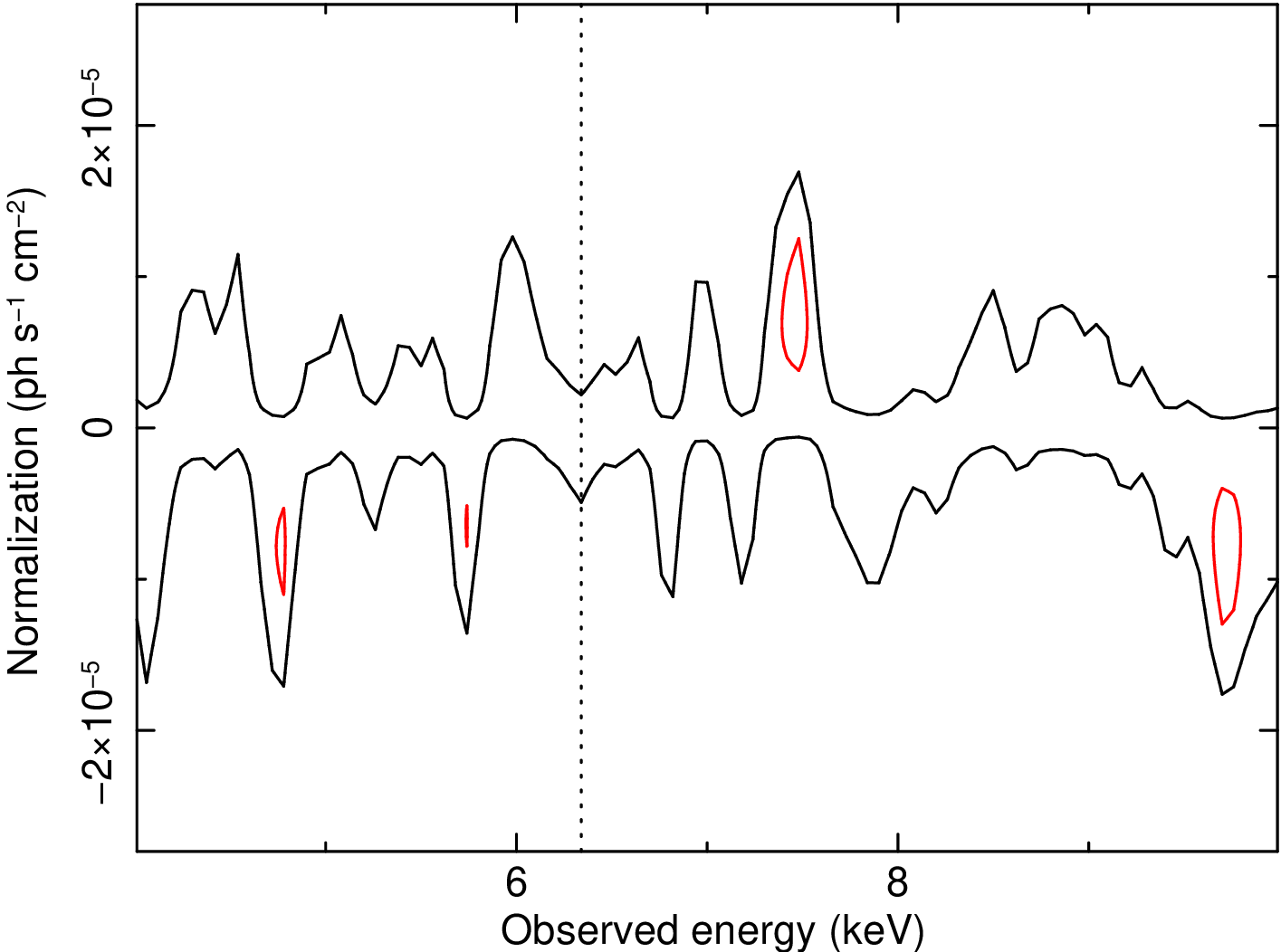}


\vspace{0.2cm}

    \includegraphics[width=3cm,height=4cm,angle=270]{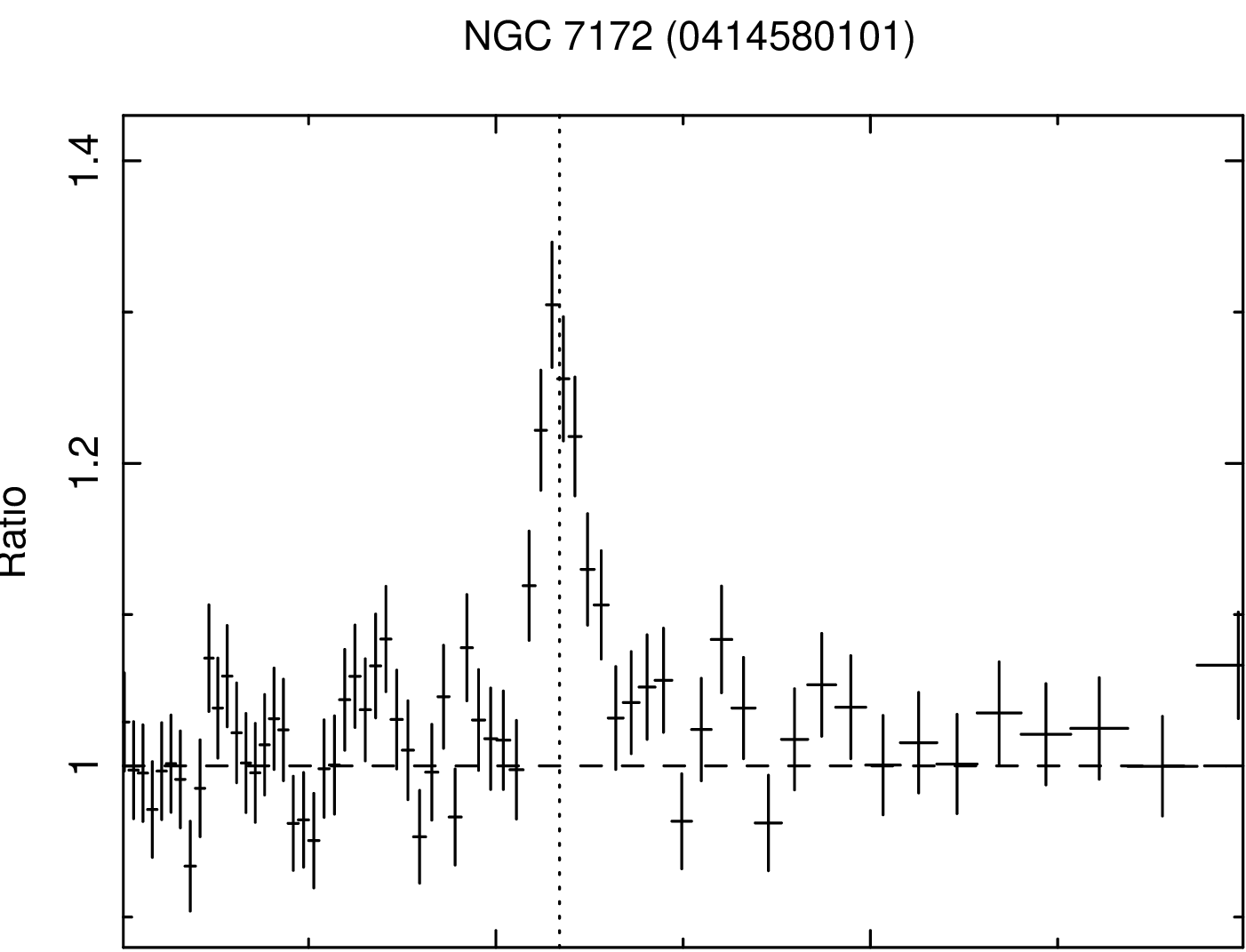}
\hspace{0.3cm}
    \includegraphics[width=3cm,height=4cm,angle=270]{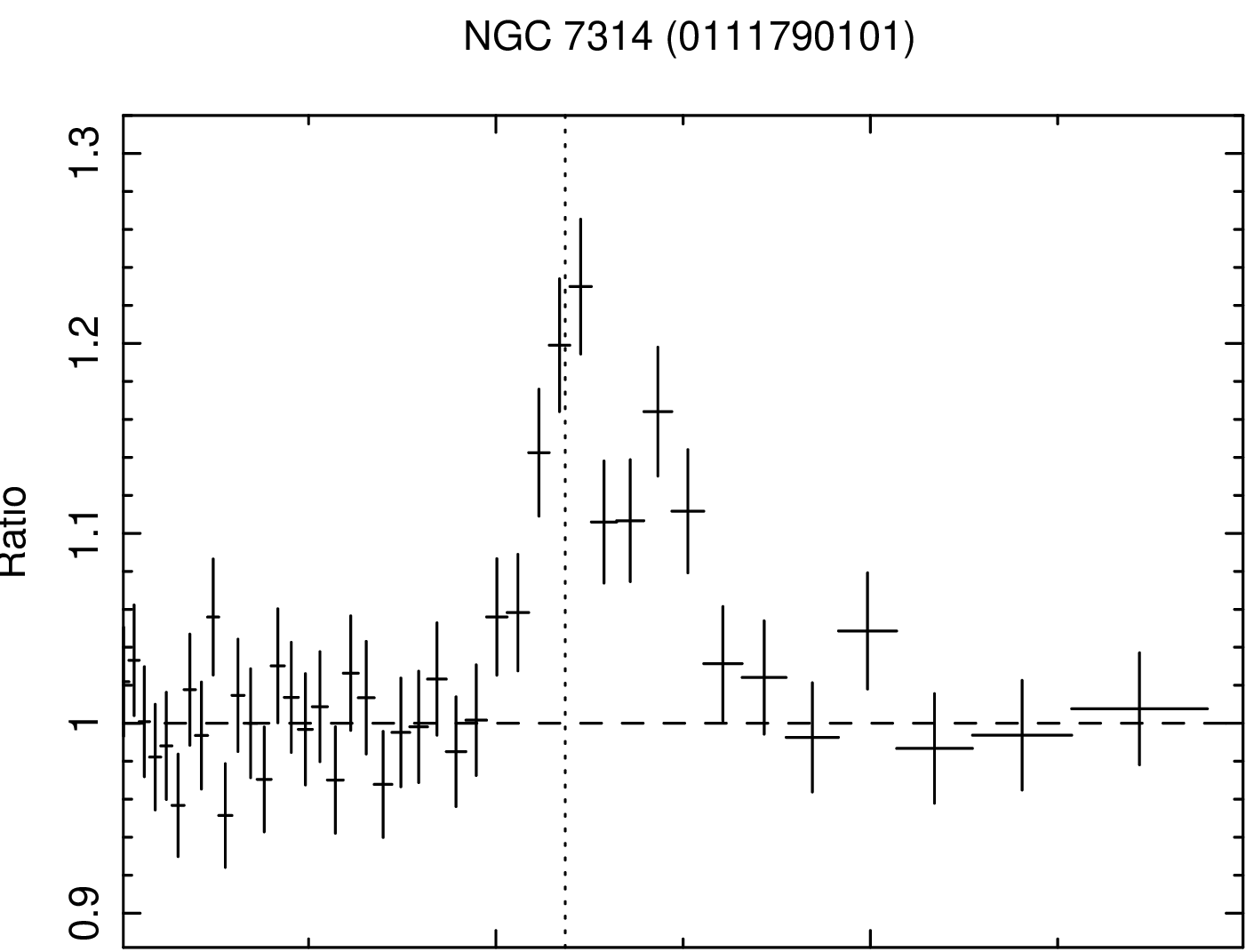}
\hspace{0.3cm}
    \includegraphics[width=3cm,height=4cm,angle=270]{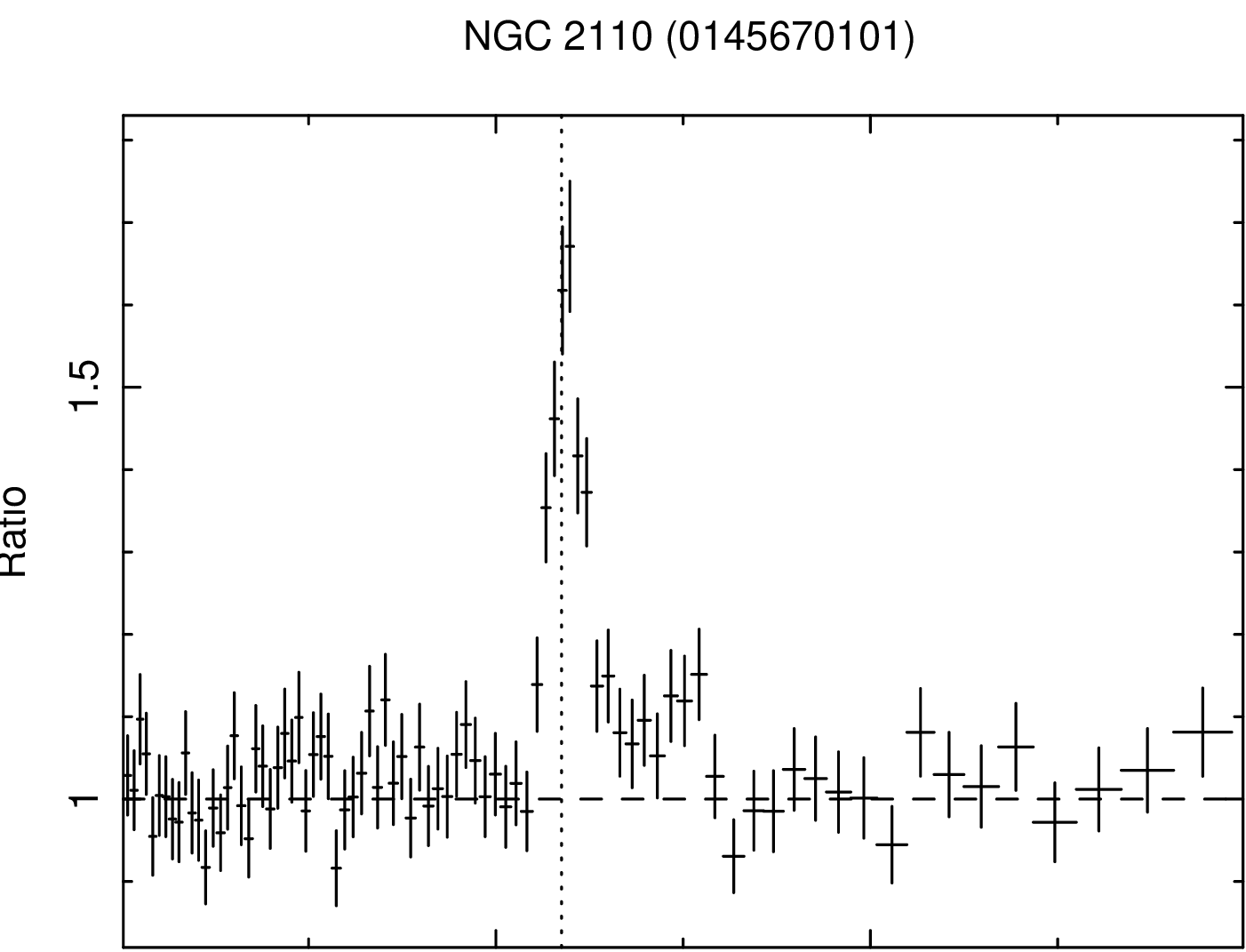}
\hspace{0.3cm}
    \includegraphics[width=3cm,height=4cm,angle=270]{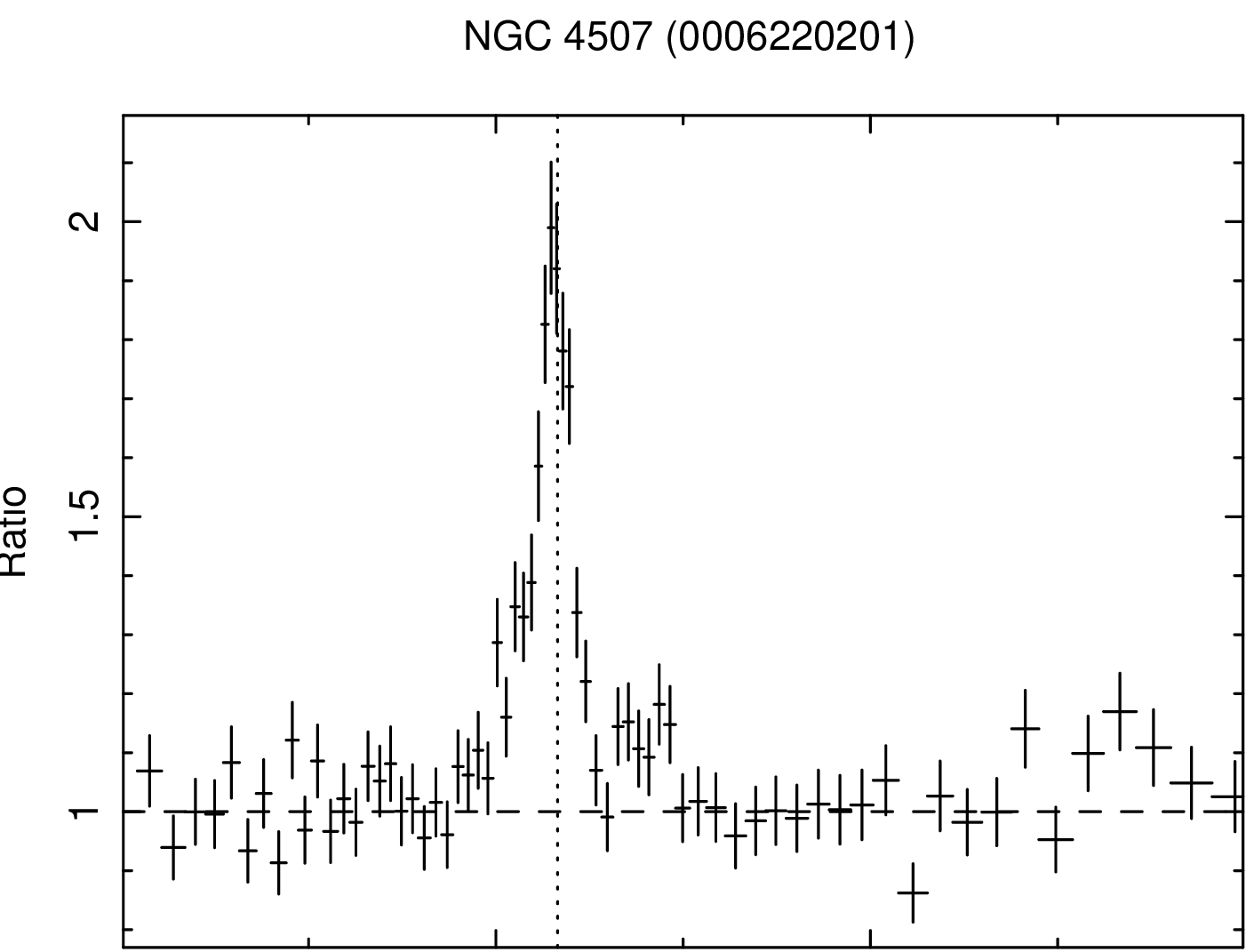}

\vspace{-0.07cm}

\hspace{-0.12cm}
    \includegraphics[width=3cm,height=4cm,angle=270]{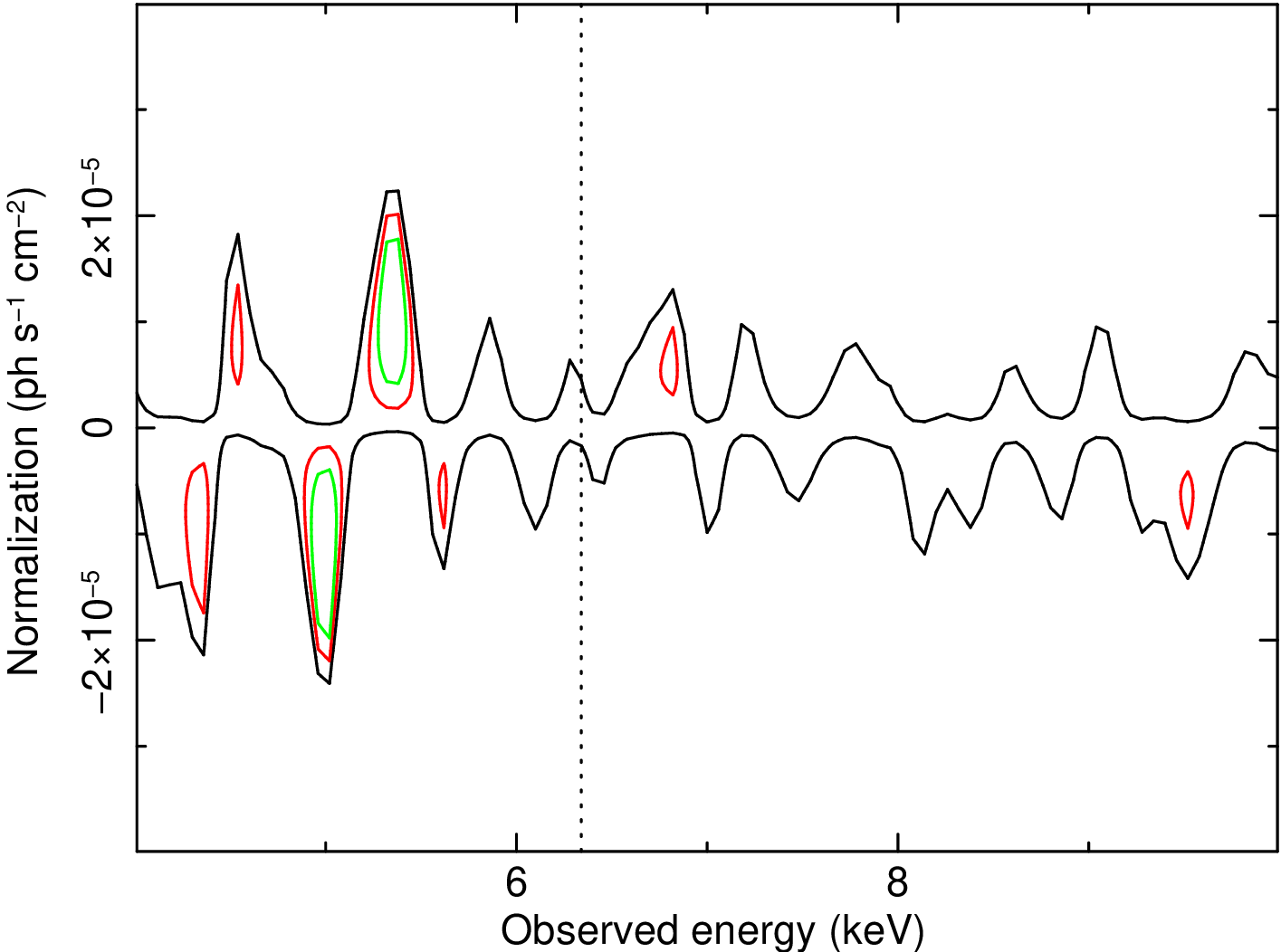}
\hspace{0.3cm}
    \includegraphics[width=3cm,height=4cm,angle=270]{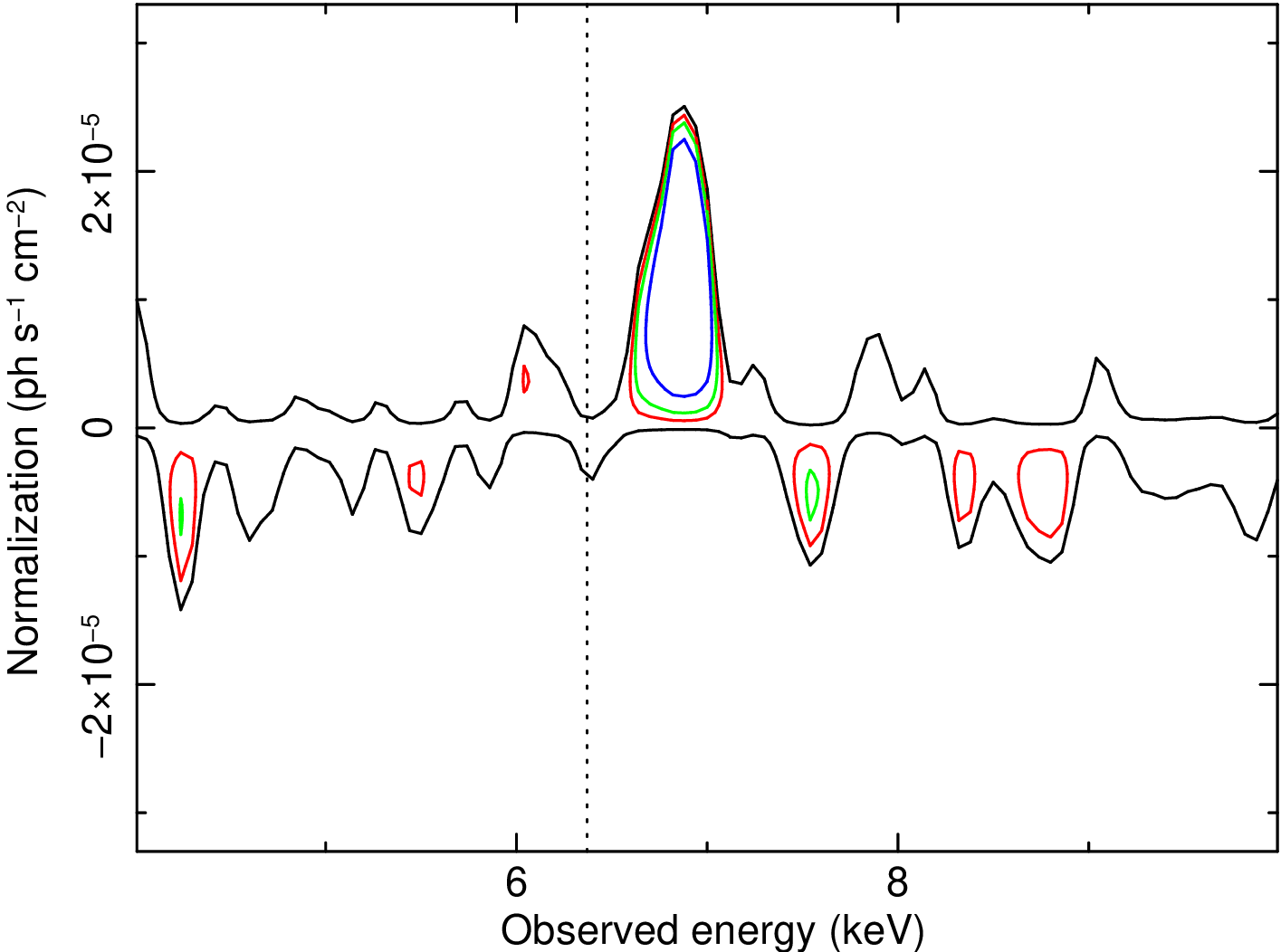}
\hspace{0.3cm}
    \includegraphics[width=3cm,height=4cm,angle=270]{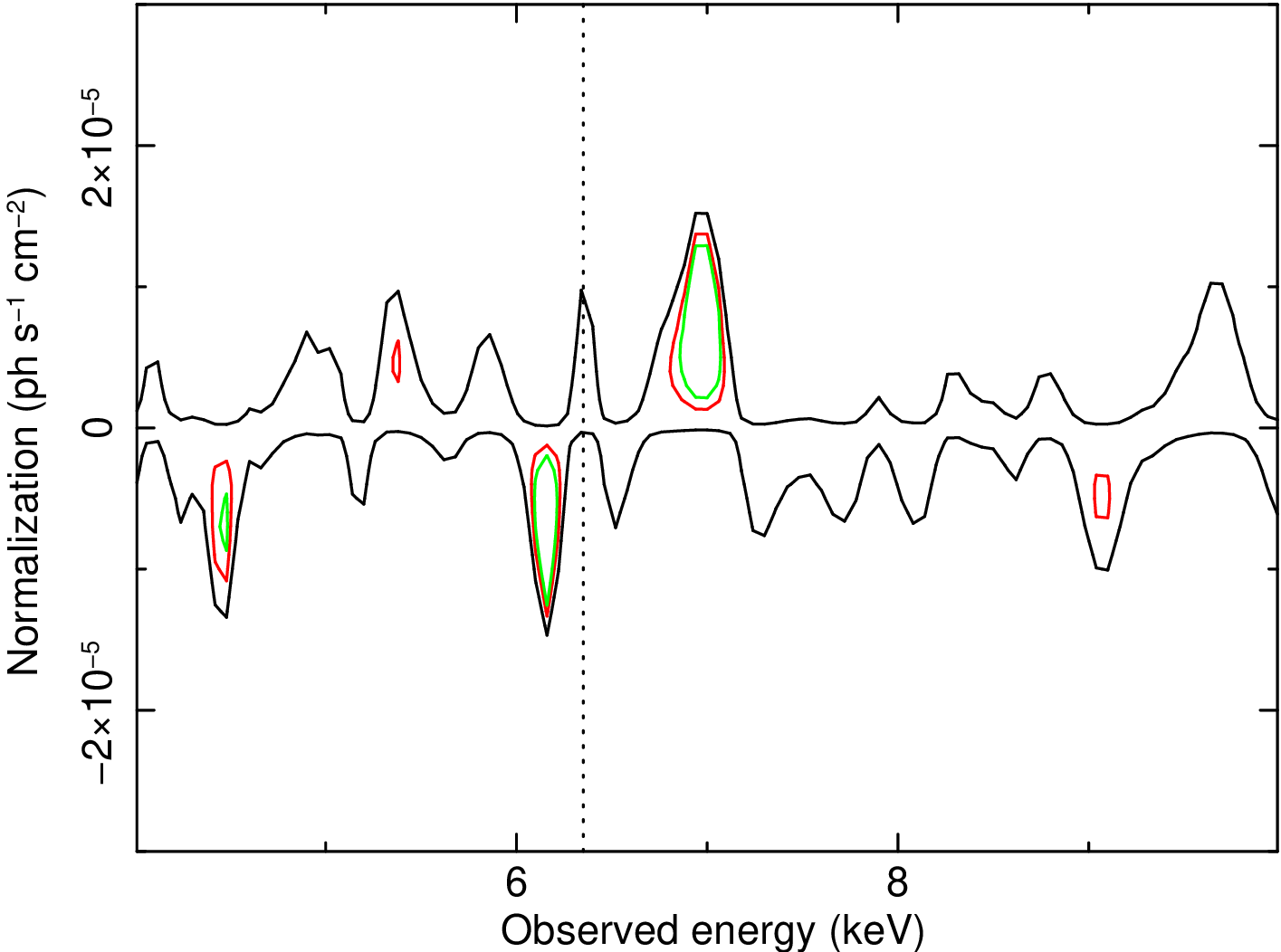}
\hspace{0.3cm}
    \includegraphics[width=3cm,height=4cm,angle=270]{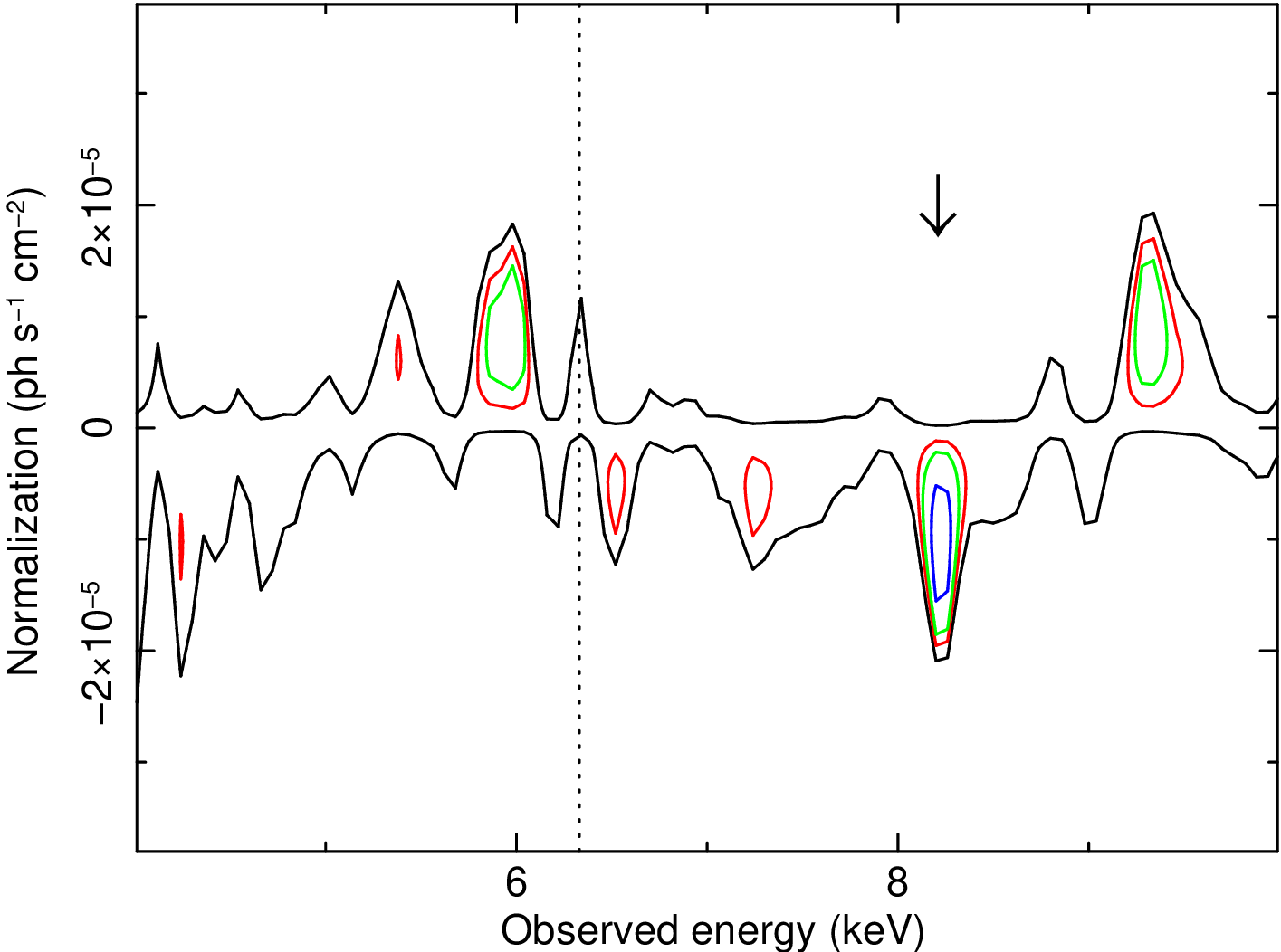}


\vspace{0.2cm}
    \includegraphics[width=3cm,height=4cm,angle=270]{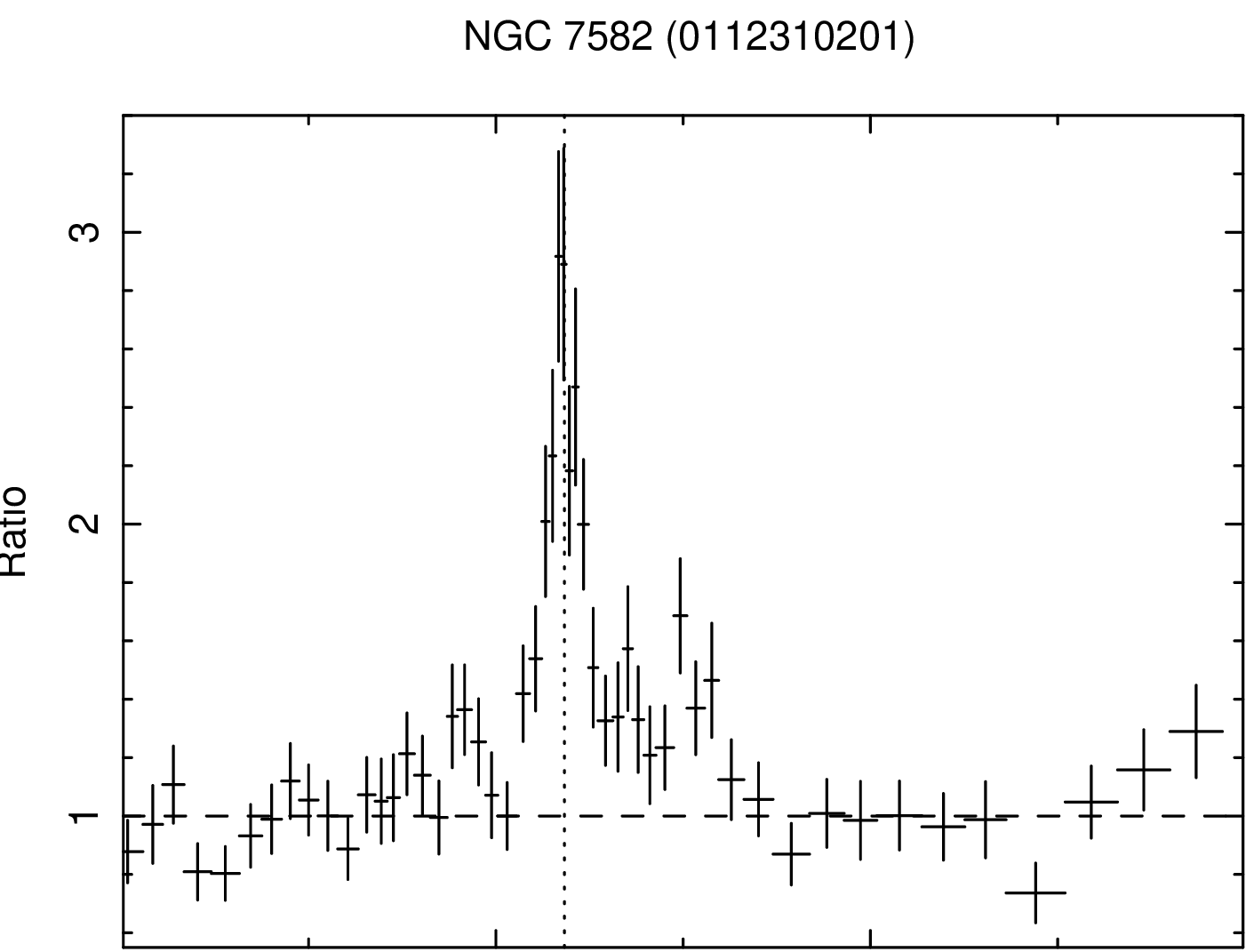}
\hspace{0.3cm}
    \includegraphics[width=3cm,height=4cm,angle=270]{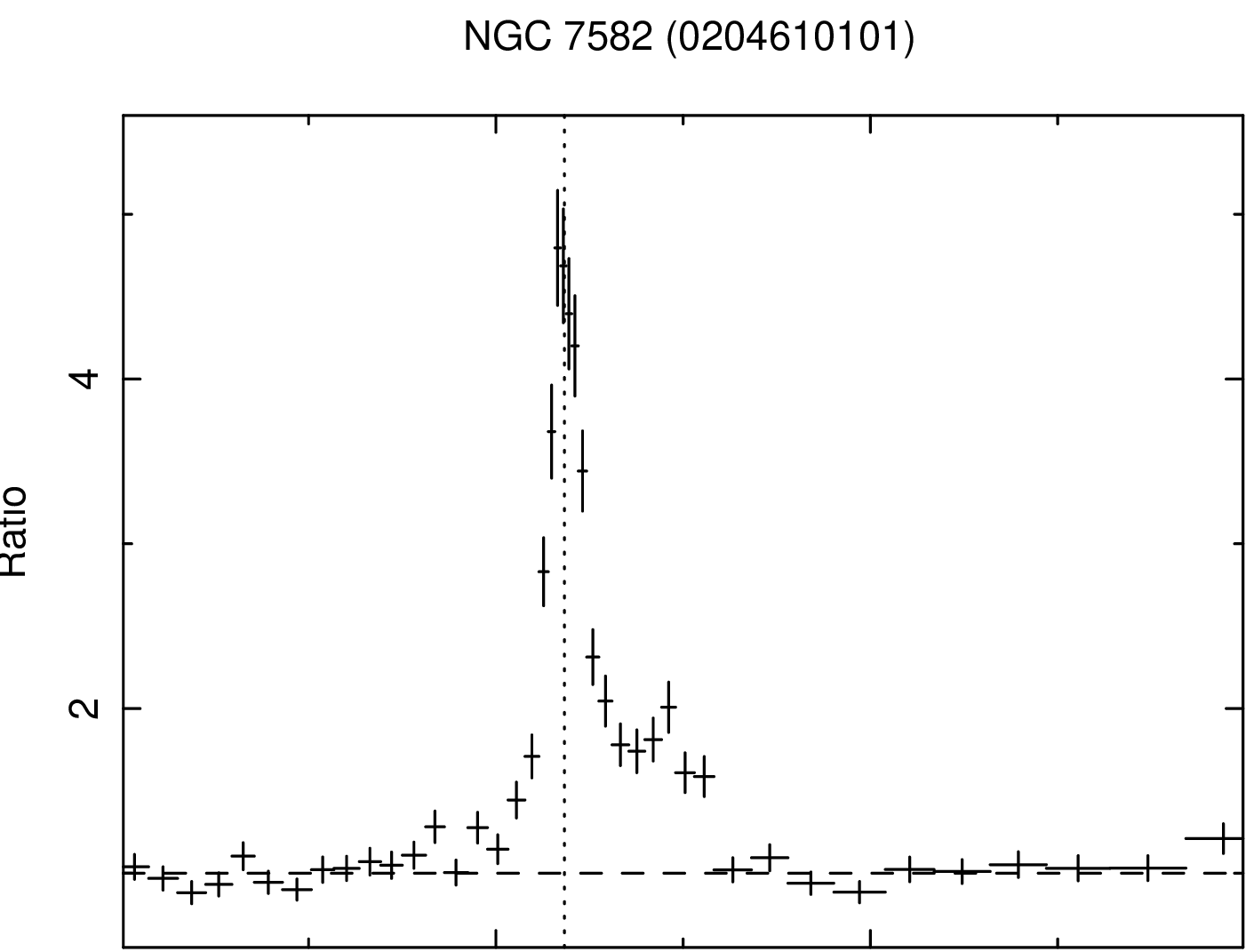}

\vspace{-0.07cm}

\hspace{-0.12cm}
    \includegraphics[width=3cm,height=4cm,angle=270]{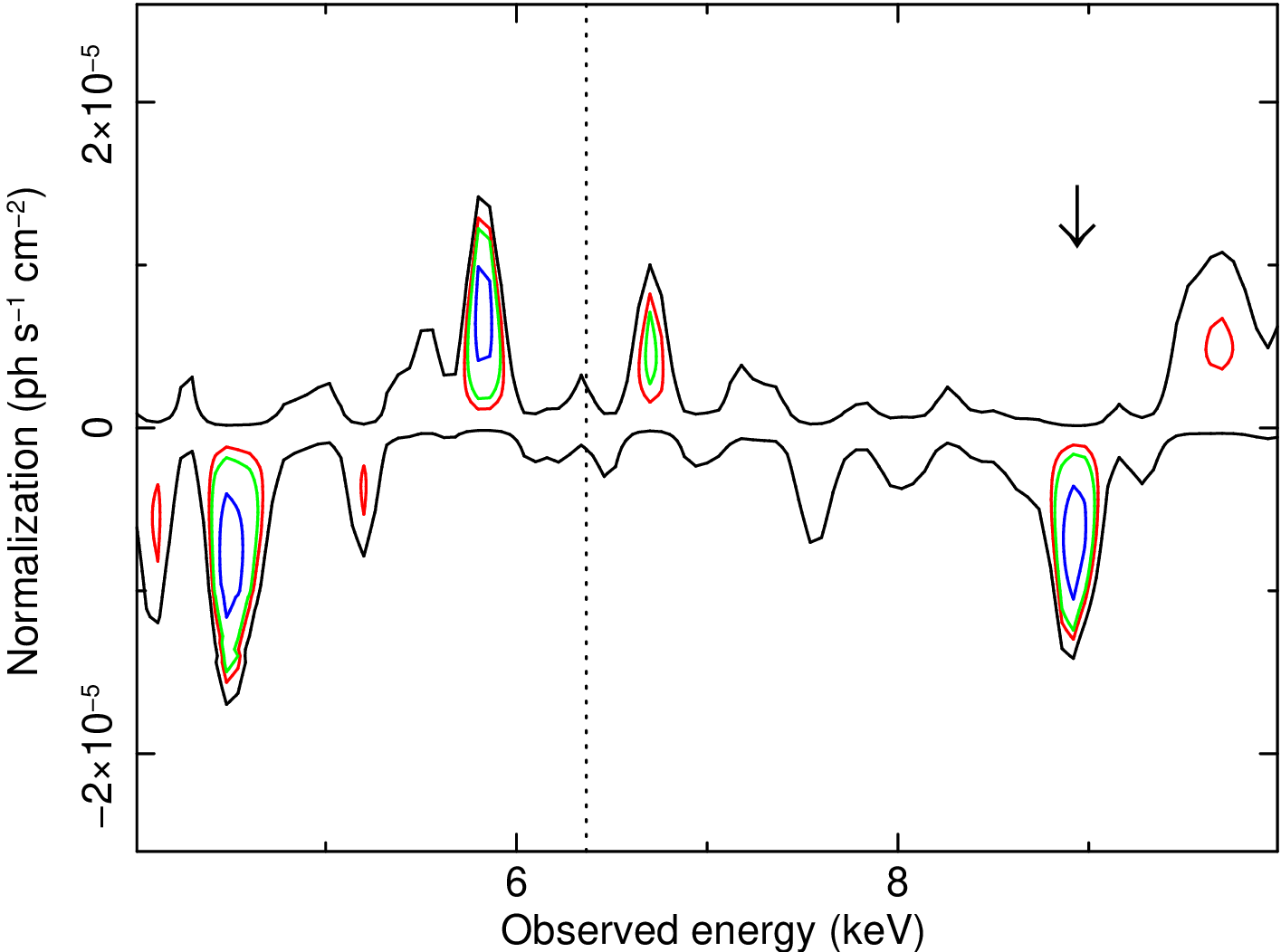}
\hspace{0.3cm}
    \includegraphics[width=3cm,height=4cm,angle=270]{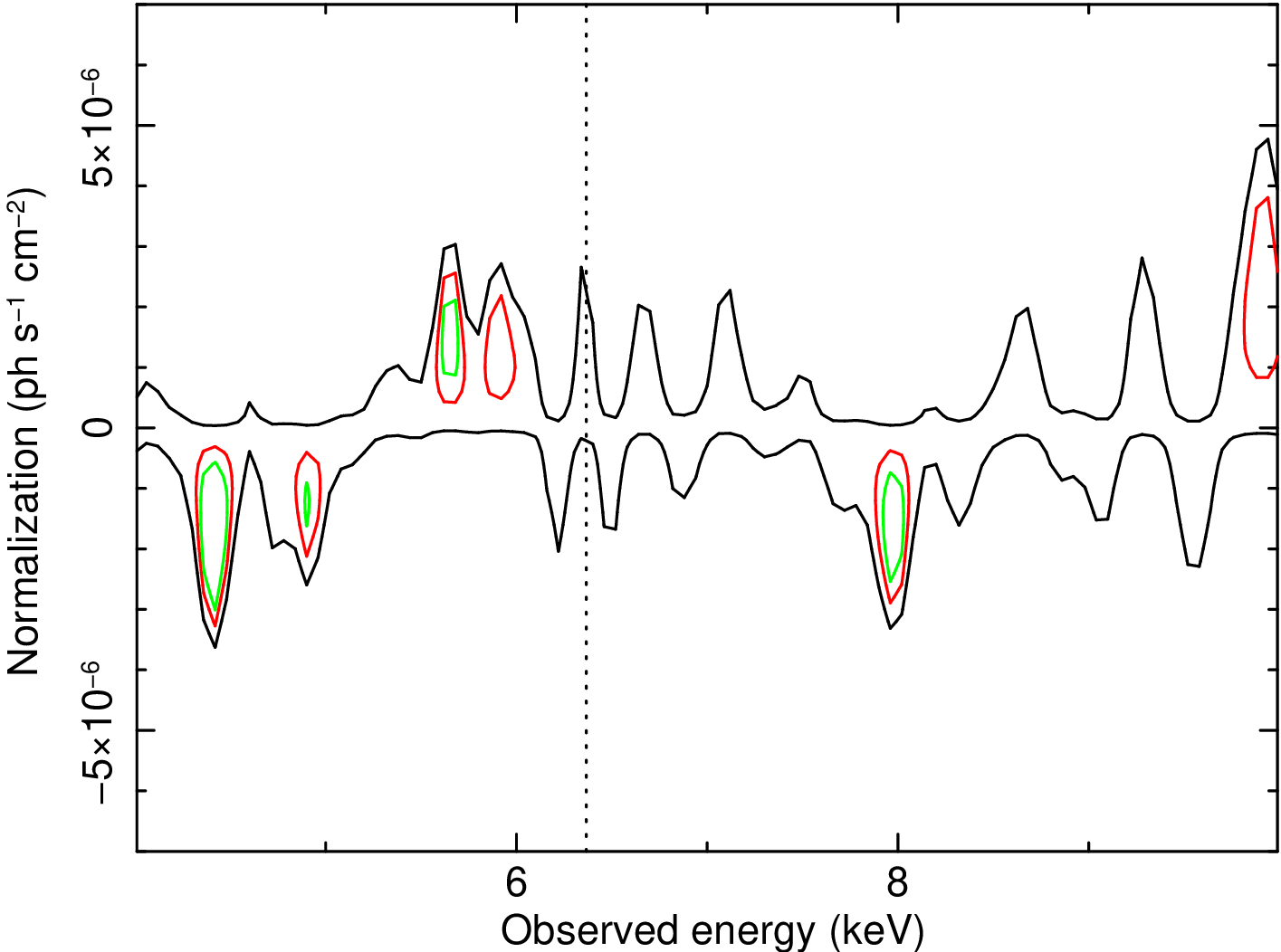}

   \caption{Ratio against the continuum (\emph{upper panel}) and contour plots with respect to the best-fit baseline model (\emph{lower panel}; 68\% (red), 90\% (green), 99\% (blue) levels) for the Type 1 sources. The Fe K absorption lines are pointed by arrows. The vertical line refer to Fe I K$\alpha$.}
    \end{figure*}

\end{appendix}

}

\end{document}